\documentclass[a4paper, oneside,reqno]{amsart}
\usepackage{graphics}
\usepackage{amssymb}
\usepackage{amsmath}
\usepackage{amscd}
\usepackage{geometry}
 \allowdisplaybreaks
\newcommand{\ZZ}{\mathbb{Z}}
\newcommand{\RR}{\mathbb{R}}
\newcommand{\FF}{\mathcal{F}}
\newcommand{\MM}{\mathcal{M}}
\newcommand{\NN}{\mathcal{N}}
\newcommand{\DD}{\mathcal{D}}
\newcommand{\OO}{\mathcal{O}}
\newcommand{\KK}{\mathcal{K}}
\newcommand{\GG}{\mathcal{G}}
\newcommand{\CC}{\mathcal{C}}
\newcommand{\PP}{\mathcal{P}}
\newcommand{\TT}{\mathbb{T}}
\newcommand{\s}{\mathcal{S}}
\newcommand{\A}{\mathcal{A}}
\newcommand{\B}{\mathcal{B}}
\newcommand{\Q}{\mathcal{Q}}
\newcommand{\e}{\epsilon}
\newcommand{\Fun}{\mathrm{Fun}}
\newcommand{\gh}{\mathrm{gh}}
\newcommand{\even}{\mathrm{even}}
\newcommand{\odd}{\mathrm{odd}}
\newcommand{\bt}{\bullet}
\newcommand{\ra}{\rightarrow}

\newcommand{\dd}{\partial}
\newcommand{\Str}{\mathrm{Str}}
\newcommand{\g}{\mathfrak{g}}
\newcommand{\m}{\mathfrak{m}}
\newcommand{\id}{\mathrm{id}}
\newcommand{\LL}{\mathcal{L}}
\newcommand{\ad}{\mathrm{ad}}
\newcommand{\tr}{\mathrm{tr}}
\newcommand{\Ber}{\mathrm{Ber}}
\newcommand{\Mes}{\mathrm{Mes}}
\newcommand{\Iter}{\mathrm{Iter}}
\newcommand{\Loop}{\mathrm{Loop}}
\newcommand{\Aut}{\mathrm{Aut}}
\newcommand{\End}{\mathrm{End}}
\newcommand{\Hom}{\mathrm{Hom}}
\newcommand{\perm}{\mathrm{perm}}
\newcommand{\sym}{\mathrm{sym}}
\newcommand{\im}{\mathrm{im}}
\newcommand{\ev}{\mathrm{ev}}
\newcommand{\Div}{\mathrm{div}\,}
\newcommand{\Span}{\mathrm{Span}}
\newcommand{\wh}{\widehat}
\newcommand{\mr}{\mathrm}
\newcommand{\mc}{\mathcal}
\newcommand{\be}{\begin{equation}}
\newcommand{\ee}{\end{equation}}
\newcommand{\car}{\circlearrowright}
\newcommand{\ora}{\overrightarrow}
\newcommand{\ola}{\overleftarrow}
\newcommand{\hra}{\hookrightarrow}
\newcommand{\xra}{\xrightarrow}
\newcommand{\aux}{\mathrm{aux}}
\newtheorem{Def}{Definition}
\newtheorem{thm}{Theorem}
\newtheorem{statement}{Statement}
\newtheorem{lemma}{Lemma}

\begin{document}

\author{P. Mn\"ev}
\address{PDMI RAS, 27, Fontanka, 191023, Saint-Petersburg, Russia}
\email{pmnev@pdmi.ras.ru}
\title{Discrete $BF$ theory}


\begin{abstract}
In this work we discuss the simplicial program for topological field theories for the case of non-abelian $BF$ theory. Discrete $BF$ theory with finite-dimensional space of fields is constructed for a triangulated manifold (or for a manifold equipped with cubical cell decomposition), that is in a sense equivalent to the topological $BF$ theory on manifold. This discrete version allows one to calculate interesting quantities from the $BF$ theory, like the effective action on cohomology, in terms of finite-dimensional integrals instead of functional integrals, as demonstrated in a series of explicit examples. We also discuss the interpretation of discrete $BF$ action as the generating function for $qL_\infty$ structure (certain ``one-loop version'' of ordinary $L_\infty$ algebra) on the cell cochains of triangulation, related to the de Rham algebra of the underlying manifold by homotopy transfer procedure. This work is a refinement of older text \cite{BF_Archive}.
\end{abstract}
\maketitle

\setcounter{tocdepth}{3}
\tableofcontents


\section{Introduction}
\label{section: intro}
This paper contains results, obtained by the author in the framework of Andrei Losev's simplicial program for topological quantum field theories. It is an extended version of older text \cite{BF_Archive} and, at the same time, English translation of author's PhD thesis (with some improvements; main additions are sections \ref{section: fin-dim argument}, \ref{section: M_gamma}, \ref{section: M_gamma coh}; also, original text contained wrong one-loop result for 2-torus in section \ref{section: 2-torus in sym gauge} which is corrected here).

The aim of simplicial program is the equivalent replacement of a topological field theory in Lagrangian formalism by a simplicial version (or, more generally, cell version). Infinite-dimensional space of fields of the topological field theory is replaced by some finite-dimensional space, associated to a triangulation (or a more general cell decomposition) of the underlying manifold. The action of topological field theory is replaced by some function (the simplicial action) on this finite-dimensional space. Observables of the topological theory should also be replaced by their simplicial counterparts. This replacement should be equivalent, i.e. the correlators of observables in topological field theory should coincide with the correlators of respective simplicial observables (and we do not suppose going to the limit of infinitely dense triangulation: any triangulation should give the exact result). Knowing the simplicial equivalent of a topological field theory, we can compute the correlators of the latter by means of finite-dimensional integrals instead of functional integrals.

One of the aims of simplicial program is to construct simplicial versions for Chern-Simons theory (yielding invariants of knots and 3-manifolds \cite{Witten}) and for Poisson-sigma model (related to Kontsevich's deformation quantization \cite{Kontsevich}, \cite{CF}). In this work we consider a simpler model of topological field theory (yet related to both models listed above): the non-abelian $BF$ theory (for abelian $BF$ theory the simplicial version was constructed in \cite{Adams}). Another significant simplification is that we do not consider observables. Instead of correlators we consider the ``effective action on de Rham cohomology of a manifold'' --- an interesting invariant of a manifold that can be computed from simplicial version of $BF$ theory (section \ref{section: category of retracts}).

Classical action for $BF$ theory on a compact orientable manifold $M$ is
$$S_{cl}=\tr\int_M B\wedge F_A$$
where $F_A=dA+A\wedge A$ is the curvature of connection $A$. Classical fields of the theory are the connection $A$ in the trivial principal $G$-bundle on $M$, and field $B$ --- the $\g$-valued $(\dim M-2)$-form on $M$. Here $G$ is a compact Lie group (the gauge group) and  $\g$ is its Lie algebra. $BF$ theory is defined for a manifold $M$ of arbitrary dimension, moreover $M$ is allowed to have boundary (switching to canonical $BF$ theory\footnote{The term ``canonical'' here has nothing to do with canonical quantization.}, introduced in section \ref{BF theory}, we also allow $M$ to be non-orientable). Classical action of $BF$ theory possesses relatively complicated gauge symmetry (reducible and open at the second stage of reducibility tower) in dimensions $\geq 4$, and to solve the problem of gauge fixing one needs to use Batalin-Vilkovisky formalism. In Batalin-Vilkovisky formalism (``BV formalism'' in the following) classical fields  $A$ and $B$ are replaced by the BV super-fields $\tilde{A}$ and $\tilde{B}$ --- two non-homogeneous $\g$-valued differential forms on $M$ (this is a convenient way to collect the original classical fields, ghosts for all stages of the reducibility tower of gauge symmetry, anti-fields for the classical fields, anti-fields for ghosts). In terms of the super-fields $\tilde{A}$, $\tilde{B}$ the master action (also called ``BV action''), is
$$S=\tr\int_M \tilde{B}\wedge (d\tilde{A}+\tilde{A}\wedge\tilde{A})$$

Simplicial equivalent of $BF$ theory is naturally constructed on the level of master action and space of BV fields (instead of the classical action and the classical space of fields). For the space of (simplicial) BV fields for a triangulation $\Xi$ of manifold $M$, one takes certain finite-dimensional space $\FF_\Xi$, constructed from the space $C^\bt(\Xi,\g)$ of $\g$-valued cell cochains of $\Xi$ (which play the role of simplicial analog of $\g$-valued differential forms on $M$). Namely, $\FF_\Xi$ is constructed as the odd cotangent bundle for the shifted space of cell cochains: $\FF_\Xi=T^*[-1](C^\bt(\Xi,\g)[1])$ (the numbers in square brackets denote shifts of grading). For the coordinate on base of $\FF_\Xi$ one uses the simplicial super-field $\omega_\Xi$ --- non-homogeneous $\g$-valued cell cochain, whose components of different degrees are assigned certain ghost numbers, so that $\deg+\gh=1$ holds; for the coordinate in the fiber one uses the second simplicial super-field $p_\Xi$ --- non-homogeneous $\g^*$-valued cell chain, whose components are also assigned ghost numbers, so that $\deg+\gh=-2$ holds. Here $\omega_\Xi$ is the simplicial analog of the BV super-field $\tilde{A}$ of topological $BF$ theory, while $p_\Xi$ is the simplicial analog of $\tilde{B}_\flat$, i.e. of the BV super-field $\tilde{B}$ of topological $BF$ theory, dualized using the pairing $\tr\int_M\bt\wedge\bt$ (the formulation of topological $BF$ theory in terms of fields $A,B_\flat$ with the master action $S=<\tilde{B}_\flat,d\tilde{A}+\frac{1}{2}[\tilde{A},\tilde{A}]>$ is called ``canonical'' $BF$ theory).

The idea of construction of the simplicial BV action is to take the effective action induced on $\FF_\Xi$. Namely, we split the infinite-dimensional space of BV fields of topological $BF$ theory on $M$ (more precisely, of its canonical version) into infrared (IR) and ultraviolet (UV) parts
$$\FF_M=\FF'\oplus\FF''$$
where the IR part is $\FF'\cong\FF_\Xi$ (hence the UV part of $\FF_M$ is infinite-dimensional). The effective action $S_\Xi$ on IR fields has to be defined as a functional integral over UV fields. This is a standard construction of quantum field theory, and it is clear in what sense it leads to an equivalent action: quantum fluctuations in UV directions are already taken into account in $S_\Xi$. However, since we are dealing with a gauge theory in BV formalism, the construction of effective action has to be modified (the standard construction would yield a perturbatively ill-defined integral over $\FF''$). Namely, one chooses a Lagrangian submanifold in the space of UV fields $\LL\subset \FF''$ and defines the effective BV action on $\FF'$ as a functional integral over $\LL$, not over the whole $\FF''$. Integrals of this kind are called ``BV integrals'' and the choice of $\LL$ is the choice of gauge fixing for BV integral. The construction of the effective BV action is discussed in section \ref{effective action: idea}. The main features of this construction are the following: first, it sends solutions of quantum master equation (QME) to solutions of QME on IR fields. Second, the dependence on the choice of $\LL$ is under control: changing $\LL$ leads to a canonical transformation of the effective action. For the case of induction of effective action for topological $BF$ theory on the space $\FF_\Xi$ of infrared BV fields, we construct the Lagrangian submanifold $\LL\subset\FF''$ from the chain homotopy operator $K_\Xi$ that contracts de Rham complex of the manifold $M$ to the subcomplex of Whitney forms of triangulation $\Xi$, which is isomorphic to the complex of cell cochains of $\Xi$ (sections \ref{section: effective action for abstract BF}, \ref{section: Whitney forms}, \ref{section: Dupont's homotopy}). The operator $K_\Xi$ is ``glued'' from certain operators (Dupont's operators), given explicitly for individual simplices of $\Xi$. An important property  of the effective action for $BF$ theory is that the corresponding BV integral is expanded as a sum over Feynman diagrams, containing only trees and one-loop diagrams  (section \ref{section: effective action for abstract BF}, Theorem \ref{thm: eff action for abstract BF}).

The construction of $K_\Xi$ or, equivalently, the choice of gauge for the BV-integral defining the induction, implies another important property of the simplicial action  $S_\Xi$ --- the simplicial locality  (section \ref{section: simplicial BF action}, Theorem \ref{thm: simplicial locality}): $S_\Xi$ is expressed as a sum of contributions of individual simplices of the triangulation $S_\Xi=\sum_{\sigma\in\Xi}\bar{S}_\sigma$. Contributions $\bar{S}_\sigma$ depend only on the restrictions of fields $\omega_\Xi,p_\Xi$ of simplicial $BF$ theory to simplex $\sigma$. Contributions $\bar{S}_\sigma$ may be recovered, if one knows the simplicial action for single simplex $\Delta^D$ with standard triangulation for every dimension $D\geq 0$. Thus, due to simplicial locality, the problem of computing the simplicial action $S_\Xi$ for any triangulation $\Xi$ of any manifold $M$ is reduced to the series of universal computations: one needs to compute the simplicial action $S_{\Delta^D}$ for the standard simplex $\Delta^D$ in every dimension $D\geq 0$.

In dimension $D=0$ the problem of computing $S_{\Delta^D}$ is trivial. In dimension $D=1$ (induction of the effective action for the interval) the problem is not quite trivial, but it can be solved exactly (section \ref{section: interval}, Theorem \ref{interval thm}). The fact that exact calculation is possible is due to the fact that the action of topological $BF$ theory on interval, restricted to the Lagrangian submanifold $\LL\subset\FF''$ over which the BV integral is taken, turns out to be quadratic. Hence the BV integral itself is Gaussian. For the simplex of dimension  $D\geq 2$ this simplification does not occur and we do not know how to obtain the explicit result. However, one can obtain the perturbative result for $S_{\Delta^D}$ (section \ref{section: simplex pert}, Theorem \ref{thm: simplex perturbative result}), i.e. one can calculate first terms of the expansion of action in powers of fields, by calculating first Feynman diagrams for the respective BV integral. In section \ref{section: simplex pert} we demonstrate the technique allowing one to compute tree Feynman diagrams for the simplex of general dimension and to recover partially the values of one-loop diagrams (also in general dimension) from trees. Explicit computation of one-loop diagrams is technically much more involved. We demonstrate such a computation in section \ref{section: q_2 on 2-simplex} for the case of simplest non-trivial one-loop diagram for $D=2$.
Knowing the perturbative result for $\Delta^D$ with certain accuracy (i.e. up to certain powers of fields), we know the simplicial $BF$ action for any triangulation $\Xi$ of any manifold $M$ with the same accuracy. And we can obtain the effective action on de Rham cohomology of $M$ with the same accuracy by computing (finite-dimensional) BV integral. The example where $M$ is the circle and $\Xi$ is its cell decomposition into two intervals and two points is discussed in section \ref{section: S on coh examples: circle, torus, sphere} (here we compute the exact result, not just perturbative, since the simplicial action is known explicitly for dimensions $D=0,1$).

In section \ref{section: BF on cubical complex} we consider the construction of discrete $BF$ theory for cubical cell decomposition $\Xi$ of manifold $M$ (i.e. all cells of $\Xi$  are cubes of some dimensions, and cells are only allowed to intersect over a face). This construction is very similar to simplicial $BF$ theory. In particular, the property of cell locality holds for the cell action (section \ref{section: cell action}, Theorem \ref{thm: cell locality}), in complete analogy with simplicial case. Thus the problem of computing  $S_\Xi$ for any cubical cell decomposition  $\Xi$ of any manifold $M$ is reduced to the series of universal computations of cell actions $S_{I^D}$ for standard cubes $I^D$ in each dimension $D\geq 0$. The new feature of cubical setting is the factorization property of Feynman diagrams for $S_{I^D}$ (section \ref{section: cube factorization}, Theorem \ref{thm: cube factorization}), greatly simplifying the perturbative computations for $S_{I^D}$. Despite this simplification, we cannot present an explicit result for $S_{I^D}$ for $D\geq 2$. However, it turns out that restrictions of the action $S_{I^D}$ to certain special subspaces in the space of cell fields (for instance, to the subspace of periodical fields) may be calculated explicitly. Thus we obtain a series of examples of manifolds $M$ with special cell decompositions $\Xi$, where the cell action can be calculated explicitly (for example: torus, cylinder, Klein bottle --- see section \ref{section: exact results for cell action}). From these examples we obtain examples of manifolds, for which the effective $BF$ action on cohomology can be calculated explicitly  (section \ref{section: S on coh examples}). Also, in section  \ref{section: S on coh properties} we prove certain properties of the effective action on cohomology, allowing one to produce more examples where it can be obtained explicitly.

Procedure of inducing the effective action for $BF$ theory, whose special cases are the transition from topological $BF$ theory on manifold $M$ to the discrete theory on a triangulation (or on a cubical cell decomposition) $\Xi$, and the transition from the the discrete theory on a triangulation to the effective theory on de Rham cohomology of $M$, has also an algebraic interpretation. Namely, the action of topological $BF$ theory may be understood as a generating function for DGLA (differential graded Lie algebra) structure on the space $\Omega^\bt(M,\g)$ of $\g$-valued differential forms on $M$ (section \ref{abstract BF theory}). Next, simplicial action on a triangulation (or on a cubical cell complex) $\Xi$ may be interpreted as generating function for ``$qL_\infty$'' structure on the space $C^\bt(\Xi,\g)$ of $\g$-valued cell cochains of $\Xi$ (section \ref{section: eff action as generating function for alg structure}). This structure is a certain natural ``one-loop'' version of $L_\infty$ algebra. Then the BV integral, defining the transition from the action of topological $BF$ theory to action $S_\Xi$, may be understood as defining the ``homotopy transfer'' of the algebraic structure from the space of differential forms $\Omega^\bt(M,\g)$ to the space of cochains $C^\bt(\Xi,\g)$. We tried to clarify this idea in section \ref{section: BF_infty}. Transition from the discrete $BF$ theory to the effective theory on cohomology, or from the original topological $BF$ theory to the effective theory on cohomology, can also be explained in these terms. The invariant of manifold $M$, given by $BF$ theory --- the effective action on cohomology modulo canonical transformations --- can be understood algebraically as the ``homotopy type of the algebra of $\g$-valued differential forms on $M$ as a $qL_\infty$ algebra''  (sections \ref{section: equivalence of qL_infty algebras}, \ref{section: category of retracts}).

An important comment is due on the logic of this work. Throughout the most part of the paper we have in mind the following logic: we formally apply abstract constructions of section \ref{section: abstract BF and effective action}, which are rigorous for finite-dimensional induction, for constructing discrete $BF$ theory and effective $BF$ theory on cohomology, hoping that results of section \ref{section: abstract BF and effective action} still apply for induction from de Rham algebra of a manifold. In particular, claims that discrete $BF$ action satisfies quantum master equation and that effective action on cohomology modulo canonical transformations is an invariant of manifold, are based on this assumption, and are not rigorously proved. Indeed, calculation of discrete $BF$ action involves computing super-traces over the space of differential forms, and these super-traces require certain regularization to make sense of them. In principle, some regularizations will give results not satisfying the QME. So one should make an independent check, whether the discrete $BF$ action, obtained using certain regularization scheme, satisfies QME. We perform such a check for the exact simplicial action for 1-simplex in section \ref{section: interval QME check} and we give a sketch of finite-dimensional proof of QME for the cell action of $D$-cube in section \ref{section: fin-dim argument}. Knowing that actions for the building blocks of discrete $BF$ theory (simplices or cubes) satisfy QME, we know that the discrete $BF$ action for any simplicial or cubical cell complex satisfies QME (by construction of gluing of $qL_\infty$ algebras, section \ref{section: gluing}). Thus we have an independent (of transfer statements of section \ref{section: abstract BF and effective action}) proof of QME for discrete $BF$ action, at least in cubical setting. In principle, one should also make finite-dimensional check of the claim that effective action on cohomology, induced from discrete theory, regarded modulo canonical transformations, is an invariant of the underlying manifold. To do this, one should analyze how the induced action on cohomology changes under local reconstructions of the cell complex (such as subdivisions and aggregations). However, we do not address this problem in present work.

\subsection{Main results}
Here is the brief review of main results of this work:
\begin{itemize}
\item Statement of simplicial locality of simplicial action (section \ref{section: simplicial BF action}, Theorem \ref{thm: simplicial locality}):
for any triangulation $\Xi$ of any manifold $M$ the simplicial $BF$ action $S_\Xi(\omega_\Xi,p_\Xi)$ is expressed as a sum over simplices of triangulation $\sigma\in\Xi$ of local contributions $\bar{S}_\sigma$ --- certain universal functions (depending only on the dimension of $\sigma$), evaluated on restrictions of simplicial fields $\omega_\Xi$, $p_\Xi$ to simplex $\sigma$. More precisely, simplicial fields are represented as
$\omega_\Xi=\sum_{\sigma\in\Xi}e_\sigma \omega^\sigma,\; p_\Xi=\sum_{\sigma\in\Xi}p_\sigma e^\sigma$, where $\{e_\sigma\}, \{e^\sigma\}$ are basis cochains and basis chains of $\Xi$, associated to simplices $\sigma\in\Xi$. The corresponding variables are $\omega^\sigma\in\g$ and $p_\sigma\in\g^*$ (they are prescribed certain ghost numbers, depending on the dimension of $\sigma$). Then simplicial locality means that the simplicial action  $S_\Xi$ is represented as
$$S_\Xi(\omega_\Xi,p_\Xi)=\sum_{\sigma\in\Xi}\bar{S}_\sigma(\{\omega^{\sigma'}\}_{\sigma'\subset\sigma},p_\sigma)$$
(cf. the more detailed discussion in section \ref{section: simplicial BF action}). Completely analogous statement holds for the cell action  $S_\Xi$ of cubical cell decomposition $\Xi$ of manifold $M$ (section \ref{section: cell action}, Theorem \ref{thm: cell locality}).
\item Explicit result for simplicial $BF$ action for 1-simplex $\Delta^1$ with standard triangulation (section \ref{section: interval}, Theorem \ref{interval thm}):
\begin{multline*}S_{\Delta^1}(\omega^0,\omega^1,\omega^{01},p_0,p_1,p_{01})
=\left<p_0,\frac{1}{2}[\omega^0,\omega^0]\right>_\g+\left<p_1,\frac{1}{2}[\omega^1,\omega^1]\right>_\g
+\\
+\left<p_{01},\frac{1}{2}[\omega^{01},\omega^0+\omega^1]+\left(\frac{\ad_{\omega^{01}}}{2}\coth\frac{\ad_{\omega^{01}}}{2}\right)\circ (\omega^1-\omega^0)\right>_\g
+\hbar\;\tr_\g\log\left(\frac{\sinh\frac{\ad_{\omega^{01}}}{2}}{\frac{\ad_{\omega^{01}}}{2}}\right)
\end{multline*}
\item Perturbative result for simplicial $BF$ action for $D$-simplex $\Delta^D$ with standard triangulation (section \ref{section: simplex pert}, Theorem \ref{thm: simplex perturbative result}):
\begin{multline*}
S_{\Delta^D}(\{\omega^\sigma\}_{\sigma\subset\Delta^D},\{p_\sigma\}_{\sigma\subset\Delta^D})=\sum_{\sigma,\sigma_1\subset\Delta^D}c^\sigma_{\sigma_1} <p_\sigma,\omega^{\sigma_1}>_\g+\\
+\frac{1}{2}\sum_{\sigma,\sigma_1,\sigma_2\subset\Delta^D}c^\sigma_{\sigma_1,\sigma_2} <p_\sigma,[\omega^{\sigma_1},\omega^{\sigma_2}]>_\g+
\frac{1}{2}\sum_{\sigma,\sigma_1,\sigma_2,\sigma_3\subset\Delta^D}c^\sigma_{\sigma_1,\sigma_2,\sigma_3} <p_\sigma,[\omega^{\sigma_1},[\omega^{\sigma_2},\omega^{\sigma_3}]]>_\g+\\
+\hbar \frac{1}{2}\sum_{\sigma_1,\sigma_2\subset\Delta^D}q_{\sigma_1,\sigma_2}\tr_\g(\ad_{\omega^{\sigma_1}}\ad_{\omega^{\sigma_2}})+
O(p\omega^4+\hbar\omega^3)
\end{multline*}
where  $<,>_\g$ is the pairing between $\g^*$ and $\g$, and $\tr_\g$ is the trace in adjoint representation of $\g$. Combinatorial coefficients $c^\sigma_{\sigma_1},\; c^\sigma_{\sigma_1,\sigma_2},\; c^\sigma_{\sigma_1,\sigma_2,\sigma_3},q_{\sigma_1,\sigma_2}$ depend on the combinatorics of intersection of faces in $\Delta^D$ and their possible values are
\begin{eqnarray*}
c^\sigma_{\sigma_1}\in\{0,\pm 1\}\\
c^\sigma_{\sigma_1,\sigma_2}\in\{0,\pm\frac{|\sigma_1|!\,|\sigma_2|!}{(|\sigma_1|+|\sigma_2|+1)!}\}\\
c^\sigma_{\sigma_1,\sigma_2,\sigma_3}\in\{0,\pm\frac{|\sigma_1|!\,|\sigma_2|!\,|\sigma_3|!}{(|\sigma_2|+|\sigma_3|+1) \cdot (|\sigma_1|+|\sigma_2|+|\sigma_3|+1)!}\}\\
q_{\sigma_1,\sigma_2}\in\{0,\hat{\mc{A}}_D+(D-1)\hat{\mc{B}}_D,\pm \hat{\mc{B}}_D\}
\end{eqnarray*}
where $|\sigma|$ denotes the dimension of simplex. Concrete value of each coefficient depends on the combinatorics of intersection of faces and on their mutual orientations (precise formulation of the result is given in section \ref{section: simplex pert}). For $\hat{\mc{A}}_D$ we obtain the formula $$\hat{\mc{A}}_D=\sum_{n=1}^D C^{n-1}_{D-1}\frac{(-1)^{n+1}}{(n+1)^2(n+2)}$$
while for $\hat{\mc{B}}_D$ only the values for lower dimensions $D\leq 3$ are known:
$$\hat{\mc{B}}_0=\hat{\mc{B}}_1=0,\;\hat{\mc{B}}_2=\frac{1}{270},\;\hat{\mc{B}}_3=\frac{1}{270}-\frac{1}{648}$$
(see the explicit computation in section \ref{section: q_2 on 2-simplex}).
\item Factorization property and perturbative result for cell $BF$ action $S_{I^D}$ for $D$-cube $I^D$ (section \ref{section: cube factorization}, Theorem \ref{thm: cube factorization}). Factorization property means that the problem of computing Feynman diagrams for $S_{I^D}$ is reduced to the problem of computing Feynman diagrams for interval $I=\Delta^1$, but with extended propagator $K^{\lambda,d\lambda}$ (section \ref{section: tensor product for ind data}) that is not just the chain homotopy for interval, but a linear combination of chain homotopy, projection to Whitney forms and identity.
\item Series of examples of explicitly calculable cell actions (section \ref{section: exact results for cell action}, Statement \ref{statement: exact cell action examples}) and examples of explicitly calculable effective actions on cohomology (section \ref{section: S on coh examples}). The most interesting examples here are the circle $M=\s^1$:
    $$S_{H^\bt(\s^1,\g)}=<p_+,\frac{1}{2}[\omega^+,\omega^+]>_\g+<p_I,[\omega^I,\omega^+]>_\g+\hbar\;\tr_\g\log \left(\frac{\sinh\frac{\ad_{\omega^I}}{2}}{\frac{\ad_{\omega^I}}{2}}\right)$$
    (indices ``$+$'' and ``$I$'' of the fields correspond to the basis $e_+=1,\; e_I=dt$ in de Rham cohomology of the circle $H^\bt(\s^1)$) and the Klein bottle $M=\mr{KB}$:
    $$S_{H^\bt(\mr{KB},\g)}
=<p_{++},\frac{1}{2}[\omega^{++},\omega^{++}]>_\g+<p_{I+},[\omega^{I+},\omega^{++}]>_\g-\hbar\;\tr_\g\log\left(\frac{\ad_{\omega^{I+}}}{2}\coth \frac{\ad_{\omega^{I+}}}{2}\right)
$$
(indices ``$++$'', ``$I+$'' correspond to the basis $e_{++}=1$, $e_{I+}=dt_1$ in $H^\bt(\mr{KB})$).
\end{itemize}

\subsection{Plan of the paper}
Sections \ref{intro to BV}, \ref{gauge fixing methods} are introductory. Sections \ref{section: abstract BF and effective action} and \ref{section: BF on simplicial complex} are central for the work: in section \ref{section: abstract BF and effective action} we introduce the necessary constructions on abstract level, in section \ref{section: BF on simplicial complex} we apply them for the construction of the simplicial version of topological $BF$ theory. In sections \ref{section: BF on cubical complex} and \ref{section: effective theory on coh} we develop the technique allowing in special cases to obtain exact results for effective $BF$ action on cohomology: in section \ref{section: BF on cubical complex} we discuss the discrete $BF$ theory on cubical cell decomposition of a manifold and the factorization property of Feynman diagrams, in section \ref{section: effective theory on coh} we discuss the effective action on cohomology and some examples where it can be obtained explicitly.

Now we will give a more extended commentary on the contents of the work, section by section.

\begin{itemize}
\item \ref{section: intro}: Introduction.
\item \ref{intro to BV}: We give a brief review of Batalin-Vilkovisky formalism and the necessary notions of super-geometry. The exposition is mostly based on the paper \cite{Schwarz} by A. Schwarz.
\item \ref{gauge fixing methods}: We discuss the three main methods of gauge fixing for gauge field theories: Faddeev-Popov method, BRST method and Batalin-Vilkovisky method. In section \ref{BF theory} we introduce the topological $BF$ theory and describe the gauge fixing for it, suggested in \cite{Wallet}, \cite{Ikemori}. For the detailed review of $BF$ theory in Batalin-Vilkovisky formalism, see \cite{CR}.
\item \ref{section: abstract BF and effective action}: We discuss in detail the construction of effective BV action on the example of a natural generalization of $BF$ theory in BV formalism --- the ``abstract $BF$ theory\footnote{It is actually at the same time a generalization and a toy model for topological $BF$ theory, since the space of fields is implicitly assumed to be finite-dimensional throughout the section \ref{section: abstract BF and effective action}.}''. We also discuss the algebraic interpretation of the construction of inducing the effective action.
\begin{itemize}
\item \ref{abstract BF theory}: We introduce the abstract $BF$ theory --- an abstract model of gauge field theory in BV formalism, associated to a unimodular differential graded Lie algebra $V$. The special case of abstract $BF$ theory for $V=\Omega^\bt(M,\g)$ corresponds to the (canonical) topological $BF$ theory on manifold $M$.
\item \ref{effective action: idea}: We discuss the general construction of effective  BV action and its most important features: solution of quantum master equation is transferred to a solution of quantum master equation; if the induction data is deformed (i.e. the gauge fixing condition for BV integral is deformed), the effective action changes by a canonical transformation; canonical transformation of the initial action leads to a canonical transformation of the effective action (Statements \ref{statement: QME for induced action}, \ref{statement: induction data deform}, \ref{statement: induced canonical transformation}).
\item \ref{section: effective action for abstract BF}: We specialize the general construction of effective BV action to the case of abstract $BF$ theory. We introduce the class of convenient gauges (Lagrangian submanifolds in the space of ultraviolet fields), associated to chain homotopies, contracting $V$ to a subcomplex, and we obtain the perturbative expansion for effective action (Theorem \ref{thm: eff action for abstract BF}). We also discuss the dependence of effective action on the choice of induction data (Statement \ref{statement: infinitesimal ind data deform for abstract BF}).
\item \ref{section: eff action as generating function for alg structure}: We give algebraic interpretation of the effective action for abstract $BF$ theory as a generating function for certain algebraic structure on a subcomplex  $V'\hra V$ --- the structure of ``$qL_\infty$ algebra'', i.e. the set of classical and quantum operations $l_{(n)}:\Lambda^n V'\ra V'$, $q_{(n)}: \Lambda^n V'\ra \RR$, satisfying two sequences of quadratic relations --- ``homotopy Jacobi identities'' and ``homotopy unimodularity relations''. A $qL_\infty$ algebra may be understood as a certain one-loop completion of an ordinary $L_\infty$ algebra (such objects appeared earlier in different context, as algebras over ``wheeled $L_\infty$''operad, see \cite{Merkulov}). We also give an equivalent description of $qL_\infty$ structure on $V'$, as a cohomological vector field $Q$ on $V'[1]$, endowed with a consistent measure $\mu$ on $V'[1]$. Reader is referred to \cite{Granaker} for operadic treatment of the subject.
\item \ref{section: BF_infty}: We introduce the class of ``$BF_\infty$ theories'', associated to $qL_\infty$ algebras in analogy with the way abstract $BF$ theories are associated to unimodular DGLAs. Concept of a $BF_\infty$ theory may be thought of as an axiomatization of the effective theory for abstract $BF$ theory. Effective action for a $BF_\infty$ theory is again an action of  $BF_\infty$ type, and we formulate the perturbation expansion for such induction (Theorem \ref{thm: eff action for BF_infty}). In the language of $qL_\infty$ algebras, transition to the effective theory is formulated as the homotopy transfer of $qL_\infty$ structure to a subcomplex $V'\hra V$. Also, in section \ref{section: equivalence of qL_infty algebras} we discuss the concept of equivalence for $qL_\infty$ algebras, where the equivalence relation is generated by canonical transformations of the respective $BF_\infty$ actions, and by the induction operation.
\end{itemize}
\item \ref{section: BF on simplicial complex}: We apply the constructions of section \ref{section: abstract BF and effective action} to construct simplicial $BF$ theory.
    \begin{itemize}
    \item \ref{section: Whitney forms}, \ref{section: Dupont's homotopy}: Here we remind two well-known constructions, allowing us to define the induction data for BV integral, defining the simplicial $BF$ action, --- the construction of Whitney forms \cite{Whitney} and the construction of Dupont's chain homotopy operator \cite{Dupont}. This exposition is based on \cite{Getzler}.
    \item \ref{section: simplicial BF action}: Here we formulate the key feature of simplicial $BF$ action --- the property of simplicial locality (Theorem \ref{thm: simplicial locality}). This property allows us to reduce the problem of computing the simplicial $BF$ action in general case to the series of universal computations for one standard simplex $\Delta^D$ (endowed with standard triangulation) in each dimension $D=0,1,2,\ldots$.
    \item \ref{section: gluing}: We discuss the abstract gluing procedure for $qL_\infty$ algebras, which generalizes the reconstruction of simplicial action for a triangulation from simplicial actions for individual simplices. In section \ref{section: gluing-induction} we prove on abstract level that, under certain consistence conditions, the procedures of induction and gluing commute. This statement is an abstract generalization of the simplicial locality property of simplicial $BF$ action for a triangulation.
    \item \ref{section: interval}: We obtain the explicit result for simplicial action for standard 1-simplex (Theorem \ref{interval thm}). De Rham parts of Feynman diagrams for the corresponding BV integral are given in terms of Bernoulli numbers, and the explicit check of classical master equation for the effective action (which holds by construction) yields non-trivial (but known, cf. \cite{AD}) quadratic relations for Bernoulli numbers --- section  \ref{section: interval QME check}.
    \item \ref{section: simplex pert}. We obtain the perturbative result for simplicial $BF$ action for the standard simplex of general dimension (Theorem \ref{thm: simplex perturbative result}). In section \ref{section: q_2 on 2-simplex} we demonstrate the explicit computation of one-loop Feynman diagram for the case of simplest non-trivial diagram and the 2-simplex.
    \end{itemize}
\item \ref{section: BF on cubical complex}: We introduce the discrete $BF$ theory on a cubical cell decomposition of manifold. This discussion is a modification of the discussion for the simplicial setting. The main difference is the factorization property of Feynman diagrams for the cell action for standard cube. This property greatly simplifies perturbative computations and leads to a series of examples where the cell action can be computed explicitly.
    \begin{itemize}
    \item \ref{section: tensor product for ind data}, \ref{section: cell action}: We discuss the construction of tensor product for induction data, which we use to construct the induction data from differential forms on cube to cell cochains of cube; and then, similarly to the simplicial case, we construct the induction data from differential forms on a manifold to cell cochains of a cubical cell decomposition. The latter induction data allows us to construct the BV integral, defining the action of the discrete $BF$ theory on a cubical cell decomposition of a manifold (the ``cell action''). In complete analogy with the simplicial setting, the cell action satisfies the cell locality property (Theorem \ref{thm: cell locality}).
    \item \ref{section: cube factorization}: We discuss the factorization property of Feynman diagrams for the cell action for cube (arising from the construction of tensor product for chain homotopies) and obtain the perturbative result  (Theorem \ref{thm: cube factorization}).
    \item \ref{section: exact results for cell action}: Using the factorization property of Feynman diagrams, we obtain some examples of exactly computable cell action: torus, cylinder. Using the gluing procedure for the cylinder, we also obtain explicit result for the Klein bottle. Our examples are summarized in Statement \ref{statement: exact cell action examples}.
    \item \ref{section: fin-dim argument}: We give a sketch of finite-dimensional proof of the fact that the cell action for $D$-cube satisfies quantum master equation. We argue that the problem can be reduced to checking certain properties (boundary factorization and closeness) for an object, living on interval (``FC form'' on cochains of interval), which is in a sense a generating object for cell actions for cubes of all dimensions $D$. These properties can be checked explicitly.
    \end{itemize}
\item \ref{section: effective theory on coh}: We discuss the action of effective $BF$ theory on de Rham cohomology of manifold, which is an interesting invariant of manifolds. We discuss the possible way of computing it via discrete $BF$ theory, some properties, allowing to compute it exactly in some cases, and explicit examples.
    \begin{itemize}
    \item \ref{section: category of retracts}: Here we present the general picture of induction as the transfer of $qL_\infty$ structure along morphisms in ``category of retracts'' and its specialization for the case of induction of effective action from topological $BF$ theory. We discuss the way of computing effective action on cohomology via discrete $BF$ theory.
    \item \ref{section: S on coh properties}: We discuss some specific properties of effective action on cohomology, allowing one to compute it explicitly in some cases.
    \item \ref{section: S on coh examples}: We give examples of explicitly computable effective action on cohomology. Of particular interest here is the pair of examples: circle and Klein  bottle.
    \end{itemize}
\end{itemize}

\subsection{Open problems} Here we list some of the questions concerning discrete $BF$ theory.
\begin{itemize}
\item It would be nice to have a finite-dimensional proof that the effective $BF$ action on de Rham cohomology of a manifold, regarded modulo canonical transformation, is a PL invariant of manifolds. To do that, one should analyze how the effective action on cohomology, induced from discrete $BF$ theory,  behaves under local reconstructions of triangulation (or cubical cell decomposition) of manifold. Namely one should check that these reconstructions induce canonical transformations on the effective action on cohomology.
\item It is interesting to understand, what sort of invariant of manifolds the effective action on cohomology is. In particular, is it a homotopy invariant? More specifically, is it strictly weaker than Massey operations on cohomology plus the fundamental group?
\item One-loop effective action on cohomology of a manifold $M$, restricted to the Maurer-Cartan set, can be viewed as a perturbative approximation to the ``torsion'' function $\det (d+A)$ (properly gauge-fixed) on the moduli space of flat connections on $M$, in the neighbourhood of zero connection. It would be interesting to calculate this torsion globally (on the whole moduli space of flat connections) in some examples. A related question: is it possible to recover one-loop effective action on cohomology non-perturbatively, just looking at singularities of the moduli space of flat connections? This seems likely from our examples of $S_D\ltimes\ZZ_2^D$-bundles over circle in section \ref{section: M_gamma coh}.
\item It would be good to have a finite-dimensional proof that the simplicial action for $D$-simplex satisfies quantum master equation, for some regularization scheme for the super-traces. We outlined such a proof for the cubical setting in section \ref{section: fin-dim argument}, but it essentially relies on the tensor product story.
\item One should study observables in discrete $BF$ theory. In particular, there should be discrete versions of observables of topological $BF$ theory, associated to knots \cite{CCRFM}.
\item A very natural idea is to study the discrete $BF$ theory, associated to a non-trivial principal bundle on a manifold, and its possible application to combinatorial formulae for characteristic classes.
\item Since the 2-dimensional $BF$ theory is a specific case of Poisson sigma model, corresponding to linear Kirillov-Kostant Poisson structure on $\g^*$, one can try to write Kontsevich's deformation for $\g^*$ via discrete $BF$ theory on a triangulated disk.
\item The major goal is to advance to discrete versions of Chern-Simons theory and of Poisson sigma model. Another interesting model, very close to Chern-Simons, is the 3-dimensional $BF$ theory with cosmological term, i.e. with classical action $\tr\int_M B\wedge F_A + B\wedge B\wedge B$.
\end{itemize}

\subsection{References}
Our main references for Batalin-Vilkovisky formalism are \cite{AKSZ},\cite{Schwarz} and the original papers \cite{BV81},\cite{BV83}. Also, more recent papers \cite{Khudaverdian},\cite{Severa} explain the half-densities picture in BV formalism. Concept of the effective BV action is used in \cite{AKLL},\cite{KL} and an alternative understanding of effective BV action is proposed in \cite{Costello}. Homotopy transfer for classical $A_\infty$ algebras is explained in \cite{KS}, for the transfer for $qL_\infty$ algebras cf. \cite{Granaker}. Historical references for $BF$ theory in Batalin-Vilkovisky formalism are \cite{Ikemori},\cite{Wallet} and a review of the subject is given in \cite{CR}. We also borrowed two important constructions from paper \cite{Getzler}: Whitney forms on simplex and Dupont's chain homotopy (the historical references are \cite{Whitney},\cite{Dupont}).

\subsection{Acknowledgements} I wish to thank my PhD advisor Ludwig Faddeev for the permanent support and numerous enlightening discussions, and Andrei Losev for statement of the problem, inspiration and ideas.   I am also grateful to  Alberto Cattaneo, Kevin Costello, Giovanni Felder, Nikolai Mn\"ev, Florian Sch\"atz, Dennis Sullivan, Bruno Vallette, Scott Wilson for valuable discussions on the subject. The idea of tensor product for chain homotopies, that I learned from discussion with Dennis Sullivan, is crucial for sections \ref{section: BF on cubical complex} and \ref{section: effective theory on coh} of this work.

This work has been
partially supported by RFBR 08-01-00638 grant, by SNF Grant 20-113439, by
the European Union through the FP6 Marie Curie RTN ENIGMA (contract
number MRTN-CT-2004-5652), and by the European Science Foundation
through the MISGAM program. I also thank Zurich University and ETH Zurich for
hospitality and partial support.

\section{Preliminaries: introduction to Batalin-Vilkovisky formalism}
\label{intro to BV}
This section is a very sketchy recollection of basics of super-geometry and BV formalism.
Our discussion of the geometrical principles of BV formalism is mostly based on A. Schwarz's paper \cite{Schwarz}, so the space of fields is supposed to be a finite-dimensional $\ZZ$-graded manifold (following the physical tradition, we are using $\ZZ$-grading here instead of $\ZZ_2$-grading). The main references are \cite{Schwarz},\cite{AKSZ},\cite{Khudaverdian},\cite{Severa}. One of the aims of this section, as well as of section \ref{gauge fixing methods}, is introducing notations and sign conventions.

\subsection{Gerstenhaber algebras and Batalin-Vilkovisky algebras}
\begin{Def} A Gerstenhaber algebra (or ``odd Poisson algebra'') is a graded commutative algebra $\CC$ endowed with an odd Poisson bracket of degree +1 (also called the anti-bracket), i.e. a bilinear map
$\{\bt,\bt\}: \CC\otimes\CC\ra\CC$ satisfying the following relations:
\begin{eqnarray}
\e(\{X,Y\})=\e(X)+\e(Y)+1 \label{antibracket degree}\\
\{X,Y\}=- (-1)^{(\e(X)+1)(\e(Y)+1)}\{Y,X\} \label{antibracket symmetry}\\
(-1)^{(\e(X)+1)(\e(Z)+1)}\{X,\{Y,Z\}\}+\mr{cycl.perm. }\,X,Y,Z = 0\label{antibracket Jacobi}\\
\{XY,Z\}=X\{Y,Z\}+(-1)^{\e(Y)\;(\e(Z)+1)}\{X,Z\}Y \label{antibracket derivation right}\\
\{X,YZ\}=\{X,Y\}Z+(-1)^{(\e(X)+1)\;\e(Y)}Y\{X,Z\} \label{antibracket derivation left}
\end{eqnarray}
for any homogeneous elements $X,Y,Z\in\CC$. Here $\e(X)\in\ZZ$ denotes the degree of $X$ in $\CC$.
\end{Def}
Relations (\ref{antibracket degree},\ref{antibracket symmetry},\ref{antibracket Jacobi}) mean that the anti-bracket endows $\CC[1]$ with the structure of graded Lie algebra, while relations (\ref{antibracket derivation right},\ref{antibracket derivation left}) mean that the anti-bracket is a biderivation of the commutative multiplication on $\CC$.

\begin{Def} A Batalin-Vilkovisky algebra is a graded commutative unital algebra $\CC$, endowed with a BV Laplacian, i.e. a linear map $\Delta: \CC\ra\CC$ satisfying the following relations:
\begin{eqnarray}
\e(\Delta X)=\e(X)+1 \\
\Delta^2=0 \\
\Delta(1)=0 \\
\Delta(XYZ)=\Delta(XY)Z+(-1)^{(\e(X)+1)\;\e(Y)}Y\Delta(XZ)+(-1)^{\e(X)}X\Delta(YZ)-\nonumber\\
-\Delta(X)YZ-(-1)^{\e(X)}X\Delta(Y)Z-(-1)^{\e(X)+\e(Y)}XY\Delta(Z)\label{7term relation}
\end{eqnarray}
for any homogeneous elements $X,Y,Z\in\CC$.
\end{Def}
Relation (\ref{7term relation}) is the Leibniz identity for a second order differential operator. A BV algebra is automatically a Gerstenhaber algebra with anti-bracket
\be\{X,Y\}=(-1)^{\e(X)}\Delta(XY)-(-1)^{\e(X)}\Delta(X)Y-X\Delta(Y)\label{antibracket from BV Laplacian}\ee
and moreover the BV Laplacian is a derivation  of this anti-bracket:
\be \Delta\{X,Y\}=\{\Delta X,Y\}+(-1)^{\e(X)+1}\{X,\Delta Y\} \ee
This implies that $\Delta$ and $\{,\}$ endow $\CC[1]$ with the structure of differential graded Lie algebra.

\textbf{Example.} If $V$ is a graded vector space, then the algebra of polynomial functions on $V\oplus V^*[-1]$, i.e.
$$\CC=S^\bt(V\oplus V^*[-1])^*=(S^\bt V^*)\otimes (S^\bt V[1])$$
has a canonical BV algebra structure. If $(x^i)$ are (homogeneous) coordinates on $V$ and $(\xi_i)$ are dual coordinates on $V^*[-1]$ then the BV Laplacian is
$$\Delta=\sum_i (-1)^{\e(x^i)}\frac{\dd}{\dd x^i}\frac{\dd}{\dd \xi_i}$$
and the corresponding anti-bracket is
$$\{f,g\}=\sum_i f\left(\frac{\overleftarrow\dd}{\dd
x^i}\frac{\overrightarrow\dd}{\dd \xi_i}-\frac{\overleftarrow\dd}{\dd
\xi_i}\frac{\overrightarrow\dd}{\dd x^i}\right)g $$
for $f,g\in \CC$.

\subsection{$\ZZ$-graded manifolds}
\label{graded manifolds}
\begin{Def}
We call a $\ZZ$-graded manifold $\MM=\bigoplus_{k\in\ZZ}\MM^k$ a sum of vector bundles  over a smooth manifold $\MM_0$ (the body of $\MM$), where we suppose that the rank of $\MM^k$ vanishes for all but finitely many values of $k$. The ring of functions on $\MM$ is defined to be the graded (super-)commutative algebra of sections
$$\Fun(\MM):=\Gamma(\MM_0,S^\bt \MM^*_\even\otimes \Lambda^\bt \MM^*_\odd)=\Gamma(\MM_0,S^\bt\MM^*_\even)\otimes_{C^\infty(\MM_0)} \Gamma(\MM_0,\Lambda^\bt\MM^*_\odd)$$
where $\MM_\even=\bigoplus_k \MM^{2k}$, $\MM_\odd=\bigoplus_k \MM^{2k+1}$ are even and odd parts of $\MM$ respectively, $S^\bt$ and $\Lambda^\bt$ denote the sums of symmetric powers and of exterior powers of the bundle respectively, and $\Gamma$ is the space of sections of bundle. Grading is defined by assigning degree $-k$ to sections of $(\MM^k)^*$.
\end{Def}
If $(x^I)$ are local coordinates on an open subset $U_0\subset\MM_0$ and $(\chi^\alpha)$ are coordinates in the fiber of $\MM$ then we say that $(x^I,\chi^\alpha)$ are local coordinates on $U=\pi^{-1}U_0\subset\MM$, where $\pi: \MM\ra \MM_0$ is the projection to base. For the ring of functions we have $$\Fun(\MM)|_U\cong C^\infty(U_0)\otimes\RR[\chi^\alpha]$$ where we treat coordinates $\chi^\alpha$ corresponding to odd directions in $\MM$ as Grassman variables.
In what follows ``graded'' will always mean ``$\ZZ$-graded''.

\begin{Def}
The odd tangent bundle $T[1]\MM$ of a graded manifold $\MM$ is a graded manifold represented by the bundle $\bigoplus_{k\in\ZZ}(T[1]\MM)^k$ over $\MM_0$ where the elements of grading are defined as
$$(T[1]\MM)^k:=
\left\{\begin{array}{ll}T\MM_0\oplus\MM^{-1}\oplus\MM^{0}&\text{ for }k=-1,\\
\MM^k\oplus\MM^{k+1}&\text{ for }k\neq -1
\end{array}\right.$$
Ring of functions on $T[1]\MM$ is defined as
$$\Fun(T[1]\MM):=\Omega^\bt(\MM_0)\otimes_{C^\infty(\MM_0)} \Fun(\MM)\otimes_{C^\infty(\MM_0)} \Fun(\MM[1])$$
where $\Omega^\bt(\MM_0)$ is the algebra of differential forms on $\MM_0$.
A $k$-form on $\MM_0$ is understood as element of degree $k$ in $\Fun(T[1]\MM)$.
\end{Def}
If $(x^i)$ are local coordinates on $\MM$ and $(\psi^i)$ are coordinates in the fiber of $T[1]\MM$ (as a bundle over $\MM$), then for the degrees of coordinates we have $\e(\psi^i)=\e(x^i)+1$.
Functions on the odd tangent bundle $T[1]\MM$ are called differential forms on $\MM$:
$$\Omega^\bt(\MM):=\Fun(T[1]\MM)$$
Vector fields on $\MM$ are understood as the derivations of algebra $\Fun(\MM)$:
$$\mr{Vect}(\MM):=\mr{Der}(\Fun(\MM))$$
An important vector field on a graded manifold is the Euler vector field $E$, acting on homogeneous functions $f\in\Fun(\MM)$ as
$$E(f):= \e(f)\cdot f$$
where $\e(f)$ is the Grassman degree of $f$.

\begin{Def}A vector field $Q$ on a graded manifold $\MM$ is called ``cohomological'' if $\e(Q)=1$ (i.e. applying $Q$ to a function increases degree by 1) and $Q^2=0$. A graded manifold endowed with a cohomological vector field $(\MM,Q)$ is called a $Q$-manifold.
\end{Def}
In particular, for any $\MM$ there is a natural cohomological vector field on the odd tangent bundle $T[1]\MM$ --- the de Rham differential on $\Omega^\bt(\MM)$. In local coordinates it is written as $$\delta=\psi^i\frac{\dd}{\dd x^i}$$

\begin{Def} The odd cotangent bundle $T^*[-1]\MM$ of a graded manifold $\MM$ is a graded manifold, represented by the bundle $\bigoplus_{k\in\ZZ}(T^*[-1]\MM)^k$ over $\MM_0$ where
$$(T^*[-1]\MM)^k:=
\left\{\begin{array}{ll}T^*\MM_0\oplus\MM^{1}\oplus(\MM^0)^*&\text{ for }k=1,\\
\MM^{k}\oplus(\MM^{1-k})^*&\text{ for }k\neq 1
\end{array}\right.$$
The ring of functions on $T^*[-1]\MM$ is
$$\Fun(T^*[-1]\MM):=\mc{V}^\bt(\MM_0)\otimes_{C^\infty(\MM_0)} \Fun(\MM)\otimes_{C^\infty(\MM_0)} \Fun(\MM^*[-1])$$
where $\mc{V}^\bt(\MM_0)$ is the space of polyvector fields on $\MM_0$, and a $k$-polyvector is regarded as an element of degree $-k$ in $\Fun(T^*[-1]\MM)$.
\end{Def}
If $(x^i)$ are local coordinates on $\MM$ and $(\xi_i)$ are coordinates in the fiber of $T^*[-1]\MM$ (as a bundle over $\MM$), then the degrees of coordinates satisfy  $\e(x^i)+\e(\xi_i)=-1$.

\begin{Def} The Berezinian bundle of a graded manifold $\MM$ is the following linear bundle over $\MM_0$:
$$\mr{Ber}(\MM)=\Lambda^{\dim\MM_0}T^*[-1]\MM_0\otimes\Lambda^{\mr{rk} \MM_\even}\MM_\even^*[-1]\otimes \Lambda^{\mr{rk} \MM_\odd}\MM_\odd$$
Its sections $\mu\in\Gamma(\MM_0,\mr{Ber}(\MM))$ are called the Berezin measures on $\MM$.
\end{Def}
Integration over a graded manifold is defined as the linear map
$$\int_\MM:\quad\Fun(\MM)\otimes_{C^\infty(\MM_0)}\Gamma(\MM_0,\Ber(\MM))\ra\RR$$
that associates to a function $f=f_\even\otimes f_\odd$ and a Berezin measure $\mu=\mu_\odd\otimes\mu_\even$ (where $f_\even\in\Fun(\MM_\even)$, $f_\odd\in \Fun(\MM_\odd)$, $\mu_\even\in\Gamma(\MM_0,\mr{Ber}(\MM_\even))$, $\mu_\odd\in\Gamma(\MM_0,\Lambda^{\mr{rk} \MM_\odd}\MM_\odd)$) the integral
$$\int_\MM f\mu:=\int_{\MM_\even}f_\even <f_\odd,\mu_\odd> \mu_\even$$
Here $<\bt,\bt>:\Lambda^\bt \MM_\odd^*\otimes \Lambda^\bt \MM_\odd\ra C^\infty(\MM_0)$ is the fiberwise canonical pairing.
We will use the notation $$\mr{Mes}(\MM):=\Fun(\MM)\otimes_{C^\infty(\MM_0)}\Gamma(\MM_0,\Ber(\MM))$$ for the space of integration measures which are allowed to be non-constant in the fiber of $\MM$ (as a bundle over $\MM_0$), so that the integral is a map $\int_\MM: \mr{Mes}(\MM)\ra\RR$.

For the Berezinian bundle of odd cotangent bundle we have
\be\Ber(T^*[-1]\MM)=\Ber(\MM)\otimes\Ber(\MM)\label{Berezinian of odd cotangent bundle}\ee

\subsection{$P$-manifolds}
On the odd cotangent bundle $T^*[-1]\NN$ of a graded manifold $\NN$ one has a natural 2-form of grade -1
\be\omega=\sum_i(-1)^{\e(x^i)}\delta x^i\wedge\delta \xi_i\label{canonical odd symplectic form}\ee where $(x^i)$ are local coordinates on $\NN$ and $(\xi_i)$ are the conjugate coordinates in fiber (``grade'' here means total degree minus the de Rham degree of the form).
\begin{Def} Graded manifold $\MM$ is called a $P$-manifold (or ``odd symplectic'', or ``anti-symplectic'' manifold) if it is endowed with a 2-form $\omega$ of grade $-1$ and $\MM$ can be covered by a system of open neighbourhoods $(U_\alpha)$, each equipped with a Darboux coordinate system $(x^i_{(\alpha)},\xi_{(\alpha)i})$, so that in each $U_\alpha$ the form $\omega$ is canonical  (\ref{canonical odd symplectic form}) and the transition maps $\phi_{\alpha\beta}$ are symplectomorphisms, i.e.  $\phi_{\alpha\beta}^*\omega=\omega$.
\end{Def}
By definition, each $P$-manifold locally looks like $T^*[-1]\NN$ for some $\NN$ (and $\NN$ can always be chosen to be purely even). Actually, a stronger global statement holds:
\begin{thm}[A. Schwarz, \cite{Schwarz}]Each $P$-manifold $\MM$ is equivalent (symplectomorphic) to the odd cotangent bundle $T^*[-1]\NN$ for some $\NN$ which can be chosen to be purely even.
\end{thm}
Note that this statement does not have an analog in ordinary (even) symplectic geometry where the problem of classifying symplectic manifolds is much harder.

The ring of functions on a $P$-manifold $\MM$ has the structure of Gerstenhaber algebra where the anti-bracket in Darboux coordinates is written as
$$\{f,g\}=\sum_i f\left(\frac{\overleftarrow\dd}{\dd
x^i}\frac{\overrightarrow\dd}{\dd \xi_i}-\frac{\overleftarrow\dd}{\dd
\xi_i}\frac{\overrightarrow\dd}{\dd x^i}\right)g $$
for $f,g\in\Fun(\MM)$. Suppose in some (non necessarily Darboux) coordinate system $(z^a)$ the odd symplectic form is
$$\omega=\frac{1}{2}\omega_{ab}(z)\,\delta z^a\wedge\delta z^b$$
then the corresponding anti-bracket is
\be\{f,g\}=f\left(\sum_{a,b}(-1)^{\e(z^a)}\omega^{ab}(z)\frac{\overleftarrow\dd}{\dd
z^a}\frac{\overrightarrow\dd}{\dd z^b}\right)g\label{BV bracket}\ee

Another important difference between $P$-manifolds and ordinary symplectic manifolds is that the former do not possess a canonical measure, while for the latter one has the Liouville measure $\frac{1}{(\dim \MM/2)!}\omega^{\dim \MM/2}$ canonically constructed from symplectic form $\omega$. In odd symplectic case this expression does not make sense since the measure is not a differential form, and also because anti-symplectic form is always nilpotent: $\omega\wedge\omega=0$.

\begin{Def}
A $QP$-manifold is a $P$-manifold  endowed with a cohomological vector field $Q$, preserving the odd symplectic form, i.e. Lie derivative of $\omega$ along $Q$ vanishes.
\end{Def}
Such $Q$ is necessarily Hamiltonian (see \cite{AKSZ}), i.e. there exists a function $S^0\in \Fun(\MM)$ of degree $\e(S^0)=0$ such that $Q=\hat{S}^0=\{S^0,\bt\}$.

Borrowing the terminology of classical (Hamiltonian) mechanics, we call the pull-back action of a symplectomorphism on functions $\phi^*:\Fun(\MM)\ra\Fun(\MM)$ the ``canonical transformation''. Canonical transformations are automorphisms of Gerstenhaber algebra $\Fun(\MM)$, i.e. they preserve grading, multiplication and anti-bracket. Infinitesimal canonical transformation acts on functions as
\be f\mapsto f+\{f,R\}\label{canonical transformation of a scalar}\ee
where the function $R\in\Fun(\MM)$ of degree $\e(R)=-1$ is the generator of infinitesimal canonical transformation.

Lagrnagian submanifolds of a $P$-manifold are introduced just as in ordinary symplectic geometry, i.e. a submanifold $\LL\subset\MM$ is called symplectic if $\omega|_\LL=0$ and $\dim \LL=\frac{1}{2}\dim\MM$.

There are two standard constructions of Lagrangian submanifolds that will be important for us: small deformation of a given Lagrangian submanifold defined by a gauge fixing fermion, and the construction of odd conormal bundle. The first construction is as follows. Suppose $\MM$ has the form of odd cotangent bundle $\MM=T^*[-1]\NN$ with standard $P$-structure and coordinates $x^i$ in base and $\xi_i$ in fiber. Then $\NN\subset\MM$ is itself a Lagrangian submanifold and its deformation in the class of Lagrangian submanifolds is defined by a function $\Psi(x)\in\Fun(\NN)$ of degree $\e(\Psi)=-1$ (the ``gauge fixing fermion'') as
\be\LL_\Psi:=\left\{(x,\xi)|\xi_i=-\frac{\dd}{\dd x^i}\Psi(x)\right\}\label{Lagrangian submanifold from gauge fixing fermion}\ee
Observe that in coordinates $({x'}^i,\xi'_i)$ related to coordinates $(x^i,\xi_i)$ by canonical transformation ${x'}^i=x^i+\{x^i,-\Psi\}=x^i,\, \xi'_i=\xi_i+\{\xi_i,-\Psi\}$ the Lagrangian submanifold $\LL_\Psi$ is given simply as $\xi'=0$.

The second construction is the following. Suppose again $\MM=T^*[-1]\NN$ and let $\KK\subset\NN$ be a submanifold of the base. Then the odd conormal bundle $N^*[-1]\KK\subset \MM$ is defined as
\be N^*[-1]\KK:=\left\{(x,\xi)|x\in\KK,\xi\bot T_x \KK\right\}\label{conormal bundle}\ee
So $N^*[-1]\KK$ is a bundle over $\KK$ where the fiber over point $x\in\KK$ is the subspace of covectors $\xi\in T^*_x[-1]\NN$, orthogonal to the tangents space to $\KK$ at point $x$.
These two constructions are in a sense enough to describe all Lagrangian submanifolds due to the following theorem.

\begin{thm}[A. Schwarz, \cite{Schwarz}] Every Lagrangian submanifold $\LL\subset T^*[-1]\NN$ may be continuously deformed to submanifold of type (\ref{conormal bundle}) for some $\KK\subset\NN$.
\end{thm}
If $\LL\subset\MM$ is a Lagrangian submanifold then locally, in neighbourhood of $\LL$, $\MM$ is equivalent to $T^*[-1]\LL$.

\subsection{$SP$-manifolds}
\label{SP manifolds}
There are two approaches to define the ``BV manifold''. The first approach is due to A. Schwarz \cite{Schwarz}: a BV manifold is an ``$SP$-manifold'' $\MM$, i.e. odd symplectic manifold endowed a measure (satisfying certain consistency condition), then from this data one constructs BV Laplacian on functions, thus making $\Fun(\MM)$ a BV algebra. The second approach, explored in \cite{Khudaverdian},\cite{Severa}, will be briefly sketched in section \ref{integrals over Lagrangian submanifolds}: here the BV manifold is just a $P$-manifold (without the choice of measure), and the BV Laplacian is canonical in every system of Darboux coordinates, but instead of functions it acts on scalar densities of weight $1/2$ (the ``half-densities'').

\begin{Def}
$SP$-manifold is a $P$-manifold $\MM$ endowed with a measure $\mu\in\Mes(\MM)$ consistent with the odd symplectic form $\omega$. Consistency means here that $\MM$ may be covered by a system of open neighbourhoods $(U_\alpha)$ such that in Darboux coordinates $(x_{(\alpha)}^i,\xi_{(\alpha)i})$ on each $U_\alpha$ the measure coincides with the coordinate Berezin measure, i.e.
$\mu=\prod_i\DD x_{(\alpha)}^i \DD \xi_{(\alpha)i}$ (and the form $\omega$ is canonical (\ref{canonical odd symplectic form})), and all transition maps are unimodular symplectomorphisms, i.e.  $\phi_{\alpha\beta}^*\omega=\omega$ and the Jacobians equal 1:
$$\mr{Jac}(\phi_{\alpha\beta})=\frac{\dd (x_{(\alpha)},\xi_{(\alpha)})}{\dd (x_{(\beta)},\xi_{(\beta)})}=1$$
\end{Def}
For an arbitrary coordinate system $(z^a)$ we call the measure $\mu_\mr{coord}=\prod_a \DD z^a$ the coordinate measure (associated to the given coordinate system). If $\mu=\rho(z)\mu_\mr{coord}$ is an arbitrary measure, we call $\rho$ the density of measure (in coordinates $(z^a)$). Under the change of coordinates $\phi$ density transforms as  $\rho\mapsto \rho'=\mr{Jac}(\phi)\cdot\phi^*\rho$, i.e. $\rho(z)\mapsto\rho'(z')=\frac{\dd z}{\dd z'}\;\rho(z)$.

Measure $\mu$ on $\MM$ defines the divergence for vector fields $\mr{div}_\mu: \mr{Vect}(\MM)\ra\Fun(\MM)$ by relation
$$\int_\MM v(f)\mu=\int_\MM \mr{div}_\mu v\cdot f\mu$$ for a given vector field $v\in\mr{Vect}(\MM)$ and arbitrary function $f$. If in local coordinates $(z^a)$ on $\MM$ the measure $\mu$ has density $\rho(z)$ then the divergence acts as
$$\mr{div}_\mu:\quad \sum_a v^a(z)\frac{\dd}{\dd z^a}\mapsto \sum_a (-1)^{(\e(v)+1)\cdot \e(z^a)} \frac{1}{\rho(z)}\,\frac{\dd}{\dd z^a}\left(\rho(z) v^a(z)\right)$$

On a $SP$-manifold $\MM$ from the measure $\mu$ one constructs the BV Laplacian $\Delta_\mu: \Fun(\MM)\ra\Fun(\MM)$ that sends a function $f$ to (up to a factor) the divergence of the Hamiltonian vector field $\hat f=\{f,\bt\}$:
$$\Delta_\mu f=(-1)^{\e(f)}\frac{1}{2}\mr{div}_\mu\hat f=(-1)^{\e(f)}\frac{1}{2}\mr{div}_\mu \{f,\bt\}$$
In local Darboux coordinates $(x^i,\xi_i)$ where measure $\mu$ has unit density (these coordinates exist by definition of $SP$-manifold), operator $\Delta_\mu$ has canonical form:
$$\Delta_\mu=\sum_i(-1)^{\e(x^i)}\frac{\dd}{\dd x^i}\frac{\dd}{\dd \xi_i}$$
which immediately implies $\Delta_\mu^2=0$. Hence $\Delta_\mu$ endows $\Fun(\MM)$ with the structure of BV algebra.
In general coordinate system $(z^a)$  the BV operator is
$$\Delta_\mu: f(z)\mapsto \frac{1}{2\rho(z)}\,\frac{\dd}{\dd z^a}\rho(z)\,\omega^{ab}(z)\frac{\dd}{\dd z^b}f(z)$$
The anti-bracket, defined on functions by (\ref{antibracket from BV Laplacian}) coincides with the anti-bracket constructed from odd symplectic form and does not depend on the choice of measure $\mu$.

If one multiplies the measure $\mu$ by a function $g\in\Fun(\MM)$ of degree $\e(g)=0$ the BV Laplacian changes as
$$\Delta_\mu\mapsto\Delta_{g\mu}=\Delta_\mu+\{\frac{1}{2}\log g,\bt\}$$
The condition that new BV Laplacian is nilpotent is a nontrivial condition on function $g$:
$$\Delta_{g\mu}^2=0\Leftrightarrow \Delta_\mu g^{1/2}=0$$

\subsection{Integrals over Lagrangian submanifolds}
\label{integrals over Lagrangian submanifolds}
Let $\MM$ be a $SP$-manifold. Then on Lagrangian submanifolds $\LL\subset\MM$ one has the induced measure $\sqrt{\mu}|_\LL$. Let $(x_i,\xi_i)$ be the coordinate system on $\MM$ where $\LL$ is given by $\xi_i=0$, and let $\rho$ be the density of measure $\mu$ in these coordinates, i.e. $\mu=\rho(x,\xi)\prod_i\DD x^i\DD\xi_i\in\mr{Mes}(\MM)$. Then the induced measure on $\LL$ may be locally written as $\sqrt{\mu}|_\LL=\rho(x,0)^{1/2}\prod_i\DD x^i\in\mr{Mes}(\LL)$. Due to the property (\ref{Berezinian of odd cotangent bundle}) of Berezinian bundle, taking square root of measure on an odd-symplectic manifold is a natural operation.

The following odd symplectic version of Stokes' theorem is the key statement in BV formalism.
\begin{thm}[Batalin-Vilkovisky \cite{BV81}, A. Schwarz \cite{Schwarz}]
\begin{enumerate}
\item Let $\LL_1,\LL_2\subset\MM$ be two Lagrangian submanifolds in $SP$-manifold $\MM$, such that $\LL_1$ can be continuously deformed to $\LL_2$ in the class of Lagrangian submanifolds. Let $f\in\Fun(\MM)$ be a function satisfying $\Delta_\mu f=0$ (i.e. $f$ is BV cocycle). Then
\be \int_{\LL_1} f\cdot \sqrt{\mu}|_{\LL_1} = \int_{\LL_2} f\cdot \sqrt{\mu}|_{\LL_2}\label{BV gauge independence}\ee
\item Let $\LL\subset\MM$ be a Lagrangian submanifold in $SP$-manifold $\MM$ and let $f\in\Fun(\MM)$ be a function. Then
\be\int_\LL \Delta_\mu f\cdot \sqrt{\mu}|_{\LL}=0\label{int of BV coboundary}\ee
\end{enumerate}
\end{thm}
Actually, in \cite{Schwarz} a stronger version of (\ref{BV gauge independence}) is proven: one can relax the requirement that $\LL_1$ can be deformed into $\LL_2$ in the class of Lagrangian submanifolds to the condition that projections of $\LL_1$ and $\LL_2$ to the body $\MM_0$ of $\MM$ are homologous.

An alternative to Schwarz's approach to BV manifolds is the following. Let now $\MM$ be a $P$-manifold (without additional $S$-structure). Define the BV Laplacian in any Darboux coordinate system as the canonical one:
$$\Delta\chi=\sum_i (-1)^{\e(x^i)}\frac{\dd}{\dd x^i}\frac{\dd}{\dd\xi_i}\chi$$
Then we cannot say that BV Laplacian acts on functions, since $\Delta f$ would not transform as a scalar under canonical transformations. Instead we have to say that $\Delta$ acts on scalar densities of weight $1/2$ (the half-densities), i.e.
$$\chi\in\mr{HalfDens}(\MM):=\Fun(\MM)\otimes \Gamma(\MM_0,\Ber(\MM)^{\otimes\frac{1}{2}})$$
Canonical transformation $\phi$ acts on half-densities as $\chi\mapsto\mr{Jac}(\phi)^{1/2}\cdot\phi^*\chi$. An infinitesimal canonical transformation with generator $R$ acts on half-density $\chi$ as
\be\chi\mapsto \chi+\{\chi,R\}+\chi\Delta R\label{canonical transformation for semi-density}\ee
An integral of a half-density over Lagrangian submanifold is understood in the following sense. Let  $(x^i,\xi_i)$ be Darboux coordinates on $\MM$ where $\LL$ is given by $\xi_i=0$ and let $\chi(x,\xi)$ be the half-density in this coordinate system. Then integral of $\chi$ over $\LL$ is understood as
$$\int_\LL \chi:=\int_\LL \chi(x,0)\prod_i \DD x^i$$
Statements (\ref{BV gauge independence},\ref{int of BV coboundary}) in the language of half-densities are:
$$\int_{\LL_1}\chi=\int_{\LL_2}\chi$$
for any half-density $\chi\in\mr{HalfDens}(\MM)$ satisfying $\Delta\chi=0$, and
$$\int_\LL\Delta\chi=0$$
for any half-density $\chi\in\mr{HalfDens}(\MM)$.

So in Schwarz's picture the BV Laplacian acts on functions on $SP$-manifold and is not canonical, but depends on the choice of measure, while in the picture of half-densities the BV Laplacian acts on half-densities on a $P$-manifold and has canonical form in every Darboux coordinate system. The correspondence between the two pictures is as follows: there is a non-canonical isomorphism between functions and half-densities on a $P$-manifold:
\be \Fun(\MM)\cong \mr{HalfDens(\MM)} \label{fun - half-dens}\ee which uses reference $S$-structure and sends $f\in\Fun(\MM)$ to $\chi=f\sqrt{\rho}\in \mr{HalfDens(\MM)}$ where $\rho$ is the density of reference measure. The obvious advantage of Schwarz's picture of $SP$-manifolds is that it defines a BV algebra structure on $\Fun(\MM)$ while the advantage of picture of half-densities is that is more canonical. A very elegant homological algebra construction of space of half-densities together with BV Laplacian on it is proposed in \cite{Severa}.

In most part of present work we will be dealing with the situation where the space of fields is just a graded vector space (with $P$-structure) and Lagrangian submanifolds are vector subspaces. Here we have the preferred measure (canonical up to constant factor) --- the Lebesgue measure, and so isomorphism (\ref{fun - half-dens}) becomes canonical (up to constants). This will allow us to forget the subtle difference between functions and half-densities.

\subsection{Master equation}
From now on we need to allow functions on graded manifolds to be formal power series in ``Planck constant'' $\hbar$ (an infinitesimal formal parameter).

Let $\MM$ be a $SP$-manifold. Consider functions on $\MM$ of form $e^{S/\hbar}$ where $S=S^0+\hbar S^1+\hbar^2 S^2+\cdots$ is a function on $\MM$, regular in $\hbar$.
We say that $S$ satisfies the quantum master equation if
\be \Delta e^{S/\hbar}=0 \label{QME} \ee
(where $\Delta$ denotes the BV Laplacian, associated to the $SP$-structure) or equivalently
\be \frac{1}{2}\{S,S\}+\hbar \Delta S=0 \label{QME MC}\ee
(note that this is just the Maurer-Cartan equation on $S/\hbar$ for the dg Lie algebra structure $(\Delta,\{\bt,\bt\})$ on $\Fun(\MM)$). A solution of quantum master equation is called the BV action (or master action). In terms of decomposition in Planck constant $S=S^0+\hbar S^1+\hbar^2 S^2+\cdots$ the quantum master equation (\ref{QME MC}) is equivalent to the system of equations
\begin{eqnarray}
\{S^0,S^0\}&=&0\label{CME}\\
\{S^0,S^1\}+\Delta S^0&=&0\label{QME1}\\
\{S^0,S^2\}+\frac{1}{2}\{S^1,S^1\}+\Delta S^1&=&0\label{QME2}\\
\vdots\nonumber
\end{eqnarray}
Equation (\ref{CME}) is called the classical master equation on $S^0$. Hamiltonian vector field
\be Q_{\MM}:=\widehat{S^0}=\{S^0,\bt\}\label{Q on BV fields}\ee
generated by a solution of classical master equation is called the BRST operator on $\MM$ (this is a BRST operator on BV manifold unlike the BRST operator that we will encounter in section \ref{BRST method}). It satisfies $Q_{\MM}^2=0$, i.e. it is a cohomological vestor field on $\MM$.

Given a solution of classical master equation $S^0$ we have a series of obstructions for extending it to a solution of quantum master equation, as seen from (\ref{QME1}), (\ref{QME2}), \ldots Namely, for equation (\ref{QME1}) on $S^1$ to be solvable, we need the class $[\Delta S^0]\in H^1_{Q_{\MM}}(\Fun(\MM))$ in cohomology of BRST operator to vanish. Then for equation (\ref{QME2}) to be solvable, we need the class $[\Delta S^1+\frac{1}{2}\{S^1,S^1\}]\in H^1_{Q_{\MM}}(\Fun(\MM))$ to vanish, and so on. In particular, if $H^1_{Q_{\MM}}(\Fun(\MM))=0$, all the obstructions vanish automatically and hence $S^0$ can be extended to a solution of QME.

An important role in BV formalism is played by another nilpotent operator constructed from a solution of quantum master equation:
$$\delta_\mr{BV}=\{S,\bt\}+\hbar \Delta=Q_{\MM}+\hbar (\widehat{S^1}+\Delta)+\hbar^2\widehat{S^2}+\hbar^3\widehat{S^3}+\cdots$$
The operator $\delta_\mr{BV}$ is a deformation of $Q_\MM$ and it is related to the operator $\hbar \Delta$ by a similarity transformation:
$$\delta_\mr{BV}=\hbar e^{-S/\hbar}\Delta e^{S/\hbar}$$
Unlike $Q_\MM$, the operator $\delta_\mr{BV}$ is of order 2 and hence cannot be geometrically interpreted as a vector field on $\MM$.

If $S$ is a solution of quantum master equation and $R\in\Fun(\MM)$ is a function of degree $\e(R)=-1$ and regular in $\hbar$, then
\be S'=S+\{S,R\}+\hbar\Delta R=S+\delta_{BV}R\ee
is also a solution of quantum master equation in lowest order in $R$, since one can equivalently write
\be e^{S'/\hbar}=e^{S/\hbar}+\Delta (e^{S/\hbar} R)\label{canonical transformation for exp(S)}\ee
The transformation (the ``canonical transformation of action'')
\be S\mapsto S'=S+\{S,R\}+\hbar\Delta R\label{canonical transformation for action}\ee
can be understood as the infinitesimal canonical transformation of function $S$ with generator $R$, plus the effect of changing the reference measure (the last term in \ref{canonical transformation for action}). I.e. the fact that $\Delta_\mu e^{S'/\hbar}=0$ (where $\mu$ is the original chosen measure on $\MM$) can be reinterpreted as $\Delta_{\mu_R} e^{S_R/\hbar} = 0$ where $S_R=S+\{S,R\}$ is the action transformed as a function and $\mu_R=e^{2\Delta R}\mu$ is the transformed measure.

In the alternative picture one says that the original manifold did not carry $S$-structure, BV Laplacian is canonical and independent of coordinate transformations, and then $S$ is not a function, but rather a ``log-half-density'' (meaning that $\chi=e^{S/\hbar}$ is a half-density; to make sense of this definition and be able to take exponentials, one still has to use isomorphism (\ref{fun - half-dens}) for some reference measure). Then one recognizes in (\ref{canonical transformation for action}) the transformation rule (\ref{canonical transformation for semi-density}) for half-densities.


Due to Batalin-Vilkovisky theorem (\ref{BV gauge independence}), for any chosen Lagrangian submanifold $\LL\subset\MM$, the integral $\int_{\LL} e^{S/\hbar}\cdot \sqrt{\mu}|_{\LL}$ is invariant under canonical transformations of action (this is obvious from formula (\ref{canonical transformation for exp(S)})).
Two solutions of quantum master equation related by a canonical transformation are considered physically equivalent BV actions.

\textbf{Gauge transformations in BV formalism.}
Classical part of BV action $S^0$ generates its own gauge symmetries in the following sense. Hamiltonian vector fields generated by derivatives of $S^0$
$$v_a=\{\frac{\dd}{\dd z^a}S^0,\bt\}$$ annihilate the classical part of action $v_a S^0=0$, as a consequence of classical master equation (\ref{CME}). Derivatives $R^0_a=\frac{\dd}{\dd z^a}S^0$ along coordinates $z^a$ of degree $\e(z^a)=+1$ may be interpreted as generators of canonical transformations, preserving $S^0$. Analogous statement holds for the full BV action $S$: infinitesimal canonical transformation with generator $R_a=\frac{\dd}{\dd z^a}S$ preserves $S$, since $S\mapsto S+\delta_\mr{BV}R_a$
and $\delta_\mr{BV}R_a=0$ as implied by (\ref{QME MC}).

\section{Preliminaries: gauge fixing}
\label{gauge fixing methods}
In sections \ref{FP method}, \ref{BRST method}, \ref{BV method} we give a brief review of three main methods of solving the problem of gauge fixing (i.e. the problem of constructing the perturbatively well-defined functional integral for a gauge theory): Faddeev-Popov method, BRST and BV methods. In section \ref{BF theory} we briefly discuss the gauge fixing for topological $BF$ theory in BV formalism.

We will call the $\ZZ$-grading on spaces of fields which will arise here the ``ghost number'' and denote it as $\gh$. We consider only purely bosonic gauge theories here and hence we need not introduce the  more complicated  $\ZZ\oplus\ZZ_2$-grading on spaces of fields.
\subsection{Gauge fixing: Faddeev-Popov method}
\label{FP method}
Let $\FF_{cl}$ be the manifold of classical fields with coordinates $(x^i)$ (classical fields have ghost number 0). Let $\GG$ be a Lie group of gauge symmetry acting on $\FF_{cl}$
$$\GG\circlearrowright\FF_{cl}$$
i.e. we have a group homomorphism from $\GG$ to group of diffeomorphisms of  $\FF_{cl}$: $$\kappa:\GG\ra\mr{Diff}(\FF_{cl})$$
where we suppose that $\kappa$ is injective (i.e. the action of $\GG$ on $\FF_{cl}$ is irreducible). The group action induces the Lie algebra homomorphism from Lie algebra $\A$ of $\GG$ to vector fields on $\FF_{cl}$:
\be\hat\kappa: e_\alpha\mapsto v_\alpha=v^i_\alpha(x) \frac{\dd}{\dd x^i}\in\mr{Vect}(\FF_{cl})\label{small gauge transformation}\ee
where $(e_\alpha)$ is a basis in $\A$. Finally, let $S_{cl}\in\Fun(\FF_{cl})$ be a $\GG$-invariant function, the classical action of gauge theory. So $S_{cl}$ can be viewed as a function on the orbit space: $S_{cl}\in\Fun(\FF_{cl}/\GG)$. The problem of gauge fixing is to make sense of integral
\be\mr{``}\;\int_{\FF_{cl}}e^{S/\hbar}\;\;\mr{"}\label{int over Fcl}\ee
so that it could be computed by stationary phase method. Gauge symmetry of $S_{cl}$ implies that the Hessian of action in any stationary point is degenerate and the perturbative expansion is not well-defined. A natural way to understand integral (\ref{int over Fcl}) would be as an integral over coset
\be\int_{\FF_{cl}/\GG}e^{S/\hbar}\label{int over coset}\ee
But this answer is not satisfactory from physical point of view where $\FF_{cl}$ and $\GG$ are usually spaces of sections of certain bundles on a manifold (the space-time), and one would like to understand (\ref{int over Fcl}) also as an integral over space of sections of some bundle on space-time.

The main idea of Faddeev-Popov method is the following. Choose a function $\phi:\FF_{cl}\ra\A$ such that each orbit of $\GG$-action intersects the surface $\phi^{-1}(0)\subset\FF_{cl}$ only once (i.e. $\phi^{-1}(0)\sim \FF_{cl}/\GG$). The integral over coset (\ref{int over coset}) is then rewritten as
\be\int_{\FF_{cl}}e^{S_{cl}/\hbar}\prod_\alpha\delta(\phi^\alpha(x))\cdot\det\left(v^i_\alpha(x)\frac{\dd}{\dd x^i}\phi^\beta(x)\right)\prod_i \DD x^i \label{FP}\ee
and the important statement is that this expression is invariant under deformations of function $\phi$. Delta functions in (\ref{FP}) localize the integral to the surface $\phi^{-1}(0)\subset\FF_{cl}$ while the determinant  ensures the invariance under deformations of $\phi$. Finally, introducing additional Grassman fields $c^\alpha$, $\bar{c}_\alpha$ with ghost numbers +1 and -1 respectively and the Lagrangian multiplier $\lambda_\alpha$ with ghost number 0, we can rewrite (\ref{FP}) as the integral
\be \int_{\FF_{FP}}e^{S_{FP}/\hbar}\prod_i \DD x^i\prod_{\alpha}\DD c^\alpha \DD\bar{c}_\alpha \DD\lambda_\alpha \label{FP1}\ee
over extended space of fields
\be\FF_{FP}=\FF_{cl}\oplus \A[1]\oplus\A^*[-1]\oplus\A^*\label{FP fields}
\ee
with coordinates $(x^i,c^\alpha,\bar{c}_\alpha,\lambda_\alpha)$ for the Faddeev-Popov action
\be S_{FP}(x,c,\bar{c},\lambda):=S_{cl}(x)+\lambda_\alpha \phi^\alpha(x)+\bar{c}_\alpha \frac{\dd \phi^\alpha(x)}{\dd x^i} v^i_\beta(x) c^\beta \label{FP action}\ee
Integral (\ref{FP1}) is perturbatively well-defined and does not depent on choice of gauge. Hence it solves the problem of gauge fixing for the integral (\ref{int over Fcl}).

\textbf{Remark.} We were implicitly assuming that the coordinate measure on $\FF_{cl}$ is invariant under $\GG$-action, or equivalently
\be \mr{div}\; v_\alpha=0 \label{div v_alpha=0}\ee

\textbf{Example: Yang-Mills theory in Lorenz gauge.} Let $M$ be a Riemannian manifold (the space-time), let $G$ be a gauge group and $\g$ its Lie algebra. Then for the Yang-Mills theory we set $\FF_{cl}=\g\otimes\Omega^1(M)$ --- the space of connections on trivial $G$-bundle on $M$. The group of gauge symmetry is $\GG=G^M$, i.e. the group of fiberwise rotations of the principal bundle. Its Lie algebra $\A=\g\otimes\Omega^0(M)$ acts on connections $A\in\FF_{cl}$ by the usual gauge trasformations $A\mapsto A+d_A\alpha$ where $\alpha\in\A$ is the generator of gauge transformation and $d_A=d+[A,\bt]$ is the covariant differential. Classical Yang-Mills action is $S_{cl}(A)=\frac{1}{4}\tr_\g\int_M *F_A\wedge F_A$ where $*:\Omega^\bt(M)\ra\Omega^{\dim(M)-\bt}(M)$ is the Hodge star, $F_A=dA+A\wedge A$ is the curvature of connection $A$, and the trace $\tr_\g$ is evaluated in the adjoint representation of $\g$. Lorenz gauge for Yang-Mills theory corresponds to choosing the gauge fixing function $\phi$ to be the Hodge operator $\phi=d^*:\g\otimes\Omega^1(M)\ra\g\otimes\Omega^0(M)$. This choice of $\phi$ produces  the Faddev-Popov action
$$S_{FP}=S_{cl}(A)+\tr\int_M \lambda\wedge d^*A+\tr\int_M \bar{c}\wedge d^*d_A c$$
where $c$, $\bar{c}$ and $\lambda$ are $\g$-valued 0-, $\dim(M)$- and $\dim(M)$-forms respectively, with ghost numbers +1, -1 and 0. We used the pairing $\tr\int\bt\wedge\bt:\A^{\otimes 2}\ra\RR$ here to identify $\A^*$ with $\A$ in (\ref{FP fields}). Without using this identification we should have written
\be S_{FP}=S_{cl}(A)+<\lambda,d^*A>+<\bar{c},d^*d_A c>\label{Yang-Mills FP action}\ee
and understand $c$ as a $\g$-valued 0-form and $\bar{c}$ and $\lambda$ as $\g^*$-valued 0-currents. Here $<\bt,\bt>$ is the canonical pairing of $\g$-valued forms and $\g^*$-valued currents.

\subsection{Gauge fixing: BRST formalism}
\label{BRST method}
The framework of BRST formalism is as follows. One embeds the manifold of classical fields $\FF_{cl}$ into 0-th degree of the $\ZZ$-graded manifold of BRST fields $\FF_\mr{BRST}$. We call grading on $\FF_\mr{BRST}$ the ghost number $\gh$. In addition, $\FF_\mr{BRST}$ is a $Q$-manifold in terminology of \cite{AKSZ}, i.e. the algebra of functions $\Fun(\FF_\mr{BRST})$ is endowed the derivation (the ``BRST operator'') $Q$, satisfying $Q^2=0$, $\gh(Q)=1$ (in the language of super-geometry $Q$ is called the cohomological vector field on $\FF_\mr{BRST}$). Also, $\FF_\mr{BRST}$ is endowed with  a $Q$-invariant measure $\mu$, i.e. $\mr{div}_\mu Q=0$. This implies that the integral of a BRST coboundary vanishes:
\be\int_{\FF_\mr{BRST}}(Qf)\mu=0\label{BRST Stokes thm}\ee
for every $f\in\Fun(\FF_\mr{BRST})$. There are two more conditions that relate the data of BRST formalism $(\FF_\mr{BRST},Q)$ to the data of classical gauge theory:
\begin{itemize}
\item classical action is a BRST cocycle: $$QS_{cl}=0$$
\item $\FF_\mr{BRST}$ is a resolution of the space of orbits of gauge symmetry $\FF_{cl}/\GG$ in the following sense: cohomology of $Q$ in ghost number 0 is isomorphic to ring of functions on $\FF_{cl}/\GG$:
    $$H^0_Q(\Fun(\FF_\mr{BRST}))\sim \Fun(\FF_{cl}/\GG)$$
\end{itemize}
For the situation of section \ref{FP method}, i.e. for a gauge theory with simple (meaning irreducible, closed) gauge symmetry, the minimal BRST resolution is constructed as \be\FF_\mr{minBRST}=\FF_{cl}\oplus\A[1]\label{minBRST fields}\ee
In coordinates $(x^i,c^\alpha)$ the BRST operator is
\be Q_\mr{min}=-c^\alpha v^i_\alpha(x)\frac{\dd}{\dd x^i}+\frac{1}{2}f^{\alpha}_{\beta\gamma}c^\beta c^\gamma\frac{\dd}{\dd c^{\alpha}}\label{Qmin}\ee
where  $f^\alpha_{\beta\gamma}$ are structure constants of the Lie algebra of gauge transformations $\A$. The condition $Q_\mr{min}S_{cl}=0$ is equivalent to the $\GG$-invariance of $S_{cl}$. The condition $Q_\mr{min}^2=0$ encodes the Jacobi identity for structure constants $f^\alpha_{\beta\gamma}$ together with the condition that (\ref{small gauge transformation}) is a homomorphism. Thus, in BRST formalism all the information on the Lie algebra of gauge symmetry $\A$ and its action on $\FF_{cl}$  is encoded in the (minimal) BRST operator $Q_\mr{min}$. If (\ref{div v_alpha=0}) holds and $\A$ is unimodular (which means $f^\alpha_{\alpha\gamma}=0$), we have $\mr{div} Q=0$ (with respect to the coordinate measure), and thus we can take $\mu=\mu_\mr{coord}=\prod_i \DD x^i \prod_\alpha \DD c^\alpha$.

Gauge fixing in BRST formalism is done as follows. Choose some function $\Psi\in\Fun(\FF_\mr{BRST})$ with $\gh(\Psi)=-1$ (the gauge fixing fermion). Then formally, due to (\ref{BRST Stokes thm}), we have
$$\int_{\FF_\mr{BRST}}e^{\frac{1}{\hbar}S_{cl}}\mu=\int_{\FF_\mr{BRST}}e^{\frac{1}{\hbar}(S_{cl}+Q\Psi)}\mu$$
and the right part does not depend on $\Psi$ (if it is defined). So in BRST formalism one has to take some gauge fixing fermion $\Psi$ such that the integral
\be\int_{\FF_\mr{BRST}}e^{\frac{1}{\hbar}(S_{cl}+Q\Psi)}\mu\label{BRST gauge fixed}\ee
is perturbatively well-defined and declare it to be the value of ill-defined expression (\ref{int over Fcl}).

\textbf{Minimal vs. full BRST (auxiliary fields).} Condition of existence of such $\Psi$ that (\ref{BRST gauge fixed}) is well-defined, forces us to extend $(\FF_\mr{minBRST},Q_\mr{min})$ to some larger BRST manifold $(\FF_\mr{fullBRST},Q_\mr{full})$. Indeed, in $\Fun(\FF_\mr{minBRST})$ there are no functions of ghost number $-1$ at all. To solve this problem, we can extend the space of BRST fields without changing the cohomology of $Q$:
\be\FF_\mr{minBRST}\mapsto \FF_\mr{fullBRST}=\FF_\mr{minBRST}\oplus T[1]V_\mr{Aux}\label{BRST space extension}\ee
where $V_\mr{Aux}$ is some graded vector space and $T[1]V_\mr{Aux}=V_\mr{Aux}\oplus V_\mr{Aux}[1]$ is its odd tangent bundle. Since $\Fun(T[1]V_\mr{Aux})$ is identified with the de Rham algebra $\Omega^\bt(V_\mr{Aux})$, we have a natural cohomological vector field on $T[1]V_\mr{Aux}$ --- the de Rham differential $d_{V_\mr{Aux}}=\lambda^I\frac{\dd}{\dd \bar{c}^I}$. Here $(\bar{c}^I,\lambda^I)$ are coordinates in the base and fiber of $T[1]V_\mr{Aux}$ respectively. We extend the minimal BRST operator as
\be Q_\mr{min}\mapsto Q_\mr{full}=Q_\mr{min}+d_{V_\mr{Aux}}\label{Q extension}\ee
Clearly, the cohomology of BRST operator after this extension gets tensor multiplied by de Rham cohomology of $V_\mr{Aux}$ (which is contractible), and hence do not change at all.
We extend the integration measure trivially as $\mu_\mr{min}\mapsto\mu_\mr{full}=\mu_\mr{min} \mu_\mr{Aux}$ where $\mu_\mr{Aux}=\prod_I \DD\bar{c}^I \DD \lambda^I$ is the Lebesgue measure on $V_\mr{Aux}$ (it is defined up to constant factor).

For the theory with irreducible gauge symmetry we can take $V_\mr{Aux}=\A^*[-1]$. Then the full space of BRST fields is $$\FF_\mr{fullBRST}=\FF_{cl}\oplus\A[1]\oplus T[1](\A^*[-1])$$
(which indeed coincides with (\ref{FP fields})) and the full BRST operator is
$$Q_\mr{full}=-c^\alpha v^i_\alpha(x)\frac{\dd}{\dd x^i}+\frac{1}{2}f^{\alpha}_{\beta\gamma}c^\beta c^\gamma\frac{\dd}{\dd c^{\alpha}}+\lambda_\alpha \frac{\dd}{\dd \bar{c}_\alpha}$$
Now we can construct the gauge fixing fermion $\Psi$ using the same object as in Faddeed-Popov method, i.e. the function $\phi:\FF_{cl}\ra \A$:
\be \Psi=\bar{c}_\alpha \phi^\alpha(x)\label{gauge fixing fermion}\ee
Then the BRST action
$$S_{cl}(x)+Q_\mr{full}\Psi=S_{cl}(x)+\bar{c}_\alpha\frac{\dd \phi^\alpha}{\dd x^i}v^i_\beta(x)c^\beta+\lambda_\alpha \phi^\alpha(x)$$
is exactly the Faddeev-Popov action (\ref{FP action}).

\textbf{Case of reducible gauge symmetry.}
BRST formalism works also in the case of reducible (but closed) gauge symmetry, i.e. when the action of the group of gauge symmetries $\kappa_1:\GG_1\ra\mr{Diff}(\FF_{cl})$ is reducible (has nonzero kernel). Then there is another group $\GG_2$ acting on $\GG_1$ by right shifts, i.e. there is a group homomorphism $\kappa_2(x):\GG_2\ra\mr{RShifts}(\GG_1)$ depending in general on $x\in\FF_{cl}$ (we denoted the group of right shifts of $\GG_1$ by $\mr{RShifts}(\GG_1)$). Using the identification $\mr{RShifts}(\GG_1)=\GG_1$ (that sends a right shift to its value on the unit element $1\in\GG_1$) we say that $\kappa_2$ is a homomorphism $\kappa_2(x):\GG_1\ra \GG_2$ depending on a point $x\in\FF_{cl}$. The exactness condition is that the image of $\kappa_2(x)$ coincides with the stabilizer of $x\in\FF_{cl}$ in
$\GG_1$: $$\mr{im}(\kappa_2(x))=\mr{Stab}_x\subset\GG_1$$
If the action $\kappa_2(x)$ is again reducible, we introduce the next stage of reducibility tower, the group $\GG_3$ and homomorphism $\kappa_3(x):\GG_3\ra\mr{RShifts}(\GG_2)=\GG_2$ with the condition $\mr{im}(\kappa_3(x))=\ker(\kappa_2(x))\in\GG_2$ etc. up to some stage $p$ where $\ker(\kappa_p)=\{1\}\subset\GG_p$ (the action $\kappa_p$ is irreducible). Thus, we have an exact sequence of group actions $$\GG_p\circlearrowright\cdots\circlearrowright\GG_2\circlearrowright\GG_1\circlearrowright\FF_{cl}$$
or, equivalently, an exact sequence of group homomorphisms
$$1\ra\GG_p\xra{\kappa_p(x)}\cdots\xra{\kappa_3(x)}\GG_2\xra{\kappa_2(x)}\GG_1\xra{\kappa_1}\mr{Diff}(\FF_{cl})$$
In terms of infinitesimal gauge transformations, we have an exact sequence of Lie algebra actions
\be\A_p\circlearrowright\cdots\circlearrowright\A_2\circlearrowright\A_1\circlearrowright\FF_{cl}\label{BRST action tower}\ee
or, equivalently, an exact sequence of Lie algebra homomorphisms
\be 0\ra\A_p\xra{\hat\kappa_p(x)}\cdots\xra{\hat\kappa_3(x)} \A_2\xra{\hat\kappa_2(x)} \A_1\xra{\hat\kappa_1}\mr{Vect}(\FF_{cl})\label{BRST symmetry tower}\ee
where homomorphisms $\hat\kappa_2(x),\ldots,\hat\kappa_p(x)$ depend on a point  $x\in\FF_{cl}$ and the exactness conditions are: $\mr{im}(\hat\kappa_2(x))=\mr{Stab}_x\in\A_1,\;\mr{im}(\hat\kappa_3(x))=\ker(\hat\kappa_2(x)),\;\ldots,\;\ker(\hat\kappa_p(x))=0$. Alternatively, one can say that the reducibility tower of infinitesimal gauge symmetry is an exact sequence of Lie algebroids over $\FF_{cl}$ (and morphisms of Lie algebroids)
$$0\ra\A_p\times \FF_{cl}\ra\cdots\ra\A_2\times \FF_{cl}\ra\A_1\times \FF_{cl}\ra T\FF_{cl}$$
Here $A_1\times\FF_{cl}$ is understood as an algebroid over $\FF_{cl}$ with anchor $\hat\kappa_1:\A_1\times \FF_{cl}\ra T\FF_{cl}$ (the anchor map relation here is equivalent to the fact that $\hat\kappa_1:\A_1\ra \mr{Vect}(\FF_{cl})$ is a Lie algebra homomorphism). Other stages $\A_2\times \FF_{cl},\;\ldots,\;\A_p\times \FF_{cl}$ are understood as algebroids over $\FF_{cl}$ with zero anchor. Maps $\hat\kappa_2,\;\ldots,\;\hat\kappa_p$ here are fiberwise Lie algebra homomorphisms.

The minimal space of BRST fields is constructed for the case of reducible gauge symmetry as
\be \FF_\mr{minBRST}=\FF_{cl}\oplus\A_1[1]\oplus\A_2[2]\oplus\cdots\oplus \A_p[p]\label{minBRST fields for reducible case}\ee
with coordinates $x^i\in\Fun(\FF_{cl})$ (the classical fields, $\gh(x^i)=0$), $c_1^{\alpha_1}\in\Fun(\A_1[1])$ (ghosts for the classical fields, or ``first ghosts'', $\gh(c_1^{\alpha_1})=1$), $c_2^{\alpha_2}\in\Fun(\A_2[2])$ (ghosts for the first ghosts, or ``second ghosts'', $\gh(c_2^{\alpha_2})=2$) etc. The BRST operator $Q_\mr{min}$ is constructed from structure constants of the algebraic structure (\ref{BRST action tower}). The full BRST space of fields is defined as before (\ref{BRST space extension}) but $V_\mr{Aux}$ is now the following (cf. \cite{BV83}):
\be V_\mr{Aux}=\bigoplus_{1\leq B\leq A\leq p}V_{AB}\label{BRST auxiliary space}\ee
where
$$V_{AB}=\left\{\begin{array}{ll} \A_A^*[-A+B-1] & \mbox{for $B$ odd}\\
\A_A[A-B] &  \mbox{for $B$ even}\end{array}\right.$$
i.e.
\begin{eqnarray*}
V_\mr{Aux}=&\A_1^*[-1]\oplus \\
&(\A_2^*[-2]\oplus\A_2[0])\oplus \\
&(\A_3^*[-3]\oplus\A_3[1]\oplus\A_3^*[-1])\oplus \\
&\vdots
\end{eqnarray*}
And, as in (\ref{Q extension}), we extend the minimal BRST operator by the de Rham operator on $V_\mr{Aux}$.

\textbf{Example: abelian $p$-form field.}
A simple example of gauge theory with reducible gauge symmetry is the free abelian $p$-form field. Here $\FF_{cl}=\Omega^p(M)$, i.e. the classical field $x$ is a $p$-form on a Riemannian manifold $M$ (the space-time). The classical action is $S_{cl}=\int_M *dx\wedge dx$ where $*$ is the Hodge star. This action possesses abelian gauge symmetry $x\mapsto x+d x_1$ with $x_1\in\A_1=\Omega^{p-1}(M)$. This symmetry is reducible if $p\geq 2$: one can shift $x_1$ by an exact form $x_1\mapsto x_1+d x_2$ where $x_2\in\A_p=\Omega^{p-2}(M)$ etc. Thus the gauge symmetry tower (\ref{BRST symmetry tower}) for this case is just a piece of de Rham complex
$$0\ra\Omega^0(M)\xra{d}\cdots\xra{d}\Omega^{p-2}(M)\xra{d}\Omega^{p-1}(M)\xra{d} \Omega^p(M)$$
We wrote $\Omega^p(M)$ instead of $\mr{Vect}(\Omega^p(M))$ as the last term since here $A_1$ acts on $\FF_{cl}$ by constant vector fields. Notice that the exactness condition does not hold here, but it is  spoiled only by finite-dimensional cohomology. Minimal BRST fields here are the classical field $x\in\Omega^p(M)$ and the tower of ghosts $c_k\in\Omega^{p-k}(M)$, $k=1,\ldots,p$ with ghost numbers $\gh(x)=0$, $\gh(c_k)=k$. The minimal BRST operator is
$$Q_\mr{min}=dc_1\frac{\dd}{\dd x}+dc_2\frac{\dd}{\dd c_1}+\cdots+dc_p\frac{\dd}{\dd c_{p-1}}$$
For the purpose of gauge fixing we introduce auxiliary fields (\ref{BRST space extension},\ref{BRST auxiliary space}): the anti-ghosts $\bar{c}_{AB}$ and the Lagrangian multipliers $\lambda_{AB}$, where $1\leq B\leq A\leq p$. For the odd $A$ we set $\bar{c}_{AB},\lambda_{AB}\in(\Omega^{p-A}(M))^*$ with ghost numbers $\gh(\bar{c}_{AB})=B-A-1$ for anti-ghosts and $\gh(\lambda_{AB})=B-A$ for Lagrangian multipliers. For even $A$ we set $\bar{c}_{AB},\lambda_{AB}\in\Omega^{p-A}(M)$ with ghost numbers $\gh(\bar{c}_{AB})=A-B$, $\gh(\lambda_{AB})=A-B+1$. The full BRST operator is
$$Q_\mr{full}=dc_1\frac{\dd}{\dd x}+dc_2\frac{\dd}{\dd c_1}+\cdots+dc_p\frac{\dd}{\dd c_{p-1}}+\sum_{1\leq B\leq A\leq p}\lambda_{AB} \frac{\dd}{\dd \bar{c}_{AB}}$$
Imposing the Lorenz gauge corresponds to choosing the following gauge fixing fermion:
\begin{eqnarray*}\Psi=&&<\bar{c}_{11},d^*x>+\\
&&<\bar{c}_{21},d^*c_1>+<\bar{c}_{11},d\bar{c}_{22}>+\\
&&<\bar{c}_{31},d^*c_2>+<\bar{c}_{21},d\bar{c}_{32}>+<\bar{c}_{33},d^*\bar{c}_{22}>+\\
&&\cdots
\end{eqnarray*}
where as in (\ref{Yang-Mills FP action}) we denote the canonical pairing of currents and forms as $<\bt,\bt>$.

\subsection{Gauge fixing: BV formalism}
\label{BV method}
In Batalin-Vilkovisky approach one embeds the manifold of classical fields $\FF_{cl}$ into the ``space of BV fields'' $\FF_\mr{BV}=T^*[-1]\FF_\mr{BRST}$, the BV manifold with canonical BV Laplacian
$$\Delta=\sum_a(-1)^{\gh(\Phi^a)}\frac{\dd}{\dd\Phi^a}\frac{\dd}{\dd \Phi^+_a}$$ where $(\Phi^a)$ are coordinates on $\FF_\mr{BRST}$ and $(\Phi^+_a)$ are conjugate coordinates in the fiber of $T^*[-1]\FF_\mr{BRST}$. Coordinates $\Phi^a$ are called fields (and may be classical fields, ghosts, anti-ghosts or Lagrangian multipliers), while $\Phi^+_a$ are called ``anti-fields''. Next, one needs a BV action (or ``master action''), i.e. a function $S\in\Fun(\FF_\mr{BV})$ regular in $\hbar$ with ghost number $\gh(S)=0$ and satisfying the quantum master equation (\ref{QME}). There are also the following additional requirements for the BV action:
\begin{itemize}
\item Consistency with the classical action:
$$S^0|_{\FF_\mr{BRST}}=S_{cl}$$
\item Properness condition: the rank of Hessian of $S$ in any stationary point of the classical action $x\in\FF_{m.s.}$ equals $\frac{1}{2}\dim\FF_\mr{BV}$.
\end{itemize}
We denoted $\FF_{m.s.}=\{x\in\FF_{cl}:\delta S_{cl}(x)=0\}\subset\FF_{cl}$ the locus of stationary points of the classical action (the ``mass shell'').

The gauge fixing in BV formalism is performed by replacing the expression (\ref{int over Fcl}) with the integral of exponential of the BV action over a Lagrangian submanifold
$\LL\subset\FF_\mr{BV}$:
\be\int_\LL e^{S/\hbar}\label{integral over Lagrangian submfd}\ee
Choice of gauge corresponds here to choosing a specific $\LL$, and the invariance of the integral under continuous deformation of $\LL$ is implied by Batalin-Vilkovisky theorem (\ref{BV gauge independence}). Replacing (\ref{int over Fcl}) with (\ref{integral over Lagrangian submfd}) is also justified by the fact that on a special Lagrangian submanifold $\LL=\FF_\mr{BRST}\subset\FF_\mr{BV}$ the integrand in (\ref{integral over Lagrangian submfd}) becomes the exponential of classical action. Properness condition on $S$ is necessary for the existence of such $\LL$ for which the integral (\ref{integral over Lagrangian submfd}) is perturbatively well-defined.

Consider a gauge theory that can be treated in BRST formalism with the space of BRST fields $\FF_\mr{BRST}$, the BRST operator $Q$, measure $\mu\in\mr{Mes}(\FF_\mr{BRST})$ of density $\rho$ in coordinates $\Phi^a$, and with a gauge fixing fermion $\Psi$. Then we set
\be\FF_\mr{BV}=T^*[-1]\FF_\mr{BRST}\label{BV fields from BRST}\ee
and construct the BV action as
\be S=S_{cl}- Q(\Phi^a) \Phi^+_a +\hbar\log\rho\label{BV action from BRST}\ee
(instead of including the term $\hbar\log\rho$ in the BV action, one can replace the canonical BV Laplacian $\Delta$ by $\Delta_{\mu^2}$). We choose $\LL$ of the form (\ref{Lagrangian submanifold from gauge fixing fermion}), i.e. a Lagrangian deformation of $\FF_\mr{BRST}\subset\FF_\mr{BV}$ defined by the gauge fixing fermion $\Psi$:
$$\Phi^+_a=-\frac{\dd}{\dd \Phi^a}\Psi$$
Then $$\int_{\LL_\Psi}e^{S/\hbar}=\int_{\LL_\Psi}e^{S|_{\LL_\Psi}/\hbar}=\int_{\FF_\mr{BRST}}e^{\frac{1}{\hbar}(S_{cl}+Q\Psi)}\mu$$
and the integral over the Lagrangian submanifold reduces to (\ref{BRST gauge fixed}). Observe that the classical master equation for (\ref{BV action from BRST}) is satisfied due to $QS_{cl}=0$, $Q^2=0$, while the for the full quantum master equation we need additionally $\mr{div}_\mu(Q)=0$.
The BRST operator $Q$ is extended to the whole space of BV fields by construction
(\ref{Q on BV fields}):
$$Q_{\FF_\mr{BV}}=\{S^0,\bt\}=Q(\Phi^a)\frac{\dd}{\dd\Phi^a}+ \left((S_{cl}- Q(\Phi^b)\Phi^+_b)\frac{\ola\dd}{\dd \Phi^a}\right)\cdot\frac{\dd}{\dd \Phi^+_a}$$
Notice that the subalgebra of functions independent of anti-fields $\Fun(\FF_\mr{BRST})\subset\Fun(\FF_\mr{BV})$ is closed under the action of operator $Q_{\FF_\mr{BV}}$, and it acts there as the usual BRST operator: $Q_{\FF_\mr{BV}}|_{\FF_\mr{BRST}}=Q$.

In particular, for a gauge theory with closed irreducible gauge symmetry $\A\circlearrowright \FF_{cl}$ the minimal space of BV fields is
$$\FF_\mr{minBV}=T^*[1](\FF_{cl}\oplus\A[1])$$
with coordinates $x^i$, $c^\alpha$ on base and $x^+_i$, $c^+_\alpha$ on fiber (the ghost numbers are respectively 0, 1, -1, -2). The minimal BV action is constructed using
(\ref{Qmin},\ref{BV action from BRST}):
$$S_\mr{min}(x,c,x^+,c^+)=S_{cl}(x)-x^+_i v^i_\alpha(x) c^\alpha + \frac{1}{2}f^\alpha_{\beta\gamma}c^+_\alpha c^\beta c^\gamma$$
As in the case of minimal BRST formalism, we have a problem with the gauge fixing fermion $\Psi$ here, i.e. we cannot find a Lagrangian deformation of $\FF_\mr{BRST}$.

To solve this problem, as in BRST case, we introduce an auxiliary vector space (space of anti-ghosts) $V_\mr{Aux}$ and
set
$$\FF_\mr{fullBV}:=\FF_\mr{minBV}\oplus T^*[-1](T[1]V_\mr{Aux})$$
and
$$S_\mr{full}:=S_\mr{min}-\bar{c}^+_I\lambda^I$$
where we use coordinates $(\bar{c}^I,\lambda^I,\bar{c}^+_I,\lambda^+_I)$ on $T^*[-1](T[1]V_\mr{Aux})$.

For a gauge theory with irreducible gauge symmetry $\A\circlearrowright\FF_{cl}$ we have
$V_\mr{Aux}=\A^*[-1]$ and
$$\FF_\mr{fullBV}=T^*[-1]\left(\FF_{cl}\oplus\A[1]\oplus T[1](\A^*[-1])\right)$$
with coordinates $(x^i,c^\alpha,\bar{c}_\alpha,\lambda_\alpha;x^+_i,c^+_\alpha,\bar{c}^{+\alpha},\bar\lambda^{+\alpha})$, and the ghost numbers are respectively 0, 1, -1, 0, -1, -2, 0, -1.
The full BV action is
$$S_\mr{full}=S_{cl}(x)-x^+_i v^i_\alpha(x) c^\alpha + \frac{1}{2}f^\alpha_{\beta\gamma}c^+_\alpha c^\beta c^\gamma-\bar{c}^{+\alpha}\lambda_\alpha$$
For the gauge fixing fermion we can take (\ref{gauge fixing fermion}), i.e. $\LL_\Psi$ is given by
$$\LL_\Psi=\{(x^i,c^\alpha,\bar{c}_\alpha,\lambda_\alpha;x^+_i,c^+_\alpha,\bar{c}^{+\alpha},\bar\lambda^{+\alpha})\quad |\quad x^+_i=-\frac{\dd\phi^\alpha(x)}{\dd x^i}\,\bar{c}_\alpha,\,c^+_\alpha=0,\,\bar{c}^{+\alpha}=-\phi^\alpha(x),\,\lambda^{+\alpha}=0\}$$
Integral (\ref{integral over Lagrangian submfd}) in this case again coincides with Faddeev-Popov integral (\ref{FP1},\ref{FP action})
\begin{multline} \int_{\LL_\Psi\subset\FF_\mr{fullBV}}e^{S_\mr{full}/\hbar}=\\ \int \prod_i \DD x^i \prod_\alpha\DD c^\alpha \DD \bar{c}_\alpha \DD\lambda_\alpha\;\exp\frac{1}{\hbar}\left(S_{cl}(x)+\bar{c}_\alpha\frac{\dd \phi^\alpha(x)}{\dd x^i}v^i_\beta(x)c^\beta+\lambda_\alpha \phi^\alpha(x)\right)\label{FP integral in BV}\end{multline}

Notice that in minimal BV formalism, despite the absence of small Lagrangian deformations of $\FF_\mr{minBRST}\subset\FF_\mr{minBV}$, there are other Lagrangian submanifolds: the conormal bundles (\ref{conormal bundle}) for submanifolds $\KK\subset\FF_\mr{BRST}$. Strictly speaking, Batalin-Vilkovisky theorem (\ref{BV gauge independence}) is not applicable to this case, since the transition to conormal bundle is not a small deformation. Let us return to the irreducible gauge symmetry case $\A\circlearrowright\FF_{cl}$. Using the function $\phi:\FF_{cl}\ra\A$ we can construct submanifold
\be\KK=\phi^{-1}(0)\oplus\A[1]=\{(x^i,c^\alpha)|\phi^\alpha(x)=0\}\subset\FF_\mr{minBRST}\label{KK from phi}\ee
and its conormal bundle
\be N^*[-1]\KK=\{(x^i,c_\alpha;x^+_i,c^+_\alpha)|\phi^\alpha(x)=0,x^+_i=-\frac{\dd \phi^\alpha(x)}{\dd x^i}\,\bar{c}_\alpha,c^+_\alpha=0\}\subset\FF_\mr{minBV}\label{conormal bundle from phi}\ee
Here anti-ghosts $\bar{c}_\alpha$ appeared as coordinates on the space of covectors, orthogonal to the surface $\phi^{-1}(0)\in\FF_{cl}$. Notice that if we evaluate the integral over the Lagrangian multiplier $\lambda_\alpha$ in (\ref{FP integral in BV}) and take into account the arising delta functions $\prod_\alpha\delta(\phi^\alpha(x))$, we obtain precisely the integral over  conormal bundle
(\ref{conormal bundle from phi}) in minimal BV formalism:
\be\int_{N^*[-1]\KK\subset\FF_\mr{minBV}}e^{S_\mr{min}/\hbar}=\int_{\LL_\Psi\subset\FF_\mr{fullBV}}e^{S_\mr{full}/\hbar}\label{BV min-full correspondence}\ee
for $\Psi$ and $\KK$ constructed from the function $\phi:\FF_{cl}\ra\A$ using (\ref{gauge fixing fermion},\ref{KK from phi}). The right hand side of (\ref{BV min-full correspondence}) does not depend on choice of  $\phi$ due to Batalin-Vilkovisky theorem (\ref{BV gauge independence}), and hence the integral over conormal bundle in minimal BV formalism is also invariant.
This means that Batalin-Vilkovisky theorem (\ref{BV gauge independence}) is also valid for some class of conormal bundles in minimal BV formalism. Full BV formalism with auxiliary fields can be regarded from this point of view as a technical tool to prove (\ref{BV gauge independence}) for such non-continuous Lagrangian deformations.

The case of gauge theory with closed reducible gauge symmetry is translated from BRST to BV formalism straightforwardly, using the construction (\ref{BV fields from BRST},\ref{BV action from BRST}). However, Batalin-Vilkovisky approach works in more complicated situations. For instance, in the case of open gauge symmetry, i.e. when the distribution of vector fields on $\FF_{cl}$ annihilating $S_{cl}$ is integrable to a Lie group action only on the mass shell $\FF_{m.s.}\subset\FF_{cl}$, not on the whole  $\FF_{cl}$. Here the space of BV fields is again constructed using (\ref{minBRST fields},\ref{BV fields from BRST}) where $\A$ is the Lie algebra of gauge symmetry on the mass shell. The BRST operator $Q$ constructed formally using (\ref{Qmin}) will be nilpotent only modulo equations of motion $\delta S_{cl}=0$, and hence the BRST method does not work, and construction  (\ref{BV action from BRST}) does not work also. Finding the BV action $S$ (which will now be nonlinear in anti-fields $\Phi^+_a$) is a non-trivial problem here. The BRST operator on space of BV fields is again defined by (\ref{Q on BV fields}), but the subalgebra of functions $\Fun(\FF_\mr{BRST})\subset\Fun(\FF_\mr{BV})$ is no longer closed under its action. An example of a gauge theory with open gauge symmetry is the Poisson sigma model, a 2-dimensional model of topological field theory, famous due to its relation to Kontsevich's deformation quantization of Poisson manifolds \cite{CF},\cite{Kontsevich}.
Another similar possible situation is when the gauge symmetry is closed but reducible and the stabilizer $\mr{Stab}_x\in\GG_1$ depends on whether $x$ is on mass-shell or not, i.e. the reducibility tower (\ref{BRST symmetry tower}) is different over different points $x\in\FF_{cl}$. Here Batalin-Vilkovisky method suggests that we again formally construct space of BV fields using (\ref{minBRST fields for reducible case},\ref{BV fields from BRST}), where $\A_1,\ldots,\A_p$ are defined by the reducibility tower for $x\in\FF_{m.s.}$. An example of this situation is provided by the topological $BF$ theory in dimension $D\geq 4$ (in dimension $D<4$ the usual BRST formalism works). We will discuss this example in section \ref{BF theory}.

Let us briefly summarize the main features of Faddeev-Popov, BRST and Batalin-Vilkovisky methods.
\begin{itemize}
\item Faddeev-Popov method. Data: the space of classical fields $\FF_{cl}$ endowed with an action of Lie algebra $\A\circlearrowright\FF_{cl}$, the classical action is an invariant function of classical fields
$S_{cl}\in \Fun(\FF_{cl})^\A$. The gauge fixing is done using the gauge fixing function $\phi:\FF_{cl}\ra\A$. Ghosts $c,\, \bar{c}$ are introduced as a tool, allowing one to raise the determinant of geometric origin into action.

\item BRST method. Data: the $\ZZ$-graded space of fields $\FF_\mr{BRST}$ endowed with a cohomological vector field $Q$ (which encodes the gauge symmetry of the original classical system). BRST action is the class of the classical action in $Q$-cohomology: $S_\mr{BRST}=[S_{cl}]\in H^0_Q(\Fun(\FF_\mr{BRST}))$. The gauge fixing is the choice of a representative for this class.

\item Batalin-Vilkovisky method. Data: the BV manifold $(\FF_\mr{BV},\Delta)$, master action $S\in\Fun(\FF_\mr{BV})$, solving the quantum master equation $\Delta e^{S/\hbar}=0$ ($S$ encodes both the classical action and its gauge symmetry). Master actions $S$ and $S'$ differing by a canonical transformation $e^{S'/\hbar}-e^{S/\hbar}=\Delta(\cdots)$ are considered physically equivalent. The gauge fixing is the choice of a Lagrangian submanifold $\LL\subset\FF_\mr{BV}$.
\end{itemize}

\subsection{Topological $BF$ theory}
\label{BF theory}
Let $M$ be a $D$-dimensional compact orientable manifold, let $G$ be a compact Lie group (the gauge group) and denote $\g:=\mr{Lie}(G)$ its Lie algebra. Let also $P$ be a principal $G$-bundle on $M$ and $\ad P$ the adjoint bundle. The classical fields of $BF$ theory are a connection on $P$ and a $(D-2)$-form with values in $\ad P$:
$$A\in\mr{Conn}(M,P),\; B\in\Omega^{D-2}(M,\ad P)$$
The classical action of $BF$ theory is
\be S_{cl}(A,B)=\tr_g\int_M B\wedge F_A\label{BF classical action}\ee
where $F_A\in \Omega^2(M,\ad P)$ is the curvature of connection $A$ and $\tr_g$ is the trace in adjoint representation of $\g$.
We will consider only the case of the trivial bundle $P$.  Here  $A\in \Omega^1(M,\g)$ is a $\g$-valued 1-form and $B\in \Omega^{D-2}(M,\g)$ is a $\g$-valued $(D-2)$-form. The classical action is
$$S_{cl}(A,B)=\tr_\g\int_M B\wedge (dA+A\wedge A)$$
and the space of classical fields is
$$\FF_{cl}=\mr{Conn}(M,P)\oplus \Omega^{D-2}(M,\ad P)=\Omega^1(M,\g)\oplus\Omega^{D-2}(M,\g)$$
The equations of motion are
\be\left\{\begin{array}{l} F_A=0 \\ d_A B=0 \end{array}\right.\label{BF eq of motion}\ee
where $d_A=d+\ad_A$ is the covariant differential defined by the connection $A$. In other words, $(A,B)$ is a stationary point of the classical action $S_{cl}$ iff the connection $A$ is flat and $B$ is a covariantly constant $(D-2)$-form.

The classical action is invariant under gauge transformations
\be \left\{\begin{array}{lll}A & \mapsto & A^g=gAg^{-1}+g\,dg^{-1} \\
B & \mapsto & gBg^{-1}+d_{A^g}\tau_1 \end{array}\right.
\ee
where $g\in \Gamma(M,P)=\Omega^0(M,G)$ is a fiberwise rotation of the bundle $P$ and $\tau_1\in\Omega^{D-3}(M,\ad P)=\Omega^{D-3}(M,\g)$. Hence the group of gauge symmetry for $BF$ thory is
$\GG_1=\Omega^0(M,G)\ltimes\Omega^{D-3}(M,\g)$. However, in the case $D\geq 4$ the action $\GG_1\circlearrowright\FF_{cl}$ is reducible on the mass shell, since one can shift $\tau_1\mapsto \tau_1+d_{A^g}\tau_2$ if the connection $A$ is flat. Here $\tau_2\in\Omega^{D-4}(M,\g)$. Therefore we have an action $\GG_2\car\GG_1\car\FF_{cl}$, where $\GG_2=\Omega^{D-4}(M,\g)$. In case $D\geq 5$ this action is again reducible, since $\tau_2$ can be shifted by a  $d_{A^g}$-exact form, etc. So we obtain the reducibility tower of the gauge symmetry on the mass shell: $\GG_{D-2}\car\cdots\car\GG_2\car\GG_1\car\FF_{cl}$ with $\GG_k=\Omega^{D-2-k}(M,\g)$ for $2\leq k\leq D-2$. Here all groups except $\GG_1$ are abelian.  Infinitesimally the gauge symmetry on the mass shell is described by a tower of Lie algebra actions
$$\A_{D-2}\car\cdots\car\A_2\car\A_1\car\FF_{cl}$$
with $\A_1=\mr{Lie}(\GG_1)=\Omega^0(M,\g)\ltimes\Omega^{D-2}(M,\g)$, with coordinates $(c,\tau_1)$ and action
$$\left\{\begin{array}{lll}A & \mapsto & A+d_A c \\
B & \mapsto & B+[B,c]+d_A \tau_1 \end{array}\right.$$
Further, $\A_k=\mr{Lie}(\GG_k)=\Omega^{D-2-k}(M,\g)$ for $2\leq k\leq D-2$, and $\A_k$ acts on $\A_{k-1}$ by shifts $\tau_{k-1}\mapsto\tau_{k-1}+d_A\tau_k$.

\textbf{Super-field formalism.} Following the general routine we introduce the (minimal) space of BRST fields
$$\FF_\mr{BRST}=\FF_{cl}\oplus \A_1[1]\oplus\A_2[2]\oplus\cdots\oplus A_{D-2}[D-2]$$
with coordinates $(A,B;c,\tau_1;\tau_2;\cdots;\tau_{D-2})$. Here we understand $c$ and $\tau_1$ as ghosts, i.e. $\gh(c)=\gh(\tau_1)=1$. Next, $\tau_2$ is the ghost for ghost $\tau_1$ and $\gh(\tau_2)=2$ etc. Classical fields are prescribed ghost number zero: $\gh(A)=\gh(B)=0$. The space of BV fields is the odd cotangent bundle $T^*[-1](\FF_\mr{BRST})$. Next we introduce the ``super-fields'' $\tilde{A}$ and $\tilde{B}$
\begin{eqnarray} \tilde{A} & := & c+A+B^++\tau_1^++\cdots+\tau_{D-2}^+ \label{super A}\\
\tilde{B} & := & \tau_{D-2}+\cdots+\tau_1+B+A^++c^+ \label{super B}
\end{eqnarray}
Super-fields are understood as non-homogeneous differential forms  $\tilde{A},\tilde{B}\subset\Omega^0(M,\g)\oplus\cdots\oplus\Omega^D(M,\g)$ where different components are prescribed different ghost numbers, so that for $\tilde{A}$ the total degree is $\gh+\deg=1$ (where $\deg$ denotes the de Rham degree of a form), and for $\tilde{B}$ the total degree is $\gh+\deg=D-2$. Here we use the identification $(\Omega^k(M,\g))^*\simeq\Omega^{D-k}(M,\g)$ using the pairing $\tr\int\bt\wedge\bt$ on $\Omega^\bt(M,\g)$. Note that $\tilde{A}$ and $\tilde{B}$ are mutually canonically conjugate. In terms of super-fields the BV action for $BF$ theory is written in elegant form:
\be S(\tilde{A},\tilde{B})=\tr\int_M \tilde{B}\wedge (d\tilde{A}+\tilde{A}\wedge\tilde{A})\label{BF action via superfields}\ee
(the integral is non-trivial only for the component of integrand of de Rham degree $D$). This result is due to Wallet \cite{Wallet} and Ikemori \cite{Ikemori}. The expression for BV action in terms of classical fields, ghosts and anti-fields is obtained from (\ref{BF action via superfields}) by substitution of decompositions (\ref{super A},\ref{super B}) for super-fields. In particular, in dimensions $D=2,3,4$ the BV action for $BF$ theory is
\begin{eqnarray}
D=2&:&S=\tr\int_M B F_A+A^+ d_A c + B^+ [B,c]+\frac{1}{2}c^+ [c,c] \label{BF action D=2}\\
D=3&:&S=\tr\int_M B F_A+A^+ d_A c + B^+ ([B,c]+d_A\tau_1)+\frac{1}{2}c^+ [c,c]+\tau_1^+[\tau_1,c] \label{BF action D=3}\\
D=4&:&S=\tr\int_M B F_A+A^+ d_A c + B^+ ([B,c]+d_A\tau_1)+\nonumber\\
&&+\frac{1}{2}c^+ [c,c]+\tau_1^+([\tau_1,c]+d_A\tau_2)+\tau_2^+[\tau_2,c]+\frac{1}{2}[B^+,B^+]\tau_2\label{BF action D=4}
\end{eqnarray}
(we are sloppy with signs here, cf. \cite{CR} for careful discussion).
Notice that for $D=2,3$ the BV action is linear in anti-fields and hence can be obtained from BRST formalism and construction (\ref{BV action from BRST}). However starting from dimension $D=4$ the BV action contains terms, quadratic in anti-fields. This indeed reflects the fact that starting from this dimension we have the ``open'' tower of reducibility of gauge symmetry, i.e. the tower depends on whether the classical fields $(A,B)$ satisfy the equations of motion (\ref{BF eq of motion}).

\textbf{Canonical $BF$ theory.} In what follows we will use the more general and fundamental version of $BF$ theory, the ``canonical'' $BF$ theory. The initial data are the same: a compact manifold $M$ (but now we do not require orientability),  a compact Lie group $G$ with $\g=\mr{Lie}(G)$ and principal $G$-bundle $P$ on $M$ which we assume to be trivial for simplicity. Classical fields are: the connection
$A\in\mr{Conn}(M,P)=\Omega^1(M,\g)$ and the de Rham 2-current $\beta\in(\Omega^2(M,\g))^*$. The classical action is
\be S_{cl}(A,\beta)=<\beta,F_A>\label{BF canonical action}\ee
Notice that unlike (\ref{BF classical action}) this action does not use the integration over $M$ and the inner product on $\g$, it uses only the canonical pairing between forms and currents $<\bt,\bt>: (\Omega^\bt(M,\g))^*\otimes \Omega^\bt(M,\g)\ra\RR$. This implies in particular that this action makes sense for non-orientable $M$. In case of orientable $M$ the field $\beta$ is related to $B$ by the index lowering operation $\flat:\Omega^\bt(M,\g)\ra(\Omega^\bt(M,\g))^*$, associated to the Poincar\'{e} pairing on forms
$(\bt,\bt)_P=\tr\int_M\bt\wedge\bt:\Omega^\bt(M,\g)^{\otimes 2}\ra\RR$, i.e. $\beta=B_\flat$.
The equations of motion are $F_A=0$, $d_A^*\beta=0$, where $d_A^*\beta:=-<\beta,d_A\bt>$.
As before one introduces the tower of reducibility of gauge symmetry on the mass shell:
$\A_{D-2}\car\cdots\car\A_2\car\A_1\car\FF_{cl}$, where now $\A_1=\Omega^0(M,\g)\ltimes(\Omega^3(M,\g))^*$ acting on classical fields by
$A\mapsto A+d_A c$, $\beta\mapsto\beta-\ad^*_c\beta+d_A^*\chi_1$; further $\A_k=(\Omega^{k+2}(M,\g))^*$ for $2\leq k\leq D-2$ and the action $\A_k\car\A_{k-1}$ is given by shifts $\chi_{k-1}\mapsto\chi_{k-1}+d_A^*\chi_k$. The ghosts, associated to these symmetries are organized together with the classical fields into super-fields
\begin{eqnarray}\omega := \tilde{A}&=&c+A+\beta^+ + \chi_1^+ +\cdots+\chi_{D-2}^+ \label{omega physical}\\
p := \tilde{B}&=&c^+ + A^+ + \beta + \chi_1 +\cdots+\chi_{D-2} \label{p physical}
\end{eqnarray}
Here we introduced the notation $(\omega,p)$ for the super-fields of canonical $BF$ theory, which will be used throughout this text. The super-field $\omega$ is again understood as a non-homogeneous differential form $\omega\in\Omega^\bt(M,\g)$ of total degree $\gh+\deg=1$. The second super-field $p$ is understood as a non-homogeneous de Rham current $p\in(\Omega^\bt(M,\g))^*$ of total degree  $\gh+\deg=-2$
(we use the convention that the de Rham  degree of a $k$-current in $(\Omega^k(M,\g))^*$ is $-k$). Notice that here we did not have to use the Poincar\'{e} pairing on forms $(\bt,\bt)_P$ to define the super-fields. The BV action for canonical $BF$ theory is
\be S(\omega,p)=<p,F_\omega>=<p,d\omega+\frac{1}{2}[\omega,\omega]>\label{BF canonical BV action}\ee
Action (\ref{BF canonical BV action}) on the space of BV fields of canonical $BF$ theory
\be\FF_\mr{BV}=T^*[-1]\mr{Maps}(T[1]M,\; \g[1])\label{BF BV fields}\ee is a special case of general constructions of $PQ$-manifolds introduced in  \cite{AKSZ}, and therefore the canonical $BF$ theory in BV formalism is sometimes called the AKSZ theory. This construction, the super-field formalism and the form of BV action  suggest that apart from ``physically'' chosen Lagrangian submanifold $\FF_\mr{BRST}\subset\FF_\mr{BV}$ (which is distinguished by Batalin-Vilkovisky construction starting from connection $A$ and 2-current $\beta$), there is another important Lagrangian submanifold $\mr{SConn}(M,\g)=\mr{Maps}(T[1]M,\; \g[1])\subset \FF_\mr{BV}$ where the super-connection $\omega$ lives. In what follows the representation  (\ref{BF BV fields}) for the space of BV fields of canonical $BF$ theory will be crucial for us, i.e. precisely the one related to Lagrangian submanifold $\mr{SConn}(M,\g)$.

Topological $BF$ theory is one of the simplest models of topological field theory. It can be defined on a manifold of arbitrary dimension, possibly non-orientable, possibly with boundary. In lower dimensions $BF$ theory  is related to other well-known topological field theories. For instance, for $\dim M=2$ the $BF$ theory is a special case of Poisson-sigma model, corresponding to the target $\g^*$ (with linear Poisson structure); for $\dim M=3$ it is the Chern-Simons theory with special gauge Lie algebra $\g\oplus\g^*$ (and the inner product pairs $\g$ to $\g^*$). Next, the 3-dimensional $BF$ theory with cosmological term, i.e.  $S_{cl}=\tr\int_M B\wedge F_A+B\wedge B\wedge B$, is closely related to Chern-Simons theory (with general gauge group $G$) \cite{CCRFM}. In particular the partition functions are related by $Z_{BF+B^3}=|Z_{CS}|^2$. For 3-dimensional $BF$ theory the observables associated to knots are constructed \cite{CCRFM}, and the vacuum expectation value for them yields the Alexander-Conway polynomial (this situation is analogous to Chern-Simons theory, where the vacuum expectation value for Wilson loops yields knot invariants \cite{Witten}).

\section{Abstract $BF$ theory and effective action for it}
\label{section: abstract BF and effective action}
In section \ref{abstract BF theory} we introduce the ``abstract $BF$ theory'': to an arbitrary unimodular differential graded Lie algebra (DGLA) $(V,d,[\bt,\bt])$ we associate a space of BV fields $\FF$ (a graded vector space with canonical BV Laplacian) and a BV action $S$ on this space, constructed from structure constants of differential and commutator in $V$, and satisfying the quantum master equation (QME). Moreover it turns out that classical part of QME is equivalent to the three quadratic relations for differential and commutator in DGLA (the Poincar\'{e}, Leibniz and Jacobi identities), while the quantum part of QME is equivalent to the unimodularity property of the Lie bracket. Topological $BF$ theory on a manifold $M$ corresponds to a special case of abstract $BF$ theory: $V=\Omega^\bt(M,\g)$ (the de Rham algebra of $M$ with coefficients in gauge Lie algebra $\g$).

In section \ref{effective action: idea} we discuss the general construction of effective (or ``induced'') action on infrared (IR) fields in BV formalism: the effective action is constructed as a BV integral over Lagrangian submanifold $\LL$ in the space of ultraviolet (UV) fields. The main property of this construction is that it transfers solutions of QME to solutions of QME (now on the space of IR fields). We also discuss the dependence of effective action on choice of $\LL$ (i.e. the choice of gauge for BV integral) and on the choice of splitting of fields into IR and UV parts: it turns that an infinitesimal change in this ``induction data'' leads to an infinitesimal canonical transformation of the effective action (moreover, one can write an explicit formula for the generator of this transformation). We also show that a canonical transformation of action leads to a canonical transformation of effective action, which allows one to speak of induction for equivalence class of action modulo canonical transformations. The construction of effective BV action was used in framework different from ours in papers \cite{KL},\cite{AKLL}.

In section \ref{section: effective action for abstract BF} we specialize the construction of effective BV action for the case of inducing effective action $S'$ for abstract $BF$ theory. We associate the space of IR fields to a subcomplex (more precisely, a deformation retract) $V'\hra V$. Splitting of fields into IR and UV fields is defined by embedding $\iota: V'\ra V$ and retraction $r: V\ra V'$, which are assumed to be chain maps. Lagrangian submanifold $\LL$ defining the BV integral is constructed from Hodge decomposition of $V$, which is in turn defined by the chain homotopy $K: V\ra V$ retracting $V$ to $V'$. This $\LL$ is a linear subspace and therefore we need not bother about measure in BV integral over $\LL$ (we use the translation-invariant measure). And hence we can forget the difference between functions and half-densities. The pay-off is that we compute the action (as a function of IR fields) modulo constants, and normalize it by $S'(0)=0$. We discuss the Feynman rules for BV integral, defining the perturbation series for effective action $S'$. It turns out that two types of diagrams contribute: binary rooted trees (the edges are oriented towards the root) and one-loop graphs with leaves (incoming external edges) and vertices of valence 3 (two incoming edges, one outgoing edge). The internal edges are decorated with ``propagator'' $-K$, leaves --- with IR fields, the 3-valent vertices are decorated with the Lie bracket in $V$. An important feature here are the rules of orientation of edges: each vertex has exactly one outgoing edge. This implies in particular the absence of multi-loop Feynman diagrams.

In section \ref{section: eff action as generating function for alg structure} we interpret the effective action for abstract $BF$ theory, associated to a subcomplex $V'\hra V$ as a generating function for certain algebraic structure on $V'$ (analogously to the way in which the initial action of abstract $BF$ theory is a generating function for unimodular DGLA structure on $V$), namely a one-loop version of $L_\infty$ structure. We call this structure the ``$qL_\infty$ algebra''. It can be defined either in super-geometric terms, as a pair of a (pointed) cohomological vector field $Q$ and a $Q$-invariant measure $\mu$ on $V'[1]$, or in terms of operations, as a set of ``classical'' operations $l_{(n)}:\Lambda^n V'\ra V'$ (generated by Taylor components of the cohomological vector field) plus a set of ``quantum operations'' $q_{(n)}:\Lambda^n V'\ra \RR$ (generated by Taylor components of logarithm of density of the measure). Cohomologicity condition for $Q$ and $Q$-invariance of $\mu$ can be reformulated as two systems of quadratic equations on operations: the system of homotopy Jacobi identities (the standard quadratic relations for a $L_\infty$ algebra) plus a system of homotopy unimodularity relations. In terms of BV formalism this set of relations is equivalent to the QME for effective action $S'$ (which is satisfied automatically by the construction of effective action via BV integral). The $qL_\infty$ algebras appeared earlier in another context (as algebras over ``wheeled $L_\infty$ operad'') in \cite{Merkulov}.

In section \ref{section: BF_infty} we introduce the notion of ``$BF_\infty$ theory'', generalizing the notion of abstract $BF$ theory. A $BF_\infty$ theory is associated to a $qL_\infty$ algebra: solution of the QME on the space of fields (constructed as for abstract $BF$ theory) is constructed from the structure constants of classical and quantum operations. In particular, abstract $BF$ theory corresponds to the case when all operations except $l_{(1)}$ and $l_{(2)}$ vanish (the case of unimodular DGLA). The effective theory for abstract $BF$ theory is a $BF_\infty$ theory. Moreover, it turns out that the class of $BF_\infty$ theories is closed under the operation of induction of effective action on a subcomplex. The induction is analogous to induction from abstract $BF$ theory, but here we have to allow vertices of arbitrary valence in Feynman diagrams, and each vertex is allowed to have at most one outgoing edge. Vertices with one outgoing edge are decorated by classical operations $l_{(n)}$, vertices with all edges incoming are decorated by quantum operations $q_{(n)}$. As before, only tree and one-loop diagrams appear here. $BF_\infty$ theories and $qL_\infty$ algebras are two equivalent languages describing the same object. In the language of $qL_\infty$ algebras, induction is the one-loop analog of homotopy transfer of a $L_\infty$ structure to a subcomplex. Then (section \ref{section: equivalence of qL_infty algebras}) we discuss the notion of equivalence for $qL_\infty$ algebras, originating from the equivalence of $BF_\infty$ theories modulo canonical transformations (we introduce the class of ``special'' canonical transformations that preserve the $BF_\infty$ ansatz for action). We also declare the induced $qL_\infty$ algebra to be equivalent to the former one. Therefore the homotopy type of a given $qL_\infty$ algebra is understood as the induced $qL_\infty$ structure on cohomology modulo canonical transformations. Next, in section \ref{section: S' via U and I} we give the construction for $L_\infty$ quasi-isomorphism between classical parts of induced and former $qL_\infty$ structures via BV integral. We also interpret the effective $BF_\infty$ action in terms of $L_\infty$ morphism and the ``torsion'' (this interpretation may be understood as a way to organize Feynman diagrams for the effective action; we will use it for computations in sections \ref{section: exact results for cell action}, \ref{section: S on coh examples}).

\subsection{Abstract $BF$ theory}
\label{abstract BF theory}
Let $V$ be a $\ZZ$-graded vector space. We construct the space of fields as the odd cotangent bundle of $V$ shifted:
\begin{equation}\FF:=T^*[-1](V[1])=V[1]\oplus V^*[-2] \label{space of fields}
\end{equation}
The ring of functions on $\FF$ is the graded symmetric algebra of $\FF^*$ (with coefficients in formal power series in Planck constant $\hbar$):
$$\Fun(\FF)=S^\bt(V^*[-1]\oplus V[2])$$
We call the grading in $V$ and $V^*$ the ``degree'' ($\deg$), and the grading in $\Fun(\FF)$ the ``ghost number'' ($\gh$).

Let $(e_i)$ be a basis $V$ and $(e^i)$ the dual basis in $V^*$. Denote $(\omega^i)\subset V^*[-1]$ and $(p_i)\subset V[2]$ the sets of coordinate functions on $V[1]$ and $V^*[-2]$, associated to this basis. We have
\begin{eqnarray}\deg(e_i)+\gh(\omega^i)=1\label{deg+gh for omega} \\ \deg(e^i)+\gh (p_i)=-2 \label{deg+gh for p}\\ \deg (e^i)=-\deg (e_i)\end{eqnarray}
and the ring of functions $\Fun(\FF)$ is understood as the ring of polynomials in variables $(\omega^i,p_i)$ with coefficients in $\RR[[\hbar,\hbar^{-1}]]$:
$$\Fun(\FF)=\RR[[\hbar,\hbar^{-1}]][\omega^i,p_i]$$
Introduce the super-fields $\omega$ and $p$ as
\begin{eqnarray}
\omega&:=& e_i \omega^i\in V\otimes V^*[-1]\subset V\otimes\Fun(\FF) \\
p&:=& p_i\, e^i\in V[2]\otimes V^*\subset \Fun(\FF)\otimes V^*
\end{eqnarray}
Therefore $\omega$ and $p$ are functions on the space of fields with values in $V$ and $V^*$ respectively: $\omega$ is the generating function for coordinate functions on base of $\FF=T^*[-1](V[1])$ and $p$ is the generating function for coordinates on the base. One can also give the definition of super-fields that is manifestly independent on choice of basis: one defines super-fields as shifted identity maps
\begin{eqnarray*}\omega:& V[1]\ra V \\ p:& V^*[-2]\ra V^* \end{eqnarray*}

In the traditional physical interpretation, $\omega$ and $p$ are elements of total degree 1 and -2 respectively in vector spaces $V\otimes(\oplus_k\RR[k])$ and $V^*\otimes(\sum_k\RR[k])$, bigraded by the pair (degree, ghost number), with $\RR$ shifted in ghost number:
\begin{eqnarray*}\omega&\in&\tilde{V}=\left[V\otimes(\oplus_k\RR[k])\right]^{\deg+\gh=1}\\
p&\in&\tilde{V}^*=\left[V^*\otimes(\oplus_k\RR[k])\right]^{\deg+\gh=-2}
\end{eqnarray*}
In other words, in this picture $\omega$ and $p$ are elements of vector spaces $V$, $V^*$, where components of different degrees are prescribed certain ghost numbers, so that the total degree is 1 and -2 respectively.

The space of fields (\ref{space of fields}), being the odd ctangent bundle, is endowed with a canonical odd symplectic form
$$\Omega_\mr{BV}=<\delta p,\delta\omega>=\sum_i(-1)^{|i|}\delta p_i\wedge\delta\omega^i$$
We use notation
$|i|=\deg e^i$ for degrees of basis elements; $\delta$ denotes the de Rham differential on $\FF$ and the canonical pairing $<\bt,\bt>:V^*\otimes V\ra\RR$ couples $e^i$ with $e_i$: $<e^i,e_j>=\delta^i_j$ (the Kronecker symbol). Pairing in inverse order has a sign: $<e_i,e^j>=(-1)^{|i|}\delta^j_i$.
Odd symplectic form $\Omega_\mr{BV}$ in turn defines the canonical anti-bracket $\{\bt,\bt\}$ on $\FF$:
$$\{f,g\}=f\left(\frac{\overleftarrow\dd}{\dd
\omega^i}\frac{\overrightarrow\dd}{\dd p_i}-\frac{\overleftarrow\dd}{\dd
p_i}\frac{\overrightarrow\dd}{\dd \omega^i}\right)g $$
where $f,g\in\Fun(\FF)$
It is also convenient to write anti-bracket in terms of super-fields
$\omega$, $p$. For this purpose we introduce the derivatives in super-fields:
$$\begin{array}{lll}\frac{\ola\dd}{\dd\omega}=\frac{\ola\dd}{\dd\omega^i}e^i &,& \frac{\ora\dd}{\dd
p}=e_i \frac{\ora\dd}{\dd p_i},\\
\frac{\ora\dd}{\dd\omega}=(-1)^{|i|}e^i\frac{\ora\dd}{\dd\omega^i} &,& \frac{\ola\dd}{\dd
p}=\frac{\ola\dd}{\dd p_i}e_i \end{array}$$
Therefore in terms of $\omega$, $p$ the anti-bracket is
$$\{f,g\}=f\left<\frac{\overleftarrow\dd}{\dd
\omega},\frac{\overrightarrow\dd}{\dd p}\right>g-(-1)^{(\gh f+1)(\gh g+1)}g\left<\frac{\overleftarrow\dd}{\dd
\omega},\frac{\overrightarrow\dd}{\dd p}\right>f= f\left(\left<\frac{\overleftarrow\dd}{\dd
\omega},\frac{\overrightarrow\dd}{\dd p}\right>-\left<\frac{\overleftarrow\dd}{\dd
p},\frac{\overrightarrow\dd}{\dd \omega}\right>\right)g$$

Since now (contrary to the situation of section \ref{SP manifolds}) $\FF$ is a vector space, it has a distinguished class of measures: the constant measures $\mu\in\Ber(\FF)$ (here we mean the Berezinian of a vector space, i.e. in terms of general definition in section \ref{graded manifolds} we treat $\FF$ as a graded manifold whose body is a single point). Therefore $\FF$ possesses a canonical BV Laplacian
$$\Delta=\sum_i (-1)^{|i|+1}\frac{\dd}{\dd\omega^i}\frac{\dd}{\dd p_i}=-\left<\frac{\ora\dd}{\dd\omega},\frac{\ora\dd}{\dd p}\right>$$

Let $V$ be endowed with structure of differential graded Lie algebra (DGLA), i.e. with a couple of linear maps $d: V\ra V$ and $[\bt,\bt]: V\otimes V\ra V$ satisfying the  following properties:
\begin{itemize}
\item differential has degree 1: \be|dx|=|x|+1\label{degree of d}\ee
\item bracket has degree 0: \be|[x,y]|=|x|+|y|\label{degree of [,]}\ee
\item bracket is graded skew-symmetric: \be [x,y]=-(-1)^{|x|\,|y|}[x,y]\ee
\item Poincar\'{e} identity: \be d^2 x=0\label{Poincare identity}\ee
\item Leibniz identity: \be d [x,y]=[dx,y]+(-1)^{|x|}[x,dy]\label{Leibniz identity}\ee
\item Jacobi identity: \be(-1)^{|x|\,|z|}[x,[y,z]]+\mr{cycl. perm.}=0\label{Jacobi identity}\ee
\end{itemize}
where $x,y,z\in V$ and we use notation $|x|:=\deg(x)$. Let us require in addition the unimodularity property for $V$:
\be\Str_V[x,\bt]=0\label{unimodularity}\ee
for every $x\in V$. Here $\Str_V$ denotes super-trace over $V$.

We define the action of abstract $BF$ theory, associated to a DGLA $(V,d,[,])$ as
\begin{eqnarray} S&=&<p,d\omega+\frac{1}{2}[\omega,\omega]>\nonumber\\ &=&\sum_{i,j}d^i_j p_i\omega^j+\frac{1}{2}\sum_{i,j,k}(-1)^{(|j|+1)|k|}f^i_{jk} p_i\omega^j\omega^k\label{abstract BF action}
\end{eqnarray}
where $d^i_j:=<e^i,de_j>$ and $f^i_{jk}:=<e^i,[e_j,e_k]>$ are the structure constants of differential and bracket respectively. Therefore $S\in\Fun(\FF)$ is given as acertain cubic polynomial in coordinates  $(\omega^i,p_i)$ on $\FF$, and its coefficients are (up to signs) the structure constants of differential and bracket. So $S$ is in a sense generating function for DGLA operations on $V$. Next, it follows immediately from (\ref{degree of d},\ref{degree of [,]}) and (\ref{deg+gh for omega},\ref{deg+gh for p}) that all monomials with nonvanishing coefficients in (\ref{abstract BF action}) have ghost number zero. Hence $\gh(S)=0$.

\begin{statement}\label{statement: QME for abstract BF} The action of abstract $BF$ theory $S=<p,d\omega+\frac{1}{2}[\omega,\omega]>$ satisfies quantum master equation, i.e.
$\frac{1}{2}\{S,S\}+\hbar\Delta S=0$.
\end{statement}
\textbf{Proof.} Since $S=S^0$ does not depend on $\hbar$, QME is equivalent to the system
\begin{eqnarray}\{S,S\}&=&0\label{CME for abstract BF}\\
\Delta S&=&0 \label{QME for abstract BF}
\end{eqnarray}
where (\ref{CME for abstract BF}) is the classical master equation (CME) and (\ref{QME for abstract BF}) is the quantum part of QME. Let us compute $\{S,S\}$:
\begin{eqnarray*}
\frac{1}{2}\{S,S\}&=&\sum_i S\frac{\ola\dd}{\dd\omega^i}\frac{\ora\dd}{\dd p_i}S \\
&=&\sum_i \left(\sum_j d^j_i p_j+\sum_{j,k}(-1)^{(|k|+1)|i|}f^j_{ki}p_j\omega^k\right)\cdot \\
&&\cdot\left(\sum_l d^i_l \omega^l+\frac{1}{2}\sum_{l,m}(-1)^{(|l|+1)|m|}f^i_{lm}\omega^l\omega^m\right)\\
&=&\sum_{i,j,l}d^j_i d^i_l p_j\omega^l+\frac{1}{2}\sum_{i,j,l,m}(-1)^{(|l|+1)|m|}d^j_i f^i_{lm}p_j\omega^l\omega^m+
\sum_{i,j,k,l}(-1)^{(|k|+1)|i|}f^j_{ki}d^i_lp_j\omega^k\omega^l+\\ &&+
\frac{1}{2}\sum_{i,j,k,l,m}(-1)^{(|k|+1)|i|+(|l|+1)|m|}f^j_{ki}f^i_{lm}p_j\omega^k\omega^l\omega^m \\
&=& \sum_{i,j}<e^i,d^2e_j>p_i\omega^j+\\ &&+\frac{1}{2}\sum_{i,j,k}((-1)^{(|j|+1)|k|}<e^i,d[e_j,e_k]>+2(-1)^{(|j|+1)(|k|+1)}<e^i,[e_j,de_k]>)p_i\omega^j\omega^k+\\
&&+\frac{1}{2}\sum_{i,j,k,l}(-1)^{(|j|+1)(|k|+|l|)+(|k|+1)|l|}<e^i,[e_j,[e_k,e_l]]>p_i\omega^j\omega^k\omega^l \\
&=&\sum_{i,j}<e^i,d^2e_j>p_i\omega^j+\\
&&+\frac{1}{2}\sum_{i,j,k}(-1)^{(|j|+1)|k|}<e^i,d[e_j,e_k]-[de_j,e_k]-(-1)^{|j|}[e_j,de_k]>p_i\omega^j\omega^k+\\
&&+\frac{1}{6}\sum_{i,j,k,l}(-1)^{(|j|+1)(|k|+|l|)+(|k|+1)|l|}\cdot \\
&&\cdot<e^i,[e_j,[e_k,e_l]]+(-1)^{|j|(|k|+|l|)}[e_k,[e_l,e_j]]+
(-1)^{(|j|+|k|)|l|}[e_l,[e_j,e_k]]>p_i\omega^j\omega^k\omega^l
\end{eqnarray*}
Here we notice that coefficients of all monomials vanish: due to identity $d^2=0$ for monomials $p_i\omega^j$, due to Leibniz identity for monomials $p_i\omega^j\omega^k$, and due to Jacobi identity for monomials $p_i\omega^j\omega^k\omega^l$. Hence (\ref{CME for abstract BF}) holds. Moreover it is clear that CME  (\ref{CME for abstract BF}) is the generating equation for quadratic relations on operations on DGLA operations --- the Poincar\'{e}, Leibniz and Jacobi identities (\ref{Poincare identity},\ref{Leibniz identity},\ref{Jacobi identity}). Let us now check (\ref{QME for abstract BF}):
\begin{eqnarray*}
\Delta S&=&\sum_i (-1)^{|i|+1}\frac{\dd}{\dd\omega^i}\frac{\dd}{\dd p_i} S\\
&=&\sum_i (-1)^{|i|+1} d^i_i+\sum_{i,k}(-1)^{(|i|+1)(|k|+1)}f^i_{ik}\omega^k\\
&=&0+\sum_{i,k} (-1)^{|i|+|k|}<e^i,[e_k,e_i]>\omega^k\\
&=&\sum_{k:\;|k|=0}(\Str_V[e_k,\bt])\;\omega^k
\end{eqnarray*}
Therefore (\ref{QME for abstract BF}) is equivalent to the unimodularity property for $V$ (\ref{unimodularity}).
\\$\Box$

\textbf{Remark.} Notice that these computations are much more elegant if done in terms of super-fields. For CME we have:
\begin{eqnarray*}
\frac{1}{2}\{S,S\}&=&S\left<\frac{\ola\dd}{\dd\omega},\frac{\ora\dd}{\dd p}\right>S\\
&=&\left<<p,d\bt+[\omega,\bt]>,d\omega+\frac{1}{2}[\omega,\omega] \right>\\
&=&<p,d(d\omega+\frac{1}{2}[\omega,\omega])+[\omega,d\omega+\frac{1}{2}[\omega,\omega]]>\\
&=&<p,d^2\omega>+<p,\frac{1}{2}d[\omega,\omega]+[\omega,d\omega]>+<p,\frac{1}{2}[\omega,[\omega,\omega]]>
\end{eqnarray*}
where the first term generates Poincar\'{e} identity, second --- the Leibniz identity, third --- the Jacobi identity. Finally, computation of (\ref{CME for abstract BF}) in super-field formalism is the following:
\begin{eqnarray*}\Delta S &=& -\left<\frac{\ora\dd}{\dd\omega},\frac{\ora\dd}{\dd p}\right>S\\
&=&-\left<\frac{\ora\dd}{\dd\omega},d\omega+\frac{1}{2}[\omega,\omega]\right>\\
&=&\Str_V d+\Str_V[\omega,\bt]=\Str_V[\omega,\bt]
\end{eqnarray*}
Thus equation $\Delta S=0$ generates the unimodularity condition for $V$. Benefits of super-field formalism are evident in this example: the computations are less messy and signs are easier to track.

\textbf{Main example.}
The ordinary topological $BF$ theory (more exactly, its canonical version), discussed in section  \ref{BF theory}, corresponds to special case of abstract $BF$ theory, where the DGLA $V$ is the algebra of differential forms on manifold $M$ with coefficients in Lia algebra $\g$:\quad $V=\Omega^\bt(M,\g)$. The super-fields $\omega$ and $p$ of algebraic $BF$ theory are then precisely the super-fields (\ref{omega physical},\ref{p physical}) of section \ref{BF theory}, constructed from classical fields, ghosts and anti-fields. Action (\ref{abstract BF action}) for this choice of $V$ is (\ref{BF canonical BV action}). For QME to hold we have to require unimodularity of gauge Lie algebra $\g$.

\subsection{Effective BV action: general idea}
\label{effective action: idea}
Let $W$ and $W'$ be a pair of $\ZZ$-graded vector spaces, and let $\iota: W'\ra W$ and $r: W\ra W'$ be the embedding and projection: a pair of linear maps of degree 0, satisfying $r\circ\iota=\id_{W'}$. Thus $W$ splits into a sum of vector subspaces
\be W=\iota(W')\oplus W''\label{W decomp}\ee
with $W''=\ker(r)\subset W$.
Next, let $\FF=T^*[-1]W=W\oplus W^*[-1]$ and $\FF'=T^*[-1]W'=W'\oplus W'^*[-1]$ be the corresponding odd cotangent bundles, equipped with canonical BV Laplacians $\Delta$ and $\Delta'$ (associated to the constant Berezinians on $\FF$ and $\FF'$). Space $\FF$ splits analogously to (\ref{W decomp}):
\be\FF=\hat\iota (\FF')\oplus \FF''\label{fields decomp}\ee
where $\hat\iota=\iota\oplus r^*: \FF'\ra\FF$ is the embedding of $\FF'$ into $\FF$ and $\hat r=r\oplus\iota^*: \FF\ra\FF'$ is the corresponding projection, and $\FF''$ is again the kernel of projection: $\FF''=\ker(\hat r)=T^*[-1]W''\subset\FF$. We call $\FF$ the space of fields, $\FF'$ the space of infrared (IR) fields, $\FF''$ the space of ultraviolet (UV) fields. We will also need the pull-backs, associated to the embedding and projection: $\hat\iota^*:\Fun(\FF)\ra\Fun(\FF')$ and $\hat{r}^*:\Fun(\FF')\ra\Fun(\FF)$. Algebra of functions on $\FF$ factorizes as $\Fun(\FF)=\hat{r}^*\Fun(\FF')\otimes\Fun(\FF'')\cong\Fun(\FF')\otimes\Fun(\FF'')$. We will use the latter isomorphism for convenience of notations, but one should keep in mind that it depends on $\iota$ and $r$. The BV Laplacian on $\FF$ splits as
\be\Delta=\Delta'+\Delta''\label{Delta decomposition}\ee
where $\Delta''$ is the canonical BV operator on $\FF''$.
Let also $S\in\Fun(\FF)$ be some BV action on $\FF$, i.e. a function satisfying the QME: $\Delta e^{S/\hbar}=0$.

We define the induced (effective) action on IR fields $S'\in\Fun(\FF')$ via an integral over Lagrangian submanifold $\LL\subset\FF''$:
\be e^{S'(z')/\hbar}=\int_\LL e^{S(\hat\iota(z')+z'')/\hbar}\label{effective action}\ee
where $z'\in\FF'$, $z''\in\FF''$. We use the measure on $\LL$ induced from the Lebesgue measure on ambient space $\FF''$:\quad $\mu_\LL=\sqrt{\mu_{\FF''}}\,|_\LL$.
Expression (\ref{effective action}) should be understood as the fiber BV integral $(\hat{r}_\LL)_* e^{S/\hbar}$ for the (fiberwise Lagrangian) sub-bundle $\hat{r}_\LL:\FF'\oplus\LL\ra \FF'$ of the bundle $\hat{r}:\FF\ra\FF'$.

\begin{statement}\label{statement: QME for induced action} Effective action $S'$ defined by (\ref{effective action}) satisfies QME on $\FF'$, i.e.
$$\Delta' e^{S'/\hbar}=0$$
\end{statement}

\textbf{Proof.}
Indeed, we have
\begin{eqnarray*}
\Delta'e^{S'/\hbar}&=&\int_\LL \Delta' e^{S/\hbar}\\
&=&\int_\LL (\Delta-\Delta'')e^{S/\hbar}\\
&=&0-\int_\LL\Delta''e^{S/\hbar}\\
&=&0
\end{eqnarray*}
where first we use the splitting for BV Laplacian $\Delta$, then the QME for $S$ (on the space $\FF$) and, lastly, we use theorem (\ref{int of BV coboundary}) for the integral of $\Delta''$-coboundary over the Lagrangian submanifold $\LL\subset\FF''$.
\\$\Box$

\begin{statement}\label{statement: induced canonical transformation} Canonical transformation of action $S$ on $\FF$: $$S\mapsto S+\{S,R\}+\hbar\Delta R$$ with the generator
$R\in\Fun(\FF)$, $\gh(R)=-1$ leads to the canonical transformation of effective action $$S'\mapsto S'+\{S',R'\}+\hbar\Delta' R'$$ with the generator of induced transformation given by
\be R'=e^{-S'/\hbar}\cdot\int_\LL e^{S/\hbar}R\label{induced canonical transformation}\ee
\end{statement}

\textbf{Proof.} Let us use formula (\ref{canonical transformation for exp(S)}) for the canonical transformation of exponential of the action on $\FF$ to compute the transformation of the integral (\ref{effective action}):
\begin{eqnarray*}e^{S'/\hbar}&\mapsto& e^{\tilde{S}'/\hbar}=\int_\LL e^{S/\hbar}+\Delta (e^{S/\hbar}R)\\
&=&e^{S'/\hbar}+\int_\LL (\Delta'+\Delta'')(e^{S/\hbar}R)\\
&=&e^{S'/\hbar}+\Delta'\int_\LL e^{S/\hbar}R+0\\
&=&e^{S'/\hbar}+\Delta'\left(e^{S'/\hbar}\left(e^{-S'/\hbar}\cdot\int_\LL e^{S/\hbar}R\right)\right)
\end{eqnarray*}
where we again use the splitting of the BV Laplacian (\ref{Delta decomposition}) and the theorem (\ref{int of BV coboundary}) for the integral of a BV-coboundary. The last expression is precisely the exponential form for canonical transformation for the effective action $S'$ with the generator (\ref{induced canonical transformation}).
\\$\Box$

The induction data for the effective action $S'$ on $\FF'$ is the triplet $(\iota,r,\LL)$ --- embedding, projection, defining the splitting (\ref{fields decomp}) and the Lagrangian submanifold $\LL\subset\FF''$ (and $\FF''\subset\FF$  is itself defined by $r$).
\begin{statement}\label{statement: induction data deform} Continuous deformation of the induction data $(\iota,r,\LL)$ leads to a canonical transformation for the effective action $S'$.
\end{statement}
\textbf{Proof.} For the proof it suffices to consider the infinitesimal deformations of the induction data $(\iota,r,\LL)$. These deformations fall into two types:
\begin{itemize}
\item Deformations of type I: $(\iota,r,\LL)\mapsto (\iota,r,\LL_\Psi)$, i.e.  deformations of the Lagrangian submanifold $\LL\mapsto \LL_\Psi$ given by gauge fixing fermions $\Psi\in\Fun(\LL)$, $\gh(\Psi)=-1$. Embedding and projection are not changed.
\item Deformations of type II: $(\iota,r,\LL)\mapsto (\iota+\delta\iota, r+\delta r,\tilde\LL)$, i.e. infinitesimal deformations of embedding and projection, supplemented by a ``minimal'' deformation of $\LL$ (we will elaborate on this further). Deformation of $\LL$ is necessary, since $\LL$ has to be a Lagrangian submanifold in $\FF''$, while the latter changes under change of $\iota,r$.
\end{itemize}
Consider type I deformations first. Let $(x''^i,\xi''_i)$ be a Darboux coordinate system on $\FF''$ for which $\LL$ is given by the equation $\xi''=0$. Then
\begin{eqnarray*}e^{S'/\hbar}&\mapsto& \int_{\LL_\Psi}e^{S/\hbar}\\
&=&\int_\LL \left(e^{S/\hbar}-\left(\frac{\dd}{\dd x''^i}\Psi(x'')\right)\frac{\dd}{\dd \xi''_i}e^{S/\hbar}\right)\\
&=&e^{S'/\hbar}+\int_\LL\Psi\cdot\Delta''e^{S/\hbar}\\
&=&e^{S'/\hbar}+\int_\LL\Psi\cdot(\Delta-\Delta')e^{S/\hbar}\\
&=&e^{S'/\hbar}-\Delta'\left(\int_\LL\Psi e^{S/\hbar}\right)
\end{eqnarray*}
Hence the deformation of $\LL$ by a gauge fixing fermion $\Psi$ leads to the canonical transformation for $S'$ with the generator
$$R'=-e^{-S'/\hbar}\int_\LL\Psi e^{S/\hbar}$$

Consider now the type II deformations. Let $\iota\mapsto \tilde\iota=\iota+\delta\iota$, $r\mapsto\tilde r=r+\delta r$ be an infinitesimal deformation of embedding and projection, where $\delta\iota: W'\ra W$ and $\delta r: W\ra W'$. The condition that (deformed) embedding and retraction still invert each other means that $(r+\delta r)\circ(\iota+\delta\iota)=\id_{W'}$ and hence $\delta\iota$ and $\delta r$ are subject to relation
$$r\circ\delta\iota+\delta r\circ\iota=0$$
The next idea is to trade deformation of $\iota$ and $r$ for a certain canonical transformation on the space of fields $\FF$, which is in turn equivalent to changing initial action to a canonically transformed one. Then we use Statement \ref{statement: induced canonical transformation} to induce the canonical transformation on the effective action.

Introduce a linear deformation of identity map $\id_W+\delta\phi: W\ra W$ with
\be\delta\phi=\delta\iota\circ r-\iota\circ\delta r\circ \PP''=\PP''\circ\delta\iota\circ r-\iota\circ\delta r\label{delta phi}\ee
where $\PP''=\id_W-\iota\circ r$ is the projector to the second subspace in splitting (\ref{W decomp}) (``UV projector''). Then the following two properties hold:
\begin{eqnarray*}
\iota+\delta\iota&=&(\id_W+\delta\phi)\circ\iota\\
r+\delta r&=& r\circ (\id_W+\delta\phi)^{-1}=r\circ (\id_W-\delta\phi)
\end{eqnarray*}
(we work in first order in deformations).

On the level of cotangent bundles $\FF,\FF'$ the picture is as follows. Embedding $\hat\iota=\iota\oplus r^*:\FF'\ra\FF$ and projection $\hat r=r\oplus\iota^*:\FF\ra\FF'$ are deformed as $\hat\iota\mapsto\hat{\tilde\iota}=\hat\iota+\delta\hat\iota$, $\hat r\mapsto\hat{\tilde r}=\hat r+\delta\hat r$, where
\begin{eqnarray*}
\delta\hat\iota&=&\delta\iota\oplus\delta r^*\\
\delta\hat r&=&\delta r\oplus \delta\iota^*
\end{eqnarray*}
And there is a linear deformation of identity map $\id_\FF+\delta\hat{\phi}:\FF\ra\FF$, where $$\delta\hat\phi=\delta\phi\oplus(-\delta\phi^*)$$
The pull-back on functions $(\id_\FF+\delta\hat{\phi})^*:\Fun(\FF)\ra\Fun(\FF)$ is the infinitesimal canonical transformation $f\mapsto f+\{f,R\}$ with generator $$R=<\xi,\delta\phi (x)>$$
where $x\in W$, $\xi\in W^*[-1]$ and $<\bt,\bt>$ is the canonical pairing.
Note that \be \Delta'' R=\Str_{W''}\delta\phi=0\label{Delta R=0}\ee since $\delta\phi$ maps from $W''$ to $\iota(W')$.
Next, $\delta\hat\phi$ has the following properties:
\begin{eqnarray*}
\hat\iota+\delta\hat\iota&=&(\id_\FF+\delta\hat\phi)\circ\hat\iota\\
\hat r+\delta\hat r&=& \hat r\circ (\id_\FF-\delta\hat\phi)
\end{eqnarray*}
The latter property means in particular that the map $\id_\FF+\delta\hat\phi$ moves $\FF''=\ker(\hat r)\subset\FF$ to $\tilde\FF''=\ker(\hat r+\delta\hat r)\subset\FF$. Therefore we define the minimal deformation of $\LL$ as $$\tilde\LL=(\id_\FF+\delta\hat\phi)\LL$$
and hence the integral (\ref{effective action}) is transformed as
\begin{eqnarray} e^{S'/\hbar}&\mapsto&\int_{\tilde\LL}e^{S(\hat{\tilde\iota}(z')+\tilde z'')/\hbar}\label{induction data deform 1}\\
&=&\int_\LL e^{S((\id_\FF+\delta\hat\phi)\circ (\hat\iota(z')+z''))/\hbar}\label{induction data deform 2}\\
&=&\int_\LL (e^{S/\hbar}+\Delta (e^{S/\hbar}R))\nonumber\\
&=&e^{S'/\hbar}+\Delta'(e^{S'/\hbar}R')\nonumber
\end{eqnarray}
where $R'=e^{-S'/\hbar}\cdot\int_\LL e^{S/\hbar}R$ and in the last step we used the Statement \ref{statement: induced canonical transformation}. Also in the transition from integral over $\tilde\LL$ (\ref{induction data deform 1}) to integral over $\LL$ (\ref{induction data deform 2}) we use (\ref{Delta R=0}) (otherwise there could be additionally a contribution of deformation of measure on the Lagrangian submanifold).

So we showed that infinitesimal deformation of the induction data $(\iota,r,\LL)\mapsto (\iota+\delta\iota,r+\delta r,\tilde\LL_\Psi)$ leads to the canonical transformation for the effective action $S'$, with generator given by
\be R'=e^{-S'/\hbar}\cdot\int_\LL e^{S/\hbar}(-\Psi+<\xi,(\delta\iota\circ r-\iota\circ\delta r\circ\PP'')(x)>)\label{R' from deforming induction data}\ee
\\$\Box$

\textbf{Remark.} It is reasonable to split type II deformations into two subtypes: the type IIa (parallel) deformations, not changing the splitting $W=\iota(W')\oplus W''$ and hence coming from automorphisms of $W'$, and the type IIb (perpendicular) deformations, shifting $\iota(W')$ only in the direction of $W''$, and twisting the splitting $W=\iota(W')\oplus W''$. Conditions on $\delta\iota$ and $\delta r$ are
$$\delta\iota_{||}\circ r+\iota\circ\delta r_{||}=0$$ for IIa deformations and
$$r\circ\delta\iota_\bot=0$$ for type IIb deformations. This classification is equivalent to representing the automorphism $\delta\phi: W\ra W$ (\ref{delta phi}) as
\be\delta\phi=\iota\circ\delta\chi_{||}\circ r+\delta\phi_\bot=\iota\circ\left(r\circ\delta \iota\right)\circ r+
\left(\PP''\circ\delta\iota\circ r-\iota\circ\delta r\circ\PP''\right)\label{IIa and IIb separation of delta phi}\ee
where $\delta\chi_{||}=r\circ\delta \iota: W'\ra W'$ is the automorphism of $W'$, generating the parallel part of deformation, ans $\delta\phi_\bot=\PP''\circ\delta\iota\circ r-\iota\circ\delta r\circ\PP'': W\ra W$ is the automorphism of $W$, generating the perpendicular part of deformation. Parallel (IIa) and perpendicular (IIb) parts of deformation affect the effective action quite differently: the parallel part just induces a change of IR variables  $z'$ in \ref{effective action}), while the perpendicular part deforms the Lagrangian submanifold of integration itself. The splitting (\ref{IIa and IIb separation of delta phi}) suggests the following form of the result (\ref{R' from deforming induction data}), where effects of all three types (I, IIa, IIb) of deformations of induction data are split manifestly:
\be R'=<\xi',r\circ\delta\iota( x')>+e^{-S'/\hbar}\cdot\int_\LL e^{S/\hbar}(-\Psi+<\xi,(\PP''\delta\iota\circ r-\iota\circ\delta r\circ\PP'')(x)>)\label{R' from deforming induction data 2}\ee
where $x'\in W'$, $\xi'\in W'^*[-1]$.

\subsection{Effective action for the abstract $BF$ theory}
\label{section: effective action for abstract BF}
Now we will apply the formalism of the section \ref{effective action: idea} to the abstract $BF$ theory, associated to a DGLA $(V,d,[\bt,\bt])$, with the space of fields (\ref{space of fields}) and the action (\ref{abstract BF action}). In notations of section \ref{effective action: idea} we have to set $W=V[1]$.

Let  $V$ be a $\ZZ$-graded vector space endowed with a DGLA structure, i.e. the differential $d$ and Lie bracket $[\bt,\bt]$. Let also $(V',d)$ be a deformation retract of the cochain complex $(V,d)$, with embedding $\iota:V'\hra V$ and projection (more adequate term in this setting is ``retraction'')  $r:V\ra V'$ --- linear maps of degree 0, satisfying $r\iota=\id_V$. In this setting $\iota$ and $r$ are additionally required to be chain maps (i.e. $d\iota=\iota d$, $d r=r d$) inducing an isomorphism on cohomology $H^\bt(V')\cong H^\bt(V)$ (i.e. $\iota, r$ are quasi-isomorphisms). Thus we have a splitting of $V$ into IR and UV parts:
$$V=\iota(V')\oplus V''$$
where $V''=\ker(r)\subset V$. Now this splitting is consistent with the differential $d$ and the UV part $V''$ is acyclic. We define IR and UV projectors as
\begin{eqnarray*}\PP'&=&\iota r\; :V\ra V \\ \PP''&=&\id_V-\PP'\;:V\ra V\end{eqnarray*}
Next we introduce the odd cotangent bundles $\FF=T^*[-1](V[1])$, $\FF'=T^*[-1](V'[1])$, $\FF''=T^*[-1](V''[1])\subset\FF$ --- the space of fields of abstract $BF$ theory, space of IR fields (on which we will construct the effective action) and the space of UV fields. On the level of spaces of fields we have the splitting
$$\FF=\hat\iota(\FF')\oplus\FF''$$
where $\hat\iota=\iota\oplus r^*:\FF'\ra\FF$ is the embedding of IR fields, $\hat r=r\oplus \iota^*:\FF\ra\FF'$ is the retraction to IR fields.

To specify the Lagrangian submanifold $\LL$ for the integral (\ref{effective action}), we need to introduce a new object --- the chain homotopy operator, contracting $V$ to $\iota(V')$, i.e. a linear map $K:V^\bt\ra V^{\bt-1}$ of degree -1 satisfying
\begin{eqnarray} K\iota=rK&=&0 \label{K property 1}\\ dK+Kd&=&\PP'' \label{K property 2}\\ K^2&=&0 \label{K property 3}
\end{eqnarray}
Condition (\ref{K property 1}) means that $K$ acts nontrivially only on $V''\subset V$ and is zero on $\iota(V')\subset V$. Define the $d$-exact and $K$-exact parts of $V''$:\quad  $V''_{d-ex}=d(V'')$, $V''_{K-ex}=K(V'')$. Then (due to (\ref{K property 2})) we have
\be V''=V''_{d-ex}\oplus V''_{K-ex}\label{V'' decomp}\ee
and $d: V''_{K-ex}\ra V''_{d-ex}$ is the isomorphism (of vector spaces), and $K: V''_{d-ex}\ra V''_{K-ex}$  is its inverse. Projectors to the first and second subspaces in the decomposition (\ref{V'' decomp}) are $\PP''_{d-ex}=dK$ and $\PP''_{K-ex}=Kd$ respectively. Hence the triplet of maps $(\iota,r,K)$ defines the Hodge decomposition for $V$:
\be V=\iota(V')\oplus V''_{d-ex}\oplus V''_{K-ex}\label{Hodge decomp}\ee
The dual decomposition for  $V^*$ is
\be V^*=r^*(V')\oplus V''^*_{K^*-ex}\oplus V''^*_{d^*-ex}\label{dual Hodge decomp}\ee
where $d^*,K^*:V^*\ra V^*$ are the dual operators for $d,K$, and $V''^*_{K^*-ex}=K^*(V''^*)$, $V''^*_{d^*-ex}=d^*(V''^*)$.
Define the Lagrangian submanifold $\LL_K\subset\FF''$ as the odd conormal bundle
\be\LL_K=N^*[-1](V''_{K-ex}[1])=V''_{K-ex}[1]\oplus V''^*_{K^*-ex}[-2]\subset\FF''\label{L via K}\ee
Notice that $\LL$ is actually a linear subspace in $V$.

Next, using the construction (\ref{effective action}) we define the effective actionе $S'\in\Fun(\FF')$ for abstract $BF$ theory as
\begin{eqnarray}e^{S'(\omega',p')/\hbar}&=&\int_{\LL_K} e^{S(\iota(\omega')+\omega'',r^*(p')+p'')/\hbar}\label{BV integral 0}\\
&=&\int_{\LL_K} \exp\left(\frac{1}{\hbar}<r^*(p')+p'',d\iota(\omega')+d\omega''+\frac{1}{2}[\iota(\omega')+\omega'',\iota(\omega')+\omega'']>\right)\nonumber
\end{eqnarray}
where IR and UV super-fields are defined as in section \ref{abstract BF theory}, i.e. as the shifted identity maps $\omega':V'[1]\ra V'$, $p':V'^*[-2]\ra V'^*$, $\omega'': V''[1]\ra V''$, $p'':V''^*[-2]\ra V''^*$. Let $(e'_i)$ be a basis in $V'$,\; $(e''_I)$ be a basis in $V''_{K-ex}$ and $(e''_{\bar{I}})$ be a basis in $V''_{d-ex}$ (the dual bases in dual spaces are denoted the same, but with raised index). Then
\begin{eqnarray*}\omega'=e'_i\omega'^i&,& p'=p'_i e'^i \\ \omega''_{\LL}=\omega''_{K-ex}=e''_I\omega''^I &,& p''_{d^*-ex}=p''_I e''^I \\
\omega''_{d-ex}=e''_{\bar{I}}\omega''^{\bar{I}} &,& p''_{\LL}=p''_{K^*-ex}=p''_{\bar{I}} e''^{\bar{I}}
\end{eqnarray*}
where we introduced the splitting of UV super-fields $\omega'', p''$ into $K$-exact и $d$-exact parts: $\omega''=\omega''_{\LL}+\omega''_{d-ex}$, $p''=p''_{d^*-ex}+p''_{\LL}$ in accord with the splitting (\ref{V'' decomp}).
Written in coordinates on $\LL_K$, the integral (\ref{BV integral 0}) is
\begin{multline}\label{BV integral}e^{S'(\omega',p')/\hbar}= \\ =\frac{1}{N}\int e^{\frac{1}{\hbar}<r^*(p')+p''_{\LL},d\iota(\omega')+d\omega''_{\LL}+\frac{1}{2}[\iota(\omega')+\omega''_{\LL},\iota(\omega')+\omega''_{\LL}]>}
\prod_I \DD\omega''^I \prod_{\bar{I}} \DD p''_{\bar{I}}
\end{multline}
where $$N=\int e^{\frac{1}{\hbar}<p''_{\LL},d\omega''_{\LL}>} \prod_I \DD\omega''^I \prod_{\bar{I}} \DD p''_{\bar{I}}$$ accounts for the normalization of measure on $\LL_K$. Integral (\ref{BV integral}) is understood perturbatively, i.e. as the sum of Feynman diagrams, arising in the stationary phase decomposition\footnote{There is some abuse of terminology here: point $\omega''_{\LL}=p''_{\LL}=0$ is in general not a stationary point of $S$} near the point $\omega''_{\LL}=p''_{\LL}=0$ on $\LL_K$. Here $S_0=<p''_{\LL},d\omega''_{\LL}>$ is interpreted as the free (Gaussian) part of the action, and the other terms in action are treated as a perturbation. I.e. we decompose the action on $\LL_K$ as
$$S|_{\LL_K}=S_0+S_{int}$$
where the free part is
$$S_0=<p''_{\LL},d\omega''_{\LL}>=\sum_{\bar{I},J}d^{\bar{I}}_Jp''_{\bar{I}}\omega''^J$$
and the perturbation (``interaction''):
\begin{eqnarray*}S_{int}&=&<p',d\omega'>+\frac{1}{2}<p',r[\iota(\omega'),\iota(\omega')]>+<p',r[\iota(\omega'),\omega''_{\LL}]> \\ &&+
\frac{1}{2}<p',r[\omega''_{\LL},\omega''_{\LL}]>+\frac{1}{2}<p''_{\LL},[\iota(\omega'),\iota(\omega')]>  \\ && +<p''_{\LL},[\iota(\omega'),\omega''_{\LL}]>+
\frac{1}{2}<p''_{\LL},[\omega''_{\LL},\omega''_{\LL}]>\\
&=&\sum_{i,j}d^i_j p'_i \omega'^j+\frac{1}{2}\sum_{i,j,k}(-1)^{(|j|+1)|k|} f^i_{jk} p'_i \omega'^j\omega'^k+
\sum_{i,j,K}(-1)^{(|j|+1)|K|}p'_i\omega'^j\omega''^K \\ &&+\frac{1}{2}\sum_{i,J,K}(-1)^{(|J|+1)|K|}p'_i\omega''^J\omega''^K+
\frac{1}{2}\sum_{\bar{I},j,k}(-1)^{(|j|+1)|k|} f^{\bar{I}}_{jk} p''_{\bar{I}} \omega'^j\omega'^k \\ &&+
\sum_{\bar{I},j,K}(-1)^{(|j|+1)|K|} f^{\bar{I}}_{jK} p''_{\bar{I}} \omega'^j\omega''^K+
\frac{1}{2}\sum_{\bar{I},J,K}(-1)^{(|J|+1)|K|} f^{\bar{I}}_{JK} p''_{\bar{I}} \omega''^J\omega''^K
\end{eqnarray*}
We use the natural notations for the structure constants of the differential and of the Lie bracket:
$d^i_j=<e^i,e_j>$, $d^{\bar{I}}_J=<e^{\bar{I}},e_J>$, $f^i_{jk}=<e^i,r[\iota(e_j),\iota(e_k)]>$, $f^i_{jK}=<e^i,r[\iota(e_j),e_K]>$, $f^i_{JK}=<e^i,r[e_J,e_K]>$, $f^{\bar{I}}_{jk}=<e^{\bar{I}},[\iota(e_j),\iota(e_k)]>$, $f^{\bar{I}}_{jK}=<e^{\bar{I}},[\iota(e_j),e_K]>$, $f^{\bar{I}}_{JK}=<e^{\bar{I}},[e_j,e_k]>$.

Introduce the notation
$$\ll f\gg_0=\frac{1}{N}\int f e^{S_0/\hbar}\prod_I \DD\omega''^I \prod_{\bar{I}} \DD p''_{\bar{I}}$$
for the average of function $f\in\Fun(\FF)$ over UV fields with the Gaussian measure defined by free part of action $S_0$. Expectation value $\ll f\gg_0\in\Fun(\FF')$ is understood as the function of IR fields $\omega',p'$. Propagator for the integral (\ref{BV integral}) is
$$\ll\omega''^I p''_{\bar{J}}\gg_0=-\hbar K^I_{\bar{J}}$$
where $K^I_{\bar{J}}=<e^I,Ke_{\bar{J}}>$ is the matrix of the chain homotopy operator $K:V''_{d-ex}\ra V''_{K-ex}$, inverse to the matrix of differential $d^{\bar{J}}_I$.
According to the standard routine, the perturbation series for (\ref{BV integral}) is generated by the expression
\begin{eqnarray}e^{S'/\hbar}&=& \ll e^{S_{int}/\hbar}\gg_0  \nonumber
\\&=&\left.\left(\exp \left(-\hbar K^I_{\bar{J}}\frac{\dd}{\dd p''_{\bar{J}}}\frac{\dd}{\dd\omega''^I}\right)\circ e^{S_{int}/\hbar}\right)\right|_{\omega''=p''=0} \label{exp Wick formula} \\
&=&\left.\left(\exp \left<\frac{\ora\dd}{\dd \omega''_\LL},\hbar K\frac{\ora\dd}{\dd p''_\LL}\right>\circ e^{S_{int}/\hbar}\right)\right|_{\omega''=p''=0} \label{exp Wick formula super}
\end{eqnarray}
Sum over Feynman diagrams arises from expanding both exponentials in right hand side of (\ref{exp Wick formula}) or (\ref{exp Wick formula super}) into Taylor series. The standard argument of perturbation theory gives $S'$ as the sum of connected Feynman diagrams. We will first demonstrate the calculation of the first terms of perturbation series for (\ref{BV integral}), and then we state the general result.
First terms of Taylor expansions for the exponentials in (\ref{exp Wick formula super}) yield
\begin{multline}e^{S'/\hbar}=\exp\frac{1}{\hbar}\left((<p',d\omega'>+\frac{1}{2}<p',r[\iota(\omega'),\iota(\omega')]>\right)\cdot\\
\cdot (1
+\left<\frac{\ora\dd}{\dd \omega''_\LL},\hbar K\frac{\ora\dd}{\dd p''_\LL}\right>
\circ\frac{1}{\hbar}<p''_\LL,[\iota(\omega'),\omega''_\LL]>
+\left<\frac{\ora\dd}{\dd \omega''_\LL},\hbar K\frac{\ora\dd}{\dd p''_\LL}\right>\circ\\
\circ\frac{1}{\hbar}<p',r[\iota(\omega'),\omega''_\LL]>\frac{1}{2\hbar}<p''_\LL,[\iota(\omega'),\iota(\omega')]>+\\
+\left<\frac{\ora\dd}{\dd \omega''_{\LL(1)}},\hbar K\frac{\ora\dd}{\dd p''_{\LL(1)}}\right>\left<\frac{\ora\dd}{\dd \omega''_{\LL(2)}},\hbar K\frac{\ora\dd}{\dd p''_{\LL(2)}}\right>\circ\\ \circ\frac{1}{2!}\frac{1}{\hbar}<p''_{\LL(1)},[\iota(\omega'),\omega''_{\LL(1)}]>\frac{1}{\hbar}<p''_{\LL(2)},[\iota(\omega'),\omega''_{\LL(2)}]>+\\
+\left<\frac{\ora\dd}{\dd \omega''_{\LL(1)}},\hbar K\frac{\ora\dd}{\dd p''_{\LL(1)}}\right>\left<\frac{\ora\dd}{\dd \omega''_{\LL(2)}},\hbar K\frac{\ora\dd}{\dd p''_{\LL(2)}}\right>\circ\\
\circ \frac{1}{2!}\frac{1}{\hbar}<p''_{\LL(1)},[\iota(\omega'),\omega''_{\LL(2)}]>\frac{1}{\hbar}<p''_{\LL(2)},[\iota(\omega'),\omega''_{\LL(1)}]>+\cdots)
\end{multline}
Here indices $(1)$, $(2)$ tell which derivatives acts on which variables. We also separated the part of $S_{int}$ independent of UV fields. To evaluate the terms of perturbation series we use the following two identities:
\be \left<\frac{\ora\dd}{\dd \omega''_{\LL(1)}},K\frac{\ora\dd}{\dd p''_{\LL(1)}}\right>\circ <f,\omega''_{\LL(1)}><p''_{\LL(1)},g>=<f,-K g>\label{op form 1}\ee
where $f\in\Fun(\FF)\otimes V^*$ and $g\in\Fun(\FF)\otimes V$, and the second identity:
\be \left<\frac{\ora\dd}{\dd \omega''_{\LL(1)}},K\frac{\ora\dd}{\dd p''_{\LL(1)}}\right>\circ <p''_{\LL (1)},\OO\omega''_{\LL (1)}>=(-1)^{\gh<p,\OO\omega>}\Str_V K\OO=-(-1)^{\gh<p,\OO\omega>}\Str_V(-K\OO)\label{op form 2}\ee
where $\OO\in\Fun(\FF)\otimes\mr{End}(V)$ is a linear operator, $\Str_V$ is the super-trace over $V$, defined by
$$\Str_V \OO=\sum_a (-1)^{|a|(1+\gh<e^a,\OO e_a>)}<e^a,\OO e_a>$$
Now we can write the first few terms of the perturbation series for (\ref{BV integral}) as
\begin{multline}e^{S'/\hbar}=\exp\frac{1}{\hbar}\left((<p',d\omega'>+\frac{1}{2}<p',r[\iota(\omega'),\iota(\omega')]>\right)\cdot (1-\;
\Str_V (-K[\iota(\omega'),\bt])+\\
+\hbar^{-1}\frac{1}{2}<p',r[\iota(\omega'),-K[\iota(\omega'),\iota(\omega')]]>+\frac{1}{2}\Str_V (-K[\iota(\omega'),\bt])\cdot \Str_V (-K[\iota(\omega'),\bt])-\\
-\frac{1}{2}\;\Str_V (-K[\iota(\omega'),-K[\iota(\omega'),\bt]])+\cdots) \label{pert series for exp}
\end{multline}
The fact that the operator $-\hbar K$ plays the role of propagator for super-fields (the ``super-propagator'') is also implied by the following observation:
$$\ll\omega''_\LL\otimes p''_\LL\gg_0=-\hbar K$$
where $\omega''_\LL\otimes p''_\LL$ is understood as an element of $\Fun(\LL)\otimes V\otimes V^*\cong \Fun(\LL)\otimes\mr{End}(V)$.
Transition to the series for effective action $S'$ itself amounts to taking logarithm of the series (\ref{pert series for exp}). Considerable portion of terms disappear (namely, the ones corresponding to non-connected Feynman diagrams):
\begin{multline}S'=<p',d\omega'>+\frac{1}{2}<p',r[\iota(\omega'),\iota(\omega')]>+\frac{1}{2}<p',r[\iota(\omega'),-K[\iota(\omega'),\iota(\omega')]]>+\\
+\frac{1}{2}<p',r[\iota(\omega'),-K[\iota(\omega'),-K[\iota(\omega'),\iota(\omega')]]]>+\\
+\frac{1}{8}<p',r[-K[\iota(\omega'),\iota(\omega')],-K[\iota(\omega'),\iota(\omega')]]>+\cdots
-\hbar\;\Str_V (-K[\iota(\omega'),\bt])-\\
-\hbar\frac{1}{2}\;\Str_V (-K[\iota(\omega'),-K[\iota(\omega',\bt)]])- \hbar\frac{1}{2}\;\Str_V (-K[-K[\iota(\omega'),\iota(\omega')],\bt])+\cdots \label{pert series for S'}
\end{multline}

To formulate the general result, it is convenient to use the notations for iterated operations. Let $T$ be a planar binary rooted tree (edges are supposed to be oriented in the direction of root). Leaves and the root are understood as external edges (with single incident vertex), all vertices have valence 3. Each internal edge $e\in T$ has a parent edge $e_p$ (such that $e_p$ and $e$ have a common incident vertex, and $e_p$ is placed closer to the root) and two children: the left $e_l$ and right $e_r$ (leaves only have a parent, the root has only children). To denote a tree we make use of the standard correspondence between trees and bracket structures.

Next, let $X$ and $Y$ be two $\ZZ$-graded vector spaces and let $\lambda_2: X\otimes X\ra X$, $\tilde{\lambda}_2: X\otimes X\ra Y$ be two bilinear maps. Then for a tree $T$ with $|T|=n$ leaves we define the $n$-linear map
$$\Iter_{T;\lambda_2;\tilde{\lambda}_2}: X^{\otimes n}\ra Y$$
by the following algorithm. Let $(x_1,\cdots,x_n)\in X^n$. Decorate $i$-th leaf (going around the tree counterclockwise, starting from the root) by $x_i$. Then we gradually decorate the other edges  $e\in T$ following the rule $x_e=\lambda_2 (x_{e_l},x_{e_r})$ for internal edges $e$ and $x_{root}=\tilde{\lambda}_2 (x_{root_l},x_{root_r})$ for the root. The value of iterated operation is defined as
$\Iter_{T;\lambda_2;\tilde{\lambda}_2}(x_1,\cdots,x_n)=x_{root}$.
For example:
$$\Iter_{((*(**))(**));\lambda_2;\tilde{\lambda}_2}(x_1,x_2,x_3,x_4,x_5)=\tilde{\lambda}_2(\lambda_2(x_1,\lambda_2(x_2,x_3)),\lambda_2(x_4,x_5))$$

We also need to introduce a notation for traces of iterated operations. Namely, let $L$ be a planar oriented graph with one cycle (``one-loop'' in terminology of Feynman diagrams), whose vertices are of valence 3 (moreover, there are always two incoming edges and one outgoing), and with $|L|=n$ external edges --- leaves (oriented as incoming).
If we cut $L$ along any edge in the cycle, we obtain a tree $T$ with $n+1$ leaves, one of which is marked (the one corresponding to the cut edge). We define $n$-linear map
$\Loop_{L;\lambda_2}: X^{\otimes n}\ra \RR$
as a super-trace over $X$
$$\Loop_{L;\lambda_2}(x_1,\ldots,x_n)=\Str_X\;\Iter_{T;\lambda_2;\lambda_2}(x_1,\cdots,x_{i-1},\bt,x_i,\cdots,x_n)$$
where $i$ is the position of the marked leaf in $T$ (we enumerate the leaves, going around $T$ counterclockwise, starting from the root). Example:
$$\Loop_{(((**)\bt)*);\lambda_2}=\Str_X\lambda_2(\lambda_2(\lambda_2(x_1,x_2),\bt),x_3)$$
where $*$ denotes non-marked leaves and $\bt$ is the marked leaf.

Let us denote $\bf{T}_\mr{Pl}$ the set of planar trees and $\bf{L}_\mr{Pl}$ the set of planar oriented one-loop graphs. We also denote $\bf{T}_\mr{nonPl}$ the set of trees without planar structure (i.e. the set $\bf{T}_\mr{Pl}$ modulo graph isomorphisms) and  denote $\bf{L}_\mr{nonPl}$ the set of one-loop graphs without planar structure. For a non-planar graph $\Gamma$ we denote $\mr{Aut}(\Gamma)$ its automorphism group, i.e. the group of permutations of vertices not changing the incidence matrix. The order of group of automorphisms is denoted $|\mr{Aut}(\Gamma)|$.
Now we can state the general perturbative result for (\ref{BV integral}).

\begin{thm}\label{thm: eff action for abstract BF} Effective action for the abstract $BF$ theory, associated to a DGLA $(V,d,[,])$, induced on the space $\FF'=T^*[-1]V'[1]$ and defined by the integral (\ref{BV integral}), has the form
\be S'(\omega',p')=S'^0(\omega',p')+\hbar S'^1(\omega')\label{S' decomp into tree and loop}\ee
where the tree part is the sum over non-planar binary rooted trees:
\be S'^0(\omega',p')=<p',d\omega'>+\sum_{T\in{\bf{T}}_\mr{nonPl},|T|\geq 2}\frac{1}{|\mr{Aut}(T)|}<p',\Iter_{T;-K[\bt,\bt];\,r[\bt,\bt]}(\iota(\omega'),\cdots,\iota(\omega'))> \label{sum over trees}\ee
and the one-loop part is the sum over non-planar connected oriented one-loop graphs:
\be S'^1(\omega')=-\sum_{L\in\bf{L}_\mr{nonPl}}\frac{1}{|\mr{Aut}(L)|}\Loop_{L;-K[\bt,\bt]}(\iota(\omega'),\cdots,\iota(\omega'))\label{sum over loops}\ee
First terms for $S'^0$ are
\begin{multline}\label{pert series for tree effective action}S'^0(\omega',p')=<p',d\omega'>+\frac{1}{2}<p',r[\iota(\omega'),\iota(\omega')]>+\frac{1}{2}<p',r[\iota(\omega'),-K[\iota(\omega'),\iota(\omega')]]>+\\
+\frac{1}{2}<p',r[\iota(\omega'),-K[\iota(\omega'),-K[\iota(\omega'),\iota(\omega')]]]>+\\
+\frac{1}{8}<p',r[-K[\iota(\omega'),\iota(\omega')],-K[\iota(\omega'),\iota(\omega')]]>+\cdots
\end{multline}
(first two terms yield the restriction of initial action on the IR fields $\hat\iota^*\circ S=S(\iota(\omega'),r^*(p'))$).  For the one-loop part $S'^1$:
\begin{multline}\label{pert series for loop effective action}S'^1(\omega')=-\;\Str_V (-K[\iota(\omega'),\bt])-\frac{1}{2}\;\Str_V (-K[\iota(\omega'),-K[\iota(\omega',\bt)]])-\\ - \frac{1}{2}\;\Str_V (-K[-K[\iota(\omega'),\iota(\omega')],\bt])
- \frac{1}{3}\;\Str_V (-K[\iota(\omega'),-K[\iota(\omega'),-K[\iota(\omega'),\bt]]])-\\
-\frac{1}{2}\;
\Str_V (-K[-K[\iota(\omega'),\iota(\omega')],-K[\iota(\omega',\bt)]])
-\frac{1}{2}\;\Str_V (-K[-K[\iota(\omega'),-K[\iota(\omega'),\iota(\omega')]],\bt])
+\cdots
\end{multline}
\end{thm}
This result is standard and essentially contained in (\ref{exp Wick formula super}), we just introduced the ``operation formalism'' for Feynman diagrams, which is convenient for this specific theory and justified by identities (\ref{op form 1},\ref{op form 2}). The important property of integral (\ref{BV integral}) is that no Feynman diagrams with more than one loop appear. This is due to the fact that edges of Feynman graphs are oriented (since the free part of action couples the fields $\omega$ and $p$, living in dual spaces) and due to the special rules of orientation of edges: each vertex has exactly one outgoing edge (since  $S_{int}$  is linear in field $p$).

Feynman rules for (\ref{BV integral}) can be formulated as follows: the propagator decorating the internal edges is $-K$. The ``incoming'' external edges are decorated with $\iota(\omega')$, ``outgoing'' --- with $r^*(p')$. In 3-valent vertices one evaluates the Lie bracket $[\bt,\bt]$. There is also a special graph --- the trivial ``tree'' with one leaf, where we decorate the edge with $d$.

The following statement is a direct consequence of the construction of effective action via the integral (\ref{BV integral 0}) and the Statement \ref{statement: QME for induced action}.
\begin{statement} Effective action $S'$ for the abstract $BF$ theory satisfies QME:
$$\Delta' e^{S'(\omega',p')/\hbar}=0$$
\end{statement}
It is informative to check the QME explicitly in lowest orders of perturbation series (\ref{pert series for S'}). Due to (\ref{S' decomp into tree and loop}) and due to independence of $S'^1$ on $p'$ (which implies $\Delta'S'^1=\{S'^1,S'^1\}=0$) the QME for $S'$ is equivalent to the system
(\ref{CME},\ref{QME1}):
\begin{eqnarray} \{S'^0,S'^0\}&=&0 \label{CME for S'}\\
\{S'^0,S'^1\}+\Delta S'^0&=&0 \label{QME for S'}
\end{eqnarray}
Let us check the CME in lowest orders in $\omega'$ using the super-field formalism:
\begin{multline*}
\frac{1}{2}\{S'^0,S'^0\}=S'^0\left<\frac{\ola\dd}{\dd\omega'},\frac{\ora\dd}{\dd p'}\right>S'^0=\\
=\left< <p',d\bt>,d\omega'\right>+\left< <p',d\bt>,\frac{1}{2}r[\iota(\omega'),\iota(\omega')]\right>+
\left<<p',r[\iota(\omega'),\iota(\bt)]>,d\omega'\right>+\\
+\left<<p',d\bt>,\frac{1}{2}r[\iota(\omega'),-K[\iota(\omega'),\iota(\omega')]]\right>+
\left<<p',r[\iota(\omega'),\iota(\bt)]>,\frac{1}{2}r[\iota(\omega'),\iota(\omega')]\right>+\\
+\left<<p',r[\iota(\omega'),-K[\iota(\omega'),\iota(\bt)]]>-\frac{1}{2}<p',r[\iota(\bt),-K[\iota(\omega'),\iota(\omega')]]>,d\omega'\right>+
\cdots= \\
=<p',d^2\omega'>+<p',r\left(\frac{1}{2}d[\iota(\omega'),\iota(\omega')]+[\iota(\omega'),d\iota(\omega')]\right)>+\\
+<p',r\left(\frac{1}{2}[\iota(\omega'),\iota\circ r [\iota(\omega'),\iota(\omega')]]
+\frac{1}{2}d[\iota(\omega'), -K [\iota(\omega'),\iota(\omega')]]+\right.\\
\left.+[\iota(\omega'), -K[\iota(\omega'),d\iota(\omega')]]-
\frac{1}{2}[d\iota(\omega'), -K [\iota(\omega'),\iota(\omega')]]\right)>+\cdots= \\
=0+0+\frac{1}{2}<p',r[\iota(\omega'),(\PP'+dK+Kd)[\iota(\omega'),\iota(\omega')]]>+\cdots=
\\=\frac{1}{2}<p',r[\iota(\omega'),[\iota(\omega'),\iota(\omega')]]>+\cdots=0+O(p'\omega'^4)
\end{multline*}
Hence, in the order $O(p'\omega')$ the CME holds due to the Poincar\'{e} identity $d^2=0$, in the order $O(p'\omega'^2)$ --- due to Leibniz identity for $V$, in higher orders --- due to Leibniz and Jacobi identities for $V$ and the defining property of the chain homotopy $\PP'+dK+Kd=\id_V$. Let us now check the quantum part of QME:
\begin{multline*}\{S'^0,S'^1\}+\Delta S'^0=S'^1\left<\frac{\ola\dd}{\dd\omega'},\frac{\ora\dd}{\dd p'}\right>S'^0-\left<\frac{\ora\dd}{\dd\omega'},\frac{\ora\dd}{\dd p'}\right>S'^0 \\=(\left<-\Str_V(-K[\iota(\bt_2),\bt_1]),d\omega'\right>+
\left<-\Str_V(-K[\iota(\bt_2),\bt_1]),\frac{1}{2}r[\iota(\omega'),\iota(\omega')]\right>+
\\ +\left<-\Str_V(-K[\iota(\omega'),-K[\iota(\bt_2),\bt_1]]),d\omega'\right>+
\left<-\Str_V(-K[-K[\iota(\omega'),\iota(\bt_2)],\bt_1]),d\omega'\right>+\cdots)+\\
+(\Str_{V'}d+\Str_{V'}r[\iota(\omega'),\iota(\bt)]+\Str_{V'}r[\iota(\omega'),-K[\iota(\omega'),\iota(\bt)]]+
\frac{1}{2}\Str_{V'}r[-K[\iota(\omega'),\iota(\omega')],\iota(\bt)]+\cdots)\\
=(-\Str_V(-K[d\iota(\omega'),\bt])-
\frac{1}{2}\Str_V(-K[\iota r[\iota(\omega'),\iota(\omega')],\bt])-
\\ -\Str_V(-K[\iota(\omega'),-K[d\iota(\omega'),\bt]])-
\Str_V(-K[-K[\iota(\omega'),d\iota(\omega')],\bt])+\cdots)+
\\+(\Str_{V}\PP'[\iota(\omega'),\bt]+\Str_{V}\PP'[\iota(\omega'),-K[\iota(\omega'),\bt]]+
\frac{1}{2}\Str_{V}\PP'[-K[\iota(\omega'),\iota(\omega')],\bt]+\cdots) \\
=\Str_V (\PP'+dK+Kd)[\iota(\omega'),\bt]+\Str_V (-K[\iota(\omega'),(\PP'+dK+Kd)[\iota(\omega'),\bt]])- \\
-\frac{1}{2}\Str_V(-K[(\PP'+dK+Kd)[\iota(\omega'),\iota(\omega')],\bt])+\cdots \\
=\Str_V [\iota(\omega'),\bt]+\Str_V (-K[\iota(\omega'),[\iota(\omega'),\bt]])
-\frac{1}{2}\Str_V(-K[[\iota(\omega'),\iota(\omega')],\bt])+\cdots=0+O(\omega'^3)
\end{multline*}
In this computation we only used the cyclic property of trace, Leibniz and Jacobi identities for $V$ and the property of chain homotopy $\PP'+dK+Kd=\id_V$. Also for the lowest order of QME we had to use the unimodularity of Lie bracket on $V$.

The data necessary to define the effective action for the abstract $BF$ theory is the triplet of maps $(\iota,r,K)$.
\begin{Def} Let $(V,d)$ and $(V',d)$ be two homotopic cochain complexes. The data of induction from $V$ to $V'$ is defined to be the triplet of linear maps $(\iota,r,K)$ (embedding, retraction, chain homotopy), where $\iota: V'\ra V$, $r:V\ra V'$, $K:V\ra V$ are subject to the following relations:
\begin{eqnarray}
d\iota&=&\iota d\label{ind data axiom1}\\
dr&=&rd\label{ind data axiom2}\\
r\iota&=&\id_{V'}\label{ind data axiom3}\\
r K=K\iota&=&0\label{ind data axiom4}\\
d K+K d&=&\id_V-\iota r \label{ind data axiom5}\\
K^2&=&0\label{ind data axiom6}
\end{eqnarray}
\end{Def}
The fact that continuous deformation of the induction data leads to the canonical transformation for effective action follows directly from Statement \ref{statement: induction data deform}. However, using formula (\ref{R' from deforming induction data 2}), we can make a more precise statement.
\begin{statement}\label{statement: infinitesimal ind data deform for abstract BF} Infinitesimal deformations of the induction data $(\iota,r,K)$ are of form $(\iota,r,K)\mapsto(\iota+\delta\iota,r+\delta r,K+\delta_0 K+\delta K)$ where the following relations should hold: $d\;\delta\iota=\delta\iota\; d$, $d\;\delta r=\delta r\; d$ ,$r\;\delta\iota+\delta r\;\iota=0$, $K\;\delta K+\delta K\; K=0$, $d\;\delta K+\delta K\; d=0$, $r\;\delta K=\delta K\; \iota=0$ and where $\delta_0 K=-K\;\delta\iota\; r-\iota\;\delta r\; K$ is the minimal deformation of $K$, associated to the deformation of embedding and retraction. Under such deformation the effective action changes by a canonical transformation
$$S'\mapsto S'+\{S',R'\}+\hbar\Delta R'$$
with generator
\begin{multline}R'(\omega',p')=<p',r\;\delta\iota(\omega')>+\\
+e^{-S'/\hbar}\int_{\LL_K}e^{S(\iota(\omega')+\omega'',r^*(p')+p'')/\hbar}\left(-<p''_\LL,d\;\delta K\omega''_\LL>+<p''_\LL,\delta\iota(\omega')>-<p',\delta r(\omega''_\LL)>\right)\label{R' for infinitesimal ind data deform}\end{multline}
\end{statement}
\textbf{Proof.} We will separately treat the deformations of type I (deforming $K$, keeping $\iota$, $r$ fixed) and of type II (deformation of $\iota$, $r$, supplemented by the minimal deformation of $K$). Type I deformation acts on the Lagrangian submanifold $\LL_K$ by $\omega''_\LL\mapsto \omega''_\LL+\delta\omega''_\LL$, $p''_\LL\mapsto p''_\LL+\delta p''_\LL$, where the deformation $\omega''_\LL$ has to satisfy the equation $(K+\delta K)(\omega''_\LL+\delta\omega''_\LL)=0$ (plus the dual equation for the deformation of $p''_\LL$). General solution can be given as $\delta\omega''_\LL=-d\;\delta K\omega''_\LL$. Hence the deformation $K\mapsto K+\delta K$ leads to the deformation of Lagrangian submanifold $\LL_K\mapsto (\LL_K)_\Psi$, with the gauge fixing fermion $\Psi=<p''_\LL,d\;\delta K\omega''_\LL>$. Type II deformations are treated as in the proof of Statement \ref{statement: induction data deform}. Minimal deformation of the chain homotopy $\delta_0 K$ under type II deformation is obtained from equations $d\;\delta_0 K+\delta_0 K\;d=-\delta\iota\; r-\iota\;\delta r$, $\delta_0 K\; K+K\;\delta_0 K=0$. The solution is $\delta_0 K=-K\;\delta\iota\; r-\iota\;\delta r\; K$.
\\$\Box$

\subsection{Effective action for $BF$ theory as a generating function for algebraic structure on the subcomplex}
\label{section: eff action as generating function for alg structure}
Let $S'\in\Fun(\FF')$ be the effective action for abstract $BF$ theory on space $\FF=T^*[-1](V[1])$, associated to the DGLA $(V,d,[\bt,\bt])$. According to the perturbative result of section \ref{section: effective action for abstract BF}, we have
$$S'(\omega',p')=S'^0(\omega',p')+\hbar S'^1(\omega')$$
with the one-loop part $S'^1$ independent of $p'$ and the tree part $S'^0$ depending on $p'$ linearly. Let us write the tree part as
$$S'^0(\omega',p')=\sum_i (-1)^{|i|+1} p'_i Q^i(\omega')$$
with $Q^i\in\Fun(V'[1])$ a collection of functions on $V'[1]$, and let us define a vector field on $V'[1]$ (a derivation of algebra of functions $\Fun(V'[1])$)
$$Q=Q^i\frac{\dd}{\dd \omega'^i}\in\mr{Vect}(V'[1])=\mr{Der}(\Fun(V'[1]))$$
Therefore
$$S'^0(\omega',p')=\sum_i (-1)^{|i|+1} p'_i Q\omega'^i=-<p',Q\omega'>$$
(in super-field formalism we mean that $Q$ acts on $\omega'^i$ and is moved through the basis elements $e'_i\in V$). The CME $\{S'^0,S'^0\}=0$ means that $Q^2=0$, i.e. $Q$ is a cohomological vector field on $V'[1]$. Notice also that the BRST operator $Q_{\FF'}=\{S^0,\bt\}$ on the full space of BV fields $\FF'=T^*[-1](V'[1])$ is the cohomological vector field on $\FF'$, tangent to the zero section $V'[1]\subset \FF'$, and $Q$ is the restriction $Q=Q_{\FF'}|_{V'[1]}$.

Let us decompose components of $Q$ into the Taylor series in variables $\omega'^i$:
$$Q^i(\omega')=\sum_{n=1}^\infty \frac{1}{n!} Q^i_{(n)}(\omega')=\sum_{n=1}^\infty\frac{1}{n!}\sum_{j_1,\ldots,j_n}Q^i_{(n)j_1\cdots j_n}\omega'^{j_1}\cdots\omega'^{j_n}$$
where the Taylor coefficients are defined as derivatives
$$Q^i_{(n)j_1\cdots j_n}=\left.\left(\frac{\dd}{\dd \omega'^{j_n}}\cdots \frac{\dd}{\dd \omega'^{j_1}}Q^i(\omega')\right)\right|_{\omega'=0}$$
Now define the set of super-antisymmetric polylinear maps (the ``classical operations'') $l_{(n)}:\Lambda^n V'\ra V'$, by fixing their values on the basis of $V'$ (i.e. by means of structure constants):
\begin{equation}
l_{(n)}(e'_{j_1},\ldots,e'_{j_n})=\sum_i e'_i l_{(n)j_1\cdots j_n}^i \\ =
\sum_i (-1)^{|i|+1} e'_i\; Q_{(n)j_1\cdots j_n}^i \e(j_1,\ldots,j_n)
\label{classical operations def}
\end{equation}
where $\e(j_1,\ldots,j_n)$ is the sign defined as
\be \e(j_1,\ldots,j_n)=(-1)^{(|j_{n-1}|+1)\cdot|j_n|+(|j_{n-2}|+1)\cdot(|j_{n-1}|+|j_n|)+\cdots+(|j_1|+1)\cdot(|j_2|+\cdots+|j_n|)} \label{sign}\ee
The fact that operations $l_{(n)}$ are really (super-)antisymmetric is checked straightforwardly:
\begin{multline}
l_{(n)}(e'_{j_1},\ldots,e'_{j_k},e'_{j_{k+1}},\ldots,e'_{j_n})=\sum_i (-1)^{|i|+1} e'_i\; Q_{(n)j_1\cdots j_k j_{k+1}\cdots j_n}^i \e(j_1,\ldots ,j_k, j_{k+1},\ldots ,j_n)\\
=\sum_i (-1)^{|i|+1} e'_i\; (-1)^{(|j_k|+1)(|j_{k+1}|+1)}Q_{(n)j_1\cdots j_{k+1} j_{k}\cdots j_n}^i (-1)^{|j_k|+|j_{k+1}|}\e(j_1,\ldots ,j_{k+1}, j_{k},\ldots ,j_n) \\
=(-1)^{1+|j_k||j_{k+1}|}l_{(n)}(e'_{j_1},\ldots,e'_{j_{k+1}},e'_{j_{k}},\ldots,e'_{j_n})
\end{multline}
Property $\gh(Q)=1$ implies $\deg (l_{(n)})=2-n$. In terms of maps $l_{(n)}$ the vector field $Q$ is
\be Q=-\left<\sum_{n=1}^\infty\frac{1}{n!}l_{(n)}(\omega',\ldots,\omega'),\frac{\dd}{\dd\omega'}\right> \label{Q via l}\ee
(we mean that $l_{(n)}$ acts on the basis elements $e'_i$, while variables $\omega'^i$ are carried through, accounting for the sign in definition (\ref{classical operations def})). The tree part of action is
\be S'^0=\left<p',\sum_{n=1}^\infty\frac{1}{n!}l_{(n)}(\omega',\ldots,\omega')\right> \label{S'^0 via l}\ee
The CME in terms of classical operations $l_{(n)}$ is the system of quadratic equations on the structure constants of operations $l_{(n)}$:
\be \sum_{r\geq 0,s\geq 1:\;r+s=n}\frac{1}{r!\;s!}\;l_{(r+1)}(\omega',\ldots,\omega',l_{(s)}(\omega',\ldots,\omega'))=0 \label{quadratic relations}\ee
for all $n\geq 1$.

Pair $(V',Q)$, with $V'$ a $\ZZ$-graded vector field and $Q$ a cohomological vector field on $V'[1]$, vanishing at point $0\in V'[1]$, is called an $L_\infty$ algebra (or a `` homotopy Lie algebra'').
Hence the tree effective action endows $V'$ with the structure of $L_\infty$ algebra. In alternative definition, the $L_\infty$ algebra is a $\ZZ$-graded vector space $V'$, endowed with a sequence of polylinear super-antsymmetric maps (the $L_\infty$ operations) $l_{(n)}:\Lambda^n V'\ra V'$ of degree $2-n$ for each $n\geq 1$, and operations are required to satisfy the system of quadratic relations (\ref{quadratic relations}) --- the (higher) homotopy Jacobi identities. The two definitions of $L_\infty$ algebra are related by (\ref{Q via l}): cohomological vector field is the ``generating function'' for operations.

Let us also write the first equations of the system (\ref{quadratic relations}) in the standard form of equations on structure constants (without the super-fields). They are obtained by applying the differential operator $\frac{\dd}{\dd\omega'^{i_1}}\cdots \frac{\dd}{\dd\omega'^{i_n}}$ to (\ref{quadratic relations}).
For $n=1$ we get
$$l_{(1)}(l_{(1)}(e'_i))=0$$
--- the Poincar\'{e} identity, i.e. $l_{(1)}$ is the coboundary operator. For $n=2$:
$$l_{(1)}(l_{(2)}(e'_i,e'_j))-l_{(2)}(l_{(1)}(e'_i),e'_j)-(-1)^{|i|}l_{(2)}(e'_i,l_{(1)}(e'_j))=0$$
--- the Leibniz identity, i.e. $l_{(1)}$ is a derivation of $l_{(2)}$.
\begin{multline} l_{(2)}(e'_i,l_{(2)}(e'_j,e'_k))+(-1)^{|i|(|j|+|k|)}l_{(2)}(e'_j,l_{(2)}(e'_k,e'_i))+
(-1)^{|k|(|i|+|j|)}l_{(2)}(e'_k,l_{(2)}(e'_i,e'_j))+\\
+(-1)^{|i|+|j|}l_{(3)}(e'_i,e'_j,l_{(1)}(e'_k))+(-1)^{(|i|+1)(|j|+|k|)}l_{(3)}(e'_j,e'_k,l_{(1)}(e'_i))+\\+
(-1)^{|i|+|k|(|i|+|j|+1)}l_{(3)}(e'_k,e'_i,l_{(1)}(e'_j))
+l_{(1)}(l_{(3)}(e'_i,e'_j,e'_k))
\end{multline}
--- the homotopy Jacobi identity. In case $l_{(3)}=0$, this identity becomes the usual Jacobi identity for $l_{(2)}$. In case $l_{(3)}=l_{(4)}=\cdots=0$ the $L_\infty$ algebra is the ordinary DGLA. The DGLA case corresponds to $Q$ at most quadratic in coordinates on $V'[-1]$. In DGLA case $l_{(1)}=d$ is the differential, $l_{(2)}=[\bt,\bt]$ is the Lie bracket, and the cohomological vector field is
$$Q=-<d\omega',\frac{\dd}{\dd\omega'}>-<[\omega',\omega'],\frac{\dd}{\dd\omega'}>$$

The perturbation series (\ref{sum over trees}) for the tree effective action gives $L_\infty$ operations on $V'$ as sums over trees:
\be l_{(n)}(\omega',\ldots,\omega')=n!\sum_{T\in{\bf{T}}_\mr{nonPl},\; |T|=n}\frac{1}{|\mr{Aut}(T)|}\;\Iter_{T;-K[\bt,\bt];\,r[\bt,\bt]}(\iota(\omega'),\cdots,\iota(\omega'))\label{induced l_n}\ee
for $n>1$ and $l_{(1)}(\omega')=d\omega'$. Structure constants of operations $l_{(n)}$ are obtained from (\ref{induced l_n}) by applying $\frac{\dd}{\dd\omega^{i_1}}\cdots \frac{\dd}{\dd\omega^{i_n}}$. In particular, for $l_{(1)}$ we have $$l_{(1)}(e'_i)=de'_i$$ --- the differential on $V'$. For $l_{(2)}$:
$$l_{(2)}(e'_i,e'_j)=r[\iota(e'_i),\iota(e'_j)]$$
--- the Lie bracket in $V$, projected to $V'$. For $l_{(3)}$ we obtain
\begin{multline}l_{(3)}(e'_i,e'_j,e'_k)=r[-K[\iota(e'_i),\iota(e'_j)],\iota(e'_k)]+\\
+(-1)^{|i|(|j|+|k|)}r[-K[\iota(e'_j),\iota(e'_k)],\iota(e'_i)]+
(-1)^{|k|(|i|+|j|)}r[-K[\iota(e'_k),\iota(e'_i)],\iota(e'_j)]
\end{multline}

Next let us write the one-loop part of effective action as $S'^1(\omega')=\log \rho(\omega')$,
with $\rho\in\Fun(V'[1])$. Then in terms of $Q$ and $\rho$ the effective action is
\be S'=-<p',Q\omega'>+\hbar\log\rho \label{S' via Q and rho}\ee
The classical part of QME (\ref{CME for S'}), as discussed above, is equivalent to $Q^2=0$. The quantum part of QME (\ref{QME for S'}) in terms of $Q$, $\rho$ is
\begin{multline}\Delta S'^0+\{S'^1,S'^0\}=\sum_i\frac{\dd}{\dd\omega'^i} (Q \omega'^i)+\sum_i(Q\omega'^i)\left(\frac{\dd}{\dd\omega'^i}\log\rho\right)=
\sum_i\frac{1}{\rho}\; \frac{\dd}{\dd\omega'^i} (\rho Q \omega'^i)=\\
=\mr{div}Q+Q\log\rho=\mr{div}_\rho Q=0\end{multline}
where $\mr{div}_\rho Q$is the divergence of vector field $Q$, with the measure $\mu=\rho\cdot \mu_\mr{coord}=\rho\cdot\prod_i \DD\omega'^i$ on $V'[1]$, i.e. the measure of density $\rho$. Therefore we have the following interpretation of QME:
\begin{eqnarray}
Q^2&=&0 \label{CME via Q}\\
\mr{div}_\rho Q&=&0 \label{QME via Q and rho}
\end{eqnarray}
i.e. $Q$ is a cohomological vector field on $V'[1]$ and $\rho$ is the density of a $Q$-invariant measure on $V'[1]$ (i.e. such that measure of a set is preserved by flow generated by $Q$).

We define ``quantum operations'' on $V'$ --- a sequence of polylinear super-antisymmetric maps $q_{(n)}:\Lambda^n V'\ra\RR$ of degree $\deg q_{(n)}=-n$ via the Taylor expansion for $S'^1(\omega')=\log\rho(\omega')$:
\be S'^1(\omega')=\log\rho(\omega')=\sum_{n=1}^\infty\frac{1}{n!}\;q_{(n)}(\omega',\ldots,\omega')\label{S'^1 via q}\ee
I.e. the structure constants of quantum operations are defined as
$$q_{(n)}(e'_{i_1},\ldots,e'_{i_n})=\e(i_1,\ldots,i_n)\left.\left(\frac{\dd}{\dd\omega'^{i_n}}
\cdots\frac{\dd}{\dd\omega'^{i_1}}\log\rho(\omega')\right)\right|_{\omega'=0}$$
The quantum part of QME (\ref{QME via Q and rho})  in terms of classical and quantum operations $l_{(n)}, q_{(n)}$ is the system
\be\frac{1}{n!}\;\Str_{V'}l_{(n+1)}(\omega',\ldots,\omega',\bt)+
\sum_{r\geq 0,s\geq 1:\;r+s=n}\frac{1}{r!s!}\;q_{(r+1)}(\omega',\ldots,\omega',l_{(s)}(\omega',\ldots,\omega'))=0 \label{relations on q and l}\ee
for $n\geq 1$ (for $n=0$ the equation is satisfied automatically: $\Str_{V'}l_{(1)}(\bt)=0$, since $\deg l_{(1)}=1$). Equations for structure constants of operations are obtained from (\ref{relations on q and l}), just as for (\ref{quadratic relations}), by applying
$\frac{\dd}{\dd\omega'^{i_1}}\cdots \frac{\dd}{\dd\omega'^{i_n}}$. In particular, for $n=1$:
$$\Str_{V'}l_{(2)}(e'_i,\bt)+q_{(1)}(l_{(1)}(e'_i))=0$$
is the homotopy version of unimodularity property (\ref{unimodularity}).
For $n=2$:
$$\Str_{V'} l_{(3)}(e'_i,e'_j,\bt)+q_{(1)}(l_{(2)}(e'_i,e'_j))+(-1)^{|i|+1}q_{(2)}(e'_i,l_{(1)}(e'_j))
+ (-1)^{(|i|+1)|j|} q_{(2)}(e'_j,l_{(1)}(e'_i))=0$$

The perturbation series for one-loop effective action (\ref{sum over loops}) gives the quantum operations on $V'$ as sums over one-loop graphs:
$$q_{(n)}(\omega',\ldots,\omega')=
-n!\sum_{L\in{\bf{L}}_\mr{nonPl},\;|L|=n}\frac{1}{|\mr{Aut}(L)|}\Loop_{L;-K[\bt,\bt]}(\iota(\omega'),\cdots,\iota(\omega'))$$
The structure constants are obtained as usual, by applying $\frac{\dd}{\dd\omega'^{i_1}}\cdots \frac{\dd}{\dd\omega'^{i_n}}$. In particular, for $n=1$ we have
$$q_{(1)}(e'_i)=-\;\Str_{V}(-K[\iota(e'_i),\bt])$$
For $n=2$:
\begin{multline*}
q_{(2)}(e'_i,e'_j)=-\;\Str_{V}(-K[-K[e'_i,e'_j],\bt])-\\
-(-1)^{|i|+1}\;\Str_V(-K[e'_i,-K[e'_j,\bt]])-
(-1)^{(|i|+1)|j|}\;\Str_V(-K[e'_j,-K[e'_i,\bt]])
\end{multline*}

Now we can summarize the algebraic structure generated by the effective action $S'$ for abstract $BF$ theory, associated to a DGLA structure on $V$, on a subcomplex (more precisely, deformation retract) $V'\stackrel{\iota}{\hookrightarrow} V$.
\begin{Def}We call a $qL_\infty$ algebra (a ``quantum'' $L_\infty$ algebra) a $\ZZ$-graded vector field $V'$, endowed with a cohomological vector field $Q$ (i.e. of degree 1 and satisfying $Q^2=0$) on $V'[1]$, vanishing at point $0\in V'[1]$, and also with a $Q$-invariant (i.e. satisfying (\ref{QME via Q and rho})) measure $\mu$ on $V'[1]$.
\end{Def}
Equivalently, in terms of operations:
\begin{Def} A $qL_\infty$ algebra is a $\ZZ$-graded vector space $V'$, endowed with two sets of polylinear super-antisymmetric maps $l_{(n)}:\Lambda^n V'\ra V'$ and $q_{(n)}:\Lambda^n V'\ra\RR$ for $n=1,2,3,\ldots$ --- the classical and quantum operations, satisfying two systems of equations: the system of homotopy Jacobi identities (\ref{quadratic relations}) and the system of homotopy unimodularity relations (\ref{relations on q and l}).
\end{Def}
Therefore the effective action $S'$ generates the structure of $qL_\infty$ algebra on $V'$ by means of (\ref{S' via Q and rho}) in terms of $Q,\rho$, or by (\ref{S'^0 via l},\ref{S'^1 via q}) in terms of operations. Notice that the effective action is the generating function for structure constants of classical and quantum operations on $V'$, precisely as the initial action of abstract $BF$ theory was the generating function for structure constants of differential and Lie bracket on $V$.

\subsection{$BF_\infty$ theory}
\label{section: BF_infty}
Let $(V,Q,\mu)$ be a $qL_\infty$ algebra. We define the corresponding $BF_\infty$ theory by the BV action
\begin{multline} S(\omega,p)=-<p,Q(\omega)\omega>+\hbar \log\rho(\omega)=\\
=\left<p,\sum_{n=1}^\infty\frac{1}{n!}l_{(n)}(\omega,\ldots,\omega)\right>+\hbar \sum_{n=1}^\infty\frac{1}{n!}q_{(n)}(\omega,\ldots,\omega)\label{BF_infty action}\end{multline}
on the space of fields
$$\FF=T^*[-1](V[1])$$
with canonical BV Laplacian $\Delta$.
We use the notations for super-fields, operations, etc. as before. The difference from situation of section \ref{abstract BF theory} is just that we construct the action not from the unimodular DGLA structure on a $\ZZ$-graded vector space $V$, but from the $qL_\infty$ structure. Abstract $BF$ theory then is the special case of $BF_\infty$ theory, with $l_{(3)}=l_{(4)}=\cdots=0$, $q_{(1)}=q_{(2)}=\cdots=0$,  i.e. only the first two classical operations are non-zero. The QME for action (\ref{BF_infty action}) is satisfied automatically due to (\ref{CME via Q},\ref{QME via Q and rho}) or, equivalently, due to (\ref{quadratic relations},\ref{relations on q and l}).

We define the effective action for $BF_\infty$ theory precisely as we did for abstract $BF$ theory in section \ref{section: effective action for abstract BF}. I.e. we need a chain complex $(V',d)$, embedding $\iota: V'\ra V$ and retraction $r: V\ra V'$ that are supposed to be quasi-isomorphisms (we understand $d=l_{(1)}$ as the differential on $V$). We also need a chain homotopy $K:V\ra V$, contracting $V$ to $\iota(V')$, and we assume the standard set of relations:
$d\iota=\iota d$, $dr=rd$, $r\;\iota=\id_{V'}$, $K\iota=rK=0$, $dK+Kd=\PP''$, $K^2=0$. The induction data --- the triplet $(\iota,r,K)$ define the Hodge decomposition (\ref{Hodge decomp}) for $V$ and the Lagrangian subspace $\LL_K$ in the space of UV fields (\ref{L via K}). The effective action on IR fields $S'\in\Fun(\FF')$ is defined as in section \ref{section: effective action for abstract BF}, i.e. by the integral
\begin{multline}e^{S'(\omega',p')/\hbar}=\int_{\LL_K}e^{S(\iota(\omega')+\omega'',r^*(p')+p'')/\hbar}\\
=\frac{1}{N}\int\prod_I\DD\omega''^I\prod_{\bar{I}}\DD p''_{\bar{I}}\;\exp\frac{1}{\hbar}
\left(\left<r^*(p')+p''_\LL,\sum_{n=1}^\infty \frac{1}{n!} l_{(n)}(\iota(\omega')+\omega''_\LL,\ldots,\iota(\omega')+\omega''_\LL)\right>+\right.\\
\left.+\hbar
\sum_{n=1}^\infty \frac{1}{n!} q_{(n)}(\iota(\omega')+\omega''_\LL,\ldots,\iota(\omega')+\omega''_\LL)\right) \label{BV integral for BF infty}
\end{multline}

To formulate the perturbative result for the integral (\ref{BV integral for BF infty}), we need the natural generalization of the $\Iter$, $\Loop$ notation of section \ref{section: effective action for abstract BF}.
Let $T$ be a planar rooted tree, not necessarily binary. Then for two sets of polylinear maps $\lambda_n:X^{\otimes n}\ra X$, $\tilde{\lambda}_n:X^{\otimes n}\ra Y$ with $n\geq 1$, we define the map $\Iter_{T;\{\lambda_n\};\{\tilde{\lambda}_n\}}: X^{\otimes |T|}\ra Y$ as follows: edges of $T$ are decorated with elements of $X$. Leaves are decorated with arguments of $\Iter$, an internal edge with $n$ children is decorated with $\lambda_n$ applied to the children, the root is decorated with  $\tilde{\lambda}_n$ applied to children. We read the value of  $\Iter$ from the root. Likewise the traces of iterated operations $\Loop$ are generalized to the case of connected one-loop graphs with vertices of arbitrary valence $n+1\geq 2$, where each vertex is required to have precisely one outgoing edge. We define $\Loop_{L;\{\lambda_n\}}:X^{\otimes |L|}\ra\RR$ as before: the graph $L$ is cut along an edge in the cycle and becomes a rooted tree $T$ with $i$-th leaf marked, and we set
$$\Loop_{L;\{\lambda_n\}}(x_1,\ldots x_{|L|})=\Str_X\; \Iter_{T;\{\lambda_n\};\{\lambda_n\}}(x_1,\ldots,x_{i-1},\bt,x_i,\ldots,x_{|L|+1})$$
For example:
\begin{eqnarray*}\Iter_{(*(**)*);\{\lambda_n\};\{\tilde{\lambda}_n\}}(x_1,x_2,x_3,x_4)
&=&\tilde{\lambda}_3(x_1,\lambda_2(x_2,x_3),x_4)\\
\Loop_{(*(*\bt)*);\{\lambda_n\}}(x_1,x_2,x_3)&=&\Str_X\;\lambda_3(x_1,\lambda_2(x_2,\bt),x_3)
\end{eqnarray*}
We denote $\bf{T}_\mr{Pl}$, $\bf{T}_\mr{nonPl}$, $\bf{L}_\mr{Pl}$, $\bf{L}_\mr{nonPl}$ the sets of planar/non-planar trees and one-loop oriented graphs respectively, with vertices of arbitrary valence $\geq 3$, and requirement that each vertex has precisely one outgoing edge.

\begin{thm}\label{thm: eff action for BF_infty} Effective action for  $BF_\infty$ theory, associated to a $qL_\infty$ algebra $(V,Q,\mu)$, induced on the space $\FF'=T^*[-1](V'[1])$ by means of integral (\ref{BV integral for BF infty}), is
$$S'(\omega',p')=S'^0(\omega',p')+\hbar S'^1(\omega')$$
where the tree part is the sum over non-planar rooted trees:
\be S'^0(\omega',p')=<p',d\omega'>+\sum_{T\in{\bf{T}}_\mr{nonPl}:\;|T|\geq 2}\frac{1}{|\mr{Aut}(T)|}<p',\Iter_{T;\{-K l_{(n)}\}_{n\geq 2};\,\{r l_{(n)}\}_{n\geq 2}}(\iota(\omega'),\ldots,\iota(\omega'))>\label{pert series for S'^0 for BF_infty}\ee
and the one-loop part is
\begin{multline}S'^1(\omega')=-\sum_{L\in{\bf{L}}_\mr{nonPl}}\frac{1}{|\mr{Aut}(L)|}\Loop_{L;\{-K l_{(n)}\}_{n\geq 2}}(\iota(\omega'),\ldots,\iota(\omega'))+\\
+\sum_{T\in{\bf{T}}_\mr{nonPl}:\;|T|\geq 1} \frac{1}{|\mr{Aut}(T)|}\Iter_{T;\{-K l_{(n)}\}_{n\geq 2};\,\{q_{(n)}\}_{n\geq 1}}(\iota(\omega'),\ldots,\iota(\omega'))\label{pert series for S'^1 for BF_infty}
\end{multline}
Effective action $S'$ satisfies QME and defines a $BF_\infty$ theory on $\FF'$, associated to the $qL_\infty$ structure on $V'$, with classical operations $l'_{(1)}=d$ and
\be l'_{(n)}(\omega',\ldots,\omega')=n!\sum_{T\in{\bf{T}}_\mr{nonPl}:\;|T|=n}\frac{1}{|\mr{Aut}(T)|}\Iter_{T;\{-K l_{(m)}\}_{m\geq 2};\,\{r l_{(m)}\}_{m\geq 2}}(\iota(\omega'),\ldots,\iota(\omega'))\label{induced l_n from qL_infty}\ee
for $n\geq 2$, and quantum operations
\begin{multline}q'_{(n)}(\omega',\ldots,\omega')=-n!\sum_{L\in{\bf{L}}_\mr{nonPl}:\;|L|=n}\frac{1}{|\mr{Aut}(L)|}\Loop_{L;\{-K l_{(m)}\}_{m\geq 2}}(\iota(\omega'),\ldots,\iota(\omega'))+\\
+n!\sum_{T\in{\bf{T}}_\mr{nonPl}:\;|T|=n} \frac{1}{|\mr{Aut}(T)|}\Iter_{T;\{-K l_{(m)}\}_{m\geq 2};\,\{q_{(m)}\}_{m\geq 1}}(\iota(\omega'),\ldots,\iota(\omega'))\label{induced q_n from qL_infty}\end{multline}
for $n\geq 1$.
\end{thm}

\textbf{Proof.} Perturbation series for (\ref{BV integral for BF infty}) is obtained following the same routine as in section \ref{section: effective action for abstract BF}. We just have to adjust the Feynman rules to include vertices of all valences $n+1\geq 3$ with one outgoing edge, corresponding to classical operations $l_{(n)}$ on $V$ or, equivalently, to terms of order $O(p\,\omega^n)$ in $S$. We also have to include root vertices with no outgoing edges of all valences $n\geq 1$, corresponding to quantum operations $q_{(n)}$ on $V$ or, equivalently, to terms of order $O(\hbar \omega^n)$ in $S$. Feynman trees containing a vertex of this second type in the root yield the second term in (\ref{pert series for S'^1 for BF_infty}). The fact that $S'$ satisfies QME is an implication of the construction (\ref{BV integral for BF infty}) and Statement \ref{statement: QME for induced action}. Since the effective action $S'$ again satisfies the ansatz (\ref{BF_infty action}), we can interpret it as the action of $BF_\infty$ theory, associated to the $qL_\infty$ structure on $V'$. Formulae (\ref{induced l_n from qL_infty},\ref{induced q_n from qL_infty}) for operations follow directly from the Feynman diagram expansion (\ref{pert series for S'^0 for BF_infty},\ref{pert series for S'^1 for BF_infty}). Equations (\ref{quadratic relations},\ref{relations on q and l}) for $\{l'_{(n)}\}$, $\{q'_{(n)}\}$ are satisfied automatically, since  $S'$ satisfies the QME.
\\$\Box$

Therefore one can understand the class of $BF_\infty$ theories as the closure of the class of abstract $BF$ theories w.r.t. operation of inducing the effective action.

\begin{Def}\label{def: induced qL_infty} Let $(V,Q,\mu)$ be a $qL_\infty$ algebra and $V'\stackrel{\iota}{\hookrightarrow}V$ a deformation retract of $V$. Then we call the $qL_\infty$ structure on $V'$ defined by (\ref{induced l_n from qL_infty},\ref{induced q_n from qL_infty}) the induced $qL_\infty$ structure, with the induction data $(\iota,r,K)$.
\end{Def}

\begin{statement}\label{statement: iterated induction} Let $(V,Q,\mu)$ be a $qL_\infty$ algebra, $V_1\stackrel{\iota_1}{\hookrightarrow}V$ a subcomplex in $V$, $(\iota_1,r_1,K_1)$ --- the induction data from $V$ to $V_1$,and let $V_2\stackrel{\iota_2}{\hookrightarrow}V_1$ be a subcomplex in $V_1$, and $(\iota_2,r_2,K_2)$ --- the induction data from $V_1$ to $V_2$. We define the composition of induction data as
\be (\iota,r,K)=(\iota_2,r_2,K_2)\circ(\iota_1,r_1,K_1):=(\iota_1\iota_2,r_2 r_1, K_1+ \iota_1 K_2 r_1)\label{composition of iota,r,K}\ee
Then the iterated induction of effective action on $V_2$ through $V_1$: $$(\FF,S)\stackrel{(\iota_1,r_1,K_1)}{\longrightarrow}(\FF_1,S_1)
\stackrel{(\iota_2,r_2,K_2)}{\longrightarrow} (\FF_2,S_2)$$
yields the same action $S_2$ on $V_2$ as the direct induction to $V$, with the composed induction data:
$$(\FF,S)\stackrel{(\iota_2,r_2,K_2)\circ(\iota_1,r_1,K_1)}{\longrightarrow}(\FF_2,S_2)$$
Or equivalently, in the language of $qL_\infty$ algebras: iterated induction
$$(V,Q,\mu)\stackrel{(\iota_1,r_1,K_1)}{\longrightarrow}(V_1,Q_1,\mu_1)
\stackrel{(\iota_2,r_2,K_2)}{\longrightarrow} (V_2,Q_2,\mu_2)$$
yields the same $qL_\infty$ structure on $V_2$ as the direct induction
$$(V,Q,\mu)\stackrel{(\iota_2,r_2,K_2)\circ(\iota_1,r_1,K_1)}{\longrightarrow}(V_2,Q_2,\mu_2)$$
\end{statement}

\textbf{Proof.}
Let us first check that(\ref{composition of iota,r,K}) satisfies the relations for induction data. Indeed, obviously $\iota_1\iota_2$ and $r_2 r_1$ are chain maps, then $r\iota=r_2 (r_1 \iota_1)\iota_2=\id_{V_2}$, $K\iota=(K_1+\iota_1 K_2 r_1)\iota_1\iota_2=(K_1\iota_1)\iota_2+\iota_1 (K_2 \iota_2)=0$, $r K=r_2 r_1 (K_1+\iota_1 K_2 r_1)=r_2 (r_1 K_1)+(r_2 K_2) r_1=0$, $d K+K d=d (K_1+\iota_1 K_2 r_1)+(K_1+\iota_1 K_2 r_1) d=(d K_1+K_1 d)+\iota_1 (d K_2+K_2 d) r_1=(\id_V-\iota_1 r_1)+\iota_1(\id_{V_1}-\iota_2 r_2)r_1=\id_V-\iota r$, $K^2=(K_1+\iota_1 K_2 r_1)(K_1+\iota_1 K_2 r_1)=K_1^2+\iota_1 K_2^2 r_1+\iota_1 K_2 (r_1 K_1)+(K_1 \iota_1) K_2 r_1=0$. Hence (\ref{composition of iota,r,K}) is the legitimate induction data.

Let us introduce the following notations for the UV parts of complexes $V,V_1$:\quad $V''_1=\ker(r_1)\subset V$, $V''_2=\ker(r_2)\subset V_1$, i.e. splitting of $V$ and $V_1$ int IR and UV parts is: $V=\iota_1(V_1)\oplus V''_1$, $V_1=\iota_2(V_2)\oplus V''_2$. Corresponding Hodge decompositions defined by chain homotopies $K_1$, $K_2$ are $V=\iota_1(V_1)\oplus V''_{1,d-ex}\oplus V''_{1,K_1-ex}$ and $V_1=\iota_2(V_2)\oplus V''_{2,d-ex}\oplus V''_{2,K_2-ex}$. At the same time, the induction data (\ref{composition of iota,r,K}) define another splitting of $V$ into IR and UV parts: $V=\iota(V_2)\oplus V''_{12}$, где $V''_{12}=\ker(r)=V''_1\oplus\iota_1(V''_2)\subset V$. Therefore the Hodge decomposition is
$V=\iota(V_2)\oplus V''_{12,d-ex}\oplus V''_{12,K-ex}$, where $V''_{12,d-ex}=V''_{1,d-ex}\oplus \iota_1 (V''_{2,d-ex})$ and $V''_{12,K-ex}=V''_{1,K_1-ex}\oplus\iota_1(V''_{2,K_2-ex})$.

According to the definition of effective action (\ref{BV integral for BF infty}), the exponential of action on $V_1$ is
\be e^{S_1(\omega_1,p_1)/\hbar}=\int_{\LL_{K_1}\subset\FF''_1}e^{S(\iota_1(\omega_1)+\omega''_1,r^*_1(p_1)+p''_1)/\hbar} \label{eff act on V_1}\ee
where $\LL_{K_1}=N^*[-1](V''_{1,K_1-ex}[1])\subset \FF''_1=T^*[-1](V''_1[1])$. Iterating the procedure, for the exponential of effective action, induced on $V_2$ from (\ref{eff act on V_1}), we obtain the following:
\begin{multline}e^{S_2(\omega_2,p_2)/\hbar}=\int_{\LL_{K_2}\subset\FF''_2}e^{S_1(\iota_2(\omega_2)+\omega''_2,r^*_2(p_2)+p''_2)/\hbar}\\
=\int_{\LL_{K_2}\subset\FF''_2}\int_{\LL_{K_1}\subset\FF''_1}e^{S(\iota_1(\iota_2(\omega_2)+\omega''_2)+\omega''_1,r^*_1(r^*_2(p_2)+p''_2)+p''_1)/\hbar}\\
=\int_{\LL_{K_1}\oplus \hat\iota_1(\LL_{K_2})}e^{S(\iota(\omega_2)+\omega''_{12},r^*(p_2)+p''_{12})/\hbar}
\end{multline}
where $\LL_{K_2}=N^*[-1](V''_{2,K_2-ex}[1])\subset \FF''_2=T^*[-1](V''_2[1])$ and we denoted $\hat\iota_1=\iota_1\oplus r_1^*:\FF_1\ra \FF$. Notice that $\LL_{K_1}\oplus \hat\iota_1(\LL_{K_2})=N^*[-1](V''_{12,K-ex})$, hence we obtained precisely the expression defining effective action on $V_2$ with the induction data (\ref{composition of iota,r,K}).
$\Box$

\subsubsection{Equivalence of $qL_\infty$ algebras}
\label{section: equivalence of qL_infty algebras}
Talking about canonical transformations for the action of $BF_\infty$ theory, we would like to restrict to the ``special'' canonical transformations, preserving the ansatz (\ref{BF_infty action}) for $BF_\infty$ action.
\begin{Def}\label{infinitesimal special canonical transform} We call the infinitesimal special canonical transformation (SCT) for a  $BF_\infty$ action $S$ on $\FF=T^*[-1](V[1])$ the infinitesimal canonical transformation $S\mapsto S+\{S,R\}+\hbar \Delta R$ with the generator of the form
\be R=<p,v(\omega)\omega>+\hbar \chi(\omega)\label{infinitesimal special transform generator}\ee
where $v(\omega)\in\mr{Vect}(V[1])$ is some vector field on $V[1]$ with ghost number $\gh(v)=0$, and $\chi(\omega)\in\Fun(V[1])$ is a function on $V[1]$ with ghost number $\gh(\chi)=-1$.  Equivalently, we call the infinitesimal SCT of a $qL_\infty$ algebra $(V,Q,\mu)$ the infinitesimal transformation of cohomological vector field and measure of form
\begin{eqnarray}Q&\mapsto& Q+[Q,v]\\
\rho&\mapsto&\rho\cdot (1+\mr{div}_\rho v+Q \chi)
\end{eqnarray}
where the bracket means the Lie bracket of vector fields.
\end{Def}
The fact that for any given $BF_\infty$ action $S$ the general infinitesimal canonical transformation, preserving the $BF_\infty$ ansatz for it, has to be of form (\ref{infinitesimal special transform generator}), is a consequence of trivial degree counting for $p$ and $\hbar$ in the formula for canonical transformation $S\ra S+\{S,R\}+\hbar \Delta R$.
We can also give the definition of finite SCTs.

\begin{Def}\label{finite special canonical transform} We call the finite special canonical transformation for a $BF_\infty$ action $S=-<p,Q\omega>+\hbar\log\rho$ на $\FF=T^*[-1](V[1])$ the finite canonical transformation of form
\be S\mapsto \tilde{S}=-<p,U^*Q(U^*)^{-1}\omega>+\hbar(U^*\log\rho+\log \mr{Jac}(U)+U^* Q\chi)\label{fin spec can transf for S}\ee
where $U:V[1]\ra V[1]$ is a diffeomorphism of ghost number 0, i.e. such that the pull-back $U^*:\Fun(V[1])\ra\Fun(V[1])$ preserves grading, and $\chi\in\Fun(V[1])$ is a function on $V[1]$ of ghost number -1. Equivalently, on the language of $qL_\infty$ algebras, we call the pair $(U,\chi)$ a finite SCT between two $qL_\infty$ structures $(Q,\mu)$ and $(\tilde{Q},\tilde{\mu})$ on $V$, if
\begin{eqnarray}
\tilde{Q}&=&U^*Q (U^*)^{-1}\label{homotopy of Q}\\
\tilde{\mu}&=&U^*(\mu\cdot e^{Q\chi})\label{homotopy of mu}
\end{eqnarray}
where the pull-back $U^*$ in (\ref{homotopy of mu}) acts as on measure (not as on function).
\end{Def}
Definition \ref{infinitesimal special canonical transform} is obtained from here by substituting $U^*=\id_{\Fun(V[1])}+v$ where $v$ is an infinitesimal vector field.

\begin{statement} \label{statement: special can transf induction}
Let $S$ be a $BF_\infty$ action on $V$ and $S'$ an effective action on $V'$ for it, induced using the data $(\iota,r,K)$.
Let also $\tilde{S}$ be an action, differing from $S$ by an infinitesimal SCT. Then the effective action $\tilde S'$ on $V'$, constructed with the same induction data $(\iota,r,K)$, differs from $S'$ by an infinitesimal SCT.
\end{statement}
\textbf{Proof.}
Obvious, since the infinitesimal canonical transformation $S\mapsto\tilde{S}$ in the class of $BF_\infty$ action, leads to a canonical transformation for the effective action again in the class of $BF_\infty$ actions. Hence $\tilde S'$ differs from $S'$ by an infinitesimal SCT.
\\$\Box$

\begin{statement}\label{statement: ind data deform for BF_infty}
Let $S$ be a $BF_\infty$ action on $V$, and let $S'_1$ and $S'_2$ be two effective actions for it on $V'$, induced using data $(\iota_1,r_1,K_1)$ and $(\iota_2,r_2,K_2)$ respectively. Then $S'_1$ and $S'_2$ differ by a SCT.
\end{statement}
To prove this we will need the following lemma.
\begin{lemma} Let $(V,d)$ and $(V',d)$ be two quasi-isomorphic chain complexes. Then the configuration space of possible induction data $(\iota,r,K)$ from $V$ to $V'$ (i.e. of triples embedding-retraction-chain homotopy) is a certain bundle over the configurations space of Hodge decompositions $\{V=W\oplus V''_{d-ex}\oplus V''_{K-ex}:\;W\sim V'\}$, with the fiber --- the configurations space of automorphisms $V'$ (as a chain complex) $\{(\lambda,\lambda^{-1},0)\}\sim \Aut(V')$. The base of this bundle is connected.
\end{lemma}
\textbf{Proof of lemma.}
Given a triplet $(\iota,r,K)$, we construct the Hodge decomposition
$V=W\oplus V''_{d-ex}\oplus V''_{K-ex}$, где $W=\iota(V')$, $V''_{d-ex}=\ker(r)\cap \mr{im}(d)$, $V''_{K-ex}=\ker(r)\cap\mr{im}(K)$.
And vice versa: given a Hodge decomposition for $V$ and an embedding $\iota:V'\ra W\subset V$, we reconstruct the retraction as $r=\iota^{-1}\PP_{W}$ (with $\PP_{W}$ the projector to the first term in Hodge decomposition), and the chain homotopy is reconstructed as $K=(d|_{V''_{K-ex}\ra V''_{d-ex}})^{-1}$, i.e. as the inverse map to the differential $V''_{K-ex}\stackrel{d}{\ra}V''_{d-ex}$. Hence the space of triplets $(\iota,r,K)$ is a principal bundle over the space of Hodge decompositions for $V$ with fiber (and the structure group) being the group $\mr{Aut}(V')$ of chain automorphisms of $V'$:
$$\mr{Aut}(V')\car\{(\iota,r,K)\}\ra \{V=W\oplus V''_{d-ex}\oplus V''_{K-ex}:\;W\sim V'\}$$
Automorphisms $\lambda\in\mr{Aut}(V')$ act on $\{(\iota,r,K)\}$ as $$(\iota,r,K)\mapsto(\lambda,\lambda^{-1},0)\circ (\iota,r,K)=(\iota\lambda,\lambda^{-1} r,K)$$

Consider the case when differential on $V'$ is zero, i.e. $V'=H^\bt(V)$ is the cohomology of $V$. For the Hodge decomposition spaces $V''_{d-ex}=\im(d)=d(V)$ and $W\oplus V''_{d-ex}=\ker(d)$ are fixed (defined by the structure of chain complex on $V$). Let us fix some linear subspace $W_0\subset \ker(d)$, such that $\ker(d)=W_0\oplus \im(d)$ and $W_0\cap \im(d)=\{0\}$. Then the general $W$, satisfying these conditions, is obtained from $W_0$ as a graph $W=\mr{graph}(\alpha)$ of some linear, degree-preserving map $\alpha\in\mr{Hom}^0(W_0\ra\im(d))$. Now let $W$ be fixed. Choose some linear subspace $V''_{K-ex,0}\subset V$, satisfying $V=\ker(d)\oplus V''_{K-ex,0}$ и $V''_{K-ex,0}\cap\ker(d)=\{0\}$. Then the general $V''_{K-ex}$, satisfying these conditions is of the form $V''_{K-ex}=\mr{graph}(\beta)$, where $\beta\in\mr{Hom}^0(V''_{K-ex,0}\ra\ker(d))$. Hence, the space of Hodge decompositions
$\{V=W\oplus V''_{d-ex}\oplus V''_{K-ex}:\;W\sim V'\}$ is a bundle over $\mr{Hom}^0(W_0\ra\im(d))$ with fiber $\mr{Hom}^0(V''_{K-ex,0}\ra\ker(d))$ (i.e. the base and the fiber are vector spaces), and hence is connected (and even retractible). Then for the case $V'=H^\bt(V)$ the lemma is proved. Note also that this space is non-empty, since the cohomology can always be embedded into $V$ as $W_0$, and it is always possible to choose a complement $V''_{K-ex,0}$ to $\ker(d)$ in $V$.

Consider now the general case, with $V'$ a general chain complex, quasi-isomorphic to $V$. Choose some induction data $(\iota_1,r_1,K_1)$ from $V'$ to the cohomology $H^\bt(V)$. Then we have a decomposition $V'=\iota_1(H(V))\oplus \tilde V''_{d-ex}\oplus \tilde V''_{K_1-ex}$. Then for a general Hodge decomposition for $V$: $V=\iota(V')\oplus V''_{d-ex}\oplus V''_{K-ex}$, generated by a triplet $(\iota,r,K)$, we have
\begin{eqnarray} \iota(\iota_1(H(V)))\oplus \im(d)&=&\ker(d)\subset V \label{lemma eq 1}\\
\iota(\tilde V''_{d-ex})\oplus V''_{d-ex}&=&\im(d)\subset V \label{lemma eq 2} \\
\iota(\tilde V''_{K-ex})\oplus V''_{K-ex}\oplus \ker(d)&=&V \label{lemma eq 3}
\end{eqnarray}
Using the argumentation as above, we find that the space of images of the  embeddings, restricted to cohomology, is $\{\iota(\iota_1(H(V)))\}=\Hom^0(\iota_0(\iota_1(H(V)))\ra\im(d))$, where $\iota_0:V'\ra V$ is some embedding. Then the space of images of embeddings, restricted to the exact part of $V'$, is the Grassmanian $\{\iota(\tilde V''_{d-ex})\}=\mr{Gr}(\im(d),\tilde V''_{d-ex})=\prod_n\mr{Gr}((\im(d))^n,(\tilde V''_{d-ex})^n)$. The space  $\{V''_{d-ex}\}$ with $\iota(\tilde V''_{d-ex})$ fixed is $\Hom^0(V''_{d-ex,0}\ra \iota(\tilde V''_{d-ex}))$ for some $V''_{d-ex,0}$, satisfying (\ref{lemma eq 2}). Next, choose some subspace $U_0\subset V$ such that $U_0\oplus \ker(d)=V$. General complement of $\ker(d)$ in $V$ has the form $U=\mr{graph}(\beta)$ with
$\beta\in\Hom^0(U_0\ra\ker(d))$. With $U$ fixed, the space of images of embeddings, restricted to the $K_1$-exact part of $V'$, is $\{\iota(\tilde V''_{K_1-ex})\}=\mr{Gr}(U,\tilde V''_{K_1-ex})$; for the space $\{V''_{K-ex}\}$ with $U$, $\iota(\tilde V''_{K_1-ex})$ fixed, we have $\{V''_{K-ex}\}=\Hom^0(V''_{K-ex,0}\ra \iota(\tilde V''_{K_1-ex}))$ for some arbitrarily chosen complement $V''_{K-ex,0}$ of $\iota(\tilde V''_{K_1-ex}))$ in $U$. Therefore we showed that the configuration space of Hodge decompositions is homotopic to a certain vector bundle over the product of Grassmanians
$$\mr{Gr}(\im(d),\tilde V''_{d-ex})\times \mr{Gr}(U_0,\tilde V''_{K_1-ex})$$
In particular, it is connected.
\\$\Box$

\textbf{Proof of Statement \ref{statement: ind data deform for BF_infty}.}
For an infinitesimal deformation of induction data $(\iota_1,r_1,K_1)\mapsto (\iota_2,r_2,K_2)$ we can use Statement \ref{statement: infinitesimal ind data deform for abstract BF}, since it never uses any specifics of abstract $BF$ theory (as compared to $BF_\infty$). Hence $S'_1$ and $S'_2$ are related by an infinitesimal canonical transformation. Since both effective actions satisfy $BF_\infty$ ansatz, this is a SCT.

Now consider the case of finite deformations $(\iota_1,r_1,K_1)\mapsto (\iota_2,r_2,K_2)$. According to Lemma, such a transformation can be represented as a composition
$$(\iota_1,r_1,K_1)\mapsto (\tilde\iota_2,\tilde r_2,\tilde K_2)\mapsto (\lambda,\lambda^{-1},0)\circ (\tilde\iota_2,\tilde r_2,\tilde K_2)=(\iota_2,r_2,K_2)$$
where $(\iota_1,r_1,K_1)$ and $(\tilde\iota_2,\tilde r_2,\tilde K_2)$ are in the same connected component of the configuration space of induction data. For the deformation $(\iota_1,r_1,K_1)\mapsto (\tilde\iota_2,\tilde r_2,\tilde K_2)$  the corresponding effective actions are related by a SCT, as a consequence of infinitesimal case. Hence we only need to consider deformations of induction data of form $(\iota_1,r_1,K_1)\mapsto (\lambda,\lambda^{-1},0)\circ (\iota_1,r_1,K_1)=(\iota_2,r_2,K_2)$. These deformations (automorphisms of $V'$) affect the effective action as $S'_2=\hat\lambda^* S'_1$ with $\hat\lambda=\lambda\oplus(\lambda^*)^{-1}$, which is obvious from the definition (\ref{BV integral for BF infty}). According to the Definition \ref{finite special canonical transform}, this is a finite special canonical deformation on $V'$ with $(U,\chi)=(\lambda,0)$.
\\$\Box$

\begin{Def}\label{def: homotopy of qL_infty algebras} We call two $qL_\infty$ algebras $(V_1,Q_1,\mu_1)$ and $(V_2,Q_2,\mu_2)$ equivalent, if their cohomologies (i.e. cohomologies of the differential $d=l_{(1)}$) are isomorphic $H^\bt(V_1)=H^\bt(V_2)$, and the induced  $qL_\infty$ structures on cohomology $(H^\bt(V),Q'_1,\mu'_1)$ and $(H^\bt(V),Q'_2,\mu'_2)$ are related by a SCT $(U,\chi)$.
\end{Def}
Due to the Statement \ref{statement: ind data deform for BF_infty}, this definition does not depend on the choice of induction data from both $qL_\infty$ algebras to the cohomology. Also, according to Statement \ref{statement: special can transf induction}, two $qL_\infty$ structures on space $V$, related by an infinitesimal SCT, are equivalent. Also a $qL_\infty$ algebra $(V,Q,\mu)$ is equivalent to any  $qL_\infty$ algebra $(V',Q',\mu')$ induced from it, as a consequence of the Statement \ref{statement: iterated induction}.

Two $BF_\infty$ theories, corresponding to two equivalent $qL_\infty$ algebras, are considered equivalent.

\textbf{Remark.} We should mention two drawbacks of our discussion of equivalence of $qL_\infty$ algebras. First, our treatment of the transfer of a canonical transformation to a subcomplex relies on Statement \ref{statement: induced canonical transformation} and on the formula (\ref{induced canonical transformation}) for transferring an infinitesimal canonical transformation. Therefore the Statement \ref{statement: special can transf induction} is formulated for infinitesimal SCTs and is straightforwardly generalized to the connected component of identity in the configuration space of finite SCTs, but not to other connected components --- such a generalization would require the description of transfer of canonical transformations in finite (not infinitesimal) terms. Therefore, in particular, we cannot guarantee that two $qL_\infty$ structures on space $V$, related by a finite SCT $(U,\chi)$, are necessarily equivalent.

Second, $L_\infty$ algebras have a nice notion of morphism. Namely, one calls an $L_\infty$ morphism between two $L_\infty$ algebras $(V_1,Q_1)$ and $(V_2,Q_2)$ a non-linear map $U:V_1[1]\ra V_2[1]$ of degree zero ,such that  $Q_1 U^*=U^* Q_2$. Two $L_\infty$ algebras are called equivalent (or homotopic), if there exist an $L_\infty$-quasi-isomorphism between them, i.e. one additionally requires that the linear part of $U$ induces an isomorphism on cohomology $H^\bt(V_1)\cong H^\bt(V_2)$. On the other hand, for the case of $qL_\infty$ algebras we could only define an invertible $qL_\infty$ morphism (or $qL_\infty$ automorphism) --- the special canonical transformation (Definition \ref{finite special canonical transform}), which is quite restrictive (in particular, it preserves the dimension). We would like to say that induction of a $qL_\infty$ algebra is another instance of $qL_\infty$ quasi-isomorphism, and our definition of equivalence of $qL_\infty$ algebras (Definition \ref{def: homotopy of qL_infty algebras}) is inspired by this idea. But we do not know how to give a general definition of $qL_\infty$ morphism, generalizing these special cases --- the $qL_\infty$ automorphism and the induction.

\subsubsection{Effective action in terms of $L_\infty$ morphism and torsion}
\label{section: S' via U and I}
\begin{statement}\label{statement: L_infty morph for induction}
Let $S$ be the $BF_\infty$ action, associated to a $qL_\infty$ algebra $(V,Q,\mu)$, and let $S'$ be the effective action, associated to the induced $qL_\infty$ algebra $(V',Q',\mu')$, with the induction data $(\iota,r,K)$. Define a nonlinear map $U: V'[1]\ra V[1]$ via the pull-back on functions $U^*:\Fun(V[1])\ra \Fun(V'[1])$, given by
\be f(\omega)\mapsto U^*(f)(\omega')=e^{-S'(\omega',p')/\hbar}\int_{\LL_K}e^{S(\iota(\omega')+\omega'',r^*(p')+p'')/\hbar} f(\omega)\label{L_infty morph via pullback}\ee
Then the perturbation series for $U$ is
\be U(\omega')=\iota(\omega')+\sum_{T\in{\bf{T}}_\mr{nonPl}:\;|T|\geq 2}\frac{1}{|\mr{Aut}(T)|}\Iter_{T;\{-Kl_{(n)}\}_{n\geq 2};\{-Kl_{(n)}\}_{n\geq 2}}(\iota(\omega'),\ldots,\iota(\omega')) \label{L_infty morph pert series}\ee
Moreover, $U$ is an $L_\infty$ quasi-isomorphism between the $L_\infty$ algebras $(V,Q)$ and $(V',Q')$, i.e. \be U^*Q=Q'U^* \label{L_infty morph condition}\ee holds, and the linear part of $U$ induces an isomorphism between cohomologies of $V$ and $V'$.
\end{statement}

\textbf{Proof.} First, we need to check that $U^*$ defined by (\ref{L_infty morph via pullback}) is a homomorphism, i.e. $U^*(fg)=U^*(f)U^*(g)$ (this implies that $U^*$ is really a pull-back for some map $U:V'[1]\ra V[1]$). This follows from the observation that field $\omega$ does not interact with itself, i.e. there are no connected Feynman diagrams with more than one outgoing 1-valent vertex (corresponding to insertions of $\omega$ in the integral), and the other vertices generated by the action $S$. Next, let us check the property (\ref{L_infty morph condition}) of $L_\infty$ morphism.
To do this. we apply the BV Laplacian $\Delta'$  to the integral $\int_{\LL_K}e^{S/\hbar} f=e^{S'/\hbar}U^*(f)$. We have
\begin{multline}\Delta'(e^{S'/\hbar}U^*(f))=\Delta'(e^{S'/\hbar})U^*(f)+\{e^{S'/\hbar},U^*(f)\}+e^{S'/\hbar}\Delta'(U^*(f)) \\
=0+\frac{1}{\hbar}e^{S'/\hbar}\{S',U^*(f)\}+0=\frac{1}{\hbar}e^{S'/\hbar}Q'U^*(f)
\end{multline}
On the other hand,
\begin{multline}\Delta'\int_{\LL_K}e^{S/\hbar} f=\int_{\LL_K} (\Delta-\Delta'')(e^{S/\hbar} f)=\int_{\LL_K}\Delta (e^{S/\hbar} f)
\\=\frac{1}{\hbar}\int_{\LL_K}e^{S/\hbar}\{S,f\}=\frac{1}{\hbar}\int_{\LL_K}e^{S/\hbar}Qf=\frac{1}{\hbar}e^{S'/\hbar}U^*(Qf)
\end{multline}
Hence $U$ is really an $L_\infty$ morphism. It is a quasi-isomorphism, because the linear part of $U$ is the embedding $\iota$, which is itself a linear quasi-isomorphism.
Perturbative expansion (\ref{L_infty morph pert series}) follows from considering Feynman diagrams for the integral (\ref{L_infty morph via pullback}).
\\$\Box$

Expression $U(\omega')$ in (\ref{L_infty morph pert series}) looks ambiguously. We understand it as defined by $U^*(f)(\omega')=f(U(\omega'))$, and therefore $U(\omega')=U^*(\omega)$, just as $\iota(\omega')=\iota^*(\omega)$, where we mean that $\iota$ acts on $e'_i$ in decomposition $\omega'=e'_i\omega'^i$, while $\iota^*$ acts on $\omega^i$ in $\omega=e_i\omega^i$. One can also interpret  $U(\omega')$ in terms of components of the $L_\infty$ morphism --- polylinear antisymmetric maps $U_{(n)}:\Lambda^n V'\ra V$, which are introduced analogously to defining the classical operations $l_{(n)}$ via Taylor series for $Q$. Namely, the structure constants of components of $U_{(n)}$ are defined from
$$U(\omega')=e_i U^*(\omega^i)=\sum_{n=1}^\infty\frac{1}{n!}e_i U^i_{(n)}(e'_{j_1},\ldots,e'_{j_n})\omega'^{j_1}\cdots\omega'^{j_n}\e(j_1,\ldots,j_n)$$
where the sign $\e$ is defined by (\ref{sign}). Therefore $U(\omega')$ is interpreted either as $U^*$ acting on coordinates $\omega^i$ on $V[1]$, or via the components $U_{(n)}$, acting on vectors $e'_j\in V'$.

It follows from comparing (\ref{L_infty morph pert series}) and (\ref{pert series for S'^0 for BF_infty}), that we can write
$$S'^0(\omega',p')=<p',\sum_{n=1}^\infty r\circ \frac{1}{n!} l_{(n)}(U(\omega'),\ldots,U(\omega'))>=S^0(U(\omega'),r^*(p'))$$
i.e. the value of tree part of the effective action in any point of $\FF'$ is given by the value of the original action in the point, shifted along $\LL_K$, and the shift is given by the $L_\infty$ morphism: $\omega'\mapsto \omega=U(\omega')=\iota(\omega')+K(\cdots)$, $p'\mapsto r^*(p')$.

\begin{statement} Effective action for $BF_\infty$ theory, defined by (\ref{BV integral for BF infty}), can be written as
\be S'(\omega',p')=S^0(U(\omega'),r^*(p'))+\hbar S^1 (U(\omega'))+\hbar\;\Str_V\log(1+K\circ I(U(\omega'))) \label{S' via U and I}\ee
where the linear $\omega$-dependent operator $I(\omega)\in\Fun(V[1])\otimes\End(V)$ is defined by
\be I(\omega)=\sum_{n=1}^\infty\frac{1}{n!}l_{(n+1)}(\omega,\ldots,\omega,\bt)\label{I}\ee
where $U$ is the $L_\infty$ morphism (\ref{L_infty morph pert series}).
\end{statement}

\textbf{Proof.} The fact that $S^0(U(\omega'),r^*(p'))$ and $S^1(U(\omega'))$ coincide with (\ref{pert series for S'^0 for BF_infty}) and the second sum (over trees) in (\ref{pert series for S'^1 for BF_infty}), follows immediately from the formula (\ref{L_infty morph pert series}) for $U$. So we just have to prove that $\Str_V\log(1+K I(U(\omega')))$ coincides with the first sum in (\ref{pert series for S'^1 for BF_infty}). Indeed,
\begin{multline} \Str_V\log(1+K I(U(\omega')))=-\sum_{m=1}^\infty \frac{1}{m}\Str_V (-K I(U(\omega')))^m\\
=-\sum_{m=1}^\infty \frac{1}{m}\Str_V\left(\sum_{T\in{\bf{T}}_\mr{Pl}:\;|T|\geq 1}\Iter_{T;\{-\frac{1}{n!}K l_{(n)}\}_{n\geq 2};\{-\frac{1}{n!}K l_{(n+1)}(*,\cdots,*,\bt)\}_{n\geq 1}}(\iota(\omega'),\ldots,\iota(\omega'))\right)^m\\
=-\sum_{L\in{\bf{L}}_\mr{Pl}}\frac{1}{[L]}\Loop_{L;\{-\frac{1}{n!}K l_{(n)}\}_{n\geq 2}}(\iota(\omega'),\ldots,\iota(\omega'))\\=
-\sum_{L\in{\bf{L}}_\mr{nonPl}}\frac{1}{|\Aut(L)|}\Loop_{L;\{-K l_{(n)}\}_{n\geq 2}}(\iota(\omega'),\ldots,\iota(\omega'))
\end{multline}
where passed to summing over planar graphs for the ease of treatment of the symmetry coefficients. We denoted $[L]$ the length of cycle in $L$; $\bf{L}_\mr{Pl}$ is to be understood as the set of planar one-loops graphs with one marked edge in the cycle.
\\$\Box$

Expression (\ref{S' via U and I}) for the effective action can also be understood as follows. Introduce the function $\tau\in\Fun(V'[1])$:
\be \tau(\omega')=\mr{Sdet}_V (1+K I(U(\omega')))\label{torsion}\ee
where we mean the super-determinant: $\mr{Sdet}_V (1+K I(U(\omega'))):=\exp\left(\Str_V \log(1+K I(U(\omega')))\right)$. Function $\tau$ depends on the induction data $(\iota,r,K)$ and may be understood as the gauge fixing for the ill-defined expression $\mr{Sdet}_V (d+I(U(\omega')))$ (the determinant of Hessian of action $\frac{\dd}{\dd\omega''_{\LL}}\frac{\dd}{\dd p''_\LL}S|_{\LL_K}$ in the stationary point on $\LL_K$). We call $\tau$ the ``torsion'',  where the term is inspired by the Ray-Singer torsion. In a sense $\tau$ plays the role of Jacobian of the $L_\infty$ morphism (cf. the definition of finite SCT (\ref{fin spec can transf for S})). Therefore, in terms of $L_\infty$ morphism $U$ and torsionя $\tau$, the effective action is
$$S'(\omega',p')=S(U(\omega'),r^*(p'))+\hbar\log\tau(\omega')$$

\section{Simplicial $BF$ theory}
\label{section: BF on simplicial complex}
Here we will apply the construction of effective action from section \ref{section: abstract BF and effective action} to the case of topological $BF$ theory, i.e. abstract $BF$ theory associated to the de Rham DGLA $V=\Omega^\bt(M,\g)$, and induction to the subcomplex $V'=C^\bt(\Xi,\g)\hookrightarrow V$  of $\g$-valued cell cochains of the simplicial complex $\Xi$ (a triangulation of $M$). Here $M$ is assumed to be a compact manifold, possibly with corners, possibly non-orientable, and we assume that $\g$ is the Lie algebra of some compact finite-dimensional Lie group $G$ (the gauge group). First we have to present the induction data $(\iota,r,K)$, i.e. embedding $\iota:C^\bt(\Xi,\g)\ra \Omega^\bt(M,\g)$, retraction $r:\Omega^\bt(M,\g)\ra C^\bt(\Xi,\g)$ and chain homotopy $K:\Omega^\bt(M,\g)\ra \Omega^{\bt-1}(M,\g)$. The idea is to set $\iota$ to be the embedding of cell cochains of $\Xi$ as Whitney forms \cite{Whitney}, for the retraction $r$ we take the integration of a differential form over simplices of $\Xi$, for $K$ we take the explicit Dupont's chain homotopy \cite{Dupont}, retracting the differential forms to Whitney forms. We use \cite{Getzler} in our exposition of these constructions.

In sections \ref{section: Whitney forms},\ref{section: Dupont's homotopy} we present the construction of Whitney forms and Dupont's operator on a standard simplex, and construct the induction data for a general simplicial complex $\Omega^\bt(M,\g)\xra{(\iota_\Xi,r_\Xi,K_\Xi)}C^\bt(\Xi,\g)$.

In section \ref{section: simplicial BF action} we introduce the notion of simplicial $BF$ action $S_\Xi$ --- the effective action for topological $BF$ theory on a manifold  $M$, induced on cell cochains of a triangulation $\Xi$ with the induction data $(\iota_\Xi,r_\Xi,K_\Xi)$. We also discuss the important property of simplicial action --- its simplicial locality (Theorem \ref{thm: simplicial locality}). Namely, $S_\Xi$ is represented as a sum of contributions (``reduced actions'') for individual simplices $\Xi$. These contributions are given by universal formulae and can be obtained from solving the problem of induction for single standard simplex (in all dimensions $D\geq 0$).

In section \ref{section: gluing} we propose the abstract construction of gluing for $qL_\infty$ algebras and prove that it is consistent with the induction procedure (section \ref{section: gluing-induction}). This construction  clarifies in a sense the simplicial locality property of simplicial $BF$ action from a more abstract point of view.

In section \ref{section: interval} we compute the reduced action for 1-simplex explicitly (Theorem \ref{interval thm}). Explicit check of the QME for the simplicial action for 1-simplex (section \ref{section: interval QME check}) leads to a non-trivial quadratic identity for Bernoulli numbers (\ref{interval CME Bernoulli relation}). Of course, this identity can also be obtained more directly \cite{AD}. Also, from the explicit check of quantum part of QME we see that the one-loop part of the simplicial action for 1-simplex can be completely reconstructed from the tree part by means of QME. Classical part of the induced $qL_\infty$ structure on cell cochains of the interval (section \ref{section: interval qL_infty structure}), generated by the tree part of simplicial action for 1-simplex, is a known result (see \cite{CG}, \cite{LS}).

In section \ref{section: simplex pert} we address the problem of computing the reduced action for the simplex of general dimension. We cannot find the explicit formula here and can only present the perturbative result, i.e. we compute the contributions of first few Feynman diagrams (Theorem \ref{thm: simplex perturbative result}). The technique we used here allows one to compute the values of tree diagrams for the simplex of arbitrary dimension. However, values of one-loop diagrams can only be partially reconstructed using the symmetry of simplex, the tree result and the QME. In section \ref{section: q_2 on 2-simplex} we demonstrate the explicit computation of the simplest non-trivial one-loop diagram for 2-simplex, and the result turns out to agree with the prediction we derived from the tree result and QME.

\subsection{Whitney forms}
\label{section: Whitney forms}
Let $\Delta^n=[0\cdots n]$ be the standard $n$-simplex with vertices labeled by integers from $0$ to $n$ and with barycentric coordinates $(t_0,\ldots,t_n)$ with the constraint $t_0+\cdots+t_n=1$ and $t_i\geq 0$ for every $i$. Let us introduce a collection of linear differential forms
\be \chi_{i_0\cdots i_k}=k!\sum_{r=0}^k (-1)^r t_{i_r} dt_{i_0}\wedge\cdots\widehat{d t_{i_r}}\cdots\wedge dt_{i_k}\in\Omega^k(\Delta^n)\label{chi def}\ee
for every sequence $i_0,\ldots,i_k$ of vertices of $\Delta^n$, hat means exclusion. Forms $\chi_{i_0\cdots i_k}$ have the following properties.
\begin{itemize}
\item Consistency with permutations of vertices: \be\chi_{i_{\pi(0)}\cdots i_{\pi(k)}}=(-1)^\pi \chi_{i_0\cdots i_k}\label{chi symmetry}\ee
for any permutation $\pi\in S_k$. Therefore the form $\chi_{i_0\cdots i_k}=\chi_\sigma$ can be understood as associated to the oriented $k$-dimensional face $\sigma=[i_0\cdots i_k]$ of  $\Delta^n$.
\item Restriction of $\chi_\sigma$ to $\sigma$ yields the standard volume form on $\sigma$:
$$(\chi_\sigma)|_\sigma=k! dt_{i_1}\wedge\cdots\wedge dt_{i_k}$$
In particular, the integral of $\chi_\sigma$ over the face $\sigma$ is \be \int_\sigma \chi_\sigma=1\label{chi int 1}\ee
\item If $\sigma$, $\sigma'$ are two faces of $\Delta^n$ and $\sigma$ is not contained in $\sigma'$, then \be(\chi_\sigma)|_{\sigma'}=0\label{chi restriction}\ee
In particular, for any face $\sigma'\neq\sigma$ we have \be \int_{\sigma'}\chi_\sigma=0\label{chi int 2}\ee
\item De Rham differential acts on forms $\chi$ as
\be d\chi_{i_0\cdots i_k}=\sum_i \chi_{i i_0\cdots i_k}\label{d of Whitney form}\ee
\end{itemize}

Forms $\chi_\sigma$ span the subcomplex $\Omega_W^\bt(\Delta^n)\subset \Omega^\bt(\Delta^n)$ in the DGA of differential forms on $\Delta^n$. Complex $\Omega_W^\bt(\Delta^n)$ is called the Whitney complex, and its elements are called Whitney forms. Isomorphism between the complex  $C^\bt(\Delta^n)$ of cell cochains of the standard triangulation of $\Delta^n$ (i.e. the one consisting of faces of $\Delta^n$ of all dimensions) and the Whitney complex $\Omega_W^\bt(\Delta^n)$ is given by the map $$\iota_{\Delta^n}:e_\sigma\mapsto\chi_\sigma$$
sending a basis cell cochain $e_\sigma$ to the Whitney form $\chi_\sigma$ for each face $\sigma\subset\Delta^n$. Therefore we have the embedding $\iota_{\Delta^n}:C^\bt(\Delta^n)\hra \Omega^\bt(\Delta^n)$. Its image is the Whitney complex $\iota_{\Delta^n}(C^\bt(\Delta^n))=\Omega_W^\bt(\Delta^n)\subset \Omega^\bt(\Delta^n)$. The fact that this embedding is a chain map follows from (\ref{d of Whitney form}). An important property of Whitney complex for a simplex is that it is  consistent with restriction to a face: $\Omega_W^\bt(\Delta^n)|_\sigma=\Omega_W^\bt(\sigma)$ for any face $\sigma\subset\Delta^n$.

Let now $\Xi$ be a triangulation of the manifold $M$.  For every simplex  $\sigma\in\Xi$ we define the corresponding basis Whitney form $\chi_\sigma$ by its restrictions to the simplices of triangulation $\sigma'\in\Xi$:
$$(\chi_\sigma)|_{\sigma'}=\left\{
\begin{array}{ll}\chi_\sigma^{(\sigma')},&\sigma\subset\sigma'\\ 0,&\sigma\not\subset\sigma'\end{array}\right.$$
where $\chi_\sigma^{(\sigma')}\in\Omega^\bt(\sigma')$ is the Whitney form on $\sigma'$, corresponding to the face $\sigma\subset\sigma'$, defined by (\ref{chi def}). Forms $\chi_\sigma$ are piecewise-linear differential forms on $M$, and their linear span $\Omega_W^\bt(\Xi)\subset\Omega^\bt(M)$ (Whitney complex for the triangulation $\Xi$) is a subcomplex,  i.e. it is closed under de Rham differential. Alternatively one can formulate the construction of $\Omega_W^\bt(\Xi)$ by gluing Whitney complexes for simplices of highest dimension in $\Xi$ over faces of codimension 1. More precisely, we are gluing Whitney complexes by restriction maps $\Omega_W^\bt(\sigma)\ra\Omega_W^\bt(\sigma)|_{\sigma'}=\Omega_W^\bt(\sigma')$, with $\sigma'\subset\sigma$, $\dim\sigma'=\dim\sigma-1$.

The embedding of cell cochains of triangulation $\Xi$ into differential forms is again given by $\iota_\Xi:e_\sigma\mapsto \chi_\sigma$, i.e. a basis cochain associated to simplex $\sigma\in\Xi$ is sent to the corresponding Whitney form on $M$. Therefore we have the chain map
$\iota_\Xi:C^\bt(\Xi)\hra \Omega^\bt(M)$, whose image is $\Omega_W^\bt(\Xi)$. We define the retraction $r_\Xi:\Omega^\bt(M)\ra C^\bt(\Xi)$ by integrals over simplices:
$$\alpha\mapsto\sum_{\sigma\in\Xi}\left(\int_\sigma\alpha\right)e_\sigma$$
for $\alpha\in\Omega^\bt(M)$. The fact that $r_\Xi$ is a chain map is implied by Stokes theorem, and the property $r_\Xi\circ\iota_\Xi=\id_{C^\bt(\Xi)}$ follows from the properties (\ref{chi int 1},\ref{chi int 2}) of Whitney forms.

\subsection{Dupont's chain homotopy}
\label{section: Dupont's homotopy}
First consider the case of standard $n$-simplex $\Delta^n=[0\cdots n]$. For every vertex $i$ of the simplex we define the dilation map
$$\phi_i:[0,1]\times \Delta^n\ra\Delta^n$$ given by
$$\phi_i(u;t_0,\ldots,t_n)=(ut_0,\ldots,u t_i+(1-u),\ldots,u t_n)$$
Let $\pi:[0,1]\times\Delta^n\ra\Delta^n$ be the projection to second factor and let $\pi_*:\Omega^\bt([0,1]\times \Delta^n)\ra\Omega^{\bt-1}(\Delta^n)$ be the integration over the first factor. Let us introduce the operators
$$h^i:\Omega^\bt(\Delta^n)\ra\Omega^{\bt-1}(\Delta^n)$$
as
$$h^i\alpha=\pi_*\phi_i^*\alpha$$
for $\alpha\in\Omega^\bt(\Delta^n)$. Let $\ev^i:\Omega^\bt(\Delta^n)\ra \RR$ be the evaluation of a form in vertex $i$ (vanishing for forms of degree $\geq 1$). Stokes' theorem implies that $h^i$ is the chain homotopy between the identity map $\id:\Omega^\bt(\Delta^n)\ra\Omega^\bt(\Delta^n)$ and $\ev^i$:
$$d h^i+h^i d=\id-\ev^i$$
Also the following holds: $$h^i h^j+h^j h^i=0$$
since $h^i h^j\alpha=\pi_* \phi^*_{ij}\alpha$, where $\phi_{ij}:[0,1]\times [0,1]\times\Delta^n\ra\Delta^n$ is the map
$$\phi_{ij}(u,v;t_0,\ldots,t_n)=(uv t_k+v (1-u)\delta_{ik}+(1-v)\delta_{jk})$$
and $\phi_{ji}(u,v)=\phi_{ij}(\tilde{v},\tilde{u})$, where $(u,v)\mapsto (\tilde{u},\tilde{v})$ is a diffeomorphism of the square $[0,1]\times [0,1]$, given by
$1-v=\tilde{u}(1-\tilde{v})$, $v (1-u)=(1-\tilde{u})$.

Dupont's chain homotpy operator $K_{\Delta^n}:\Omega^\bt(\Delta^n)\ra\Omega^{\bt-1}(\Delta^n)$ is defined as
\begin{eqnarray}K_{\Delta^n}&=&\sum_{k=0}^{n-1}(-1)^k\sum_{0\leq i_0<\cdots< i_k\leq n}\chi_{i_0\cdots i_k}h^{i_k}\cdots h^{i_0} \label{K Dupont}\\
&=&\sum_{k=0}^{n-1}(-1)^k\sum_{i_0,\ldots,i_k=0}^n t_{i_0}dt_{i_1}\cdots dt_{i_k} h^{i_k}\cdots h^{i_0}\nonumber
\end{eqnarray}
and satisfies the following properties.
\begin{enumerate}
\item Consistency with restrictions to faces: for a face $\sigma\subset\Delta^n$ we have
\be (K_{\Delta^n}\alpha)|_\sigma=K_\sigma (\alpha|_\sigma) \label{K restriction to face}\ee
for any form $\alpha\in\Omega^\bt(\Delta^n)$
\item Dupont's operator vanishes on Whitney forms \be K_{\Delta^n} \iota_{\Delta^n}=0\label{iota K Dupont}\ee
and the integrals over faces for the image of $K$ also vanish
\be r_{\Delta^n} K_{\Delta^n}=0 \label{rK Dupont}\ee
\item The ``simplicial de Rham theorem'': operator $K_{\Delta^n}$ is a chain homotopy between the identity map $\id:\Omega^\bt(\Delta^n)\ra\Omega^\bt(\Delta^n)$ and the projection to Whitney forms:
    \be d K_{\Delta^n}+K_{\Delta^n} d=\id-\iota_{\Delta^n} r_{\Delta^n} \label{dK+Kd Dupont}\ee
\item The property
\be (K_{\Delta^n})^2=0 \label{K^2=0 on simplex}\ee
(see \cite{Getzler} for the proof).
\end{enumerate}
Property (\ref{K restriction to face}) follows from the property (\ref{chi restriction}) for Whitney forms. Namely, it implies that upon restriction to $\sigma$ all terms in (\ref{K Dupont}) vanish except those, where all $i_0,\ldots,i_k$ are vertices of the face $\sigma$. We also use the fact that dilation $\phi_i$ for a vertex $i\in\sigma\subset\Delta^n$ sends points of $\sigma$ to points of $\sigma$. Also (\ref{K restriction to face}) implies in particular that the restriction $(K_{\Delta^n}\alpha)|_\sigma$ does not have a component of degree $\dim\sigma$, which implies (\ref{rK Dupont}). Property (\ref{iota K Dupont}) is proved by a straightforward computation, which we will present here.

\textbf{Proof of (\ref{iota K Dupont})} We have to prove that $K\chi_{i_0\cdots i_k}=0$. Using $S_n$-symmetry of the simplex $\Delta^n$, let us set $[i_0\cdots i_k]=[0\cdots k]$. First we consider the action of $h^i$ on $\chi_{0\cdots k}$. For $i>k$ we have
\begin{eqnarray*}
\phi_i^*\chi_{0\cdots k}&=&k! \sum_{r=0}^k (-1)^r u t_r (t_0 du+ u dt_0)\cdots\widehat{(t_r du+ u dt_r)}\cdots (t_k du+ u dt_k)\\
&=&k! u^k du\left(\sum_{r=0}^k\sum_{s=0}^{r-1}(-1)^{r+s} t_r t_s dt_0\cdots\widehat{dt_s}\cdots\widehat{dt_r}\cdots dt_k\right.+\\
&&+\left.\sum_{r=0}^k\sum_{s=r+1}^{k}(-1)^{r+s+1} t_r t_s dt_0\cdots\widehat{dt_r}\cdots\widehat{dt_s}\cdots dt_k\right)+u^{k+1}\chi_{0\cdots k}
\end{eqnarray*}
upon transposition of $r$ and $s$ the first sum in brackets becomes the second sum with opposite sign, hence the expression in brackets vanishes and $\phi_i^*\chi_{0\cdots k}=u^{k+1}\chi_{0\cdots k}$ is a form of degree 0 on the interval $[0,1]$. Hence $$i>k\Rightarrow h^i\chi_{0\cdots k}=0$$
Now consider the case $i\leq k$.
\begin{multline}\phi_i^*\chi_{0\cdots k}  =  k!\sum_{r=0}^k (-1)^r (u t_r+(1-u)\delta_{ir})\prod_{0\leq j\leq k,\;j\neq r} ((t_j-\delta_{ij})du+u dt_j)\\
=  k!u^{k-1}du\left(\sum_{0\leq s<r\leq k} (-1)^{r+s} (u t_r+(1-u)\delta_{ir}) (t_s-\delta_{is}) dt_0\cdots\wh{dt_s}\cdots\wh{dt_r}\cdots dt_k\right.+\\
+\left.\sum_{0\leq r<s\leq k} (-1)^{r+s+1} (u t_r+(1-u)\delta_{ir}) (t_s-\delta_{is}) dt_0\cdots\wh{dt_r}\cdots\wh{dt_s}\cdots dt_k\right)+\\
+\left(u^{k+1}\chi_{0\cdots k}+(-1)^i k! (1-u)u^k dt_0\cdots\wh{dt_i}\cdots dt_k\right)\\
= k!u^k du\left(\sum_{0\leq s<r\leq k}(-1)^{r+s} t_r t_s dt_0\cdots\wh{dt_s}\cdots\wh{dt_r}\cdots dt_k
+\right.\\
\left.+\sum_{0\leq r<s\leq k} (-1)^{r+s+1} dt_r dt_s dt_0\cdots\wh{dt_r}\cdots\wh{dt_s}\cdots dt_k\right)-\\
-k!u^k du \left(\sum_{i<r\leq k}(-1)^{i+r}t_r dt_0\cdots\wh{dt_i}\cdots\wh{dt_r}\cdots dt_k+\sum_{0\leq r<i}(-1)^{i+r+1}t_r dt_0\cdots\wh{dt_r}\cdots\wh{dt_i}\cdots dt_k\right)+\\
+k! (1-u)u^{k-1} du\left(\sum_{0\leq s<i}(-1)^{i+s}t_s dt_0\cdots\wh{dt_s}\cdots\wh{dt_i}\cdots dt_k+\right.\\
\left.+\sum_{i<s\leq k}(-1)^{i+s+1} t_s dt_0\cdots\wh{dt_i}\cdots\wh{dt_s}\cdots dt_k\right)+\\
+\left(u^{k+1}\chi_{0\cdots k}+(-1)^i k! (1-u)u^k dt_0\cdots\wh{dt_i}\cdots dt_k\right)\\
=\left(k! u^k du\cdot 0+(-1)^i k u^k du \chi_{0\cdots\wh{i}\cdots k}+(-1)^i k (1-u) u^{k-1}du \chi_{0\cdots\wh{i}\cdots k}\right)+\\
+\left(u^{k+1}\chi_{0\cdots k}+(-1)^i k! (1-u)u^k dt_0\cdots\wh{dt_i}\cdots dt_k\right)\\
=(-1)^i k u^{k-1}du\;\chi_{0\cdots\wh{i}\cdots k}+\left(u^{k+1}\chi_{0\cdots k}+(-1)^i k! (1-u)u^k dt_0\cdots\wh{dt_i}\cdots dt_k\right)
\label{phi of chi}
\end{multline}
Only the first term in the final expression is a degree 1 form in variable $u$. Therefore we obtain the elegant result
$$h^i\chi_{0\cdots k}=\left\{\begin{array}{ll}(-1)^i \chi_{0\cdots\wh{i}\cdots k},& 0\leq i\leq k\\ 0,& i>k \end{array}\right.$$
Or for the Whitney form, associated to an arbitrary face $\sigma=[j_0\cdots j_k]\subset\Delta^n$:
$$h^i\chi_{j_0\cdots j_k}=\left\{\begin{array}{ll}(-1)^r \chi_{j_0\cdots\wh{j_r}\cdots j_k},& i=j_r
\\ 0,& i\not\in[j_0\cdots j_k] \end{array}\right.$$
Using this we can compute $K_{\Delta^n}\chi_{0\cdots k}$:
\begin{multline}K_{\Delta^n}\chi_{0\cdots k}=\sum_{l=0}^{n-1} (-1)^l\sum_{0\leq i_0<\cdots< i_l\leq n}\chi_{i_0\cdots i_l}h^{i_l}\cdots h^{i_0}\chi_{0\cdots k}\\
=\sum_{l=0}^{k-1} (-1)^l\sum_{0\leq i_0<\cdots< i_l\leq k}(-1)^{i_0+(i_1+1)+\cdots+(i_l+l)}\chi_{i_0\cdots i_l}\chi_{0\cdots\wh{i_0}\cdots\wh{i_l}\cdots k}\\
=\sum_{l=0}^{k-1} (-1)^l\sum_{0\leq i_0<\cdots< i_l\leq k}\sum_{0\leq r\leq l}\sum_{0\leq s\leq k,\;s\neq i_j
}(-1)^{r+s+\sharp\{j:\,i_j<s\}} t_r t_s
dt_{i_l}\cdots\wh{dt_{i_r}}\cdots dt_{i_0}\cdot \\
 \cdot dt_0\cdots\,\wh{dt_{i_0}}\cdots\wh{dt_s}\cdots\wh{dt_{i_l}}\cdots dt_k\\
=\sum_{l=0}^{k-1} (-1)^l\sum_{0\leq i_0<\cdots< i_l\leq k}\left(\sum_{0\leq r\leq l}\sum_{0\leq s< i_r,\;s\neq i_j}
(-1)^{r+s+i_r+\sharp\{j:\,i_j<s\}+\sharp\{j:\,s<i_j<i_r\}}\;\cdot\right.\\
\left.\cdot\; t_r t_s dt_0\cdots \wh{dt_s}\cdots\wh{dt_{i_r}}\cdots dt_k+\right.\\
+\left.\sum_{0\leq r\leq l}\sum_{i_r<s\leq k:\;s\neq i_j}
(-1)^{r+s+i_r+\sharp\{j:\,i_j<s\}+\sharp\{j:\,i_r<i_j<s\}}t_r t_s dt_0\cdots \wh{dt_{i_r}}\cdots\wh{dt_{s}}\cdots dt_k\right)\\
=\sum_{l=0}^{k-1} (-1)^l \binom{k-1}{l}\cdot\left(\sum_{0\leq s<q\leq k}(-1)^{s+q}t_q t_s dt_0\cdots\wh{dt_s}\cdots\wh{dt_q}\cdots dt_k+\right.\\
\left.+
\sum_{0\leq q<s\leq k}(-1)^{s+q+1}t_q t_s dt_0\cdots\wh{dt_q}\cdots\wh{dt_s}\cdots dt_k\right)=0 \label{K of chi calculation}
\end{multline}
The last expression vanishes, since the expression in brackets is zero. Here $\binom{k-1}{l}$ denotes the binomial coefficients.
\\$\Box$

Notice that in computation (\ref{K of chi calculation}) we proved the following important property for Whitney forms: for any face
$\sigma=[i_0\cdots i_k]\subset\Delta^n$ and any $0\leq l\leq k-1$ we have the quadratic relation for Whitney forms:
\be \sum_{j_0<\cdots<j_l,\, j_{l+1}<\cdots<j_k:\;\{j_0\cdots j_l\}\cup\{j_{l+1}\cdots j_k\}=\{i_0\cdots i_k\}}(-1)^{(j_0\cdots j_k)}\chi_{j_0\cdots j_l}\wedge \chi_{j_{l+1}\cdots j_k}=0 \label{chi quadratic relations}\ee
where we sum over ways to split the set of vertices of $\sigma$ into two subsets with $l+1$ and $k-l$ elements, respectively. The sign $(-1)^{(j_0\cdots j_k)}$ is the sign of permutation $[j_0\cdots j_k]\ra [i_0\cdots i_k]$. For example, for the face $\sigma=[012]$ and $l=0$ the relation (\ref{chi quadratic relations}) is
$$\chi_0 \chi_{12}-\chi_1 \chi_{02}+\chi_2 \chi_{01}=0$$
Let us also give the proof of the simplicial de Rham theorem (\ref{dK+Kd Dupont}).

\textbf{Proof of (\ref{dK+Kd Dupont})}
Let us compute the super-sommutator $[d,K_{\Delta^n}]=dK+Kd$:
\begin{multline*}
[d,K_{\Delta^n}]=\sum_{k=0}^n (-1)^k \sum_{i_0,\ldots,i_k=0}^n \left(dt_{i_0}\cdots dt_{i_k} h^{i_k}\cdots h^{i_0}+\right.\\
\left.+
\sum_{r=0}^k (-1)^r t_{i_0}dt_{i_1}\cdots dt_{i_k} h^{i_k}\cdots [d,h^{i_r}]\cdots h^{i_0}\right)\\
=\sum_{k=0}^n (-1)^k \sum_{i_0,\ldots,i_k=0}^n \left(dt_{i_0}\cdots dt_{i_k} h^{i_k}\cdots h^{i_0}+
\sum_{r=0}^k (-1)^r t_{i_0}dt_{i_1}\cdots dt_{i_k} h^{i_k}\cdots \wh{h^{i_r}}\cdots h^{i_0}\right.-\\
-\left.\sum_{r=0}^k (-1)^r t_{i_0}dt_{i_1}\cdots dt_{i_k} h^{i_k}\cdots \ev^{i_r}\cdots h^{i_0}\right)\\
=\sum_{k=0}^n (-1)^k \left(\sum_{i_0,\ldots,i_k=0}^n dt_{i_0}\cdots dt_{i_k} h^{i_k}\cdots h^{i_0}+
\sum_{i_1,\ldots,i_k=0}^n dt_{i_1}\cdots dt_{i_k} h^{i_k}\cdots h^{i_1}\right)-\\
-\sum_{k=0}^{n}\sum_{i_0,\ldots,i_k=0}^n t_{i_0}dt_{i_1}\cdots dt_{i_k}\ev^{i_k}h^{i_{k-1}}\cdots h^{i_0}\\
=\id-\sum_{k=0}^{n}\sum_{i_0,\ldots,i_k=0}^n t_{i_0}dt_{i_1}\cdots dt_{i_k}\ev^{i_k}h^{i_{k-1}}\cdots h^{i_0}
\end{multline*}
Now we only have to prove that \be\ev^{i_k}h^{i_{k-1}}\cdots h^{i_0}\alpha=\int_{[i_0\cdots i_k]}\alpha\label{int over face}\ee --- the integral over the respective face of $\Delta^n$, for any $k$-form $\alpha\in\Omega^k(\Delta^n)$ (it is clear that for a form of degree $\neq k$ left and right hand sides of (\ref{int over face}) vanish automatically).

Denote $f(x)=(h^{i_{k-1}}\cdots h^{i_0}\alpha)(x)$ the function on $\Delta^n$ and $x\in\Delta^n$ a point. We can write $$f=\pi_* \phi^*_{i_0\cdots i_{k-1}}\alpha$$ where $$\phi_{i_0\cdots i_{k-1}}=\phi_{i_0}\cdots \phi_{i_{k-1}}:[0,1]^k\times \Delta^n\ra\Delta^n$$
and $\pi_*$ is the integration over cube $[0,1]^k$. Instead of computing $f(x)$ as the integral over cube of the pull-back of $\alpha$ by the map $\phi_{i_0\cdots i_{k-1}}$, we can compute the integral of $\alpha$ itself over the image of the cube in the simplex (this image depends on the point $x$):
$$f(x)=\int_{\phi_{i_0\cdots i_{k-1}}([0,1]^k\times\{x\})\subset\Delta^n}\alpha$$
Since $\phi_{i_0\cdots i_{k-1}}$ is the iterated dilation --- towards the vertex $[i_{k-1}]$, then towards $[i_{k-2}]$ etc., it is clear that image of the cube $[0,1]^k$ is the convex hull of the set of vertices $[i_0],\cdots,[i_{k-1}]$ and the point $x$:
$$\phi_{i_0\cdots i_{k-1}}([0,1]^k\times\{x\})=\mr{Conv}([i_0],\cdots,[i_{k-1}],x)$$
i.e. a $k$-simplex inside $\Delta^n$, containing the $(k-1)$-face $[i_0\cdots i_{k-1}]$ and with a vertex in point $x$.  We denote the geometric vertices with square brackets to distinguish them from labels. Therefore we showed that for any $k$-form $\alpha\in\Omega^k(\Delta^n)$, any $(k-1)$-face $[i_0\cdots i_{k-1}]\subset\Delta^n$ and any point $x\in\Delta^n$, we have
$$(h^{i_{k-1}}\cdots h^{i_0}\alpha)(x)=\int_{\mr{Conv}([i_0],\cdots,[i_{k-1}],x)}\alpha$$
In particular, setting $x=[i_k]$, we recover (\ref{int over face}).
\\$\Box$

Now suppose we have a manifold $M$ with triangulation $\Xi$. We define the chain homotopy $K_\Xi:\Omega^\bt(M)\ra\Omega^\bt(M)$ by restrictions to simplices $\Xi$, i.e. for every $\sigma\in\Xi$ we set
\be (K_\Xi\alpha)|_\sigma=K_\sigma (\alpha|_\sigma)\label{K on Xi}\ee
for any differential form $\alpha\in\Omega^\bt(M)$. Due to the property (\ref{K restriction to face}), definition (\ref{K on Xi}) is self-consistent. This definition also implies that properties (\ref{iota K Dupont},\ref{rK Dupont},\ref{dK+Kd Dupont},\ref{K^2=0 on simplex}) of Dupont's operator on a simplex globalize to the triangulation straightforwardly:
\begin{eqnarray*}
K_\Xi\iota_\Xi&=&0\\
r_\Xi K_\Xi&=&0\\
d K_\Xi+K_\Xi d&=&\id_{\Omega^\bt(M)}-\iota_\Xi r_\Xi\\
(K_\Xi)^2&=&0
\end{eqnarray*}

\subsection{Simplicial $BF$ action}
\label{section: simplicial BF action}
We are interested in the topological $BF$ theory, i.e. the abstract $BF$ theory, associated to the DGLA of $\g$-valued differential forms on the manifold $M$: $$V=\Omega^\bt(M,\g)=\g\otimes\Omega^\bt(M)$$
and in the effective action $S_\Xi$ on the space of fields $\FF_\Xi=T^*[-1](V_\Xi[1])$, constructed canonically from the complex of $\g$-valued cell cochains of the triangulation $\Xi$ of $M$ :
$$V_\Xi=C^\bt(\Xi,\g)=\g\otimes C^\bt(\Xi)$$
For the induction data we choose the triplet of maps $(\iota_\Xi,r_\Xi,K_\Xi)$, constructed in sections \ref{section: Whitney forms}, \ref{section: Dupont's homotopy}. Space $V$ splits into IR and UV parts
$$\g\otimes\Omega^\bt(M)=\g\otimes \iota_\Xi(C^\bt(\Xi))\oplus\g\otimes\Omega''^\bt(M)$$
where IR part is the complex of $\g$-valued Whitney forms on triangulation $\g\otimes\Omega_W^\bt(\Xi)$ and UV part consists of $\g$-valued forms $\alpha''\in\g\otimes\Omega''^\bt(M)$ whose integrals over all simplices of the triangulation vanish:
$$\int_\sigma \alpha''=0$$
The space of UV forms is in turn split into $d$-exact and $K_\Xi$-exact parts: $$\g\otimes\Omega''^\bt(M)=\g\otimes\Omega''^\bt_{d-ex}(M)\oplus \g\otimes\Omega''^\bt_{K-ex}(M)$$

It is important to note the two specific properties of this situation (as compared to the general case of induction of effective action for abstract $BF$ theory). First, the maps $(\iota_\Xi,r_\Xi,K_\Xi)$ act nontrivially only on the de Rham part of forms, i.e. on the second factor in $\g\otimes\Omega^\bt(M)$, and are trivial in coefficients $\g$. Second, the induction data $(\iota_\Xi,r_\Xi,K_\Xi)$ are consistent with restrictions to the simplices of triangulation $\Omega^\bt(M)\ra\Omega^\bt(\sigma)$, $C^\bt(\Xi)\ra C^\bt(\sigma)$.

Let $(e_\sigma)$ be the basis cochains on $\Xi$, corresponding to the simplices of triangulation, and let $(e^\sigma)$ be the basis chains. Let also $(T_a)$ be some basis in Lie algebra $\g$ and $(T^a)$ be the dual basis in $\g^*$. Then, according to general formalism, the coordinates on the space of IR fields $\FF_\Xi=T^*[-1](V_\Xi[1])$ are $(\omega^{\sigma a},p_{\sigma a})$, with ghost numbers $\gh(\omega^{\sigma a})=1-|\sigma|$, $\gh(p_{\sigma a})=-2+|\sigma|$, where $|\sigma|$ denotes the dimension of simplex $\sigma$. Therefore the super-fields are
$$\omega_\Xi=\sum_{a,\sigma}T_a e_\sigma \omega^{\sigma a},\;p_\Xi=\sum_{a,\sigma}p_{\sigma a} T^a e^\sigma$$
However, for this situation it is more convenient to pass to the system of $\g$-valued coordinates $\omega^\sigma=\sum_a T_a \omega^{\sigma a}: \FF_\Xi\ra\g$ and $\g^*$-valued coordinates $p_\sigma=\sum_a p_{\sigma a}T^a: \FF_\Xi\ra\g^*$. The ghost numbers are again $\gh(\omega^\sigma)=1-|\sigma|$, $\gh(p_\sigma)=-2+|\sigma|$. In terms of coordinates $(\omega^\sigma,p_\sigma)$ the super-fields are
$$\omega_\Xi=\sum_\sigma e_\sigma \omega^\sigma,\;p_\Xi=\sum_{\sigma} p_\sigma e^\sigma$$
One can understand $\omega_\Xi$ as a $\g$-valued cochain on $\Xi$ of total degree $\gh+\deg=1$ (we assume that basis cochains and chains have de Rham degree $\deg(e_\sigma)=-\deg(e^\sigma)=|\sigma|$), and $p_\Xi$ --- as a $\g^*$-valued chain of total degree $\gh+\deg=-2$. Also $\omega^\sigma$ is understood as the value of cochain $\omega_\Xi$ on simplex $\sigma\in\Xi$, and $p_\sigma$ --- as the value of chain $p_\Xi$ on $\sigma$.

We call the action $S_\Xi(\omega_\Xi,p_\Xi)$ of effective $BF_\infty$ theory on the space $\FF_\Xi$, defined by perturbation series (\ref{sum over trees},\ref{sum over loops}), the ``simplicial $BF$ action'' for triangulation $\Xi$.

\begin{thm}[Simplicial locality of the simplicial $BF$-action $S_\Xi$]
\label{thm: simplicial locality}
There exists a sequence of universal functions
$$\bar{S}_{\Delta^n}(\{\omega^\sigma\}_{\sigma\subset\Delta^n},p_{\Delta^n})\in\Fun(g\otimes C^\bt(\Delta^n)[1]\oplus\g^*\otimes C_n(\Delta^n)[-2])$$
parameterized by non-negative integers $n$ (each $\bar{S}$ is a certain function of values of the cochain $\omega$ on all faces of $\Delta^n$ and value of $p$ on the bulk cell only), such that for any triangulation $\Xi$ of any manifold $M$ the simplicial action $S_\Xi$ can be represented as a sum of contributions of individual simplices of $\Xi$ (of all dimensions)
\be S_\Xi(\omega_\Xi,p_\Xi)=\sum_{\sigma\in\Xi}\bar{S}_\sigma(\{\omega^{\sigma'}\}_{\sigma'\subset\sigma},p_\sigma) \label{simplicial locality decomp}\ee
I.e. the contributions of simplices are given by universal functions, do not depend on the combinatorics of $\Xi$ and depend only on the restriction of fields to the simplex.
\end{thm}
\textbf{Proof.}
Consider first the tree part of effective action on $\Xi$:
\begin{multline*}S^0_\Xi=\sum_{\sigma,\sigma_1\in\Xi}<p_\sigma e^\sigma,d e_{\sigma_1} \omega^{\sigma_1}>
+\sum_{T\in{\bf{T}}_\mr{nonPl}:\,|T|\geq 2}\sum_{\sigma,\sigma_1,\ldots,\sigma_{|T|}\in\Xi}\frac{1}{|\Aut(T)|}\cdot\\
\cdot
<p_\sigma e^\sigma,r_\Xi\circ\Iter_{T;-K_\Xi[\bt,\bt];[\bt,\bt]}(\iota_\Xi(e_{\sigma_1} \omega^{\sigma_1}),\ldots,\iota_\Xi(e_{\sigma_{|T|}} \omega^{\sigma_{|T|}}))>\\
=\sum_{\sigma,\sigma_1\in\Xi} d^\sigma_{\sigma_1}<p_\sigma,\omega^{\sigma_1}>_\g+\\
+
\sum_{T\in{\bf{T}}_\mr{nonPl}:\,|T|\geq 2}\sum_{\sigma,\sigma_1,\ldots,\sigma_{|T|}\in\Xi}\frac{1}{|\Aut(T)|}\;\left<p_\sigma,\int_\sigma
\Iter_{T;-K_\Xi[\bt,\bt];[\bt,\bt]}(\chi_{\sigma_1} \omega^{\sigma_1},\ldots,\chi_{\sigma_{|T|}} \omega^{\sigma_{|T|}})\right>_\g
\end{multline*}
where $d^\sigma_{\sigma_1}=<e^\sigma,de_{\sigma_1}>$ is the matrix of differential on $C^\bt(\Xi)$, i.e. $d^\sigma_{\sigma_1}=\pm 1$ if $\sigma_1$ is a face of codimension 1 in $\sigma$ and $d^\sigma_{\sigma_1}=0$ otherwise. We denote $<\bt,\bt>_\g:\;\g^*\otimes\g\ra\RR$ the canonical pairing between the Lie algebra and its dual. Operator $K_\Xi$ has the following property: non-vanishing of the restriction $(K_\Xi\alpha)|_\sigma\neq 0$ for a form $\alpha\in\Omega^\bt(M)$ implies $\alpha|_\sigma\neq 0$. Also non-vanishing of the restriction of product of Whitney forms $(\chi_{\sigma_1}\wedge\chi_{\sigma_2})|_\sigma\neq 0$ implies that $\sigma_1$ and $\sigma_2$ are faces of $\sigma$; and $(\alpha\wedge\chi_{\sigma_1})|_{\sigma}\neq 0$ implies that $\sigma_1$ is a face of $\sigma$ and $\alpha|_\sigma\neq 0$. It follows from these observations that in the expression
$$\sum_{\sigma_1,\ldots,\sigma_{|T|}}\int_\sigma
\Iter_{T;-K_\Xi[\bt,\bt];[\bt,\bt]}(\chi_{\sigma_1} \omega^{\sigma_1},\ldots,\chi_{\sigma_{|T|}} \omega^{\sigma_{|T|}})$$
only those terms contribute where all simplices $\sigma_1,\ldots,\sigma_{|T|}$ are faces of the simplex $\sigma$. Therefore we can write
$$S^0_\Xi=\sum_{\sigma\in\Xi}\bar{S}^0_\sigma (\{\omega^{\sigma'}\}_{\sigma'\subset\sigma},p_\sigma)$$
where we define the summand $\bar{S}^0_\sigma$ as a sum over trees
\begin{multline*}\bar{S}^0_\sigma=\left<p_\sigma,\sum_{\sigma_1}d^\sigma_{\sigma_1}\omega^{\sigma_1}+\right.\\
\left.+
\sum_{T\in{\bf{T}}_\mr{nonPl}:\,|T|\geq 2}\sum_{\sigma_1,\ldots,\sigma_{|T|}\subset\sigma}\frac{1}{|\Aut(T)|}\int_\sigma
\Iter_{T;-K_\sigma[\bt,\bt];[\bt,\bt]}(\chi_{\sigma_1}\omega^{\sigma_1},\ldots,
\chi_{\sigma_{|T|}}\omega^{\sigma_{|T|}})\right>_\g
\end{multline*}
Here we understand Whitney forms $\chi_{\sigma_j}=\chi_{\sigma_j}^{(\sigma)}\in\Omega^\bt(\sigma)$ as differential forms on the simplex $\sigma$.

Now consider the one-loop part of the effective action on $\Xi$
$$S^1_\Xi=-\sum_{L\in{\bf{L}}_\mr{nonPl}}\sum_{\sigma_1,\ldots,\sigma_{|L|}\in\Xi}\frac{1}{|\Aut(L)|}\Loop_{L;-K_\Xi[\bt,\bt]}
(\chi_{\sigma_1}\omega^{\sigma_1},\ldots,\chi_{\sigma_{|L|}}\omega^{\sigma_{|L|}})$$
We would like to split the super-traces over $\Omega^\bt(M,\g)$ according to the following splitting of the space of differential forms:
\be \Omega^\bt(M)=\bigoplus_{\sigma\in\Xi}\Omega^\bt_0(\sigma) \label{Omega splitting}\ee
where $\Omega^\bt_0(\sigma)$ denotes the space of differential forms on the simplex $\sigma$ with zero restriction to the boundary $\dd\sigma$:
$$\Omega^\bt_0(\sigma)=\{\alpha\in\Omega^\bt(\sigma):\;\alpha|_{\dd\sigma}=0\}$$
In the splitting (\ref{Omega splitting}) we assume that the projection $\Omega^\bt(M)\ra \Omega^\bt_0(\sigma)$ is given just by restriction to the corresponding simplex; the embedding $\Omega^\bt_0(\sigma)\ra\Omega^\bt(M)$ for a simplex $\sigma$ of top dimension is given by extension of a form $\alpha\in\Omega^\bt_0(\sigma)$ by zero to the complement of $\sigma$ in $M$. For a simplex of dimension $\dim\sigma<\dim M$ we embed $\alpha\in\Omega^\bt_0(\sigma)$ into $\Omega^\bt(M)$ by means of a smearing function $\rho_\sigma\in C^\infty(M)$ supported in the small neighborhood $U_\sigma\subset M$ of the face $\sigma$ (i.e. $U_\sigma$ is a ``thickening'' of $\sigma$), and equal to 1 on the face $\sigma$ itself. I.e. the embedding $\Omega^\bt_0(\sigma)\ra\Omega^\bt(M)$ is given by $\alpha\mapsto \rho_\sigma\cdot\pi_\sigma^*\alpha$, where $\pi_\sigma:U_\sigma\ra \sigma$ is a projection to the face.

Using (\ref{Omega splitting}), we split the super-traces over $\Omega^\bt(M)$ as
$$\Loop_{L;-K_\Xi[\bt,\bt];\Omega^\bt(M,\g)}(\chi_{\sigma_1}\omega^{\sigma_1},\ldots,\chi_{\sigma_{|L|}}\omega^{\sigma_{|L|}})=
\sum_{\sigma\in\Xi}\Loop_{L;-K_\sigma[\bt,\bt];\Omega^\bt_0(\sigma,\g)}(\chi_{\sigma_1}\omega^{\sigma_1},\ldots,\chi_{\sigma_{|L|}}\omega^{\sigma_{|L|}})$$
(we are now indicating the space over which the super-trace is taken in the $\Loop$ notation). According to the arguments we used for the tree part of $S_\Xi$,\quad $\Loop_{L;-K_\sigma[\bt,\bt];\Omega^\bt_0(\sigma,\g)}(\chi_{\sigma_1}\omega^{\sigma_1},\ldots,\chi_{\sigma_{|L|}}\omega^{\sigma_{|L|}})$
may be non-vanishing only if all the simplices $\sigma_1,\ldots,\sigma_{|L|}$ are faces of $\sigma$. Hence
$$S^1_\Xi=\sum_{\sigma}\bar{S}^1_\sigma (\{\omega^{\sigma'}\}_{\sigma'\subset\sigma})$$
where
$$\bar{S}^1_\sigma=-\sum_{L\in{\bf{L}}_\mr{nonPl}}\sum_{\sigma_1,\ldots,\sigma_{|L|}\subset\sigma}\frac{1}{|\Aut(L)|}
\Loop_{L;-K_\sigma[\bt,\bt];\Omega^\bt_0(\sigma,\g)}(\chi_{\sigma_1}\omega^{\sigma_1},\ldots,\chi_{\sigma_{|L|}}\omega^{\sigma_{|L|}})$$

Therefore, (\ref{simplicial locality decomp}) is proved, and moreover we have
\begin{multline}\label{reduced action}\bar{S}_\sigma=\bar{S}^0_\sigma(\{\omega^{\sigma'}\}_{\sigma'\subset\sigma},p_\sigma)+
\hbar \bar{S}^1_\sigma(\{\omega^{\sigma'}\}_{\sigma'\subset\sigma})
=\left<p_\sigma,\sum_{\sigma_1}d^\sigma_{\sigma_1}\omega^{\sigma_1}+\right.\\
\left.+
\sum_{T\in{\bf{T}}_\mr{nonPl}:\,|T|\geq 2}\sum_{\sigma_1,\ldots,\sigma_{|T|}\subset\sigma}\frac{1}{|\Aut(T)|}\int_\sigma
\Iter_{T;-K_\sigma[\bt,\bt];[\bt,\bt]}(\chi_{\sigma_1}\omega^{\sigma_1},\ldots,
\chi_{\sigma_{|T|}}\omega^{\sigma_{|T|}})\right>_\g-\\
-\hbar \sum_{L\in{\bf{L}}_\mr{nonPl}}\sum_{\sigma_1,\ldots,\sigma_{|L|}\subset\sigma}\frac{1}{|\Aut(L)|}
\Loop_{L;-K_\sigma[\bt,\bt];\Omega^\bt_0(\sigma,\g)}(\chi_{\sigma_1}\omega^{\sigma_1},\ldots,\chi_{\sigma_{|L|}}\omega^{\sigma_{|L|}})
\end{multline}
$\Box$

Simplicial locality property (\ref{simplicial locality decomp}) implies that to find simplicial action $S_\Xi$ for any triangulation of any manifold, it suffices to know the sequence of universal functions $\bar{S}_{\Delta^n}$ for $n\geq 0$. The latter can be recovered from simplicial actions $S_{\Delta^n}$ for $M=\Delta^n$ the standard simplex (with standard triangulation, consisting of all faces) for each dimension $n\geq 0$:
\be\bar{S}_{\Delta^n}=\sum_{\sigma\subset\Delta^n}(-1)^{n-|\sigma|}S_\sigma \label{Sbar via S}\ee
Therefore, to write down the general action of simplicial $BF$-теории, we need to make a series of universal computations: we need to compute $S_{\Delta^n}$ for $n=0,1,2,\ldots$

Notice also that functions $\bar{S}_{\Delta^n}$ do not satisfy QME, since the spaces $g\otimes C^\bt(\Delta^n)[1]\oplus\g^*\otimes C_n(\Delta^n)[-2]$ where they act do not have a canonical BV structure. However, their sum (\ref{simplicial locality decomp}) over faces of any simplicial complex $\Xi$ satisfies QME by construction.

We call $\bar{S}_{\Delta^n}$ the ``reduced'' simplicial $BF$ action for simplex $\Delta^n$.

\textbf{Remark.} We obtained the action $S_\Xi$ on a simlicial complex by inducing it as effective action for the topological $BF$ theory on a manifold, with some special choice of the induction data $(\iota_\Xi,r_\Xi,K_\Xi)$, and discovered that the result (\ref{simplicial locality decomp}) in a sense is built up from results for individual simplices. It turns out that one can take a different approach and obtain $S_\Xi$ by means of a certain universal gluing procedure for $BF_\infty$ theories, applied to simplicial actions on individual simplices, i.e. we do not need the manifold $M$ in this approach and we are not inducing an effective action, but instead we construct a new solution $S_\Xi$ of QME from a sequence of ``standard'' solutions $\{S_{\Delta^n}\}_{n\geq 0}$. In particular, $\Xi$ is not required to be a triangulation of some manifold here.
We will come to this point in a more abstract setting in section \ref{section: gluing-induction}.

\subsection{Gluing procedure for $qL_\infty$ algebras}
\label{section: gluing}
In this section we describe the abstract construction of gluing for $qL_\infty$ algebras.

\begin{Def}\label{def: gluing data} We call the following set of data the ``gluing data'':
\begin{itemize}
\item two $qL_\infty$ algebras $(V,Q_V,\rho_V)$ and $(W,Q_W,\rho_W)$
\item two projections $\pi_1,\pi_2: V\ra W$
\item two embeddings $\iota_1,\iota_2: W\ra V$
\end{itemize}
We assume that the following axioms are satisfied:
\begin{eqnarray}
\pi_1 \iota_1=\pi_2 \iota_2=\id_W \label{gluing axioms 1} \\
\pi_1 \iota_2=\pi_2\iota_1=0 \label{gluing axioms 2} \\
Q_V\pi_1^*=\pi_1^* Q_W \label{gluing axioms 3}\\
Q_V\pi_2^*=\pi_2^* Q_W \label{gluing axioms 4}
\end{eqnarray}
where $\pi_{1,2}^*:\Fun(W[1])\ra\Fun(V[1])$ are the pull-backs by $\pi_{1,2}$.
\end{Def}
Properties (\ref{gluing axioms 1},\ref{gluing axioms 2}) mean that the projections $\pi_{1,2}$ invert the respective embeddings $\iota_{1,2}$ and that the images of embeddings $\iota_{1,2}$ in $V$ do not intersect. Properties (\ref{gluing axioms 3},\ref{gluing axioms 4}) mean that $\pi_{1,2}$ are linear $L_\infty$ morhisms between $L_\infty$ algebras $(V,Q_V)$ and $(W,Q_W)$. In terms of operations this property is written as
\begin{eqnarray}
\pi_1 l_{V(n)}(\underbrace{\omega_V,\ldots,\omega_V}_n)=l_{W(n)}(\underbrace{\pi_1\omega_V,\ldots,\pi_1\omega_V}_n)\label{gluing axiom 3'}\\
\pi_2 l_{V(n)}(\underbrace{\omega_V,\ldots,\omega_V}_n)=l_{W(n)}(\underbrace{\pi_2\omega_V,\ldots,\pi_2\omega_V}_n)\label{gluing axiom 4'}
\end{eqnarray}
for $n\geq 1$. Notice that for $n=1$ this means that $\pi_{1,2}$ are chain maps. Embeddings $\iota_{1,2}$ are not required to be consistent with algebraic structure on $V,W$. In particular, they are not required to be chain maps. Therefore, $\iota_{1,2}$ are just two linear maps of vector spaces.

Due to (\ref{gluing axioms 1},\ref{gluing axioms 2}), we have the following decomposition for vector space $V$:
\be V=U\oplus\iota_1(W)\oplus\iota_2(W)\label{V=UWW}\ee
где $U=\ker(\iota_1 \pi_1+\iota_2 \pi_2)=\ker(\pi_1)\cap\ker(\pi_2)\subset V$.
Let us introduce the following linear combinations of $\pi_{1,2}$ and $\iota_{1,2}$:
\be \pi_-=\pi_2-\pi_1,\quad \pi_+=\frac{\pi_1+\pi_2}{2},\quad \iota_-=\frac{\iota_2-\iota_1}{2},\quad \iota_+=\iota_1+\iota_2 \label{gluing pi_pm iota_pm}\ee
(notice that $\pi_\pm$ are not $L_\infty$ morphisms). They automatically satisfy $$\pi_-\iota_-=\pi_+\iota_+=\id_W,\quad \pi_-\iota_+=\pi_+\iota_-=0$$ and we have the twisted version of decomposition (\ref{V=UWW}):
\be V=U\oplus\iota_+(W)\oplus\iota_-(W)=V'\oplus V'' \label{V=V'+V''}\ee
where we denoted
$$V'=\ker(\pi_-)=U\oplus\iota_+(W),\quad V''=\iota_-(W)$$
Let $\iota_{V',V}:V'\ra V$ and $\pi_{V,V'}:V\ra V'$ be the embedding and projection defined by (\ref{V=V'+V''}). Also denote $$\pi=\pi_1 \iota_{V',V}: V'\ra W$$
the projection from $V'$ to $W$ (notice that restrictions of $\pi_1$ and $\pi_2$ to $V'$ coincide and one could define $\pi=\pi_2\iota_{V',V}$).
\begin{Def}\label{def: glued qL_infty alg} We define the ``glued'' $qL_\infty$ algebra from the gluing data $((V,Q_V,\rho_V),(W,Q_W,\rho_W),\pi_{1,2},\iota_{1,2})$ as the triplet $(V',Q_{V'},\rho_{V'})$:
\begin{eqnarray}V'=\ker\pi_-\subset V \label{glued alg 1}\\
Q_{V'}=\iota_{V',V}^* Q_V \pi_{V,V'}^* \label{glued alg 2} \\
\rho_{V'}=\frac{\iota_{V',V}^*\rho_V}{\pi^*\rho_W}\label{rho_V'}\label{glued alg 3}
\end{eqnarray}
where $\iota_{V',V}^*:\Fun(V[1])\ra\Fun(V'[1])$ and $\pi_{V,V'}^*:\Fun(V'[1])\ra\Fun(V[1])$ are the pull-backs by $\iota_{V',V}$ and $\pi_{V,V'}$ respectively. Equivalently (\ref{glued alg 2},\ref{glued alg 3}) can be formulated in terms of operations:
\begin{eqnarray}
l_{V'(n)}(\underbrace{\omega_{V'},\ldots,\omega_{V'}}_n)=\pi_{V,V'} l_{V(n)}(\underbrace{\iota_{V',V}\omega_{V'},\ldots,\iota_{V',V}\omega_{V'}}_n) \label{glued l_n}\\
q_{V'(n)}(\underbrace{\omega_{V'},\ldots,\omega_{V'}}_n)= q_{V(n)}(\underbrace{\iota_{V',V}\omega_{V'},\ldots,\iota_{V',V}\omega_{V'}}_n)- q_{W(n)}(\underbrace{\pi\omega_{V'},\ldots,\pi\omega_{V'}}_n)\label{glued q_n}
\end{eqnarray}
for $n\geq 1$.
\end{Def}

\begin{statement}\label{gluing statement}
\begin{enumerate}
\item Triplet $(V',Q_{V'},\rho_{V'})$ satisfies the relations of $qL_\infty$ algebra, i.e.
\begin{eqnarray}
Q_{V'}^2&=&0 \label{gluing CME}\\
Q_{V'}(\rho_{V'})+\rho_{V'}\Div Q_{V'}&=&0\label{gluing QME}
\end{eqnarray}
\item\label{gluing statement part 2} The $L_\infty$ part of the glued algebra $(V',Q_{V'})$ is a $L_\infty$ subalgebra in $(V,Q_V)$, i.e. the cohomological vector field $Q_V$ is tangent to the subspace $V'[1]\subset V[1]$ and $Q_{V'}$ is the restriction $Q_{V'}=Q_V|_{V'[1]}$. In other words, embedding $\iota_{V',V}$ and projection $\pi_{V,V'}$ are linear $L_\infty$ morphisms.
\end{enumerate}
\end{statement}
First we prove the lemma, generalizing (\ref{gluing statement part 2}).
\begin{lemma}\label{gluing lemma} Let $\MM$ and $\NN$ be two $Q$-manifolds, with cohomological vector fields $Q_\MM$ and $Q_\NN$ respectively. Let $u_{1,2}:\MM\ra \NN$ be a pair of maps, satisfying $Q_\MM u_{1,2}^*=u_{1,2}^* Q_\NN$. Then $Q_\MM$ is tangent to the submanifold $\MM'=\{x\in \MM\,|\,u_1(x)=u_2(x)\}\subset \MM$ and restriction $Q_{\MM'}=(Q_\MM)|_{\MM'}$ is a cohomological vector field on $\MM'$.
\end{lemma}

\textbf{Proof of Lemma \ref{gluing lemma}.} The ring of functions $\Fun(\MM')$ is canonically identified with the quotient $\Fun(\MM)/(\im(u_2^*-u_1^*))$. There is also the canonical embedding $\iota_{\MM',\MM}:\MM'\ra \MM$ (pull-back $\iota_{\MM',\MM}^*$ is precisely the canonical projection to the quotient $\Fun(\MM)\ra\Fun(\MM')$). Let us introduce temporarily some projection $\pi_{\MM,\MM'}:\MM\ra \MM'$ satisfying $\pi_{\MM,\MM'}\circ\iota_{\MM',\MM}=\id_{\MM'}$.
Define the vector filed $Q_{\MM'}$ on $\MM'$ as
\be Q_{\MM'}=\iota_{\MM',\MM}^* Q_\MM \pi_{\MM,\MM'}^*\ee
Statement that $Q_\MM$ is tangent to $\MM'$ means that this $Q_{\MM'}$ does not depend on projection $\pi_{\MM,\MM'}^*$.

Indeed, let $\tilde{\pi}_{\MM,\MM'}$ be another projection and $f\in\Fun(\MM')$ a test function. Then $(\tilde\pi_{\MM,\MM'}^*-\pi_{\MM,\MM'}^*)(f)$ is mapped to zero by $\iota_{\MM',\MM}^*$ (the projection to quotient) and therefore lies in the ideal $(\im(u_2^*-u_1^*))$. Hence we have $(\tilde\pi_{\MM,\MM'}^*-\pi_{\MM,\MM'}^*)(f)=\sum_i g_i\cdot (u_2^*-u_1^*)(h_i)$ for some functions $g_i\in\Fun(\MM)$ and some $h_i\in\Fun(\NN)$. Apply $Q_\MM$ to this decomposition:
\begin{eqnarray} Q_\MM(\tilde\pi_{\MM,\MM'}^*-\pi_{\MM,\MM'}^*)(f)&=&\sum_i Q_\MM(g_i)\cdot (u_2^*-u_1^*)(h_i)+ g_i\cdot Q_\MM(u_2^*-u_1^*)(h_i) \\
&=&\sum_i Q_\MM(g_i)\cdot (u_2^*-u_1^*)(h_i)+ g_i\cdot (u_2^*-u_1^*)Q_\NN(h_i)
\end{eqnarray}
In the first step used the Leibniz identity, in the second --- that $u_{1,2}^*$ commute with $Q$. Therefore $Q_\MM(\tilde\pi_{\MM,\MM'}^*-\pi_{\MM,\MM'}^*)(f)$ lies in the ideal $(\im(u_2^*-u_1^*))$, and therefore $\iota_{\MM',\MM}^*Q_\MM(\tilde\pi_{\MM,\MM'}^*-\pi_{\MM,\MM'}^*)=0$. Therefore $Q_{\MM'}$ does not depend on the choice of projection $\pi_{\MM,\MM'}$.

Next, it is easy to show that $Q_{\MM'}^2=0$. Indeed,
\be Q_{\MM'}^2=\iota_{\MM',\MM}^*Q_\MM^2\pi_{\MM,\MM'}^*-\iota_{\MM',\MM}^* Q_\MM (1-\pi_{\MM,\MM'}^*\iota_{\MM',\MM}^*) Q_\MM \pi_{\MM,\MM'}^*\ee
First term is zero, since $Q_\MM^2=0$. Second term is zero, since $1-\pi_{\MM,\MM'}^*\iota_{\MM',\MM}^*$ is the projection to ideal $(\im(u_2^*-u_1^*))\subset\Fun(\MM)$, and as we checked above this ideal is closed under action of $Q_\MM$. Therefore $Q_{\MM'}^2=0$.
\\$\Box$

\textbf{Proof of Statement \ref{gluing statement}.}
Part (\ref{gluing statement part 2}) of the Statement is the linear case of Lemma \ref{gluing lemma}, with $\MM=V[1]$, $\NN=W[1]$ being graded vector spaces, with $Q_\MM=Q_V$, $Q_\NN=Q_W$; maps $u_{1,2}$ are the linear $L_\infty$ morphisms $\pi_{1,2}$. Their are linearity implies that the submanifold $\MM'=V'[1]$ is a linear subspace. That $\pi_{V,V'}$ is an $L_\infty$ morphism, follows from construction (\ref{glued alg 2}), while the fact that the embedding $\iota_{V',V}$ is an $L_\infty$ morphism follows from Lemma \ref{gluing lemma}.

Classical part of QME (\ref{gluing CME}) follows from Lemma \ref{gluing lemma}. Let us check the quantum part of QME (\ref{gluing QME}): $Q_{V'}(\log\rho_{V'})+\Div Q_{V'}=0$. Since the embedding $\iota_{V',V}:V'\ra V$ is a (linear) $L_\infty$ morphism, the projection $\pi=\pi_1\iota_{V',V}: V'\ra W$ is also a (linear) $L_\infty$ morphism. Hence
\begin{eqnarray}
Q_{V'}(\log\rho_{V'})&=&Q_{V'}(\iota_{V',V}^*\log\rho_V-\pi^*\log\rho_W)\\
&=&\iota_{V',V}^*Q_V\log\rho_V-\pi^* Q_W\log\rho_W\\
&=&-\iota_{V',V}^*\Div Q_V + \pi^* \Div Q_W
\end{eqnarray}
where in the last step we used QME for $V$ and $W$. So we only need to show that
\be (\Div Q_V)|_{V'[1]}=\Div Q_{V'}+\pi^*\Div Q_W\label{div decomp}\ee

Let us use the decomposition of $Q_V$ in $L_\infty$ operations $l_{V(n)}:\Lambda^n V\ra V$:
\be Q_V=-\left<\sum_{n=1}^\infty \frac{1}{n!}l_{V(n)}(\omega,\cdots,\omega),\frac{\dd}{\dd\omega}\right> \label{gluing Q via l} \ee
Let us split $\omega$ into components, tangent and normal to $V'$: $\omega=\omega'+\omega''$, where $\omega'=(\id_V-\iota_- \pi_-)\omega$ and $\omega''=\iota_- \pi_- \omega$. Compute the divergence $(\Div Q_V)|_{V'[1]}$:
\begin{eqnarray}
(\Div Q_V)|_{V'[1]}&=&-\left.\left(\Str_V\sum_{n=1}^\infty\frac{1}{(n-1)!}l_{V(n)}(\omega,\cdots,\omega,\bt)\right)\right|_{V'[1]} \\
&=&-\Str_{V'}\sum_{n=1}^\infty\frac{1}{(n-1)!}l_{V(n)}(\omega',\cdots,\omega',\bt)-\\ & & -\Str_{V''}\sum_{n=1}^\infty\frac{1}{(n-1)!}l_{V(n)}(\omega',\cdots,\omega',\bt)\nonumber\\
&=&\Div Q_{V'}-\Str_{V''}\sum_{n=1}^\infty\frac{1}{(n-1)!}l_{V(n)}(\omega',\cdots,\omega',\bt)
\end{eqnarray}
(this is just the splitting of divergence into tangential and normal parts). We have to compute the last term, i.e. the normal part of divergence $(\Div''Q_V)|_{V'[1]}$:
\begin{eqnarray}
\pi_- l_{V(n)}(\omega',\cdots,\omega',\iota_- x)&=&(\pi_2-\pi_1) l_{V(n)}(\omega',\cdots,\omega',\iota_- x)\label{eq18}\\
&=&l_{W(n)}(\pi\omega',\cdots,\pi\omega',\pi_-\iota_- x)\label{eq19}\\
&=&l_{W(n)}(\pi\omega',\cdots,\pi\omega',x)\label{eq20}
\end{eqnarray}
where $x\in W$,\quad $l_{W(n)}$ are the operations on $W$, and we used the fact that projections $\pi_{1,2}$ are  $L_\infty$ morphisms and therefore can be carried through the operations. It follows immediately that $\Str_{V''}l_{V(n)}(\omega',\cdots,\omega',\bt)=\Str_W l_{W(n)}(\pi\omega',\cdots,\pi\omega',\bt)$. So we showed that $(\Div''Q_V)|_{V'[1]}=\pi^*\Div Q_W$, which implies (\ref{div decomp}).
\\$\Box$

Note that the key point in the proof that forces us to introduce the embeddings $\iota_{1,2}$ with non-intersecting images is the step (\ref{eq19})-(\ref{eq20}), where we use $\pi_-\iota_-=\id_W$.

\textbf{Split case.} The most important case of gluing is when the initial $qL_\infty$ algebra is the direct sum of two $qL_\infty$ algebras:
$V=V_1\oplus V_2$, $Q_V=Q_{V_1}\otimes \id+\id\otimes Q_{V_2}$, $\rho_V=\rho_{V_1}\otimes \rho_{V_2}$, and the embeddings and projections act like $\iota_1,\pi_1: W\rightleftarrows V_1$ and $\iota_2,\pi_2: W\rightleftarrows V_2$ (of course, we require that projections are $L_\infty$ morphisms). Therefore we have two separate splittings $V_1=U_1\oplus \iota_1(W)$, $V_2=U_2\oplus\iota_2(W)$. The glued  $qL_\infty$ structure arises on the space $V'=U_1\oplus U_2\oplus\iota_+(W)$.

\textbf{Main example: gluing induced $qL_\infty$ algebras on simplicial complexes.} For the simplicial $BF$ theory this means the following: let $\Xi_1$, $\Xi_2$ and $F$ be three simplicial complexes, where $F$ is embedded into $\Xi_1$ and $\Xi_2$ as a simplicial subcomplex. Then for the cochain complexes $V_{1,2}=\g\otimes C^\bt(\Xi_{1,2})$, $W=\g\otimes C^\bt(F)$ we have the natural gluing data --- embeddings and projections. Namely, let $\{\sigma^{1}\}$, $\{\sigma^{2}\}$ be sets of simplices of $\Xi_1$ and $\Xi_2$ not lying in images of $F$ in $\Xi_1$ and $\Xi_2$ respectively, and let $\{\sigma^{1F}\}$ и $\{\sigma^{2F}\}$ be the sets of simplices of images of $F$ in $\Xi_1$ and $\Xi_2$, respectively. Denote also $\{\sigma^F\}$ the set of simplices of $F$. Then embeddings $\iota_{1,2}$ and projections $\pi_{1,2}$ are given by:
\begin{eqnarray*}\iota_{1,2}:&e_{\sigma^F}\mapsto e_{\sigma^{1,2 F}}\\
\pi_{1,2}:&\left\{\begin{array}{lll}e_{\sigma^{1,2F}}&\mapsto& e_{\sigma^{F}}\\
e_{\sigma^{1,2}}&\mapsto& 0 \end{array}\right.
\end{eqnarray*}
The glued space with the basis $\{e_{\sigma^1}\}\cup\{e_{\sigma^2}\}\cup\{e_{\sigma^{+F}}=e_{\sigma^{1F}}+e_{\sigma^{2F}}\}$
is naturally identified with the space of cochains $C^\bt(\Xi')$ of the simplicial complex, glued from $\Xi_1$ and $\Xi_2$ along $F$, with the set of simplices
$\{\sigma^1\}\cup\{\sigma^2\}\cup\{\sigma^{+F}\}$. The fact that projections $\pi_{1,2}$ are $L_\infty$ morphisms for the respective $L_\infty$ algebras, follows from (\ref{simplicial locality decomp}). Indeed,
\begin{multline*}\pi_1 l_{\Xi_1(n)}\left(\sum_{\sigma^1}e_{\sigma^1}\omega^{\sigma^1}+\sum_{\sigma^{1F}}e_{\sigma^{1F}}\omega^{\sigma^{1F}},
\ldots,\sum_{\sigma^1}e_{\sigma^1}\omega^{\sigma^1}+\sum_{\sigma^{1F}}e_{\sigma^{1F}}\omega^{\sigma^{1F}}\right)\\=
l_{F(n)}\left(\sum_{\sigma^{1F}}\pi_1(e_{\sigma^{1F}})\omega^{\sigma^{1F}},\ldots,\sum_{\sigma^{1F}}\pi_1(e_{\sigma^{1F}})\omega^{\sigma^{1F}}\right)
\end{multline*}
i.e. the projection is carried through classical operations --- the defining property of linear $L_\infty$ morphism, and analogously for $\pi_2$.
If the initial $qL_\infty$ structures on $\g\otimes C^\bt(\Xi_{1,2})$ are
\begin{eqnarray*}Q_{\Xi^1}&=&\sum_{n=1}^\infty \frac{1}{n!}\sum_{\sigma,\sigma_1,\ldots,\sigma_n\in \Xi_1} (-1)^{|\sigma|+1} l_{\Xi_1(n)}^\sigma \left(e_{\sigma_1}\omega^{\sigma_1},\ldots,e_{\sigma_n}\omega^{\sigma_n}\right) \frac{\dd}{\dd \omega^\sigma}\\
Q_{\Xi^2}&=&\sum_{n=1}^\infty \frac{1}{n!}\sum_{\sigma,\sigma_1,\ldots,\sigma_n\in \Xi_2} (-1)^{|\sigma|+1} l_{\Xi_2(n)}^\sigma \left(e_{\sigma_1}\omega^{\sigma_1},\ldots,e_{\sigma_n}\omega^{\sigma_n}\right) \frac{\dd}{\dd \omega^\sigma}\\
\log\rho_{\Xi_1}&=&\sum_{n=1}^\infty \frac{1}{n!}\sum_{\sigma_1,\ldots,\sigma_n\in \Xi_1} q_{\Xi_1(n)} \left(e_{\sigma_1}\omega^{\sigma_1},\ldots,e_{\sigma_n}\omega^{\sigma_n}\right)\\
\log\rho_{\Xi_2}&=&\sum_{n=1}^\infty \frac{1}{n!}\sum_{\sigma_1,\ldots,\sigma_n\in \Xi_2} q_{\Xi_2(n)} \left(e_{\sigma_1}\omega^{\sigma_1},\ldots,e_{\sigma_n}\omega^{\sigma_n}\right)
\end{eqnarray*}
then, according to the general construction of gluing, the $qL_\infty$ structure on $\g\otimes C^\bt(\Xi')$ is
\begin{eqnarray*}Q_{\Xi'}&=&\sum_{n=1}^\infty \frac{1}{n!}\sum_{\sigma^1\in \Xi_1\backslash F;\;\sigma_1,\ldots,\sigma_n\in \Xi_1\subset \Xi'} (-1)^{|\sigma^1|+1} l_{\Xi_1(n)}^{\sigma^1} \left(e_{\sigma_1}\omega^{\sigma_1},\ldots,e_{\sigma_n}\omega^{\sigma_n}\right) \frac{\dd}{\dd \omega^{\sigma^1}}+\\
&&+\sum_{n=1}^\infty \frac{1}{n!}\sum_{\sigma^2\in \Xi_2\backslash F;\;\sigma_1,\ldots,\sigma_n\in \Xi_2\subset \Xi'} (-1)^{|\sigma^2|+1} l_{\Xi_2(n)}^{\sigma^2} \left(e_{\sigma_1}\omega^{\sigma_1},\ldots,e_{\sigma_n}\omega^{\sigma_n}\right) \frac{\dd}{\dd \omega^{\sigma^2}}+\\
&&+\sum_{n=1}^\infty \frac{1}{n!}\sum_{\sigma^{+F},\sigma_1^{+F},\ldots,\sigma_n^{+F}\in F\subset \Xi'} (-1)^{|\sigma^{+F}|+1} l_{F(n)}^{\sigma^{+F}} \left(e_{\sigma_1^{+F}}\omega^{\sigma_1^{+F}},\ldots,e_{\sigma_n}^{+F}\omega^{\sigma_n^{+F}}\right) \frac{\dd}{\dd \omega^{\sigma^{+F}}}\\
\log\rho_{\Xi'}&=&\sum_{n=1}^\infty \frac{1}{n!}\sum_{\sigma_1,\ldots,\sigma_n\in \Xi_1\subset \Xi'}  q_{\Xi_1(n)} \left(e_{\sigma_1}\omega^{\sigma_1},\ldots,e_{\sigma_n}\omega^{\sigma_n}\right)+\\
&&+\sum_{n=1}^\infty \frac{1}{n!}\sum_{\sigma_1,\ldots,\sigma_n\in \Xi_2\subset \Xi'} q_{\Xi_2(n)} \left(e_{\sigma_1}\omega^{\sigma_1},\ldots,e_{\sigma_n}\omega^{\sigma_n}\right) -\\
&&-\sum_{n=1}^\infty \frac{1}{n!}\sum_{\sigma_1^{+F},\ldots,\sigma_n^{+F}\in F\subset \Xi'} q_{F(n)} \left(e_{\sigma_1^{+F}}\omega^{\sigma_1^{+F}},\ldots,e_{\sigma_n}^{+F}\omega^{\sigma_n^{+F}}\right)
\end{eqnarray*}
Or, in the shorter notation,
\begin{eqnarray*}
Q_{\Xi'}&=&Q_{\Xi_1}|_{\omega^{\sigma^{1F}}\mapsto\omega^{\sigma^{+F}},\frac{\dd}{\dd\omega^{\sigma^{1F}}}\mapsto \frac{\dd}{\dd\omega^{\sigma^{+F}}}} +Q_{\Xi_2}|_{\omega^{\sigma^{2F}}\mapsto\omega^{\sigma^{+F}},\frac{\dd}{\dd\omega^{\sigma^{2F}}}\mapsto \frac{\dd}{\dd\omega^{\sigma^{+F}}}}-\\ &&\qquad -Q_F|_{\omega^{\sigma^{F}}\mapsto\omega^{\sigma^{+F}},\frac{\dd}{\dd\omega^{\sigma^{F}}}\mapsto \frac{\dd}{\dd\omega^{\sigma^{+F}}}}\\
\log\rho_{\Xi'}&=&\log\rho_{\Xi_1}|_{\omega^{\sigma^{1F}}\mapsto\omega^{\sigma^{+F}}} +\log\rho_{\Xi_2}|_{\omega^{\sigma^{2F}}\mapsto\omega^{\sigma^{+F}}} -\log\rho_F|_{\omega^{\sigma^{F}}\mapsto\omega^{\sigma^{+F}}}
\end{eqnarray*}
The glued $BF_\infty$ action on $\g\otimes C^\bt(\Xi')$ can be written as
\be S_{\Xi'}=S_{\Xi_1}|_{\omega^{\sigma^{1F}}\mapsto\omega^{\sigma^{+F}},\; p_{\sigma^{1F}}\mapsto p_{\sigma^{+F}}}+S_{\Xi_2}|_{\omega^{\sigma^{2F}}\mapsto\omega^{\sigma^{+F}},\; p_{\sigma^{2F}}\mapsto p_{\sigma^{+F}}}- S_F|_{\omega^{\sigma^{F}}\mapsto\omega^{\sigma^{+F}},\; p_{\sigma^{F}}\mapsto p_{\sigma^{+F}}} \label{gluing S}\ee

Therefore, iterating this construction we can ``glue'' the simplicial action $S_\Xi$ for an arbitrary simplicial complex $\Xi$ from the actions for individual simplices: we start with any simplex and gradually glue the others one by one. The resulting action
$S_\Xi$ satisfies QME by construction (Statement \ref{gluing statement}). The fact that $S_\Xi$ does not depend on the order in which we glue the simplices follows from the fact that actions for individual simplices can be written as
$$S_\sigma=\sum_{\sigma'\subset\sigma} \bar{S}_\sigma'$$
Then for $S_\Xi$, using (\ref{gluing S}), we obtain the expression (\ref{simplicial locality decomp}), manifestly independent on the order of gluing. Here we never use the construction of effective action, nor the manifold $M$, geometrically realizing the simplicial complex $\Xi$. Therefore the gluing construction gives a more general variant of simplicial $BF$ theory, where $\Xi$ is allowed to be any simplicial complex, not necessarily a triangulation of some manifold.

\textbf{Example: circle.} Let us also give an example of gluing, where $V$ is not of form $V_1\oplus V_2$: the gluing of the interval into the circle. The general construction successfully works in this situation. We have $V= \g e_0\oplus \g e_1\oplus \g e_{01}$ --- the cochain complex of the (triangulated) interval $[0,1]$, where $e_{0,1,01}$ are the basis cochains. Also $W=\g e_A$ (A - is the label of the gluing point). Projections $\pi_{1,2}$ and embeddings $\iota_{1,2}$ are given by
\begin{eqnarray}\pi_1:x e_0+ y e_1+z e_{01}&\mapsto& x e_A\\
\pi_2:x e_0+ y e_1+z e_{01}&\mapsto& y e_A\\
\iota_1: x e_A & \mapsto & x e_0\\
\iota_2: x e_A & \mapsto & x e_1
\end{eqnarray}
We obtain $V=V'\oplus V''$, where $V'=\ker \pi_-=\g\otimes e_+\oplus\g\otimes e_{01}$ (here $e_+=\iota_+(e_A)=e_1+e_2$) is the cochain complex of the circle, embedded into the cochain complex of the interval, and $V''=\g\otimes e_-$, where $e_-=\iota_- (e_A)=\frac{e_2-e_1}{2}$ is the cochain complex of point $A$, embedded into $V$ as a linear subspace (not a subcomplex). We will return to this example again in section \ref{section: gluing, bc examples}.

\subsubsection{Imposing a boundary condition}
\label{section: gluing, boundary condition}
Let us mention another construction for $qL_\infty$ algebras, resembling the gluing (but simpler) --- the construction of imposing a boundary condition. Let $V$ and $W$ be two $qL_\infty$ algebras and  $\pi_1:V\ra W$ be a linear $L_\infty$ morphism. Then the space $V'=\ker \pi_1\subset V$ has the $qL_\infty$ structure $Q_{V'}=(Q_V)|_{V'}$, $\rho_{V'}=(\rho_V)|_{V'}$. The CME holds due to Lemma \ref{gluing lemma}, and the quantum part of QME is proved by checking that the normal part of divergence of $Q_V$ vanishes: $\Div'' Q_V=0$, by the argument analogous to (\ref{eq18})-(\ref{eq20}).

A geometrical example is the following: let  $\Xi$ be a simplicial complex and $F$ its simplicial subcomplex, then $V=\g\otimes C^\bt(\Xi)$, $W=\g\otimes C^\bt(F)$, the projection is constructed from the geometrical embedding. Then we have the $qL_\infty$ structure on $V'=\g\otimes C^\bt(\Xi')$, where $\Xi'=\Xi\backslash F$ (it is not a true simplicial complex, but rather a relative one).

For example, we can set $\Xi$ to be the interval $[0,1]$ and $F$ to be a point $[A]$, embedded as the right boundary point $[1]$ of the interval. Then $\Xi'$ is the interval without the right boundary point, and it is not a true cell complex, but the cochain complex is well-defined:
$V'=\g\otimes e_{0}\oplus \g\otimes e_{01}$ and it has the relict $qL_\infty$ structure.

We will present the explicit result for this example and for the gluing of interval into a circle in section \ref{section: gluing, bc examples}.

\subsubsection{Consistency of operations of gluing and induction}
\label{section: gluing-induction}
In this section we will prove the statement that (under certain conditions on the gluing and induction data) operations of gluing and induction for $qL_\infty$ algebras commute. In particular this allows to interpret the gluing of effective $BF_\infty$ actions as effective action for some other $BF_\infty$ theory (which is useful as an indirect method of computing effective actions, since technically gluing is much simpler than induction). This statement should be considered as an abstract version of Theorem \ref{thm: simplicial locality}.

Let $(V,Q_V,\rho_V)$, $(W,Q_W,\rho_W)$ be two $qL_\infty$ algebras and let $\pi_{1,2}:V\ra W$, $\iota_{1,2}:W\ra V$ be the gluing data. Then $V'=\ker\pi_-$ is endowed with the glued $qL_\infty$ structure $(V',Q_{V'},\rho_{V'})$ (all notations are as in Definitions \ref{def: gluing data},\ref{def: glued qL_infty alg}). Let also $\bar V\hra V$, $\bar W\hra W$ be two deformation retracts and  $V\xra{(\iota_V,r_V,K_V)} \bar V$ and $W\xra{(\iota_W,r_W,K_W)}\bar W$ be two sets of induction data (arrows denote the direction in which the $qL_\infty$ structure is transferred). Then $\bar V$, $\bar W$ are endowed with induced $qL_\infty$ structures (by means of the formulae of homotopy transfer (\ref{induced l_n from qL_infty},\ref{induced q_n from qL_infty}), see Definition \ref{def: induced qL_infty}): $(\bar V, Q_{\bar V},\rho_{\bar V})$ and $(\bar W, Q_{\bar W},\rho_{\bar W})$. Therefore we have the following set of data:
\be \begin{CD}
(V,Q_V,\rho_V),\; (W,Q_W,\rho_W)@>{(\iota_V,r_V,K_V),\;(\iota_W,r_W,K_W)}>\text{induction}> (\bar V, Q_{\bar V},\rho_{\bar V}),\;(\bar W, Q_{\bar W},\rho_{\bar W})\\
@V(\pi_{1,2},\iota_{1,2})V\text{gluing}V \\
(V',Q_{V'},\rho_{V'})
\end{CD} \label{gluing-induction diagram}\ee
\begin{Def}[Consistent gluing and induction data, induced gluing data, glued induction data] We say that two sets of induction data $(\iota_V,r_V,K_V)$, $(\iota_W,r_W,K_W)$ are consistent with the gluing data $(\pi_{1,2},\iota_{1,2})$, if the following holds:
\begin{eqnarray}
\iota_W \bar \pi_1=\pi_1 \iota_V,\quad \iota_W \bar\pi_2=\pi_2 \iota_V\label{induction-gluing agreement axiom 1}\\
r_W\pi_1=\bar\pi_1 r_V,\quad r_W\pi_2=\bar\pi_2 r_V\label{induction-gluing agreement axiom 2}\\
K_W \pi_1=\pi_1 K_V,\quad K_W \pi_2=\pi_2 K_V\label{induction-gluing agreement axiom 3}
\end{eqnarray}
where we introduced two projections $\bar\pi_{1,2}: \bar V\ra\bar W$:
\be \bar\pi_1=r_W \pi_1 \iota_V,\quad \bar\pi_2=r_W \pi_2 \iota_V\label{gluing pi induced}\ee
If properties (\ref{induction-gluing agreement axiom 1},\ref{induction-gluing agreement axiom 2},\ref{induction-gluing agreement axiom 3}) hold, we call the ``induced gluing data'' the collection of maps  $\bar\pi_{1,2}: \bar V\ra\bar W$, $\bar\iota_{1,2}:\bar W\ra\bar V$, where the projections are given by (\ref{gluing pi induced}) and the embeddings are
\be \bar\iota_1=r_V\iota_1 \iota_W,\quad \bar\iota_2=r_V\iota_2 \iota_W \label{gluing iota induced}\ee
Then we introduce linear combinations of projections and embeddings $\bar\pi_{\pm}: \bar V\ra\bar W$, $\bar\iota_\pm: \bar W\ra \bar V$ as in (\ref{gluing pi_pm iota_pm}) and define the glued space $\bar V'=\ker\bar\pi_-\subset \bar V$, with embedding $\iota_{\bar V',\bar V}:\bar V'\ra \bar V$ and projection $\pi_{\bar V,\bar V'}:\bar V\ra\bar V'$. We define the ``glued induction data''  $(\iota_{V'},r_{V'},K_{V'})$ as
\begin{eqnarray}
\iota_{V'}=\pi_{V,V'} \iota_V\iota_{\bar V',\bar V}:\quad \bar V'\ra V' \label{glued ind data 1}\\
r_{V'}= \pi_{\bar V,\bar V'}r_V\iota_{V',V}:\quad V'\ra\bar V' \label{glued ind data 2}\\
K_{V'}= \pi_{V,V'} K_V\iota_{V',V}:\quad V'\ra V' \label{glued ind data 3}
\end{eqnarray}
\end{Def}

\begin{statement}\label{statement: induction-gluing} Let $(\iota_V,r_V,K_V)$, $(\iota_W,r_W,K_W)$ be the induction data consistent with the gluing data $(\pi_{1,2},\iota_{1,2})$, i.e. properties  (\ref{induction-gluing agreement axiom 1}--\ref{induction-gluing agreement axiom 3}) hold. Then:
\begin{enumerate}
\item\label{statement: induction-gluing 1} Maps (\ref{gluing pi induced},\ref{gluing iota induced}) satisfy the axioms of gluing data
(\ref{gluing axioms 1}--\ref{gluing axioms 4}).
\item\label{statement: induction-gluing 2} Maps (\ref{glued ind data 1}--\ref{glued ind data 3}) satisfy the axioms of induction data (\ref{ind data axiom1}--\ref{ind data axiom6}).
\item\label{statement: induction-gluing 3} The glued $qL_\infty$ structure on $\bar V'$, constructed from the gluing data $(\bar\pi_{1,2},\bar\iota_{1,2})$ coincides with induced $qL_\infty$ structure, constructed using the induction data $(\iota_{V'},r_{V'},K_{V'})$.
\end{enumerate}
In other words, the diagram (\ref{gluing-induction diagram}) is completed to a commutative square
\be \begin{CD}
(V,Q_V,\rho_V),\; (W,Q_W,\rho_W)@>{(\iota_V,r_V,K_V),\;(\iota_W,r_W,K_W)}>\text{induction}> (\bar V, Q_{\bar V},\rho_{\bar V}),\;(\bar W, Q_{\bar W},\rho_{\bar W})\\
@V(\pi_{1,2},\iota_{1,2})V\text{gluing}V   @V(\bar\pi_{1,2},\bar\iota_{1,2})V\text{induced gluing}V\\
(V',Q_{V'},\rho_{V'})@>(\iota_{V'},r_{V'},K_{V'})>\text{glued induction}> (\bar V',Q_{\bar V'},\rho_{\bar V'})
\end{CD} \label{gluing-induction square}
\ee
\end{statement}

\textbf{Proof. } Let us first prove (\ref{statement: induction-gluing 1}). Axioms (\ref{gluing axioms 1},\ref{gluing axioms 2}) of gluing are checked trivially using (\ref{induction-gluing agreement axiom 1},\ref{induction-gluing agreement axiom 2}):
\begin{eqnarray*}\bar\pi_1\bar\iota_1=\bar\pi_1 r_V\iota_1\iota_W=r_W\pi_1\iota_1\iota_W=r_W\id_W\iota_W=\id_{\bar W}\\
\bar\pi_1\bar\iota_2=\bar\pi_1 r_V\iota_2\iota_W=r_W\underbrace{\pi_1\iota_2}_{=0}\iota_W=0
\end{eqnarray*}
and analogously one checks $\bar\pi_2\bar\iota_2=\id_{\bar{W}}$, $\bar\pi_2\bar\iota_1=0$. Then we have to check that $\bar\pi_{1,2}$ are $L_\infty$ morphisms, i.e. that
$$\bar\pi_1 l_{\bar V(n)}(\omega_{\bar V},\ldots,\omega_{\bar V})=l_{\bar W(n)}(\bar\pi_1\omega_{\bar V},\ldots,\bar\pi_1\omega_{\bar V}),\quad
\bar\pi_2 l_{\bar V(n)}(\omega_{\bar V},\ldots,\omega_{\bar V})=l_{\bar W(n)}(\bar\pi_2\omega_{\bar V},\ldots,\bar\pi_2\omega_{\bar V})$$
for $n\geq 1$. Induced $L_\infty$ operations $l_{\bar V(n)}:\Lambda^n\bar V\ra\bar V$,\quad $l_{\bar W(n)}:\Lambda^n\bar W\ra\bar W$ are constructed by means of (\ref{induced l_n from qL_infty}) as sums over trees:
\begin{eqnarray*}l_{\bar V(n)}(\omega_{\bar V},\ldots,\omega_{\bar V})&=&n!\sum_{T\in{\bf{T}}_\mr{nonPl}:\;|T|=n}\frac{1}{|\Aut(T)|}\cdot\\
&&\qquad\cdot \Iter_{T;\{-K_V l_{V(m)}\}_{m\geq 2}; \{r_V l_{V(m)}\}_{m\geq 2}}(\iota_V(\omega_{\bar V}),\ldots,\iota_V(\omega_{\bar V}))\\
l_{\bar W(n)}(\omega_{\bar W},\ldots,\omega_{\bar W})&=&n!\sum_{T\in{\bf{T}}_\mr{nonPl}:\;|T|=n}\frac{1}{|\Aut(T)|}\cdot\\
&&\qquad\cdot\Iter_{T;\{-K_W l_{W(m)}\}_{m\geq 2}; \{r_W l_{W(m)}\}_{m\geq 2}}(\iota_W(\omega_{\bar W}),\ldots,\iota_W(\omega_{\bar W}))
\end{eqnarray*}
where $n\geq 2$ (for $n=1$ the unary operations are differentials of the corresponding complexes $l_{\bar V(1)}=d_{\bar V}$, $l_{\bar W(1)}=d_{\bar W}$).
Using (\ref{induction-gluing agreement axiom 1}--\ref{induction-gluing agreement axiom 3}) and (\ref{gluing axiom 3'}), we obtain
\begin{multline*}
\bar \pi_1 l_{\bar V(n)}(\omega_{\bar V},\ldots,\omega_{\bar V})\\
=n!\sum_{T\in{\bf{T}}_\mr{nonPl}:\;|T|=n}\frac{1}{|\Aut(T)|} \Iter_{T;\{-K_V l_{V(m)}\}_{m\geq 2}; \{\underbrace{\bar\pi_1 r_V}_{=r_W\pi_1} l_{V(m)}\}_{m\geq 2}}(\iota_V(\omega_{\bar V}),\ldots,\iota_V(\omega_{\bar V}))\\
=n!\sum_{T\in{\bf{T}}_\mr{nonPl}:\;|T|=n}\frac{1}{|\Aut(T)|} \Iter_{T;\{-K_W l_{W(m)}\}_{m\geq 2}; \{r_W l_{W(m)}\}_{m\geq 2}}(\underbrace{\pi_1\iota_V(\omega_{\bar V})}_{=\iota_W\bar\pi_1\omega_{\bar V}},\ldots, \underbrace{\pi_1\iota_V(\omega_{\bar V})}_{=\iota_W\bar\pi_1\omega_{\bar V}})\\
=l_{\bar W(n)}(\bar\pi_1\omega_{\bar V},\ldots,\bar\pi_1\omega_{\bar V})
\end{multline*}
for $n\geq 2$. For $n=1$ we have to check that projection $\bar\pi_1$ is a chain map, which is obvious since by definition (\ref{gluing pi induced}) is is a composition of chain maps. So we checked that $\bar\pi_1$ is an $L_\infty$ morphism. For $\bar\pi_2$ the check is analogous. Therefore, we proved the part (\ref{statement: induction-gluing 1}) of the Statement.

Let us check the part (\ref{statement: induction-gluing 2}). The fact that $\iota_{V'},r_{V'}$ are chain maps (axioms (\ref{ind data axiom1},\ref{ind data axiom2})) follows from part (\ref{gluing statement part 2}) of the Statement \ref{gluing statement}: $\iota_{V',V},\pi_{V,V'}$ and $\iota_{\bar V',\bar V},\pi_{\bar V,\bar V'}$ are $L_\infty$ morphisms and, in particular, chain maps. Therefore the glued embedding $\iota_{V'}$ (\ref{glued ind data 1}) and glued retraction $r_{V'}$ (\ref{glued ind data 2}) are compositions of chain maps and hence are chain maps themselves. Next, to check axiom (\ref{ind data axiom3}), let us use the following representation for the projection to the glued space:
$$\iota_{V',V}\pi_{V,V'}=\id_V-\iota_-\pi_-$$
(obvious from construction of the glued space $V'=\ker\pi_-$). This implies
$$r_{V'}\iota_{V'}=\pi_{\bar V,\bar V'}r_V\underbrace{\iota_{V',V}\pi_{V,V'}}_{=\id_V-\iota_-\pi_-}\iota_V\iota_{\bar V',\bar V}=
\id_{\bar V'}-\pi_{\bar V,\bar V'}r_V\iota_-\underbrace{\pi_-\iota_V}_{=\iota_W\bar\pi_-}\iota_{\bar V',\bar V}=\id_{\bar V'}-0$$
The last term vanishes, since the projection $\bar\pi_-$ vanishes on the image of embedding $\iota_{\bar V',\bar V}$. We check axiom (\ref{ind data axiom4}) analogously:
\begin{multline*}
r_{V'}K_{V'}=\pi_{\bar V,\bar V'}r_V\underbrace{\iota_{V',V}\pi_{V,V'}}_{=\id_V-\iota_-\pi_-}K_V\iota_{V',V}=
\pi_{\bar V,\bar V'}\underbrace{r_V K_V}_{=0}\iota_{V',V}-\pi_{\bar V,\bar V'}r_V\iota_-\pi_-K_V\iota_{V',V}=\\=
0-\pi_{\bar V,\bar V'}r_V\iota_- K_W\underbrace{\pi_-\iota_{V',V}}_{=0}=0\\
K_{V'}\iota_{V'}=\pi_{V,V'}K_V\underbrace{\iota_{V',V}\pi_{V,V'}}_{=\id_V-\iota_-\pi_-}\iota_V\iota_{\bar V',\bar V}=
\pi_{V,V'}\underbrace{K_V\iota_V}_{=0}\iota_{\bar V',\bar V}-\pi_{V,V'}K_V\iota_-\pi_-\iota_V \iota_{\bar V',\bar V}=\\
=0-\pi_{V,V'}K_V\iota_-\iota_W\underbrace{\bar\pi_- \iota_{\bar V',\bar V}}_{=0}=0
\end{multline*}
Next, let us check the property (\ref{ind data axiom5}) of chain homotopy:
\begin{multline*}
d_{V'}K_{V'}+K_{V'}d_{V'}=\pi_{V,V'}(\underbrace{d_V K_V+K_V d_V}_{=\id_V-\iota_V r_V})\iota_{V',V}=\id_{V'}-\pi_{V,V'}\iota_V (\underbrace{\iota_{\bar V',\bar V} \pi_{\bar V,\bar V'}+\bar\iota_-\bar\pi_-}_{=\id_{\bar V}}) r_V\iota_{V',V}=\\
=\id_{V'}-\iota_{V'}r_{V'}-\pi_{V,V'}\iota_V\bar\iota_- r_W\underbrace{\pi_-\iota_{V',V}}_{=0}=\id_{V'}-\iota_{V'}r_{V'}
\end{multline*}
Finally, check (\ref{ind data axiom6}):
\begin{multline*}
(K_{V'})^2=\pi_{V,V'}K_V\underbrace{\iota_{V',V}\pi_{V,V'}}_{=\id_V-\iota_-\pi_-}K_V\iota_{V',V}=\pi_{V,V'}\underbrace{(K_V)^2}_{=0} \iota_{V',V}-
\pi_{V,V'}K_V\iota_-K_W\underbrace{\pi_-\iota_{V',V}}_{=0}=0
\end{multline*}
Therefore, the part (\ref{statement: induction-gluing 2}) of the Statement is proved.

Let us turn to part (\ref{statement: induction-gluing 3}). The classical operations of $qL_\infty$ structure on $V'$, glued from $(\bar V,Q_{\bar V},\rho_{\bar V}),\;(\bar W,Q_{\bar W},\rho_{\bar W})$, according to (\ref{glued l_n}), are
\begin{multline}
l_{\bar V' (n)}(\omega_{\bar V'},\ldots,\omega_{\bar V'})=\pi_{\bar V,\bar V'}l_{\bar V (n)}(\iota_{\bar V',\bar V} \omega_{\bar V'},\ldots, \iota_{\bar V',\bar V} \omega_{\bar V'})=\\
=n!\sum_{T\in{\bf{T}}_\mr{nonPl}:\;|T|=n}\frac{1}{|\Aut(T)|} \Iter_{T;\{-K_V l_{V(m)}\}_{m\geq 2}; \{\pi_{\bar V,\bar V'} r_V l_{V(m)}\}_{m\geq 2}}(\iota_V\iota_{\bar V',\bar V}\omega_{\bar V'},\ldots,\iota_V\iota_{\bar V',\bar V}\omega_{\bar V'})\label{gluing-induction eq2}
\end{multline}
Let us use that
\be\pi_{\bar V,\bar V'} r_V=\pi_{\bar V,\bar V'} r_V(\iota_{V',V}\pi_{V,V'}+\iota_-\pi_-) =r_{V'}\pi_{V,V'}+\pi_{\bar V,\bar V'} r_V \iota_- \pi_-\label{gluing-induction eq1} \ee
and that $\pi_{V,V'}$ и $\pi_{1,2}$ are $L_\infty$ morphisms:
\begin{multline}\Iter_{T;\{-K_V l_{V(m)}\}_{m\geq 2}; \{\pi_{\bar V,\bar V'} r_V l_{V(m)}\}_{m\geq 2}}(\iota_V\iota_{\bar V',\bar V}\omega_{\bar V'},\ldots,\iota_V\iota_{\bar V',\bar V}\omega_{\bar V'})=\\
=\Iter_{T;\{-K_V l_{V(m)}\}_{m\geq 2}; \{(r_{V'}\pi_{V,V'}+\pi_{\bar V,\bar V'} r_V \iota_- \pi_-) l_{V(m)}\}_{m\geq 2}}(\iota_V\iota_{\bar V',\bar V}\omega_{\bar V'},\ldots,\iota_V\iota_{\bar V',\bar V}\omega_{\bar V'})=\\
=\Iter_{T;\{-K_V l_{V(m)}\}_{m\geq 2}; \{r_{V'}l_{V'(m)}(\pi_{V,V'}\bt,\ldots,\pi_{V,V'}\bt)\}_{m\geq 2}}(\iota_V\iota_{\bar V',\bar V}\omega_{\bar V'},\ldots,\iota_V\iota_{\bar V',\bar V}\omega_{\bar V'})+\\
+\Iter_{T;\{-K_W l_{W(m)}\}_{m\geq 2}; \{\pi_{\bar V,\bar V'} r_V \iota_- l_{W(m)}\}_{m\geq 2}}(\iota_W\underbrace{\bar\pi_2\iota_{\bar V',\bar V}}_{=\bar\pi}\omega_{\bar V'},\ldots,\iota_W\underbrace{\bar\pi_2\iota_{\bar V',\bar V}}_{=\bar\pi}\omega_{\bar V'})-\\
-\Iter_{T;\{-K_W l_{W(m)}\}_{m\geq 2}; \{\pi_{\bar V,\bar V'} r_V \iota_- l_{W(m)}\}_{m\geq 2}}(\iota_W\underbrace{\bar\pi_1\iota_{\bar V',\bar V}}_{=\bar\pi}\omega_{\bar V'},\ldots,\iota_W\underbrace{\bar\pi_1\iota_{\bar V',\bar V}}_{=\bar\pi}\omega_{\bar V'})=\\
=\Iter_{T;\{-K_V l_{V(m)}\}_{m\geq 2}; \{r_{V'}l_{V'(m)}(\pi_{V,V'}\bt,\ldots,\pi_{V,V'}\bt)\}_{m\geq 2}}(\iota_V\iota_{\bar V',\bar V}\omega_{\bar V'},\ldots,\iota_V\iota_{\bar V',\bar V}\omega_{\bar V'})
\label{gluing-induction eq3}
\end{multline}
where $\bar\pi=\bar\pi_1\iota_{\bar V',\bar V}=\bar\pi_2\iota_{\bar V',\bar V}: \bar V'\ra\bar W$. So the second term in (\ref{gluing-induction eq1}) does not contribute to (\ref{gluing-induction eq2}). Now we want to show that analogous manipulation of carrying the projection $\pi_{V,V'}$ through a vertex of the tree can be performed not just in the root of tree $T$, but in any vertex. For this reason consider a sub-tree $\tilde{T}\subset T$, having the vertex of $T$ in question as its root, and containing all its descendants. We have
\begin{multline*}
\Iter_{\tilde{T};\{-K_V l_{V(m)}\}_{m\geq 2}; \{-\pi_{V,V'}K_V l_{V(m)}\}_{m\geq 2}}(\iota_V\iota_{\bar V',\bar V}\omega_{\bar V'},\ldots,\iota_V\iota_{\bar V',\bar V}\omega_{\bar V'})=\\
=\Iter_{\tilde{T};\{-K_V l_{V(m)}\}_{m\geq 2}; \{(-K_{V'}\pi_{V,V'}-\pi_{V,V'}K_V\iota_-\pi_- )l_{V(m)}\}_{m\geq 2}}(\iota_V\iota_{\bar V',\bar V}\omega_{\bar V'},\ldots,\iota_V\iota_{\bar V',\bar V}\omega_{\bar V'})=\\
=\Iter_{\tilde{T};\{-K_V l_{V(m)}\}_{m\geq 2}; \{-K_{V'}l_{V'(m)}(\pi_{V,V'}\bt,\ldots,\pi_{V,V'}\bt)\}_{m\geq 2}}(\iota_V\iota_{\bar V',\bar V}\omega_{\bar V'},\ldots,\iota_V\iota_{\bar V',\bar V}\omega_{\bar V'})+\\
+\Iter_{\tilde{T};\{-K_W l_{W(m)}\}_{m\geq 2}; \{-\pi_{V,V'}K_V\iota_- l_{W(m)}\}_{m\geq 2}}(\iota_W\underbrace{\bar\pi_2\iota_{\bar V',\bar V}}_{=\bar\pi}\omega_{\bar V'},\ldots,\iota_W\underbrace{\bar\pi_2\iota_{\bar V',\bar V}}_{\bar\pi}\omega_{\bar V'})-\\
-\Iter_{\tilde{T};\{-K_W l_{W(m)}\}_{m\geq 2}; \{-\pi_{V,V'}K_V\iota_- l_{W(m)}\}_{m\geq 2}}(\iota_W\underbrace{\bar\pi_1\iota_{\bar V',\bar V}}_{=\bar\pi}\omega_{\bar V'},\ldots,\iota_W\underbrace{\bar\pi_1\iota_{\bar V',\bar V}}_{\bar\pi}\omega_{\bar V'})=\\
=\Iter_{\tilde{T};\{-K_V l_{V(m)}\}_{m\geq 2}; \{-K_{V'}l_{V'(m)}(\pi_{V,V'}\bt,\ldots,\pi_{V,V'}\bt)\}_{m\geq 2}}(\iota_V\iota_{\bar V',\bar V}\omega_{\bar V'},\ldots,\iota_V\iota_{\bar V',\bar V}\omega_{\bar V'})
\end{multline*}
From this we obtain by induction
\begin{multline}
\Iter_{\tilde{T};\{-K_V l_{V(m)}\}_{m\geq 2}; \{-\pi_{V,V'}K_V l_{V(m)}\}_{m\geq 2}}(\iota_V\iota_{\bar V',\bar V}\omega_{\bar V'},\ldots,\iota_V\iota_{\bar V',\bar V}\omega_{\bar V'})=\\
=\Iter_{\tilde{T};\{-K_{V'} l_{V'(m)}\}_{m\geq 2}; \{-K_{V'} l_{V'(m)}\}_{m\geq 2}}(\underbrace{\pi_{V,V'}\iota_V\iota_{\bar V',\bar V}}_{=\iota_{V'}} \omega_{\bar V'},\ldots,\underbrace{\pi_{V,V'}\iota_V\iota_{\bar V',\bar V}}_{=\iota_{V'}}\omega_{\bar V'})
\label{gluing-induction eq4}
\end{multline}
Therefore, using (\ref{gluing-induction eq3},\ref{gluing-induction eq4}), we can rewrite the glued $L_\infty$ operations on $\bar V'$ (\ref{gluing-induction eq2}) as
\begin{multline*}
l_{\bar V' (n)}(\omega_{\bar V'},\ldots,\omega_{\bar V'})=\\
=
n!\sum_{T\in{\bf{T}}_\mr{nonPl}:\;|T|=n}\frac{1}{|\Aut(T)|} \Iter_{T;\{-K_{V'} l_{V'(m)}\}_{m\geq 2}; \{ r_{V'} l_{V'(m)}\}_{m\geq 2}}(\iota_{V'}(\omega_{\bar V'}),\ldots,\iota_{V'}(\omega_{\bar V'}))
\end{multline*}
So we obtained precisely the expression for the $L_\infty$ operations on $\bar V'$, induced from the $L_\infty$ algebra $(V',Q_{V'})$. Therefore we proved part (\ref{statement: induction-gluing 3}) on the level of classical operations. We also have to consider the case of unary operation $n=1$ separately, since representation of induced operations as sums over trees works only for $n\geq 2$. But checking consistency on the level of glued unary operation (differential) $l_{\bar V'(1)}=d_{\bar V'}$ is checking its consistency with the induction data $(\iota_{V'},r_{V'},K_{V'})$, or more precisely checking axioms (\ref{ind data axiom1},\ref{ind data axiom2}). We already made this check proving part (\ref{statement: induction-gluing 2}) of the Statement.

Let us now turn to proving part (\ref{statement: induction-gluing 3}) on the level of quantum operations $q_{\bar V',(n)}$. The quantum operations on $\bar V'$, glued from $\bar V$, $\bar W$, are according to (\ref{glued q_n}):
\begin{multline}q_{\bar V'(n)}(\omega_{\bar V'},\ldots,\omega_{\bar V'})=q_{\bar V(n)}(\iota_{\bar V',\bar V}\omega_{\bar V'},\ldots,\iota_{\bar V',\bar V}\omega_{\bar V'})- q_{\bar W(n)}(\bar\pi\omega_{\bar V'},\ldots,\bar\pi\omega_{\bar V'})=\\
=-n!\sum_{L\in{\bf{L}}_\mr{nonPl}:\;|L|=n}\frac{1}{|\Aut(L)|}\left(\Loop_{L;\{-K_V l_{V(m)}\}_{m\geq 2};V}(\iota_V\iota_{\bar V',\bar V}\omega_{\bar V'},\ldots,\iota_V\iota_{\bar V',\bar V}\omega_{\bar V'})-\right.\\
\left.-\Loop_{L;\{-K_W l_{W(m)}\}_{m\geq 2};W}(\iota_W\bar\pi\omega_{\bar V'},\ldots,\iota_W \bar\pi \omega_{\bar V'})\right)+\\
+n!\sum_{T\in{\bf{T}}_\mr{nonPl}:\;|T|=n}\frac{1}{|\Aut(T)|}\left(\Iter_{T;\{-K_V l_{V(m)}\}_{m\geq 2};\{q_{V(m)}\}_{m\geq 1}} (\iota_V\iota_{\bar V',\bar V}\omega_{\bar V'},\ldots,\iota_V\iota_{\bar V',\bar V}\omega_{\bar V'})-\right.\\
\left.-\Iter_{T;\{-K_W l_{W(m)}\}_{m\geq 2};\{q_{W(m)}\}_{m\geq 1}} (\iota_W\bar\pi\omega_{\bar V'},\ldots,\iota_W\bar\pi\omega_{\bar V'})\right)
\label{gluing-induction eq5}
\end{multline}
(the third index of $\Loop$ denotes the space over which the super-trace is taken). On the other hand, the quantum operations induced from $V'$ (let us temporarily denote them $\tilde{q}_{\bar V'(n)}$) are
\begin{multline}
\tilde q_{\bar V'(n)}(\omega_{\bar V'},\ldots,\omega_{\bar V'})=
-n!\sum_{L\in{\bf{L}}_\mr{nonPl}:\;|L|=n}\frac{1}{|\Aut(L)|}\Loop_{L;\{-K_{V'}l_{V'(m)}\}_{m\geq 2};V'}(\iota_{V'}\omega_{\bar V'},\ldots, \iota_{V'}\omega_{\bar V'})+\\
+n!\sum_{T\in{\bf{T}}_\mr{nonPl}:\;|T|=n}\frac{1}{|\Aut(T)|}\Iter_{T;\{-K_{V'}l_{V'(m)}\}_{m\geq 2};\{q_{V'(m)}\}_{m\geq 1}} (\iota_{V'}\omega_{\bar V'},\ldots, \iota_{V'}\omega_{\bar V'})
\label{gluing-induction eq6}
\end{multline}
Denote $\hat L$ the tree, obtained by cutting one-loop diagram $L$ along some edge of the cycle, and assume for convenience that we chose such a planar representative for $L$, that the cut edge in $L$ (i.e. the marked leaf in $\hat{L}$) is the last leaf of $\hat{L}$ if we are going around in counterclockwise direction, starting from the root. Let us compute the super-traces over $V$ in (\ref{gluing-induction eq5}), using the splitting $V=\iota_{V',V}(V')\oplus \iota_-(W)$:
\begin{multline}
\Loop_{L;\{-K_V l_{V(m)}\}_{m\geq 2};V}(\iota_V\iota_{\bar V',\bar V}\omega_{\bar V'},\ldots,\iota_V\iota_{\bar V',\bar V}\omega_{\bar V'})=\\
=\Str_{V}\Iter_{\hat L;\{-K_V l_{V(m)}\}_{m\geq 2};\{-K_V l_{V(m)}\}_{m\geq 2}}(\iota_V\iota_{\bar V',\bar V}\omega_{\bar V'},\ldots,\iota_V\iota_{\bar V',\bar V}\omega_{\bar V'},\bt)=\\
=\Str_{V'}\Iter_{\hat L;\{-K_V l_{V(m)}\}_{m\geq 2};\{-\pi_{V,V'}K_V l_{V(m)}\}_{m\geq 2}}(\iota_V\iota_{\bar V',\bar V}\omega_{\bar V'},\ldots,\iota_V\iota_{\bar V',\bar V}\omega_{\bar V'},\iota_{V',V}\bt)+\\
+\Str_{W}\Iter_{\hat L;\{-K_V l_{V(m)}\}_{m\geq 2};\{-\pi_-K_V l_{V(m)}\}_{m\geq 2}}(\iota_V\iota_{\bar V',\bar V}\omega_{\bar V'},\ldots,\iota_V\iota_{\bar V',\bar V}\omega_{\bar V'},\iota_-\bt)
\label{gluing-induction eq7}
\end{multline}
To compute the first term here, we use (\ref{gluing-induction eq4}) (or rather a trivial modification, where the last leaf is decorated with $\iota_{V',V}\bt$):
\begin{multline}\Str_{V'}\Iter_{\hat L;\{-K_V l_{V(m)}\}_{m\geq 2};\{-\pi_{V,V'}K_V l_{V(m)}\}_{m\geq 2}}(\iota_V\iota_{\bar V',\bar V}\omega_{\bar V'},\ldots,\iota_V\iota_{\bar V',\bar V}\omega_{\bar V'},\iota_{V',V}\bt)=\\
=\Iter_{\hat L;\{-K_{V'} l_{V'(m)}\}_{m\geq 2};\{-K_{V'} l_{V'(m)}\}_{m\geq 2}}(\iota_{V'}\omega_{\bar V'},\ldots,\iota_{V'}\omega_{\bar V'},\bt)=\\
=\Loop_{L;\{-K_{V'} l_{V'(m)}\}_{m\geq 2};V'}(\iota_{V'}\omega_{\bar V'},\ldots,\iota_{V'}\omega_{\bar V'})
\label{gluing-induction eq8}
\end{multline}
To compute the second term in (\ref{gluing-induction eq7}), we carry the projection $\pi_-$ through to the leaves, using (\ref{induction-gluing agreement axiom 1},\ref{induction-gluing agreement axiom 3}) and that $\pi_{1,2}$ are $L_\infty$ morphisms (\ref{gluing axiom 3'},\ref{gluing axiom 4'}):
\begin{multline}
\Str_{W}\Iter_{\hat L;\{-K_V l_{V(m)}\}_{m\geq 2};\{-\pi_-K_V l_{V(m)}\}_{m\geq 2}}(\iota_V\iota_{\bar V',\bar V}\omega_{\bar V'},\ldots,\iota_V\iota_{\bar V',\bar V}\omega_{\bar V'},\iota_-\bt)=\\
=\Str_W\Iter_{\hat L;\{-K_W l_{W(m)}\}_{m\geq 2};\{-K_W l_{W(m)}\}_{m\geq 2}}(\iota_W\underbrace{\bar\pi_2\iota_{\bar V',\bar V}}_{=\bar\pi}\omega_{\bar V'},\ldots,\iota_W\underbrace{\bar\pi_2\iota_{\bar V',\bar V}}_{=\bar\pi}\omega_{\bar V'},\underbrace{\pi_2\iota_-}_{\frac{1}{2}\id_W}\bt)-\\
-\Str_W\Iter_{\hat L;\{-K_W l_{W(m)}\}_{m\geq 2};\{-K_W l_{W(m)}\}_{m\geq 2}}(\iota_W\underbrace{\bar\pi_1\iota_{\bar V',\bar V}}_{=\bar\pi}\omega_{\bar V'},\ldots,\iota_W\underbrace{\bar\pi_1\iota_{\bar V',\bar V}}_{=\bar\pi}\omega_{\bar V'},\underbrace{\pi_1\iota_-}_{-\frac{1}{2}\id_W}\bt)=\\
=\Loop_{L;\{-K_W l_{W(m)}\}_{m\geq 2};W}(\iota_W\bar\pi\omega_{\bar V'},\ldots,\iota_W\bar\pi\omega_{\bar V'})
\label{gluing-induction eq9}
\end{multline}
Substituting (\ref{gluing-induction eq7}) together with (\ref{gluing-induction eq8},\ref{gluing-induction eq9}) into (\ref{gluing-induction eq5}), we see that the first term (the sum over one-loop graphs) in (\ref{gluing-induction eq5}) coincides with the first term in (\ref{gluing-induction eq6}). Now we only have to compare sums over trees in (\ref{gluing-induction eq5}) and (\ref{gluing-induction eq6}). Let us compute the contributions of trees in (\ref{gluing-induction eq6}) using (\ref{gluing-induction eq4}):
\begin{multline*}
\Iter_{T;\{-K_{V'}l_{V'(m)}\}_{m\geq 2};\{q_{V'(m)}\}_{m\geq 1}} (\iota_{V'}\omega_{\bar V'},\ldots, \iota_{V'}\omega_{\bar V'})=\\
=\Iter_{T;\{-K_{V}l_{V(m)}\}_{m\geq 2};\{q_{V'(m)}(\pi_{V,V'}\bt,\ldots, \pi_{V,V'}\bt)\}_{m\geq 1}} (\iota_{V}\iota_{\bar V',\bar V}\omega_{\bar V'},\ldots, \iota_{V}\iota_{\bar V',\bar V}\omega_{\bar V'})=\\
=\Iter_{T;\{-K_{V}l_{V(m)}\}_{m\geq 2};\{q_{V(m)}(\underbrace{\iota_{V',V}\pi_{V,V'}}_{=\id_V-\iota_-\pi_-}\bt,\ldots, \underbrace{\iota_{V',V}\pi_{V,V'}}_{=\id_V-\iota_-\pi_-}\bt)\}_{m\geq 1}} (\iota_{V}\iota_{\bar V',\bar V}\omega_{\bar V'},\ldots, \iota_{V}\iota_{\bar V',\bar V}\omega_{\bar V'})-\\
-\Iter_{T;\{-K_{V}l_{V(m)}\}_{m\geq 2};\{q_{W(m)}(\underbrace{\pi\pi_{V,V'}}_{=\pi_1}\bt,\ldots, \underbrace{\pi\pi_{V,V'}}_{=\pi_1}\bt)\}_{m\geq 1}} (\iota_{V}\iota_{\bar V',\bar V}\omega_{\bar V'},\ldots, \iota_{V}\iota_{\bar V',\bar V}\omega_{\bar V'})
\end{multline*}
In the decompositions of projections $\iota_{V',V}\pi_{V,V'}=\id_V-\iota_-\pi_-$ the second term is negligible, since
\begin{multline*}
\Iter_{\tilde T;\{-K_{V}l_{V(m)}\}_{m\geq 2};\{-\pi_-K_{V}l_{V(m)}\}_{m\geq 2}} (\iota_{V}\iota_{\bar V',\bar V}\omega_{\bar V'},\ldots, \iota_{V}\iota_{\bar V',\bar V}\omega_{\bar V'})=\\
=\Iter_{\tilde T;\{-K_{W}l_{W(m)}\}_{m\geq 2};\{-K_{W}l_{W(m)}\}_{m\geq 2}} (\iota_{W}\underbrace{\bar\pi_2\iota_{\bar V',\bar V}}_{=\bar\pi}\omega_{\bar V'},\ldots, \iota_{W}\underbrace{\bar\pi_2\iota_{\bar V',\bar V}}_{=\bar\pi_2}\omega_{\bar V'})-\\
-\Iter_{\tilde T;\{-K_{W}l_{W(m)}\}_{m\geq 2};\{-K_{W}l_{W(m)}\}_{m\geq 2}} (\iota_{W}\underbrace{\bar\pi_1\iota_{\bar V',\bar V}}_{=\bar\pi}\omega_{\bar V'},\ldots, \iota_{W}\underbrace{\bar\pi_1\iota_{\bar V',\bar V}}_{=\bar\pi_2}\omega_{\bar V'})=0
\end{multline*}
(analogous to the cancellation in (\ref{gluing-induction eq3})). Therefore,
\begin{multline*}
\Iter_{T;\{-K_{V'}l_{V'(m)}\}_{m\geq 2};\{q_{V'(m)}\}_{m\geq 1}} (\iota_{V'}\omega_{\bar V'},\ldots, \iota_{V'}\omega_{\bar V'})=\\
=\Iter_{T;\{-K_{V}l_{V(m)}\}_{m\geq 2};\{q_{V(m)}\}_{m\geq 1}} (\iota_{V}\iota_{\bar V',\bar V}\omega_{\bar V'},\ldots, \iota_{V}\iota_{\bar V',\bar V}\omega_{\bar V'})-\\
-\Iter_{T;\{-K_{W}l_{W(m)}\}_{m\geq 2};\{q_{W(m)}\}_{m\geq 1}} (\iota_{W}\bar\pi\omega_{\bar V'},\ldots, \iota_{W}\bar\pi\omega_{\bar V'})
\end{multline*}
And we see that contributions of trees in (\ref{gluing-induction eq5}) and (\ref{gluing-induction eq6}) coincide. This concludes the proof of part (\ref{statement: induction-gluing 3}) of the Statement.
\\$\Box$

\textbf{Main example: gluing de Rham algebras vs. gluing induced $qL_\infty$ structures on triangulations.} The following case of Statement \ref{statement: induction-gluing} is important for the simplicial $BF$ theory. Let $M$ be a compact manifold with boundary and let $N$ be another compact manifold (possibly with boundary) endowed with two embeddings into boundary of $M$:
$$\mr{emb}_1,\mr{emb}_2:N\ra \dd M$$
such that images of $\mr{emb}_1,\mr{emb}_2$ do not intersect
$$\mr{emb}_1(N)\cap \mr{emb}_2(N)=\varnothing$$
Then we have two pull-backs for differential forms:
$$\pi_{1,2}=\mr{emb}_{1,2}^*:\quad \Omega^\bt(M,\g)\ra \Omega^\bt(M,\g)$$
Let $U_1, U_2$ be small neighborhoods (thickenings) of $\mr{emb}_1(N)\subset M$ and $\mr{emb}_2(N)\subset M$, respectively, and let $\rho_{1,2}\in C^\infty(M)$ be two smearing functions supported on $U_1$ and $U_2$, respectively, having value 1 on $\mr{emb}_1(N)$ and $\mr{emb}_2(N)$, respectively:
$$\rho_1\circ \mr{emb}_1=\rho_2\circ\mr{emb}_2=1$$
Denote also $\pi_\bot^1: U_1\ra N,\quad \pi_\bot^2: U_2\ra N$ the projections from $U_1,U_2$ to $N$ (a point of $U_1$ is sent to the nearest (in some metric) point of the image $\mr{emb}_1(N)$, value of $\pi_\bot^1$ is the preimage of this point in $N$; analogously for $\pi_\bot^2$).
Embeddings $\iota_{1,2}$ for differential forms are introduced as
$$\iota_{1,2}=\rho_{1,2}\cdot (\pi_\bot^{1,2})^*:\quad \Omega^\bt(N,\g)\ra\Omega^\bt(M,\g)$$
Projections $\pi_{1,2}$ and embeddings $\iota_{1,2}$ are the gluing data for $qL_\infty$ algebras (for this case, just DGLA) of differential forms $V=\Omega^\bt(M,\g)$, $W=\Omega^\bt(N,\g)$. Axioms (\ref{gluing axioms 1},\ref{gluing axioms 2}) of gluing are obvious and axioms (\ref{gluing axioms 3},\ref{gluing axioms 4}) follow from the fact that $\pi_{1,2}$ are DGLA homomorphisms and, hence, linear $L_\infty$ morphisms. Next, the glued $qL_\infty$ algebra (which is again a  DGLA)
$V'$ is identified with the algebra of differential forms $\Omega^\bt(M',\g)$ on the manifold, obtained from $M$  by gluing together two components of the boundary, $\mr{emb}_1(N)$ and $\mr{emb}_2(N)$:
$$M'=M/\{\mr{emb}_1(x)\sim\mr{emb}_2(x)|\;x\in N\}$$
Notice that to identify $V'$ with $\Omega^\bt(M',\g)$, we have to require only the tangent component of differential forms to the surface of gluing $N\hra M'$ to be continuous, while the normal (in some metric) component is allowed to have a jump when passing across the surface of gluing.

Now suppose that we have a triangulation $\Xi$ of the manifold $M$ and a triangulation $F$ of $N$. Then the cochains $C^\bt(\Xi,\g)$ are endowed with the  $qL_\infty$ structure induced from $\Omega^\bt(M,\g)$, an the cochains $C^\bt(F,\g)$ are endowed with $qL_\infty$ structure induced from $\Omega^\bt(N,\g)$. We denote the respective standard induction data (embedding cochains as Whitney forms, retraction by integrals over simplices, Dupont's chain homotopy) by $(\iota_\Xi,r_\Xi,K_\Xi)$ and $(\iota_F,r_F,K_F)$. Assume also that embeddings $\mr{emb}_{1,2}$ are consistent with triangulation (i.e. they map simplices into simplices) and embed $F$ into $\Xi$ as a simplicial subcomplex in two ways: $\overline{\mr{emb}}_{1,2}:F\ra \Xi$. Then the consistency conditions (\ref{induction-gluing agreement axiom 1},\ref{induction-gluing agreement axiom 2},\ref{induction-gluing agreement axiom 3}) for gluing and induction are satisfied automatically (namely, (\ref{induction-gluing agreement axiom 1}) follows from the consistency of Whitney forms on a simplex with face restrictions, (\ref{induction-gluing agreement axiom 3}) follows from (\ref{K restriction to face}) and \ref{induction-gluing agreement axiom 2} is obvious). Induced gluing data $\bar\pi_{1,2}: C^\bt(\Xi,\g)\ra C^\bt(F,\g),\quad \bar\iota_{1,2}:C^\bt(F,\g)\ra C^\bt(\Xi,\g)$ is simply given by restrictions of cochains on $\Xi$ to $F$ (induced from simplicial embeddings $\overline{\mr{emb}}_{1,2}$), and by the respective embeddings of cochains on $F$ into cochains on $\Xi$ (supported on $\overline{\mr{emb}}_{1,2}(F)$, respectively). Indeed, unlike the case of differential forms, we do not need smearing functions for embeddings here. Glued induction data $\Omega^\bt(M',\g)\xra{(\iota_{\Xi'},r_{\Xi'},K_{\Xi'})} C^\bt(\Xi',\g)$ turn out to be the standard induction data for triangulation $\Xi'$. Therefore, due to Statement \ref{statement: induction-gluing}, the action glued from $S_{\Xi}$, $S_F$ coincides with the effective action $S_{\Xi'}$, induced from topological $BF$ theory on $M'$ with the standard induction data. Diagram (\ref{gluing-induction square}) here takes the form
$$ \begin{CD}
\Omega^\bt(M,\g),\; \Omega^\bt(N,\g)@>{(\iota_{\Xi},r_\Xi,K_\Xi),\;(\iota_F,r_F,K_F)}>\text{standard induction}> (C^\bt(\Xi,\g),S_\Xi),\;(C^\bt(F,\g),S_F)\\
@V(\pi_{1,2},\iota_{1,2})V\text{gluing}V   @V(\bar\pi_{1,2},\bar\iota_{1,2})V\text{induced gluing}V\\
\Omega^\bt(M',\g)@>(\iota_{\Xi'},r_{\Xi'},K_{\Xi'})>\text{standard induction}> (C^\bt(\Xi',\g),S_{\Xi'})
\end{CD}
$$

Another possible case is when the cell complex $\Xi'$ is not a ``honest'' triangulation. For example, for gluing the interval with standard triangulation into the circle (gluing the two boundary points) the glued cell complex is $\{[+],[01]\}$. Gluing data (for differential forms) are obviously consistent with the standard induction data for interval. Hence the glued simplicial action for interval really gives the effective action for circle. The glued induction data are:
\begin{eqnarray}
\iota_{\s^1}:\quad x e_+ +y e_{01}&\mapsto& x+y dt \label{circle ind data 1}\\
r_{\s^1}:\quad f+g dt&\mapsto& f(0) e_+ + \left( \int_{\s^1} g(\tilde t) d\tilde t \right) e_{01} \label{circle ind data 2}\\
K_{\s^1}:\quad f+g dt&\mapsto& \int_0^t g(\tilde t)d\tilde t-t\left(\int_{\s^1} g(\tilde t) d\tilde t\right) \label{circle ind data 3}
\end{eqnarray}

Of particular interest is the case when the manifold $M=M_1\sqcup M_2$ is a disjoint union and $\mr{emb}_1: N\ra \dd M_1$, $\mr{emb}_2: N\ra \dd M_2$; also $M_{1,2}$ and $N$ are endowed with triangulations $\Xi_{1,2},F$ and embeddings $\mr{emb}_{1,2}$ are consistent with triangulations. Then we describe the gluing of two manifolds $M_{1,2}$ along a part of boundary into a new manifold $M'$, and it turns out that the simplicial action for $M'$ with glued triangulation $\Xi'$ coincides with the action, glued from simplicial actions on $\Xi_{1,2}$ and $F$. Hence, starting from gluing two simplices along a face and gradually adding new simplices, we can obtain Theorem \ref{thm: simplicial locality} from the Statement \ref{statement: induction-gluing}.

\textbf{Remark.} The possibility to glue from the standard induction data for individual simplices $\Omega^\bt(\sigma)\xra{(\iota_\sigma,r_\sigma,K_\sigma)}C^\bt(\sigma)$ the induction data for a simplicial complex $\Omega^\bt(\Xi)\xra{(\iota_\Xi,r_\Xi,K_\Xi)}C^\bt(\Xi)$ is due to two facts. First, due to consistency of standard induction data for a simplex with restriction of differential forms and cochains to a face $\sigma'\subset \sigma$, which means that the three following diagrams commute:
$$\begin{CD}\Omega^\bt(\sigma) @<\iota_\sigma<< C^\bt(\sigma) \\
@V\bt|_{\sigma'}VV @V\bt|_{\sigma'}VV \\
\Omega^\bt(\sigma') @<\iota_{\sigma'}<< C^\bt(\sigma')
\end{CD}\qquad
\begin{CD}\Omega^\bt(\sigma) @>r_\sigma>> C^\bt(\sigma) \\
@V\bt|_{\sigma'}VV @V\bt|_{\sigma'}VV \\
\Omega^\bt(\sigma') @>r_{\sigma'}>> C^\bt(\sigma')
\end{CD}\qquad
\begin{CD}\Omega^\bt(\sigma) @>K_\sigma>> \Omega^\bt(\sigma) \\
@V\bt|_{\sigma'}VV @V\bt|_{\sigma'}VV \\
\Omega^\bt(\sigma') @>K_{\sigma'}>> \Omega^\bt(\sigma')
\end{CD}
$$
for any face $\sigma'\subset\sigma$ (first diagram is a property of Whitney forms, second is an obvious property of integrals over faces, third is a property of Dupont's chain homotopy). The second fact is the consistency of standard induction data for a simplex with the action of group of permutations of vertices of simplex $S_{n+1}$ on the differential $\Omega^\bt(\Delta^n)$ and cochains $C^\bt(\Delta^n)$:
\be
\begin{CD}\Omega^\bt(\Delta^n) @<\iota_{\Delta^n}<< C^\bt(\Delta^n) \\
@V\pi VV @V\pi VV \\
\Omega^\bt(\Delta^n) @<\iota_{\Delta^n}<< C^\bt(\Delta^n)
\end{CD}\qquad
\begin{CD}\Omega^\bt(\Delta^n) @>r_{\Delta^n}>> C^\bt(\Delta^n) \\
@V\pi VV @V\pi VV \\
\Omega^\bt(\Delta^n) @>r_{\Delta^n}>> C^\bt(\Delta^n)
\end{CD}\qquad
\begin{CD}\Omega^\bt(\Delta^n) @>K_{\Delta^n}>> \Omega^\bt(\Delta^n) \\
@V\pi VV @V\pi VV \\
\Omega^\bt(\Delta^n) @>K_{\Delta^n}>> \Omega^\bt(\Delta^n)
\end{CD}
\label{simplex ind data symmetry}
\ee
for any permutation of vertices $\pi\in S_{n+1}$. We mean the action on differential forms by $t_i\mapsto t_{\pi(i)},\;dt_i\mapsto dt_{\pi(i)}$ and on the cochains by $e_{i_0\cdots i_k}\mapsto e_{\pi(i_0)\cdots\pi(i_k)}$ where we assume $e_{\pi'(i_0)\cdots \pi'(i_k)}=(-1)^{\pi'}e_{i_0\cdots i_k}$ for ``internal'' permutations of vertices of the face $\pi':(i_0,\ldots,i_k)\mapsto (i_0,\ldots,i_k)$. Consistency of induction data with the symmetry of simplex (\ref{simplex ind data symmetry}) is important, since otherwise we had to glue simplicial complexes from simplices with enumerated vertices and take care of consistency of numeration with gluing.

\subsection{Simplicial $BF$ action for the interval}
\label{section: interval}
As we showed in section \ref{section: simplicial BF action}, the problem of computing the simplicial $BF$ action $S_\Xi$ for an arbitrary simplicial complex $\Xi$ reduces to a series of universal computations for $\Xi=\Delta^D$ --- standard simplex in dimension $D$ with standard triangulation.

\textbf{Preliminary example: simplicial action for 0-simplex.} For $D=0$ the problem is trivial, since  0-simplex $\Delta^0=[0]$ is a points ($[0]$ is the label of vertex) and the algebra of $\g$-valued differential forms $\Omega^\bt(\Delta^0,\g)$ coincides with the complex of $\g$-valued cell cochains $C^\bt(\Delta^0,\g)=\g e_0\cong\g$. I.e. the space of UV forms vanishes here: $\Omega''^\bt(\Delta^0)=\{0\}$ and the problem of induction is trivial, i.e. the simplicial action coincides with the initial action of abstract $BF$ theory for Lie algebra $\Omega^\bt(\Delta^0,\g)\cong \g$:
\be S_{\Delta^0}=<p_0,\frac{1}{2}[\omega^0,\omega^0]>_\g\label{S_Delta^0}\ee
where $<\bt,\bt>_\g:\g^*\otimes\g\ra\RR$ is the canonical pairing between $\g$ and $\g^*$.
This action is a function on the space of fields $\FF_{\Delta^0}=T^*[-1](\g[1])=\g[1]\oplus\g^*[-2]$; field $\omega^0$ is the $\g$-valued coordinate function on the first term and $p_0$ is the $\g^*$-valued coordinate function on the second term, $\gh(\omega^0)=1$, $\gh(p_0)=-2$.
Action (\ref{S_Delta^0}) is the $BF_\infty$ action, corresponding to the natural $qL_\infty$ structure on $\g$-valued cochains on a point
$C^\bt(\Delta^0,\g)\cong\g$: there is only one classical operation $l_{(2)}(x,y)=[x,y]$ for $x,y\in\g$, all other operations vanish: $l_{(1)}=l_{(3)}=l_{(4)}=\cdots=0$, $q_{(1)}=q_{(2)}=\cdots=0$. Or, in terms of cohomological vector field and density of measure:
$$Q_{\Delta^0}=-\left<\frac{1}{2}[\omega^0,\omega^0],\frac{\dd}{\dd \omega^0}\right>_\g=-\sum_{a,b,c}\frac{1}{2}f^a_{bc}\omega^{0b}\omega^{0c}\frac{\dd}{\dd\omega^{0a}}$$
--- is the Chevalley-Eilenberg differential on $\Fun(\g[1])$ --- the cochain complex of Lie algebra $\g$ (we denoted $f^a_{bc}=<T^a,[T_b,T_c]>_\g$ the structure constant of the Lie bracket in $\g$), and the density of measure is trivial:
$$\rho_{\Delta^0}=1$$
i.e. the measure $\mu_{\Delta^0}=\DD \omega^0=\prod_a \DD\omega^{0a}$  is the coordinate Berezin measure on $\g[1]$. Reduced action for 0-simplex (\ref{Sbar via S}) coincides with the full simplicial $BF$ action:
\be \label{Sbar Delta^0}\bar{S}_{\Delta^0}=S_{\Delta^0}=<p_0,\frac{1}{2}[\omega^0,\omega^0]>_\g\ee

\textbf{Case $D=1$: induction data, Hodge decomposition for forms.} Let us now turn to case $D=1$, i.e. to the problem of computing $S_{\Delta^1}$ --- the simplicial $BF$ action for the interval $\Delta^1=[0,1]$.
Cochain complex of the standard triangulation with coefficients in $\g$ is $C^\bt(\Delta^1,\g)=\g e_0\oplus \g e_1\oplus \g e_{01}$ where $e_{0},e_{1},e_{01}$ are the basis cochains, corresponding to the left and right end-points of the interval and to the bulk (the top-dimension cell). Basis Whitney forms  $(t_0,t_1)$ are written in barycentric coordinates as
$\chi_0=t_0, \chi_1=t_1, \chi_{01}=t_0 dt_1-t_1 dt_0$. Or in terms of a single coordinate $t=t_1$ (i.e. resolving the constraint on second coordinate $t_0$):
$$\chi_0=1-t, \chi_1=t, \chi_{01}=dt$$
Therefore the embedding of cochains into differential forms $\iota_{\Delta^1}:C^\bt(\Delta^1,\g)\ra \Omega^\bt(\Delta^1,\g)$ is
$$\iota_{\Delta^1}: \alpha^{0}e_0+\alpha^1 e_1+\alpha^{01} e_{01}\mapsto \alpha^0 (1-t)+\alpha^1 t+\alpha^{01}dt$$
and the retractions $r_{\Delta^1}:\Omega^\bt(\Delta^1,\g)\ra C^\bt(\Delta^1,\g)$ is
$$r_{\Delta^1}:\alpha=f(t)+g(t)dt\mapsto f(0)e_0+f(1) e_1+ \left(\int_0^1 g(t)dt\right) e_{01}$$
where we decomposed differential form $\alpha$ on the interval into components of degrees 0 and 1; $f,g$ is a pair of functions on the interval. Splitting of $\Omega^\bt(\Delta^1,\g)$ into IR and UV parts is $\Omega^\bt(\Delta^1,\g)=\Omega^\bt_W(\Delta^1,\g)\oplus \Omega''^\bt(\Delta^1,\g)$ where the IR part (the Whitney complex) is $$\Omega^\bt_W(\Delta^1,\g)=\{\alpha^0 (1-t)+\alpha^1 t+\alpha^{01} dt|\;\alpha^{0,1,01}\in\g\}$$ --- linear 0-forms and constant 1-forms on the interval, and the UV part:
$$\Omega''^\bt(\Delta^1,\g)=\{f(t)+g(t)dt|\;f(0)=f(1)=0,\int_0^1 g(t)dt=0\}$$
i.e. 0-forms, vanishing on the end-points of the interval, and 1-forms with vanishing integral over the interval.
Projectors to IR and UV forms are
\begin{eqnarray*}
\PP'=\iota_{\Delta^n}\circ r_{\Delta^n}:\; f(t)+g(t)dt& \mapsto & f(0)\cdot(1-t)+f(1)\cdot t+\left(\int_0^1 g(\tilde{t})d\tilde{t}\right)\cdot dt\\
\PP''=\id-\PP':\; f(t)+g(t)dt& \mapsto & \left(f(t)-f(0)\cdot(1-t)-f(1)\cdot t\right)+\left(g(t)-\int_0^1 g(\tilde{t})d\tilde{t}\right)\cdot dt
\end{eqnarray*}
Next, the chain homotopy, defined by (\ref{K Dupont}), acts as
\begin{multline*}K_{\Delta^1}=\chi_0 h^0+\chi_1 h^1: f(t)+g(t)dt\mapsto (1-t) \int_0^1 g(ut)t du+t \int_0^1 g(ut-u+1)(t-1)du\\
=(1-t)\int_0^t g(\tilde{t})d\tilde{t}-t\int_{t}^1 g(\tilde{t})d\tilde{t}=\int_0^t g(\tilde{t})d\tilde{t}-t\int_0^1 g(\tilde{t})d\tilde{t}
\end{multline*}
I.e. $K_{\Delta^1}$ sends 0-forms to zero (since a chain homotopy lowers the degree of a form by one), and acts on 1-forms as the integral operator
$$K_{\Delta^1}: g(t)dt\mapsto \int_0^t g(\tilde{t})d\tilde{t}-t\int_0^1 g(\tilde{t})d\tilde{t}$$
The kernel of this operator is
\be K_{\Delta^1}(t,\tilde{t})=\theta(t-\tilde{t})-t \label{interval K kernel}\ee
where $$\theta(x)=\left\{\begin{array}{ll}1,\;x\geq 0 \\ 0,\; x<0 \end{array}\right.$$ is the unit step function. Next, the $d$-exact part of $\Omega''^\bt(\Delta^1)$ obviously coincides with the space of UV 1-forms (since any 1-form on the interval with zero integral is the differential of some function, vanishing on the end-points): $\Omega''^\bt_{d-ex}(\Delta^1)=\Omega''^1(\Delta^1)$. Therefore the $K$-exact part of $\Omega''^\bt(\Delta^1)$ coincides with the space of UV 0-forms (since the differential $d:\Omega''^0(\Delta^1)\ra\Omega''^1(\Delta^1)$ is invertible, and $K_{\Delta^1}$ is the inverse): $\Omega''^\bt_{K-ex}(\Delta^1)=\Omega''^0(\Delta^1)$. Therefore the Hodge decomposition (\ref{Hodge decomp}) for cochains on the interval is
\begin{eqnarray*}
\Omega^\bt(\Delta^1,\g)&=&\Omega^\bt_W(\Delta^1,g)\oplus \Omega''^1(\Delta^1,\g)\oplus \Omega''^0(\Delta^1,\g)\\
&=&\{\alpha^0 (1-t)+\alpha^1 t+\alpha^{01}dt\}\oplus \{g(t)dt|\;\int_0^1 g(t)dt=0\}\oplus\{f(t)|\;f(0)=f(1)=0\}
\end{eqnarray*}

Simplicial action $S_{\Delta^1}$ is a function on the space
$$\FF_{\Delta^1}=T^*[-1](C^\bt(\Delta^1,\g)[1])=T^*[-1]((\g e_0\oplus \g e_1\oplus \g e_{01})[1])\cong \g[1]\oplus\g[1]\oplus\g[0]\oplus \g^*[-2]\oplus \g^*[-2]\oplus \g^*[-1]$$
with $\g$-valued coordinates $\omega^0,\omega^1,\omega^{01}$ and $\g^*$-valued coordinates $p_0,p_1,p_{01}$. The ghost numbers for coordinates are: $\gh(\omega^0)=\gh(\omega^1)=1, \gh(\omega^{01})=0,\gh(p_{0})=\gh(p_1)=-2,\gh(p_{01})=-1$.

\begin{thm}[Simplicial $BF$ action for the interval]
\label{interval thm}
The reduced simplicial $BF$ action for the interval is
$$\bar{S}_{\Delta^1}=\bar{S}_{\Delta^1}^0+\hbar \bar{S}_{\Delta^1}^1$$
with the tree part given by
\begin{eqnarray}\bar{S}_{\Delta^1}^0(\omega^0,\omega^1,\omega^{01},p_{01})&=&
\left<p_{01},(\omega^1-\omega^0)+\frac{1}{2}[\omega^{01},\omega^0+\omega^1]+\right.\\
&&\qquad\left.+\sum_{n=1}^\infty \frac{B_{2n}}{(2n)!}\underbrace{[\omega^{01},[\omega^{01},\cdots,[\omega^{01}}_{2n},\omega^1-\omega^0]\cdots]\right>_\g \nonumber\\
&=&\left<p_{01},\frac{1}{2}[\omega^{01},\omega^0+\omega^1]+\left(
\frac{\ad_{\omega^{01}}}{2}\coth\frac{\ad_{\omega^{01}}}{2}\right)\circ (\omega^1-\omega^0)\right>_\g \label{interval thm eq1}
\end{eqnarray}
where $B_n$ are Bernoulli numbers: $B_0=1,\; B_1=-1/2,\; B_2=1/6,\; B_3=0,\; B_4=-1/30,\;\ldots$ and $\ad_{\omega^{01}}=[\omega^{01},\bt]$ is the adjoint action; one-loop part of $\bar{S}_{\Delta^1}$ is
\begin{eqnarray}\bar{S}_{\Delta^1}^1(\omega^{01}) & = & \sum_{n=1}^\infty \frac{B_{2n}}{(2n)\cdot (2n)!}\;\tr_\g \underbrace{[\omega^{01},[\omega^{01},\cdots,[\omega^{01}}_{2n},\bt]\cdots ]\nonumber \\
&=&\tr_\g\log\left(\frac{\sinh\frac{\ad_{\omega^{01}}}{2}}{\frac{\ad_{\omega^{01}}}{2}}\right) \label{interval thm eq2}
\end{eqnarray}
The full simplicial $BF$ action for the interval is the sum of contributions of left and right end-points and the bulk:
\begin{multline}\label{interval thm eq3}S_{\Delta^1}(\omega^0,\omega^1,\omega^{01},p_0,p_1,p_{01})=
\bar{S}_{[0]}(\omega^0,p_0)+\bar{S}_{[1]}(\omega^1,p_1)+\bar{S}_{[01]}(\omega^0,\omega^1,\omega^{01},p_{01})\\
=\left<p_0,\frac{1}{2}[\omega^0,\omega^0]\right>_\g+\left<p_1,\frac{1}{2}[\omega^1,\omega^1]\right>_\g
+\\
+\left<p_{01},\frac{1}{2}[\omega^{01},\omega^0+\omega^1]+\left(\frac{\ad_{\omega^{01}}}{2}\coth\frac{\ad_{\omega^{01}}}{2}\right)\circ (\omega^1-\omega^0)\right>_\g
+\hbar\;\tr_\g\log\left(\frac{\sinh\frac{\ad_{\omega^{01}}}{2}}{\frac{\ad_{\omega^{01}}}{2}}\right)
\end{multline}
\end{thm}
Recall that the Bernoulli numbers $B_n$ are defined by the generating function
$$\sum_{n=0}^\infty\frac{B_n}{n!}x^n=\frac{x}{e^x-1}$$
We will also need the Bernoulli polynomials $B_n(t)$ defined by
$$\sum_{n=0}^\infty\frac{B_n(t)}{n!}x^n=\frac{x e^{xt}}{e^x-1}$$
The first Bernoulli polynomials are $B_0(t)=1,\;B_1(t)=t-\frac{1}{2},\;B_2(t)=t^2-t+\frac{1}{6},\;B_3(t)=t^3-\frac{3}{2}t^2+\frac{1}{2}t,\; B_4(t)=t^4-2t^3+t^2-\frac{1}{30}$ etc.
To prove the theorem we need the two following lemmas.
\begin{lemma}\label{interval lemma 1} For every $n\geq 1$ we have
\be \left(K_{\Delta^1}(\chi_{01}\wedge\bt)\right)^n\circ \chi_1=-\left(K_{\Delta^1}(\chi_{01}\wedge\bt)\right)^n\circ \chi_0=\frac{B_{n+1}(t)-B_{n+1}}{(n+1)!} \label{interval lemma 1 eq1}\ee
and
\be \int_0^1 \chi_{01}\wedge \left(K_{\Delta^1}(\chi_{01}\wedge\bt)\right)^n\circ \chi_1=
-\int_0^1 \chi_{01}\wedge \left(K_{\Delta^1}(\chi_{01}\wedge\bt)\right)^n\circ \chi_0=-\frac{B_{n+1}}{(n+1)!}\label{interval lemma 1 eq2}\ee
where $B_n(t)$ are the Bernoulli polynomials.
\end{lemma}

\begin{lemma}\label{interval lemma 2} For every $n\geq 2$ we have
\be \Str_{\Omega^0(\Delta^1)}\left(K(\chi_{01}\wedge\bt)\right)^n=-\frac{B_n}{n!} \label{interval lemma 2 eq1}\ee
\end{lemma}

\textbf{Proof of Lemma \ref{interval lemma 1}.}
Let us introduce the generating function
\be f(x,t)=\sum_{n=0}^\infty x^n (K_{\Delta^1}(\chi_{01}\wedge\bt))^n\circ\chi_1\label{interval lemma 1 gen function}\ee
Applying the differential $d_t=dt\wedge\frac{\dd}{\dd t}$ to both sides and using the property of chain homotopy $d_t K_{\Delta^1}+K_{\Delta^1} d_t=\PP''$, we obtain
$$dt\wedge\frac{\dd}{\dd t} f(x,t)=dt+ x  dt\wedge\left(f(x,t)-\int_0^1 f(x,\tilde{t}) d\tilde{t}\right)$$
and hence
$$\frac{\dd}{\dd t}f(x,t)=x f(x,t)+C(x)$$
where $C(x)$ does not depend on $t$. Solving this differential equation with boundary conditions $f(x,0)=0$, $f(x,1)=1$ (only the $n=0$ term in (\ref{interval lemma 1 gen function}) contributes to values of $f$ in the end-points of interval), we obtain the unique solution
\be f(x,t)=\frac{e^{xt}-1}{e^x-1}=\frac{1}{x}\left(\frac{xe^{xt}}{e^x-1}-\frac{x}{e^x-1}\right)=\sum_{n=0}^\infty \frac{B_{n+1}(t)-B_{n+1}}{(n+1)!}\;x^n \label{interval lemma 1 eq3}\ee
Next, since $K_{\Delta^1}(\chi_{01}\wedge(\chi_0+\chi_1))=K_{\Delta^1}(\chi_{01})=0$, we have $\left(K_{\Delta^1}(\chi_{01}\wedge\bt)\right)^n\circ \chi_1=-\left(K_{\Delta^1}(\chi_{01}\wedge\bt)\right)^n\circ \chi_0$ for $n\geq 1$. Therefore (\ref{interval lemma 1 eq1}) is proved. Finally, (\ref{interval lemma 1 eq2}) is obtained from (\ref{interval lemma 1 eq3}) immediately by integration over $t$ (or equivalently, (\ref{interval lemma 1 eq2}) follows from (\ref{interval lemma 1 eq1}) and the property $\int_0^1 B_n(t)dt=0$ of Bernoulli polynomials for $n\geq 1$).
\\$\Box$

\textbf{Proof of Lemma \ref{interval lemma 2}.} Since the question of computing the trace of an operator over infinite-dimensional space $\Omega^0(\Delta^1)$ is a subtle one, we propose three independent computations in three different natural bases on $\Omega^0(\Delta^1)$: in the basis of monomials $\{t^n\}$, in the basis of delta functions $\{\delta(t-t_0)\}$ and in the basis of exponentials $\{e^{2\pi i n t}-1\}$. And we check that all three bases produce the same result (\ref{interval lemma 2 eq1}).

\textit{Computation in the basis of monomials.} Introduce the notation $\kappa=K_{\Delta^1}(\chi_{01}\wedge\bt):\Omega^0(\Delta^1) \ra \Omega^0 (\Delta^1)$
$$
\kappa:\;f(t)\mapsto \int_0^t f(\tilde{t})d\tilde{t}-t\int_0^1 f(\tilde{t})d\tilde{t}
$$
We want to compute $\Str_{\Omega^0(\Delta^1)}\kappa^n$ directly by summing diagonal matrix elements of the operator $\kappa^n$ in the basis of monomials $\{t^m\}_{m=0}^\infty$. We use Dirac's bra-ket notation for matrix elements:
$$\kappa^n\circ t^m=\sum_{m'}<t^{m'}|\kappa^n|t^m>\cdot t^{m'}$$
Notice that the monomial $t^0=1$ does not contribute to the trace, since $\kappa\circ 1=0$.
Introduce the generating function
$$f_m(x,t)=\sum_{n=0}^\infty x^n \kappa^n\circ t^m$$
for $m\geq 1$. Differentiating $f_m$ in variable $t$, analogously to the argument in the proof of Lemma \ref{interval lemma 1}, we obtain the differential equation
$$\frac{\dd}{\dd t}f_m(x,t)=x f_m(x,t)+m t^{m-1}+C_m(x)$$
where $C_m(x)$ does not depend on $t$. Solving this equation with boundary conditions $f_m(x,0)=0$, $f_m(x,1)=1$, we obtain the unique solution
\begin{eqnarray}f_m(x,t)&=&\frac{e^{xt}-1}{e^x-1}\left(1-e^x\int_0^1 d\tilde{t}\; m \tilde{t}^{m-1}e^{-x\tilde{t}} \right)+
e^{xt}\int_0^t d\tilde{t}\; m \tilde{t}^{m-1}e^{-x\tilde{t}}\\
&=&\frac{e^{xt}-1}{e^x-1}\sum_{k=0}^{m-1}\frac{m!}{(m-k)!}x^{-k}-\sum_{k=1}^{m-1}\frac{m!}{(m-k)!}t^{m-k}x^{-k} \label{interval lemma 2 eq2}
\end{eqnarray}
Let us write $f_m(x,t)$ as a power series in variable $t$: $f_m(x,t)=\sum_{k=1}^\infty f_{m,k}(x)t^k$. Then the coefficient $f_{m,m}(x)$ is the generating function for diagonal matrix elements of powers of operator $\kappa$, corresponding to monomial $t^m$:
\be f_{m,m}(x)=\sum_{n=0}^\infty x^n <t^m|\kappa^n|t^m> \label{interval lemma 2 eq3} \ee
Explicit formula (\ref{interval lemma 2 eq2}) implies that
$$f_{m,m}(x)=1-\frac{x}{e^x-1}\sum_{k=m}^\infty \frac{x^k}{(k+1)!}$$
Term $1$ corresponds to the matrix element of identity operator, i.e. to the $n=0$ term in (\ref{interval lemma 2 eq3}). Notice also that $f_{m,m}(x)=1+O(x^m)$ implies that diagonal matrix elements $<t^m|\kappa^n|t^m>$ may be non-vanishing only for $m\leq n$. Therefore the matrix of operator $\kappa^n$ in monomial basis has only finitely many nonzero matrix elements on the diagonal.
Finally, summing over $m$ in (\ref{interval lemma 2 eq3}) and subtracting the contribution of identity operator, we obtain
\begin{multline*}
\sum_{n=1}^\infty x^n \Str_{\Omega^0(\Delta^1)}\kappa^n =\sum_{m=1}^\infty (f_{m,m}(x)-1)=-\frac{x}{e^x-1}\sum_{m=1}^\infty\sum_{k=m}^\infty \frac{x^k}{(k+1)!}\\
=-\frac{x}{e^x-1}\sum_{k=1}^\infty\frac{k}{(k+1)!} x^k=1-x-\frac{x}{e^x-1}=-\frac{1}{2}x-\sum_{n=2}^\infty \frac{B_n}{n!} x^n
\end{multline*}
which implies (\ref{interval lemma 2 eq1}): $\Str_{\Omega^0(\Delta^1)}\kappa^n=-\frac{B_n}{n!}$ for $n\geq 2$.

\textit{Computation in coordinate representation (in the basis of delta functions).} Another natural idea of computing $\Str_{\Omega^0(\Delta^1)}\kappa^n$ is to use the basis of delta functions $\{\delta(t-t_0)\}_{0\leq t_0\leq 1}$, i.e. use the representation of the super-trace as a convolution
\begin{multline}\Str_{\Omega^0(\Delta^1)}\kappa^n=\\
=\int_0^1 dt_1 \int_0^1 dt_2 \cdots \int_0^1 dt_n <\delta(t-t_1)|\kappa|\delta(t-t_2)><\delta(t-t_2)|\kappa|\delta(t-t_3)>\cdots <\delta(t-t_n)|\kappa|\delta(t-t_1)>\label{interval contraction}
\end{multline}
where \be \kappa(t_1,t_2)=<\delta(t-t_1)|\kappa|\delta(t-t_2)>=\theta(t_1-t_2)-t_1 \label{interval lemma 2 eq4}\ee
is the kernel (\ref{interval K kernel}) of the operator $\kappa$. Let us introduce the generating function for super-traces of all powers of $\kappa$:
$$g(x)=\sum_{n=1}^\infty \frac{1}{n} x^n\;\Str_{\Omega^0(\Delta^1)}\kappa^n$$
and rewrite it using (\ref{interval contraction},\ref{interval lemma 2 eq4}) as
\begin{multline*}
g(x)
=\left(x\int_0^1 dt_1 \theta(t_1-t_1)+\frac{1}{2}x^2 \int_0^1 dt_1 \int_0^1 dt_2 \theta(t_1-t_2)\theta(t_2-t_1)\right.+\\
+\frac{1}{3}\left.x^3 \int_0^1 dt_1 \int_0^1 dt_2 \int_0^1 dt_3 \theta(t_1-t_2)\theta(t_2-t_3)\theta(t_3-t_1)+\cdots\right)+\\
+\sum_{k=1}^\infty \frac{(-1)^k}{k}\left(x\int_0^1 dt_1\, t_1+x^2\int_0^1 dt_1\int_0^1 dt_2 \theta(t_1-t_2) t_2\right.+\\
+\left.x^3 \int_0^1 dt_1 \int_0^1 dt_2 \int_0^1 dt_3 \theta(t_1-t_2)\theta(t_2-t_3)t_3+\cdots \right)^k
\end{multline*}
Notice that
$$\int_0^1 dt_1 \int_0^1 dt_2 \theta(t_1-t_2)\theta(t_2-t_1)=\int_0^1 dt_1 \int_0^1 dt_2 \int_0^1 dt_3 \theta(t_1-t_2)\theta(t_2-t_3)\theta(t_3-t_1)=\cdots=0$$
since the integrand is supported on a set of zero measure. Next,
$$\int_0^1 dt_1 \cdots\int_0^1 dt_i \theta(t_1-t_2)\cdots \theta(t_{i-1}-t_i)t_i=\int_{1\geq t_1\geq \cdots \geq t_i\geq 0}t_i=\frac{1}{(i+1)!} $$
Finally, the integral
$$\int_0^1 dt_1 \theta(t_1-t_1)$$
is ill-defined and requires a regularization: we have to specify the value of distribution $\theta(t)$ in point $t=0$. Let us choose the symmetric regularization: $\theta(0)=\frac{1}{2}(\theta(-0)+\theta(+0))=\frac{1}{2}$. Notice that this choice affects only the coefficient of $x^1$ in $g(x)$, i.e. only the value of super-trace $\Str_{\Omega^0(\Delta^1)}\kappa$. Now we can finish the computation of $g(x)$:
\begin{multline*}g(x)=\frac{1}{2}\;x+\sum_{k=1}^\infty \frac{(-1)^k}{k} \left(\sum_{i=1}^\infty \frac{1}{(i+1)!}\;x^i\right)^k=
\frac{1}{2}\;x+\sum_{k=1}^\infty \frac{(-1)^k}{k} \left(\frac{e^x-1-x}{x}\right)^k\\
=\frac{1}{2}\;x-\log\frac{e^x-1}{x}=-\log\frac{\sinh(x/2)}{x/2}=\int_0^x \frac{d\tilde{x}}{\tilde{x}}\left(1-\frac{\tilde{x}}{2}-\frac{\tilde{x}}{e^{\tilde{x}}-1}\right)
=-\sum_{n=2}^\infty \frac{B_n}{n\; n!}\;x^n
\end{multline*}
Again we came ti the result (\ref{interval lemma 2 eq1}).

\textit{Computation in ``momentum representation'' (in the basis of exponentials).} Let us use the fact that operator  $\kappa$ takes values in UV functions, i.e. ones that vanish in points $t=0,1$. Therefore
$$\Str_{\Omega^0(\Delta^1)}\kappa^n=\Str_{\Omega''^0(\Delta^1)}\kappa^n$$
Introduce the basis $\{e^{2\pi i m t}-1\}_{m\in\ZZ,m\neq 0}$ in $\Omega''^0(\Delta^1)$. It is extremely simple to compute super-traces of operators $\kappa^n$ in this basis, since it is the eigenbasis for $\kappa$:
$$\kappa:\;(e^{2\pi i m t}-1)\mapsto \frac{1}{2\pi i m}(e^{2\pi i m t}-1)$$
Therefore for $n\geq 1$
$$\Str_{\Omega''^0(\Delta^1)}\kappa^n=\sum_{m\in\ZZ,\, m\neq 0}\left(\frac{1}{2\pi i m}\right)^n=
\left\{\begin{array}{ll}\frac{2}{(2\pi i)^n}\zeta(n)&\text{ для чётного }n\\ 0 &\text{ для нечётного }n\end{array}\right.$$
where $\zeta(n)$ is the Riemann zeta function. Using Euler's formula values of zeta function in even integer points, we again obtain (\ref{interval lemma 2 eq1}).

Thus all three bases gave the same result for $\Str_{\Omega^0(\Delta^1)}\kappa^n$ for $n\geq 2$. Notice that the case $n=1$ is ambiguous: in coordinate representation and in the basis of exponentials we obtained $\Str_{\Omega^0(\Delta^1)}\kappa=0$, however we needed a regularization: in coordinate representation we had to specify the value of $\theta(0)$, in exponential basis we have to choose the order of summation for the conditionally convergent sum over eigenvalues $\sum_{m\in\ZZ,\, m\neq 0}\frac{1}{2\pi i m}$. In the monomial basis we even obtained the wrong value $\Str_{\Omega^0(\Delta^1)}\kappa=\frac{1}{2}$. Notice also that the computation in monomial basis, despite its elegance (only finitely many diagonal matrix elements of $\kappa^n$ are non-zero), is the least transparent: basis of monomials $\{t^m\}$ does not have a well-defined dual basis on the interval w.r.t. pairing $(f,g)=\int_0^1 f g\; dt $, while the basis of delta functions is orthogonal (self-dual) and the basis of exponential is orthogonal too.
\\$\Box$

\textbf{Proof of theorem \ref{interval thm}.} Let us compute the reduced action on the interval $\bar{S}_{\Delta^1}$ using the series (\ref{reduced action}). Notice that in the case of interval the majority of Feynman diagrams vanish. Namely, any Feynman diagram containing a vertex incident to three internal edges vanishes, since   $K_{\Delta^1}[K_{\Delta^1}\alpha,K_{\Delta^1}\beta]=0$ for any forms $\alpha,\beta\in\Omega^\bt(\Delta^1,\g)$ (since the operation $K_{\Delta^1}[K_{\Delta^1}\alpha,K_{\Delta^1}\beta]$ decreases the degree by 3). Also any tree containing a vertex, incident to the root and two internal edges vanishes, since $r_{\Delta^1}[K_{\Delta^1}\alpha,K_{\Delta^1}\beta]=0$ (since $K_{\Delta^1}\alpha$ and $K_{\Delta^1}\beta$ are functions on the interval, vanishing on the end-points, hence their commutator is again a UV function and is sent by the retraction $r_{\Delta^1}$ to zero). These observations imply that only Feynman trees of type $(*\cdots(*(*(**)))\cdots)$ (the ``branches'') and one-loop Feynman graphs of type $(*\cdots(*(*(*\bt)))\cdots)$ (``wheels'') contribute. Therefore series (\ref{reduced action}) reduces to the following:
\begin{multline*}\bar{S}_{\Delta^1}=\left<p_{01},(\omega^1-\omega^0)+\sum_{n=0}^\infty\int_0^1 [\chi_{01}\omega^{01},
(-K_{\Delta^1}[\chi_{01}\omega^{01},\bt])^n\circ(\chi_0\omega^0+\chi_1 \omega^1)] \right>_\g-\\
-\hbar \sum_{n=1}^\infty \frac{1}{n}\Str_{\g\otimes\Omega^0(\Delta^1)}(-K_{\Delta^1}[\chi_{01}\omega^{01},\bt])^n\\
=\left<p_{01},(\omega^1-\omega^0)+
\sum_{n=0}^\infty \left(\int_0^1 \chi_{01}\wedge (-K_{\Delta^1}(\chi_{01}\wedge\bt))^n\circ\chi_0\right)\left((\ad_{\omega^{01}})^n\circ \omega^0\right)\right.+\\
+\left.\sum_{n=0}^\infty \left(\int_0^1 \chi_{01}\wedge (-K_{\Delta^1}(\chi_{01}\wedge\bt))^n\circ\chi_1\right)\left((\ad_{\omega^{01}})^n\circ \omega^1\right) \right>_\g-\\
-\hbar\sum_{n=1}^\infty\frac{1}{n}\left(\Str_{\Omega^0(\Delta^1)}(-K_{\Delta^1}(\chi_{01}\wedge\bt))^n\right)\cdot(\tr_g\, (\ad_{\omega^{01}})^n)
\end{multline*}
where we separated the de Rham part of values of Feynman diagrams from the trivial expressions in  $\g$-coefficients. Finally, using (\ref{interval lemma 1 eq2},\ref{interval lemma 2 eq1}) and that $\int_0^1\chi_{01}\wedge \chi_0=\int_0^1\chi_{01}\wedge \chi_1=\frac{1}{2}$, we come to expressions (\ref{interval thm eq1},\ref{interval thm eq2}). Notice that we did not need the value of the ill-defined super-trace $\Str_{\Omega^0(\Delta^1)}K_{\Delta^1}(\chi_{01}\wedge\bt)$, since in the expression for $\bar{S}_{\Delta^1}$ it comes with the vanishing (due to unimodularity of $\g$) factor $\tr_\g\ad(\omega^{01}) =\tr_\g[\omega^{01},\bt]$.
\\$\Box$

\subsubsection{Checking QME for $S_{\Delta^1}$ explicitly}
\label{section: interval QME check}
The fact that simplicial action  $S_{\Delta^1}$ for the interval (\ref{interval thm eq3}) satisfies QME follows from its construction via BV integral (Statement \ref{statement: QME for induced action}). However we can check QME for $S_{\Delta^1}$ explicitly. This check is an important evidence for the self-consistency of the whole construction and correctness of computation of $S_{\Delta^1}$ (in particular, of the computation of the super-traces (\ref{interval lemma 2 eq1})).

First let us check CME $\{S_{\Delta^1}^0,S_{\Delta^1}^0\}=0$. Introduce the notation
$$\tilde{B}_n=\left\{\begin{array}{ll}B_n&\text{for even  }\;n \\ 0&\text{for odd  }\;n \end{array}\right.$$
Write the tree part of the simplicial action on the interval (\ref{interval thm eq3}) as
$$S^0_{\Delta^1}=\left<p_0,\frac{1}{2}[\omega^0,\omega^0]\right>_\g+\left<p_1,\frac{1}{2}[\omega^1,\omega^1]\right>_\g+
\left<p_{01},\frac{1}{2}[\omega^{01},\omega^0+\omega^1]+\sum_{n=0}^\infty\frac{\tilde{B}_n}{n!}(\ad_{\omega^{01}})^n\circ (\omega^1-\omega^0)\right>_\g$$
Compute the anti-bracket $\{S_{\Delta^1}^0,S_{\Delta^1}^0\}$:
\begin{multline}
\frac{1}{2}\{S_{\Delta^1}^0,S_{\Delta^1}^0\}=S_{\Delta^1}^0
\left(\left<\frac{\ola\dd}{\dd\omega^0},\frac{\ora\dd}{\dd p_0}\right>_\g+
\left<\frac{\ola\dd}{\dd\omega^1},\frac{\ora\dd}{\dd p_1}\right>_\g+
\left<\frac{\ola\dd}{\dd\omega^{01}},\frac{\ora\dd}{\dd p_{01}}\right>_\g \right)S_{\Delta^1}^0\\
=<p_0,\frac{1}{2}[\omega^0,[\omega^0,\omega^0]]>_\g+<p_1,\frac{1}{2}[\omega^1,[\omega^1,\omega^1]]>_\g+\\
+
<p_{01},\frac{1}{4}[\omega^{01},[\omega^0,\omega^0]]+\frac{1}{4}[\omega^{01},[\omega^1,\omega^1]]-
\frac{1}{4}[[\omega^{01},\omega^0+\omega^1],\omega^0+\omega^1]>_\g+\\
+<p_{01},\frac{1}{2}\sum_{n=0}^\infty\frac{\tilde{B}_n}{n!}(\ad_{\omega^{01}})^n\circ([\omega^1,\omega^1]-[\omega^0,\omega^0])-
\frac{1}{2}[\sum_{n=0}^\infty\frac{\tilde{B}_n}{n!}(\ad_{\omega^{01}})^n\circ (\omega^1-\omega^0),\omega^0+\omega^1]>_\g+\\
+\left<\frac{1}{2}[\omega^{01},\omega^0+\omega^1]+\sum_{n=0}^\infty\frac{\tilde{B}_n}{n!}(\ad_{\omega^{01}})^n\circ (\omega^1-\omega^0),\frac{\dd}{\dd\omega^{01}}\right>_\g\circ
\left<p_{01},\sum_{n=0}^\infty\frac{\tilde{B}_n}{n!}(\ad_{\omega^{01}})^n\circ (\omega^1-\omega^0)\right>_\g
\label{interval CME eq3}
\end{multline}
First two terms vanish due to Jacobi identity. The remainder is a sum of expressions of type
\be (\ad_x)^a[(\ad_x)^b\circ y,(\ad_x)^c\circ z] \label{interval CME eq1}\ee
with $x,y,z\in\g$, $a,b,c\geq 0$. This expressions are not independent for different $(a,b,c)$: there are relations between them due to Jacobi identity. In particular,
\be (\ad_x)^a[(\ad_x)^b\circ y,(\ad_x)^c\circ z]=\sum_{k=0}^a \binom{a}{k} \;[(\ad_x)^{b+k}\circ y,(\ad_x)^{a+c-k}\circ z]\label{interval CME eq2}\ee
where $\binom{a}{k}$ are binomial coefficients. We can use this to transform expressions of type (\ref{interval CME eq1}) to the  ``canonical form'' --- sums of expressions of type $[(\ad_x)^a\circ y,(\ad_x)^b\circ z]$ (they are not anymore related by Jacobi identity). Using (\ref{interval CME eq2}) we can write
\begin{multline*}\left<\sum_{i=0}^\infty f_i (\ad_x)^i\circ y,\frac{\dd}{\dd x}\right>_\g\circ \left(\sum_{j=1}^\infty g_j (\ad_x)^j\circ z\right)
=\sum_{i=0}^\infty \sum_{j=1}^\infty \sum_{k=0}^{j-1} f_i g_j (\ad_x)^{k}[(\ad_x)^i\circ y,(\ad_x)^{j-k-1}\circ z]\\
=\sum_{i=0}^\infty \sum_{j=1}^\infty \sum_{k=0}^{j-1} \sum_{l=0}^{k} \binom{k}{l} f_i g_j [(\ad_x)^{i+l}\circ y,(\ad_x)^{j-1-l}\circ z]\\
=\sum_{I=0}^\infty \sum_{J=0}^\infty \left(\sum_{r=0}^I \binom{J+r+1}{r+1} f_{I-r}g_{J+r+1}\right)  [(\ad_x)^I\circ y,(\ad_x)^J\circ z]
\end{multline*}
Now we can continue the computation (\ref{interval CME eq3}):
\begin{multline}
\frac{1}{2}\{S_{\Delta^1}^0,S_{\Delta^1}^0\}=\left<p_{01},\frac{1}{4}[[\omega^{01},\omega^1-\omega^0],\omega^1-\omega^0]\right.+\\
+\sum_{I,J=0}^\infty \frac{1}{2}\frac{\tilde{B}_{I+J}}{(I+J)!}\binom{I+J}{I} ([(\ad_{\omega^{01}})^I\circ\omega^1,(\ad_{\omega^{01}})^J\circ\omega^1]-
[(\ad_{\omega^{01}})^I\circ\omega^0,(\ad_{\omega^{01}})^J\circ\omega^0])-\\
-\sum_{I=0}^\infty \frac{1}{2}\frac{\tilde{B}_I}{I!}[(\ad_{\omega^{01}})^I\circ (\omega^1-\omega^0),\omega^0+\omega^1]-\\
-\sum_{I=1}^\infty\sum_{J=0}^\infty\frac{1}{2} \frac{\tilde{B}_{I+J}}{(I+J)!}\binom{I+J}{I}[(\ad_{\omega^{01}})^I\circ (\omega^0+\omega^1),
(\ad_{\omega^{01}})^J\circ (\omega^1-\omega^0)]-\\
-\left.\sum_{I,J=0}^\infty\left(\sum_{r=0}^I\frac{\tilde{B}_{I-r}}{(I-r)!}\;\frac{\tilde{B}_{J+r+1}}{(J+r+1)!}\;\binom{J+r+1}{r+1}\right)
[(\ad_{\omega^{01}})^I\circ (\omega^1-\omega^0),(\ad_{\omega^{01}})^J\circ (\omega^1-\omega^0)] \right>_\g
\label{interval CME eq4}
\end{multline}
Third and fourth terms together yield
\begin{multline*}-\sum_{I=0}^\infty \frac{1}{2}\frac{\tilde{B}_I}{I!}[(\ad_{\omega^{01}})^I\circ (\omega^1-\omega^0),\omega^0+\omega^1]-\\
-\sum_{I=1}^\infty\sum_{J=0}^\infty\frac{1}{2} \frac{\tilde{B}_{I+J}}{(I+J)!}\binom{I+J}{I}[(\ad_{\omega^{01}})^I\circ (\omega^0+\omega^1),
(\ad_{\omega^{01}})^J\circ (\omega^1-\omega^0)]\\
=-\sum_{I,J=0}^\infty\frac{1}{2} \frac{\tilde{B}_{I+J}}{(I+J)!}\binom{I+J}{I}[(\ad_{\omega^{01}})^I\circ (\omega^0+\omega^1),
(\ad_{\omega^{01}})^J\circ (\omega^1-\omega^0)]\\=
\sum_{I,J=0}^\infty\frac{1}{2} \frac{\tilde{B}_{I+J}}{(I+J)!}\binom{I+J}{I}([(\ad_{\omega^{01}})^I\circ \omega^0,
(\ad_{\omega^{01}})^J\circ \omega^0]-[(\ad_{\omega^{01}})^I\circ \omega^1,(\ad_{\omega^{01}})^J\circ \omega^1])
\end{multline*}
i.e. they cancel the second term in (\ref{interval CME eq4}). Therefore
\begin{multline}
\frac{1}{2}\{S_{\Delta^1}^0,S_{\Delta^1}^0\}=\left<p_{01},\frac{1}{4}[[\omega^{01},\omega^1-\omega^0],\omega^1-\omega^0]-\right.\\
-\left.
\sum_{I,J=0}^\infty\left(\sum_{r=0}^I\frac{\tilde{B}_{I-r}}{(I-r)!}\;\frac{\tilde{B}_{J+r+1}}{(J+r+1)!}\;\binom{J+r+1}{r+1}\right)
[(\ad_{\omega^{01}})^I\circ (\omega^1-\omega^0),(\ad_{\omega^{01}})^J\circ (\omega^1-\omega^0)] \right>_\g\\
=<p_{01},\sum_{I,J=0}^\infty\A_{IJ}[(\ad_{\omega^{01}})^I\circ (\omega^1-\omega^0),(\ad_{\omega^{01}})^J\circ (\omega^1-\omega^0)]>_\g
\label{interval CME eq5}
\end{multline}
where coefficients are
\begin{multline}\label{interval CME Bernoulli relation}
\A_{IJ}:
=\frac{1}{8}(\delta_{I,1}\delta_{J,0}+\delta_{I,0}\delta_{J,1})-\\
-\frac{1}{2}
\left(\sum_{r=0}^I\frac{\tilde{B}_{I-r}}{(I-r)!}\;\frac{\tilde{B}_{J+r+1}}{(J+r+1)!}\;\binom{J+r+1}{r+1}+
\sum_{s=0}^J\frac{\tilde{B}_{J-s}}{(J-s)!}\;\frac{\tilde{B}_{I+s+1}}{(I+s+1)!}\;\binom{I+s+1}{s+1}\right)=0
\end{multline}
--- a two-parametric family of quadratic relations for Bernoulli numbers \cite{AD}.

Let us now check the quantum part of QME $S_{\Delta^1}$:
\begin{multline}
\label{interval QME eq1}
\Delta S^0_{\Delta^1}+\{S^0_{\Delta^1},S^1_{\Delta^1}\}\\
=\left(-\left<\frac{\dd}{\dd\omega^0},\frac{\dd}{\dd p_0}\right>_\g-
\left<\frac{\dd}{\dd\omega^1},\frac{\dd}{\dd p_1}\right>_\g+\left<\frac{\dd}{\dd\omega^{01}},\frac{\dd}{\dd p_{01}}\right>_\g\right)S^0_{\Delta^1}+
S^0_{\Delta^1}\left<-\frac{\ola\dd}{\dd p_{01}},\frac{\ora\dd}{\dd\omega^{01}}\right>_\g S^1_{\Delta^1}\\
=\left(\tr_g\ad_{\omega^0}+\tr_g\ad_{\omega^1}-\frac{1}{2}\tr_g\ad_{\omega^0+\omega^1}-\sum_{n=2}^\infty\frac{\tilde{B}_n}{n!}
\sum_{k=0}^{n-1}\tr_g\; (\ad_{\omega^{01}})^k[(\ad_{\omega^{01}})^{n-k-1}\circ (\omega^1-\omega^0),\bt]\right)+\\
+\left<\frac{1}{2}[\omega^{01},\omega^0+\omega^1]+\sum_{n=0}^\infty \frac{\tilde{B}_n}{n!}(\ad_{\omega^{01}})^n\circ(\omega^1-\omega^0),\frac{\dd}{\dd\omega^{01}}\right>_\g\circ
\left(\sum_{m=2}^\infty\frac{\tilde{B}_m}{m\cdot m!}\tr_\g(\ad_{\omega^{01}})^m\right)\\
=\tr_g\ad_{\omega^0}+\tr_g\ad_{\omega^1}-\frac{1}{2}\tr_g\ad_{\omega^0+\omega^1}-\sum_{n=2}^\infty\frac{\tilde{B}_n}{n!}
\sum_{k=0}^{n-1}\tr_g\; [(\ad_{\omega^{01}})^{n-k-1}\circ (\omega^1-\omega^0),(\ad_{\omega^{01}})^k\bt]+\\
+\sum_{m=2}^\infty \frac{1}{2}\frac{\tilde{B}_m}{m!}\tr_g\; [[\omega^{01},\omega^1-\omega^0],(\ad_{\omega^{01}})^{m-1}\bt]+
\sum_{n=0}^\infty\sum_{m=2}^\infty\frac{\tilde{B}_n}{n!}\frac{\tilde{B}_m}{m!}
\tr_g\;[(\ad_{\omega^{01}})^n\circ(\omega^1-\omega^0),(\ad_{\omega^{01}})^{m-1}\bt]
\end{multline}
Notice that the first three terms vanish due to unimodularity of $\g$. To simplify the rest of the expression, it is convenient to use the following relation:
$$\tr_\g \;[[x,y],(\ad_x)^a\bt]=\tr_g\; \ad_x [y,(\ad_x)^a\bt]-\tr_g\;[y,(\ad_x)^{a+1}\bt]=0$$
for any $x,y\in\g$ and $a\geq 0$, where we use Jacobi identity and cyclic property of the trace. This implies
$$\tr_g\;[(\ad_x)^a\circ y,(\ad_x)^b\bt]=0$$
for $a\geq 1$, $b\geq 0$. Therefore to the fourth term in (\ref{interval QME eq1}) only terms with $k=n-1$ contribute, and to the sixth --- only terms with $n=0$, and fifth term vanishes. Therefore
\be \Delta S^0_{\Delta^1}+\{S^0_{\Delta^1},S^1_{\Delta^1}\}=
-\sum_{n=2}^\infty\frac{\tilde{B}_n}{n!}\tr_g\; [\omega^1-\omega^0,(\ad_{\omega^{01}})^{n-1}\bt]+
\sum_{m=2}^\infty\frac{\tilde{B}_m}{m!}
\tr_g\;[\omega^1-\omega^0,(\ad_{\omega^{01}})^{m-1}\bt]=0 \label{interval QME eq2}\ee

Thus we checked QME for the action (\ref{interval thm eq3}) explicitly.

\textbf{Remark (indirect recovering of one-loop result via QME).} Notice that computation (\ref{interval QME eq1},\ref{interval QME eq2}) shows that the one-loop part of simplicial action on the interval can be recovered completely from the tree part, i.e. the values of super-traces (\ref{interval lemma 2 eq1}) can be indirectly calculated using QME from the tree part of action (i.e. from values of integrals (\ref{interval lemma 1 eq2})), and the result coincides with the one we obtained explicitly, using different bases in the space of differential forms on the interval. Checking QME for the simplicial action is a perfectly unambiguous finite-dimensional computation, unlike the explicit computation of super-traces over  $\Omega^0(\Delta^1)$, and therefore is in a sense a more rigorous proof of the result (\ref{interval lemma 2 eq1}).

On the other hand , the CME for $S^0_{\Delta^1}$, as our computation shows, is equivalent to a nontrivial identity for Bernoulli numbers (\ref{interval CME Bernoulli relation}).
Notice also that the computation (\ref{interval CME eq3},\ref{interval CME eq4}) explicitly shows that the reduced action itself for the interval does not satisfy CME: the contribution of end-points of the interval plays an essential role in the cancellation.

\subsubsection{Induced $qL_\infty$ structure on $C^\bt(\Delta^1,\g)$}
\label{section: interval qL_infty structure}
 It follows from the formula (\ref{interval thm eq3}) for $S_{\Delta^1}$ that the $qL_\infty$ structure on cochains of the standard triangulation of the interval $C^\bt(\Delta^1,\g)=\g e_0\oplus \g e_1\oplus \g e_{01}$ is given by the cohomological vector field
\begin{multline} Q_{\Delta^1}=-\left<\frac{1}{2}[\omega^0,\omega^0],\frac{\dd}{\dd\omega^0}\right>_\g-
\left<\frac{1}{2}[\omega^1,\omega^1],\frac{\dd}{\dd\omega^1}\right>_\g+\\
+\left<\frac{1}{2}[\omega^{01},\omega^0+\omega^1]+\left(\frac{\ad_{\omega^{01}}}{2}\coth \frac{\ad_{\omega^{01}}}{2}\right)\circ (\omega^1-\omega^0),\frac{\dd}{\dd\omega^{01}}\right>_\g \label{interval Q}
\end{multline}
on the space of cochains with shifted grading $C^\bt(\Delta^1,\g)[1]\cong\g[1]\oplus\g[1]\oplus\g$, and by the density
$$\rho_{\Delta^1}=\exp\left( \tr_\g\log\frac{\sinh\frac{\ad_{\omega^{01}}}{2}}{\frac{\ad_{\omega^{01}}}{2}}\right)=
\mr{det}_\g \left(\frac{\sinh\frac{\ad_{\omega^{01}}}{2}}{\frac{\ad_{\omega^{01}}}{2}}\right)$$
of $Q_{\Delta^1}$-invariant measure
$$\mu_{\Delta^1}=\rho_{\Delta^1}\DD\omega^0 \DD\omega^1 \DD\omega^{01}=\prod_a \DD\omega^{0a}\cdot\prod_b \DD\omega^{0b}\cdot\mr{det}_\g \left(\frac{\sinh\frac{\ad_{\omega^{01}}}{2}}{\frac{\ad_{\omega^{01}}}{2}}\right)\prod_ c\DD\omega^{01c}$$
where indices $a,b,c$ run over the basis of $\g$. Thus the measure $\mu_{\Delta^1}$ is the product of coordinate Berezin measure on $\g[1]\oplus\g[1]$ (i.e. on the cochains concentrated on the end-points of the interval) and the invariant measure on Lie algebra
$\g$ (i.e. on cochains concentrated on the bulk of the interval):
$$\mr{det}_\g \left(\frac{\sinh\frac{\ad_{\omega^{01}}}{2}}{\frac{\ad_{\omega^{01}}}{2}}\right)\prod_ c\DD\omega^{01c}=\exp^*\mu_G$$
--- the pull-back of Haar measure on Lie group $G$ by the exponential map $\exp: \g\ra G$, see \cite{Kirillov}.

The classical and quantum operations $\{l_{(n)}\}$, $\{q_{(n)}\}$ are written in terms of the super-field $\omega=e_0\omega^0+e_1\omega^1+e_{01}\omega^{01}$
as
\begin{eqnarray*}
l_{(2)}(\omega,\omega)&=&e_0[\omega^0,\omega^0]+e_1[\omega^1,\omega^1]+e_{01}[\omega^{01},\omega^0+\omega^1]\\
l_{(n)}(\underbrace{\omega,\ldots,\omega}_n)&=& n B_{n-1}\; e_{01}\left((\ad_{\omega^{01}})^n\circ(\omega^1-\omega^0)\right),\;\text{ для }n\neq 2\\
q_{(n)}(\underbrace{\omega,\cdots,\omega}_n)&=&\frac{B_n}{n}\tr_g(\ad_{\omega^{01}})^n
\end{eqnarray*}
In particular, operations $l_{(4)}=l_{(6)}=l_{(8)}=\cdots=0$ and $q_{(3)}=q_{(5)}=q_{(7)}=\cdots=0$ vanish, since the respective coefficients (the Bernoulli numbers) vanish. Also  $q_{(1)}=0$ due to unimodularity of $\g$. Equivalently one can say that polylinear super-antisymmetric operations
$l_{(n)}:\Lambda^n C^\bt(\Delta^1,\g)\ra C^\bt(\Delta^1,\g)$ and $q_{(n)}:\Lambda^n C^\bt(\Delta^1,\g)\ra\RR$ act on $\g$-valued cochains $\alpha=e_0\alpha^0+e_1\alpha^1+e_{01}\alpha^{01}\in C^\bt(\Delta^1,\g)$ as
\begin{eqnarray*}
l_{(2)}(\alpha_1,\alpha_2)&=&e_0[\alpha_1^0,\alpha_2^0]+e_1 [\alpha_1^1,\alpha_2^1]+e_{01}\left(\frac{1}{2}[\alpha_1^{01},\alpha_2^0+\alpha_2^1]+
\frac{1}{2}[\alpha_1^0+\alpha_1^1,\alpha_2^{01}]\right),\\
l_{(n)}(\alpha_1,\ldots,\alpha_n)&=&\frac{B_{n-1}}{(n-1)!}\;e_{01}\sum_{k=1}^n (-1)^{n-k}\cdot\\
&&\quad\cdot\sum_{\left(\pi:(1\cdots\wh{k}\cdots n)\mapsto (1\cdots\wh{k}\cdots n)\right)\in S_{n-1}}\ad_{\alpha^{01}_{\pi(1)}}\cdots\wh{\ad_{\alpha^{01}_{\pi(k)}}}\cdots \ad_{\alpha^{01}_{\pi(n)}}\circ (\alpha^1_k-\alpha^0_k)\\ &&\text{ for }n\neq 2,\\
q_{(n)}(\alpha_1,\ldots,\alpha_n)&=&\frac{B_n}{n\cdot n!}\sum_{\left(\pi:(1\cdots n)\mapsto (1\cdots n)\right)\in S_n}
\tr_g\left(\ad_{\alpha^{01}_{\pi(1)}}\cdots \ad_{\alpha^{01}_{\pi(n)}}\right)
\end{eqnarray*}
where the lower index for $\alpha$ is the number of the cochain (the upper index, as usual, is the simplex of the triangulation), sums over $\pi$ are sums over permutations and $S_n$ is the symmetric group.

Another way to formulate the result is as follows. In terms of basis $e_{\sigma a}= T_a e_\sigma$ on $C^\bt(\Delta^1,\g)$ (where $T_a$ is the basis in $\g$) we represent cochains as $\alpha=\sum_a (e_{0a}\alpha^{0a}+e_{1a}\alpha^{1a}+e_{01a}\alpha^{01a})\in C^\bt(\Delta^1,\g)$, where $\alpha^{0a},\alpha^{1a},\alpha^{01a}\in\RR$, and we can write the operations in terms of structure constants:
\begin{eqnarray*}l_{(n)}(\alpha_1,\ldots,\alpha_n)&=&\sum_{\sigma,\sigma_1,\ldots,\sigma_n\in\{[0],[1],[01]\}}\sum_{a,a_1,\ldots,a_n}
e_{\sigma a} l^{\sigma a}_{(n)\sigma_1 a_1,\ldots,\sigma_n a_n}\alpha_1^{\sigma_1 a_1}\cdots \alpha_n^{\sigma_n a_n}\\
q_{(n)}(\alpha_1,\ldots,\alpha_n)&=&\sum_{\sigma_1,\ldots,\sigma_n\in\{[0],[1],[01]\}}\sum_{a_1,\ldots,a_n}
q_{(n)\sigma_1 a_1,\ldots,\sigma_n a_n}\alpha_1^{\sigma_1 a_1}\cdots \alpha_n^{\sigma_n a_n}
\end{eqnarray*}
where the structure constants are
\begin{multline*}
l^{01a}_{(1)1b}=-l^{01a}_{(1)0b}=\delta^a_b,\\
l^{0a}_{(2) 0b,0c}=l^{1a}_{(2) 1b,1c}=f^a_{bc},\;l^{01a}_{(2) 01b,0c}=l^{01a}_{(2) 01b,1c}=l^{01a}_{(2) 0b,01c}=l^{01a}_{(2) 1b,01c}=\frac{1}{2}f^a_{bc},\\
l^{01a}_{(n)01 a_1,\ldots,01 a_{k-1},1 a_k,01 a_{k+1},\ldots, 01 a_n}=-l^{01a}_{(n)01 a_1,\ldots,01 a_{k-1},1 a_k,01 a_{k+1},\ldots, 01 a_n}
=(-1)^{n-k}\frac{B_{n-1}}{(n-1)!}\;\cdot\\
\cdot\sum_{\left(\pi:(1\cdots\wh{k}\cdots n)\mapsto (1\cdots\wh{k}\cdots n)\right)\in S_{n-1}}
\sum_{b_1,\ldots,\wh{b_k},\ldots,b_{n-1}}
f^a_{a_{\pi(1)}b_1}f^{b_1}_{a_{\pi(2)}b_2}\cdots f^{b_{k-2}}_{a_{\pi(k-1)}b_{k-1}}f^{b_{k-1}}_{a_{\pi(k+1)}b_{k+1}}\cdots f^{b_{n-1}}_{a_{\pi(n)}a_k}\\
\text{ для }n\geq 2 \text{ и } 1\leq k\leq n,\\
q_{(n)01a_1,\ldots, 01a_n}=\frac{B_n}{n\cdot n!}\sum_{\left(\pi:(1\cdots n)\mapsto (1\cdots n)\right)\in S_n}
\sum_{b_1,\ldots,b_n}f^{b_n}_{a_{\pi(1)}b_1}f^{b_1}_{a_{\pi(2)}b_2}\cdots f^{b_{n-1}}_{a_{\pi(n)}b_n}
\end{multline*}
All the rest structure constants vanish. We used the notation $f^a_{bc}=<T^a,[T_b,T_c]>_\g$ for structure constants of Lie algebra $\g$.

From the general construction (Statement \ref{statement: L_infty morph for induction}), we know that $L_\infty$ structure (\ref{interval Q}) on the space $C^\bt(\Delta^1,\g)$ is homotopic to de Rham algebra of the interval (with coefficients in $\g$) $\Omega^\bt(\Delta^1,\g)$. Moreover it is easy to compute the $L_\infty$ quasi-isomorphism (\ref{L_infty morph pert series}) $U_{\Delta^1}:C^\bt(\Delta^1,\g)[1]\ra \Omega^\bt(\Delta^1,\g)[1]$.
\begin{statement}
The $L_\infty$ quasi-isomorphism $U_{\Delta^1}:C^\bt(\Delta^1,\g)[1]\ra \Omega^\bt(\Delta^1,\g)[1]$ between the  $L_\infty$ algebra $(C^\bt(\Delta^1,\g),Q_{\Delta^1})$ and the DGLA of $\g$-valued differential forms on the interval $(\Omega^\bt(\Delta^1,\g),d,[\bt,\bt])$ is
\begin{eqnarray}U_{\Delta^1}(e_0\omega^0+e_1\omega^1+e_{01}\omega^{01})&=&
\omega^0 t+\omega^1 (1-t)+\omega^{01}dt+\\
&&\qquad+\sum_{n=1}^\infty (-1)^n\frac{B_{n+1}(t)-B_{n+1}}{(n+1)!}(\ad_{\omega^{01}})^n\circ (\omega^1-\omega^0)\nonumber\\
&=&\omega^0+\left(\frac{1-e^{-t\,\ad_{\omega^{01}}}}{1-e^{-\ad_{\omega^{01}}}}\right)\circ(\omega^1-\omega^0)+\omega^{01}dt \label{interval U}
\end{eqnarray}
\end{statement}
\textbf{Proof.} The fact that only trees of type  $(*(*\cdots(*(**))\cdots))$ contribute to (\ref{L_infty morph pert series}) follows from the argument from the proof of Theorem \ref{interval thm}. Therefore the series (\ref{L_infty morph pert series}) is
$$U_{\Delta^1}(e_0\omega^0+e_1\omega^1+e_{01}\omega^{01})=\chi_0\omega^0+\chi_1\omega^1+\chi_{01}\omega^{01}+\sum_{n=1}^\infty
(-K[\chi_{01}\omega^{01},\bt])^n\circ (\chi_0 \omega^0+\chi_1 \omega^1)$$
Using the result (\ref{interval lemma 1 eq1}), we immediately obtain (\ref{interval U}). Notice also that we can write (\ref{interval U}) in a more symmetrical form:
$$U_{\Delta^1}(e_0\omega^0+e_1\omega^1+e_{01}\omega^{01})=\left(\frac{1-e^{t_0\,\ad_{\omega^{01}}}}{1-e^{\ad_{\omega^{01}}}}\right)\circ\omega^0
+\left(\frac{1-e^{-t_1\,\ad_{\omega^{01}}}}{1-e^{-\ad_{\omega^{01}}}}\right)\circ\omega^1+\omega^{01} dt$$
where  $t_1=t, t_0=1-t$ are the barycentric coordinates on the interval.
\\$\Box$

\subsubsection{Examples for constructions of section \ref{section: gluing}: gluing two intervals into one, gluing interval into circle, tearing end-point off the interval}
\label{section: gluing, bc examples}
Now, having the result (\ref{interval thm eq3}) at our disposal, we can write explicit results for some simplest examples for constructions of section \ref{section: gluing}.

\textbf{Gluing two intervals into one.} Consider gluing two intervals into one by gluing right end-point of one interval with left end-point of the other. Let $\Xi_1=\{[0],[1'],[01']\}$ и $\Xi_2=\{[1''],[2],[1''2]\}$  be the standard triangulations for two intervals $[01']$ and $[1''2]$, and let $F=\{[A]\}$ be the simplicial complex, consisting of a single point. Embeddings
$$F\hra\Xi_1:\;[A]\mapsto [1']$$
of $F$ into $\Xi_1$ as the right end-point and
$$F\hra\Xi_2:\;[A]\mapsto [1'']$$
of $F$ into $\Xi_2$ as the left end-point induce the embeddings $\iota_{1,2}$ and projections $\pi_{1,2}$ for cochain complexes $V_1=C^\bt(\Xi_1,\g)=\g e_0\oplus\g e_{1'}\oplus \g e_{01'}$, $V_2=C^\bt(\Xi_2,\g)=\g e_{1''}\oplus \g e_2\oplus \g e_{1''2}$, $W=C^\bt(F,\g)=\g e_A$:
\begin{eqnarray*}\iota_1:W\ra V_1,\; x^A e_A\mapsto x^A e_{1'}\\
\iota_2:W\ra V_2,\;x^A e_A\mapsto x^A e_{1''}\\
\pi_1: V_1\ra W,\; x^0 e_0+x^{1'} e_{1'}+x^{01'}e_{01'}\mapsto x^{1'} e_{A}\\
\pi_2: V_2\ra W,\; x^{1''} e_{1''}+x^{2} e_{2}+x^{1''2}e_{1''2}\mapsto x^{1''} e_{A}
\end{eqnarray*}
The space of the glued $qL_\infty$ algebra is constructed as $V'=\ker\pi_-\subset V_1\oplus V_2$, hence here it is
$$V'=\g e_0\oplus \g e_{01'}\oplus \g e_2 \oplus \g e_{1''2}\oplus \g e_{1^+}$$
where $e_{1^+}=e_{1'}+e_{1''}$, and $V'$ is interpreted as the space of cochains $V'=C^\bt(\Xi',\g)$ on the glued simplicial complex $\Xi'=\{[0],[1^+],[2],[01'],[1''2]\}$, where the right end-point of the interval $[01']$ is identified with the 0-simplex $[1^+]$ and with the left end-point of the interval $[1''2]$. This identification does not affect $V'$ as a vector space, but is related to the differential (operation $l_{(1)}$) of the glued $qL_\infty$ structure on $V'$. For the glued  $qL_\infty$ structure on $V'$, using the known results for the point and for the interval, we obtain:
\begin{eqnarray*}Q_{\Xi'}=Q_{\Xi_1}+Q_{\Xi_2}-Q_F&=& -\left<\frac{1}{2}[\omega^0,\omega^0],\frac{\dd}{\dd \omega^0}\right>_\g
-\left<\frac{1}{2}[\omega^{1^+},\omega^{1^+}],\frac{\dd}{\dd \omega^{1^+}}\right>_\g-\left<\frac{1}{2}[\omega^2,\omega^2],\frac{\dd}{\dd \omega^2}\right>_\g+\\
&&+\left<\frac{1}{2}[\omega^{01'},\omega^0+\omega^{1^+}]+\left(\frac{\ad_{\omega^{01'}}}{2}
\coth\frac{\ad_{\omega^{01'}}}{2}\right)\circ (\omega^{1^+}-\omega^0),\frac{\dd}{\dd\omega^{01'}}\right>_\g+\\
&&+\left<\frac{1}{2}[\omega^{1''2},\omega^{1^+}+\omega^{2}]+\left(\frac{\ad_{\omega^{1''2}}}{2}
\coth\frac{\ad_{\omega^{1''2}}}{2}\right)\circ (\omega^{2}-\omega^{1^+}),\frac{\dd}{\dd\omega^{1''2}}\right>_\g\\
\rho_{\Xi'}=\frac{\rho_{\Xi_1}\rho_{\Xi_2}}{\rho_{F}}&=&\mr{det}_\g\left(\frac{\sinh\frac{\ad_{\omega^{01'}}}{2}}{\frac{\ad_{\omega^{01'}}}{2}}\right)
\cdot \mr{det}_\g\left(\frac{\sinh\frac{\ad_{\omega^{1''2}}}{2}}{\frac{\ad_{\omega^{1''2}}}{2}}\right)
\end{eqnarray*}
Super-fields for $\FF'=T^*[-1](V'[1])$ are $\omega'=e_0\omega^0+e_{1^+}\omega^{1^+}+e_2\omega^2+e_{01'}\omega^{01'}+e_{1''2}\omega^{1''2}$,
$p'=p_0 e^0+p_{1^+} e^{1^+}+p_2 e^2+p_{01'}e^{01'}+p_{1''2}e^{1''2}$, where $\omega^{1^+}=\frac{1}{2}(\omega^{1'}+\omega^{1''})$ and $p_{1^+}=p_{1'}+p_{1''}$. The glued action on $\FF'$ is
\begin{multline*}S_{\Xi'}=S_{\Xi_1}+S_{\Xi_2}-S_F=\left<p_0,\frac{1}{2}[\omega^0,\omega^0]\right>_\g
+\left<p_{1^+},\frac{1}{2}[\omega^{1^+},\omega^{1^+}]\right>_\g+\left<p_2,\frac{1}{2}[\omega^2,\omega^2]\right>_\g+\\
+\left<p_{01'},\frac{1}{2}[\omega^{01'},\omega^0+\omega^{1^+}]+\left(\frac{\ad_{\omega^{01'}}}{2}
\coth\frac{\ad_{\omega^{01'}}}{2}\right)\circ (\omega^{1^+}-\omega^0)\right>_\g+\\
+\left<p_{1''2},\frac{1}{2}[\omega^{1''2},\omega^{1^+}+\omega^{2}]+\left(\frac{\ad_{\omega^{1''2}}}{2}
\coth\frac{\ad_{\omega^{1''2}}}{2}\right)\circ (\omega^{2}-\omega^{1^+})\right>_\g+\\
+\hbar\left(\tr_\g\log\left(\frac{\sinh\frac{\ad_{\omega^{01'}}}{2}}{\frac{\ad_{\omega^{01'}}}{2}}\right)+
\tr_\g\log\left(\frac{\sinh\frac{\ad_{\omega^{1''2}}}{2}}{\frac{\ad_{\omega^{1''2}}}{2}}\right)\right)
\end{multline*}

\textbf{Circle glued from the interval.} For the gluing of the interval into the circle, we have $\Xi=\{[0],[1],[01]\}$ --- the standard triangulation of the interval, $F=\{[A]\}$ --- the ``triangulation'' of the point. Embeddings of
$F$ into $\Xi$ as left or right end-point
\begin{eqnarray*}
F\hra \Xi:&[A]\mapsto [0],\\
F\hra \Xi:&[A]\mapsto [1]
\end{eqnarray*}
induce embeddings and retractions for the cochain complexes $V=C^\bt(\Xi,\g)=\g e_0\oplus \g e_1\oplus \g e_{01}$, $W=C^\bt(F,\g)=\g e_A$:
\begin{eqnarray*}\iota_1:W\ra V,\; x^A e_A\mapsto x^A e_{0}\\
\iota_2:W\ra V,\;x^A e_A\mapsto x^A e_{1}\\
\pi_1: V\ra W,\; x^0 e_0+x^{1} e_{1}+x^{01}e_{01}\mapsto x^{0} e_{A}\\
\pi_2: V\ra W,\; x^0 e_0+x^{1} e_{1}+x^{01}e_{01}\mapsto x^{1} e_{A}
\end{eqnarray*}
The space of the glued $qL_\infty$ algebra is $V'=\ker\pi_-=\g e_+\oplus \g e_{01}\subset V$, where $e_+=e_0+e_1$. We identify $V'$ with the cochain complex $V'=C^\bt(\Xi',\g)$ for the triangulated circle $\Xi'=\{[+],[01]\}$ (here $[+]$ is the label of the 0-simplex, glued from the end-points $[0]$ and $[1]$ of the former interval). The glued $qL_\infty$ structure on $V'$ is
\begin{eqnarray*}
Q_{\Xi'}=Q_\Xi-Q_F&=&-\left<\frac{1}{2}[\omega^+,\omega^+],\frac{\dd}{\dd \omega^+}\right>_\g
+\left<[\omega^{01},\omega^{+}],\frac{\dd}{\dd\omega^{01}}\right>_\g\\
\rho_{\Xi'}=\frac{\rho_{\Xi}}{\rho_{F}}&=&\mr{det}_\g\left(\frac{\sinh\frac{\ad_{\omega^{01}}}{2}}{\frac{\ad_{\omega^{01}}}{2}}\right)
\end{eqnarray*}
Super-fields for $\FF'=T^*[-1](V'[1])$ are $\omega'=e_+\omega^+ +e_{01}\omega^{01}$, $p'=p_+ e^+ + p_{01} e^{01}$, where $\omega^+=\frac{1}{2}(\omega^0+\omega^1)$ и $p_+=p_0+p_1$. The glued action on $\FF'$ is
\be S_{\Xi'}=S_\Xi-S_F=\left<p_+,\frac{1}{2}[\omega^+,\omega^+]\right>_\g+\left<p_{01},[\omega^{01},\omega^{+}]\right>_\g
+\hbar\;\tr_\g\log\left(\frac{\sinh\frac{\ad_{\omega^{01}}}{2}}{\frac{\ad_{\omega^{01}}}{2}}\right) \label{circle}\ee
As we know from the arguments of section \ref{section: gluing-induction}, this action is the true effective $BF_\infty$ action, induced from the topological $BF$ theory on the circle with the induction data (\ref{circle ind data 1}--\ref{circle ind data 3}).

\textbf{Interval with an end-point torn off.} Finally, let us return to the example of tearing off the end-point of the interval from section \ref{section: gluing, boundary condition}. Let $\Xi=\{[0],[1],[01]\}$ be the triangulated interval and $F=\{[A]\}$ a point. Embedding $F\hra \Xi:\; [A]\mapsto [1]$ induces the retraction for cochain complexes $V=C^\bt(\Xi,\g)=\g e_0\oplus \g e_1\oplus \g e_{01}$, $W=C^\bt(F,\g)=\g e_A$:
$$\pi_1: V\ra W,\; x^0 e_0+x^{1} e_{1}+x^{01}e_{01}\mapsto x^{1} e_{A}$$
(we do not need the embedding $W\ra V$ for the construction of imposing the boundary condition). Then $V'=\ker\pi_1=\g e_0\oplus \g e_{01}\subset V$. We can understand $V'$ as the vector space of maps  $V'=C^\bt(\Xi',\g):=\g^{\Xi'}$ from the set of simplices $\Xi'=\{[0],[01]\}$ (this is not a simplicial complex, since it is not closed under the boundary operator) into the Lie algebra $\g$. Then $V'$ has the relict $qL_\infty$ structure
\begin{eqnarray*}Q_{\Xi'}=Q_\Xi|_{\Xi'}&=&-\left<\frac{1}{2}[\omega^0,\omega^0],\frac{\dd}{\dd \omega^0}\right>_\g+
\left<\frac{1}{2}[\omega^{01},\omega^0]-\left(\frac{\ad_{\omega^{01}}}{2}
\coth\frac{\ad_{\omega^{01}}}{2}\right)\circ \omega^0,\frac{\dd}{\dd\omega^{01}}\right>_\g\\
\rho_{\Xi'}=\rho_\Xi|_{\Xi'}&=&\mr{det}_\g\left(\frac{\sinh\frac{\ad_{\omega^{01}}}{2}}{\frac{\ad_{\omega^{01}}}{2}}\right)
\end{eqnarray*}
and the corresponding action is
\begin{multline*}S_{\Xi'}=S_\Xi-S_F=\left<p_0,\frac{1}{2}[\omega^0,\omega^0]\right>_\g+\left<p_{01},\frac{1}{2}[\omega^{01},\omega^0]-\left(\frac{\ad_{\omega^{01}}}{2}
\coth\frac{\ad_{\omega^{01}}}{2}\right)\circ \omega^0\right>_\g+\\
+\hbar\;\tr_\g\log\left(\frac{\sinh\frac{\ad_{\omega^{01}}}{2}}{\frac{\ad_{\omega^{01}}}{2}}\right)\end{multline*}

Notice also that if we tear both end-points off the interval, i.e. take $\Xi=\{[0],[1],[01]\}$, $F=\{[A],[B]\}$, where $[A]$ is embedded as $[0]$ and $[B]$ --- as $[1]$, we obtain $V'=\g e_{01}=C^\bt(\Xi',\g)$, where $\Xi'=\{[01]\}$ is just the bulk of the interval, and the $qL_\infty$ structure is
\begin{eqnarray*}Q_{\Xi'}&=&0\\
\rho_{\Xi'}&=&\mr{det}_\g\left(\frac{\sinh\frac{\ad_{\omega^{01}}}{2}}{\frac{\ad_{\omega^{01}}}{2}}\right)
\end{eqnarray*}
I.e. all classical operations vanish, but the quantum operations are nontrivial. The corresponding action consists of one-loop part only:
$$S_{\Xi'}=
\hbar\;\tr_\g\log\left(\frac{\sinh\frac{\ad_{\omega^{01}}}{2}}{\frac{\ad_{\omega^{01}}}{2}}\right)$$

\subsection{Perturbative results for the simplex of dimension $D\geq 2$}
\label{section: simplex pert}

We successfully solved the problem of computing the reduced simplicial action on the standard simplex $\bar{S}_{\Delta^D}$ for dimensions $D=0,1$ and obtained explicit results (\ref{Sbar Delta^0}) and (\ref{interval thm eq1},\ref{interval thm eq2}). Now we address the case of simplex $\Delta^D$ of higher dimension $D\geq 2$. Unfortunately, we cannot obtain the explicit result here and can only compute firs Feynman diagrams in expansion (\ref{reduced action}) for $\bar{S}_{\Delta^D}$.

\textbf{Splitting of values of Feynman diagrams into de Rham part and $\g$-part.} Introduce notation $\bar{S}_{\Delta^D,\Gamma}$ for the contribution of Feynman diagram $\Gamma\in\bf{T}_\mr{nonPl}\cup \bf{L}_\mr{nonPl}$ in the expansion
(\ref{reduced action}) for $\bar{S}_{\Delta^D}$, i.e.
\be\bar{S}_{\Delta^D}=\sum_{T\in\bf{T}_\mr{nonPl}}\bar{S}_{\Delta^D,T}+\hbar\sum_{L\in\bf{L}_\mr{nonPl}}\bar{S}_{\Delta^D,L} \label{Sbar Feynman diagrams}\ee
For convenience reason we introduce the tree with single leaf $(*)$, and its contribution to $\bar{S}_{\Delta^D}$ is the first term in (\ref{reduced action}):
$$\bar{S}_{\Delta^D,(*)}=\sum_{k=0}^D (-1)^k <p_{\Delta^D},\omega^{0\cdots\wh{k}\cdots D}>_\g$$
(this sum should be understood as a sum over faces of codimension 1, i.e. over $\sigma=[0\cdots\wh{k}\cdots D]\subset\Delta^D$).
Let us split values of Feynman diagrams $\bar{S}_{\Delta^D,\Gamma}$ into the de Rham part and the part in $\g$-coefficients
\begin{multline*}\bar{S}_{\Delta^D,T}=\frac{1}{|\Aut(T)|}\sum_{\sigma_1,\ldots,\sigma_{|T|}\subset\Delta^D}\left<p_{\Delta^D},\int_{\Delta^D}
\Iter_{T;-K_{\Delta^D}[\bt,\bt];[\bt,\bt]}(\chi_{\sigma_1}\omega^{\sigma_1},\ldots,\chi_{\sigma_{|T|}}\omega^{\sigma_{|T|}})\right>_\g\\
=\frac{1}{|\Aut(T)|}\sum_{\sigma_1,\ldots,\sigma_{|T|}\subset\Delta^D}\int_{\Delta^D}
\Iter_{T;-K_{\Delta^D}(\bt\wedge\bt);(\bt\wedge\bt)}(\chi_{\sigma_1},\ldots,\chi_{\sigma_{|T|}})\cdot\\
\cdot<p_{\Delta^D},\Iter_{T;[\bt,\bt];[\bt,\bt]}(\omega^{\sigma_1},\ldots,\omega^{\sigma_{|T|}})>_\g \e_T(|\sigma_1|,\ldots,|\sigma_{|T|}|)
\end{multline*}
where we mean that for every non-planar tree $T$ we choose some planar structure and use it to evaluate $\Iter_T$. Sign $\e_T(|\sigma_1|,\ldots,|\sigma_{|T|}|)=\pm 1$ is defined from
\begin{multline*}\Iter_{T;-K_{\Delta^D}[\bt,\bt];[\bt,\bt]}(\chi_{\sigma_1}\omega^{\sigma_1},\ldots,\chi_{\sigma_{|T|}}\omega^{\sigma_{|T|}})\\
=\Iter_{T;-K_{\Delta^D}(\bt\wedge\bt);(\bt\wedge\bt)}(\chi_{\sigma_1},\ldots,\chi_{\sigma_{|T|}})
\Iter_{T;[\bt,\bt];[\bt,\bt]}(\omega^{\sigma_1},\ldots,\omega^{\sigma_{|T|}})
\e_T(|\sigma_1|,\ldots,|\sigma_{|T|}|)
\end{multline*}
and arises from the permutation of variables $\omega^\sigma$ of parity $(1-|\sigma|)\mod 2$ with forms $\chi_\sigma$ of parity $|\sigma|\mod 2$  and odd operators $K_{\Delta^D}$. Obviously the sign $\e_T$ depends only on tree $T$ and dimensions of faces $\sigma_1,\ldots,\sigma_{|T|}$. In particular,
\begin{eqnarray*}\e_{(*)}(|\sigma_1|)&=&+1\\\e_{(**)}(|\sigma_1|,|\sigma_2|)&=&(-1)^{(|\sigma_1|+1)|\sigma_2|}\\
\e_{(*(**))}(|\sigma_1|,|\sigma_2|,|\sigma_3|)&=&(-1)^{(|\sigma_2|+1)|\sigma_3|+(|\sigma_1|+1)(|\sigma_2|+|\sigma_3|+1)}
\end{eqnarray*}
We split values of one-loop diagrams analogously:
\begin{multline*} \bar{S}_{\Delta^D,L}=-\frac{1}{|\Aut(L)|}\sum_{\sigma_1,\ldots,\sigma_{|L|}\subset\Delta^D}\Loop_{L;-K_{\Delta^D}[\bt,\bt];\Omega_0^\bt(\Delta^D,\g)}
(\chi_{\sigma_1}\omega^{\sigma_1},\ldots,\chi_{\sigma_{|L|}}\omega^{\sigma_{|L|}})\\
=-\frac{1}{|\Aut(L)|}\sum_{\sigma_1,\ldots,\sigma_{|L|}\subset\Delta^D}\Loop_{L;-K_{\Delta^D}(\bt\wedge\bt);\Omega_0^\bt(\Delta^D)}
(\chi_{\sigma_1},\ldots,\chi_{\sigma_{|L|}})\cdot\\
\cdot \Loop_{L;[\bt,\bt];\g}(\omega^{\sigma_1},\ldots,\omega^{\sigma_{|L|}})
\e_L(|\sigma_1|,\ldots,|\sigma_{|L|}|)
\end{multline*}
where we again fix some planar structure for every non-planar graph $L$ (and mark some edge in the cycle). Sign $\e_L=\pm 1$ is determined by
\begin{multline*}\Loop_{L;-K_{\Delta^D}[\bt,\bt];\Omega_0^\bt(\Delta^D,\g)}
(\chi_{\sigma_1}\omega^{\sigma_1},\ldots,\chi_{\sigma_{|L|}}\omega^{\sigma_{|L|}})\\
=\Loop_{L;-K_{\Delta^D}(\bt\wedge\bt);\Omega_0^\bt(\Delta^D)}
(\chi_{\sigma_1},\ldots,\chi_{\sigma_{|L|}})\cdot \Loop_{L;[\bt,\bt];\g}(\omega^{\sigma_1},\ldots,\omega^{\sigma_{|L|}})
\e_L(|\sigma_1|,\ldots,|\sigma_{|L|}|)
\end{multline*}
Similarly to $\e_T$ it depends only on the graph $L$  and on the dimensions of faces. For example,
\begin{eqnarray*}
\e_{(*\bt)}(|\sigma_1|)&=&+1\\
\e_{((**)\bt)}(|\sigma_1|,|\sigma_2|)&=&(-1)^{(|\sigma_1|+1)|\sigma_2|}\\
\e_{(*(*\bt))}(|\sigma_1|,|\sigma_2|)&=&(-1)^{(|\sigma_1|+1)(|\sigma_2|+1)}
\end{eqnarray*}
Notice that if the length of the cycle in one-loop diagram is $[L]=1$, then $\bar{S}_{\Delta^D,L}=0$, since $\Loop_{L;[\bt,\bt];\g}(\omega^{\sigma_1},\ldots,\omega^{\sigma_{|L|}})=\tr_\g[\cdots,\bt]=0$ due to unimodularity of $\g$. Therefore, only one-loop diagrams $L$ with cycle of length $[L]\geq 2$ contribute to (\ref{Sbar Feynman diagrams}), e.g. $L=(*(*\bt)),(*(*(*\bt))),(*((**)\bt)),\ldots$.

Let us introduce notations for de Rham parts of diagrams:
$$C_{\Delta^D,T}(\sigma_1,\ldots,\sigma_{|T|})=\int_{\Delta^D}
\Iter_{T;-K_{\Delta^D}(\bt\wedge\bt);(\bt\wedge\bt)}(\chi_{\sigma_1},\ldots,\chi_{\sigma_{|T|}})$$
for planar trees $T$ and
$$C_{\Delta^D,L}(\sigma_1,\ldots,\sigma_{|L|})=
\Loop_{L;-K_{\Delta^D}(\bt\wedge\bt);\Omega^\bt_0(\Delta^D)}(\chi_{\sigma_1},\ldots,\chi_{\sigma_{|L|}})$$
for planar one-loop graphs $L$. Thus $C_{\Delta^D,\Gamma}:\{\sigma\subset\Delta^D\}^{|\Gamma|}\ra \RR$ is a map from $|\Gamma|$-tuples of faces of the simplex $\Delta^D$ to numbers. For any Feynman graph, $C_{\Delta^D,\Gamma}(\sigma_1,\ldots,\sigma_{|\Gamma|})$ depends on combinatorics of relative arrangement of the faces $\sigma_1,\ldots,\sigma_{|\Gamma|}$. In terms of $C_{\Delta^D,\Gamma}$ contributions $\bar{S}_{\Delta^D,\Gamma}$ of Feynman diagrams to the reduced action $\bar{S}_{\Delta^D}$ for the simplex are
\begin{multline*}\bar{S}_{\Delta^D,T}=\frac{1}{|\Aut(T)|}\sum_{\sigma_1,\ldots,\sigma_{|T|}\subset\Delta^D}\e_T(|\sigma_1|,\ldots,|\sigma_{|T|}|)\cdot\\
\cdot C_{\Delta^D,T}(\sigma_1,\ldots,\sigma_{|T|})<p_{\Delta^D},\Iter_{T;[\bt,\bt],[\bt,\bt]}(\omega^{\sigma_1},\ldots,\omega^{\sigma_{|T|}})>_\g \end{multline*}
for trees and
$$\bar{S}_{\Delta^D,L}=-\frac{1}{|\Aut(L)|}\sum_{\sigma_1,\ldots,\sigma_{|L|}\subset\Delta^D}\e_L(|\sigma_1|,\ldots,|\sigma_{|L|}|)
C_{\Delta^D,L}(\sigma_1,\ldots,\sigma_{|L|})\Loop_{L;[\bt,\bt];\g}(\omega^{\sigma_1},\ldots,\omega^{\sigma_{|L|}})$$
for one-loop graphs. Thus the non-trivial part of the perturbative computation of $\bar{S}_{\Delta^D}$ is computation of numbers $C_{\Delta^D,\Gamma}$. Notice that the de Rham part of the contribution of a Feynman tree is expressed in terms of multiple integrals (due to the construction of Dupont's operator (\ref{K Dupont}) on $\Delta^D$), while the for a one-loop diagram one has to evaluate a super-trace of certain integral operator over the infinite-dimensional space of differential forms $\Omega^\bt_0(\Delta^D)$. So one-loop computation is technically much more involved and might in principle contain divergencies. However, there is an alternative indirect way to recover certain part of the one-loop result, without explicitly computing super-traces, but using QME and the tree result. Another feature, simplifying the perturbative computations for simplex, is that result has to be symmetric w.r.t. permutations of vertices of $\Delta^D$.

\textbf{Symmetries of de Rham parts of Feynman diagrams $C_{\Delta^D,\Gamma}$.}
If $\pi:(0\cdots D)\ra(0\cdots D)$ is a permutation of vertices of $\Delta^D$, $\pi\in S_{n+1}$, we will denote its action on faces of $\Delta^D$
by
$$\perm_\pi:[i_0\cdots i_k]\mapsto [\pi(i_0)\cdots\pi(i_k)]$$
or just by $\pi[i_0\cdots i_k]:=[\pi(i_0)\cdots\pi(i_k)]$. Permutations also act on differential forms on $\Delta^D$ by homomorphisms $\perm_{\pi}^*:\Omega^\bt(\Delta^D)\ra \Omega^\bt(\Delta^D)$ and send $t_i\mapsto t_{\pi(i)}$, $dt_i\mapsto dt_{\pi(i)}$. The following properties of the action of permutations on forms are important for us
\begin{eqnarray*}\int_{\Delta^D}\perm_\pi^*\alpha&=&(-1)^\pi\int_{\Delta^D}\alpha\\
K_{\Delta^D}(\perm_\pi^*\alpha)&=&\perm_\pi^*\alpha\\
\perm_\pi^* \chi_\sigma&=&\chi_{\pi\sigma}
\end{eqnarray*}
for any permutation $\pi:(0\cdots D)\ra(0\cdots D)$, form $\alpha\in\Omega^\bt(\Delta^D)$ and face $\sigma\subset\Delta^D$.

A straightforward consequence of these properties is the ``external'' symmetry of de Rham parts of Feynman diagrams (symmetry w.r.t permutations of vertices of $\Delta^D$):
\begin{eqnarray}
C_{\Delta^D,T}(\pi\sigma_1,\ldots,\pi\sigma_{|T|})&=&(-1)^\pi C_{\Delta^D,T}(\sigma_1,\ldots,\sigma_{|T|})\label{external symmetry T}\\
C_{\Delta^D,L}(\pi\sigma_1,\ldots,\pi\sigma_{|L|})&=&C_{\Delta^D,L}(\sigma_1,\ldots,\sigma_{|L|})\label{external symmetry L}
\end{eqnarray}
for any $\pi: (0\cdots n)\ra (0\cdots n)$, $T\in \bf{T}_\mr{Pl}$, $L\in \bf{L}_\mr{Pl}$, $\sigma_i\subset\Delta^D$. Difference in the behavior of trees and one-loop diagrams (presence vs. absence of sign $(-1)^\pi$) is due to the fact that for a tree diagram
$$C_{\Delta^D,T}(\pi\sigma_1,\ldots,\pi\sigma_{|T|})=
\int_{\Delta^D}
\perm^*_\pi\Iter_{T;-K_{\Delta^D}(\bt\wedge\bt);(\bt\wedge\bt)}(\chi_{\sigma_1},\ldots,\chi_{\sigma_{|T|}})=
(-1)^\pi C_{\Delta^D,T}(\sigma_1,\ldots,\sigma_{|T|})$$
--- the sign comes from the pairing of permuted form with non-permuted fundamental class of the simplex, while
\begin{multline*}C_{\Delta^D,L}(\pi\sigma_1,\ldots,\pi\sigma_{|L|})=\Str_{\Omega^\bt_0(\Delta^D)}\left(\perm_\pi^*\circ\Iter_{L;-K_{\Delta^D}(\bt\wedge\bt);-K_{\Delta^D}(\bt\wedge\bt)}
(\chi_{\sigma_1},\ldots,\chi_{\sigma_{|L|}})\circ \perm_\pi^*\right)\\
=C_{\Delta^D,L}(\sigma_1,\ldots,\sigma_{|L|})
\end{multline*}
There is no sign here, since permutation acts on the operator under super-trace by similarity transformation, and hence does not change the value of super-trace.

Property (\ref{chi symmetry}) of Whitney forms implies the ``internal'' symmetry of $C_{\Delta^D,\Gamma}$ (consistency with permutations of vertices inside faces):
\begin{eqnarray}
C_{\Delta^D,\Gamma}(\pi_1\sigma_1,\ldots,\pi_{|\Gamma|}\sigma_{|\Gamma|})&=&(-1)^{\pi_1}\cdots (-1)^{\pi_{|\Gamma|}} C_{\Delta^D,\Gamma}(\sigma_1,\ldots,\sigma_{|\Gamma|})\label{internal symmetry}
\end{eqnarray}
where every $\pi_j:\sigma_j\ra\sigma_j$ is a permutation of vertices of $\sigma_j$.

There is the third kind of symmetry of $C_{\Delta^D,\Gamma}$ --- symmetry w.r.t isomorphisms of graphs $\kappa:\Gamma\ra\Gamma'$:
$$C_{\Delta^D,\Gamma'}(\sigma_{\kappa(1)},\ldots,\sigma_{\kappa(|\Gamma|)})=
\e_\kappa(|\sigma_1|,\ldots,|\sigma_{|\Gamma|}|) C_{\Delta^D,\Gamma}(\sigma_1,\ldots,\sigma_{|\Gamma|})$$
where we mean that $\kappa$ sends leaves of $\Gamma$ into leaves of $\Gamma'$. Sign $\e_\kappa(|\sigma_1|,\ldots,|\sigma_{|\Gamma|}|)$ depends on $\kappa$ and dimensions of faces only (not on the combinatorics of their arrangement).
For example, for $\kappa: (*_1 (*_2 *_3))\ra ((*_3 *_2) *_1)$ (index say which leaves go to which ones), we have
\begin{multline*}C_{\Delta^D,((**)*)}(\sigma_3,\sigma_2,\sigma_1)=
\int_{\Delta^D}-K_{\Delta^D}(\chi_{\sigma_3}\wedge\chi_{\sigma_2})\wedge\chi_{\sigma_1}\\
=(-1)^{|\sigma_2| |\sigma_3|+|\sigma_1| (|\sigma_2|+|\sigma_3|+1)} \int_{\Delta^D}\chi_{\sigma_1}\wedge\left(-K_{\Delta^D}(\chi_{\sigma_2}\wedge\chi_{\sigma_3})\right)\\
=(-1)^{|\sigma_2| |\sigma_3|+|\sigma_1| (|\sigma_2|+|\sigma_3|+1)} C_{\Delta^D,(*(**))}(\sigma_1,\sigma_2,\sigma_3)
\end{multline*}
Restrictions on values of $C_{\Delta^D,\Gamma}$ arise from the case when $\kappa:\Gamma\ra\Gamma$ is an automorphisms of the planar graph. For example, for $\kappa: (*_1(*_2 *_3))\ra (*_1 (*_3 *_2))$ we have
\begin{multline}
C_{\Delta^D,(*(**))}(\sigma_1,\sigma_3,\sigma_2)=
\int_{\Delta^D}\chi_{\sigma_1}\wedge\left(-K_{\Delta^D}(\chi_{\sigma_3}\wedge\chi_{\sigma_2})\right)\\
=(-1)^{|\sigma_2||\sigma_3|}\int_{\Delta^D}\chi_{\sigma_1}\wedge\left(-K_{\Delta^D}(\chi_{\sigma_2}\wedge\chi_{\sigma_3})\right)=
(-1)^{|\sigma_2||\sigma_3|}C_{\Delta^D,(*(**))}(\sigma_1,\sigma_2,\sigma_3)\label{C symmetry under graph automorphism}
\end{multline}

Now we can formulate the explicit perturbative result for $\bar{S}_{\Delta^D}$ for the simplex of arbitrary dimension.

\begin{thm}
\label{thm: simplex perturbative result}
For the standard simplex of arbitrary dimension $D\geq 0$ first terms of perturbative expansion for the reduced action are
$$\bar{S}_{\Delta^D}=\bar{S}_{\Delta^D,(*)}+\bar{S}_{\Delta^D,(**)}+\bar{S}_{\Delta^D,(*(**))}+\hbar \bar{S}_{\Delta^D,(*(*\bt))}+O(p\omega^4+\hbar\omega^3)$$
where
\begin{eqnarray}
\bar{S}_{\Delta^D,(*)}&=&\sum_{\sigma_1\subset \Delta^D,\,|\sigma_1|=D-1}\eta_{\Delta^D,(*)}(\sigma_1)<p_{\Delta^D},\omega^{\sigma_1}>_\g\label{simplex pert result 1}\\
\bar{S}_{\Delta^D,(**)}&=&\frac{1}{2}
\sum_{\sigma_1,\sigma_2\subset\Delta^D,\,|\sigma_1|+|\sigma_2|=D}(-1)^{(|\sigma_1|+1)|\sigma_2|}\eta_{\Delta^D,(**)}(\sigma_1,\sigma_2)\cdot\nonumber\\
&&\cdot\frac{|\sigma_1|!\,|\sigma_2|!}{(|\sigma_1|+|\sigma_2|+1)!}<p_{\Delta^D},[\omega^{\sigma_1},\omega^{\sigma_2}]>_\g\label{simplex pert result 2}\\
\bar{S}_{\Delta^D,(*(**))}&=&
\frac{1}{2}\sum_{\sigma_1,\sigma_2,\sigma_3\subset\Delta^D,\,|\sigma_1|+|\sigma_2|+|\sigma_3|=D+1}
(-1)^{(|\sigma_1|+1)(|\sigma_2|+|\sigma_3|+1)+(|\sigma_2|+1)|\sigma_3|}\cdot\nonumber\\
&&\cdot\eta_{\Delta^D,(*(**))}(\sigma_1,\sigma_2,\sigma_3)\cdot\frac{|\sigma_1|!\,|\sigma_2|!\,|\sigma_3|!}{(|\sigma_2|+|\sigma_3|+1)\cdot (|\sigma_1|+|\sigma_2|+|\sigma_3|+1)!}\cdot \nonumber\\
&&\cdot <p_{\Delta^D},[\omega^{\sigma_1},[\omega^{\sigma_2},\omega^{\sigma_3}]]>_\g\label{simplex pert result 3}\\
\bar{S}_{\Delta^D,(*(*\bt))}&=&\frac{1}{2}\left(\A_D\sum_{0\leq i<j\leq D}\tr_\g(\ad_{\omega^{ij}})^2
+\B_D\sum_{0\leq i<j<k\leq D}\tr_g(\ad_{\omega^{jk}-\omega^{ik}+\omega^{ij}})^2\right)\label{simplex pert result 4}
\end{eqnarray}
Coefficients $\eta_{\Delta^D,T}\in\{\pm1,0\}$ depend on combinatorics of arrangement of faces and are defined as:
\begin{itemize}
\item if $\sigma_1=[0\cdots \wh{k}\cdots D]$, then
$$\eta_{\Delta^D,(*)}(\sigma_1)=(-1)^k$$
\item if the intersection of faces $\sigma_1=[i_0\cdots i_{|\sigma_1|}]$ and $\sigma_2=[j_0\cdots j_{|\sigma_2|}]$ is a single vertex $\sigma_1\cap\sigma_2=[i_r]=[j_s]$, then
    $$\eta_{\Delta^D,(**)}(\sigma_1,\sigma_2)=(-1)^s (-1)^{(i_0\cdots i_{|\sigma_1|}j_0\cdots\wh{j_s}\cdots j_{|\sigma_2|})}$$
    otherwise $\eta_{\Delta^D,(**)}(\sigma_1,\sigma_2)=0$
\item if faces $\sigma_1=[i_0\cdots i_{|\sigma_1|}]$, $\sigma_2=[j_0\cdots j_{|\sigma_2|}]$, $\sigma_3=[k_0\cdots k_{|\sigma_3}]$ satisfy
$\sigma_2\cap\sigma_3=[j_r]=[k_s]$, $\sigma_1\cap\sigma_2=[i_u i_q]=[j_v j_r]$, $\sigma_1\cap\sigma_3=[i_q]=[k_s]$, then
\be \eta_{\Delta^D,(*(**))}(\sigma_1,\sigma_2,\sigma_3)=(-1)^{r+s+v+\theta(v-r)+1}(-1)^{(i_0\cdots i_{|\sigma_1|}j_0\cdots\wh{j_v}\cdots\wh{j_r}\cdots j_{|\sigma_2|}k_0\cdots \wh{k_s}\cdots k_{|\sigma_3|})} \label{eta1}\ee
if $\sigma_2\cap\sigma_3=[j_r]=[k_s]$,$\sigma_1\cap\sigma_2=[i_q]=[j_r]$, $\sigma_1\cap\sigma_3=[i_u i_q]=[k_v k_s]$, then
\be \eta_{\Delta^D,(*(**))}(\sigma_1,\sigma_2,\sigma_3)=
(-1)^{r+s+v+\theta(v-s)+1+|\sigma_2|}(-1)^{(i_0\cdots i_{|\sigma_1|}j_0\cdots\wh{j_r}\cdots j_{|\sigma_2|}k_0\cdots\wh{k_v}\cdots \wh{k_s}\cdots k_{|\sigma_3|})}\label{eta2}\ee
otherwise $\eta_{\Delta^D,(*(**))}(\sigma_1,\sigma_2,\sigma_3)=0$.
\end{itemize}
Coefficients $\A_D$, $\B_D$ depend on the dimension $D$ of the simplex only, moreover
\be \A_D=\frac{(-1)^{D+1}}{(D+1)^2(D+2)} \label{simplex A value}\ee
\end{thm}
Before starting to prove the theorem, we first need two intermediate results.

\begin{lemma}
\label{lemma: chi product}
Let $\sigma_1=[i_0\cdots i_{|\sigma_1|}]$ and $\sigma_2=[j_0\cdots j_{|\sigma_2|}]$ be two faces of $\Delta^D$. Then if the intersection of $\sigma_1$ and $\sigma_2$ is a simplex of dimension $|\sigma_1\cap\sigma_2|\geq 1$, then
\be \chi_{\sigma_1}\wedge\chi_{\sigma_2}=0 \label{chi product 1}\ee
If intersection of $\sigma_1$ and $\sigma_2$ is a single vertex $[i_r]=[j_s]$, then
\be\chi_{\sigma_1}\wedge\chi_{\sigma_2}=(-1)^{s}\frac{|\sigma_1|!\;|\sigma_2|!}{(|\sigma_1|+|\sigma_2|)!}t_{j_s}\chi_{i_0\cdots i_{|\sigma_1|} j_0\cdots\wh{j_s}\cdots j_{|\sigma_2|}}\label{chi product 2}\ee
For the integral of product of two Whitney forms over $\Delta^D$ we have
\be\int_{\Delta^D}\chi_{\sigma_1}\wedge\chi_{\sigma_2}=
\left\{\begin{array}{l}(-1)^{s}(-1)^{(i_0\cdots i_{|\sigma_1|} j_0\cdots\wh{j_s}\cdots j_{|\sigma_2|})}\frac{|\sigma_1|!\;|\sigma_2|!}{(|\sigma_1|+|\sigma_2|+1)!},\\\qquad \text{ if }|\sigma_1|+|\sigma_2|=D \text{ and }\sigma_1\cap\sigma_2=[i_r]=[j_s]\\
0,\qquad \text{ otherwise}
\end{array}\right.\label{chi product 3}\ee
The second sign is the sign of permutation $(0\cdots D)\mapsto (i_0\cdots i_{|\sigma_1|} j_0\cdots\wh{j_s}\cdots j_{|\sigma_2|})$.
\end{lemma}

\textbf{Proof of Lemma \ref{lemma: chi product}.}
Let us prove (\ref{chi product 1}). Using the symmetry $S_{D+1}\car\Delta^D$, we can choose $\sigma_1=[0\cdots a]$, $\sigma_2=[a-c,a+b-c]$ without loss of generality, where $a=|\sigma_1|, b=|\sigma_2|, c=|\sigma_1\cap\sigma_2|\geq 1$. In case $c\geq 2$:
$$\chi_{\sigma_1}\wedge\chi_{\sigma_2}=a!b!\sum_{0\leq p\leq a,\,a-c\leq q\leq a+b-c}(-1)^{p+q+a+c} t_p t_q dt_0\cdots \wh{dt_p}\cdots
dt_{a}dt_{a-c}\cdots\wh{dt_q}\cdots dt_{a+b-c}=0$$
since every term in this expression contains $dt_j\wedge dt_j=0$ for some $a-c\leq j \leq a$. For the case $c=1$:
$$\chi_{\sigma_1}\wedge\chi_{\sigma_2}=a!b!\sum_{0\leq p\leq a,\,a-1\leq q\leq a+b-1}(-1)^{p+q+a+1} t_p t_q
dt_0\cdots \wh{dt_p}\cdots dt_{a}dt_{a-1}\cdots\wh{dt_q}\cdots dt_{a+b-1}$$
Only two terms contribute here: $p=a-1$, $q=a$ and $p=a$, $q=a-1$, and they cancel each other. Thus (\ref{chi product 1}) is proved.

Now consider the case $c=0$ (i.e. $\sigma_1$ and $\sigma_2$ intersect over a 0-simplex $[a]$). We have
\begin{multline*}\chi_{\sigma_1}\wedge\chi_{\sigma_2}=a!b!\sum_{0\leq p\leq a,\,a\leq q\leq a+b}(-1)^{p+q+a} t_p t_q
dt_0\cdots \wh{dt_p}\cdots dt_{a}dt_{a}\cdots\wh{dt_q}\cdots dt_{a+b}\\
=a!b!\left(\sum_{0\leq p\leq a}(-1)^{p} t_p t_a
dt_0\cdots \wh{dt_p}\cdots dt_{a}dt_{a+1}\cdots dt_{a+b}+\right.\\
\left.+
\sum_{a+1\leq q\leq a+b}(-1)^{q} t_a t_q
dt_0\cdots dt_{a-1}dt_{a}\cdots\wh{dt_q}\cdots dt_{a+b}\right)\\
=a!b!t_a\sum_{0\leq p\leq a+b}(-1)^p t_p dt_0\cdots\wh{dt_p}\cdots dt_{a+b}=
\frac{a!\;b!}{(a+b)!}t_a \chi_{0\cdots (a+b)}
\end{multline*}
Using the symmetry of simplex $\Delta^D$, we obtain (\ref{chi product 2}) from this.

For the integral (\ref{chi product 3}) to be non-zero, it is necessary that $\chi_{\sigma_1}\wedge\chi_{\sigma_2}$ is a $D$-form, i.e. $|\sigma_1|+|\sigma_2|=D$. Therefore $\chi_{\sigma_1}$ and $\chi_{\sigma_2}$ have to intersect. If the intersection is a simplex of dimension $>0$, the integral vanishes due to (\ref{chi product 1}). If the intersection of $\chi_{\sigma_1}$ and $\chi_{\sigma_2}$ is 0-simplex, we hit the case (\ref{chi product 2}). Therefore, using the symmetry of $\Delta^D$, we reduce (\ref{chi product 3}) to the integral
\begin{multline}\int_{\Delta^D}t_0 \chi_{0\cdots D}=D!\int_{\Delta^D}\left((t_0)^2 dt_1\cdots dt_D + \sum_{p=1}^D (-1)^p t_0 t_p dt_0\cdots\wh{dt_p}\cdots dt_D\right)\\
=D!\left(\frac{2!}{(D+2)!}+D\frac{1!\;1!}{(D+1+1)!}\right)=\frac{1}{D+1}\label{chi product 4}
\end{multline}
We exploit the following useful formula here (see \cite{Getzler}), for the integral of a monomial over simplex:
\be\int_{\Delta^D}t_1^{a_1}\cdots t_D^{a_D} dt_1\cdots dt_D=\frac{a_1!\cdots a_D!}{(a_1+\cdots+a_D+D)!}\label{int of monomial over simplex}\ee
Thus for the integral (\ref{chi product 3}), using (\ref{chi product 2}) and  (\ref{chi product 4}), we obtain
\begin{multline*}\int_{\Delta^D}\chi_{i_0\cdots i_{|\sigma_1|}}\wedge \chi_{j_0\cdots j_{|\sigma_2|}}=
(-1)^{s}\frac{|\sigma_1|!\;|\sigma_2|!}{(|\sigma_1|+|\sigma_2|)!}\int_{\Delta^D}t_{j_s}\chi_{i_0\cdots i_{|\sigma_1|} j_0\cdots\wh{j_s}\cdots j_{|\sigma_2|}}\\
=(-1)^{s}\frac{|\sigma_1|!\;|\sigma_2|!}{(|\sigma_1|+|\sigma_2|)!}(-1)^{(i_0\cdots i_{|\sigma_1|} j_0\cdots\wh{j_s}\cdots j_{|\sigma_2|})}\frac{1}{D+1}=(-1)^{s}(-1)^{(i_0\cdots i_{|\sigma_1|} j_0\cdots\wh{j_s}\cdots j_{|\sigma_2|})}\frac{|\sigma_1|!\;|\sigma_2|!}{(|\sigma_1|+|\sigma_2|+1)!}
\end{multline*}
$\Box$

\begin{lemma}
\label{lemma: K of chi product}
Let $\sigma_1=[i_0\cdots i_{|\sigma_1|}]$ and $\sigma_2=[j_0\cdots j_{|\sigma_2|}]$ be two faces of $\Delta^D$. If $\sigma_1$ and $\sigma_2$ intersect over a simplex of dimension $\geq 1$ or do not intersect at all, then
\be K_{\Delta^D}(\chi_{\sigma_1}\wedge\chi_{\sigma_2})=0\label{K of chi product 1}\ee
If they intersect over a single vertex $[i_r]=[j_s]$, then
\be K_{\Delta^D}(\chi_{\sigma_1}\wedge\chi_{\sigma_2})=
(-1)^{r+s}\frac{|\sigma_1|!\;|\sigma_2|!}{(|\sigma_1|+|\sigma_2|+1)\cdot(|\sigma_1|+|\sigma_2|-1)!}t_{j_s}\chi_{i_0\cdots\wh{i_r}\cdots i_{|\sigma_1|} j_0\cdots\wh{j_s}\cdots j_{|\sigma_2|}}\label{K of chi product 2}\ee
\end{lemma}

\textbf{Proof of Lemma \ref{lemma: K of chi product}.} Case $|\sigma_1\cap\sigma_2|\geq 1$ follows immediately from (\ref{chi product 1}). Consider the case $\sigma_1\cap\sigma_2=\varnothing$. Using the symmetry of $\Delta^D$, we set $\sigma_1=[0\cdots a]$, $\sigma_2=[a+1,\cdots a+b+1]$.
As implied by the computation (\ref{phi of chi}) and the fact that $\phi_i^*$ is a homomorphism,
$$\phi_i^*(\chi_{0\cdots a}\wedge\chi_{(a+1)\cdots (a+b+1)})=u^{a+b+2} \chi_{0\cdots a}\wedge\chi_{(a+1)\cdots (a+b+1)}$$ for $i>a+b+1$. This is a form of degree zero in $u$ and hence does not contribute to $K_{\Delta^D}(\chi_{\sigma_1}\wedge\chi_{\sigma_2})$. Therefore we can write
\begin{multline}K_{\Delta^D}(\chi_{0\cdots a}\wedge\chi_{(a+1)\cdots (a+b+1)})
=\sum_{-1\leq k<a,-1\leq l<b}(-1)^{k+l+1}\cdot\\
\cdot \sum_{0\leq i_0<\cdots<i_k\leq a,\; a+1\leq j_0<\cdots< j_l\leq a+b+1}\chi_{i_0\cdots i_k j_0\cdots j_l}h^{j_l}\cdots h^{j_0}h^{i_k}\cdots h^{i_0}(\chi_{0\cdots a}\wedge\chi_{(a+1)\cdots (a+b+1)})
\label{K of chi product 3}
\end{multline}
We allow values $k=-1$ and $l=-1$ to account for the situation when one of sets $\{i_0,\ldots,i_k\}$ or $\{j_0,\ldots,j_l\}$ is empty.
To calculate the action of dilations on Whitney forms, we again use (\ref{phi of chi}):
\begin{multline}
\chi_{0\cdots a}\wedge\chi_{(a+1)\cdots (a+b+1)}\xra{\phi_{i_0}^*}
(-1)^{i_0}a u_0^{a+b}\,du_0\, \chi_{0\cdots\wh{i_0}\cdots a}\wedge\chi_{(a+1)\cdots (a+b+1)}\xra{\phi_{i_1}^*}\cdots\\
\xra{\phi_{i_k}^*}(-1)^{i_0+(i_1+1)\cdots+(i_k+k)}a(a-1)\cdots (a-k)u_0^{a+b}\cdots u_k^{a+b-k}du_k\,
\chi_{0\cdots\wh{i_0}\cdots\wh{i_k}\cdots a}\wedge\chi_{(a+1)\cdots (a+b+1)}\xra{\phi_{j_0}^*}\cdots\\
\xra{\phi_{j_l}^*}(-1)^{i_0+(i_1+1)\cdots+(i_k+k)+(j_0+k+1)+\cdots+(j_l+k+l+1)}a(a-1)\cdots (a-k) b(b-1)\cdots (b-l)\cdot\\
\cdot u_0^{a+b}\cdots u_k^{a+b-k}du_k v_0^{a+b-k-1}dv_0\cdots v_l^{a+b-k-l-1}dv_l\,
\chi_{0\cdots\wh{i_0}\cdots\wh{i_k}\cdots a}\wedge\chi_{(a+1)\cdots\wh{j_0}\cdots\wh{j_l}\cdots (a+b+1)}
\label{K of chi product 4}
\end{multline}
Here we omit the terms that are forms of non-top degree in variables $u_0,\ldots,u_k,v_0,\ldots,v_l$, since the next step is to take the integral $\pi_*$ over the cube $[0,1]^{k+l+2}$, parameterized with the dilation parameters $u_0,\ldots,u_k,v_0,\ldots,v_l$:
\begin{multline}
h^{j_l}\cdots h^{j_0}h^{i_k}\cdots h^{i_0}(\chi_{0\cdots a}\wedge\chi_{(a+1)\cdots (a+b+1)})=
\pi_* \phi^*_{j_l}\cdots\phi^*_{j_0}\phi^*_{i_k}\cdots\phi^*_{i_0}
(\chi_{0\cdots a}\wedge\chi_{(a+1)\cdots (a+b+1)})\\
=(-1)^{i_0+\cdots i_k+j_0+\cdots j_l+\frac{1}{2}(k+l+1)(k+l+2)}\frac{a}{a+b+1}\cdots\frac{a-k}{a+b-k+1}\cdot\frac{b}{a+b-k}\cdots\frac{b-l}{a+b-k-l}\cdot\\
\cdot\chi_{0\cdots\wh{i_0}\cdots\wh{i_k}\cdots a}\wedge\chi_{(a+1)\cdots\wh{j_0}\cdots\wh{j_l}\cdots (a+b+1)}\\
=(-1)^{i_0+\cdots i_k+j_0+\cdots j_l+\frac{1}{2}(k+l+1)(k+l+2)}\frac{a!b!(a+b-l-k-1)!}{(a-k-1)!(b-l-1)!(a+b+1)!}\cdot\\
\cdot\chi_{0\cdots\wh{i_0}\cdots\wh{i_k}\cdots a}\wedge\chi_{(a+1)\cdots\wh{j_0}\cdots\wh{j_l}\cdots (a+b+1)}
\label{K of chi product 5}
\end{multline}
To finish the computation of (\ref{K of chi product 3}), let us use the following observation:
\begin{multline}\chi_{i_0\cdots i_k j_0\cdots j_l}
=(k+l+1)!\left(
\sum_{p=0}^k (-1)^p t_{i_p} dt_{i_0}\cdots\wh{dt_{i_p}}\cdots dt_{i_k} dt_{j_0}\cdots dt_{j_l}+\right.\\
\left.+
\sum_{q=0}^l (-1)^{q+k+1} t_{j_q} dt_{i_0}\cdots dt_{i_k}dt_{j_0}\cdots\wh{dt_{j_q}}\cdots dt_{j_l}\right)\\
=\frac{(k+l+1)!}{k!}\chi_{i_0\cdots i_k}\wedge dt_{j_0}\cdots dt_{j_l}+
(-1)^{k+1}\frac{(k+l+1)!}{l!}dt_{i_0}\cdots dt_{i_k}\wedge \chi_{j_0\cdots j_l}\label{K of chi product 6}
\end{multline}
Hence
\begin{multline*}
K_{\Delta^D}(\chi_{0\cdots a}\wedge\chi_{(a+1)\cdots (a+b+1)})\\
=\sum_{-1\leq k<a,-1\leq l<b}(-1)^{k+l+1 + a(l+1)}\frac{a!b!(a+b-l-k-1)!(k+l+1)!}{(a-k-1)!(b-l-1)!(a+b+1)!k!}\cdot\\
\cdot\left(\sum_{0\leq i_0<\cdots< i_k\leq a}(-1)^{i_0+(i_1+1)\cdots+(i_k+k)}\chi_{i_0\cdots i_k}\wedge\chi_{0\cdots\wh{i_0}\cdots\wh{i_k}\cdots a}\right)\cdot\\
\cdot\left(\sum_{a+1\leq j_0<\cdots< j_l\leq a+b+1}(-1)^{j_0+(j_1+1)+\cdots+(j_l+l)}dt_{j_0}\cdots dt_{j_l}\wedge
\chi_{(a+1)\cdots\wh{j_0}\cdots\wh{j_l}\cdots (a+b+1)}\right)+\\
+\sum_{-1\leq k<a,-1\leq l<b}(-1)^{k+l+1 + a l}\frac{a!b!(a+b-l-k-1)!(k+l+1)!}{(a-k-1)!(b-l-1)!(a+b+1)!l!}\cdot\\
\cdot\left(\sum_{0\leq i_0<\cdots< i_k\leq a}(-1)^{i_0+(i_1+1)\cdots+(i_k+k)}dt_{i_0}\cdots dt_{i_k}\wedge\chi_{0\cdots\wh{i_0}\cdots\wh{i_k}\cdots a}\right)\cdot\\
\cdot\left(\sum_{a+1\leq j_0<\cdots< j_l\leq a+b+1}(-1)^{j_0+(j_1+1)+\cdots+(j_l+l)}\chi_{j_0\cdots j_l}\wedge
\chi_{(a+1)\cdots\wh{j_0}\cdots\wh{j_l}\cdots (a+b+1)}\right)
\end{multline*}
In the first term first sum in brackets vanishes due to quadratic relation (\ref{chi quadratic relations}) for Whitney forms, and the second sum in second term vanishes for the same reason. Hence $K_{\Delta^D}(\chi_{0\cdots a}\wedge\chi_{(a+1)\cdots (a+b+1)})=0$ and (\ref{K of chi product 1}) is proved.

To prove (\ref{K of chi product 2}), we use (\ref{chi product 2}) and the symmetry of simplex $\Delta^D$, to reduce to the case
$\sigma_1=[0]$, $\sigma_2=[0\cdots a]$. Let us split $K_{\Delta^D}(t_0\chi_{0\cdots a})$ into two parts:
\begin{multline*}
K_{\Delta^D}(t_0\chi_{0\cdots a})\\=\sum_{k=0}^{a-1}(-1)^k\left(\sum_{1\leq i_0<\cdots i_k\leq a}\chi_{i_0\cdots i_k}h^{i_k}\cdots h^{i_0}
(t_0\chi_{0\cdots a})+\sum_{1\leq i_1<\cdots i_k\leq a}\chi_{0 i_1\cdots i_k}h^{i_k}\cdots h^{i_1} h^0
(t_0\chi_{0\cdots a})\right)
\end{multline*}
The second part here corresponds to the case  $i_0=0$, the first --- to $i_0>0$. Computation of $h^{i_k}\cdots h^{i_0}(t_0\chi_{0\cdots a})$ is analogous to (\ref{K of chi product 4},\ref{K of chi product 5}):
\begin{multline*}
t_0\chi_{0\cdots a}\xra{\phi^*_{i_0}} (-1)^{i_0} a u_0^a du_0\, t_0 \chi_{0\cdots\wh{i_0}\cdots a}\xra{\phi^*_{i_1}}\cdots\\
\xra{\phi^*_{i_k}}
(-1)^{i_0+(i_1+1)+\cdots+ (i_k+k)} a (a-1)\cdots (a-k) u_0^a du_0 \cdots u_k^{a-k} du_k\, t_0 \chi_{0\cdots \wh{i_0}\cdots\wh{i_k}\cdots a}\\
\xra{\pi_*} (-1)^{i_0+(i_1+1)+\cdots+ (i_k+k)}\frac{a}{a+1}\cdots\frac{a-k}{a-k+1}\,t_0\chi_{0\cdots \wh{i_0}\cdots\wh{i_k}\cdots a}
\\=(-1)^{i_0+(i_1+1)+\cdots+ (i_k+k)}\frac{a-k}{a+1}\,t_0\chi_{0\cdots \wh{i_0}\cdots\wh{i_k}\cdots a}
\end{multline*}
Also $h^{i_k}\cdots h^{i_1} h^0(t_0\chi_{0\cdots a})$ is computed similarly:
\begin{multline*}
t_0\chi_{0\cdots a}\xra{\phi^*_{0}} a u_0^{a-1} du_0\, (u_0 t_0-u_0+1) \chi_{1\cdots a}\xra{\phi^*_{i_1}}
(-1)^{i_1+1} a (a-1)\cdot\\ \cdot u_0^{a-1} du_0\, u_1^{a-2} du_1\, (u_0 u_1 t_0-u_0+1) \chi_{1\cdots\wh{i_1}\cdots a}\xra{\phi^*_{i_2}}\cdots\\
\xra{\phi^*_{i_k}}(-1)^{(i_1+1)+\cdots+(i_k+1)} a (a-1)\cdots (a-k)\cdot\\ \cdot u_0^{a-1} du_0\, u_1^{a-2} du_1\cdots u_k^{a-k-1} du_k\,
(u_0 u_1\cdots u_k t_0-u_0+1) \chi_{1\cdots\wh{i_1}\cdots\wh{i_k}\cdots a}\\
\xra{\pi_*}(-1)^{(i_1+1)+\cdots+(i_k+1)}\left(\frac{a-k}{a+1}t_0-\frac{a}{a+1}+1\right) \chi_{1\cdots\wh{i_1}\cdots\wh{i_k}\cdots a}\\=
(-1)^{(i_1+1)+\cdots+(i_k+1)}\left(\frac{a-k}{a+1}t_0+\frac{1}{a+1}\right) \chi_{1\cdots\wh{i_1}\cdots\wh{i_k}\cdots a}
\end{multline*}
Therefore
\begin{multline}
K_{\Delta^D}(t_0\chi_{0\cdots a})=t_0 \sum_{k=0}^{a-1}(-1)^k\frac{a-k}{a+1}\sum_{0\leq i_0<\cdots<i_k\leq a}
(-1)^{i_0+(i_1+1)+\cdots+(i_k+k)}\chi_{i_0\cdots i_k}\wedge\chi_{0\cdots\wh{i_0}\cdots \wh{i_k}\cdots a}+\\
+\sum_{k=0}^{a-1}(-1)^k\frac{1}{a+1}\sum_{1\leq i_1<\cdots<i_k\leq a}
(-1)^{(i_1+1)+\cdots+(i_k+k)}\chi_{0 i_1\cdots i_k}\wedge\chi_{1\cdots\wh{i_1}\cdots \wh{i_k}\cdots a}\\
=t_0 \sum_{k=0}^{a-1}(-1)^k\frac{a-k}{a+1}\sum_{0\leq i_0<\cdots<i_k\leq a}
(-1)^{i_0+(i_1+1)+\cdots+(i_k+k)}\chi_{i_0\cdots i_k}\wedge\chi_{0\cdots\wh{i_0}\cdots \wh{i_k}\cdots a}-\\
-dt_0\wedge\sum_{k=0}^{a-1}(-1)^k\frac{k}{a+1}\sum_{1\leq i_1<\cdots<i_k\leq a}
(-1)^{(i_1+1)+\cdots+(i_k+k)}\chi_{i_1\cdots i_k}\wedge\chi_{1\cdots\wh{i_1}\cdots \wh{i_k}\cdots a}+\\
+t_0 \sum_{k=0}^{a-1}(-1)^k\frac{k!}{a+1}\sum_{1\leq i_1<\cdots<i_k\leq a}
(-1)^{(i_1+1)+\cdots+(i_k+k)}dt_{i_1}\cdots dt_{i_k}\wedge\chi_{1\cdots\wh{i_1}\cdots \wh{i_k}\cdots a}
\label{K of chi product 7}
\end{multline}
First and second terms vanish due to (\ref{chi quadratic relations}). We also used (\ref{K of chi product 6}) for a special case:
$$\chi_{0 i_1\cdots i_k}=k!t_0 dt_{i_1}\cdots dt_{i_k}-k dt_0\wedge\chi_{i_1\cdots i_k}$$
The last term in (\ref{K of chi product 7}) gives:
\begin{multline*}
K_{\Delta^D}(t_0\chi_{0\cdots a})=
t_0 \sum_{k=0}^{a-1}(-1)^k\frac{k!(a-k-1)!}{a+1}\sum_{1\leq i_1<\cdots<i_k\leq a}\sum_{1\leq p\leq a,\,p\neq i_j}
(-1)^{(i_1+1)+\cdots+(i_k+k)}\cdot\\
\cdot (-1)^{p+1+\sharp\{j:i_j<p\}} t_p dt_{i_1}\cdots dt_{i_k}dt_1\cdots\wh{dt_{i_1}}\cdots\wh{dt_p}\cdots\wh{dt_{i_k}}\cdots dt_a\\
=t_0 \sum_{k=0}^{a-1}(-1)^k\frac{k!(a-k-1)!}{a+1}\sum_{1\leq i_1<\cdots<i_k\leq a}\sum_{1\leq p\leq a,\,p\neq i_j}
(-1)^{p+1+\sharp\{j:i_j<p\}+\sharp\{j:i_j>p\}} t_p dt_1\cdots\wh{dt_p}\cdots dt_a\\
=t_0 \sum_{k=0}^{a-1}\frac{k!(a-k-1)!}{a+1}\sum_{p=1}^a \sum_{1\leq i_1<\cdots<i_k\leq a,\,p\neq i_j}(-1)^{p+1}
t_p dt_1\cdots\wh{dt_p}\cdots dt_a\\
=t_0 \sum_{k=0}^{a-1}\frac{k!(a-k-1)!}{a+1} C^{k}_{a-1}\frac{1}{(a-1)!}\chi_{1\cdots a}=\frac{a}{a+1}t_0\chi_{1\cdots a}
\end{multline*}
Using the symmetry of $\Delta^D$ and (\ref{chi product 2}), we obtain the general case of (\ref{K of chi product 2}):
\begin{multline*}
K_{\Delta^D}(\chi_{i_0\cdots i_{|\sigma_1|}}\wedge\chi_{j_0\cdots j_{|\sigma_2|}})=(-1)^s\frac{|\sigma_1|!|\sigma_2|!}{(|\sigma_1|+|\sigma_2|)!}
K_{\Delta^D}(t_{j_s}\chi_{i_0\cdots i_{|\sigma_1|}j_0\cdots\wh{j_s}\cdots j_{|\sigma_2|}})\\
=(-1)^{r+s}\frac{|\sigma_1|!|\sigma_2|!}{(|\sigma_1|+|\sigma_2|)!}\;\frac{|\sigma_1|+|\sigma_2|}{|\sigma_1|+|\sigma_2|+1}t_{j_s}
\chi_{i_0\cdots\wh{i_r}\cdots i_{|\sigma_1|}j_0\cdots\wh{j_s}\cdots j_{|\sigma_2|}}\\
=(-1)^{r+s}\frac{|\sigma_1|!|\sigma_2|!}{(|\sigma_1|+|\sigma_2|+1)\cdot(|\sigma_1|+|\sigma_2|-1)!}t_{j_s}
\chi_{i_0\cdots\wh{i_r}\cdots i_{|\sigma_1|}j_0\cdots\wh{j_s}\cdots j_{|\sigma_2|}}
\end{multline*}
\\$\Box$

\textbf{Proof of Theorem \ref{thm: simplex perturbative result}.}
Result (\ref{simplex pert result 1}) for $\bar{S}_{\Delta^D,(*)}$ is obvious, since we have for the de Rham part
$$\eta_{\Delta^D,(*)}(\sigma_1)=C_{\Delta^D,(*)}(\sigma_1)=\int_{\Delta^D}d\chi_{\sigma_1}=\pm 1$$
if $\sigma_1$ is a face of codimension 1 in $\Delta^D$ (otherwise $C_{\Delta^D,(*)}(\sigma_1)=0$), and the sign depends on whether the orientations of $\sigma_1$ and $\Delta^D$ are consistent. Next, the result (\ref{simplex pert result 2}) for $\bar{S}_{\Delta^D,(**)}$ immediately follows from (\ref{chi product 3}), since the de Rham part of this contribution is $$C_{\Delta^D,(**)}(\sigma_1,\sigma_2)=\int_{\Delta^D}\chi_{\sigma_1}\wedge\chi_{\sigma_2}$$

Result (\ref{simplex pert result 3}) for $\bar{S}_{\Delta^D,(*(**))}(\sigma_1,\sigma_2,\sigma_3)$ follows from (\ref{K of chi product 1},\ref{K of chi product 2}): for the de Rham part
$$C_{\Delta^D,(*(**))}=-\int_{\Delta^D} \chi_{\sigma_1}\wedge K_{\Delta^D}(\chi_{\sigma_2}\wedge\chi_{\sigma_3})$$
to be non-zero, faces $\sigma_2$ and $\sigma_3$ have to intersect in a single vertex (as implied by (\ref{K of chi product 1})). Let $\sigma_1=[i_0\cdots i_a]$, $\sigma_2=[j_0\cdots j_b]$, $\sigma_3=[k_0\cdots k_c]$ and $\sigma_2$ intersects $\sigma_3$ in the vertex $[j_r]=[k_s]$. Then
\begin{multline*}C_{\Delta^D,(*(**))}=-\int_{\Delta^D} \chi_{i_0\cdots i_a}\wedge K_{\Delta^D}(\chi_{j_0\cdots j_b}\wedge\chi_{k_0\cdots k_c})\\=
(-1)^{r+s+1}\frac{b!\;c!}{(b+c+1)\cdot(b+c-1)!}\int_{\Delta^D }t_{k_s}\chi_{i_0\cdots i_a}\wedge\chi_{j_0\cdots \wh{j_r}\cdots j_b k_0\cdots
\wh{k_s}\cdots k_c}
\end{multline*}
For the product of Whitney forms in the integrand to be non-zero, it is necessary that $\sigma_1$ intersects with $\sigma_2\cup\sigma_3\backslash (\sigma_2\cap\sigma_3)=[j_0\cdots \wh{j_r}\cdots j_b k_0\cdots
\wh{k_s}\cdots k_c]$ exactly in one vertex (we treat set-theoretic operations $\cup$, $\backslash$ as acting on the sets of vertices of simplices). Suppose this vertex $[i_u]=[j_v]$ is in the simplex $\sigma_2$ (case when it is in $\sigma_3$ is reduced to this one by the symmetry (\ref{C symmetry under graph automorphism})). Due to the constraint on dimensions $a+b+c=D+1$, simplex $\sigma_1$ also has to contain the vertex
$[i_q]=[j_r]=[k_s]$, which is therefore a common intersection point for all three faces. Using (\ref{chi product 2}), we can write
\begin{multline*}C_{\Delta^D,(*(**))}=(-1)^{r+s+1}\frac{b!\;c!}{(b+c+1)\cdot(b+c-1)!}\cdot\\
\cdot(-1)^{v+\theta(v-r)}\frac{a!\;(b+c-1)!}{(a+b+c-1)!}\int_{\Delta^D }t_{k_s} t_{j_v}
\chi_{i_0\cdots i_a j_0\cdots\wh{j_v}\cdots \wh{j_r}\cdots j_b k_0\cdots
\wh{k_s}\cdots k_c}\\
=(-1)^{r+s+v+\theta(v-r)+1}\frac{a!\;b!\;c!}{(b+c+1)\cdot (a+b+c-1)!}(-1)^{(i_0\cdots i_a j_0\cdots\wh{j_v}\cdots \wh{j_r}\cdots j_b k_0\cdots
\wh{k_s}\cdots k_c)}\int_{\Delta^D}t_0 t_1 \chi_{0\cdots D}\\
=(-1)^{r+s+v+\theta(v-r)+1}(-1)^{(i_0\cdots i_a j_0\cdots\wh{j_v}\cdots \wh{j_r}\cdots j_b k_0\cdots
\wh{k_s}\cdots k_c)}\frac{a!\;b!\;c!}{(b+c+1)\cdot (a+b+c+1)!}
\end{multline*}
where we used the symmetry of $\Delta^D$ and (\ref{int of monomial over simplex}). Thus the result (\ref{simplex pert result 3}) for $\Bar{S}_{\Delta^D,(*(**))}$ is proved.

Next, (\ref{simplex pert result 4}) is proved by the following argument. Write $\bar{S}_{\Delta^D,(*(*\bt))}$ as
\be\bar{S}_{\Delta^D,(*(*\bt))}=-\frac{1}{2}\sum_{\sigma_1,\sigma_2\subset \Delta^D}\e_{(*(*\bt))}(|\sigma_1|,|\sigma_2|)C_{\Delta^D,(*(*\bt))}(\sigma_1,\sigma_2)\;\tr_g (\ad_{\omega^{\sigma_1}}\ad_{\omega^{\sigma_2}})\label{simplex pert result 5}\ee
where the sign is $\e_{(*(*\bt))}(|\sigma_1|,|\sigma_2|)=(-1)^{(|\sigma_1|+1)(|\sigma_2|+1)}$ and the de Rham part is
$$C_{\Delta^D,(*(*\bt))}(\sigma_1,\sigma_2)=\Str_{\Omega^\bt_0(\Delta^D)}K_{\Delta^D}(\chi_{\sigma_1}\wedge K_{\Delta^D}(\chi_{\sigma_2}\wedge\bt))$$
For the super-trace to be non-zero, it is necessary for the operator $K_{\Delta^D}(\chi_{\sigma_1}\wedge K_{\Delta^D}(\chi_{\sigma_2}\wedge\bt))$ to be of degree 0, i.e. we have a constraint on dimensions of faces $|\sigma_1|+|\sigma_2|=2$. The possible variants are: $|\sigma_1|=|\sigma_2|=1$ or $|\sigma_1|=0$, $|\sigma_2|=2$ or $|\sigma_1|=2$, $|\sigma_2|=0$. Two last variants are equivalent, since due to the cyclic property of the trace we have
$$C_{\Delta^D,(*(*\bt))}(\sigma_2,\sigma_1)=(-1)^{(|\sigma_1|+1)(|\sigma_2|+1)} C_{\Delta^D,(*(*\bt))}(\sigma_1,\sigma_2)$$
Suppose $|\sigma_1|=2$, $|\sigma_2|=0$. Under the interchange of two vertices in $\sigma_1\backslash \sigma_2$ the super-trace $C_{\Delta^D,(*(*\bt))}(\sigma_1,\sigma_2)$ has to change sign, due to (\ref{internal symmetry}). On the other hand, it should not change value, due to (\ref{external symmetry L}). Hence $C_{\Delta^D,(*(*\bt))}(\sigma_1,\sigma_2)=0$ in this case and we only have to consider the case $|\sigma_1|=|\sigma_2|=1$. If simplices  $\sigma_1$ and $\sigma_2$ do not intersect, the super-trace vanishes by the same argument (interchanging two vertices of $\sigma_1$ should change the sign of super-trace on one hand, and should not on the other hand). So we are left with the cases when  $\sigma_1$ and $\sigma_2$ either coincide or intersect in one vertex. Therefore
\be C_{\Delta^D,(*(*\bt))}(\sigma_1,\sigma_2)=
\left\{\begin{array}{ll}
\tilde{\A}_D,&\text{ if }\sigma_1=\sigma_2=[ij]\\
\tilde{\B}_D,&\text{ if }\sigma_1=[ij],\;\sigma_2=[jk]\text{ or }\sigma_1=[jk],\;\sigma_2=[ij]\\
-\tilde{\B}_D,&\text{ if }\sigma_1=[ik],\;\sigma_2=[jk]\text{ or }\sigma_1=[ij],\;\sigma_2=[ik]\\
0,&\text{ otherwise}
\end{array}\right.\label{simplex pert result 6}\ee
All pairs of coinciding 1-simplices are transferred to each other by permutations of vertices of $\Delta^D$, and all pairs of 1-simplices, intersecting in one vertex, are transferred into each other by a composition of permutation of vertices of $\Delta^D$ and permutation of vertices inside the 1-simplices. Therefore $C_{\Delta^D,(*(*\bt))}$ depends only on two independent values: $\tilde{\A}_D$ and $\tilde{\B}_D$. Substituting (\ref{simplex pert result 6}) into (\ref{simplex pert result 5}), we obtain (\ref{simplex pert result 4}), where we set
\be \A_D=-\tilde{\A}_D+(D-1)\tilde{\B}_D,\; \B_D=-\tilde{\B}_D \label{simplex pert result 8}\ee

Finally, the formula (\ref{simplex A value}) for $\A_D$ follows from the QME and the tree result (\ref{simplex pert result 3}).
To use QME, we have to pass from the reduced action $\bar{S}_{\Delta^D}$ for simplex to the full simplicial action
$$S_{\Delta^D}=\sum_{\sigma\subset\Delta^D}\bar{S}_{\Delta^D}=\sum_{T\in\bf{T}_\mr{nonPl}}\sum_{\sigma\subset\Delta^D}\bar{S}_{\sigma,T}+
\hbar \sum_{L\in\bf{L}_\mr{nonPl}}\sum_{\sigma\subset\Delta^D}\bar{S}_{\sigma,L}$$
Let us check the quantum part of QME in lower orders in $\omega$:
\begin{multline*}
0=\Delta S^0_{\Delta^D}+\{S^0_{\Delta^D},S^1_{\Delta^D}\}=\sum_{\sigma\subset\Delta^D}(-1)^{|\sigma|+1}\left<\frac{\dd}{\dd\omega^\sigma},\frac{\dd}{\dd p_\sigma}\right>_\g S^0_{\Delta^D}-\sum_{\sigma\subset\Delta^D} S^0_{\Delta^D}\left<\frac{\ola\dd}{\dd p_\sigma},\frac{\ora\dd}{\dd \omega^\sigma}\right>_\g S^1_{\Delta^D}\\
=\sum_{T\in\bf{T}_\mr{nonPl}}\sum_{\sigma,\sigma'\subset\Delta^D} (-1)^{|\sigma|+1}\left<\frac{\dd}{\dd\omega^\sigma},\frac{\dd}{\dd p_\sigma}\right>_\g \bar{S}^0_{\sigma',T}-\sum_{T\in{\bf{T}}_\mr{nonPl},\,L\in{\bf{L}}_\mr{nonPl}}\sum_{\sigma,\sigma',\sigma''\subset\Delta^D} \bar{S}^0_{\sigma',T}\left<\frac{\ola\dd}{\dd p_\sigma},\frac{\ora\dd}{\dd \omega^\sigma}\right>_\g \bar{S}^1_{\sigma'',L}\\
=\sum_{T\in\bf{T}_\mr{nonPl}}\sum_{\sigma\subset\Delta^D} (-1)^{|\sigma|+1}\left<\frac{\dd}{\dd\omega^\sigma},\frac{\dd}{\dd p_\sigma}\right>_\g \bar{S}^0_{\sigma,T}-\sum_{T\in{\bf{T}}_\mr{nonPl},\,L\in{\bf{L}}_\mr{nonPl}}\sum_{\sigma\subset\sigma'\subset\Delta^D} \bar{S}^0_{\sigma,T}\left<\frac{\ola\dd}{\dd p_\sigma},\frac{\ora\dd}{\dd \omega^\sigma}\right>_\g \bar{S}^1_{\sigma',L}\\
=\sum_{\sigma\subset\Delta^D}(-1)^{|\sigma|+1}\left<\frac{\dd}{\dd\omega^\sigma},\frac{\dd}{\dd p_\sigma}\right>_\g \bar{S}_{\sigma,(*)}+
\sum_{\sigma\subset\Delta^D}(-1)^{|\sigma|+1}\left<\frac{\dd}{\dd\omega^\sigma},\frac{\dd}{\dd p_\sigma}\right>_\g \bar{S}_{\sigma,(**)}+\\
+\sum_{\sigma\subset\Delta^D}(-1)^{|\sigma|+1}\left<\frac{\dd}{\dd\omega^\sigma},\frac{\dd}{\dd p_\sigma}\right>_\g \bar{S}_{\sigma,(*(**))}
-\sum_{\sigma\subset\sigma'\subset\Delta^D} \bar{S}^0_{\sigma,(*)}\left<\frac{\ola\dd}{\dd p_\sigma},\frac{\ora\dd}{\dd \omega^\sigma}\right>_\g \bar{S}^1_{\sigma',(*(*\bt))}+O(\omega^3)
\end{multline*}
We substitute here the expressions (\ref{simplex pert result 1}--\ref{simplex pert result 4}):
\begin{multline}
0=\Delta S^0_{\Delta^D}+\{S^0_{\Delta^D},S^1_{\Delta^D}\}\\
=\sum_{\sigma\subset\Delta^D}(-1)^{|\sigma|+1}\eta_{\sigma,(*)}(\sigma)\tr_g 1+
\sum_{\sigma_1\subset\sigma\subset\Delta^D,\,|\sigma|+|\sigma_1|=|\sigma|}(-1)^{|\sigma|+|\sigma_1|}\eta_{\sigma,(**)}(\sigma,\sigma_1)
\frac{|\sigma_1|!\;|\sigma|!}{(|\sigma_1|+|\sigma|+1)!}\tr_g \ad_{\omega^{\sigma_1}}+\\
+\frac{1}{2}\sum_{\sigma_2,\sigma_3\subset\sigma\subset\Delta^D,\,|\sigma|+|\sigma_2|+|\sigma_3|=|\sigma|+1}
(-1)^{(|\sigma|+1)(|\sigma_2|+|\sigma_3|)+(|\sigma_2|+1)|\sigma_3|+1}
\eta_{\sigma,(*(**))}(\sigma,\sigma_2,\sigma_3)\cdot\\
\cdot\frac{|\sigma|!\,|\sigma_2|!\,|\sigma_3|!}{(|\sigma_2|+|\sigma_3|+1)\cdot(|\sigma|+|\sigma_2|+|\sigma_3|+1)!}
\tr_g\ad_{[\omega^{\sigma_2},\omega^{\sigma_3}]}+\\
+\sum_{\sigma_1,\sigma_2\subset\sigma\subset\Delta^D,\,|\sigma_1|+|\sigma_2|+|\sigma|=|\sigma|+1}
(-1)^{|\sigma_1| |\sigma_2|+|\sigma|}\eta_{\sigma,(*(**))}(\sigma_1,\sigma_2,\sigma)\cdot\\
\cdot\frac{|\sigma_1|!\,|\sigma_2|!\,|\sigma|!}{(|\sigma_2|+|\sigma|+1)\cdot(|\sigma_1|+|\sigma_2|+|\sigma|+1)!}
\tr_\g(\ad_{\omega^{\sigma_1}}\ad_{\omega^{\sigma_2}})+\\
+\sum_{[ij]\subset\sigma\subset\Delta^D}\A_{|\sigma|}\tr_g (\ad_{\omega^j-\omega^i}\ad_{\omega^{ij}})+
\sum_{[ijk]\subset\sigma\subset \Delta^D}\B_{|\sigma|}\tr_g
(\ad_{(\omega^k-\omega^j)-(\omega^k-\omega^i)+(\omega^j-\omega^i)}\ad_{\omega^{jk}-\omega^{ik}+\omega^{ij}})+ O(\omega^3)
\label{simplex pert result 7}
\end{multline}
Notice that the first term vanishes, since $\eta_{\sigma,(*)}(\sigma)=0$, second and third vanish due to unimodularity of $\g$, and the last one vanishes, since $(\omega^k-\omega^j)-(\omega^k-\omega^i)+(\omega^j-\omega^i)=0$. Only fourth and fifth terms are nontrivial, moreover only combinations of simplices of type $\sigma_1=[ij],\sigma_2=[i]\subset\sigma\subset\Delta^D$ or $\sigma_1=[ij],\sigma_2=[j]\subset\Delta^D$ contribute to the fourth term (otherwise $\eta_{\sigma,(*(**))}(\sigma_1,\sigma_2,\sigma)=0$). Also (\ref{eta2}) implies that the sign for the combination of first type is $\eta_{\sigma,(*(**))}([ij],[i],\sigma)=-1$, and for the combination of second type $\eta_{\sigma,(*(**))}([ij],[j],\sigma)=+1$, and the combinatorial coefficient is
$$\frac{|\sigma_1|!\,|\sigma_2|!\,|\sigma|!}{(|\sigma_2|+|\sigma|+1)\cdot(|\sigma_1|+|\sigma_2|+|\sigma|+1)!}=\frac{1}{(|\sigma|+1)^2(|\sigma|+2)}$$
Therefore, continuing the computation (\ref{simplex pert result 7}), we obtain
\begin{multline*}
0=\Delta S^0_{\Delta^D}+\{S^0_{\Delta^D},S^1_{\Delta^D}\}\\
=\sum_{[ij]\subset\sigma\subset\Delta^D}\left(\frac{(-1)^{|\sigma|}}{(|\sigma|+1)^2(|\sigma|+2)}+\A_{|\sigma|}\right)\tr_g (\ad_{\omega^j-\omega^i}\ad_{\omega^{ij}})+O(\omega^3)
\end{multline*}
Since the right hand side has to be zero function on $\FF_{\Delta^D}$, we obtain formula (\ref{simplex A value}) for $\A_D$.
\\$\Box$

Notice that the status of tree results (\ref{simplex pert result 1},\ref{simplex pert result 2},\ref{simplex pert result 3}) and of the one-loop result (\ref{simplex pert result 4},\ref{simplex A value}) is different: tree results are obtained by direct evaluation of multiple integrals for de Rham parts of the respective Feynman diagrams, while the one-loop result is obtained indirectly from the symmetry argument, giving the ansatz (\ref{simplex pert result 4}) for $\bar{S}_{\Delta^D,(*(*\bt))}$, and an indirect computation of $\A_D$  from QME and the already-known tree result. I.e. we obtain the one-loop result, without performing the direct computation of super-trace over the space $\Omega^\bt_0(\Delta^D)$. We should also note that while in dimension $D=1$ one-loop part of the reduced action can be exactly recovered from the tree part (section \ref{section: interval QME check}), in higher dimension we can only recover a part of one-loop answer: we could recover $\A_D$, but not $\B_D$. Also the value of $\B_D$ is in a sense less interesting than the value of $\A_D$, since under the (special) canonical transformation
$$S_{\Delta^D}\mapsto S_{\Delta^D}+\{S_{\Delta^D},R\}+\hbar\Delta R$$
with generator
$$R=\mr{const}\cdot\hbar\sum_{0\leq i<j<k\leq D}\tr_\g(\ad_{\omega^{jk}-\omega^{ik}+\omega^{ij}}\ad_{\omega^{ijk}})$$
the coefficient of the term
$$\hbar\sum_{0\leq i<j<k\leq D}\tr_g (\ad_{\omega^{jk}-\omega^{ik}+\omega^{ij}})^2$$
in $S_{\Delta^D}$ is shifted by a constant. Thus, if we are interested in the simplicial action modulo equivalence (i.e. modulo canonical transformations), then values of coefficients $\B_D$ are not essential.

For the full simplicial action for $\Delta^D$ the result of Theorem \ref{thm: simplex perturbative result} means the following:
\begin{multline*}
S_{\Delta^D}=\sum_{\sigma_1\subset\sigma\subset\Delta^D,\,|\sigma_1|=|\sigma|-1}\eta_{\sigma,(*)}(\sigma_1)<p_\sigma,\omega^{\sigma_1}>_\g+\\
+\frac{1}{2}\sum_{\sigma_1,\sigma_2\subset\sigma\subset\Delta^D,\,|\sigma_1|+|\sigma_2|=|\sigma|}(-1)^{(|\sigma_1|+1)|\sigma_2|}
\eta_{\sigma,(**)}(\sigma_1,\sigma_2)\frac{|\sigma_1|!\,|\sigma_2|!}{(|\sigma_1|+|\sigma_2|+1)!}<p_\sigma,[\omega^{\sigma_1},\omega^{\sigma_2}]>_\g
+\\
+\frac{1}{2}\sum_{\sigma_1,\sigma_2,\sigma_3\subset\sigma\subset\Delta^D,\,|\sigma_1|+|\sigma_2|+|\sigma_3|=|\sigma|+1}
(-1)^{(|\sigma_1|+1)(|\sigma_2|+|\sigma_3|+1)+(|\sigma_2|+1)|\sigma_3|}\eta_{\sigma,(*(**))}(\sigma_1,\sigma_2,\sigma_3)\cdot\\
\cdot\frac{|\sigma_1|!\,|\sigma_2|!\,|\sigma_3|!}{(|\sigma_2|+|\sigma_3|+1)\cdot (|\sigma_1|+|\sigma_2|+|\sigma_3|+1)!}<p_{\sigma},[\omega^{\sigma_1},[\omega^{\sigma_2},\omega^{\sigma_3}]]>_\g+\\
+\hbar \frac{1}{2}
\left(\hat{\A}_D\sum_{0\leq i<j\leq D}\tr_\g(\ad_{\omega^{ij}})^2
+\hat{\B}_D\sum_{0\leq i<j<k\leq D}\tr_g(\ad_{\omega^{jk}-\omega^{ik}+\omega^{ij}})^2\right)+O(p\omega^4+\hbar\omega^3)
\end{multline*}
where
\begin{eqnarray*}
\hat{A}_D&=&\sum_{[01]\subset\sigma\subset\Delta^D}\A_{|\sigma|}=\sum_{n=1}^D \binom{D-1}{n-1} \A_n\\
\hat{B}_D&=&\sum_{[012]\subset\sigma\subset\Delta^D}\B_{|\sigma|}=\sum_{n=2}^D \binom{D-2}{n-2} \B_n
\end{eqnarray*}
and for $\hat{\A}_D$ we know the explicit value from (\ref{simplex A value}):
$$\hat{A}_D=\sum_{n=1}^D \binom{D-1}{n-1}\frac{(-1)^{n+1}}{(n+1)^2(n+2)}$$

Another equivalent formulation of Theorem \ref{thm: simplex perturbative result} is in terms of first few operations of the $qL_\infty$ structure on $C^\bt(\Delta^D,\g)=\bigoplus_{\sigma\subset\Delta^D}\g e_\sigma$, induced from de Rham algebra of the simplex $\Omega^\bt(\Delta^D,\g)$. In terms of the super-field $\omega=\sum_{\sigma\subset\Delta^D}e_\sigma \omega^{\sigma}$ the operations $l_{(1)}$, $l_{(2)}$, $l_{(3)}$, $q_{(2)}$ are
\begin{eqnarray*}
l_{(1)}(\omega)&=&\sum_{\sigma_1\subset\sigma\subset\Delta^D,\,|\sigma_1|=|\sigma|-1}\eta_{\sigma,(*)}(\sigma_1)e_\sigma \omega^{\sigma_1}\\
l_{(2)}(\omega,\omega)&=&
\sum_{\sigma_1,\sigma_2\subset\sigma\subset\Delta^D,\,|\sigma_1|+|\sigma_2|=|\sigma|}(-1)^{(|\sigma_1|+1)|\sigma_2|}
\eta_{\sigma,(**)}(\sigma_1,\sigma_2)\frac{|\sigma_1|!\,|\sigma_2|!}{(|\sigma_1|+|\sigma_2|+1)!}\;e_\sigma [\omega^{\sigma_1},\omega^{\sigma_2}]\\
l_{(3)}(\omega,\omega,\omega)&=&
3\sum_{\sigma_1,\sigma_2,\sigma_3\subset\sigma\subset\Delta^D,\,|\sigma_1|+|\sigma_2|+|\sigma_3|=|\sigma|+1}
(-1)^{(|\sigma_1|+1)(|\sigma_2|+|\sigma_3|+1)+(|\sigma_2|+1)|\sigma_3|}\eta_{\sigma,(*(**))}(\sigma_1,\sigma_2,\sigma_3)\cdot \\
&&\cdot\frac{|\sigma_1|!\,|\sigma_2|!\,|\sigma_3|!}{(|\sigma_2|+|\sigma_3|+1)\cdot (|\sigma_1|+|\sigma_2|+|\sigma_3|+1)!}\;e_{\sigma}[\omega^{\sigma_1},[\omega^{\sigma_2},\omega^{\sigma_3}]]\\
q_{(2)}(\omega,\omega)&=&\hat{\A}_D\sum_{0\leq i<j\leq D}\tr_\g(\ad_{\omega^{ij}})^2
+\hat{\B}_D\sum_{0\leq i<j<k\leq D}\tr_g(\ad_{\omega^{jk}-\omega^{ik}+\omega^{ij}})^2
\end{eqnarray*}
Also one can say that the polylinear super-antisymmetric operations $l_{(1,2,3)}:C^\bt(\Delta^D,\g)^{\otimes 1,2,3}\ra C^\bt(\Delta^D,\g)$ and
$q_{(2)}:C^\bt(\Delta^D,\g)^{\otimes 2}\ra \RR$ act on $\g$-valued cochains $\alpha=\sum_{\sigma\subset\Delta^D}e_\sigma \alpha^\sigma$ on the simplex $\Delta^D$ as
\begin{eqnarray*}
l_{(1)}(\alpha_1)&=&\sum_{\sigma_1\subset\sigma\subset\Delta^D,\,|\sigma_1|=|\sigma|-1}\eta_{\sigma,(*)}(\sigma_1)e_\sigma \alpha_1^{\sigma_1}\\
l_{(2)}(\alpha_1,\alpha_2)&=&
\sum_{\sigma_1,\sigma_2\subset\sigma\subset\Delta^D,\,|\sigma_1|+|\sigma_2|=|\sigma|}
\eta_{\sigma,(**)}(\sigma_1,\sigma_2)\frac{|\sigma_1|!\,|\sigma_2|!}{(|\sigma_1|+|\sigma_2|+1)!}\;e_\sigma [\alpha_1^{\sigma_1},\alpha_2^{\sigma_2}]\\
l_{(3)}(\alpha_1,\alpha_2,\alpha_3)&=&
\sum_{\sigma_1,\sigma_2,\sigma_3\subset\sigma\subset\Delta^D,\,|\sigma_1|+|\sigma_2|+|\sigma_3|=|\sigma|+1}
\frac{|\sigma_1|!\,|\sigma_2|!\,|\sigma_3|!}{(|\sigma_1|+|\sigma_2|+|\sigma_3|+1)!}
\cdot \\
&&\cdot\; e_{\sigma}\left((-1)^{|\sigma_1|+1} \eta_{\sigma,(*(**))}(\sigma_1,\sigma_2,\sigma_3) \;\frac{1}{|\sigma_2|+|\sigma_3|+1}\; [\alpha_1^{\sigma_1},[\alpha_2^{\sigma_2},\alpha_3^{\sigma_3}]]+\right.\\
&&+(-1)^{|\sigma_1|(|\sigma_2|+|\sigma_3|) +|\sigma_2|+1} \eta_{\sigma,(*(**))}(\sigma_2,\sigma_3,\sigma_1) \;\frac{1}{|\sigma_3|+|\sigma_1|+1}\; [\alpha_2^{\sigma_2},[\alpha_3^{\sigma_3},\alpha_1^{\sigma_1}]]+\\
&&\left.+(-1)^{|\sigma_3|(|\sigma_1|+|\sigma_2|)+|\sigma_3|+1} \eta_{\sigma,(*(**))}(\sigma_3,\sigma_1,\sigma_2) \;\frac{1}{|\sigma_1|+|\sigma_2|+1}\; [\alpha_3^{\sigma_3},[\alpha_1^{\sigma_1},\alpha_2^{\sigma_2}]]\right)
\\
q_{(2)}(\alpha_1,\alpha_2)&=&\hat{\A}_D\sum_{0\leq i<j\leq D}\tr_\g(\ad_{\alpha_1^{ij}}\ad_{\alpha_2^{ij}})
+\hat{\B}_D\sum_{0\leq i<j<k\leq D}\tr_g(\ad_{\alpha_1^{jk}-\alpha_1^{ik}+\alpha_1^{ij}})(\ad_{\alpha_2^{jk}-\alpha_2^{ik}+\alpha_2^{ij}})
\end{eqnarray*}
Equivalently, in terms of the basis $(T_a e_\sigma)$ on $\g\otimes C^\bt(\Delta^D)$ the structure constants of operations are
\begin{eqnarray*}
l_{(1)\sigma_1 a_1}^{\sigma a}&=&\eta_{\sigma,(*)}(\sigma_1)\delta_{a_1}^a\\
l_{(2)\sigma_1 a_1,\,\sigma_2 a_2}^{\sigma a}&=&\eta_{\sigma,(**)}(\sigma_1,\sigma_2)\frac{|\sigma_1|!\,|\sigma_2|!}{(|\sigma_1|+|\sigma_2|+1)!}\;
f^a_{a_1 a_2}\\
l_{(3)\sigma_1 a_1,\,\sigma_2 a_2,\,\sigma_3 a_3}^{\sigma a}&=&
\frac{|\sigma_1|!\,|\sigma_2|!\,|\sigma_3|!}{(|\sigma_1|+|\sigma_2|+|\sigma_3|+1)!}
\cdot \\
&&\cdot\sum_b\left((-1)^{|\sigma_1|+1} \eta_{\sigma,(*(**))}(\sigma_1,\sigma_2,\sigma_3) \;\frac{1}{|\sigma_2|+|\sigma_3|+1}\;
f^a_{a_1 b} f^b_{a_2,a_3}+\right.\\
&&+(-1)^{|\sigma_1|(|\sigma_2|+|\sigma_3|) +|\sigma_2|+1} \eta_{\sigma,(*(**))}(\sigma_2,\sigma_3,\sigma_1) \;\frac{1}{|\sigma_3|+|\sigma_1|+1}\;
f^a_{a_2 b} f^b_{a_3 a_1}+\\
&&\left.+(-1)^{|\sigma_3|(|\sigma_1|+|\sigma_2|)+|\sigma_3|+1} \eta_{\sigma,(*(**))}(\sigma_3,\sigma_1,\sigma_2) \;\frac{1}{|\sigma_1|+|\sigma_2|+1}\; f^a_{a_3 b} f^b_{a_1 a_2}\right)\\
q_{(2)\sigma_1 a_1,\sigma_2 a_2}&=&\left\{\begin{array}{l}
(\hat{A}_D+(D-1)\hat{B}_D)\sum_{b,c} f^b_{a_1 c}f^c_{a_2 b},\quad\text{ if }\sigma_1=\sigma_2=[ij]\\
\hat{B}_D\sum_{b,c} f^b_{a_1 c}f^c_{a_2 b},\quad\text{ if }\sigma_1=[ij],\sigma_2=[jk]\text{ or }\sigma_1=[jk],\sigma_2=[ij]\\
-\hat{B}_D\sum_{b,c} f^b_{a_1 c}f^c_{a_2 b},\quad\text{ if }\sigma_1=[ik],\sigma_2=[jk]\text{ or }\sigma_1=[ij],\sigma_2=[ik]\\
0,\quad\text{ otherwise}
\end{array}\right.
\end{eqnarray*}
where $f^a_{bc}$ are the structure constants of $\g$.

\subsubsection{Explicit calculation of the super-trace $C_{\Delta^2,(*(*\bt))}$ on 2-simplex in coordinate representation}
\label{section: q_2 on 2-simplex}
The problem of explicit computation of the super-trace
\be C_{\Delta^D,(*(*\bt))}=\Str_{\Omega^\bt_0(\Delta^D)}K_{\Delta^D}(\chi_{\sigma_1}\wedge K_{\Delta^D}(\chi_{\sigma_2}\wedge\bt))\label{str1}\ee
on $D$-simplex can be approached in the following way. Introduce the basis
$$|t^0_1,\ldots,t^0_D;i_1,\ldots,i_n >=dt_{i_1}\cdots dt_{i_n}\delta(t_1-t_1^0)\cdots\delta(t_D-t_D^0)$$
on the space $\Omega^\bt_0(\Delta^D)$, enumerated by points inside the simplex $t_1,\ldots,t_D>0$, $t_1+\cdots+t_D<1$ and by sequences of integers $1\leq i_1<\cdots < i_n\leq D$, where $0\leq n\leq D$. The dual basis can be written as
$$<t^0_1,\ldots,t^0_D;i_1,\ldots,i_n |=(-1)^{n(D-n)}*(dt_{i_1}\cdots dt_{i_n})\,\delta(t_1-t_1^0)\cdots\delta(t_D-t_D^0) $$
where
$$*(dt_{i_1}\cdots dt_{i_n})=(-1)^{(i_1\cdots i_n 1\cdots\wh{i_1}\cdots\wh{i_n}\cdots D)} dt_1\cdots \wh{dt_{i_1}}\cdots \wh{dt_{i_n}}\cdots dt_D$$
is the standard Hodge star. Then we have the orthogonality property
\begin{multline*}
<t^0_1,\ldots,t^0_D;i_1,\ldots,i_n |\tilde{t}^0_1,\ldots,\tilde{t}^0_D;\tilde{i}_1,\ldots,\tilde{i}_n >\\
=\int_{\Delta^D}(-1)^{n(D-n)}*(dt_{i_1}\cdots dt_{i_n})\,\delta(t_1-t_1^0)\cdots\delta(t_D-t_D^0)
dt_{\tilde{i}_1}\cdots dt_{\tilde{i}_n}\delta(t_1-\tilde{t}_1^0)\cdots\delta(t_D-\tilde{t}_D^0)\\
=\delta_{i_1\tilde{i}_1}\cdots\delta_{i_n\tilde{i}_n}\delta(t_1^0-\tilde{t}_1^0)\cdots\delta(t_D^0-\tilde{t}_D^0)
\end{multline*}
and the completeness
\begin{equation*}\sum_{n=0}^D(-1)^{nD} \sum_{1\leq i_1<\cdots<i_n\leq D}\int_{\Delta^D}
dt_1^0\cdots dt_D^0\;|t^0_1,\ldots,t^0_D;i_1,\ldots,i_n ><t^0_1,\ldots,t^0_D;i_1,\ldots,i_n |=\id_{\Omega_0^\bt(\Delta^D)}
\end{equation*}
Let us write the super-trace (\ref{str1}) as
\begin{multline}
C_{\Delta^D,(*(*\bt))}(\sigma_1,\sigma_2)\\
=\sum_{n=0}^{D-1}(-1)^{n}\sum_{1\leq i_1<\cdots<i_n\leq D}\sum_{a,b=0}^{D-1}(-1)^{a+b}\sum_{0\leq j_0<\cdots<j_a\leq D,\; 0\leq k_0<\cdots<k_b \leq D}\int_{\Delta^D}dt_1^0\cdots dt_D^0\cdot\\
\cdot <t^0_1,\ldots,t^0_D;i_1,\ldots,i_n |
\chi_{j_0\cdots j_a} h^{j_a}\cdots h^{j_0}\chi_{\sigma_1}\chi_{k_0\cdots k_b}h^{k_b}\cdots h^{k_0}
\chi_{\sigma_2}|t^0_1,\ldots,t^0_D;i_1,\ldots,i_n >\\
=\sum_{n=0}^{D-1}\sum_{1\leq i_1<\cdots<i_n\leq D}\sum_{\bar{\sigma_1},\bar{\sigma_2}\subset\Delta^D}(-1)^{(|\bar{\sigma}_1|+|\bar{\sigma}_2|+1)n+|\bar{\sigma}_1| |\bar{\sigma}_2|+1}
\cdot\\ \cdot
\pi_*\int_{\Delta^D}dt_1^0\cdots dt_D^0 <t^0_1,\ldots,t^0_D;i_1,\ldots,i_n |
\chi_{\bar{\sigma}_1}\phi^*_{\bar{\sigma}_1}\chi_{\sigma_1}\chi_{\bar{\sigma}_2}\phi^*_{\bar{\sigma}_2}
\chi_{\sigma_2}|t^0_1,\ldots,t^0_D;i_1,\ldots,i_n >
\label{str2}
\end{multline}
where we introduced notations $\phi^*_{\bar{\sigma}_1}=\phi^*_{j_a}\cdots \phi^*_{j_0}$, $\bar{\sigma}_1=[j_0\cdots j_a]$,
$\phi^*_{\bar{\sigma}_2}=\phi^*_{k_b}\cdots \phi^*_{k_0}$, $\bar{\sigma}_2=[k_0\cdots k_b]$, and $\pi_*$ is the integral over auxiliary variables (the dilation parameters) $u_0,\ldots, u_a,v_0,\ldots,v_b$. Thus we propose first to compute the super-trace over the space of differential forms on simplex with fixed dilation parameters, and then to integrate the result over dilation parameters. In this way we are avoiding the direct treatment of the very singular kernel of Dupont's operator in coordinate representation.

Introduce the notation $\nu(\sigma_1,\sigma_2;\bar{\sigma}_1,\bar{\sigma}_2)$ for the summand in (\ref{str2}):
\begin{multline*}\nu(\sigma_1,\sigma_2;\bar{\sigma}_1,\bar{\sigma}_2)=
\sum_{n=0}^{D-1}\sum_{1\leq i_1<\cdots<i_n\leq D}
(-1)^{(|\bar{\sigma}_1|+|\bar{\sigma}_2|+1)n+|\bar{\sigma}_1| |\bar{\sigma}_2|+1}\cdot\\
\cdot\pi_*\int_{\Delta^D}dt_1^0\cdots dt_D^0 <t^0_1,\ldots,t^0_D;i_1,\ldots,i_n |
\chi_{\bar{\sigma}_1}\phi^*_{\bar{\sigma}_1}\chi_{\sigma_1}\chi_{\bar{\sigma}_2}\phi^*_{\bar{\sigma}_2}
\chi_{\sigma_2}|t^0_1,\ldots,t^0_D;i_1,\ldots,i_n >
\end{multline*}
Thus
\be C_{\Delta^D,(*(*\bt))}(\sigma_1,\sigma_2)=\sum_{\bar{\sigma_1},\bar{\sigma_2}\subset\Delta^D}
\nu(\sigma_1,\sigma_2;\bar{\sigma}_1,\bar{\sigma}_2) \label{str3}\ee
There is a collection of arguments, allowing to state a priori that some terms in (\ref{str3}) coincide and some vanish:
\begin{enumerate}
\item\label{str arg1} ``external'' symmetry of $\Delta^D$:
$$\nu(\pi\sigma_1,\pi\sigma_2;\pi\bar{\sigma}_1,\pi\bar{\sigma}_2)=\nu(\sigma_1,\sigma_2;\bar{\sigma}_1,\bar{\sigma}_2)$$
where $\pi$ is a permutation of vertices of $\Delta^D$
\item\label{str arg2} ``internal'' symmetry of $\sigma_1,\sigma_2,\bar{\sigma}_1,\bar{\sigma}_2$:
$$\nu(\pi_1\sigma_1,\pi_2\sigma_2;\bar{\pi}_1\bar{\sigma}_1,\bar{\pi}_2\bar{\sigma}_2)=(-1)^{\pi_1}(-1)^{\pi_2}
\nu(\sigma_1,\sigma_2;\bar{\sigma}_1,\bar{\sigma}_2)$$
where $\pi_1,\pi_2,\bar{\pi}_1,\bar{\pi}_2$ are the permutations of vertices in the respective simplices
\item\label{str arg3} cyclic symmetry of the trace:
$$\nu(\sigma_2,\sigma_1;\bar{\sigma}_2,\bar{\sigma}_1)=\nu(\sigma_1,\sigma_2;\bar{\sigma}_1,\bar{\sigma}_2)$$
\item\label{str arg4} if $\sigma_1\subset\bar{\sigma}_2$ or $\sigma_2\subset\bar{\sigma}_1$, then
$$\nu(\sigma_1,\sigma_2;\bar{\sigma}_1,\bar{\sigma}_2)=0$$
since the product of one of the pairs of Whitney forms vanishes
\item\label{str arg5} if $\bar{\sigma}_1\cup \bar{\sigma}_2\neq \Delta^D$ (recall that notation $\cup$ is used for the set-theoretic operation on the sets of vertices), then
\be \nu(\sigma_1,\sigma_2;\bar{\sigma}_1,\bar{\sigma}_2)=0 \label{str8}\ee
since in this case the point $(t_1^0,\cdots,t_D^0)$ cannot be mapped into itself by the sequence of dilations $\phi^*_{\bar{\sigma}_1}\phi^*_{\bar{\sigma}_2}$ (unless all the dilation parameters are equal to 1 simultaneously).
\end{enumerate}
Notice that the last property is in a sense a regularization: point $(t_1^0,\cdots,t_D^0)$ cannot be mapped into itself by a sequence of dilations, but it can be mapped into arbitrarily close points in certain sector of directions.

Let us now restrict to the dimension $D=2$. To obtain values of constants $\tilde{\A}_D$ and $\tilde{\B}_D$ from (\ref{simplex pert result 6}), it suffices to compute $C_{\Delta^2,(*(*\bt))}(\sigma_1,\sigma_2)$ for two cases: $\sigma_1=\sigma_2=[12]$ and $\sigma_1=[01],\sigma_2=[12]$. As properties (\ref{str arg1}--\ref{str arg5}) imply, for the first case there are only two independent non-vanishing contributions:
\be C_{\Delta^2,(*(*\bt))}([12],[12])=4\nu([12],[12];[01],[2])+2\nu([12],[12];[02],[01]) \label{str6}\ee
and for the second case --- four contributions:
\begin{multline} C_{\Delta^2,(*(*\bt))}([01],[12])=2\nu([01],[12];[01],[2])+2\nu([01],[12];[02],[1])+\\ +2\nu([01],[12];[01],[02])+\nu([01],[12];[01],[12]) \label{str7}\end{multline}
Let us demonstrate the computation of $\nu([12],[12];[01],[2])$ in detail:
\be \nu([12],[12];[01],[2])=-\sum_{i=1}^2\pi_*\int_{\Delta^2}dt_1^0 dt_2^0 <t_1^0,t_2^0;i|\chi_{01}\phi^*_1\phi^*_0\chi_{12}\chi_2\phi^*_2\chi_{12}|t_1^0,t_2^0;i> \label{str4}\ee
(we use the obvious observation that only diagonal matrix elements for basis 1-forms contribute). Then
\begin{multline*}|t_1^0,t_2^0;1>=dt_1 \delta(t_1-t_1^0)\delta(t_2-t_2^0)\xra{\chi_{12}\wedge}-t_1 dt_1 dt_2  \delta(t_1-t_1^0)\delta(t_2-t_2^0)\\
\xra{\phi^*_2}-v_1^2 t_1 dv_1(t_1 dt_2-(t_2-1)dt_1)  \delta(v_1 t_1-t_1^0)\delta(v_1 t_2-v_1+1-t_2^0)\\
\xra{\chi_2\wedge}-v_1^2 t_1 t_2 dv_1(t_1 dt_2-(t_2-1)dt_1)\delta(v_1 t_1-t_1^0)\delta(v_1 t_2-v_1+1-t_2^0)\\
\xra{\chi_{12}\wedge}-v_1^2 t_1^2 t_2 dv_1 dt_1 dt_2\delta(v_1 t_1-t_1^0)\delta(v_1 t_2-v_1+1-t_2^0)\\
\xra{\phi^*_0}-v_1^2u_2^4 t_1^2 t_2 dv_1 du_2 (t_1 dt_2-t_2 dt_1)\delta(v_1 u_2 t_1-t_1^0)\delta(v_1 u_2 t_2-v_1+1-t_2^0)\\
\xra{\phi^*_1}-v_1^2u_2^4 u_1 (u_1 t_1-u_1+1)^2 t_2^2 dv_1 du_2 du_1\delta(v_1u_2(u_1t_1-u_1+1)-t_1^0)\delta(v_1
u_2u_1t_2-v_1+1-t_2^0)\\
\xra{\chi_{01}\wedge}-v_1^2u_2^4 u_1 (u_1 t_1-u_1+1)^2 t_2^2 ((1-t_2)dt_1+t_1dt_2) dv_1 du_2 du_1\cdot\\ \cdot \delta(v_1u_2(u_1t_1-u_1+1)-t_1^0)\delta(v_1
u_2u_1t_2-v_1+1-t_2^0)
\end{multline*}
We write only the terms that are top-degree forms in auxiliary variables $u_1,u_2,v_1$. Therefore the matrix element in the integral over simplex in (\ref{str4}) for $i=1$ is
\begin{multline}
<t_1^0,t_2^0;1|\chi_{01}\phi^*_1\phi^*_0\chi_{12}\chi_2\phi^*_2\chi_{12}|t_1^0,t_2^0;1>\\=
-v_1^2u_2^4 u_1 (u_1 t_1^0-u_1+1)^2 (t^0_2)^2 (1-t_2^0) dv_1 du_2 du_1\;\delta(v_1u_2(u_1t_1^0-u_1+1)-t_1^0)\;\delta(v_1
u_2u_1t_2^0-v_1+1-t_2^0)\label{str5}
\end{multline}
Integral over $\Delta^2$ is evaluated using the delta functions, and we are left with non-trivial integral over dilation parameters $u_1,u_2,v_1$:
\begin{multline*}
\pi_*\int_{\Delta^2}dt_1^0 dt_2^0 <t_1^0,t_2^0;1|\chi_{01}\phi^*_1\phi^*_0\chi_{12}\chi_2\phi^*_2\chi_{12}|t_1^0,t_2^0;1>\\
=\int_0^1 du_1\int_0^1 du_2 \int_0^1 dv_1
\frac{1}{(1-u_1u_2v_1)^2}\left(-\frac{u_1 u_2^4 v_1^3 (1-u_1)^2 (1-u_1u_2)(1-v_1)^2}{(1-u_1u_2v_1)^5}\right)=-\frac{1}{360}
\end{multline*}
Factor $\frac{1}{(1-u_1u_2v_1)^2}$ in the integrand is the Jacobian arising from delta-functions in (\ref{str5}). Analogously, for the $i=2$ term in (\ref{str4}) we obtain
\begin{multline*}
\chi_{01}\phi^*_1\phi^*_0\chi_{12}\chi_2\phi^*_2\chi_{12}|t_1^0,t_2^0;2>\\=
-v_1u_2^3u_1t_2^2(u_1t_1-u_1+1)(v_1u_2u_1t_2-v_1+1)((1-t_2)dt_1+t_1dt_2)dv_1du_2du_1\cdot\\
\cdot\delta(v_1u_2(u_1t_1-u_1+1)-t_1^0)\;\delta(v_1
u_2u_1t_2-v_1+1-t_2^0)\\
\Rightarrow<t_1^0,t_2^0;2|\chi_{01}\phi^*_1\phi^*_0\chi_{12}\chi_2\phi^*_2\chi_{12}|t_1^0,t_2^0;2>
=-v_1u_2^3u_1t_1^0(t^0_2)^2(u_1t_1^0-u_1+1)(v_1u_2u_1t_2^0-v_1+1) dv_1du_2du_1\cdot\\
\cdot\delta(v_1u_2(u_1t^0_1-u_1+1)-t_1^0)\;\delta(v_1
u_2u_1t^0_2-v_1+1-t_2^0)\\
\Rightarrow\pi_*\int_{\Delta^2}dt_1^0 dt_2^0 <t_1^0,t_2^0;2|\chi_{01}\phi^*_1\phi^*_0\chi_{12}\chi_2\phi^*_2\chi_{12}|t_1^0,t_2^0;2>\\
=\int_0^1 du_1\int_0^1 du_2 \int_0^1 dv_1
\frac{1}{(1-u_1u_2v_1)^2}\left(-\frac{u_1u_2^4v_1^2(1-u_1)^2(1-v_1)^3}{(1-u_1u_2v_1)^5}\right)=-\frac{1}{360}
\end{multline*}
And therefore
\begin{multline*}\nu([12],[12];[01],[2])\\=-\sum_{i=1}^2\pi_*\int_{\Delta^2}dt_1^0 dt_2^0 <t_1^0,t_2^0;i|\chi_{01}\phi^*_1\phi^*_0\chi_{12}\chi_2\phi^*_2\chi_{12}|t_1^0,t_2^0;i>=-\left(-\frac{1}{360}-\frac{1}{360}\right)=\frac{1}{180}
\end{multline*}
Other terms in (\ref{str6},\ref{str7}) are computed in the same manner:
\begin{multline*}
<t_1^0,t_2^0;1| \chi_{02}\phi^*_2\phi^*_0\chi_{12}\chi_{01}\phi_1^*\phi_0^*\chi_{12} |t_1^0,t_2^0;1>\\
=-v_2^2u_2^3u_1(t^0_1)^2t_2^0(v_1u_2u_1t_1^0-v_1+1)(u_1t_2^0-u_1+1)dv_2dv_1du_2du_1\cdot\\
\cdot \delta(v_2(v_1u_2u_1t_1^0-v_1+1)-t_1^0)\;\delta(v_2u_2u_2(u_1 t_2^0-u_1+1)-t_2^0)\\
<t_1^0,t_2^0;2| \chi_{02}\phi^*_2\phi^*_0\chi_{12}\chi_{01}\phi_1^*\phi_0^*\chi_{12} |t_1^0,t_2^0;2>\\
=-v_2^2v_1u_2^4u_1(t^0_1)^2(1-t_1^0)(u_1t_2^0-u_1+1)^2dv_2dv_1du_2du_1\cdot\\
\cdot \delta(v_2(v_1u_2u_1t_1^0-v_1+1)-t_1^0)\;\delta(v_2u_2u_2(u_1 t_2^0-u_1+1)-t_2^0)
\displaybreak[0]\\
\Rightarrow \nu([12],[12];[02],[01])=-\sum_{i=1}^2\pi_*\int_{\Delta^2}dt_1^0dt_2^0<t_1^0,t_2^0;i| \chi_{02}\phi^*_2\phi^*_0\chi_{12}\chi_{01}\phi_1^*\phi_0^*\chi_{12} |t_1^0,t_2^0;i>\\
=-\left(\int_0^1du_1\int_0^1du_2\int_0^1dv_1\int_0^1dv_2\frac{1}{(1-u_1u_2v_1v_2)^2}
\left(-\frac{u_1(1-u_1)^2u_2^4v_1(1-v_1)^3v_2^5}{(1-u_1u_2v_1v_2)^5}\right)+\right.\\
+\int_0^1du_1\int_0^1du_2\int_0^1dv_1\int_0^1dv_2\frac{1}{(1-u_1u_2v_1v_2)^2}\\ \left.
\left(-\frac{u_1(1-u_1)^2u_2^4v_1(1-v_1)^2v_2^4(1-v_2+v_1v_2-u_1u_2v_1v_2)}{(1-u_1u_2v_1v_2)^5}\right)\right)\\
=-\left(-\frac{1}{2160}-\frac{1}{2160}\right)=\frac{1}{1080}
\displaybreak[1]\\
<t_1^0,t_2^0;1| \chi_{02}\phi^*_2\phi^*_0\chi_{01}\chi_{1}\phi_1^*\chi_{12} |t_1^0,t_2^0;1>\\=u_1u_2^2v_1(t_1^0)^2t_2^0 (1-v_1+u_1u_2v_1t_1^0)(1-u_2+u_1u_2t_0^0)dv_1du_2du_1\cdot\\
\cdot \delta(1-v_1+u_1u_2v_1t_1^0-t_1^0)\;\delta(u_2v_1(1-u_1+u_1t_2^0)-t_2^0)\\
<t_1^0,t_2^0;2| \chi_{02}\phi^*_2\phi^*_0\chi_{01}\chi_{1}\phi_1^*\chi_{12} |t_1^0,t_2^0;2>\\=
u_1u_2^3v_1^2(t_1^0)^2(1-t_1^0)(1-u_1+u_1t_2^0)(1-u_2+u_1u_2t_0^0)dv_1du_2du_1\cdot\\
\cdot \delta(1-v_1+u_1u_2v_1t_1^0-t_1^0)\;\delta(u_2v_1(1-u_1+u_1t_2^0)-t_2^0)
\displaybreak [0]
\\
\Rightarrow \nu([01],[12];[02],[1])=-\sum_{i=1}^2\pi_*\int_{\Delta^2}dt_1^0dt_2^0<t_1^0,t_2^0;i| \chi_{02}\phi^*_2\phi^*_0\chi_{01}\chi_{1}\phi_1^*\chi_{12} |t_1^0,t_2^0;i>\\
=-\left(\int_0^1du_1\int_0^1du_2\int_0^1dv_1\frac{1}{(1-u_1u_2v_1)^2}\cdot
\frac{u_1(1-u_1)u_2^3(1-u_2)v_1^2(1-v_1)^3}{(1-u_1u_2v_1)^5}+\right.\\
\left.+\int_0^1du_1\int_0^1du_2\int_0^1dv_1\frac{1}{(1-u_1u_2v_1)^2}\cdot
\frac{u_1(1-u_1)u_2^3(1-u_2)(1-u_1u_2)v_1^3(1-v_1)^2}{(1-u_1u_2v_1)^5}\right)\\
=-\left(\frac{1}{720}+\frac{1}{720}\right)=-\frac{1}{360}
\displaybreak[1]\\
<t_1^0,t_2^0;1| \chi_{01}\phi^*_1\phi^*_0\chi_{01}\chi_{02}\phi_2^*\phi_0^*\chi_{12} |t_1^0,t_2^0;1>\\
=u_2^3v_1v_2^2(1-u_1+u_1t_1^0)^2t_2^0(1-t_2^0)(1-u_2+u_1u_2t_0^0)dv_2dv_1du_2du_1\cdot\\
\cdot \delta(u_2v_1v_2(1-u_1+u_1t_1^0)-t_1^0)\;\delta(v_2(1-v_1+u_1u_2v_1t_2^0)-t_2^0)\\
<t_1^0,t_2^0;2| \chi_{01}\phi^*_1\phi^*_0\chi_{01}\chi_{02}\phi_2^*\phi_0^*\chi_{12} |t_1^0,t_2^0;2>
\\=u_2^2v_2^2t_1^0t_2^0(1-u_1+u_1t_1^0)(1-v_1+u_1u_2v_1t_2^0)(1-u_2+u_1u_2t_0^0)dv_2dv_1du_2du_1\cdot\\
\cdot \delta(u_2v_1v_2(1-u_1+u_1t_1^0)-t_1^0)\;\delta(v_2(1-v_1+u_1u_2v_1t_2^0)-t_2^0)
\displaybreak[0]\\
\Rightarrow  \nu([01],[12];[01],[02])=-\sum_{i=1}^2\pi_*\int_{\Delta^2}dt_1^0dt_2^0
<t_1^0,t_2^0;i| \chi_{01}\phi^*_1\phi^*_0\chi_{01}\chi_{02}\phi_2^*\phi_0^*\chi_{12} |t_1^0,t_2^0;i>\\
=-\left(\int_0^1du_1\int_0^1du_2\int_0^1dv_1\int_0^1dv_2\frac{1}{(1-u_1u_2v_1v_2)^2}\cdot\right.\\
\cdot\frac{(1-u_1)^2u_2^3v_1(1-v_1)v_2^3(1-u_2+u_1u_2-u_1u_2v_2)(1-v_2+v_1v_2-u_1u_2v_1v_2)}{(1-u_1u_2v_1v_2)^5}+\\
\left.+\int_0^1du_1\int_0^1du_2\int_0^1dv_1\int_0^1dv_2\frac{1}{(1-u_1u_2v_1v_2)^2}\cdot
\frac{(1-u_1)^2u_2^3v_1(1-v_1)^2v_2^4(1-u_2+u_1u_2-u_1u_2v_2)}{(1-u_1u_2v_1v_2)^5}\right)\\
=-\left(\frac{1}{864}+\frac{1}{1440}\right)=-\frac{1}{540}
\displaybreak[1]\\
<t_1^0,t_2^0;1| \chi_{01}\phi^*_1\phi^*_0\chi_{01}\chi_{12}\phi_2^*\phi_1^*\chi_{12} |t_1^0,t_2^0;1>=\\=
-u_2^2v_2(1-u_1+u_1t_1^0)(1-v_2+u_2v_1v_2-u_1u_2v_1v_2+u_1u_2v_1v_2t_1^0)t_2^0(1-t_2^0)(1-u_2+u_1u_2t_0^0)dv_2dv_1du_2du_1\cdot\\
\cdot\delta(1-v_2+u_2v_1v_2-u_1u_2v_1v_2+u_1u_2v_1v_2t_1^0-t_1^0)\;\delta(v_2(1-v_1+u_1u_2v_1t_2^0)-t_2^0)\\
<t_1^0,t_2^0;2| \chi_{01}\phi^*_1\phi^*_0\chi_{01}\chi_{12}\phi_2^*\phi_1^*\chi_{12} |t_1^0,t_2^0;2>=\\
=-u_2^2v_2^2t_1^0(1-u_1+u_1t_1^0)t_2^0(1-v_1+u_1u_2v_1t_2^0)(1-u_2+u_1u_2t_0^0)dv_2dv_1du_2du_1\cdot\\
\cdot\delta(1-v_2+u_2v_1v_2-u_1u_2v_1v_2+u_1u_2v_1v_2t_1^0-t_1^0)\;\delta(v_2(1-v_1+u_1u_2v_1t_2^0)-t_2^0)
\displaybreak[0]\\
\Rightarrow  \nu([01],[12];[01],[12])=-\sum_{i=1}^2\pi_*\int_{\Delta^2}dt_1^0dt_2^0
<t_1^0,t_2^0;i| \chi_{01}\phi^*_1\phi^*_0\chi_{01}\chi_{12}\phi_2^*\phi_1^*\chi_{12} |t_1^0,t_2^0;i>\\
=-\left(\int_0^1du_1\int_0^1du_2\int_0^1dv_1\int_0^1dv_2\frac{1}{(1-u_1u_2v_1v_2)^2}\cdot\right.\\
\cdot\left(-\frac{u_2^2(1-u_2)(1-v_1)v_2^2(1-u_1v_2)(1-v_2+v_1v_2-u_1u_2v_1v_2)(1-v_2+u_2v_1v_2-u_1 u_2v_1v_2)}{(1-u_1u_2v_1v_2)^5}\right)+\\
+\int_0^1du_1\int_0^1du_2\int_0^1dv_1\int_0^1dv_2\frac{1}{(1-u_1u_2v_1v_2)^2}\cdot\\
\left.\cdot\left(-\frac{u_2^2(1-u_2)(1-v_1)^2v_2^3(1-u_1v_2)(1-v_2+u_2v_1v_2-u_1 u_2v_1v_2)}{(1-u_1u_2v_1v_2)^5}\right)\right)\\
=-\left(-\frac{1}{288}-\frac{1}{480}\right)=\frac{1}{180}
\end{multline*}
Finally the matrix elements $<t_1^0,t_2^0;i| \chi_{01}\phi^*_1\phi^*_0\chi_{01}\chi_{2}\phi_2^*\chi_{12} |t_1^0,t_2^0;i>=0$ vanish for $i=1,2$ and hence
$$\nu([01],[12];[01],[2])=0$$
and we obtain the following values for the super-traces (\ref{str6},\ref{str7}):
$$C_{\Delta^2,(*(*\bt))}([12],[12])=4\nu([12],[12];[01],[2])+2\nu([12],[12];[02],[01])=4\cdot\frac{1}{180}+2\cdot\frac{1}{1080}=\frac{13}{540}$$
and
\begin{multline*}
C_{\Delta^2,(*(*\bt))}([01],[12])=2\nu([01],[12];[01],[2])+2\nu([01],[12];[02],[1])+\\
+2\nu([01],[12];[01],[02])+\nu([01],[12];[01],[12])\\
=2\cdot 0+2\cdot\left(-\frac{1}{360}\right)+2\cdot\left(-\frac{1}{540}\right)+\frac{1}{180}=-\frac{1}{270}
\end{multline*}
Comparing this to (\ref{simplex pert result 6},\ref{simplex pert result 8}), we conclude that
\be \A_2=-\frac{1}{36},\;\B_2=\frac{1}{270} \label{A and B for 2-simplex}\ee
Notice that the value of $\A_2$ obtained from the explicit computation of the super-trace coincides with the value (\ref{simplex A value}) predicted by QME.

Analogous but more tedious computation of the super-trace $C_{\Delta^3,(*(*\bt))}$ for 3-simplex in coordinate representation yields
$$\A_3=\frac{1}{80},\;\B_3=-\frac{1}{648}$$
(value of $\A_3$ again coincides with the prediction (\ref{simplex A value})).

The scheme of computation of $C_{\Delta^2,(*(*\bt))}$ we employed here is the most simple and convenient among those we know. The key points here are interchanging the order of taking super-trace and integration over dilation parameters (\ref{str2}), and the use of property (\ref{str8}), playing the role of regularization. This regularization can be equivalently formulated as follows: the dilation parameters take values in the interval $[0,1-\e]$ and $\e$ is taken to zero. A remarkable feature of this scheme is that we never encounter divergent quantities on intermediate stages of the computation.

We also computed $C_{\Delta^2,(*(*\bt))}$ using two other schemes. One method is to compute in coordinate representation, but without interchanging the super-trace and the integral over dilation parameters. Here the problem reduces to computing convolutions of (singular) kernels of Dupont's operator. As a regularization we used the point-splitting: one computes not the diagonal matrix elements, but matrix elements between nearby points (plus one should average over the direction of splitting). Here in the intermediate stages one encounters logarithmic divergencies. Nevertheless, in the final result divergencies cancel out and one comes again to the result (\ref{A and B for 2-simplex}).
Another, more cumbersome scheme is to use the basis of monomials. As a regularization one uses restricting the total degree of monomials by a large number $N\ra\infty$. In the intermediate stages one again encounters divergencies $\sim\log N$, but they cancel out in the end and the result coincides with (\ref{A and B for 2-simplex}).

Using the general perturbative result of Theorem \ref{thm: simplex perturbative result} and the value of $\B_2$, obtained from the explicit calculation of the super-trace (\ref{A and B for 2-simplex}), we can write down the simplicial $BF$ action on 2-simplex up to terms of order $O(p\omega^4+\hbar\omega^3)$:
\begin{multline}
S_{\Delta^2}=\bar{S}_0+\bar{S}_1+\bar{S}_2+\bar{S}_{01}+\bar{S}_{12}+\bar{S}_{20}+\bar{S}_{012}=
\\=<p_0,\frac{1}{2}[\omega^0,\omega^0]>_g+<p_1,\frac{1}{2}[\omega^1,\omega^1]>_g+<p_2,\frac{1}{2}[\omega^2,\omega^2]>_g+\\
+\left(<p_{01},(\omega^1-\omega^0)+\frac{1}{2}[\omega^{01},\omega^0+\omega^1]+\frac{1}{12}[\omega^{01},[\omega^{01},\omega^1-\omega^0]]>_\g+
\hbar\frac{1}{24}\tr_g(\ad_{\omega^{01}})^2\right)+\\
+\left(<p_{12},(\omega^2-\omega^1)+\frac{1}{2}[\omega^{12},\omega^1+\omega^2]+\frac{1}{12}[\omega^{12},[\omega^{12},\omega^2-\omega^1]]>_\g+
\hbar\frac{1}{24}\tr_g(\ad_{\omega^{12}})^2\right)+\\
+\left(<p_{20},(\omega^0-\omega^2)+\frac{1}{2}[\omega^{20},\omega^2+\omega^0]+\frac{1}{12}[\omega^{20},[\omega^{20},\omega^0-\omega^2]]>_\g+
\hbar\frac{1}{24}\tr_g(\ad_{\omega^{20}})^2\right)+\\
+\left(<p_{012},(\omega^{01}+\omega^{12}+\omega^{20})+\frac{1}{3}[\omega^{012},\omega^0+\omega^1+\omega^2]
+\frac{1}{6}([\omega^{01},\omega^{12}]+[\omega^{12},\omega^{20}]+[\omega^{20},\omega^{01}]+\right.\\
+\frac{1}{72}([\omega^{01},[\omega^{01},\omega^{01}+\omega^{12}+\omega^{20}]]+[\omega^{12},[\omega^{12},\omega^{01}+\omega^{12}+\omega^{20}]]
+[\omega^{20},[\omega^{20},\omega^{01}+\omega^{12}+\omega^{20}]])+\\
+\frac{1}{24}([\omega^{012},[\omega^{01},\omega^1-\omega^0]]+[\omega^{012},[\omega^{12},\omega^2-\omega^1]]+
[\omega^{012},[\omega^{20},\omega^0-\omega^2]])+\\
+\frac{1}{36}([\omega^{01},[\omega^{012},\omega^1-\omega^0]]+[\omega^{12},[\omega^{012},\omega^2-\omega^1]]+
[\omega^{20},[\omega^{012},\omega^0-\omega^2]])>_\g+\\
\left.+\hbar\left(-\frac{1}{72}(\tr_g(\ad_{\omega^{01}})^2+\tr_g(\ad_{\omega^{12}})^2+\tr_g(\ad_{\omega^{20}})^2)
+\frac{1}{540}\tr_g(\ad_{\omega^{01}+\omega^{12}+\omega^{20}})^2\right)\right)+\\+O(p\omega^4+\hbar\omega^3)
\label{S on triangle}
\end{multline}
Here we split $S_{\Delta^2}$ into the contributions of three vertices, three edges and the bulk.

\section{Effective $BF$ theory on a cubical complex}
\label{section: BF on cubical complex}

Here we discuss a modification of the setting of section \ref{section: BF on simplicial complex}: we consider the effective action for topological $BF$ theory, induced on cochains of cubical cell decomposition $\Xi$ of manifold $M$. The induction data for a cubical CW-complex  are constructed from standard induction data for individual cubes (analogously to the construction of induction data for a triangulation from standard induction data for simplices in section \ref{section: BF on simplicial complex}). In turn, the standard induction data for the cube are constructed from standard induction data for the interval, using the ``tensor product construction'' (section \ref{section: tensor product for ind data}). Next, in complete analogy with the simplicial situation, we introduce the cell action on $\Xi$, which features the cell locality property (section \ref{section: cell action}). Hence we come to the problem of computing the cell action for one standard cube (in each dimension). Specific construction of chain homotopy for the standard cube leads to ``factorization'' of Feynman diagrams for the cell action (section \ref{section: cube factorization}): de Rham parts of Feynman diagrams for the cell action for $D$-cube are decomposed into sums of $D$-fold products of Feynman diagrams for interval, where edges of the factors are decorated with either chain homotopy for interval, or identity, or the IR projector for interval and we sum over allowed decorations (in an equivalent formulation, we decorate edges of the factor with the ``extended propagator'' for the interval, depending on auxiliary parameters, and instead of summing over decorations, we integrate over the auxiliary parameters). This property greatly simplifies the problem of computing Feynman diagrams for the cube, and we can easily write a closed formula for any particular Feynman diagram for the cube of general dimension (Theorem \ref{thm: cube factorization}). However, we cannot present a closed formula for the cell action for the cube of dimension $\geq 2$, and can only give perturbative results. Next, it turns out that restrictions of the cell action for cube to certain subspaces of cochains of the cube can be calculated explicitly (only special Feynman diagrams contribute and we can explicitly sum up the contributions), which leads to some examples of manifolds with certain cell decompositions, where the cell action can be calculated explicitly (section \ref{section: exact results for cell action}). We regard these examples as preparation for the explicit examples of effective action on de Rham cohomology, which will be discussed in section \ref{section: S on coh examples}.

Section \ref{section: fin-dim argument} is a digression where we give a sketch of finite-dimensional argument, why the cell action for cube satisfies quantum master equation (which by gluing construction automatically implies that cell action for any cubical CW-complex solves QME).

\subsection{Tensor product of induction data}
\label{section: tensor product for ind data}
Let $(V_1,d_1)$ and $(V_2,d_2)$ be two cochain complexes, $V'_1 \hra V_1$ and $V'_2\hra V_2$ --- two retracts. Let also $(\iota_1,r_1,K_1)$ be induction data from $V_1$ to $V'_1$ and $(\iota_2,r_2,K_2)$ be induction data from $V_2$ to $V'_2$.  Then we may define two sets of induction data from tensor product of complexes $(V_1\otimes V_2, d_1\otimes\id+\id\otimes d_2)$ to the tensor product of retracts $(V'_1\otimes V'_2, d_1\otimes \id+\id\otimes d_2)$:
\begin{eqnarray}(\iota_1,r_1,K_1)\otimes_L (\iota_2,r_2,K_2)&=&(\iota_1\otimes\iota_2,r_1\otimes r_2,K_1\otimes\id+\PP'_1\otimes K_2)
\label{tensor L}\\
(\iota_1,r_1,K_1)\otimes_R (\iota_2,r_2,K_2)&=&(\iota_1\otimes\iota_2,r_1\otimes r_2,\id\otimes K_2+K_1\otimes \PP'_2)\label{tensor R}
\end{eqnarray}
where $\PP'_1=\iota_1 r_1$, $\PP'_2=\iota_2 r_2$ are IR projectors for $V_1$ and $V_2$ (we will also need the UV projectors $\PP''_1=\id-\PP'_1$, $\PP''_2=\id-\PP'_2$). We call (\ref{tensor L}) and (\ref{tensor R}) the left and right tensor product of induction data, respectively. Left tensor product may be understood as contracting first $V_1$ and then $V_2$, i.e. as the composition of inductions
$$V_1\otimes V_2\xra{(\iota_1,r_1,K_1)\otimes\id}V'_1\otimes V_2\xra{\id\otimes(\iota_2,r_2,K_2)}V'_1\otimes V'_2$$
and the composition (\ref{composition of iota,r,K}) is
$$\left(\id\otimes(\iota_2,r_2,K_2)\right)\circ\left((\iota_1,r_1,K_1)\otimes\id\right)=(\iota_1\otimes\iota_2,r_1\otimes r_2,
K_1\otimes\id+\iota_1\circ r_1\otimes K_2)=(\iota_1,r_1,K_1)\otimes_L (\iota_2,r_2,K_2)$$
According to the Statement \ref{statement: iterated induction}, this definition  implies that the tensor product of induction data automatically satisfies axioms (\ref{ind data axiom1}--\ref{ind data axiom6}). Analogously, the right tensor product is interpreted as contracting first $V_2$ and then $V_1$:
$$V_1\otimes V_2\xra{\id\otimes(\iota_2,r_2,K_2)}V_1\otimes V'_2\xra{(\iota_1,r_1,K_1)\otimes\id}V'_1\otimes V'_2$$
and the composition
$$\left((\iota_1,r_1,K_1)\otimes\id\right)\circ\left(\id\otimes(\iota_2,r_2,K_2)\right)=(\iota_1\otimes\iota_2,r_1\otimes r_2,\id\otimes K_2+K_1\otimes \iota_2\circ r_2)=(\iota_1,r_1,K_1)\otimes_R (\iota_2,r_2,K_2)$$
also automatically satisfies (\ref{ind data axiom1}--\ref{ind data axiom6}).

For the case $(V_1,d_1)=(V_2,d_2)$, $(V'_1,d_1)=(V'_2,d_2)$ it is reasonable to define also the symmetric tensor square
\begin{multline*}(\iota,r,K)^{\otimes_\sym 2}=(\iota\otimes \iota,r\otimes r,K\otimes\frac{\id+\PP'}{2}+ \frac{\id+\PP'}{2}\otimes K)=\\=
(\iota\otimes \iota,r\otimes r,K\otimes(\PP'+\frac{1}{2}\PP'')+ (\PP'+\frac{1}{2}\PP'')\otimes K)\end{multline*}
Since this is a linear combination (half-sum) of the left and right tensor products, axioms (\ref{ind data axiom1}--\ref{ind data axiom5}) are automatically satisfied by $(\iota,r,K)\otimes_\sym (\iota,r,K)$. Axiom (\ref{ind data axiom6}) is satisfied, since operators $K,\PP',\PP''$ (building blocks of the symmetric tensor square for chain homotopy) mutually commute:
\begin{eqnarray*}(K^{\otimes_\sym 2})^2&=&\left(K\otimes(\PP'+\frac{1}{2}\PP'')+ (\PP'+\frac{1}{2}\PP'')\otimes K\right)
\left(K\otimes(\PP'+\frac{1}{2}\PP'')+ (\PP'+\frac{1}{2}\PP'')\otimes K\right)\\&=&
K^2\otimes (\PP'+\frac{1}{2}\PP'')^2+(\PP'+\frac{1}{2}\PP'')^2\otimes K^2+\\
&&+K(\PP'+\frac{1}{2}\PP'')\otimes(\PP'+\frac{1}{2}\PP'')K-
(\PP'+\frac{1}{2}\PP'')K\otimes K(\PP'+\frac{1}{2}\PP'')\\&=&0
\end{eqnarray*}

Similarly, given a collection of complexes $\{(V_i,d_i)\}_{i=1}^n$, retracts $\{(V'_i,d_i)\}_{i=1}^n$ and induction data
$\{(\iota_i,r_i,K_i)\}_{i=1}^n$, we can define the tensor product of induction data
$$V_1\otimes\cdots\otimes V_n\xra{(\iota_1\otimes\cdots\otimes\iota_n,r_1\otimes\cdots\otimes r_n,(K_1\otimes\cdots\otimes K_n)_\pi)} V_1\otimes\cdots\otimes V_n$$
where we have $n!$ variants for the product of chain homotopies $(K_1\otimes\cdots\otimes K_n)_\pi$, depending on the order $\pi^{-1}\in S_n$ in which we contract factors in $V_1\otimes\cdots\otimes V_n$. Namely, if $\pi:(1\cdots n)\mapsto (1\cdots n)$ is a permutation, then
\begin{multline}(K_1\otimes\cdots\otimes K_n)_\pi
=\sum_{i=1}^n \underbrace{(\PP'+\theta(\pi(1)-\pi(i))\PP'')\otimes \cdots\otimes
(\PP'+\theta(\pi(i-1)-\pi(i))\PP'')}_{i-1}\otimes K\otimes\\
\otimes \underbrace{(\PP'+\theta(\pi(i+1)-\pi(i))\PP'')\otimes\cdots\otimes (\PP'+\theta(\pi(n)-\pi(i))\PP'')}_{n-i}
\label{tensor general}
\end{multline}
where $\theta$ stands for the unit step function. Of particular importance is the symmetric case, when all the complexes $(V_i,d_i)$ coincide, retracts $(V'_i,d_i)$ coincide and induction data $\{(\iota_i,r_i,K_i)\}_{i=1}^n$ coincide. Then we define the symmetric tensor power of the chain homotopy by averaging over all $n!$ orders of retracting the factors:
\be K^{\otimes_\sym n}=\frac{1}{n!}\sum_{\pi\in S_n}(K\otimes\cdots\otimes K)_\pi \label{tensor averaging}\ee
For instance, for $n=2$:
\be K^{\otimes_\sym 2}=K\otimes\PP'+\PP'\otimes K+\frac{1}{2}(K\otimes\PP''+\PP''\otimes K) \label{K sym square}\ee
for $n=3$:
\begin{multline}
K^{\otimes_\sym 3}=K\otimes\PP'\otimes\PP'+\PP'\otimes K\otimes\PP'+\PP'\otimes\PP'\otimes K+\\
+\frac{1}{2}(K\otimes\PP'\otimes\PP''+K\otimes\PP''\otimes\PP'+\PP'\otimes K\otimes\PP''+\PP''\otimes K\otimes\PP'+
\PP'\otimes\PP''\otimes K+\PP''\otimes\PP'\otimes K)+\\
+\frac{1}{3}(K\otimes\PP''\otimes\PP''+\PP''\otimes K\otimes\PP''+\PP''\otimes\PP''\otimes K) \label{K sym cube}
\end{multline}
Definitions (\ref{tensor general},\ref{tensor averaging}) and the following elementary property of averaging over permutations:
$$\frac{1}{n!}\sum_{\pi\in S_n}\theta(\pi(1)-\pi(2))\cdots\theta(\pi(1)-\pi(i))=\frac{1}{i}$$
imply that the product of $K$, $n-i-1$ copies of $\PP'$ and $i$ copies of $\PP''$ (in arbitrary order) enters the expression for $K^{\otimes_\sym n}$ with coefficient $\frac{1}{i+1}$. This implies the following representation for $K^{\otimes_\sym n}$:
$$K^{\otimes_\sym n}=\int_0^1 (\PP'+(1-\lambda)\PP''+d \lambda\; K)\otimes\cdots\otimes(\PP'+(1-\lambda)\PP''+ d \lambda\; K)$$
where we introduced the auxiliary parameter $\lambda\in[0,1]$ and the integrand is a differential form on the interval $[0,1]$. Therefore we introduce the ``extended chain homotopy''
\be K^{\lambda,d\lambda}=(1-\lambda)\; \id+\lambda\;\PP'+d\lambda\; K\; \in\;\Omega^\bt([0,1])\otimes \mr{End}(V)  \label{extended K}\ee
depending on $\lambda,d\lambda$. In terms of $K^{\lambda,d\lambda}$ the symmetric tensor power of the chain homotopy is simply
\be K^{\otimes_\sym n}=\int_0^1 (K^{\lambda,d\lambda})^{\otimes n} \label{K^sym n}\ee
where tensor power in the integrand $(K^{\lambda,d\lambda})^{\otimes n}\in\Omega^\bt([0,1])\otimes \mr{End}(V^{\otimes n})$ is the tensor power in endomorphisms of $V$ and $n$-fold wedge product in forms on $[0,1]$.

\textbf{Remark.} One can also write $K^{\lambda,d\lambda}$ in the elegant exponential form (a kind of heat kernel):
$$K^{\lambda,d\lambda}=e^{(d_\tau-[d,\bt])\circ (\tau K)}$$
where the new parameter $\tau\in [0,+\infty]$ (the ``proper time'') is related to $\lambda$ by $\lambda=1-e^{-\tau}$ and we denoted the de Rham operator on the ray $[0,+\infty]$ by $d_\tau$ to distinguish it from the differential $d$ in $V$. Note that operator $d_\tau-[d,\bt]$ appearing in the exponential is a differential for the graded associative algebra $\Omega^\bt([0,+\infty])\otimes \mr{End}(V)$.

\subsection{Induction data for a cubical complex, cell $BF$ action on a cubical complex and its cell locality property}
\label{section: cell action}
Let $I=[0,1]$ be the unit interval and let $I^n=\underbrace{I\times\cdots\times I}_n$ be the $n$-dimensional cube. The de Rham complex $\Omega^\bt(I^n)$ of $n$-cube can be represented as the tensor power of the de Rham complex of interval:
$$\Omega^\bt(I^n)=\underbrace{\Omega^\bt(I)\otimes\cdots\otimes\Omega^\bt(I)}_n=(\Omega^\bt(I))^{\otimes n}$$
The standard cell decomposition of the cube $I^n$, consisting of its faces of all dimensions (including the bulk cell), is the (Cartesian) power of the standard cell complex of the interval:
$$\Xi_{I^n}= \underbrace{\{0,1,I\}\times\cdots\times \{0,1,I\}}_n$$
We will denote faces of $n$-cube by sequences $\zeta=\zeta^1\cdots\zeta^n$ with $\zeta^i\in\{0,1,I\}$. Dimension of a face is $|\zeta^1\cdots\zeta^n|=|\zeta^1|+\cdots+|\zeta^n|$. The cochain complex of standard cell decomposition of the cube is the tensor power of the cochain complex for interval:
$$C^\bt(I^n)=(C^\bt(I))^{\otimes n}=(\RR e_0\oplus\RR e_1\oplus\RR e_{01})^{\otimes n}$$
The respective notation for basis cochains on the cube is: $e_{\zeta^1\cdots\zeta^n}=e_{\zeta^1}\otimes\cdots\otimes e_{\zeta^n}$.

We define the induction data from $\Omega^\bt(I^n)$ to $C^\bt(I^n)$ as the symmetric tensor power of standard induction data for the interval (sections \ref{section: Whitney forms},\ref{section: Dupont's homotopy}):
\begin{eqnarray}
\iota_{I^n}&=&\underbrace{\iota_I\otimes\cdots\otimes\iota_I}_n\\
r_{I^n}&=&\underbrace{r_I\otimes\cdots\otimes r_I}_n\\
K_{I^n}&=&K_I^{\otimes_\sym n}=\int_{I_\aux} ((1-\lambda)\;\id+\lambda\;\PP'_I+d\lambda\; K_I)^{\otimes n}\label{cube K}
\end{eqnarray}
where $I_\aux=[0,1]$ denotes the unit interval where the auxiliary parameter $\lambda$ lives.
Thus embedding $\iota_{I^n}$ maps basis cochains to products of Whitney forms (the ``cubical Whitney forms''):
\be \iota_{I^n}:e_{\zeta^1\cdots\zeta^n}\mapsto \chi_{\zeta^1\cdots\zeta^n}=\chi_{\zeta^1}\otimes\cdots\otimes\chi_{\zeta^n} \label{cube chi}\ee
For instance, for the square $I^2$ with coordinates $0\leq t_1,t_2\leq 1$ the cubical Whitney forms are
\begin{multline*}
\chi_{00}=(1-t_1)(1-t_2),\;\chi_{01}=(1-t_1)t_2,\;\chi_{10}=t_1(1-t_2),\;\chi_{11}=t_1t_2,\\
\chi_{I0}=(1-t_2)dt_1,\;\chi_{I1}=t_2dt_1,\;\chi_{0I}=(1-t_1)dt_2,\;\chi_{1I}=t_1dt_2,\;\chi_{II}=dt_1 dt_2
\end{multline*}
Retraction $r_{I^n}$ acts by integrals over faces of the cube:
$$r_{I^n}: \alpha\mapsto\sum_{\zeta^1,\cdots,\zeta^n\in\{0,1,I\}}e_{\zeta^1\cdots\zeta^n}\int_{\zeta^1\cdots\zeta^n}\alpha$$
Chain homotopy $K_{I^n}$ acts on a product of $n$ forms on the interval $\alpha_i\in\Omega^\bt(I)$ as
\begin{multline*}
K_{I^n}: \alpha_1(t_1)\wedge\cdots\wedge\alpha_n(t_n)\mapsto \\
\int_{I_\aux}
\left((1-\lambda)\;\alpha_1(t_1)+\lambda\left(\alpha_1(0)(1-t_1)+\alpha_1(1)t_1+\int_0^1\alpha_1(\tilde{t}_1)\cdot dt_1\right)+\right.\\
\left.+d\lambda \left(\int_0^{t_1}\alpha_1(\tilde{t}_1)-t_1\int_0^{1}\alpha_1(\tilde{t}_1)\right)\right)\wedge\cdots\wedge\\
\wedge\left((1-\lambda)\;\alpha_n(t_n)+\lambda\left(\alpha_n(0)(1-t_n)+\alpha_n(1)t_n+\int_0^1\alpha_n(\tilde{t}_n)\cdot dt_n\right)+\right.\\
\left.+
d\lambda \left(\int_0^{t_n}\alpha_n(\tilde{t}_n)-t_n\int_0^{1}\alpha_n(\tilde{t}_n)\right)\right)
\end{multline*}

Induction data $(\iota_{I^n},r_{I^n},K_{I^n})$ are consistent with the action of $S_n\ltimes\ZZ_2^n$ (the symmetry group of $n$-cube) on differential forms $\Omega^\bt(I^n)$ and cochains $C^\bt(I^n)$. We mean that the generator of $i$-th copy of $\ZZ_2$ inverts the orientation of $i$-th interval, acting on forms by  $\mr{inv}_i^*:\Omega^\bt(I^n)\ra\Omega^\bt(I^n)$, where $\mr{inv}_i:t_i\mapsto 1-t_i$ and $t_j\mapsto t_j$ for $j\neq i$, and on cochains of $i$-th interval by $e_0\mapsto e_1, e_1\mapsto e_0, e_I\mapsto -e_I$. The $S_n$-symmetry permutes factors in $\Omega^\bt(I^n)=\Omega^\bt(I)\otimes\cdots\otimes \Omega^\bt(I)$ and $C^\bt(I^n)=C^\bt(I)\otimes\cdots\otimes C^\bt(I)$. Consistency of induction data with the action of $S_n$ is implied by construction of $(\iota_{I^n},r_{I^n},K_{I^n})$ as the symmetric tensor power of $(\iota_{I},r_{I},K_{I})$.

Next, the induction data $(\iota_{I^n},r_{I^n},K_{I^n})$ are consistent with restriction  of forms on cochains to faces: $\Omega^\bt(I^n)\ra\Omega^\bt(\zeta)$, $C^\bt(I^n)\ra C^\bt(\zeta)$. I.e. the following three properties hold:
\begin{eqnarray}
\left.\left(\iota_{I^n}\left(\sum_{\zeta'}e_{\zeta'}
\alpha^{\zeta'}\right)\right)\right|_{\zeta}&=&
\iota_{\zeta}\left(\left.\left(\sum_{\zeta'}e_{\zeta'}
\alpha^{\zeta'}\right)\right|_{\zeta}\right)\label{cube ind data 1}\\
(r_{I^n}\alpha)|_{\zeta}&=&r_{\zeta}(\alpha|_{\zeta})\label{cube ind data 2}\\
(K_{I^n}\alpha)|_{\zeta}&=&K_{\zeta}(\alpha|_{\zeta})\label{cube ind data 3}
\end{eqnarray}
for any face $\zeta\subset I^n$, differential form $\alpha\in\Omega^\bt(I^n)$ and cochain $\sum_{\zeta'}e_{\zeta'}
\alpha^{\zeta'}\in C^\bt(I^n)$. Properties (\ref{cube ind data 1},\ref{cube ind data 2}) obviously follow from the corresponding properties for the interval. Let us check (\ref{cube ind data 3}). Using the $S_n\ltimes\ZZ_2^n$-symmetry, we transform the face $\zeta$ to standard form
$\underbrace{I\cdots I}_m\underbrace{0\cdots 0}_{n-m}$ for some $m$ (dimension of the face). Next, forms $\alpha\in\Omega^\bt(I)$ on the interval satisfy
$$(\PP'_I\alpha)|_0=\alpha(0),\;(\PP''_I\alpha)|_0=0,\;(K_I\alpha)|_0=0$$
Therefore $(K_{I^n}\bt)|_{I\cdots I 0\cdots 0}$ acts on the factorized forms $\alpha_1\otimes\cdots\otimes\alpha_n\in\Omega^\bt(I^n)$ as
\begin{multline*}(K_{I^n}(\alpha_1\otimes\cdots\otimes\alpha_n))|_{I\cdots I 0\cdots 0}\\
=\int_{I_\aux} ((1-\lambda)\;\alpha_1+\lambda\;\PP'_I\alpha_1+d\lambda\; K_I\alpha_1)\otimes\cdots\otimes ((1-\lambda)\;\alpha_m+\lambda\;\PP'_I\alpha_m+d\lambda\; K_I\alpha_m)
\otimes \alpha_{m+1}(0)\otimes\cdots\otimes\alpha_{n}(0)
\end{multline*}
Since the last $n-m$ factors yield just a multiplication by constant, this coincides with $K_{I\cdots I 0\cdots 0}((\alpha_1\otimes\cdots\otimes\alpha_n)|_{I\cdots I 0\cdots 0})=K_{I\cdots I 0\cdots 0}(\alpha_1\otimes\cdots\otimes\alpha_m)\cdot \alpha_{m+1}(0)\cdots\alpha_n(0)$. Finally (\ref{cube ind data 3}) is extended to all (not necessary factorizing) forms by linearity.

Now, using properties (\ref{cube ind data 1}--\ref{cube ind data 3}), we construct the induction data $(\iota_\Xi,r_\Xi,K_\Xi)$ for any cell decomposition $\Xi$ of any manifold $M$, where all cells are cubes. Namely, we define the action of embedding $\iota_\Xi:C^\bt(\Xi)\ra\Omega^\bt(\Xi)$ on a cell cochain $\sum_{\zeta\in\Xi}e_\zeta\alpha^\zeta\in C^\bt(\Xi)$ via restrictions to faces:
$$\left.\left(\iota_\Xi\left(\sum_{\zeta'\in\Xi}e_{\zeta'}\alpha^{\zeta'}\right)\right)\right|_{\zeta}
=\sum_{\zeta'\in\zeta}\chi^{(\zeta)}_{\zeta'}\alpha^{\zeta'}$$
where $\chi^{(\zeta)}_{\zeta'}$ is the cubical Whitney form on cube $\zeta$, associated to the face $\zeta'\subset\zeta$. It is natural to call the images of basis cochains $e_\zeta$ in $\Omega^\bt(\Xi)$ the basis cubical Whitney forms for the cell decomposition: $\iota_\Xi(e_\zeta)=\chi^{(\Xi)}_\zeta\in\Omega^{|\zeta|}(M)$. Contrary to the case of simplicial complex, these Whitney forms are not piecewise-linear, but piecewise-polynomial. Retraction $r_\Xi:\Omega^\bt(M)\ra C^\bt(\Xi)$ is given by integration over cells
$$r_\Xi\alpha=\sum_{\zeta\in \Xi}e_\zeta \int_\zeta\alpha$$
Chain homotopy $K_\Xi$ is defined via restrictions to faces:
$$(K_\Xi\alpha)|_\zeta=K_\zeta(\alpha|_\zeta)$$
for all cells $\zeta\in\Xi$ and any form $\alpha\in\Omega^\bt(M)$. Axioms of induction data (\ref{ind data axiom1}--\ref{ind data axiom6}) are satisfied for $(\iota_\Xi,r_\Xi,K_\Xi)$, since they are satisfied on every cell $\Xi$.

We extend the induction data $(\iota_\Xi,r_\Xi,K_\Xi)$ to the case of induction from $\Omega^\bt(M,\g)$ to $C^\bt(\Xi,\g)$ by $\g$-linearity.

We define the cell $BF$ action $S_\Xi\in\Fun(\FF_\Xi)$ on the cubical cell complex $\Xi$ as the effective $BF_\infty$ action, induced from the topological $BF$ theory (i.e. the abstract $BF$ theory, associated to the DGLA structure on $\Omega^\bt(M,\g)$) on the space $\FF_\Xi=T^*[-1](C^\bt(\Xi,\g)[1])$ and given by perturbation expansion (\ref{sum over trees},\ref{sum over loops}).

Similarly to the simplicial $BF$ theory, we can either use the real-valued coordinates
$(\omega^{\zeta a},p_{\zeta a})$ on the space of fields $\FF_\Xi=T^*[-1](C^\bt(\Xi,\g)[1])$, or the $\g$-
and $\g^*$-valued coordinates $\omega^\zeta=\sum_a T_a\omega^{\zeta a}, p_\zeta=\sum_a T^a p_{\zeta a}$ (recall that we $(T_a)$ is the basis in $\g$ and $(T^a)$ is the dual basis in $\g^*$). Ghost numbers for the coordinates are $\gh(\omega^\zeta)=\gh(\omega^{\zeta a})=1-|\zeta|$, $\gh(p_\zeta)=\gh(p_{\zeta a})=-2+|\zeta|$. Super-fields of the cell $BF$ theory are
$$\omega_\Xi=\sum_{\zeta,a}T_a e_{\zeta}\omega^{\zeta a}=\sum_{\zeta}e_{\zeta}\omega^{\zeta},\quad
p_\Xi=\sum_{\zeta,a}p_{\zeta a}T^a e^\zeta=\sum_\zeta p_\zeta e^\zeta$$

Cell action $S_\Xi(\omega_\Xi,p_\Xi)$ features the property, completely analogous to the Theorem \ref{thm: simplicial locality}.
\begin{thm}[Cell locality of the cell $BF$ action]
\label{thm: cell locality}
There is a family of universal functions (reduced $BF$ actions for the cubes)
$$\bar{S}_{I^n}(\{\omega^\zeta\}_{\zeta\subset I^n},p_{I_n})\in\Fun(\g\otimes C^\bt(I^n)[1]\oplus\g^*\otimes C_n (I^n)[-2])$$
for $n\geq 0$, such that for any cubical cell decomposition $\Xi$ of any manifold $M$ the cell action $S_\Xi$ is decomposed into contributions of individual cells:
$$S_\Xi(\omega_\Xi,p_\Xi)=\sum_{\zeta\in\Xi}\bar{S}_\zeta (\{\omega^{\zeta'}\}_{\zeta'\subset\zeta},p_\zeta)$$
Perturbation expansion for $\bar{S}_{I^n}$ is
\begin{multline*}
\bar{S}_{I^n}=\left<p_{I^n},\sum_{\zeta_1\subset I^n}d^{I^n}_{\zeta_1}\omega^{\zeta_1}+
\sum_{T\in{\bf{T}_\mr{nonPl}}:\,|T|\geq 2}\sum_{\zeta_1,\ldots,\zeta_{|T|}\subset I^n}\frac{1}{|\Aut(T)|}\cdot\right.\\ \left.\cdot \int_{I^n}
\Iter_{T;-K_{I^n}[\bt,\bt];[\bt,\bt]}(\chi_{\zeta_1} \omega^{\zeta_1},\ldots,\chi_{\zeta_{|T|}} \omega^{\zeta_{|T|}})\right>_\g-\\
-\hbar
\sum_{L\in{\bf{L}_\mr{nonPl}}}\sum_{\zeta_1,\ldots,\zeta_{|L|}\subset I^n}\frac{1}{|\Aut(L)|}\Loop_{L;-K_{I^n}[\bt,\bt];\Omega^\bt_0(I^n,\g)}
(\chi_{\zeta_1} \omega^{\zeta_1},\ldots,\chi_{\zeta_{|L|}} \omega^{\zeta_{|L|}})
\end{multline*}
where $\Omega^\bt_0(I^n)=\{\alpha\in\Omega^\bt(I^n):\,\alpha|_{\dd I^n}=0\}$
\end{thm}

The proof literally follows the proof of Theorem \ref{thm: simplicial locality} for simplicial case, since it uses only the consistency of induction data with face restrictions.

Thus we again come to the problem of calculating the family of universal functions $\bar{S}_{I^D}$ for $D\geq 0$. Note that, since in dimensions $D=0,1$ the cube coincides with the simplex, we know exact answers for $\bar{S}_{I^D}$ with $D=0,1$ from (\ref{Sbar Delta^0}) and (\ref{interval thm eq1},\ref{interval thm eq2}).

\subsection{Factorization of Feynman diagrams, perturbative result for the $D$-cube}
\label{section: cube factorization}
Following the lines of section \ref{section: simplex pert} we decompose the reduced action for the cube $I^D$ into contributions of Feynman diagrams:
$$\bar{S}_{I^D}=\sum_{T\in\bf{T}_\mr{nonPl}}\bar{S}_{I^D,T}+\hbar \sum_{L\in\bf{L}_\mr{nonPl}}\bar{S}_{I^D,L}$$
and split the contribution of each diagram into the de Rham part and the $\g$-part:
\begin{multline*}\bar{S}_{I^D,T}=\frac{1}{|\Aut(T)|}\sum_{\zeta_1,\ldots,\zeta_{|T|}\subset I^D}
\left<p_{I^D},\int_{I^D}\Iter_{T;-K_{I^D}[\bt,\bt];[\bt,\bt]}
(\chi_{\zeta_1} \omega^{\zeta_1},\ldots,\chi_{\zeta_{|T|}} \omega^{\zeta_{|T|}})\right>_\g\\
=\frac{1}{|\Aut(T)|}\sum_{\zeta_1,\ldots,\zeta_{|T|}\subset I^D}\int_{I^D}
\Iter_{T;-K_{I^D}(\bt\wedge\bt);(\bt\wedge\bt)}(\chi_{\zeta_1} ,\ldots,\chi_{\zeta_{|T|}})\cdot\\
\cdot <p_{I^D},\Iter_{T;[\bt,\bt];[\bt,\bt]}(\omega^{\zeta_1},\ldots,\omega^{\zeta_{|T|}})>_\g \e_{T}(|\zeta_1|,
\ldots,|\zeta_{|T|}|)\\=
\frac{1}{|\Aut(T)|}\sum_{\zeta_1,\ldots,\zeta_{|T|}\subset I^D}\e_{T}(|\zeta_1|,
\ldots,|\zeta_{|T|}|)\cdot C_{I^D,T}(\zeta_1,\ldots,\zeta_{|T|})\cdot \\ \cdot <p_{I^D},\Iter_{T;[\bt,\bt];[\bt,\bt]}(\omega^{\zeta_1},\ldots,\omega^{\zeta_{|T|}})>_\g\\
\bar{S}_{I^D,L}=-\frac{1}{|\Aut(L)|}\sum_{\zeta_1,\ldots,\zeta_{|L|}\subset I^D}\Loop_{L;-K_{I^D}[\bt,\bt];\Omega^\bt_0(I^D,\g)}
(\chi_{\zeta_1} \omega^{\zeta_1},\ldots,\chi_{\zeta_{|L|}} \omega^{\zeta_{|L|}})\\
=-\frac{1}{|\Aut(L)|}\sum_{\zeta_1,\ldots,\zeta_{|L|}\subset I^D}\Loop_{L;-K_{I^D}(\bt\wedge\bt);\Omega^\bt_0(I^D)}
(\chi_{\zeta_1},\ldots,\chi_{\zeta_{|L|}} )
\cdot \\ \cdot \Loop_{L;[\bt,\bt];\g}(\omega^{\zeta_1},\ldots,\omega^{\zeta_{|L|}})\e_L(|\zeta_1|,\ldots,|\zeta_{|L|}|)\\
=-\frac{1}{|\Aut(L)|}\sum_{\zeta_1,\ldots,\zeta_{|L|}\subset I^D}
\e_L(|\zeta_1|,\ldots,|\zeta_{|L|}|)\cdot
C_{I^D,L}(\zeta_1,\ldots,\zeta_{|L|})\cdot
\Loop_{L;[\bt,\bt];\g}(\omega^{\zeta_1},\ldots,\omega^{\zeta_{|L|}})
\end{multline*}
where we introduced the notation for de Rham parts of Feynman diagrams:
\begin{eqnarray*}
 C_{I^D,T}(\zeta_1,\ldots,\zeta_{|T|})&=&
 \int_{I^D}\Iter_{T;-K_{I^D}(\bt\wedge\bt);(\bt\wedge\bt)}(\chi_{\zeta_1} ,\ldots,\chi_{\zeta_{|T|}})\\
 C_{I^D,L}(\zeta_1,\ldots,\zeta_{|L|})&=& \Loop_{L;-K_{I^D}(\bt\wedge\bt);\Omega^\bt_0(I^D)}
(\chi_{\zeta_1},\ldots,\chi_{\zeta_{|L|}})
\end{eqnarray*}
Signs $\e_{T},\e_L=\pm 1$ arise from permuting variables $\omega^\zeta$ with cubical Whitney forms $\chi_\zeta$ and operators $K_{I^D}$,
and, hence, depend on dimensions of faces only and coincide with signs $\e_T,\e_L$ introduced in the section
\ref{section: simplex pert}.

\textbf{Calculating de Rham parts of trees.}
Contribution of the ``tree'' with one leaf $(*)$ is understood as
\be C_{I^D,(*)}(\zeta)=d^{I^D}_\zeta=\int_{I^D}d\chi_\zeta=\left\{\begin{array}{ll}
(-1)^{i+1},&\text{ if }\zeta=\underbrace{I\cdots I}_{i-1} 1 \underbrace{I\cdots I }_{D-i}\\
(-1)^{i},&\text{ if }\zeta=\underbrace{I\cdots I}_{i-1} 0 \underbrace{I\cdots I }_{D-i}\\
0,&\text{ otherwise}
\end{array}\right. \label{cube d matrix}\ee

Next let us calculate the contribution of the next simplest tree $C_{I^D,(**)}$:
\begin{multline*}
C_{I^D,(**)}(\zeta_1,\zeta_2)=\int_{I^D}\chi_{\zeta_1}\wedge\chi_{\zeta_2}=
\int_{I^D}(\chi_{\zeta_1^1}\otimes\cdots\otimes\chi_{\zeta_1^D})\wedge (\chi_{\zeta_2^1}\otimes\cdots\otimes\chi_{\zeta_2^D})\\
=(-1)^{|\zeta_1^D|\cdot(|\zeta_2^1|+\cdots+|\zeta_2^{D-1}|)\;+\;|\zeta_1^{D-1}|\cdot(|\zeta_2^1|+\cdots+|\zeta_2^{D-2}|)\;+\;\cdots\;+\;|\zeta_1^2|\cdot |\zeta_2^1|}
\int_{I^D}(\chi_{\zeta_1^1}\wedge\chi_{\zeta_2^1})\otimes\cdots\otimes(\chi_{\zeta_1^D}\wedge\chi_{\zeta_2^D})\\
=(-1)^{\sum_{1\leq j<i\leq D}|\zeta_1^i|\cdot |\zeta_2^j|}\prod_{i=1}^D\int_I \chi_{\zeta_1^i}\wedge \chi_{\zeta_2^i}
\end{multline*}
Therefore the value of $C_{I^D,(**)}$ for a pair of faces $\zeta_1=\zeta_1^1\cdots\zeta_1^D,\;\zeta_2=\zeta_2^1\cdots\zeta_2^D$ of the $D$-cube is represented (up to sign due to reshuffling of the tensor product) as the product of values of $C_{I,(**)}$ on the interval, evaluated on projections $\zeta_1^i,\zeta_2^i$ of the initial faces to $i$-th interval:
$$C_{I^D,(**)}(\zeta_1,\zeta_2)=(-1)^{\sum_{1\leq j<i\leq D}|\zeta_1^i|\cdot |\zeta_2^j|}\prod_{i=1}^D C_{I,(**)}(\zeta_1^i,\zeta_2^i)$$
The problem of calculating $C_{I,(**)}(\zeta_1^i,\zeta_2^i)=\int_I\chi_{\zeta_1^i}\wedge\chi_{\zeta_2^i}$ on the interval is in turn solved straightforwardly by considering all pairs of faces of the interval:
\be C_{I,(**)}(0,I)=C_{I,(**)}(I,0)=C_{I,(**)}(1,I)=C_{I,(**)}(I,1)=\frac{1}{2} \label{cube (**) values}\ee
and other values of $C_{I,(**)}$ are zero for dimensional reasons (total form degree has to be $\deg\chi_{\zeta_1^i}+\deg\chi_{\zeta_2^i}=1$).

Next let us calculate $C_{I^D,(*(**))}$:
\begin{multline*}
C_{I^D,(*(**))}(\zeta_1,\zeta_2,\zeta_3)=-\int_{I^D}\chi_{\zeta_1}\wedge K_{I^D}(\chi_{\zeta_2}\wedge\chi_{\zeta_3})\\
=-\int_{I^D}(\chi_{\zeta_1^1}\otimes\cdots\otimes\chi_{\zeta_1^D})\wedge
\int_{I_\aux} ((1-\lambda)\;\id+\lambda\;\PP'_I+d\lambda\; K_I)^{\otimes D}
((\chi_{\zeta_2^1}\otimes\cdots\otimes\chi_{\zeta_2^D})\wedge(\chi_{\zeta_3^1}\otimes\cdots\otimes\chi_{\zeta_3^D})) \\
=-(-1)^{|\zeta_1|+D}(-1)^{\sum_{1\leq j<i\leq D}|\zeta_1^i|\cdot|\zeta_2^j|+|\zeta_1^i|\cdot|\zeta_3^j|+|\zeta_2^i|\cdot|\zeta_3^j|}\cdot\\ \cdot
\int_{I_\aux} \int_{I^D}
(\chi_{\zeta_1^1}\wedge K_I^{\lambda,d\lambda}(\chi_{\zeta_2^1}\wedge\chi_{\zeta_3^1}))\otimes\cdots\otimes
(\chi_{\zeta_1^D}\wedge K_I^{\lambda,d\lambda}(\chi_{\zeta_2^D}\wedge\chi_{\zeta_3^D}))\\
=-(-1)^{|\zeta_1|+D}(-1)^{\sum_{1\leq j<i\leq D}|\zeta_1^i|\cdot|\zeta_2^j|+|\zeta_1^i|\cdot|\zeta_3^j|+|\zeta_2^i|\cdot|\zeta_3^j|}
\int_{I_\aux} \prod_{i=1}^D \int_I \chi_{\zeta_1^i}\wedge K_I^{\lambda,d\lambda}(\chi_{\zeta_2^i}\wedge\chi_{\zeta_3^i})
\end{multline*}
Sign $(-1)^{|\zeta_1|+D}$ comes from moving $\int_{I_\aux}$ to the left, and the second sign is due to the reshuffling of tensor product. Denote
$$C_{I,(*(**))}^{\lambda,d\lambda}(\zeta_1^i,\zeta_2^i,\zeta_3^i)= \int_I\chi_{\zeta_1^i}\wedge K_I^{\lambda,d\lambda}(\chi_{\zeta_2^i}\wedge\chi_{\zeta_3^i})$$
i.e. $C_{I,(*(**))}^{\lambda,d\lambda}$ is a linear function of $\lambda,d\lambda$, depending on the triple of faces of the interval.
In terms of $C_{I,(*(**))}^{\lambda,d\lambda}$ the de Rham part of the contribution of Feynman tree $(*(**))$ in the restricted action for $D$-cube is
\begin{multline*}C_{I^D,(*(**))}(\zeta_1,\zeta_2,\zeta_3)\\=
-(-1)^{|\zeta_1|+D}(-1)^{\sum_{1\leq j<i\leq D}|\zeta_1^i|\cdot|\zeta_2^j|+|\zeta_1^i|\cdot|\zeta_3^j|+|\zeta_2^i|\cdot|\zeta_3^j|}
\int_{I_\aux} \prod_{i=1}^D C_{I,(*(**))}^{\lambda,d\lambda}(\zeta_1^i,\zeta_2^i,\zeta_3^i)
\end{multline*}
Therefore $C_{I^D,(*(**))}$ again factorizes (up to sign and integration over $\lambda$) into certain universal differential forms $C_{I,(*(**))}^{\lambda,d\lambda}$ in $\lambda$, depending on the triple of faces of the interval. These universal expressions are calculated straightforwardly, considering all triples of faces of the interval and using explicit expressions for Whitney forms on the interval $\chi_0=1-t,\,\chi_1=t,\,\chi_I=dt$ and the explicit formula for $K_{I}^{\lambda,d\lambda}$
\begin{multline*}
K_{I}^{\lambda,d\lambda}(f(t)+g(t)dt)=((1-\lambda)\;\id +\lambda\;\PP'_I+d\lambda\; K_I)\circ(f(t)+g(t)dt)\\
=\left((1-\lambda)\; f(t)+\lambda\;f(0)\cdot (1-t)+\lambda\; f(1)\cdot t\right)+\\
+
\left((1-\lambda)\; g(t)dt+(\lambda\;dt-d\lambda\; t)\int_0^1 g(\tilde{t})d\tilde{t}+d\lambda\;\int_0^t g(\tilde{t})d\tilde{t}\right)
\end{multline*}
for any pair of functions $f(t),g(t)$ on the interval. We obtain the following result for $C_{I,(*(**))}^{\lambda,d\lambda}$:
\begin{multline}
C_{I,(*(**))}^{\lambda,d\lambda}(I,0,0)=C_{I,(*(**))}^{\lambda,d\lambda}(I,1,1)=\frac{1}{3}+\frac{1}{6}\lambda,\quad C_{I,(*(**))}^{\lambda,d\lambda}(I,0,1)=C_{I,(*(**))}^{\lambda,d\lambda}(I,1,0)=\frac{1}{6}-\frac{1}{6}\lambda,\\
C_{I,(*(**))}^{\lambda,d\lambda}(0,0,I)=C_{I,(*(**))}^{\lambda,d\lambda}(0,I,0)=C_{I,(*(**))}^{\lambda,d\lambda}(1,1,I)=
C_{I,(*(**))}^{\lambda,d\lambda}(1,I,1)=\frac{1}{3}-\frac{1}{12}\lambda,\\
C_{I,(*(**))}^{\lambda,d\lambda}(0,1,I)=C_{I,(*(**))}^{\lambda,d\lambda}(0,I,1)=C_{I,(*(**))}^{\lambda,d\lambda}(1,0,I)=
C_{I,(*(**))}^{\lambda,d\lambda}(1,I,0)=\frac{1}{6}+\frac{1}{12}\lambda,\\
C_{I,(*(**))}^{\lambda,d\lambda}(I,0,I)=C_{I,(*(**))}^{\lambda,d\lambda}(I,I,0)=\frac{1}{12}\;d\lambda,\quad
C_{I,(*(**))}^{\lambda,d\lambda}(I,1,I)=C_{I,(*(**))}^{\lambda,d\lambda}(I,I,1)=-\frac{1}{12}\;d\lambda
\label{cube (*(**)) values}
\end{multline}
and for all other triples of faces $C_{I,(*(**))}^{\lambda,d\lambda}$ vanishes.

Continuing this argument to any Feynman tree $T$, we obtain
\begin{multline*}
C_{I^D,T}(\zeta_1,\ldots,\zeta_{|T|})=(-1)^{|T|}\int_{I^D}\Iter_{T;K_{I^D}(\bt\wedge\bt),(\bt\wedge\bt)}(\chi_{\zeta_1},\ldots,\chi_{\zeta_{|T|}})\\
=(-1)^{|T|}\int_{I^D}\Iter_{T;\int_0^1 (K_{I}^{\lambda_k,d\lambda_k})^{\otimes D} (\bt\wedge\bt);(\bt\wedge\bt)}(\chi_{\zeta_1^1}\otimes\cdots\otimes\chi_{\zeta_1^D},\ldots, \chi_{\zeta_{|T|}^1}\otimes\cdots\otimes\chi_{\zeta_{|T|}^D})\\
=\hat\e_{I^D,T}(\zeta_1,\ldots,\zeta_{|T|})\int_{I_\aux^{|T|-2}}
\prod_{i=1}^D C_{I,T}^{\lambda_1,d\lambda_1;\cdots;\lambda_{|T|-2},d\lambda_{|T|-2}}(\zeta_1^i,\ldots,\zeta_{|T|}^i)
\end{multline*}
where we assume that we enumerated the internal edges of the tree (the standard enumeration is the one coming from reading the bracket structure representing the tree from left to right) and associated to every edge its own auxiliary variable $\lambda_k$; $\int_{I_\aux^{|T|-2}}$ means integration over all auxiliary parameters. Sign $\hat\e_{I^D,T}(\zeta_1,\ldots,\zeta_{|T|})=\pm 1$ comes from extracting integrals over auxiliary parameters to the left, from the reshuffling of tensor product and contains the sign $(-1)^{|T|}$ (arising due to changing the propagator $-K_{I^D}$ to $K_{I^D}$). For instance,
\begin{eqnarray*}\hat\e_{I^D,(**)}(\zeta_1,\zeta_2)&=&(-1)^{\sum_{1\leq j<i\leq D}|\zeta_1^i|\cdot |\zeta_2^j|}\\
\hat\e_{I^D,(*(**))}(\zeta_1,\zeta_2,\zeta_3)&=&-(-1)^{|\zeta_1|+D}(-1)^{\sum_{1\leq j<i\leq D}|\zeta_1^i|\cdot|\zeta_2^j|+|\zeta_1^i|\cdot|\zeta_3^j|+|\zeta_2^i|\cdot|\zeta_3^j|}
\end{eqnarray*}
(we assume that orientation on $I_\aux^{|T|-2}$ is given by the volume form $d\lambda_1\cdots d\lambda_{|T|-2}$).
Factors \\ $C_{I,T}^{\lambda_1,d\lambda_1;\cdots;\lambda_{|T|-2},d\lambda_{|T|-2}}(\zeta_1^i,\ldots,\zeta_{|T|}^i)$, into which $C_{I^D,T}(\zeta_1,\ldots,\zeta_{|T|})$ factorizes, depend only on the tree $T$ and a $|T|$-tuple of faces of the interval, and are differential forms in auxiliary variables $\lambda_1,\ldots,\lambda_{|T|-2}$, with which we decorate the internal edges of $T$:
$$C_{I,T}^{\lambda_1,d\lambda_1;\cdots;\lambda_{|T|-2},d\lambda_{|T|-2}}(\zeta_1^i,\ldots,\zeta_{|T|}^i)=
\int_I\Iter_{T;K^{\lambda_k,d\lambda_k}_I(\bt\wedge\bt);(\bt\wedge\bt)}(\chi_{\zeta_1^i},\ldots,\chi_{\zeta_{|T|}^i})$$
For every given tree $T$ one can compute expressions $C_{I,T}^{\{\lambda_k,d\lambda_k\}}$ for all $|T|$-tuples of faces of the interval.

\textbf{Calculating de Rham parts of one-loop diagrams.} Feynman graphs $L$ with cycle of length 1 do not contribute to $\bar{S}_{I^D}$ due to the argument of section \ref{section: simplex pert}: $\g$-parts of these diagrams vanish $\Loop_{L;[\bt,\bt];\g}(\omega^{\zeta_1},\ldots,\omega^{\zeta_{|L|}})=0$ due to unimodularity of $\g$. So let us consider the simplest diagram with cycle of length 2: $L=(*(*\bt))$. Factorization occurs here by the same mechanism as for tree diagrams, but we have to use also the general factorization property of super-traces:
$$\Str_{V_1\otimes\cdots\otimes V_n}\OO_1\otimes\cdots\otimes\OO_n=(\Str_{V_1}\OO_1)\cdots (\Str_{V_n}\OO_n)$$
for a collection of graded vector spaces $V_1,\ldots,V_n$ and a collection of degree 0 endomorphisms $\OO_1\in\End(V_1),\ldots,\OO_n\in\End(V_n)$. Notice also that the space of differential forms on the cube with zero boundary condition also factorizes:
$$\Omega^\bt_0(I^D)=\underbrace{\Omega^\bt_0(I)\otimes\cdots\otimes\Omega^\bt_0(I)}_D$$
Therefore
\begin{multline*}C_{I^D,(*(*\bt))}(\zeta_1,\zeta_2)=\Str_{\Omega^\bt_0(I^D)}K_{I^D}(\chi_{\zeta_1}\wedge K_{I^D}(\chi_{\zeta_2}\wedge\bt))\\
=\Str_{(\Omega^\bt_0(I))^{\otimes D}}\int_{I_\aux} (K_I^{\lambda_1,d\lambda_1})^{\otimes D}\cdot\\ \cdot
\left((\chi_{\zeta_1^1}\otimes\cdots\otimes\chi_{\zeta_1^D})\wedge \int_{I_\aux} (K_I^{\lambda_2,d\lambda_2})^{\otimes D}
((\chi_{\zeta_2^1}\otimes\cdots\otimes\chi_{\zeta_2^D})\wedge\bt)\right)\\
=(-1)^{|\zeta_1|+1}(-1)^{\sum_{1\leq j<i\leq D}|\zeta_1^i|\cdot |\zeta_2^j|}\int_{I_\aux^2}\prod_{i=1}^D\Str_{\Omega^\bt_0(I)}K_I^{\lambda_1,d\lambda_1}(\chi_{\zeta_1^i}\wedge K_I^{\lambda_2,d\lambda_2}(\chi_{\zeta_2^i}\wedge\bt))\\
=(-1)^{|\zeta_1|+1}(-1)^{\sum_{1\leq j<i\leq D}|\zeta_1^i|\cdot |\zeta_2^j|}\int_{I_\aux^2}\prod_{i=1}^D C_{I,(*(*\bt))}^{\lambda_1,d\lambda_1;\lambda_2,d\lambda_2}(\zeta_1^i,\zeta_2^i)
\end{multline*}
where the sign comes from extracting the integration over $\lambda_2$ to the left and from the reshuffling of tensor product. Factors $C_{I,(*(*\bt))}^{\lambda_1,d\lambda_1;\lambda_2,d\lambda_2}(\zeta_1^i,\zeta_2^i)$ depend only on the pair of faces the interval and are differential forms in auxiliary variables:
$$C_{I,(*(*\bt))}^{\lambda_1,d\lambda_1;\lambda_2,d\lambda_2}(\zeta_1^i,\zeta_2^i)=\Str_{\Omega^\bt_0(I)}K_I^{\lambda_1,d\lambda_1}(\chi_{\zeta_1^i}\wedge K_I^{\lambda_2,d\lambda_2}(\chi_{\zeta_2^i}\wedge\bt))$$
To calculate $C_{I,(*(*\bt))}^{\lambda_1,d\lambda_1;\lambda_2,d\lambda_2}$, we will need the following properties of super-traces over the space of differential forms on the interval.

First, the super-trace of multiplication by a function $f(t)\in C^\infty(I)$ is
\be \Str_{\Omega^\bt(I)}f(t)\wedge\bt=\frac{f(0)+f(1)}{2} \label{str of multiplication}\ee
This result is fixed uniquely by the following conditions:
\begin{itemize}
\item linearity in $f(t)$,
\item invariance under diffeomorphisms of the interval: for a diffeomorphism $\phi:[0,1]\ra[0,1]$ we should have
$$\Str_{\Omega^\bt(I)}\phi^*f\wedge\bt=\Str_{\Omega^\bt(I)}f\wedge\bt$$
\item consistency with the symmetry of the interval:
$$\Str_{\Omega^\bt(I)}f(1-t)\wedge\bt=\Str_{\Omega^\bt(I)}f(t)\wedge\bt$$
\item normalization condition: super-trace of unity equals the Euler characteristic of $\Omega^\bt(I)$, i.e. the Euler characteristic of the interval
$$\Str_{\Omega^\bt(I)}1\wedge\bt=1$$
\end{itemize}
Analogous result for the super-trace over the space of forms on interval with zero boundary condition is
\be\Str_{\Omega^\bt_0(I)}f(t)\wedge\bt=-\frac{f(0)+f(1)}{2}\label{str of multiplication 1}\ee
since all the conditions are the same except the normalization: Euler characteristic of $\Omega^\bt_0(I)$ (or of the interval without end-points) is -1. Alternatively, one can obtain
(\ref{str of multiplication 1}) from (\ref{str of multiplication}) as
\begin{multline*}\Str_{\Omega^\bt_0(I)}f(t)\wedge\bt=\left(\Str_{\Omega^\bt_0(I)\oplus\Omega^\bt(\{0\})\oplus\Omega^\bt(\{1\})}f(t)\wedge\bt\right)
-\left(\Str_{\Omega^\bt(\{0\})}f(t)\wedge\bt\right)-\left(\Str_{\Omega^\bt(\{1\})}f(t)\wedge\bt\right)\\
=\frac{f(0)+f(1)}{2}-f(0)-f(1)=-\frac{f(0)+f(1)}{2}
\end{multline*}

The second property we need is:
\be \Str_{\Omega^\bt_0(I)}K_I(f(t)dt\wedge\bt)=\int_0^1 \left(\frac{1}{2}-t\right)f(t)dt \label{short loop}\ee
It can be obtained straightforwardly:
\begin{multline*}
g(t)\xra{K_I(f(t)dt\wedge\bt)}\int_0^t f(\tilde{t})g(\tilde{t})d\tilde{t}-t\int_0^1 f(\tilde{t})g(\tilde{t})d\tilde{t}=
\int_0^1 d\tilde{t}\cdot(\theta(t-\tilde{t})-t)f(\tilde{t})g(\tilde{t})\\
\Rightarrow\Str_{\Omega^\bt_0(I)}K_I(f(t)dt\wedge\bt)=\int_0^1 (\theta(0)-t)f(t)dt
\end{multline*}
using symmetric regularization for the unit step function $\theta(0)=\frac{1}{2}$, we obtain (\ref{short loop}).
Alternatively, one can deduce (\ref{short loop}) indirectly, by the following argument. Consider another super-trace
$$\Str_{\Omega^\bt_0(I)}\PP''_I(h(t)\wedge\bt)$$
One one hand it is calculated using (\ref{str of multiplication 1}) as
\begin{multline}\Str_{\Omega^\bt_0(I)}\PP''_I(h(t)\wedge\bt)=\Str_{\Omega^\bt_0(I)}h(t)\wedge\bt-\Str_{\Omega^\bt_0(I)}\PP'_I(h(t)\wedge\bt)=
\Str_{\Omega^\bt_0(I)}h(t)\wedge\bt+\int_0^1 h(t)dt\\
=-\frac{h(0)+h(1)}{2}+\int_0^1 h(t)dt \label{short loop 1}
\end{multline}
(we calculated the super-trace containing the IR projector $\PP'_I$ directly, as the super-trace over 1-dimensional space of IR forms on interval with zero boundary conditions $\Omega'^\bt_0(I)=\Omega'^1(I)=\RR dt$). On the other hand
\begin{multline}\Str_{\Omega^\bt_0(I)}\PP''_I(h(t)\wedge\bt)=\Str_{\Omega^\bt_0(I)}(d K_I+K_I d)(h(t)\wedge\bt)\\=
\Str_{\Omega^\bt_0(I)}(d K_I(h(t)\wedge\bt)+K_I((d h(t))\wedge\bt)+K_I(h(t)\wedge d\bt))\\
=\Str_{\Omega^\bt_0(I)}(d K_I(h(t)\wedge\bt)+K_I((d h(t))\wedge\bt)-d K_I(h(t)\wedge \bt))=
\Str_{\Omega^\bt_0(I)}K_I((d h(t))\wedge\bt)\label{short loop 2}
\end{multline}
where we use the defining property of chain homotopy $dK_I+K_Id=\PP''_I$, Leibniz identity and the cyclic property of super-trace. Comparing (\ref{short loop 2}) with (\ref{short loop 1}), we obtain
$$\Str_{\Omega^\bt_0(I)}K_I((d h(t))\wedge\bt)=-\frac{h(0)+h(1)}{2}+\int_0^1 h(t)dt$$
Substituting $h(t)=\int_0^t f(\tilde{t})d\tilde{t}$, we obtain (\ref{short loop}) by integration by parts.

Another useful property:
$$\Str_{\Omega^\bt_0(I)}\PP'_I\OO_1\PP'_I\OO_2\cdots\PP'_I\OO_n=-\left(\int_0^1 \OO_1\circ dt\right)\cdots\left(\int_0^1 \OO_n\circ dt\right)$$
for a collection of operators $\OO_1,\ldots,\OO_n:\Omega^\bt_0(I)\ra \Omega^\bt_0(I)$,--- follows from the fact that only the diagonal matrix element of $dt$ contributes to this super-trace, and from the explicit formula for the projector
$$\PP'(f(t)dt)=dt \int_0^1 f(\tilde{t})d\tilde{t}$$

Let us now demonstrate the calculation of values of $C_{I,(*(*\bt))}^{\lambda_1,d\lambda_1;\lambda_2,d\lambda_2}$ for some pairs of faces of the interval.
\begin{multline*}
C_{I,(*(*\bt))}^{\lambda_1,d\lambda_1;\lambda_2,d\lambda_2}(0,0)=\Str_{\Omega^\bt_0(I)}((1-\lambda_1)\;\id+\lambda_1\;\PP'_I)\circ
\left(\chi_0\wedge((1-\lambda_2)\;\id+\lambda_2\;\PP'_I)\circ(\chi_0\wedge\bt)\right)\\
=\Str_{\Omega^\bt_0(I)}((1-\lambda_1)(1-\lambda_2)\;\chi_0\chi_0\wedge\bt+(1-\lambda_1)\lambda_2\;\chi_0\wedge\PP'_I(\chi_0\wedge\bt)+\\
+\lambda_1(1-\lambda_2)\;\PP'_I(\chi_0\chi_0\wedge\bt)+\lambda_1 \lambda_2\;
\PP'_I(\chi_0\wedge\PP'_I(\chi_0\wedge\bt)))\\
=(1-\lambda_1)(1-\lambda_2)\cdot\Str_{\Omega^\bt_0(I)}\chi_0\chi_0\wedge\bt+((1-\lambda_1)\lambda_2+\lambda_1(1-\lambda_2))\cdot
\Str_{\Omega^\bt_0(I)}\PP'_I(\chi_0\chi_0\wedge\bt)+\\
+\lambda_1 \lambda_2\cdot
\Str_{\Omega^\bt_0(I)}\PP'_I(\chi_0\wedge\PP'_I(\chi_0\wedge\bt))\\
=-\frac{1}{2}(1-\lambda_1)(1-\lambda_2)-\frac{1}{3}((1-\lambda_1)\lambda_2+\lambda_1(1-\lambda_2))-\frac{1}{4}\lambda_1 \lambda_2=
-\frac{1}{2}+\frac{1}{6}\lambda_1+\frac{1}{6}\lambda_2-\frac{1}{12}\lambda_1\lambda_2,\\
C_{I,(*(*\bt))}^{\lambda_1,d\lambda_1;\lambda_2,d\lambda_2}(0,1)=\Str_{\Omega^\bt_0(I)}((1-\lambda_1)\;\id+\lambda_1\;\PP'_I)\circ
\left(\chi_0\wedge((1-\lambda_2)\;\id+\lambda_2\;\PP'_I)\circ(\chi_1\wedge\bt)\right)\\
=\Str_{\Omega^\bt_0(I)}((1-\lambda_1)(1-\lambda_2)\;\chi_0\chi_1\wedge\bt+(1-\lambda_1)\lambda_2\;\chi_0\wedge\PP'_I(\chi_1\wedge\bt)+\\
+\lambda_1 (1-\lambda_2)\;\PP'_I(\chi_0\chi_1\wedge\bt)+\lambda_1 \lambda_2\;
\PP'_I(\chi_0\wedge\PP'_I(\chi_1\wedge\bt)))\\
=0\cdot(1-\lambda_1)(1-\lambda_2)-\frac{1}{6}(1-\lambda_1)\lambda_2-\frac{1}{6}\lambda_1(1-\lambda_2)-\frac{1}{4}\lambda_1\lambda_2=
-\frac{1}{6}\lambda_1-\frac{1}{6}\lambda_2+\frac{1}{12}\lambda_1\lambda_2 ,\\
C_{I,(*(*\bt))}^{\lambda_1,d\lambda_1;\lambda_2,d\lambda_2}(I,0)=
\Str_{\Omega^\bt_0(I)}((1-\lambda_1)\;\id+\lambda_1\;\PP'_I+d\lambda_1\; K_I)\circ
\left(\chi_I\wedge((1-\lambda_2)\;\id+\lambda_2\;\PP'_I+d\lambda_2\; K_I)\circ(\chi_0\wedge\bt)\right)\\
=\Str_{\Omega^\bt_0(I)}\left(((1-\lambda_1)\;\id+\lambda_1\;\PP'_I)\circ
\left(\chi_I\wedge d\lambda_2\; K_I\circ(\chi_0\wedge\bt)\right)+\right.\\
\left.+d\lambda_1\; K_I\circ
\left(\chi_I\wedge((1-\lambda_2)\;\id+\lambda_2\;\PP'_I)\circ(\chi_0\wedge\bt)\right)\right)\\
=\Str_{\Omega^\bt_0(I)}\left(-(1-\lambda_1)d\lambda_2\;\chi_I\wedge K_I(\chi_0\wedge\bt)-
\lambda_1 d\lambda_2\;\PP'_I(\chi_I\wedge K_I(\chi_0\wedge\bt))+\right.\\
\left.+d\lambda_1\;(1-\lambda_2)\; K_I(\chi_I\chi_0\wedge\bt)+
d\lambda_1\; \lambda_2\; K_I(\chi_I\wedge\PP'_I(\chi_0\wedge\bt))
\right)\\
=\Str_{\Omega^\bt_0(I)}\left(((1-\lambda_1)\;d\lambda_2+d\lambda_1\;(1-\lambda_2))\cdot K_I(\chi_0\chi_I\wedge\bt)-\lambda_1\;d\lambda_2\;\PP'_I(\chi_I\wedge K_I(\chi_0\wedge\bt))+\right. \\ \left.+d\lambda_1\;\lambda_2\;\PP'_I(\chi_0\wedge K_I(\chi_I\wedge\bt))\right)\\
=((1-\lambda_1)\;d\lambda_2+d\lambda_1\;(1-\lambda_2))\cdot\int_0^1 (\frac{1}{2}-t)(1-t)dt+\lambda_1\;d\lambda_2\cdot\int_0^1 dt\; K_I((1-t)dt)+
d\lambda_1 \;\lambda_2\cdot 0\\
=\frac{1}{12}((1-\lambda_1)\;d\lambda_2+d\lambda_1\;(1-\lambda_2))+\frac{1}{12}\lambda_1\;d\lambda_2=\frac{1}{12}d\lambda_1+\frac{1}{12}d\lambda_2-\frac{1}{12}d\lambda_1\;\lambda_2,\\
C_{I,(*(*\bt))}^{\lambda_1,d\lambda_1;\lambda_2,d\lambda_2}(I,I)=\Str_{\Omega^\bt_0(I)}d\lambda_1\; K_I
(\chi_I\wedge d\lambda_2\; K_I(\chi_I\wedge\bt))=d\lambda_1\; d\lambda_2\;\Str_{\Omega^\bt_0(I)}K_I
(\chi_I\wedge K_I(\chi_I\wedge\bt))\\
=d\lambda_1\;d\lambda_2\;\int_0^1 dt\int_0^1 d\tilde{t}\; (\theta(t-\tilde{t})-t)(\theta(\tilde{t}-t)-\tilde{t})=-\frac{1}{12}d\lambda_1\;d\lambda_2
\end{multline*}
Using the symmetry of interval and the cyclic property of super-trace, we obtain all values of $C_{I,(*(*\bt))}^{\lambda_1,d\lambda_1;\lambda_2,d\lambda_2}$ from these four. The result is:
\begin{multline}
C_{I,(*(*\bt))}^{\lambda_1,d\lambda_1;\lambda_2,d\lambda_2}(0,0)=C_{I,(*(*\bt))}^{\lambda_1,d\lambda_1;\lambda_2,d\lambda_2}(1,1)
=-\frac{1}{2}+\frac{1}{6}\lambda_1+\frac{1}{6}\lambda_2-\frac{1}{12}\lambda_1\lambda_2,\\
C_{I,(*(*\bt))}^{\lambda_1,d\lambda_1;\lambda_2,d\lambda_2}(0,1)=C_{I,(*(*\bt))}^{\lambda_1,d\lambda_1;\lambda_2,d\lambda_2}(1,0)
=-\frac{1}{6}\lambda_1-\frac{1}{6}\lambda_2+\frac{1}{12}\lambda_1\lambda_2,\\
C_{I,(*(*\bt))}^{\lambda_1,d\lambda_1;\lambda_2,d\lambda_2}(I,0)=-C_{I,(*(*\bt))}^{\lambda_1,d\lambda_1;\lambda_2,d\lambda_2}(I,1)
=\frac{1}{12}d\lambda_1+\frac{1}{12}d\lambda_2-\frac{1}{12}d\lambda_1\;\lambda_2,\\
C_{I,(*(*\bt))}^{\lambda_1,d\lambda_1;\lambda_2,d\lambda_2}(0,I)=-C_{I,(*(*\bt))}^{\lambda_1,d\lambda_1;\lambda_2,d\lambda_2}(1,I)
=\frac{1}{12}d\lambda_1+\frac{1}{12}d\lambda_2-\frac{1}{12}\lambda_1\;d\lambda_2,\\
C_{I,(*(*\bt))}^{\lambda_1,d\lambda_1;\lambda_2,d\lambda_2}(I,I)=-\frac{1}{12}d\lambda_1\;d\lambda_2
\label{cube (*(* bt)) values}
\end{multline}

Similarly, for a general one-loop Feynman diagram $L$ we have:
\begin{multline*}
C_{I^D,L}(\zeta_1,\ldots,\zeta_{|L|})\\=(-1)^{|L|}\Loop_{L;\int_{I_\aux} (K_I^{\lambda_k,d\lambda})^{\otimes D}(\bt\wedge\bt); (\Omega^\bt_0(I))^{\otimes D}}(\chi_{\zeta_1^1}\otimes\cdots\otimes\chi_{\zeta_1^D},\ldots,\chi_{\zeta_{|L|}^1} \otimes\cdots\otimes\chi_{\zeta_{|L|}^D})\\
=\hat\e_{I^D,L}(\zeta_1,\ldots,\zeta_{|L|})\int_{I_\aux^{|L|}} \prod_{i=1}^D C_{I,L}^{\lambda_1,d\lambda_1;\cdots;\lambda_{|L|},d\lambda_{|L|}}(\zeta_1^i,\ldots,\zeta_{|L|}^i)
\end{multline*}
where the sign $\hat\e_{I^D,L}(\zeta_1,\ldots,\zeta_{|L|})=\pm 1$ comes from extracting integral over auxiliary parameters to the left and from the reshuffling of tensor product, and also contains the sign $(-1)^{|L|}$. For instance:
$$\hat\e_{I^D,(*(*\bt))}(\zeta_1,\zeta_2)=(-1)^{|\zeta_1|+1}(-1)^{\sum_{1\leq j<i\leq D}|\zeta_1^i|\cdot |\zeta_2^j|}$$
Factors $C_{I,L}^{\lambda_1,d\lambda_1;\cdots;\lambda_{|L|},d\lambda_{|L|}}(\zeta_1^i,\ldots,\zeta_{|L|}^i)$, into which $C_{I^D,L}(\zeta_1,\ldots,\zeta_{|L|})$ factorizes, depend only on the one-loop diagram $L$ and on an $|L|$-tuple of faces of the interval and are differential forms in auxiliary variables $\lambda_1,\cdots,\lambda_{|L|}$, with which we decorate the internal edges of $L$:
$$C_{I,L}^{\lambda_1,d\lambda_1;\cdots;\lambda_{|L|},d\lambda_{|L|}}(\zeta_1^i,\ldots,\zeta_{|L|}^i)=
\Loop_{L;K_I^{\lambda_k,d\lambda_k}(\bt\wedge\bt);\Omega^\bt_0(I)}(\chi_{\zeta_1^i},\ldots,\chi_{\zeta_{|L|}^i})$$
For any given diagram $L$ one can calculate $C_{I,L}^{\{\lambda_k,d\lambda_k\}}$ explicitly for all $|L|$-tuples of faces of the interval.

Let us summarize the discussion of this section.
\begin{thm}[Factorization of Feynman diagrams for the reduced cell $BF$ action for cube]
\label{thm: cube factorization}
The reduced cell $BF$ action for $D$-cube is expanded as a sum over Feynman diagrams:
\begin{multline*}\bar{S}_{I^D}
=\sum_{T\in{\bf{T}}_\mr{nonPl}}\frac{1}{|\Aut(T)|}\sum_{\zeta_1,\ldots,\zeta_{|T|}\subset I^D}\e_T(|\zeta_1|,\ldots,|\zeta_{|T|}|)
\cdot C_{I^D,T}(\zeta_1,\ldots,\zeta_{|T|})\cdot \\ \cdot <p_{I^D},\Iter_{T;[\bt,\bt];[\bt,\bt]}(\omega^{\zeta_1},\ldots,\omega^{\zeta_{|T|}})>_\g-\\
-\hbar\sum_{L\in{\bf{L}}_\mr{nonPl}}\frac{1}{|\Aut(L)|}\sum_{\zeta_1,\ldots,\zeta_{|L|}\subset I^D}\e_L(|\zeta_1|,\ldots,|\zeta_{|L|}|)
\cdot C_{I^D,L}(\zeta_1,\ldots,\zeta_{|L|})\cdot \Loop_{L;[\bt,\bt];\g}(\omega^{\zeta_1},\ldots,\omega^{\zeta_{|L|}})
\end{multline*}
Here $C_{I^D,(*)}(\zeta_1)=d^{I^D}_{\zeta_1}$ is given by structure constants of differential on cell cochains of $I^D$ (\ref{cube d matrix}), while de Rham parts of all other diagrams factorize:
\begin{eqnarray*}
C_{I^D,T}(\zeta_1,\ldots,\zeta_{|T|})&=&\hat{\e}_{I^D,T}(\zeta_1,\ldots,\zeta_{|T|})
\int_{I_\aux^{|T|-2}}\prod_{i=1}^D C_{I,T}^{\lambda_1,d\lambda_1;\cdots;\lambda_{|T|-2},d\lambda_{|T|-2}}(\zeta_1^i,\ldots, \zeta_{|T|}^i),\\
C_{I^D,L}(\zeta_1,\ldots,\zeta_{|L|})
&=&\hat{\e}_{I^D,L}(\zeta_1,\ldots,\zeta_{|L|})
\int_{I_\aux^{|L|}} \prod_{i=1}^D C_{I,L}^{\lambda_1,d\lambda_1;\cdots;\lambda_{|L|},d\lambda_{|L|}}(\zeta_1^i,\ldots, \zeta_{|L|}^i)
\end{eqnarray*}
where the factors $C_{I,T}^{\lambda_1,d\lambda_1;\cdots;\lambda_{|T|-2},d\lambda_{|T|-2}}$, $C_{I,L}^{\lambda_1,d\lambda_1;\cdots;\lambda_{|L|},d\lambda_{|L|}}$ depend on the Feynman diagram $\Gamma$, the tuple of faces of the interval (with which the leaves of $\Gamma$ are decorated), and are differential forms in auxiliary variables $(\lambda_k)$ (with which internal edges of $\Gamma$ are decorated). Feynman rules for them are: vertices are decorated with the operation of wedge product $\bt\wedge\bt$, edge number $k$ is decorated with the extended propagator
$$K_I^{\lambda_k,d\lambda_k}=(1-\lambda)\;\id+\lambda_k\; \PP'_I+d\lambda_k\; K_I$$
where $\PP'_I:f+g dt\mapsto (1-t)f(0)+tf(1)+dt(\int_0^1 g(\tilde{t})d\tilde{t})$, $\PP''_I=\id-\PP'_I$ и $K_I:f+gdt\mapsto \int_0^t g(\tilde{t})d\tilde{t}-t\int_0^1 g(\tilde{t})d\tilde{t}$ are the standard projectors and standard chain homotopy for the interval; leaf number $j$ is decorated with a Whitney form on the interval $\chi_{\zeta_j^i}$, the root is decorated by the fundamental class of the interval, for one-loop diagram one evaluates the super-trace over the space $\Omega^\bt_0(I)$ of differential forms, vanishing in the end-points. I.e.
\begin{eqnarray*}
C_{I,T}^{\lambda_1,d\lambda_1;\cdots;\lambda_{|T|-2},d\lambda_{|T|-2}}(\zeta_1^i,\ldots, \zeta_{|T|}^i)&=&
\int_I\Iter_{T;K_I^{\lambda_k,d\lambda_k}(\bt\wedge\bt);(\bt\wedge\bt)}(\chi_{\zeta_1^i},\ldots,\chi_{\zeta_{|T|}^i})\\
C_{I,L}^{\lambda_1,d\lambda_1;\cdots;\lambda_{|L|},d\lambda_{|L|}}(\zeta_1^i,\ldots, \zeta_{|L|}^i)&=&
\Loop_{L;K_I^{\lambda_k,d\lambda_k}(\bt\wedge\bt);\Omega^\bt_0(I)}(\chi_{\zeta_1^i},\ldots,\chi_{\zeta_{|L|}^i})
\end{eqnarray*}
Signs $\e_\Gamma=\pm 1$ depend only on the dimensions of faces $\zeta_1,\ldots,\zeta_{|\Gamma|}$ and come from moving the coordinates on the space of fields $\omega^{\zeta_j}$ to the right. Signs $\hat\e_{I^D,\Gamma}=\pm 1$ depend on combinatorics of the tuple of faces and come from the reshuffling of tensor product, moving the integral over auxiliary variables to the left, and include sign $-1$ for each internal edge of $\Gamma$.

In particular, the first terms of perturbation series for $\bar{S}_{I^D}$ are:
\begin{multline*}
\bar{S}_{I^D}=\sum_{\zeta_1\subset I^D:\,|\zeta_1|=D-1}d^{I^D}_{\zeta_1}<p_{I^D},\omega^{\zeta_1}>_\g+\\
+\frac{1}{2}
\sum_{\zeta_1,\zeta_2\subset I^D:\,|\zeta_1|+|\zeta_2|=D}(-1)^{(|\zeta_1|+1)\cdot |\zeta_2|}
(-1)^{\sum_{1\leq j<i\leq D}|\zeta_1^i|\cdot |\zeta_2^j|}\cdot\\ \cdot\left(\prod_{i=1}^D C_{I,(**)}(\zeta_1^i,\zeta_2^i)\right)<p_{I^D},[\omega^{\zeta_1},\omega^{\zeta_2}]>_\g+\\
+\frac{1}{2}\sum_{\zeta_1,\zeta_2,\zeta_3\subset I^D:\,|\zeta_1|+|\zeta_2|+|\zeta_3|=D+1}
(-1)^{(|\zeta_1|+1)\cdot(|\zeta_2|+|\zeta_3|+1)+(|\zeta_2|+1)\cdot|\zeta_3|}\cdot\\ \cdot
(-1)^{1+|\zeta_1|+D+\sum_{1\leq j<i\leq D}(|\zeta_1^i|\cdot|\zeta_2^j|+|\zeta_1^i|\cdot|\zeta_3^j|+|\zeta_2^i|\cdot|\zeta_3^j|)}\cdot\\
\cdot\left(\int_{I_\aux} \prod_{i=1}^D C_{I,(*(**))}^{\lambda,d\lambda}(\zeta_1^i,\zeta_2^i,\zeta_3^i)\right)
<p_{I^D},[\omega^{\zeta_1},[\omega^{\zeta_2},\omega^{\zeta_3}]]>_\g-\\
-\hbar\; \frac{1}{2}\sum_{\zeta_1,\zeta_2\subset I^D;\,|\zeta_1|+|\zeta_2|=2}(-1)^{(|\zeta_1|+1)\cdot (|\zeta_2|+1)}
(-1)^{1+|\zeta_1|+\sum_{1\leq j<i\leq D}|\zeta_1^i|\cdot|\zeta_2^j|}\cdot\\
\cdot \left(\int_{I_\aux^2} \prod_{i=1}^D C_{I,(*(*\bt))}^{\lambda_1,d\lambda_1;\lambda_2,d\lambda_2}(\zeta_1^i,\zeta_2^i)\right)
\tr_g (\ad_{\omega^{\zeta_1}}\ad_{\omega^{\zeta_2}})+O(p\omega^4+\hbar \omega^3)
\end{multline*}
Values of factors $C_{I,(**)}$, $C_{I,(*(**))}^{\lambda,d\lambda}$, $C_{I,(*(*\bt))}^{\lambda_1,d\lambda_1;\lambda_2,d\lambda_2}$ are calculated in (\ref{cube (**) values},\ref{cube (*(**)) values},\ref{cube (*(* bt)) values}).
\end{thm}

Using this result we can, for instance, write explicitly (with integrals over auxiliary parameters evaluated) the perturbative result for the square $I^2$:
\begin{multline}
\bar{S}_{I^2}
=<p_{II},(\omega^{I0}+\omega^{1I}-\omega^{I1}-\omega^{0I})+\\+
\frac{1}{4}([\omega^{I0},\omega^{0I}]+[\omega^{I0},\omega^{1I}]+[\omega^{I1},\omega^{0I}]+[\omega^{I1},\omega^{1I}])+
\frac{1}{4}[\omega^{00}+\omega^{01}+\omega^{10}+\omega^{11},\omega^{II}]+\\
+\frac{7}{288}([\omega^{I0},[\omega^{I0},\omega^{0I}-\omega^{1I}]]+[\omega^{I1},[\omega^{I1},\omega^{0I}-\omega^{1I}]]+
[\omega^{0I},[\omega^{0I},\omega^{I0}-\omega^{I1}]]+[\omega^{1I},[\omega^{1I},\omega^{I0}-\omega^{I1}]])+\\
+\frac{5}{288}([\omega^{I0},[\omega^{I1},\omega^{0I}-\omega^{1I}]]+[\omega^{I1},[\omega^{I0},\omega^{0I}-\omega^{1I}]]+
[\omega^{0I},[\omega^{1I},\omega^{I0}-\omega^{I1}]]+[\omega^{1I},[\omega^{0I},\omega^{I0}-\omega^{I1}]])+\\
+\frac{5}{144}([\omega^{II},[\omega^{I0},\omega^{00}-\omega^{10}]]+[\omega^{II},[\omega^{I1},\omega^{01}-\omega^{11}]]+
[\omega^{II},[\omega^{0I},\omega^{01}-\omega^{00}]]+[\omega^{II},[\omega^{1I},\omega^{11}-\omega^{10}]])+\\
+\frac{1}{144}([\omega^{II},[\omega^{I0},\omega^{01}-\omega^{11}]]+[\omega^{II},[\omega^{I1},\omega^{00}-\omega^{10}]]+
[\omega^{II},[\omega^{0I},\omega^{11}-\omega^{10}]]+[\omega^{II},[\omega^{1I},\omega^{01}-\omega^{00}]])+\\
+\frac{7}{288}([\omega^{I0},[\omega^{II},\omega^{00}-\omega^{10}]]+[\omega^{I1},[\omega^{II},\omega^{01}-\omega^{11}]]+
[\omega^{0I},[\omega^{II},\omega^{01}-\omega^{00}]]+[\omega^{1I},[\omega^{II},\omega^{11}-\omega^{10}]])+\\
+\frac{5}{288}([\omega^{I0},[\omega^{II},\omega^{01}-\omega^{11}]]+[\omega^{I1},[\omega^{II},\omega^{00}-\omega^{10}]]+
[\omega^{0I},[\omega^{II},\omega^{11}-\omega^{10}]]+[\omega^{1I},[\omega^{II},\omega^{01}-\omega^{00}]])>_\g+\\
+\hbar \left(-\frac{17}{1152}(\tr_\g(\ad_{\omega^{I0}}\ad_{\omega^{I0}})+\tr_\g(\ad_{\omega^{I1}}\ad_{\omega^{I1}})+
\tr_\g(\ad_{\omega^{0I}}\ad_{\omega^{0I}})+\tr_\g(\ad_{\omega^{1I}}\ad_{\omega^{1I}}))-\right.\\
-\frac{7}{576}(\tr_\g(\ad_{\omega^{I0}}\ad_{\omega^{I1}})+\tr_\g(\ad_{\omega^{0I}}\ad_{\omega^{1I}}))+\\
\left.+\frac{1}{192}(\tr_\g(\ad_{\omega^{I0}}\ad_{\omega^{0I}})+\tr_\g(\ad_{\omega^{I1}}\ad_{\omega^{1I}})-
\tr_\g(\ad_{\omega^{I0}}\ad_{\omega^{1I}})-\tr_\g(\ad_{\omega^{I1}}\ad_{\omega^{0I}}))\right)+O(p\omega^4+\hbar\omega^3)
\label{Sbar on square}
\end{multline}

Notice that for the cube we can compute one-loop diagrams explicitly in any dimension: for each given diagram (but arbitrary dimension of the cube) the calculation reduces to some finite calculation for the interval. This situation is drastically different from the case of simplex (section \ref{section: simplex pert}) where we can make explicit computations (quite cumbersome and requiring regularization) for super-traces in lower dimensions, but in general dimension we can only indirectly reconstruct certain part of the one-loop result from tree diagrams, using QME. Also calculating the  perturbative result for the cube is technically much easier than for the simplex (due to the more convenient expression for chain homotopy), but the final results for the cube are more cumbersome (c.f. (\ref{Sbar on square}) vs. (\ref{S on triangle})), since the number of combinatorial types of $n$-tuples of faces of the cube (i.e. of $n$-tuples  $\zeta_1,\ldots,\zeta_n\subset I^D$ modulo diagonal action of the symmetry group of cube $S_D\ltimes \ZZ_2^D$) increases faster with $n$ than the number of combinatorial types of $n$-tuples of faces of simplex ($n$-tuples $\sigma_1,\cdots,\sigma_n\subset\Delta^D$ modulo diagonal action of  $S_{D+1}$).

\textbf{Using $\ZZ_2^D$-symmetric basis for cochains of cube.} For some applications the following modification of the formalism of this section is useful: instead of using the basis of faces $\{e_0,e_1,e_I\}^D$ for cochains on the $D$-cube, one can use the eigenbasis of the $\ZZ_2^D$-symmetry:  $\{e_+,e_-,e_I\}^D$, where
\be e_+=e_0+e_1,\quad e_-=\frac{1}{2}(e_1-e_0),\quad e_I=e_I \label{+-I basis}\ee Statement of the Theorem \ref{thm: cube factorization} is still true, where we allow indices of $\zeta^i$ run over $\{+,-,I\}$ (the set, indexing $\ZZ_2$-symmetric combinations of faces) instead of $\{0,1,I\}$ (the set, indexing faces of the interval themselves). Corresponding linear combinations of Whitney forms on the interval are
$$\chi_+=\chi_0+\chi_1=1,\;\chi_-=\frac{1}{2}(\chi_1-\chi_0)=t-\frac{1}{2},\;\chi_I=dt$$
Values of the factors $C_{I,(**)}$, $C_{I,(*(**))}^{\lambda,d\lambda}$, $C_{I,(*(*\bt))}^{\lambda_1,d\lambda_1;\lambda_2,d\lambda_2}$ in the symmetric basis for the cochains are obtained from values in the basis of faces (\ref{cube (**) values},\ref{cube (*(**)) values},\ref{cube (*(* bt)) values}) by passing to symmetric linear combinations:
\begin{multline*}
C_{I,(**)}(I,+)=C_{I,(**)}(+,I)=1;\\
C_{I,(*(**))}^{\lambda,d\lambda}(I,+,+)=C_{I,(*(**))}^{\lambda,d\lambda}(+,+,I)=C_{I,(*(**))}^{\lambda,d\lambda}(+,I,+)=1,\qquad
C_{I,(*(**))}^{\lambda,d\lambda}(I,-,-)=\frac{1}{12}+\frac{1}{6}\lambda,\\
C_{I,(*(**))}^{\lambda,d\lambda}(-,-,I)=C_{I,(*(**))}^{\lambda,d\lambda}(-,I,-)=\frac{1}{12}-\frac{1}{12}\lambda,\qquad
C_{I,(*(**))}^{\lambda,d\lambda}(I,I,-)=C_{I,(*(**))}^{\lambda,d\lambda}(I,-,I)=-\frac{1}{12}d\lambda;\\
C_{I,(*(*\bt))}^{\lambda_1,d\lambda_1;\lambda_2,d\lambda_2}(+,+)=-1,\qquad
C_{I,(*(*\bt))}^{\lambda_1,d\lambda_1;\lambda_2,d\lambda_2}(-,-)=-\frac{1}{4}+\frac{1}{6}\lambda_1+\frac{1}{6}\lambda_2-\frac{1}{12}\lambda_1\lambda_2,\\
C_{I,(*(*\bt))}^{\lambda_1,d\lambda_1;\lambda_2,d\lambda_2}(I,-)=-\frac{1}{12}d\lambda_1-\frac{1}{12}d\lambda_2+\frac{1}{12}d\lambda_1\;\lambda_2,\qquad
C_{I,(*(*\bt))}^{\lambda_1,d\lambda_1;\lambda_2,d\lambda_2}(-,I)=-\frac{1}{12}d\lambda_1-\frac{1}{12}d\lambda_2+\frac{1}{12}\lambda_1\; d\lambda_2,\\
C_{I,(*(*\bt))}^{\lambda_1,d\lambda_1;\lambda_2,d\lambda_2}(I,I)=-\frac{1}{12}d\lambda_1\;d\lambda_2
\end{multline*}
(all the other values of  $C_{I,(**)}$, $C_{I,(*(**))}^{\lambda,d\lambda}$, $C_{I,(*(*\bt))}^{\lambda_1,d\lambda_1;\lambda_2,d\lambda_2}$ are zero).
In the basis $\{+,-,I\}$ the result for interval (Theorem \ref{interval thm}) is written as
$$\bar{S}_I=\left<p_I,\omega^- + [\omega^I,\omega^+]+\left(\frac{\ad_{\omega^I}}{2}\coth \frac{\ad_{\omega^I}}{2}\right)\circ\omega^-\right>_\g+
\hbar\;\tr_\g\log\left(\frac{\sinh\frac{\ad_{\omega^I}}{2}}{\frac{\ad_{\omega^I}}{2}}\right)$$
where fields $\omega^+,\omega^-,\omega^I$ are related to the fields in basis of faces by $\omega^+=\frac{1}{2}(\omega^0+\omega^1)$, $\omega^-=\omega^1-\omega^0$, $\omega^I=\omega^I$ (i.e. by dual transformation for (\ref{+-I basis})). Perturbative result for the square (\ref{Sbar on square}) in basis
$\{+,-,I\}\times\{+,-,I\}$ is
\begin{multline*}
\bar{S}_{I^2}=<p_{II},\omega^{-I}-\omega^{I-}+[\omega^{I+},\omega^{+I}]+[\omega^{++},\omega^{II}]-\\
-\frac{1}{12}[\omega^{I+},[\omega^{I+},\omega^{-I}]]-\frac{1}{12}[\omega^{+I},[\omega^{+I},\omega^{I-}]]
-\frac{1}{288}[\omega^{-I}-\omega^{I-},[\omega^{-I},\omega^{I-}]]-\\
-\frac{1}{12}[\omega^{II},[\omega^{I+},\omega^{-+}]]+\frac{1}{12}[\omega^{II},[\omega^{+I},\omega^{+-}]]+
\frac{1}{72}[\omega^{II},[\omega^{-I}-\omega^{I-},\omega^{--}]]-\\
-\frac{1}{12}[\omega^{I+},[\omega^{II},\omega^{-+}]]+\frac{1}{12}[\omega^{+I},[\omega^{II},\omega^{+-}]]+
\frac{1}{288}[\omega^{-I}-\omega^{I-},[\omega^{II},\omega^{--}]]>_\g+\\
+\hbar\left(-\frac{1}{24}\tr_g(\ad_{\omega^{I+}}\ad_{\omega^{I+}})-\frac{1}{24}\tr_g(\ad_{\omega^{+I}}\ad_{\omega^{+I}})
-\frac{5}{1152}\tr_g(\ad_{\omega^{I-}}\ad_{\omega^{I-}})-\frac{5}{1152}\tr_g(\ad_{\omega^{-I}}\ad_{\omega^{-I}})\right.\\
\left.+\frac{1}{192}\tr_g(\ad_{\omega^{I-}}\ad_{\omega^{-I}})\right)+O(p\omega^4+\hbar\omega^3)
\end{multline*}
where $\omega^{++}=\frac{1}{4}(\omega^{00}+\omega^{01}+\omega^{10}+\omega^{11})$, $\omega^{+-}=\frac{1}{2}(-\omega^{00}+\omega^{01}-\omega^{10}+\omega^{11})$, $\omega^{-+}=\frac{1}{2}(-\omega^{00}-\omega^{01}+\omega^{10}+\omega^{11})$,
$\omega^{--}=\omega^{00}-\omega^{01}-\omega^{10}+\omega^{11}$, $\omega^{+I}=\frac{1}{2}(\omega^{0I}+\omega^{1I})$, $\omega^{I+}=\frac{1}{2}(\omega^{I0}+\omega^{I1})$, $\omega^{-I}=\omega^{1I}-\omega^{0I}$, $\omega^{I-}=\omega^{I1}-\omega^{I0}$, $\omega^{II}=\omega^{II}$.

Note that, although in the symmetric basis in cochains on cube the expressions for $\bar{S}_{I^D}$ look less cumbersome, this basis is not suited as good as the basis of faces for gluing the cell action on a cubical complex (namely, in the basis of faces, for gluing face $\zeta$ to face $\zeta'$, we just identify $\omega^\zeta$ and $\omega^{\zeta'}$, while in basis $\{+,-,I\}$ we have to impose more involved relations).

\textbf{Remark.} Here we were using representation (\ref{cube K}) for the chain homotopy which gives value of the de Rham part of a Feynman diagram for $D$-cube as an integral over auxiliary parameters of the $D$-fold product of Feynman diagrams for interval with edges decorated by the extended propagator (\ref{extended K}) (these Feynman diagrams for the interval take values in differential forms in auxiliary parameters). This formalism is well suited for computing simple Feynman diagrams in arbitrary dimension $D$. However, for low dimensions another version of formalism might be useful, where one uses explicit expansions (\ref{K sym square},\ref{K sym cube}) for the symmetric power of chain homotopy for interval instead of introducing auxiliary parameters. Here one expresses de Rham part of a Feynman diagram $\Gamma$ for $D$-cube as a sum of $D$-fold products of Feynman diagrams for interval with edges decorated with either of operators $\PP'_I$, $\PP''_I$, $K_I$ (or one can choose instead the triple $\id$, $\PP'_I$, $K_I$), and we sum over all decorations, such that each edge of $\Gamma$ is decorated by $K_I$ in exactly one factor and by projectors in all other factors.

\subsection{Examples of exactly computable cell $BF$ action: torus, cylinder, Klein bottle, $S_D\ltimes \ZZ_2^D$-bundles over circle with fiber $\TT^D$}
\label{section: exact results for cell action}

\subsubsection{Torus $\TT^2$ in symmetric gauge}
\label{section: 2-torus in sym gauge}
Already in dimension $D=2$ we cannot write an explicit formula for the cell action for the square, and can only give the perturbative result. However, it turns out that to write the glued cell action for the torus $\TT^2=\s^1\times\s^1$, obtained by gluing edge  $I1$ to $I0$, and edge $1I$ to $0I$ in the square, we only need to know certain part of the cell action for square $S_{I^2}$ (namely, the restriction to the cochain complex of torus, embedded into the cochain complex of the square, according to the general construction of section \ref{section: gluing}), which can be calculated explicitly.

Consider the torus $\TT^2$ with standard (cubical) cell decomposition consisting of one 0-cell ($++$), two 1-cells ($+I$ and $I+$) and one 2-cell $(II)$. We denote the basis cochains $e_{++},e_{+I},e_{I+},e_{II}\in C^\bt(\TT^2)$, and the embedding $C^\bt(\TT^2)\hra C^\bt(I^2)$ is
$$e_{++}=e_{00}+e_{01}+e_{10}+e_{11},\;e_{+I}=e_{0I}+e_{1I},\;e_{I+}=e_{I0}+e_{I1},\;e_{II}=e_{II}$$
Corresponding Whitney forms written in coordinates $t_1,t_2$ on the square are
$$\chi_{++}=1,\;\chi_{+I}=dt_2,\;\chi_{I+}=dt_1,\;\chi_{II}=dt_1 dt_2$$
Their linear span $\Span(\chi_{++},\chi_{+I},\chi_{I+},\chi_{II}) =\RR\chi_{++}\oplus\RR\chi_{+I}\oplus\RR\chi_{I+}\oplus\RR\chi_{II}\subset \Omega^\bt(\TT^2)\subset\Omega^\bt(I^2)$ is the periodic part of the Whitney complex for square and, which is very important for us, it is closed under exterior product and has zero differential.
We want to compute the cell $BF$ action for the torus:
\begin{multline*}
S_{\TT^2}=\sum_{T\in{\bf{T}}_\mr{nonPl}}\frac{1}{|\Aut(T)|}\cdot\\
\cdot\sum_{\zeta,\zeta_1,\ldots,\zeta_{|T|}\in\{++,+I,I+,II\}}
<p_\zeta,\int_\zeta\Iter_{T;-K_{I^2}[\bt,\bt];[\bt,\bt]}(\chi_{\zeta_1}\omega^{\zeta_1},\ldots,\chi_{\zeta_{|T|}}\omega^{\zeta_{|T|}})>_\g-\\
-\hbar\sum_{L\in{\bf{L}}_\mr{nonPl}}\frac{1}{|\Aut(L)|}\sum_{\zeta_1,\ldots,\zeta_{|L|}\in\{++,+I,I+,II\}}
\Loop_{L;-K_{I^2}[\bt,\bt];\Omega^{\bt}(\TT^2,\g)}(\chi_{\zeta_1}\omega^{\zeta_1},\ldots,\chi_{\zeta_{|L|}}\omega^{\zeta_{|L|}})
\end{multline*}
Here the super-traces are evaluated over the space $\Omega^{\bt}(\TT^2,\g)$ of forms on torus (periodic forms on square), without condition of vanishing on the boundary of the square, contrary to the case of reduced cell action $\bar{S}_{I^2}$.

Notice that, since the space of IR forms here is closed under multiplication, every Feynman diagram containing an internal edge not in the cycle vanishes (since $K_{I^2}$ vanishes on IR forms). Thus  the only diagrams that contribute are the tree $(**)$ (tree $(*)$ does not contribute, since the differential vanishes on IR forms) and one-loop graphs of type $(*(*\bt)),\;(*(*(*\bt))),\;(*(*(*(*\bt)))),\ldots$ (``wheels''). Hence the tree part of $S_{\TT^2}$ is:
\begin{multline}S_{\TT^2}^0=\frac{1}{2}\sum_{\zeta,\zeta_1,\zeta_2\in\{++,+I,I+,II\}}
<p_\zeta,\int_{\zeta}[\chi_{\zeta_1}\omega^{\zeta_1},\chi_{\zeta_2}\omega^{\zeta_2}]>_\g\\=
\frac{1}{2}\sum_{\zeta,\zeta_1,\zeta_2\in\{++,+I,I+,II\}}(-1)^{(|\zeta_1|+1)\cdot |\zeta_2|}<p_\zeta,[\omega^{\zeta_1},\omega^{\zeta_2}]>_\g
\int_\zeta \chi_{\zeta_1}\wedge\chi_{\zeta_2}\\
=\frac{1}{2}<p_{++},[\omega^{++},\omega^{++}]>_\g+<p_{I+},[\omega^{I+},\omega^{++}]>_\g+
<p_{+I},[\omega^{+I},\omega^{++}]>_\g+\\+<p_{II},[\omega^{I+},\omega^{+I}]+[\omega^{II},\omega^{++}]>_\g
\label{torus tree part}
\end{multline}

\textbf{One-loop part.} Now let us proceed to the one-loop part.  We introduce the new notation for multiplication operators: $\m(\chi_\zeta)=\chi_\zeta\wedge\bt$. First observe that for the super-trace
$$W(\zeta_1,\ldots,\zeta_n)=\Str_{\Omega^\bt(\TT^2)}\prod_{i=1}^n K_{I^2}\m(\chi_{\zeta_i})$$
to be non-vanishing, all faces $\zeta_1,\ldots,\zeta_{|L|}$ have to be one-dimensional. Indeed, the operator under super-trace has to preserve the degree of form, therefore the sum of degrees of incoming forms is the number of leaves $|\zeta_1|+\cdots+|\zeta_{|L|}|=|L|$. Thus, unless all $|\zeta_i|=1$, there is at least one face of dimension 0, i.e. $\zeta_i=++$ for some $i$. Since $\chi_{++}=1$ and $(K_{I^2})^2=0$, such a super-trace vanishes. Therefore $S_{\TT^2}^1$ is represented as
$$S_{\TT^2}^1=-\sum_{n=2}^\infty (-1)^n \frac{1}{n}\sum_{\zeta_1,\ldots,\zeta_n\in\{I+,+I\}}\tr_\g(\ad_{\omega^{\zeta_1}}\cdots\ad_{\omega^{\zeta_n}})\cdot
\Str_{\Omega^\bt(\TT^2)}\prod_{i=1}^n K_{I^2}\m(\chi_{\zeta_i})$$
To evaluate super-traces $\Str_{\Omega^\bt(\TT^2)}\prod_{i=1}^n K_{I^2}\m(\chi_{\zeta_i})$ we use the factorization of the space of differential forms on torus $\Omega^\bt(\TT^2)=\Omega^\bt(\s^1)\otimes \Omega^\bt(\s^1)$, factorization of Whitney forms and the formula for chain homotopy
$$K_{I^2}=K_I\otimes\frac{\id+\PP'_I}{2}+\frac{\id+\PP'_I}{2}\otimes K_I$$
(we do not use the auxiliary variables of section \ref{section: cube factorization} here). Denote
$$K_1=K_I\otimes\frac{\id+\PP'_I}{2},\quad K_2=\frac{\id+\PP'_I}{2}\otimes K_I$$
In case  $\zeta_1=\zeta_2=\cdots=\zeta_n=I+$ the super-trace $W$ is
\begin{multline*}
W(I+,\ldots,I+)
=\Str_{\Omega^\bt(\s^1)\otimes \Omega^\bt(\s^1)}\left(\left(K_1+K_2\right)\m(\chi_{I+})\right)^n\\
=\Str_{\Omega^\bt(\s^1)\otimes \Omega^\bt(\s^1)}\left(K_1 \m(\chi_{I+})\right)^n=
\Str_{\Omega^\bt(\s^1)}(K_I\m(\chi_I))^n\cdot
\Str_{\Omega^\bt(\s^1)}\left(\frac{\id+\PP'_I}{2}\m(\chi_0)\right)^n\\
=\Str_{\Omega^\bt(\s^1)}(K_I\m(\chi_I))^n\cdot\Str_{\Omega^\bt(\s^1)}\left(\PP'_I+\frac{1}{2}\PP''_I\right)^n=
\Str_{\Omega^\bt(\s^1)}(K_I\m(\chi_I))^n\cdot\Str_{\Omega^\bt(\s^1)}\left(\PP'_I+\frac{1}{2^n}\PP''_I\right)\\
=\Str_{\Omega^\bt(\s^1)}(K_I\m(\chi_I))^n\cdot\left(\chi(\Omega'^\bt(\s^1))+\frac{1}{2^n}\chi(\Omega''^\bt(\s^1))\right)=0
\end{multline*}
where we use the fact that super-trace of projector to a subspace is the Euler characteristic of the subspace and that Euler characteristics of spaces IR forms and UV forms on the circle are zero: for IR forms, since the Euler characteristic of the circle is zero, and for UV forms, since the contracted subcomplex is always acyclic. We also use that in the expression for the chain homotopy $K_I\otimes\frac{\id+\PP'_I}{2}+\frac{\id+\PP'_I}{2}\otimes K_I$ only the first term contributes to this super-trace, since both components, into which the operator $(K_{I^2}\m(\chi_{I+}))^n$ factorizes, have to be of degree 0. Analogous argument shows that for the case $\zeta_1=\zeta_2=\cdots=\zeta_n=+I$ the super-trace is also zero.

In the case when not all $\zeta_1,\ldots,\zeta_n$ coincide, we use cyclic property to transform the super-trace to
\begin{multline*}
W(\underbrace{+I,\ldots,+I}_{b_m},\underbrace{I+,\ldots,I+}_{a_m},\quad\ldots,\quad\underbrace{+I,\ldots,+I}_{b_1},\underbrace{I+,\ldots,I+}_{a_1})=\\
=\Str_{\Omega^\bt(\TT^2)}(K_{I^2}\m(\chi_{+I}))^{b_m}(K_{I^2}\m(\chi_{I+}))^{a_m}\cdots
(K_{I^2}\m(\chi_{+I}))^{b_1}(K_{I^2}\m(\chi_{I+}))^{a_1}
\end{multline*}
for some $m\geq 1$ and $a_i,b_i\geq 1$ for $i=1,\ldots,m$. Degree counting argument shows that
\begin{eqnarray*}\m(\chi_{I+})(K_{I^2}\m(\chi_{I+}))^{a-1}=\m(\chi_{I+})\left(K_1\m(\chi_{I+})\right)^{a-1}\\
\m(\chi_{+I})(K_{I^2}\m(\chi_{+I}))^{b-1}=\m(\chi_{+I})\left(K_2\m(\chi_{+I})\right)^{b-1}
\end{eqnarray*}
Hence
\begin{multline}
W(\underbrace{+I,\ldots,+I}_{b_m},\underbrace{I+,\ldots,I+}_{a_m},\quad\ldots,\quad\underbrace{+I,\ldots,+I}_{b_1},\underbrace{I+,\ldots,I+}_{a_1})=\\
=\Str_{\Omega^\bt(\TT^2)}
K_{I^2}\m(\chi_{+I})
\left(K_2\m(\chi_{+I})\right)^{b_m-1}K_{I^2}\m(\chi_{I+})\left(K_1\m(\chi_{I+})\right)^{a_m-1}\cdots\\
\cdots K_{I^2}\m(\chi_{+I})
\left(K_2\m(\chi_{+I})\right)^{b_1-1}K_{I^2}\m(\chi_{I+})\left(K_1\m(\chi_{I+})\right)^{a_1-1}
\label{torus 1}
\end{multline}
Next, notice that
$$K_1\m(\chi_{+I})
\left(K_2\m(\chi_{+I})\right)^{b-1}K_1=0$$
since in the first factor in $\Omega^\bt(\s^1)\otimes \Omega^\bt(\s^1)$ we encounter either the structure $(K_I)^2$, or $K_I\PP'_I$. Similarly
$$K_2\m(\chi_{I+})
\left(K_1\m(\chi_{I+})\right)^{a-1}K_2=0$$
--- here in the second factor in $\Omega^\bt(\s^1)\otimes \Omega^\bt(\s^1)$ we encounter $(K_I)^2$ or $K_I\PP'_I$.
This implies for (\ref{torus 1}) the following:
\begin{multline*}
W(\underbrace{+I,\ldots,+I}_{b_m},\underbrace{I+,\ldots,I+}_{a_m},\quad\ldots,\quad\underbrace{+I,\ldots,+I}_{b_1},\underbrace{I+,\ldots,I+}_{a_1})=\\
=\Str_{\Omega^\bt(\TT^2)}
K_2\m(\chi_{+I})
\left(K_2\m(\chi_{+I})\right)^{b_m-1}K_1\m(\chi_{I+})\left(K_1\m(\chi_{I+})\right)^{a_m-1}\cdots\\
\cdots K_2\m(\chi_{+I})
\left(K_2\m(\chi_{+I})\right)^{b_1-1}K_1\m(\chi_{I+})\left(K_1\m(\chi_{I+})\right)^{a_1-1}+\\
+\Str_{\Omega^\bt(\TT^2)}
K_1\m(\chi_{+I})
\left(K_2\m(\chi_{+I})\right)^{b_m-1}K_2\m(\chi_{I+})\left(K_1\m(\chi_{I+})\right)^{a_m-1}\cdots\\
\cdots K_1\m(\chi_{+I})
\left(K_2\m((\chi_{+I})\right)^{b_1-1}K_2\m(\chi_{I+})\left(K_1\m(\chi_{I+})\right)^{a_1-1}=\\
=\Str_{\Omega^\bt(\TT^2)}
\left(K_2\m(\chi_{+I})\right)^{b_m}\left(K_1\m(\chi_{I+})\right)^{a_m}\cdots
\left(K_2\m(\chi_{+I})\right)^{b_1}\left(K_1\m(\chi_{I+})\right)^{a_1}-\\
-\Str_{\Omega^\bt(\TT^2)}
\left(\m(\chi_{+I}) K_2\right)^{b_m}\left(\m(\chi_{I+}) K_1\right)^{a_m}\cdots
\left(\m(\chi_{+I}) K_2\right)^{b_1}\left(\m(\chi_{I+}) K_1\right)^{a_1}=\\
=\Str_{\Omega^\bt(\s^1)}
\left(\frac{\id+\PP'_I}{2}\m(\chi_{+})\right)^{b_m}\left(K_I\m(\chi_{I})\right)^{a_m}\cdots
\left(\frac{\id+\PP'_I}{2}\m(\chi_{+})\right)^{b_1}\left(K_I\m(\chi_{I})\right)^{a_1}\cdot\\
\cdot \Str_{\Omega^\bt(\s^1)}
\left(K_I\m(\chi_{I})\right)^{b_m}\left(\frac{\id+\PP'_I}{2}\m(\chi_{+})\right)^{a_m}\cdots
\left(K_I\m(\chi_{I})\right)^{b_1}\left(\frac{\id+\PP'_I}{2}\m(\chi_{+})\right)^{a_1}-\\
-\Str_{\Omega^\bt(\s^1)}
\left(\m(\chi_{+})\frac{\id+\PP'_I}{2}\right)^{b_m}\left(\m(\chi_{I})K_I\right)^{a_m}\cdots
\left(\m(\chi_{+})\frac{\id+\PP'_I}{2}\right)^{b_1}\left(\m(\chi_{I})K_I\right)^{a_1}\cdot\\
\cdot \Str_{\Omega^\bt(\s^1)}
\left(\m(\chi_{I})K_I\right)^{b_m}\left(\m(\chi_{+})\frac{\id+\PP'_I}{2}\right)^{a_m}\cdots
\left(\m(\chi_{I})K_I\right)^{b_1}\left(\m(\chi_{+})\frac{\id+\PP'_I}{2}\right)^{a_1}=\\
=\frac{1}{2^{b_1+\cdots+b_m}}\;\Str_{\Omega^\bt(\s^1)}\left(K_I\m(\chi_{I})\right)^{a_1+\cdots+a_m}\cdot
\frac{1}{2^{a_1+\cdots+a_m}}\;\Str_{\Omega^\bt(\s^1)}\left(K_I\m(\chi_{I})\right)^{b_1+\cdots+b_m}-\\
-\frac{1}{2^{b_1+\cdots+b_m}}\;\left(-\Str_{\Omega^\bt(\s^1)}\left(K_I\m(\chi_{I})\right)^{a_1+\cdots+a_m}\right)\cdot
\frac{1}{2^{a_1+\cdots+a_m}}\;\left(-\Str_{\Omega^\bt(\s^1)}\left(K_I\m(\chi_{I})\right)^{b_1+\cdots+b_m}\right)=\\
=0
\end{multline*}
Thus de Rham parts of wheel diagrams vanish, since the two possible decorations of edges with pieces of propagator $K_1$, $K_2$ cancel each other. Therefore one-loop part of cell action for torus $\TT^2$ vanishes:
\be S_{\TT^2}^1=0 \label{torus loop Ksym} \ee

\subsubsection{Torus $\TT^D$ in asymmetric gauge}
\label{section: T^D in asym gauge}
Computation for 2-torus in section \ref{section: 2-torus in sym gauge}, leading to the result  (\ref{torus tree part},\ref{torus loop Ksym}),  was based on using the  $S_2\ltimes(\ZZ_2)^2$-symmetric chain homotopy for the square
$$K_{I^2}=K_I^{\otimes_\sym 2}=K_I\otimes\frac{\id+\PP'_I}{2}+\frac{\id+\PP'_I}{2}\otimes K_I$$
However we could alternatively use asymmetric chain homotopy $$K_{I^2,L}=K_I\otimes\id+\PP'_I\otimes K_I$$ or $$K_{I^2,R}=\id\otimes K_I+K_I\otimes\PP_I'$$
Choose for instance the chain homotopy $K_{I^2,L}$ (case of $K_{I^2,R}$ is completely analogous). The argument from section \ref{section: 2-torus in sym gauge} that the only contributing diagrams are the tree $(**)$ and the  ``wheels'' applies here as well, thus the tree part of cell action coincides with (\ref{torus tree part}) --- this computation is independent of the choice of chain homotopy.

To evaluate the super-trace
$$W_L(\zeta_1,\ldots,\zeta_n)=\Str_{\Omega^\bt(\TT^2)}\prod_{i=1}^n \left(K_{I^2,L}\m(\chi_{\zeta_i})\right)$$
(i.e. the de Rham part of the ``wheel'') we use the following observation.

\begin{lemma}\label{lemma: str on s^1 with P',K} Let $\zeta_1,\dots,\zeta_n,\bar{\zeta}_1,\ldots,\bar{\zeta}_n\in\{+,I\}$ be two tuples of cells of the standard cell decomposition of circle $\s^1$. Denote $\Phi_I=K_I$, $\Phi_+=\PP'_I$. Then
$$\Str_{\Omega^\bt(\s^1)}\prod_{i=1}^n \Phi_{\bar{\zeta}_i}\m(\chi_{\zeta_i})=\left\{\begin{array}{ll}
-\frac{\bar{B}_n}{n!},&\text{ if }\bar{\zeta}_1=\cdots=\bar{\zeta}_n=\zeta_1=\cdots=\zeta_n=I,\\
0,&\text{ otherwise }\end{array}\right.$$
\end{lemma}
(We use notation $\bar{B}_n$ from section \ref{section: interval QME check}: $\bar{B}_n=B_n$ for $n\neq 1$ and $\bar{B}_1=0$).

\textbf{Proof.} Case $\bar{\zeta}_1=\cdots=\bar{\zeta}_n=\zeta_1=\cdots=\zeta_n=I$ reduces to Lemma \ref{interval lemma 2}. Otherwise the operator under super-trace contains one of the following structures:
\begin{eqnarray*}\m(\chi_I)\PP'_I\m(\chi_+)\PP'_I\m(\chi_+)\cdots\PP'_I\m(\chi_+)\PP'_I\m(\chi_I)=0\\
K_I\m(\chi_+)K_I=0\\
\PP'_I\m(\chi_+)K_I=0\\
K_I \m(\chi_+)\PP'_I=0
\end{eqnarray*}
and another possibility is $\bar{\zeta}_1=\cdots=\bar{\zeta}_n=\zeta_1=\cdots=\zeta_n=+$, and then
$$\Str_{\Omega^\bt(\s^1)}(\PP'_I(\chi_+\wedge\bt))^n=\Str_{\Omega^\bt(\s^1)}\PP'_I=\chi(\s^1)=0$$
$\Box$

Applying this lemma to the super-trace arising in $W_L$ on the first circle, we see $W_L$ that could be non-vanishing only for the tuple $\zeta_1=\cdots=\zeta_n=I+$, but then
$$W_L(I+,\ldots,I+)=\Str_{\Omega^{\bt}(\s^1)}(K_I(\chi_I\wedge\bt))^n\cdot \Str_{\Omega^{\bt}(\s^1)}(\chi_+)^n=0$$
since the second super-trace is the Euler characteristic of the circle.
Hence the one-loop part of the cell action for torus for the propagator $K_{I^2,L}$ is
$$S^1_{\TT^2;K_{I^2,L}}=0$$
analogously for the propagator $K_{I^2,R}$
$$S^1_{\TT^2;K_{I^2,R}}=0$$
(we included the propagator in the notation for effective action). Choosing specific chain homotopy is the choice of gauge (the Lagrangian submanifold) for BV integral defining the effective action for $BF$ theory. Changing the chain homotopy, according to general theory (Statement \ref{statement: ind data deform for BF_infty}), leads to a special canonical transformation of the effective action. The fact that chain homotopies $K_{I^2,L}$, $K_{I^2,R}$ and the symmetric one $K_{I^2}$ give precisely the same result for $S_{\TT^2}$ (not just up to a SCT) is a miracle.

The latter argument (for asymmetric chain homotopy) can be generalized to the case of torus of higher dimension $\TT^D=\underbrace{\s^1\times\cdots\times \s^1}_D$, glued from the cube $I^D$ by identifying opposite faces of codimension 1. Basis in the space of cell cochains $C^\bt(\TT^D)$ is $\{e_+,e_I\}^{\times D}$ and the embedding into the cochains of cube $C^\bt(I^D)$ is the tensor power of the embedding of cochains of the circle into cochains of the interval: $\e_+=e_0+e_1$, $e_I=e_I$. Next, basis cochains of torus are embedded into differential forms on torus as coordinate forms $e_{\zeta^0\cdots\zeta^D}\mapsto \chi_{\zeta^0\cdots\zeta^D}$ where $\zeta^i\in\{+,I\}$. Thus IR forms comprise a subalgebra with zero differential in $\Omega^\bt(\TT^D)$, and hence, as in case $D=2$, only the tree $(**)$ and the wheels $(*(*\cdots(*\bt)\cdots))$ contribute to $S_{\TT^D}$. Tree part does not depend on choice of the chain homotopy and yields
\begin{multline}
S^0_{\TT^D}=\frac{1}{2}\sum_{\zeta,\zeta_1,\zeta_2\in\{+,I\}^D}<p_\zeta,\int_\zeta[\chi_{\zeta_1} \omega^{\zeta_1},\chi_{\zeta_2}\omega^{\zeta^2}]>_g\\
=\frac{1}{2}\sum_{\zeta,\zeta_1,\zeta_2\in\{+,I\}^D}(-1)^{(|\zeta_1|+1)\cdot |\zeta_2|}<p_\zeta,[\omega^{\zeta_1},\omega^{\zeta_2}]>_\g
\int_\zeta \chi_{\zeta_1}\wedge\chi_{\zeta_2}\\
=\frac{1}{2}\sum_{\zeta,\zeta_1,\zeta_2\in\{+,I\}^D}(-1)^{(|\zeta_1|+1)\cdot |\zeta_2|}c^\zeta_{\zeta_1,\zeta_2}<p_\zeta,[\omega^{\zeta_1},\omega^{\zeta_2}]>_\g
\label{S^0 for torus}
\end{multline}
where we denoted $c^\zeta_{\zeta_1,\zeta_2}=\int_\zeta \chi_{\zeta_1}\wedge\chi_{\zeta_2}\in\{\pm1,0\}$ the combinatorial factor taking value $\pm1$ if face $\zeta$ is product of faces $\zeta_1$ and $\zeta_2$, and value 0 otherwise.

Next, consider one-loop part of the action $S^1_{\TT^D}$ for the chain homotopy that consecutively contracts the first factor, then the second etc. in $\Omega^\bt(\TT^D)=\Omega^\bt(\s^1)\otimes\cdots\otimes \Omega^\bt(\s^1)$:
$$K_{\TT^D,1\cdots D}=K_I\otimes\underbrace{\id\otimes\cdots\otimes\id}_{D-1}+\PP'_I\otimes K_I\otimes \underbrace{\id\otimes\cdots\otimes \id}_{D-2}+\;\cdots\;+ \underbrace{\PP'_I\otimes\cdots\otimes\PP'_I}_{D-1}\otimes K_I$$
Due to the same argument as for $D=2$, only 1-cochains can contribute to $S^1_{\TT^D}$.
Therefore
\begin{multline} S^1_{\TT^D;K_{\TT^D,1\cdots D}}=\\
=-\sum_{n=2}^\infty (-1)^n\frac{1}{n}\sum_{\zeta_1,\ldots,\zeta_n\in\{I+\cdots+,\,\ldots,\,+\cdots+I\}}
\tr_\g(\ad_{\omega^{\zeta_1}}\cdots \ad_{\omega^{\zeta_n}})\cdot
\Str_{\Omega^\bt(\TT^D)}\prod_{i=1}^D (K_{\TT^D,1\cdots D}\m(\chi_{\zeta_i})) \label{torus 2}\end{multline}
Super-trace over $\Omega^\bt(\TT^D)$ factorizes into super-traces over $\Omega^\bt(\s^1)$, moreover for the first circle edges of the wheel are either decorated by $K_I$ or by the projector $\PP'_I$, and leaves are decorated by periodic Whitney forms $\chi_+=1,\;\chi_I=dt$. Such diagrams contribute only if all leaves are decorated with $\chi_I$ and all edges are decorated with $K_I$ (Lemma \ref{lemma: str on s^1 with P',K}). Therefore only the tuple $\zeta_1=\cdots=\zeta_n=I+\cdots+$ can contribute the sum in (\ref{torus 2}). But then super-traces over other circles vanish:
$$\Str_{\Omega^\bt(\TT^D)}(K_{\TT^D,1\cdots D}(\chi_{I+\cdots+}\wedge\bt))^D=\Str_{\Omega^\bt(\s^1)}(K_I\m(\chi_I))^n\cdot
\left(\Str_{\Omega^\bt(\s^1)}\m(\chi_+)^n\right)^{D-1}=0$$
Therefore we obtain
$$S^1_{\TT^D;K_{\TT^D,1\cdots D}}=0$$
Similarly, for any asymmetric chain homotopy, obtained by consecutive contraction of factors in $\Omega^\bt(\TT^D)=\Omega^\bt(\s^1)\otimes\cdots\otimes \Omega^\bt(\s^1)$ in the order determined by a permutation $\pi$ (\ref{tensor general}), we have
$$S^1_{\TT^D;K_{\TT^D,\pi}}=0$$

\subsubsection{Cylinders $I\times\s^1$, $I\times\TT^D$}
Consider the cylinder, obtained from the square $I\times I$ by gluing edge $I0$ to edge $I1$. This cylinder $I\times \s^1$ comes with the cell decomposition $\{0,1,I\}\times\{+,I\}$, with two 0-cells $0+,\, 1+$, three 1-cells $0I,\, 1I,\, I+$ and one 2-cell $II$. The space of cell cochains is
\begin{eqnarray*}C^\bt(I\times \s^1)=C^\bt(I)\otimes C^\bt(\s^1)=\Span(e_0,e_1,e_I)\otimes \Span(e_+,e_I)=\\
=\Span(e_{0+},e_{1+},e_{0I}, e_{1I}, e_{I+}, e_{II})\end{eqnarray*}
The embedding of cochains into $\Omega^\bt(I\times \s^1)$ is
\begin{eqnarray*}
e_{0+}\mapsto \chi_{0+}=1-t_1,\;e_{1+}\mapsto \chi_{1+}=t_1,\;e_{0I}\mapsto \chi_{0I}=(1-t_1)dt_2,\\
e_{1I}\mapsto \chi_{1I}=t_1dt_2,\; e_{I+}\mapsto \chi_{I+}=dt_1,\;e_{II}\mapsto \chi_{II}=dt_1 dt_2
\end{eqnarray*}
For the sake of simplicity we will use the asymmetric chain homotopy
$$K_{I^2,R}=\id\otimes K_I  +K_I\otimes \PP'_I$$
to calculate the effective action on cochains. Thus we have to analyze Feynman diagrams for $S_{I\times \s^1}$ with edges decorated by parts $\id\otimes K_I$, $K_I\otimes \PP'_I$ of chain homotopy.
First, notice that since Whitney forms on circle $\RR\chi_+\oplus\RR \chi_I$ are closed under multiplication, all diagrams containing an internal edge not in the cycle, decorated with $\id\otimes K_I$, vanish. Therefore all internal edges not in the cycle are decorated with $K_I\otimes\PP'_I$. Hence the only possibly non-vanishing tree diagrams are ``branches'' $(*(*\cdots (**)\cdots))$ (due to the argument we used in the proof of Theorem \ref{interval thm} for the interval). Moreover, for trees with $\geq 3$ leaves in $S^0_{I\times \s^1}$ only the following structures are possible:
\begin{multline*}
(-1)^{n+1}S_{I\times \s^1,\underbrace{(*(*\cdots(**}_{n+1})\cdots))}
\\=<p_{I+},\int_{I+}\ad_{\chi_{I+}\omega^{I+}} ((K_I\otimes\PP'_I) \ad_{\chi_{I+}\omega^{I+}})^{n-1}\circ (\chi_{1+}\omega^{1+}+\chi_{0+}\omega^{0+})>_\g+\\
+<p_{II},\int_{II}\ad_{\chi_{I+}\omega^{I+}} ((K_I\otimes\PP'_I) \ad_{\chi_{I+}\omega^{I+}})^{n-1}\circ (\chi_{1I}\omega^{1I}+\chi_{0I}\omega^{0I})>_\g+\\
+<p_{II},\int_{II}\ad_{\chi_{II}\omega^{II}} ((K_I\otimes\PP'_I) \ad_{\chi_{I+}\omega^{I+}})^{n-1}\circ (\chi_{1+}\omega^{1+}+\chi_{0+}\omega^{0+})>_\g+\\
+\sum_{k=2}^n<p_{II},\int_{II} \ad_{\chi_{I+}\omega^{I+}}((K_I\otimes\PP'_I) \ad_{\chi_{I+}\omega^{I+}})^{k-2}(K_I\otimes\PP'_I) \ad_{\chi_{II}\omega^{II}}\cdot\\
\cdot((K_I\otimes\PP'_I) \ad_{\chi_{I+}\omega^{I+}})^{n-k}
\circ (\chi_{1+}\omega^{1+}+\chi_{0+}\omega^{0+})>_\g
\end{multline*}
for $n\geq 2$ (sign $(-1)^{n+1}$ take into account that we decorate edges with $K_{I^2,R}$ instead of $-K_{I^2,R}$).
Thus the values of tree diagrams are calculated straightforwardly, using factorization and (\ref{interval lemma 1 eq2}):
\begin{multline*}
S_{I\times \s^1,\underbrace{(*(*\cdots(**}_{n+1})\cdots))}=
\frac{B_n}{n!}<p_{I+},(\ad_{\omega^{I+}})^n\circ (\omega^{1+}-\omega^{0+})>_\g+\\
+\frac{B_n}{n!}<p_{II},(\ad_{\omega^{I+}})^n\circ (\omega^{1I}-\omega^{0I})+\sum_{k=1}^n(\ad_{\omega^{I+}})^{k-1}
\ad_{\omega^{II}}(\ad_{\omega^{I+}})^{n-k}\circ (\omega^{1+}-\omega^{0+})>_\g
\end{multline*}
Tree diagrams $(*)$ and $(**)$ are calculated separately:
$$S_{I\times \s^1,(*)}=<p_{I+},\omega^{1+}-\omega^{0+}>_\g+<p_{II},\omega^{1I}-\omega^{0I}>_\g$$
--- generating function for the differential on cell cochains of the cylinder,
\begin{multline*}S_{I\times \s^1,(**)}=\frac{1}{2}<p_{0+},[\omega^{0+},\omega^{0+}]>_\g+\frac{1}{2}<p_{1+},[\omega^{1+},\omega^{1+}]>_\g+
<p_{0I},[\omega^{0I},\omega^{0+}]>_\g+\\
+<p_{1I},[\omega^{1I},\omega^{1+}]>_\g+\frac{1}{2}<p_{I+},[\omega^{I+},\omega^{0+}+\omega^{1+}]]>_\g+
\frac{1}{2}<p_{II},[\omega^{I+},\omega^{0I}+\omega^{1I}]+[\omega^{II},\omega^{0+}+\omega^{1+}]>_\g
\end{multline*}
--- generating function for the product on cochains  (projected product of Whitney forms on cylinder).

\textbf{One-loop cell action for $I\times\s^1$.} Next, to calculate one-loop cell action for the cylinder, we use the representation of $S^1_{I\times \s^1}$ via $L_\infty$ morphism (\ref{S' via U and I}):
\be S^1_{I\times \s^1}=-\sum_{n=2}^\infty (-1)^n\frac{1}{n}\;\Str_{\Omega^\bt(I\times \s^1,\g)}\left(K_{I^2,R}[U_{I\times \s^1}(\omega),\bt]\right)^n \label{cyl 1}\ee
where the $L_\infty$ morphism is computed analogously to the tree part of action, the only change is that now we decorate root  with the propagator $-K_{I^2,R}$ (instead of with the retraction to cochains):
\begin{multline*}
U_{I\times \s^1}(\omega)=\chi_{0+}\omega^{0+}+\chi_{1+}\omega^{1+}+\chi_{0I}\omega^{0I}+\chi_{1I}\omega^{1I}+\chi_{I+}\omega^{I+}+\chi_{II}\omega^{II}+\\
+\sum_{m=1}^\infty (-1)^m((K_I\otimes\PP'_I)\ad_{\chi_{I+}\omega^{I+}})^m\circ (\chi_{1+}\omega^{1+}+\chi_{0+}\omega^{0+})+\\
+\sum_{m=1}^\infty (-1)^m((K_I\otimes\PP'_I)\ad_{\chi_{I+}\omega^{I+}})^m\circ (\chi_{1I}\omega^{1I}+\chi_{0I}\omega^{0I})+\\
+\sum_{m=1}^\infty (-1)^m\sum_{k=1}^m ((K_I\otimes\PP'_I)\ad_{\chi_{I+}\omega^{I+}})^{k-1}(K_I\otimes\PP'_I)\ad_{\chi_{II}\omega^{II}} ((K_I\otimes\PP'_I)\ad_{\chi_{I+}\omega^{I+}})^{m-k}\circ\\
\circ (\chi_{1+}\omega^{1+}+\chi_{0+}\omega^{0+})\\
=\chi_{0+}\omega^{0+}+\chi_{1+}\omega^{1+}+\chi_{0I}\omega^{0I}+\chi_{1I}\omega^{1I}+\chi_{I+}\omega^{I+}+\chi_{II}\omega^{II}-\\
-\sum_{m=1}^\infty \frac{B_{m+1}(t_1)-B_{m+1}}{(m+1)!}(\ad_{\omega^{I+}})^m\circ (\omega^{1+}-\omega^{0+})
-\sum_{m=1}^\infty\frac{B_{m+1}(t_1)-B_{m+1}}{(m+1)!}\;dt_2\cdot\\ \cdot \left(\sum_{k=1}^m (\ad_{\omega^{I+}})^{k-1}\ad_{\omega^{II}}(\ad_{\omega^{I+}})^{m-k}
\circ (\omega^{1+}-\omega^{0+})+(\ad_{\omega^{I+}})^m\circ (\omega^{1I}-\omega^{0I})\right)
\end{multline*}
Notice that $U_{I\times \s^1}(\omega)$ has the following structure:
$$U_{I\times \s^1}(\omega)=f(\omega,t_1)\otimes \chi_+ + (\omega^{0I}\chi_{0}+\omega^{1I}\chi_{1}+g(\omega,t_1))\otimes \chi_I+\omega^{II}\chi_{II}$$
where $f$ and $g$ are certain functions of $t_1$ (constructed from Bernoulli polynomials), depending on the cochain $\omega$, and also $g|_{t_1=0}=g|_{t_1=1}=0$ holds. Therefore (\ref{cyl 1}) is expressed as a sum of products of super-traces over $\Omega^\bt(I)$ and over $\Omega^\bt(\s^1)$, moreover, for the second factor Lemma \ref{lemma: str on s^1 with P',K} can be applied. Therefore we have
\begin{multline*}S^1_{I\times \s^1}=-\sum_{n=2}^\infty (-1)^n\frac{1}{n}\Str_{\Omega^\bt(I\times \s^1,\g)}((\id\otimes K_I)
[(\omega^{0I}\chi_{0}+\omega^{1I}\chi_{1}+g(\omega,t_1))\otimes \chi_I,\bt])^n\\
=-\sum_{n=2}^\infty (-1)^n\frac{1}{n}\Str_{\Omega^\bt(I,\g)}([\omega^{0I}\chi_{0}+\omega^{1I}\chi_{1}+g(\omega,t_1),\bt])^n\cdot
\Str_{\Omega^\bt(\s^1)}(K_I(\chi_I\wedge\bt))^n\\
=\sum_{n=2}^\infty \frac{B_n}{n\cdot n!}\Str_{\Omega^\bt(I,\g)}([\omega^{0I}\chi_{0}+\omega^{1I}\chi_{1}+g(\omega,t_1),\bt])^n
\end{multline*}
We evaluate the super-trace over $\Omega^\bt(I,\g)$ using (\ref{str of multiplication}) (and using the fact that $g$ vanishes in end-points of the interval):
$$S^1_{I\times \s^1}=\sum_{n=2}^\infty \frac{B_n}{n\cdot n!}\;\frac{1}{2}(\tr_\g(\ad_{\omega^{0I}})^n+ \tr_\g(\ad_{\omega^{1I}})^n)=
\frac{1}{2}\tr_\g\log\left(\frac{\sinh\frac{\ad_{\omega^{0I}}}{2}}{\frac{\ad_{\omega^{0I}}}{2}}\right)+
\frac{1}{2}\tr_\g\log\left(\frac{\sinh\frac{\ad_{\omega^{1I}}}{2}}{\frac{\ad_{\omega^{1I}}}{2}}\right)$$
Thus we obtained the following result for the cylinder with chain homotopy $K_{I^2,R}=\id\otimes K_I+K_I\otimes\PP'_I$:
\begin{multline*}
S_{I\times\s^1}\\=<p_{I+},\omega^{1+}-\omega^{0+}>_\g+<p_{II},\omega^{1I}-\omega^{0I}>_\g
+\frac{1}{2}<p_{0+},[\omega^{0+},\omega^{0+}]>_\g+\frac{1}{2}<p_{1+},[\omega^{1+},\omega^{1+}]>_\g+\\
+<p_{0I},[\omega^{0I},\omega^{0+}]>_\g
+<p_{1I},[\omega^{1I},\omega^{1+}]>_\g+\frac{1}{2}<p_{I+},[\omega^{I+},\omega^{0+}+\omega^{1+}]]>_\g+\\+
\frac{1}{2}<p_{II},[\omega^{I+},\omega^{0I}+\omega^{1I}]+[\omega^{II},\omega^{0+}+\omega^{1+}]>_\g
+\sum_{n=2}^\infty\frac{B_n}{n!}<p_{I+},(\ad_{\omega^{I+}})^n\circ (\omega^{1+}-\omega^{0+})>_\g+\\
+\sum_{n=2}^\infty\frac{B_n}{n!}<p_{II},(\ad_{\omega^{I+}})^n\circ (\omega^{1I}-\omega^{0I})+\sum_{k=1}^n(\ad_{\omega^{I+}})^{k-1}
\ad_{\omega^{II}}(\ad_{\omega^{I+}})^{n-k}\circ (\omega^{1+}-\omega^{0+})>_\g+\\
+\hbar\sum_{n=2}^\infty \frac{B_n}{n\cdot n!}\;\frac{1}{2}(\tr_\g(\ad_{\omega^{0I}})^n+ \tr_\g(\ad_{\omega^{1I}})^n)
\end{multline*}
Or, evaluating sums with Bernoulli numbers:
\begin{multline}
S_{I\times\s^1}=\frac{1}{2}<p_{0+},[\omega^{0+},\omega^{0+}]>_\g+\frac{1}{2}<p_{1+},[\omega^{1+},\omega^{1+}]>_\g+\\+
<p_{0I},[\omega^{0I},\omega^{0+}]>_\g
+<p_{1I},[\omega^{1I},\omega^{1+}]>_\g+\\+\frac{1}{2}<p_{I+},[\omega^{I+},\omega^{0+}+\omega^{1+}]]>_\g+
\frac{1}{2}<p_{II},[\omega^{I+},\omega^{0I}+\omega^{1I}]+[\omega^{II},\omega^{0+}+\omega^{1+}]>_\g+\\
+\int_0^1 <p_{II}+d\nu\cdot p_{I+},\left(\frac{\ad_{\omega^{I+}+d\nu\cdot \omega^{II}}}{2}\coth\frac{\ad_{\omega^{I+}+d\nu\cdot\omega^{II}}}{2}\right)\circ
((\omega^{1+}-\omega^{0+})+d\nu\cdot(\omega^{1I}-\omega^{0I}))>_\g+\\
+\hbar\;\left(\frac{1}{2}\tr_\g\log\left(\frac{\sinh\frac{\ad_{\omega^{0I}}}{2}}{\frac{\ad_{\omega^{0I}}}{2}}\right)+
\frac{1}{2}\tr_\g\log\left(\frac{\sinh\frac{\ad_{\omega^{1I}}}{2}}{\frac{\ad_{\omega^{1I}}}{2}}\right)\right)
\label{cyl result}
\end{multline}
where the integral is evaluated over the auxiliary variable $\nu\in[0,1]$.

\textbf{Cylinder $I\times \TT^D$ for $D\geq 2$ in asymmetric gauge.} Now consider the higher-dimensional cylinder $I\times \TT^D =I\times \underbrace{\s^1\times\cdots\times \s^1}_D$ with cell decomposition
$\{0,1,I\}\times \{0,+\}\times\cdots\times \{0,+\}$ and the asymmetric chain homotopy
\be K_{I\times\TT^D}=\underbrace{\id\otimes\cdots\otimes\id}_D\otimes K_I+\underbrace{\id\otimes\cdots\otimes\id}_{D-1}\otimes K_I\otimes\PP'_I+
\cdots+K_I\otimes\underbrace{\PP'_I\otimes\cdots\otimes\PP'_I}_D \label{K for IxT^D}\ee
i.e. we first contract de Rham complex of the last circle, then next to last etc., finally we contract de Rham complex of the interval (indeed, one can choose any other order of contraction; it is only important that first we contract one of the circles, not the interval).

Similarly to the case of $I\times \s^1$, since Whitney forms on the circle are closed under multiplication, all internal edges of Feynman diagrams not belonging to the cycle have to be decorated with the part $K_I\otimes\PP'_I\otimes\cdots\otimes\PP'_I$ of the propagator. Therefore among tree diagrams only ``branches'' contribute, and we obtain for the tree part of action
\begin{multline}S^0_{I\times \TT^D}= \sum_{\zeta\in\{+,I\}^D}<p_{I\zeta},\omega^{1\zeta}-\omega^{0\zeta}>_\g  +
\frac{1}{2}\sum_{\zeta,\zeta_1,\zeta_2\in\{+,I\}^D}(-1)^{(|\zeta_1|+1)\cdot|\zeta_2|}c^\zeta_{\zeta_1,\zeta_2} \cdot\\ \cdot\left(<p_{0\zeta},[\omega^{0\zeta_1},\omega^{0\zeta_2}]>_\g+
<p_{1\zeta},[\omega^{1\zeta_1},\omega^{1\zeta_2}]>_\g+(-1)^{|\zeta_2|} <p_{I\zeta},[\omega^{I\zeta_1},\omega^{0\zeta_2}+\omega^{1\zeta_2}]>_\g\right)+\\+
\sum_{n=2}^\infty\frac{B_n}{n!}\sum_{\zeta,\zeta_1,\ldots,\zeta_{n+1}\in \{+,I\}^D} (-1)^{\sum_{1\leq i<j\leq n+1}|\zeta_i|\cdot |\zeta_j|}
c^\zeta_{\zeta_1,\ldots,\zeta_{n+1}}
<p_{I\zeta},\prod_{i=1}^n\ad_{\omega^{I\zeta_i}}\circ(\omega^{1\zeta_{n+1}}-\omega^{0\zeta_{n+1}})>_\g \label{S^0 for cylinder}
\end{multline}
where we denoted
$$c^\zeta_{\zeta_1,\ldots,\zeta_{n}}=\int_\zeta \chi_{\zeta_1}\wedge\cdots\wedge\chi_{\zeta_n}\in\{\pm1,0\}$$
One-loop part vanishes for $D\geq 2$ by the following argument. Due to Lemma \ref{lemma: str on s^1 with P',K}, all edges in the cycle are decorated with the part $\id\otimes\cdots\otimes\id\otimes K_I$ of the propagator. Hence, analyzing the factorized de Rham part of the diagram, we notice, that on each circle but the last one we are computing the super-trace of multiplication by some function and $\Str_{\Omega^\bt(\s^1)}f(t)\wedge\bt=0$ for any $f(t)$ (as implied by (\ref{str of multiplication})). Therefore
$$S^1_{I\times\s^D}=0$$
for $D\geq 2$, and for the choice (\ref{K for IxT^D}) for the chain homotopy.

\textbf{Cylinder $I\times \TT^D$ in gauge $K=\id\otimes K_{\TT^D}+K_I\otimes \PP'_{\TT^D}$.}
The discussions above for cylinder $I\times \s^1$ and for $I\times \TT^D$ with gauge (\ref{K for IxT^D}) can be generalized to the following statement:
cell action for cylinder $I\times \TT^D$ (with $D\geq 1$) with standard cell decomposition and chain homotopy
\be K_{I\times \TT^D}=\id\otimes K_{\TT^D}+K_I\otimes \PP'_{\TT^D} \label{cylinder K}\ee (where $K_{\TT^D}$ may be any chain homotopy for the torus; in particular we can take the symmetric one) is
\be S_{I\times\TT^D}=S_{I\times\TT^D}^0+\frac{1}{2}\hbar (S^1_{\{0\}\times \TT^D}+S^1_{\{1\}\times \TT^D}) \label{cylinder cell action}\ee
where $S_{I\times\TT^D}^0$ is given by (\ref{S^0 for cylinder}) and $S^1_{\{0\}\times \TT^D}$, $S^1_{\{1\}\times \TT^D}$ denote the one-loop action $S^1_{\TT^D}$ for torus $\TT^D$ (for the gauge $K_{\TT^D}$), evaluated on the restriction of cell fields $\omega, p$ of the cylinder to either of the bounding tori $\{0\}\times \TT^D, \{1\}\times \TT^D \subset I\times \TT^D$:
$$S^1_{\{0\}\times \TT^D}=S^1_{\TT^D}|_{\omega^\zeta\mapsto \omega^{0\zeta}},\quad S^1_{\{1\}\times \TT^D}=S^1_{\TT^D}|_{\omega^\zeta\mapsto \omega^{1\zeta}} $$
The argument is as follows. Since $C^\bt(\TT^D)$ is embedded into $\Omega^\bt(\TT^D)$ as a subalgebra, all edges of Feynman diagrams not belonging to the cycle have to be decorated with part $K_I\otimes \PP'_{\TT^D}$ of the propagator. This implies that the tree part of action does not depend on choice of $K_{\TT^D}$ and hence coincides with (\ref{S^0 for cylinder}). Also, $L_\infty$ morphism satisfies the ansatz
\be U_{I\times \TT^D}(\omega)=\sum_{\zeta\in \{+,I\}^D}(\omega^{0\zeta}\chi_{0\zeta}+\omega^{1\zeta}\chi_{1\zeta}+\omega^{I\zeta}\chi_{I\zeta}+f_\zeta(t_0,\omega)\otimes \chi_\zeta) \label{U for cylinder}\ee
where $t_0\in[0,1]$ is the coordinate on the interval and $f_\zeta(t_0,\omega)$ are some functions, vanishing at $t_0=0,1$. Now we have to invoke an argument similar to Lemma \ref{lemma: str on s^1 with P',K}:
\begin{lemma} \label{lm8}
Super-trace $$\Str_{\Omega^\bt(\TT^D)}\prod_{i=1}^n\OO_i\m(\chi_{\zeta_i})$$ for $\OO_i\in\{K_{\TT^D},\PP'_{\TT^D}\}$, $\zeta_i\in \{+,I\}^D$ (cells of $\TT^D$) vanishes unless all $\OO_i=K_{\TT^D}$ and all $\zeta_i$ are 1-dimensional.
\end{lemma}
\textbf{Proof.} Suppose some $\zeta_i=+\cdots +$. Then the super-trace vanishes unless $\OO_i=\OO_{i+1}=\PP'_{\TT^D}$, since $\chi_{+\cdots +}=1$ and  $K^2_{\TT^D}=\PP'_{\TT^D}K_{\TT^D}=K_{\TT^D}\PP'_{\TT^D}=0$. On the other hand, by degree counting we need $\#\{i: \OO_i=\PP'_{\TT^D}\}\leq \#\{i: \zeta_i=+\cdots +\}$. Thus the only possibly non-vanishing decoration containing the insertion of $+\cdots +$ is $\OO_i=\PP'_{\TT^D}$, $\zeta_i=+\cdots +$ for all $i$. But this vanishes, since the Euler characteristic of torus vanishes. Therefore for all decorations containing at least one $\zeta$ of dimension zero the super-trace vanishes. This (together with degree counting) leaves the only option for non-vanishing super-trace: $\OO_i=K_{\TT^D}$ for all $i$ and all $\zeta_i$ are 1-dimensional.
$\Box$

This lemma together with (\ref{U for cylinder}) implies that to the one-loop action
$$S^1_{I\times \TT^D}=-\sum_{n=2}^\infty (-1)^n\frac{1}{n}\;\Str_{\Omega^\bt(I\times \TT^D,\g)}\left(K_{I\times\TT^D}[U_{I\times \TT^D}(\omega),\bt]\right)^n 
$$
only the decoration of in-cycle edges with the part $\id\otimes K_{\TT^D}$ of chain homotopy contributes. Therefore in the first factor of $I\times \TT^D$ we are evaluating the super-trace of multiplication operator. Therefore (see \ref{str of multiplication}) only first two terms in the expression (\ref{U for cylinder}) for $L_\infty$ morphism are relevant for computing $S^1_{I\times \TT^D}$, moreover the only contributing diagrams are wheels with leaves decorated either only with $\sum_{\zeta\in \{+,I\}^D}\omega^{0\zeta}\chi_{0\zeta}$ or only with $\sum_{\zeta\in \{+,I\}^D}\omega^{1\zeta}\chi_{1\zeta}$ and all edges decorated with $\id\otimes K_{\TT^D}$. This immediately gives the result (\ref{cylinder cell action}), with factors $1/2$ coming from $\Str_{\Omega^\bt(I)}(\m(\chi_0))^n=\Str_{\Omega^\bt(I)}(\m(\chi_1))^n=\frac{1}{2}$.

\subsubsection{Klein bottle}
\label{section: KB cell}
Cell action for the Klein bottle is obtained from cell action for the cylinder (\ref{cyl result}) using the gluing construction (section \ref{section: gluing}). Namely, we set $V=C^\bt(I\times \s^1,\g)$ --- cochains of the cylinder with basis $\{e_{0+},e_{1+},e_{0I},e_{1I},e_{I+},e_{II}\}$, and $W=C^\bt (\s^1,\g)$ --- cochains of the circle  with basis $\{\bar{e}_+,\bar{e}_I\}$, and define the pair of projections $\pi_{1,2}:V\ra W$ and pair of embeddings $\iota_{1,2}:W\ra V$ as
\begin{eqnarray*}\pi_1:\quad x^{0+}e_{0+}+x^{1+}e_{1+}+x^{0I}e_{0I}+x^{1I}e_{1I}+x^{I+}e_{I+}+x^{II}e_{II}&\mapsto& x^{0+}\bar{e}_+ +x^{0I}\bar{e}_I\\
\pi_2:\quad x^{0+}e_{0+}+x^{1+}e_{1+}+x^{0I}e_{0I}+x^{1I}e_{1I}+x^{I+}e_{I+}+x^{II}e_{II}&\mapsto& x^{1+}\bar{e}_+ -x^{1I}\bar{e}_I\\
\iota_1:\quad y^+\bar{e}_+ +y^I \bar{e}_I&\mapsto& y^{+} e_{0+}+ y^{I} e_{0I}\\
\iota_2:\quad y^+\bar{e}_+ +y^I \bar{e}_I&\mapsto& y^{+} e_{1+}- y^{I} e_{1I}
\end{eqnarray*}
Since $\pi_{1,2}$ come from two geometric embeddings of circle into cylinder as a cell subcomplex, they are a priori linear $L_\infty$ morphisms, as implied by cell locality property of the cell action (alternatively, one can see this directly from the explicit formula (\ref{cyl result})). Therefore the gluing construction for $BF$ actions is applicable. The glued space is
$$V'=\ker\pi_-=\g\otimes\Span( e_{++},e_{-I}, e_{I+},e_{II})\subset V$$
where $e_{++}=e_{1+}+e_{0+}$ and $e_{-I}=e_{1I}-e_{0I}$. Space $V'$ is identified with the space of $\g$-valued cochains of the Klein bottle: $V'=C^\bt(\mr{KB},\g)$ with cell decomposition $\{++,-I,I+,II\}$
(notice that $e_{-I}$ is defined without the factor $1/2$ here, so that it is a properly normalized basis cochain for the cell $-I$).
The glued action is
\begin{multline*}S_\mr{KB}=S_{I\times\s^1}|_{\omega^{0+}\mapsto\omega^{++},\,\omega^{1+}\mapsto\omega^{++},\,\omega^{0I}\mapsto-\omega^{-I},\,
\omega^{1I}\mapsto\omega^{-I},\,p_{0+}\mapsto p_{++},\,p_{1+}\mapsto p_{++}, \,p_{0I}\mapsto -p_{-I},\,p_{1I}\mapsto p_{-I}}-\\
-S_{\s^1}|_{\bar{\omega}^+\mapsto\omega^{++},\,\bar{\omega}^I\mapsto\omega^{-I},\,\bar{p}_+\mapsto p_{++},\,\bar{p}_I\mapsto p_{-I}}=\\
=\frac{1}{2}<p_{++},[\omega^{++},\omega^{++}]>_\g+<p_{-I},[\omega^{-I},\omega^{++}]>_\g+<p_{I+},[\omega^{I+},\omega^{++}]>_\g+\\+
<p_{II},[\omega^{II},\omega^{++}]>_\g+2<p_{II},\left(\frac{\ad_{\omega^{I+}}}{2}\coth \frac{\ad_{\omega^{I+}}}{2}\right)\circ\omega^{-I}>_\g
\end{multline*}
Note that one-loop part of action cancelled between the cylinder and the circle.

\subsubsection{$S_D\ltimes \ZZ_2^D$-bundles over circle with fiber $\TT^D$}
\label{section: M_gamma}
Our calculation of the cell action for the Klein bottle can be generalized straightforwardly to bundles $M_\gamma$ over circle with fiber $\TT^D$ and structure group $S_D\ltimes \ZZ_2^D$. Namely, for any element $\gamma\in S_D\ltimes \ZZ_2^D$ we define $M_\gamma$ to be the manifold obtained from the cylinder $I\times \TT^D$ by gluing the bounding torus $\{1\}\times \TT^D$ to the bounding torus $\{0\}\times \TT^D$, twisted by $\gamma$ (we assume the standard action of the symmetry group of $D$-cube on the torus $\TT^D$):
$$M_\gamma=I\times \TT^D\; /\; (\{1\}\times \TT^D \sim \{0\}\times \gamma\circ \TT^D)$$
The cell decomposition for $M_\gamma$ is obtained from standard cell decomposition for the cylinder $I\times\TT^D$ by gluing and has cells $\{I\zeta,\bar 1\zeta \}_{\zeta\in \{+,I\}^D}$ where $\zeta$ runs over the set of cells of the torus $\TT^D$. We denoted $\bar{1}\zeta$ the cell of $M_\gamma$, obtained by gluing cell $1\zeta$ of the cylinder to cell $0\zeta$, twisted by $\gamma$. The basis cochains of $M_\gamma$ are embedded into the cochains of the cylinder as
$$e_{I\zeta}\mapsto e_{I\zeta}\, ,\quad  e_{\bar{1}\zeta}\mapsto e_{1\zeta}+\gamma\circ e_{0\zeta}$$
Here we assume that $\gamma$ acts on cochains as follows: for $\gamma=(\pi;o_1,\ldots,o_D)$ with $\pi\in S_D$ a permutation and $(o_1,\ldots,o_D)\in \ZZ_2^D$ (with $o_i=\pm 1$ for each $i$) we set $\gamma\circ e_{0\zeta}=\pm e_{0 \pi(\zeta)}$ where $\pi$ acts on faces of torus $\zeta\in \{+,I\}^D$ by permuting factors and the sign is
\be\int_{\pi(\zeta)} \gamma\circ \chi_{\zeta}=o_1^{|\zeta^1|}\cdots o_D^{|\zeta^D|}\int_{\pi(\zeta)} \pi\circ \chi_{\zeta}= \pm 1 \label{sec545 sign}\ee
(with $\chi_\zeta$ the Whitney form on torus).

The cell action for $M_\gamma$ is obtained from the action (\ref{cylinder cell action}) by gluing construction:
\be S_{M_\gamma}=\left.S_{I\times \TT^D}^0\right|_{\omega^{1\zeta}\mapsto \omega^{\bar{1}\zeta},\omega^{0\zeta}\mapsto\gamma\circ \omega^{\bar{1}\zeta},p_{1\zeta}\mapsto p_{\bar{1}\zeta}, p_{0\zeta}\mapsto \gamma\circ p_{\bar{1}\zeta}}-\left. S_{\TT^D}^0\right|_{\omega^\zeta\mapsto \omega^{\bar{1}\zeta}, p_\zeta\mapsto p_{\bar{1}\zeta}} \label{M_gamma cell action}\ee
where $\gamma$ acts on $\omega^{\bar{1}\zeta}$ and $p_{\bar{1}\zeta}$ by permutation on $\zeta$ and with the sign (\ref{sec545 sign}). This result is obtained for the chain homotopy (\ref{cylinder K}) for the cylinder $I\times \TT^D$ where we are forced to use the symmetric chain homotopy $K_{\TT^D}$ for the torus (otherwise the induction data would not be consistent with gluing for general $\gamma\in S_D\ltimes \ZZ_2^D$). Notice also that the one-loop part of the cell action cancels due to (\ref{cylinder cell action}). Therefore  $S_{M_\gamma}$ does not depend on the one-loop cell action for the torus with symmetric chain homotopy (which we did not calculate for $D\geq 3$) and is obtained only from the known tree results for cylinder and torus (\ref{S^0 for cylinder},\ref{S^0 for torus}).

Result (\ref{M_gamma cell action}) for manifolds $M_\gamma$ reduces to the result for Klein bottle\footnote{To make the notation consistent with section \ref{section: KB cell}, we have to identify the cell labels as $\bar{1}+ :=++$, $\bar{1}I := -I$.} if we choose $\gamma=((1);-1)\in S_1\ltimes \ZZ_2$ and reduces to the result\footnote{Actually, in this way we obtain the result for torus $\TT^{D+1}$ with the chain homotopy (\ref{cylinder K}) which differs from chain homotopy $K_{\TT^{D+1},\pi}$ discussed in section \ref{section: T^D in asym gauge}, but the result is the same: one-loop part of the cell action vanishes, while the tree part is obviously independent on choice of $K$ (since cell cochains are embedded into differential forms as a subalgebra).} for $(D+1)$-torus if we take $\gamma=((1,2,\ldots,D);1,1,\ldots,1)\in S_D\ltimes \ZZ_2^D$ --- the unit of the group.

Let us summarize the results of this section:
\begin{statement}
\label{statement: exact cell action examples}
\begin{itemize}
\item For the torus $\TT^2$ with cell decomposition $\{+,I\}\times\{+,I\}$ and symmetric chain homotopy $K=K_I\otimes\frac{\id+\PP'_I}{2}+\frac{\id+\PP'_I}{2}\otimes K_I$ the cell $BF$ action is
\begin{multline}
S_{\TT^2,\mr{sym}}=\frac{1}{2}<p_{++},[\omega^{++},\omega^{++}]>_\g+<p_{I+},[\omega^{I+},\omega^{++}]>_\g+\\
+<p_{+I},[\omega^{+I},\omega^{++}]>_\g+<p_{II},[\omega^{I+},\omega^{+I}]+[\omega^{II},\omega^{++}]>_\g\\
\label{torus sym cell result}
\end{multline}
\item For the torus $\TT^D$ with cell decomposition $\{+,I\}^{D}$ and asymmetric chain homotopy (\ref{tensor general}) (circles are contracted in arbitrary order) the cell $BF$ action for  $D\geq 2$ is
\be
S_{\TT^D}=\frac{1}{2}\sum_{\zeta,\zeta_1,\zeta_2\in\{+,I\}^D}(-1)^{(|\zeta_1|+1)\cdot |\zeta_2|}c^\zeta_{\zeta_1,\zeta_2}<p_\zeta,[\omega^{\zeta_1},\omega^{\zeta_2}]>_\g
\label{torus asym cell result}\ee
where $c^\zeta_{\zeta_1,\zeta_2}=\int_\zeta\chi_{\zeta_1}\wedge\chi_{\zeta_2}\in\{\pm1,0\}$.
\item For the cylinder $I\times \s^1$ with cell decomposition $\{0,1,I\}\times\{+,I\}$ and asymmetric chain homotopy
$K=\id\otimes K_I+K_I\otimes \PP'_I$ the cell $BF$ action is
\begin{multline*}S_{I\times\s^1}=\frac{1}{2}<p_{0+},[\omega^{0+},\omega^{0+}]>_\g+\frac{1}{2}<p_{1+},[\omega^{1+},\omega^{1+}]>_\g+\\+
<p_{0I},[\omega^{0I},\omega^{0+}]>_\g
+<p_{1I},[\omega^{1I},\omega^{1+}]>_\g+\\+\frac{1}{2}<p_{I+},[\omega^{I+},\omega^{0+}+\omega^{1+}]]>_\g+
\frac{1}{2}<p_{II},[\omega^{I+},\omega^{0I}+\omega^{1I}]+[\omega^{II},\omega^{0+}+\omega^{1+}]>_\g+\\
+\int_0^1 <p_{II}+d\nu \cdot p_{I+},\left(\frac{\ad_{\omega^{I+}+d\nu \cdot \omega^{II}}}{2}\coth\frac{\ad_{\omega^{I+}+d\nu \cdot\omega^{II}}}{2}\right)\circ
((\omega^{1+}-\omega^{0+})+d\nu \cdot(\omega^{1I}-\omega^{0I}))>_\g+\\
+\hbar\;\left(\frac{1}{2}\tr_\g\log\left(\frac{\sinh\frac{\ad_{\omega^{0I}}}{2}}{\frac{\ad_{\omega^{0I}}}{2}}\right)+
\frac{1}{2}\tr_\g\log\left(\frac{\sinh\frac{\ad_{\omega^{1I}}}{2}}{\frac{\ad_{\omega^{1I}}}{2}}\right)\right)
\end{multline*}
where $\nu\in[0,1]$ is an auxiliary variable.
\item For the cylinder $I\times \TT^D$ with cell decomposition $\{0,1,I\}\times\{+,I\}^D$ and asymmetric chain homotopy (\ref{tensor general}) (first we contract one of the circles and then other circles and the interval in arbitrary order) the cell $BF$ action for $D\geq 2$ is
\begin{multline}S_{I\times \TT^D}= \sum_{\zeta\in\{+,I\}^D}<p_{I\zeta},\omega^{1\zeta}-\omega^{0\zeta}>_\g  +
\frac{1}{2}\sum_{\zeta,\zeta_1,\zeta_2\in\{+,I\}^D}(-1)^{(|\zeta_1|+1)\cdot|\zeta_2|}c^\zeta_{\zeta_1,\zeta_2} \cdot\\ \cdot\left(<p_{0\zeta},[\omega^{0\zeta_1},\omega^{0\zeta_2}]>_\g+
<p_{1\zeta},[\omega^{1\zeta_1},\omega^{1\zeta_2}]>_\g+(-1)^{|\zeta_2|} <p_{I\zeta},[\omega^{I\zeta_1},\omega^{0\zeta_2}+\omega^{1\zeta_2}]>_\g\right)+\\+
\sum_{n=2}^\infty\frac{B_n}{n!}\sum_{\zeta,\zeta_1,\ldots,\zeta_{n+1}\in \{+,I\}^D} (-1)^{\sum_{1\leq i<j\leq n+1}|\zeta_i|\cdot |\zeta_j|}
c^\zeta_{\zeta_1,\ldots,\zeta_{n+1}}
<p_{I\zeta},\prod_{i=1}^n\ad_{\omega^{I\zeta_i}}\circ(\omega^{1\zeta_{n+1}}-\omega^{0\zeta_{n+1}})>_\g
\label{cylinder cell result}
\end{multline}
where $c^\zeta_{\zeta_1,\ldots,\zeta_{n}}=\int_\zeta \chi_{\zeta_1}\wedge\cdots\wedge\chi_{\zeta_n}\in\{\pm1,0\}$.
\item For the Klein bottle with cell decomposition $\{++,-I,I+,II\}$ and asymmetric chain homotopy $K=\id\otimes K_I+K_I\otimes \PP'_I$ the cell $BF$ action is
\begin{multline}
S_\mr{KB}=\frac{1}{2}<p_{++},[\omega^{++},\omega^{++}]>_\g+<p_{-I},[\omega^{-I},\omega^{++}]>_\g+<p_{I+},[\omega^{I+},\omega^{++}]>_\g+\\+
<p_{II},[\omega^{II},\omega^{++}]>_\g+2<p_{II},\left(\frac{\ad_{\omega^{I+}}}{2}\coth \frac{\ad_{\omega^{I+}}}{2}\right)\circ\omega^{-I}>_\g
\label{KB cell result}
\end{multline}
\item For the $S_D\ltimes \ZZ_2^D$-bundle $M_\gamma$ glued from the cylinder $I\times\TT^D$, with cells $\{I\zeta,\bar{1}\zeta\}_{\zeta\in\{+,I\}^D}$ and chain homotopy $\id\otimes K_{\TT^D}+K_I\otimes \PP'_{\TT^D}$ (where $K_{\TT^D}$ denotes the symmetric chain homotopy for torus) the cell action is
\be S_{M_\gamma}=\left.S_{I\times \TT^D}\right|_{\omega^{1\zeta}\mapsto \omega^{\bar{1}\zeta},\omega^{0\zeta}\mapsto\gamma\circ \omega^{\bar{1}\zeta},p_{1\zeta}\mapsto p_{\bar{1}\zeta}, p_{0\zeta}\mapsto \gamma\circ p_{\bar{1}\zeta}}-\left. S_{\TT^D}\right|_{\omega^\zeta\mapsto \omega^{\bar{1}\zeta}, p_\zeta\mapsto p_{\bar{1}\zeta}} \label{M_gamma cell action 1}\ee
with $S_{I\times\TT^D}$ given by (\ref{cylinder cell result}) and $S_{\TT^D}$ given by (\ref{torus asym cell result}).
\end{itemize}
\end{statement}

\subsection{Why does the cell action for cube satisfy quantum master equation? Sketch of finite-dimensional argument}
\label{section: fin-dim argument}
It turns out that, apart from simplifying perturbative computations and providing explicit examples, cubical setting has another advantage over simplicial setting: one can give purely finite-dimensional argument, why the cell action for standard cube satisfies quantum master equation. Indeed, formally the induced action always has to satisfy QME by Statement \ref{statement: QME for induced action}. But applying this result to the induction from field theory with infinite-dimensional space of fields, given by functional BV integral and requiring some regularization, is not rigorous. Thus it would be nice to perform independent, purely finite-dimensional check of QME for the result (due to construction of section \ref{section: gluing}, it is sufficient to check QME only for the building blocks of discrete theory --- simplices or cubes, from that point QME for discrete action for any cell complexes follows automatically). In section \ref{section: interval QME check} we performed such a check for the interval. We could also check the perturbative result for $D$-simplex up to some order in fields, but we do not know how to prove that a particular regularization scheme for Feynman diagrams yields a solution to QME (indeed, one can easily find regularization schemes which give the result not satisfying QME, even for dimension $D=1$).

It turns out that in cubical setting the situation is better: one can make a finite-dimensional check of the result for $D$-cube, essentially due to factorization of Feynman diagrams. The idea is to understand the $qL_\infty$ structure on cochains $C^\bt(I^D,\g)$ of $D$-cube as coming from certain structure on interval (not just a $qL_\infty$ structure, but a richer object that we call the ``FC form\footnote{From ``factorization on boundary'' and ``closeness''.} on configuration space of graphs''), so that quadratic relations for $qL_\infty$ operations on $C^\bt(I^D,\g)$ follow automatically from certain condition (closeness and factorization on the boundary) on this structure for interval, which may be checked directly.

Here we will give a sketchy outline of this argument, with some details omitted and we will not be careful with signs.

\textbf{Configuration space of graphs, boundary strata.} Let $\Gamma_{n\ra 1}$ be the configuration space of planar rooted trees with $n$ leaves\footnote{The notation comes from ``$n$ inputs, 1 output''.}, whose internal edges have lengths $\lambda\in [0,1]$ (different edges have independent lengths, a more correct term would be ``proper times''). $\Gamma_{n\ra 1}$ is a $(n-2)$-dimensional manifold with corners, naturally equipped with cubical cell decomposition, with cells associated to combinatorial types of trees (i.e. with lengths forgotten). In particular, top-dimension cells are associated to binary trees. Boundary strata of codimension 1 of $\Gamma_{n\ra 1}$ correspond to one edge having length 1 (lengths of other edges are not fixed). Thus $$\dd \Gamma_{n\ra 1}=\bigcup_{1< k< n,\; 1\leq i\leq k} \dd_{k,i}\Gamma_{n\ra 1}$$
with
\be \dd_{k,i}\Gamma_{n\ra 1}\simeq \Gamma_{n-k+1\ra 1}\times \Gamma_{k\ra 1} \label{Gamma stratum 1}\ee
the codimension 1 boundary stratum, corresponding to the tree with $k$ leaves, whose $i$-th leaf is attached by an edge of length 1 to the root of a tree with $n-k+1$ leaves. As a cell complex $\Gamma_{n\ra 1}$ can be viewed as the pair cell complex of Stasheff's associahedron of dimension $n-2$.

Let also $\Gamma_{m,n\ra 0}$ be the configuration space of planar one-loop graphs with $m$ leaves outside the cycle and $n$ leaves inside the cycle (the cycle splits the plane the graph is drawn on into ``outside'' part and ``inside part'', thus splitting the set of leaves into two subsets), whose internal edges have lengths $\lambda\in [0,1]$, and we assume that vertices have valence $\geq 3$ and edges are oriented, so that for each vertex one edge is outgoing and all other edges are incoming (leaves are oriented as incoming). We also assume that one-loop graphs are endowed with cyclic enumerations of outer leaves and of inner leaves (these two enumerations are separate and independent), i.e. we mark one outer leaf as outer leaf number 1 and one inner leaf as inner leaf number 1. The configuration space $\Gamma_{m,n\ra 0}$ is a $(m+n)$-dimensional manifold with corners, equipped with natural cell decomposition, with cells associated to combinatorial types of one-loop graphs, and top-dimension cells correspond to purely trivalent graphs. Space $\Gamma_{m,n\ra 0}$ is equipped with the action of group\footnote{We use the convention $\ZZ_0=\ZZ_1=\{\bf{1}\}$ for cyclic groups.} $\ZZ_m\times\ZZ_n$, corresponding to cyclic relabeling of outer and inner leaves (separately).  Codimension 1 boundary strata correspond to the situation when one of internal edges has length 1. When this edge does not belong to the cycle, it splits the graph into one-loop graph and a tree (which might be either inside the cycle, or outside the cycle). Otherwise, if the edge belongs to the cycle, cutting it we obtain a tree. Thus
$$\dd \Gamma_{m,n\ra 0}=\left(\bigcup_{1\leq k<m,\; 1\leq i\leq m}\dd_{k,i}^{0,1} \Gamma_{m,n\ra 0}\right) \cup \left(\bigcup_{1\leq k<n,\; 1\leq j\leq n}\dd_{k,j}^{1,0} \Gamma_{m,n\ra 0}\right)\cup\left(\bigcup_{1\leq i\leq m,\; 1\leq j\leq n}\dd^0_{ij} \Gamma_{m,n\ra 0}\right) $$
with
\be \dd_{k,i}^{0,1} \Gamma_{m,n\ra 0}\simeq \Gamma_{m-k+1\ra 1}\times \Gamma_{k,n\ra 0},\quad
\dd_{k,i}^{1,0} \Gamma_{m,n\ra 0}\simeq \Gamma_{m,k\ra 0}\times \Gamma_{n-k+1\ra 1},\quad
\dd^0_{ij} \Gamma_{m,n\ra 0}\simeq \Gamma_{m+n+1\ra 1} \label{Gamma stratum 2}\ee
Here upper indices of $\dd$ correspond to number of loops in the components the graph is split into by the edge of length 1. Namely, $\dd^{0,1}_{k,i}$ corresponds to boundary strata where the graph is obtained by connecting first outer leaf of one-loop graph with $k$ outer leaves and $n$ inner leaves by an edge of length 1 to the root of a tree with $m-k+1$ leaves and then shifting the enumeration of outer leaves by $i\mod m$. Boundary strata $\dd^{1,0}_{k,j}$ are constructed similarly and are given by one-loop graphs with $m$ outer leaves and $k$ inner leaves, with the first inner leaf attached by an edge of length 1 to the root of a tree with $n-k+1$ leaves and afterwards one shifts the enumeration of inner leaves by $j\mod n$. Boundary strata $\dd^0_{ij}$ are obtained by connecting the root of a tree with $m+n+1$ leaves to the leaf number $m+1$, and relabeling the leaves as
\begin{eqnarray*}
l\mapsto l+i \mod m\qquad\mr{if}\;1\leq l\leq m \\
l\mapsto l-m-1+j \mod n\qquad\mr{if}\;m+2\leq l\leq m+n
\end{eqnarray*}

Let us also denote the total configuration space of graphs by
$$\Gamma_{\bt\ra\bt}=\left(\bigcup_{n\geq 2}\Gamma_{n\ra 1}\right)\cup\left(\bigcup_{m\geq 0,\; n\geq 0,\; m+n\geq 1}\Gamma_{m,n\ra 0}\right)$$

An important subtlety about configuration spaces $\Gamma_{m,0\ra 0}$, $\Gamma_{0,n\ra 0}$ is that in the situation of the loop consisting of single edge of length zero we are forced to treat graphs differing by cyclic relabeling of leaves as equivalent. Otherwise the  spaces $\Gamma_{m,0\ra 0}$, $\Gamma_{0,n\ra 0}$ would have anomalous ``ultraviolet'' boundary strata, corresponding to graphs containing the loop consisting of single edge of length zero.

\textbf{FC forms.} Next, let $V$ be a complex with differential $d$, and let
$$TV=\RR\oplus V\oplus V^{\otimes 2}\oplus V^{\otimes 3}\oplus\cdots$$
be its tensor algebra with differential
$$0\oplus d \oplus (d\otimes\id+\id\otimes d)\oplus(d\otimes\id\otimes\id+\id\otimes d\otimes \id+\id\otimes\id\otimes d)\oplus\cdots$$
which we denote also by $d$, by abuse of notation. We define ``FC form''
$$\alpha\in \Omega^\bt(\Gamma_{\bt\ra\bt})\otimes \End(TV)$$
as a collection $\alpha=\{\alpha_{n\ra 1}\}_{n\geq 2}\cup\{\alpha_{m,n\ra 0}\}_{m,n\geq 0,\; m+n\geq 1}$ of differential forms, such that
$$\alpha_{n\ra 1}\in \Omega^\bt(\Gamma_{n\ra 1})\otimes\Hom(V^{\otimes n},V)$$
is a differential form on $\Gamma_{n\ra 1}$ with values in $n$-ary operations on $V$ for each $n\geq 2$, and
$$\alpha_{m,n\ra 0}\in[\Omega^\bt(\Gamma_{m,n\ra 0})\otimes \Hom(V^{\otimes m}\otimes V^{\otimes n},\RR)]^{\ZZ_m\times\ZZ_n}$$
is a differential form on $\Gamma_{m,n\ra 0}$ with values in poly-linear functions of $m+n$ variables in $V$ (where first $m$ are called ``outer'' variables and last $n$ are ``inner'' variables) for each pair $m,n\geq 0$ except $m=n=0$. Here we assume that $\alpha_{m,n\ra 0}$ is invariant w.r.t $\ZZ_m\times\ZZ_n$ action on $\Gamma_{m,n\ra 0}$ accompanied by (independent) cyclic permutation of outer and inner variables. We additionally require two properties to hold for $\alpha$:
\begin{itemize}
\item First, $\alpha$ is required to factorize on boundary strata of $\Gamma_{\bt\ra\bt}$ of codimension 1:
\begin{multline}
\left.\alpha_{n\ra 1}\right|_{\dd_{k,i}\Gamma_{n\ra 1}}  =  \pi^*_{\dd_{k,i}\Gamma_{n\ra 1}\ra \Gamma_{k\ra 1}}(\alpha_{k\ra 1})\circ_i \pi^*_{\dd_{k,i}\Gamma_{n\ra 1}\ra \Gamma_{n-k+1\ra 1}}(\alpha_{n-k+1\ra 1}), \\
\left.\alpha_{m,n\ra 0}\right|_{\dd^{0,1}_{k,i}\Gamma_{m,n\ra 0}}  =  \mr{cycl}_i^\mr{out}\left(\pi^*_{\dd^{0,1}_{k,i}\Gamma_{m,n\ra 0}\ra \Gamma_{k,n\ra 0}}(\alpha_{k,n\ra 0})\circ_1 \pi^*_{\dd^{0,1}_{k,i}\Gamma_{m,n\ra 0}\ra \Gamma_{m-k+1\ra 1}}(\alpha_{m-k+1\ra 1})\right), \\
\left.\alpha_{m,n\ra 0}\right|_{\dd^{1,0}_{k,j}\Gamma_{m,n\ra 0}}  =  \mr{cycl}_j^\mr{in}\left(\pi^*_{\dd^{1,0}_{k,j}\Gamma_{m,n\ra 0}\ra \Gamma_{m,k\ra 0}}(\alpha_{m,k\ra 0})\circ_{m+1} \pi^*_{\dd^{1,0}_{k,j}\Gamma_{m,n\ra 0}\ra \Gamma_{n-k+1\ra 1}}(\alpha_{n-k+1\ra 1})\right), \\
\left.\alpha_{m,n\ra 0}\right|_{\dd^{0}_{ij}\Gamma_{m,n\ra 0}}(x_1,\ldots,x_m; y_1,\ldots, y_n)  =
\\ =
\pi^*_{\dd^0_{ij}\Gamma_{m,n\ra 0}\ra \Gamma{m+n+1\ra 1}}(\Str_V\;\alpha_{m+n+1\ra 1}(x_{1+i},\ldots,x_{m+i},\bt,y_{1+j},\ldots,y_{n+j}))
\label{factorization on boundary}
\end{multline}
where $\pi^*_{\cdots}$ are pull-backs by projections to the factors in (\ref{Gamma stratum 1},\ref{Gamma stratum 2}), $\circ_i$ is $i$-th composition in $\End(TV)$ (namely, $A\circ_i B$ means ``plug $B$ into $A$ as $i$-th input'') accompanied by wedge product in differential forms on configuration space of graphs; $x_1,\ldots, x_m,y_1,\ldots, y_n$ denote elements of $V$, with indices of $x$ and $y$ variables understood as defined modulo $m$ and modulo $n$ respectively; $\mr{cycl}^\mr{out}_i$ and $\mr{cycl}^\mr{in}_i$ denote the operation of cyclic permutation of outer or inner arguments by $i$ positions.  Informally, this property means that if an edge in graph has length 1, the value of $\alpha$ on it is the composition (in $\End(TV)$) of values of $\alpha$ on the two subgraphs, the original graph is split into by the marked edge (accompanied by wedge multiplication of forms on $\Gamma_{\bt\ra\bt}$). The case when edge of length 1 belongs to the cycle is special and leads to super-trace of $\alpha$ for the corresponding tree.
\item Second, $\alpha$ is required to be closed w.r.t. the total differential on $\Omega^\bt(\Gamma_{\bt\ra\bt})\otimes\End(TV)$:
\be (d_\Gamma-[d,\bt])\alpha=0 \label{alpha is closed}\ee
where $d_\Gamma$ is the de Rham differential on configuration space of graphs $\Gamma_{\bt\ra\bt}$ and $d$ is the differential on $V$ extended to $TV$.
\end{itemize}

\textbf{$qA_\infty$ structure from an FC form.} Suppose $\alpha$ is a FC form. Then we can define operations
\be \A_{n}=\int_{\Gamma_{n\ra 1}}\alpha_{n\ra 1} \in \Hom(V^{\otimes n},V),\quad
\Q_{m,n}=\int_{\Gamma_{m,n\ra 0}}\alpha_{m,n\ra 0} \in \Hom(V^{\otimes m}\otimes V^{\otimes n},\RR) \label{induced qA_infty}\ee
as integrals of $\alpha$ over components of configuration space $\Gamma_{\bt\ra\bt}$. Properties (\ref{factorization on boundary},\ref{alpha is closed}) imply a set of quadratic relations for operations $\A_{n}, \Q_{m,n}$. Namely, if we integrate (\ref{alpha is closed}) over a component of $\Gamma_{\bt\ra\bt}$, the second term yields homotopy differential of the corresponding operation, while the first term by Stokes' theorem reduces to integral over boundary, which can be evaluated using the known behaviour of $\alpha$ on the boundary (\ref{factorization on boundary}). Thus we obtain the relations
\begin{eqnarray}
[d,\A_{n}]+\sum_{k=2}^{n-1}\sum_{i=1}^k\A_k\circ_i \A_{n-k+1} = 0, \label{qA_infty rel 1}\\
{[ d,\Q_{m,n} ]}+\sum_{k=1}^{m-1}\sum_{i=1}^m \mr{cycl}^\mr{out}_i (\Q_{k,n}\circ_1 \A_{m-k+1})+\sum_{k=1}^{n-1}\sum_{j=1}^n\mr{cycl}^\mr{in}_j\Q_{m,k}\circ_{m+1}\A_{n-k+1}+ \nonumber \\
+\sum_{i=1}^m\sum_{j=1}^n \mr{cycl}^\mr{out}_i \mr{cycl}^\mr{in}_j(\Str_V^{(m+1)}\; \A_{m+n+1})=0 \label{qA_infty rel 2}
\end{eqnarray}
where we denoted $\Str_V^{(m+1)}\A_{m+n+1}$ the super-trace of the operation regarded as an operator acting on $m+1$-st argument (other arguments are treated as parameters):
$$(\Str_V^{(m+1)}\A_{m+n+1})(x_1,\ldots,x_m;y_1,\ldots,y_n):=\Str_V \A_{m+n+1}(x_1,\ldots,x_m,\bt,y_1,\ldots,y_n)$$
Relations (\ref{qA_infty rel 1}) are the usual quadratic relations of an $A_\infty$ algebra with operations $d,\A_2,\A_3,\ldots$ on $V$, while relations (\ref{qA_infty rel 2}) extend this structure to what we call the structure of ``$qA_\infty$ algebra'' on $V$ (in analogy with $qL_\infty$ algebras) with classical operations $d,\{\A_n\}_{n\geq 2}$ and quantum operations $\{\Q_{m,n}\}_{m,n\geq 0,\; m+n\geq 1}$. Notice that unlike in $qL_\infty$ setting, here we specify the number of outer arguments and number of inner arguments for quantum operations. Another difference is that $qA_\infty$ operations are in general not supposed to be symmetric w.r.t. permutations of inputs (other than bicyclic permutations for quantum operations).

An important remark is that relation (\ref{qA_infty rel 2}) for cases $(m,n)=(1,0)$ and $(m,n)=(0,1)$ actually does not follow from  FC property, since the corresponding components of the configuration space of graphs $\Gamma_{1,0\ra 0}$, $\Gamma_{0,1\ra 0}$ each possess an ``anomalous'' boundary stratum corresponding to the only edge contracting to zero length. This stratum violates corresponding quadratic relations uncontrollably. So we do not include relation (\ref{qA_infty rel 2}) for these values of $(m,n)$ in the definition of $qA_\infty$ algebra.

\textbf{FC form, induced from DGA.} The next point is that FC forms may be constructed by certain induction procedure.  Namely, let $V$ is a DGA (differential graded associative algebra) with differential $d$ and associative product $m$ (i.e. satisfying Poincar\'{e}, Leibniz and associativity relations) and let $V'$ be a deformation retract of $V$ with $V\xra{(\iota,r,K)}V'$ some triplet of embedding--retraction--chain homotopy. Then we can construct a FC form $\alpha\in\Omega^\bt(\Gamma_{\bt\ra\bt})\otimes\End(TV')$ ``induced'' from the DGA $(V,d,m)$ as follows: for $T$ a planar binary rooted tree with $n$ leaves we denote $\Gamma_T\subset\Gamma_{n\ra 1}$ the top-dimension cell of $\Gamma_{n\ra 1}$, corresponding to combinatorial structure $T$. Then we define $\alpha_{n\ra 1}$ via its restrictions to top-dimension cells as
\be \left. \alpha_{n\ra 1} \right|_{\Gamma_T}(x_1,\ldots, x_n)=r\circ\Iter_{T;K^{\lambda_k;d\lambda_k}m(\bt,\bt);m(\bt,\bt)}(\iota(x_1),\ldots, \iota(x_n)) \label{alpha induction 1}\ee
Here $\{\lambda_k\}$ are lengths of internal edges of $T$, index $k$ enumerates these edges, $K^{\lambda,d\lambda}=(1-\lambda) \id+ \lambda \PP'+d\lambda\cdot K$ is the extended chain homotopy (\ref{extended K}); $x_1,\ldots,x_n$ are elements of $V'$. Similarly, for purely trivalent planar one-loop graph $L$ with $m$ leaves outside the cycle and $n$ leaves inside, we denote $\Gamma_L\subset \Gamma_{m,n\ra 0}$ the corresponding top-dimension cell of configuration space and set
\be \left. \alpha_{m,n\ra 0} \right|_{\Gamma_L}(x_1,\ldots, x_m;y_1,\ldots,y_n)=\Loop_{L;K^{\lambda_k;d\lambda_k}m(\bt,\bt)}(\iota(x_1),\ldots, \iota(x_m);\iota(y_1),\ldots,\iota(y_n)) \label{alpha induction 2}\ee
Then first, these restrictions to top-dimension cells actually glue into a well-defined differential form $\alpha$ on $\Gamma_{\bt\ra\bt}$ (up to a subtlety we will mention below). This is due to the property $K^{\lambda;d\lambda}|_{\lambda=0}=\id$ of extended chain homotopy and due to associativity of $m$. Then the restrictions of $\alpha$ agree on interfaces of top cells, which correspond to some edge in the graph acquiring length 0. Second, the boundary factorization properties (\ref{factorization on boundary}) are satisfied by construction, due to the property $K^{\lambda;d\lambda}|_{\lambda=1}=\PP'$ of extended chain homotopy. Third, closeness (\ref{alpha is closed}) is implied by the property $(d_\lambda-[d,\bt])K^{\lambda;d\lambda}=0$ of extended chain homotopy (where $d_\lambda=d\lambda\wedge \frac{\dd}{\dd\lambda}$ is de Rham differential in variable $\lambda$) and by Leibniz identity on $V$ (which can be written as $[d,m]=0$ with $d$ denoting the extension of differential to $TV$).
Thus $\alpha$ is a well-defined FC form. We can proceed to integrate it over components of the configuration space as in (\ref{induced qA_infty}) to construct the induced $qA_\infty$ structure on $V'$:
\begin{eqnarray}
\A_{n}^{V'}(x_1,\ldots,x_n)=\sum_{{T\in \bf{T}_\mr{Pl}}:\;|T|=n}r\circ\Iter_{T;K\, m(\bt,\bt);m(\bt,\bt)}(\iota(x_1),\ldots, \iota(x_n)) \label{induced A_n}\\
\Q_{m,n}^{V'}(x_1\ldots,x_m;y_1,\ldots,y_n)=\sum_{{L\in \bf{L}_\mr{Pl}}:\;|L|^\mr{out}=m, |L|^\mr{in}=n} \Loop_{L;K\, m(\bt,\bt)}(\iota(x_1),\ldots,\iota(x_m);\iota(y_1),\ldots,\iota(y_n)) \label{induced Q_m,n}
\end{eqnarray}
where for one-loop graphs $|L|^\mr{out}$, $|L|^\mr{in}$ denotes the numbers of outer and inner leaves, respectively. Sums over trivalent graphs arise as contributions of top-dimension cells into integrals over configuration space (\ref{induced qA_infty}). Edges become decorated with $K$ due to property $\int_0^1 K^{\lambda;d\lambda}=K$ of extended chain homotopy. As we are summing over planar graphs, no symmetry coefficients $\frac{1}{|\Aut(T)|}$, $\frac{1}{|\Aut(L)|}$ appear. Formula (\ref{induced A_n}) reproduces the well-known formula for $A_\infty$ structure, induced on a subcomplex of DGA, while (\ref{induced Q_m,n}) provides its completion with quantum operations. Quadratic relations (\ref{qA_infty rel 1},\ref{qA_infty rel 2}) are satisfied by construction.
Notice also that for the induction of FC form (\ref{alpha induction 1},\ref{alpha induction 2}) one is not forced to use the extended chain homotopy (\ref{extended K}) that is linear in $\lambda$. One can actually use any form $K^{\lambda,d\lambda}\in \Omega^\bt([0,1])\otimes \End(V)$ obeying properties
$$K^{\lambda,d\lambda}|_{\lambda=0}=\id,\quad K^{\lambda,d\lambda}|_{\lambda=1}=\PP',\quad \int_0^1 K^{\lambda,d\lambda}=K,\quad (d_\lambda-[d,\bt])K^{\lambda,d\lambda}=0 $$

The subtlety we referred to above, concerning gluing the form $\alpha$ from its restrictions to top cells, is the following. We have to suppose that original DGA has the property
$$\Str_V m(K^{\lambda,d\lambda}m(x,y),\bt)=\Str_V m(K^{\lambda,d\lambda}m(y,x),\bt)$$
for any $x,y\in V$ (in particular, this property holds automatically if the multiplication on $V$ is commutative). If this property is violated, the form (\ref{alpha induction 2}) is not continuous on the codimension 1 cells of $\Gamma_{m,0\ra 0}$ or $\Gamma_{0,n\ra 0}$, corresponding to one-loop graphs with cycle consisting of single edge of length 0.

\textbf{Tensor product of FC forms.} The next observation is that FC forms can be tensor multiplied. Namely, if $V_1$, $V_2$ are two complexes and $\alpha_1\in\Omega^\bt(\Gamma_{\bt\ra\bt})\otimes\End(TV_1)$, $\alpha_2\in\Omega^\bt(\Gamma_{\bt\ra\bt})\otimes\End(TV_2)$ are two FC forms, then we can construct the form
$$\alpha_1\otimes\alpha_2\in \Omega^\bt(\Gamma_{\bt\ra\bt})\otimes\End(T (V_1\otimes V_2))$$
where we take tensor product in endomorphisms and wedge product in differential forms on configuration space. It is easy to check that this form satisfies properties (\ref{factorization on boundary},\ref{alpha is closed}) and thus is a FC form. In particular, we can consider tensor powers of FC forms. If $V$ is a DGA, $V'$ its retract and $V\xra{(\iota,r,K)} V'$ --- induction data, we may consider the following construction: induction of FC form $\alpha \in \Omega^\bt(\Gamma_{\bt\ra\bt})\otimes\End(T V')$ followed by raising to tensor power  $\alpha^{\otimes N} \in \Omega^\bt(\Gamma_{\bt\ra\bt})\otimes\End(T (V'^{\otimes N}))$ and then followed by integration over configuration space $\Gamma_{\bt\ra\bt}$ to yield $qA_\infty$ structure on $V'^{\otimes N}$. Then one can notice that the result coincides with the $qA_\infty$ structure obtain from $V^{\otimes N}$ by transfer formulae (\ref{induced A_n},\ref{induced Q_m,n}) with chain homotopy $K^{\otimes_\mr{sym} N}$ (\ref{K^sym n}). In other words, we have the following picture:
$$\vspace{10pt}  \begin{CD}\mr{DGAs} @. V @>>> V^{\otimes N} \\
@V\mr{induction}VV @VK^{\lambda,d\lambda}VV @V(K^{\lambda,d\lambda})^{\otimes N}VV \\
\mr{FC\; forms} @. \alpha\in \Omega^\bt(\Gamma_{\bt\ra\bt})\otimes\End(T V') @>>> \alpha^{\otimes N}\in \Omega^\bt(\Gamma_{\bt\ra\bt})\otimes\End(T (V'^{\otimes N}))\\
@V\mr{integration}VV @V\int_{\Gamma_{\bt\ra\bt}}VV @V\int_{\Gamma_{\bt\ra\bt}}VV \\
qA_\infty\; \mr{algebras}\qquad @. \mr{operations }\; \{\A^{V'}_{n}\},\{\Q^{V'}_{m,n}\}\;\mr{ on }\; V' @. \mr{operations }\; \{\A^{V'^{\otimes N}}_{n}\},\{\Q^{V'^{\otimes N}}_{m,n}\}\;\mr{ on }\;V'^{\otimes N}
\end{CD}  \vspace{10pt} $$
Horizontal arrows are operations of raising to $N$-th tensor power. Composition of two arrows in the rightmost column yields the induction with chain homotopy $K^{\otimes_\mr{sym} N}$.
Notice that the extended chain homotopy $(K^{\lambda,d\lambda})^{\otimes N}$, defining the induction of FC form $\alpha^{\otimes N}$ from DGA structure on $V^{\otimes N}$, is not linear in $\lambda$.

\textbf{Commutative case.} Next, suppose $V$ is a (super-)commutative DGA (``cDGA'') and $\alpha\in\Omega^\bt(\Gamma_{\bt\ra\bt})\otimes\End(TV')$ is the induced FC form for subcomplex $V'\hra V$, as in (\ref{alpha induction 1},\ref{alpha induction 2}). Commutativity of $V$ implies certain ``commutativity'' properties of $\alpha$. Namely, if $\sigma: T\ra T'$ is a graph isomorphism of planar binary rooted trees with $n$ leaves, then it induces an isomorphism of corresponding cells of configuration space of graphs $\sigma: \Gamma_{T}\ra \Gamma_{T'} $ and the pull-back map on differential forms $\sigma^*: \Omega^\bt (\Gamma_{T'})\ra \Omega^\bt(\Gamma_T)$. Then for each tree isomorphism $\sigma$ we have a commutativity property, relating restrictions of $\alpha$ to cells $\Gamma_T$, $\Gamma_{T'}$ (which may coincide if $\sigma\in\Aut(T)$):
\be \left.\alpha_{n\ra 1}\right|_{\Gamma_T}(x_1,\ldots,x_n)=\pm\sigma^*\left(\left.\alpha_{n\ra 1}\right|_{\Gamma_{T'}}(x_{\sigma(1)},\ldots,x_{\sigma(n)})\right) \label{alpha com 1}\ee
where $\sigma(1),\ldots,\sigma(n)$ denotes the permutation of the order of leaves, induced by $\sigma$ (recall that we enumerate leaves counterclockwise, starting from root). Similarly, if $\sigma: L\ra L'$ is a graph isomorphism of planar trivalent one-loop graphs with $L$ having $m$ outer and $n$ inner leaves and $L'$ having $m'$ outer and $n'$ inner leaves (with $m+n=m'+n'$), then
\be \left.\alpha_{m,n\ra 0}\right|_{\Gamma_L}(x_1,\ldots,x_m;x_{m+1},\ldots,x_{m+n})=\pm\sigma^*\left(\left.\alpha_{m',n'\ra 0}\right|_{\Gamma_{L'}}(x_{\sigma(1)},\ldots,x_{\sigma(m')};x_{\sigma(m'+1)},\ldots,x_{\sigma(m+n)})\right) \label{alpha com 2}\ee

In turn, commutativity properties (\ref{alpha com 1},\ref{alpha com 2}) imply certain commutativity properties for the $qA_\infty$ operations $\{\A_n\}$, $\{\Q_{m,n}\}$, obtained by integrating  $\alpha$ over components of configurations space of graphs. One way to state these properties is the following. Let $\mr{FreeLie}(W)$ denote the free Lie algebra generated by arbitrary graded vector space $W$. Then there is a canonical embedding $\mr{FreeLie}(W)\hra TW$ which maps $[a,b]\mapsto a\otimes b\mp b\otimes a$. Therefore, there is a canonical embedding $\mr{FreeLie}((V'[1])^*)\otimes V'\hra \Hom(T(V'[1]),V')\simeq \Hom(TV',V')$ (with the second isomorphism given by suspension of factors, up to signs). Commutativity property for classical $A_\infty$ operations states that
\be \A_n\in \mr{FreeLie}((V'[1])^*)\otimes V' \label{qC_infty 1}\ee
Equivalently, we may say that for arbitrary (non-graded) Lie algebra $\g$ with basis $\{T_a\}$ the object
\begin{multline}l_n(\omega,\ldots,\omega)=\sum_{i_1,a_1,\ldots,i_n,a_n}\pm \omega^{i_1 a_1}\cdots \omega^{i_n a_n} T_{a_1}\cdots T_{a_n} \A_n(e_{i_1},\ldots,e_{i_n})\in \\ \in\Hom(S^n(\g\otimes V'[1]),\g\otimes V')\subset
\Hom(S^n(\g\otimes V'[1]),U(\g)\otimes V')
\label{qC_infty 2}
\end{multline}
is expressed in commutators of generators $T_a$ only (i.e. we understand product of $T_a$-s as the product in universal enveloping algebra $U(\g)$, but the result actually lies in the image of $\g$ in $U(\g)$ under canonical embedding). Here $\{e_i\}$ is a basis in $V'$ and super-commutative variables $\omega^{ia}$ are coordinates on $\g\otimes V'[1]$,\; $\omega=\sum_{i,a}T_a e_i \omega^{ia}$ is the super-field for the space $\g\otimes V'$. In other words, object $l_n$ defined by (\ref{qC_infty 2}) can be written as
\begin{multline}
l_n(\omega,\ldots,\omega)=\\
=\sum_{T\in {\bf{T}_\mr{nonPl}}:\; |T|=n}\underbrace{\frac{1}{|\Aut(T)|}\quad\sum_{i_1,\ldots,i_n,j}\pm\A_{T;i_1\cdots i_n}^j\; e_j\; \Iter_{T;[\bt,\bt];[\bt,\bt]}\left(\sum_{a_1} T_{a_1} \omega^{i_1 a_1},\ldots, \sum_{a_n} T_{a_n} \omega^{i_n a_n}\right)}_{l_{T}} \label{qL_infty from qC_infty 1}
\end{multline}
for some numbers $\A_{T;i_1\cdots i_n}$.

Commutativity condition for quantum operations is that the object
\begin{multline} q_n(\omega,\ldots,\omega)= \\
=\sum_{p+q=n} \quad \sum_{i_1,a_1,\ldots,i_n,a_n} \pm \omega^{i_1 a_1}\cdots \omega^{i_n a_n} \left(\Str_\g\; T_{a_1}\cdots T_{a_p} \bt T_{a_{p+1}}\cdots T_{a_n}\right) \Q_{p,q}(e_{i_1},\ldots,e_{i_p};e_{i_{p+1}},\ldots,e_{i_n}) \in \\
\in S^n (\g\otimes V'[1])^*
\label{qC_infty 3}
\end{multline}
is well-defined for any Lie algebra $\g$, meaning that expression under super-trace is expressed in terms of commutators of $T_a$-s and the argument $\bt\in\g$.
In other words, $q_n$ as defined by (\ref{qC_infty 3}) can be written as
\be q_n(\omega,\ldots,\omega)=\sum_{L\in{\bf{L}_\mr{nonPl}}:\; |L|=n} \underbrace{\frac{1}{|\Aut(L)|} \quad\sum_{i_1,\ldots,i_n}\pm\Q_{L;i_1\cdots i_n}\Loop_{L;[\bt,\bt];\g}\left(\sum_{a_1} T_{a_1} \omega^{i_1 a_1},\ldots, \sum_{a_n} T_{a_n} \omega^{i_n a_n}\right)}_{q_L} \label{qL_infty from qC_infty 2}\ee
where $\Q_{L;i_1\cdots i_n}$ are some numbers.


It would be natural to call a $qA_\infty$ algebra satisfying these commutativity conditions for operations $\{\A_n\}, \{\Q_{m,n}\}$ the ``$qC_\infty$ algebra'', since the commutativity condition for classical operations (\ref{qC_infty 1}) reproduces the relation on operations  of $C_\infty$ algebra, cf. \cite{CG}. Numbers $\A^j_{T;i_1\cdots i_n}$, $\Q_{L;i_1\cdots i_n}$ in (\ref{qL_infty from qC_infty 1},\ref{qL_infty from qC_infty 2}) are the  structure constants of $qC_\infty$ algebra and they are symmetric w.r.t. the action of graph automorphisms of $T$, $L$ (as non-planar graphs), which permute indices $i_1,\ldots,i_n$ as they permute the leaves.

The important feature of $qC_\infty$ structure on $V'$ is that it can be tensor multiplied by a unimodular Lie algebra $\g$ to produce $qL_\infty$ structure on space $\g\otimes V'$. Operations $\{l_n\}$, $\{q_n\}$ of this structure are given by (\ref{qC_infty 2},\ref{qC_infty 3}) (thus, we essentially defined $qC_\infty$ algebra as such a $qA_\infty$ algebra, which can be tensor multiplied by $\g$). Unimodularity of $\g$ is required, so that ``anomalous'' 1-valent quantum operations, violating quadratic relations (\ref{qA_infty rel 2}) do not contribute to $qL_\infty$ operations on $\g\otimes V'$, which is needed for the whole set of quadratic relations of $qL_\infty$ algebra to be satisfied.
The $qL_\infty$ structure on $\g\otimes V'$ obtained by this procedure from the $qC_\infty$ structure on $V'$, induced from commutative DGA $V$ with some induction data $V\xra{(\iota,r,K)} V'$,
is given by formulae (\ref{qL_infty from qC_infty 1},\ref{qL_infty from qC_infty 2}) with numbers $\A_{T;i_1\cdots i_n}^j$, $\Q_{L;i_1\cdots i_n}$ (the ``de Rham parts of Feynman diagrams'') given by
$$\A_{T;i_1\cdots i_n}^j=<e^j,r\circ\Iter_{T;K m(\bt,\bt);m(\bt,\bt)}(\iota(e_{i_1}),\ldots,\iota(e_{i_n}))>,
\quad \Q_{L;i_1\cdots i_n}=\Loop_{L;K m(\bt,\bt);V}(\iota(e_{i_1}),\ldots,\iota(e_{i_n}))$$
Therefore this algebra coincides with the $qL_\infty$ structure on $\g\otimes V'$, obtained directly by induction from DGLA $\g\otimes V$ with same induction data, extended trivially into $\g$-coefficients $\g\otimes V\xra{(\iota,r,K)}\g\otimes V'$:
$$\vspace{10pt}
\begin{CD}
\mr{cDGA}\quad V @>\g\otimes>> \mr{DGLA}\quad \g\otimes V \\
@V(\iota,r,K)VV @V(\iota,r,K)VV \\
qC_\infty\;\mr{structure\; on}\;\; V' @>\g\otimes>> qL_\infty\;\mr{structure\; on}\;\; \g\otimes V'
\end{CD}\vspace{10pt}
$$

On the level of FC form $\alpha\in \Omega^\bt(\Gamma_{\bt\ra\bt})\otimes \End(TV')$ the construction of tensor multiplying by $\g$ is as follows. We construct a new form $\beta$ on $\Gamma_{\bt\ra\bt}$ with components $\beta_{n\ra 1}\in \Omega^\bt(\Gamma_{n\ra 1})\otimes \Hom((\g\otimes V')^{\otimes n},U(\g)\otimes V')$, $\beta_{m,n\ra 0}\in \Omega^\bt(\Gamma_{m,n\ra 0})\otimes \Hom((\g\otimes V')^{\otimes (m+n)},\RR)$ defined by
\begin{eqnarray*}\beta_{n\ra 1}(T_{a_1}x_{1},\ldots,T_{a_n} x_{n})&=&T_{a_1}\cdots T_{a_n} \alpha_{n\ra 1}(x_{1},\ldots,x_{n}), \\
\beta_{m,n\ra 0}(T_{a_1}x_{1},\ldots,T_{a_m} x_{m};T_{a_{m+1}}x_{m+1},\ldots,T_{a_{m+n}} x_{m+n})&=&\\
=\tr_\g \PP_\g (T_{a_1}\cdots T_{a_m}\bt T_{a_{m+1}}\cdots T_{a_{m+n}}) &\cdot&
\alpha_{m,n\ra 0}(x_{1},\ldots,x_{m};x_{m+1},\ldots,x_{m+n})
\end{eqnarray*}
Here $x_i$ are arbitrary elements of $V'$ and $\PP_\g: U(\g)\ra \g$ is some projection, consistent with canonical embedding $\g\hra U(\g)$. Then contributions of non-planar graphs to (\ref{qL_infty from qC_infty 1},\ref{qL_infty from qC_infty 2}) are given by sums of integrals of $\beta$ over top-dimension cells of $\Gamma_{\bt\ra\bt}$ corresponding to isomorphic graphs:
$$l_T=\sum_{T'\sim T}\int_{\Gamma_{T'}}\beta, \quad q_L=\sum_{L'\sim L}\int_{\Gamma_{L'}}\beta$$
where we sum over planar realizations $T'$ of non-planar tree $T$ and over planar realizations $L'$ of non-planar one-loop graph $L$.

Another remark is that tensor product of FC forms preserves commutativity properties (\ref{alpha com 1},\ref{alpha com 2}).

\textbf{``Finite-dimensional argument'' for $qL_\infty$ structure on $C^\bt(I^D,\g)$.} The argument for the $qL_\infty$ structure $\g$-valued cell cochains of standard $D$-cube (induced from de Rham algebra $\Omega^\bt(I^D,\g)$) goes as follows. Due to discussion above, this structure on cochains may be obtained in another way: first construct the induced FC form $\alpha\in \Omega^\bt(\Gamma_{\bt\ra\bt})\otimes\End(T C^\bt(I))$ on (real-valued) cochains on interval, induced from de Rham algebra $\Omega^\bt(I)$ of interval (without $\g$-coefficients). Next we raise this form to tensor power $D$ to obtain the form $\alpha^{\otimes D}\in \Omega^\bt(\Gamma_{\bt\ra\bt})\otimes\End(T C^\bt(I^D))$ on cochains of $D$-cube. Then we integrate over configuration space of graphs to produce $qC_\infty$ structure on $C^\bt(I^D)$ which we afterwards tensor multiply by Lie algebra $\g$ to produce the desired $qL_\infty$ structure on $C^\bt(I^D,\g)$. So the picture is:
\be
\vspace{10pt}
\begin{CD}
\mr{cDGA}\; \Omega^\bt(I)  @.  \mr{cDGA}\; \Omega^\bt(I^D) @. \mr{DGLA}\; \Omega^\bt(I^D,\g) \\
@V K^{\lambda,d\lambda}_I VV @. @. \\
\mr{FC\; form}\; \alpha \;\mr{on}\;C^\bt(I) @>(\cdots)^{\otimes D}>> \mr{FC\; form}\;\alpha^{\otimes D} \;\mr{on}\; C^\bt(I^D) @. \\
@. @V\int_{\Gamma_{\bt\ra\bt}}VV @. \\
@. qC_\infty \;\mr{structure\; on}\; C^\bt(I^D) @>\g\otimes>> qL_\infty\;\mr{structure\; on}\; C^\bt(I^D,\g)
\end{CD}
\vspace{10pt}
\label{fin-dim argument}
\ee
This is essentially what we did in section \ref{section: cube factorization}: the differential forms $C_{I,T}^{\lambda_1,d\lambda_1;\cdots;\lambda_{|T|-2},d\lambda_{|T|-2}}(\zeta_1^i,\ldots,\zeta_{|T|}^i)$, $C_{I,L}^{\lambda_1,d\lambda_1;\cdots;\lambda_{|L|},d\lambda_{|L|}}(\zeta_1^i,\ldots,\zeta_{|L|}^i)$ we were computing there are just values of FC form $\alpha$ restricted to different cells of configuration space $\Gamma_{\bt\ra\bt}$, evaluated on tuples of basis vectors of $C^\bt(I)$; variables $\lambda$ are the coordinates on cells of $\Gamma_{\bt\ra\bt}$. The ``de Rham parts of Feynman diagrams'' $C_{I^D,T}(\zeta_1,\ldots,\zeta_{|T|})$, $C_{I^D,L}(\zeta_1,\ldots,\zeta_{|L|})$ are the structure constants $\A_T$, $\Q_L$  of $qC_\infty$ structure on cochains of cube, obtained by integrating $\alpha^{\otimes D}$ over $\Gamma_{\bt\ra\bt}$ (to be precise, this discussion has to be corrected to take into account the fact that in section \ref{section: cube factorization} we were computing reduced effective action, not the whole one). From this point we arrived to the effective action (or to the desired $qL_\infty$ structure) on $C^\bt(I^D,\g)$ by introducing Lie algebra factors, as in (\ref{qL_infty from qC_infty 1},\ref{qL_infty from qC_infty 2}).

Now, observe that all arrows in (\ref{fin-dim argument}) preserve the quadratic relations, and all but the first one (induction of $\alpha$ from $\Omega^\bt(I)$) deal with finite-dimensional spaces only. Thus, instead of checking quadratic relations for $qL_\infty$ structure on $C^\bt(I^D,\g)$ (which is the same as checking QME for the cell action for cube), it is sufficient to check that $\alpha$, defined using conventions of section \ref{section: cube factorization} for super-traces over $\Omega^\bt(I)$, is a true FC form, i.e. satisfies boundary factorization properties (\ref{factorization on boundary}) and is closed (\ref{alpha is closed}).

Boundary factorization properties for $\alpha$ are satisfied trivially: for all boundary strata except $\dd^0_i \Gamma_{m,n\ra 0}$ (corresponding to one of edges belonging to the loop acquiring length 1) factorization property does not depend on definition of super-trace at all and follows directly from $\left.K^{\lambda,d\lambda}_I\right|=\PP'_I$. For $\dd^0_i \Gamma_{m,n\ra 0}$ strata factorization occurs, if the super-trace satisfies $\Str_{\Omega^\bt(I)}\PP'_I \OO = \Str_{C^\bt(I)}r\OO\iota$ for any operator $\OO\in\End(\Omega^\bt(I))$, and this holds indeed for conventions of section \ref{section: cube factorization}.

As for closeness of $\alpha$, for top-dimension cells of $\Gamma_{\bt\ra\bt}$ corresponding to trees, it follows directly from Leibniz identity for wedge multiplication of forms on interval and from property $(d_\lambda-[d,\bt])K^{\lambda,d\lambda}=0$ of extended chain homotopy. For cells, corresponding to one-loop graphs, the problem is reduced to checking the following property: for every $n\geq 1$ we define the map from $n$-tuples of forms on interval to forms on the $n$-cube
$$w_n:(\psi_1,\ldots,\psi_n)\mapsto\Str_{\Omega^\bt(I)}K^{\lambda_1,d\lambda_1}_I\m(\psi_1)\cdots K^{\lambda_n,d\lambda_n}_I\m(\psi_n) $$
where $\psi_i\in \Omega^\bt(I)$ are differential forms and $\m(\psi_i)=\psi_i\wedge$ denotes the operator of multiplication by $\psi_i$ (thus $w_n$ is the wheel graph with leaves decorated by general forms on interval, not necessarily Whitney forms). Then $w_n$ has to satisfy
\be d_\lambda w_n(\psi_1,\ldots,\psi_n)=\sum_{i=1}^n (-1)^i w_n(\psi_1,\ldots,\psi_{i-1},d\psi_i,\psi_{i+1},\ldots,\psi_n) \label{w_n property}\ee
where $d_\lambda$ (essentially, the same as $d_\Gamma$ in (\ref{alpha is closed})) is the de Rham differential in variables $\lambda_1,\ldots,\lambda_n$.
Property (\ref{w_n property}) can be checked explicitly. For example, for $n=1$, evaluating the super-trace using prescriptions of section \ref{section: cube factorization}, we obtain
\begin{multline*}
w_1(\underbrace{f+dt\cdot g}_\psi)=\Str_{\Omega^\bt(I)}((1-\lambda)\id+\lambda \PP'_I+d\lambda\; K_I)\m_{f+dt\cdot g}=\\
=\lambda\Str_{C^\bt(I)}(r_I\,\m_f\,\iota_I) +(1-\lambda)\Str_{\Omega^\bt(I)}\m_f+d\lambda\; \Str_{\Omega^\bt(I)} (K_I\m_{dt\cdot g})=\\
=\lambda\left(f(0)+f(1)-\int_0^1 dt\cdot f(t)\right)+(1-\lambda)\frac{f(0)+f(1)}{2}+d\lambda\int_0^1 dt\left(\frac{1}{2}-t\right) g(t)
\end{multline*}
where we decomposed $\psi=f+dt\cdot g$ with $f$ and $g$ two functions on interval. Checking (\ref{w_n property}) for $w_1$, we have
$$
d_\lambda w_1(\psi)= d\lambda \left(\frac{f(0)+f(1)}{2}-\int_0^1 dt\cdot f(t)\right)=-\int_0^1 dt\left(\frac{1}{2}-t\right) f'(t)=-w_1(d\psi)
$$
For $n=2$ we compute
\begin{multline*}
w_2(\underbrace{f_1+dt\cdot g_1}_{\psi_1},\underbrace{f_2+dt\cdot g_2}_{\psi_2})=\\
=\Str ((1-\lambda_1)\id+\lambda_1 \PP'_I+d\lambda_1\; K_I)\m_{f_1+dt\cdot g_1} ((1-\lambda_2)\id+\lambda_2 \PP'_I+d\lambda_2\; K_I)\m_{f_2+dt\cdot g_2}=\\
=(1-\lambda_1)(1-\lambda_2)\Str\; \m_{f_1}\m_{f_2}+(1-\lambda_1)\lambda_2 \Str\; \m_{f_1} \PP'_I \m_{f_2}+\lambda_1 (1-\lambda_2)\Str\; \PP'_I \m_{f_1}\m_{f_2}+ \lambda_1\lambda_2 \Str\; \PP'_I \m_{f_1} \PP'_I \m_{f_2}-\\
-(1-\lambda_1) d\lambda_2 \Str\;(\m_{f_1}K_I\m_{dt\cdot g_2}+\m_{dt\cdot g_1} K_I \m_{f_2})-\lambda_1 d\lambda_2 \Str\; \PP'_I (\m_{f_1}K_I\m_{dt\cdot g_2}+\m_{dt\cdot g_1}K_I \m_{f_2})+\\
+d\lambda_1\,(1-\lambda_2)\Str\;(K_I\m_{f_1}\m_{dt\cdot g_2}+K\m_{dt\cdot g_1}\m_{f_2})+d\lambda_1\, \lambda_2\, \Str\; (K_I\m_{f_1}\PP'_I\m_{dt\cdot g_2}+K_I\m_{dt\cdot g_1}\PP'_I\m_{f_2})+\\
+d\lambda_1\, d\lambda_2\,\Str\; K_I\m_{dt\cdot g_1} K_I \m_{dt\cdot g_2}=\\
=(1-\lambda_1)(1-\lambda_2)\frac{f_1(0)f_2(0)+f_1(1)f_2(1)}{2}+\\
+((1-\lambda_1)\lambda_2+ \lambda_1 (1-\lambda_2)) \left(f_1(0)f_2(0)+f_1(1)f_2(1)-\int_0^1 dt f_1(t) f_2(t)\right) +\\
+\lambda_1 \lambda_2 \left(f_1(0)f_2(0)+f_1(1)f_2(1)-\left(\int_0^1 dt f_1(t)\right) \left(\int_0^1 dt f_2(t)\right)\right)+\\
+(d\lambda_1\, (1-\lambda_2)+(1-\lambda_1)\, d\lambda_2)\int_0^1 dt \left(\frac{1}{2}-t\right)(-f_1(t) g_2(t)+g_1(t)f_2(t))+\\
+\lambda_1\, d\lambda_2\,\int_0^1 dt \int_0^1 dt' g_1(t)\;(\theta(t-t')-t)\;f_2(t')- d\lambda_1\,\lambda_2\, \int_0^1 dt \int_0^1 dt' g_2(t)\;(\theta(t-t')-t)\;f_1(t')+\\
+d\lambda_1 \,d\lambda_2\, \int_0^1 dt \int_0^1 dt' \;(\theta(t'-t)-t')\;(\theta(t-t')-t)\; g_1(t)g_2(t')
\end{multline*}
Where $\Str$ everywhere means super-trace over $\Omega^\bt(I)$, and we treat $\psi=f+dt\cdot g$ as a purely odd object, so that $f$ is odd and $g$ is even. Then we go on with checking (\ref{w_n property}) for $w_2$:
\begin{multline}
d_\lambda w_2(\psi_1,\psi_2)=(d\lambda_1\, (1-\lambda_2)+(1-\lambda_1)\, d\lambda_2) \frac{f_1(0)f_2(0)+f_1(1)f_2(1)}{2} - \\
-(d\lambda_1\, (1-2\lambda_2)+(1-2\lambda_1)\, d\lambda_2) \int_0^1 dt\, f_1(t) f_2(t)- (d\lambda_1\,\lambda_2+\lambda_1\, d\lambda_2)\left(\int_0^1 dt f_1(t)\right) \left(\int_0^1 dt f_2(t)\right)+\\
+d\lambda_1\, d\lambda_2\,\int_0^1 dt \int_0^1 dt' (\theta(t-t')-t)\;(g_1(t)\,f_2(t')+g_2(t)\, f_1(t')) \label{w_2 1}
\end{multline}
On the other hand
\begin{multline}
-w_2(d\psi_1,\psi_2)+w_2(\psi_1,d\psi_2)=-(d\lambda_1\,(1-\lambda_2)+(1-\lambda_1)\,d\lambda_2)\, \int_0^1 dt \left(\frac{1}{2}-t\right) (f'_1(t) f_2(t)+f_1(t) f'_2(t))-\\
-\lambda_1\, d\lambda_2\int_0^1 dt\int_0^1 dt' f'_1(t)\, (\theta(t-t')-t)\, f_2(t')+d\lambda_1\,\lambda_2\int_0^1 dt \int_0^1 dt' f'_2(t)\, (\theta(t-t')-t)\, f_1(t')+\\
+d\lambda_1\, d\lambda_2 \int_0^1 dt \int_0^1 dt' (\theta(t'-t)-t')\,(\theta(t-t')-t)\, (f'_1(t)\, g_2(t')+g_1(t)\, f'_2(t')) \label{w_2 2}
\end{multline}
Integrating by parts, we obtain that (\ref{w_2 2}) equals (\ref{w_2 1}), and thus we checked (\ref{w_n property}) for $n=2$. We see that for these checks the conventions (\ref{str of multiplication},\ref{short loop}) of section \ref{section: cube factorization} for calculating super-traces are substantial.

Property (\ref{w_n property}) can be checked for general $n$, using the following idea: possibly problematic decorations of the wheel with pieces of $K^{\lambda,d\lambda}_I$ only appear in components of $w_n$ of degrees $0,1$ (as a differential form in variables $\lambda_i$) --- decorations either with $n$ identities, or with $n-1$ identities and one $K_I$, while for all other decorations the operator under super-trace has continuous kernel in coordinate representation, and thus the super-trace is unambiguous. Therefore, one proves that (\ref{w_n property}) holds in de Rham degrees $>2$ (in variables $\lambda_i$) in general way, evaluating super-traces in coordinate representation. For lower degrees one has to check (\ref{w_n property}) explicitly. Here the conventions (\ref{str of multiplication},\ref{short loop}) for resolving ambiguous super-traces become important.

\section{Effective $BF$ action on de Rham cohomology of manifold}
\label{section: effective theory on coh}
An interesting object from the perspective of algebraic topology is the effective action on the de Rham cohomology of a manifold $M$, induced from topological $BF$ theory on $M$. More precisely, the invariant of a manifold is the effective action on cohomology modulo canonical transformations (or equivalently, on the language of $qL_\infty$ algebras, the homotopy type of de Rham algebra $\Omega^\bt(M,\g)$ as a $qL_\infty$ algebra). Tree part of the effective action on cohomology is the generating function for Massey operations on cohomology and therefore is related to the rational homotopy type of $M$ (see \cite{Sullivan},\cite{BG}). An interesting question arises: does one-loop part of the action on cohomology contain some additional information about the manifold? In other words, is the homotopy type of the de Rham algebra as a $qL_\infty$ algebra a stronger invariant of the manifold than the homotopy type of de Rham algebra as the ordinary $L_\infty$ algebra? The answer is positive: we can provide a pair of manifolds with the same classical Massey operations on cohomology, but distinguished by quantum operations on cohomology --- circle and Klein bottle (section \ref{section: S on coh examples}).

In section \ref{section: category of retracts} we discuss the general picture of induction for topological $BF$ theory (equivalently, for the de Rham algebra as a $qL_\infty$ algebra). In particular, we discuss the construction of effective action on cohomology via discrete $BF$ theory, i.e. via simplicial $BF$ action on a triangulation or cell action on a cubical cell decomposition (\ref{ret 1}). In this way the calculation of effective action on cohomology is reduced to certain finite-dimensional BV integral, provided that we know the action for building blocks of the discrete $BF$ theory --- simplices or cubes (e.g. to some fixed order of perturbation expansion).

In section \ref{section: S on coh properties} we discuss some specific properties of the effective action on cohomology: cyclic property of the tree part of action for the case of induction directly from the topological $BF$ theory (not via discretization) with the ``Hodge induction data'', consistent with Poincar\'{e}  duality. Next we discuss the general estimates on the degrees of arguments for classical and quantum operations on cohomology. The most important property among those is that quantum operations depend on 1-cohomologies only. These estimates also imply that for a manifold with $H^1(M)=0$ the effective action on cohomology can always be calculated explicitly: all quantum operations vanish, while the complexity (number of inputs) of classical ones is bounded by dimension of the manifold (hence the action on cohomology is the sum of contributions of finitely many tree Feynman diagrams). Finally, we present a result, allowing one to compute the effective action on cohomology of the product of manifolds in some cases.

In section \ref{section: S on coh examples} we discuss some examples of manifolds for which the effective action on cohomology can be computed explicitly. A particularly interesting example here is the pair circle -- Klein bottle (it is, to our knowledge, the simplest instance of the situation, when manifolds are indistinguishable by classical Massey operations on cohomology, but are distinguished by quantum operations). We also present the result for a higher-dimensional generalization of Klein bottle --- the bundle $M_\gamma$ over circle with $D$-torus as the fiber and the transition function defined by arbitrary element $\gamma\in S_D\ltimes \ZZ_2^D$ of the group of symmetry of $D$-cube.

\subsection{Category of retracts}
\label{section: category of retracts}

\begin{Def}
Let $(V,d)$ be a cochain complex. The category of retracts $\mr{Ret}_V$ is the category whose objects are subcomplexes $V'\hra V$ quasi-isomorphic to $V$, regarded up to isomorphism. A morphism in $\mr{Ret}_V$ between objects $V_1,V_2$ is a set of induction data $V_1\xra{(\iota,r,K)} V_2$, i.e. a triplet of linear maps: embedding $\iota: V_2\ra V_1$, retraction $r: V_1\ra V_2$, chain homotopy $K: V_1\ra V_1$, such that axioms (\ref{ind data axiom1}--\ref{ind data axiom6}) are satisfied.
\end{Def}
Composition of morphisms is defined by (\ref{composition of iota,r,K}).
Objects $\mr{Ret}_V$ form a partially ordered set, with morphisms going from a larger object to a smaller one (not strictly smaller, since there are automorphisms). Also, $\mr{Ret}_V$ contains two important objects: the maximal object --- cochain complex $V$ itself, and the minimal object --- its cohomology $H^\bt(V)$.

Now suppose that  $V$ is endowed with a $qL_\infty$ structure $(Q_V,\rho_V)$ (or equivalently, with a $BF_\infty$ action on the space of fields $\FF_V=T^*[-1](V[1])$), such that the unary classical operation $l_{V(1)}$ coincides with the differential $d$ of the complex. Then this structure is transported to all objects of $\mr{Ret}_V$ along morphisms. I.e. for each $V'\hra V$ an induced $qL_\infty$ structure $(Q_{V'},\rho_{V'})$ arises, and it is unique (does not depend on the particular morphism or chain of morphisms from $V$ to $V'$, along which the structure is transported) up to special canonical transformations, due to Statement \ref{statement: ind data deform for BF_infty}. The transport of $qL_\infty$ structure along morphisms is consistent with composition of morphisms due to Statement \ref{statement: iterated induction}. All $qL_\infty$ structures on all retracts $V'$, induced from a given $qL_\infty$ structure on $V$, are equivalent in the sense of Definition \ref{def: homotopy of qL_infty algebras}, and in particular are equivalent to the induced $qL_\infty$ structure on $H^\bt(V)$.

We are interested in the category of retracts for $V=\Omega^\bt(M,\g)$ --- the complex of $\g$-valued differential forms on manifold $M$, with standard DGLA structure (regarded as a special case of $qL_\infty$ structure). The corresponding $BF_\infty$ theory is the topological $BF$ theory on $M$ with gauge Lie algebra $\g$. The $qL_\infty$ structure induced on cohomology of $V$ (which is just the de Rham cohomology of $M$ with coefficients $H^\bt(M,\g)$) contains the complete information on the homotopy type of the de Rham algebra of $M$ as a $qL_\infty$ algebra, and yields an interesting invariant of manifolds. Since the tree part of induction formula coincides with the usual formula for the homotopy transfer of a (classical) $L_\infty$ algebra, the induced classical operations  $l_{(n)}$ on $H^\bt(M,\g)$ are the Massey operations on de Rham cohomology.
The full $qL_\infty$ structure on cohomology, with quantum operations $q_{(n)}$, is a strictly stronger invariant of manifolds than the classical $L_\infty$ part: we will discuss in the section \ref{section: S on coh examples} an example of a pair of manifolds with the same Massey operations on cohomology, distinguished by quantum operations on cohomology --- circle and Klein bottle.

If the manifold $M$ is compact, orientable, has no boundary and is endowed with Riemann metric, then one constructs from metric the Hodge star $*$ and the operator
 $d^*=-*d*:\Omega^{\bt}(M)\ra\Omega^{\bt-1}(M)$. Thus one gets the classical Hodge decomposition for de Rham complex
$$\Omega^\bt(M)=\mr{Harm}^\bt(M)\oplus \Omega^\bt_{d-ex}(M)\oplus \Omega^\bt_{d^*-ex}(M)$$
into exact, co-exact and harmonic $\mr{Harm}^\bt(M)=\ker(d\,d^*+d^*d)$ forms. We call the Hodge induction data
$$\Omega^\bt(M)\xra{(\iota_\mr{Hodge},r_\mr{Hodge},K_\mr{Hodge})} H^\bt(M)$$
the embedding of cohomology into $\Omega^\bt(M)$ as harmonic forms, projection onto harmonic forms in Hodge decomposition and the chain homotopy
$$K_\mr{Hodge}=d^*/(d\, d^*+d^* d)$$
(we assume that this expression vanishes on harmonic forms). The special feature of Hodge induction data is that they are consistent with the Poincar\'{e} pairing on
$$(\alpha,\beta)_P=\int_M \alpha\wedge\beta$$
in the following sense: first, subspaces of IR forms $\mr{Harm}^\bt(M)$ and UV forms $\Omega^\bt_{d-ex}(M)\oplus \Omega^\bt_{d^*-ex}(M)$ are mutually orthogonal w.r.t. the pairing $(\bt,\bt)_P$. Second, Hodge chain homotopy is a self-adjoint operator w.r.t. $(\bt,\bt)_P$:
$$(K_\mr{Hodge}\alpha,\beta)_P=(-1)^{|\alpha|}(\alpha,K_\mr{Hodge}\beta)_P$$
in particular, this implies that subspace $\Omega^\bt_{d^*-ex}(M)$ is isotropic.

In the category of retracts $\mr{Ret}_{\Omega^\bt(M,\g)}$ we are interested only in subcomplexes  $V'\hra V=\Omega^\bt(M,\g)$ of form $V'=\g\otimes C^\bt$, where $C^\bt\hra \Omega^\bt(M)$ and differential on $V'$ is trivial in coefficients $\g$. Moreover, we only consider morphisms $(\iota,r,K)$ that are trivial in coefficients $\g$. Apart from the largest object $\Omega^\bt(M,\g)$ and the smallest object $H^\bt(M,\g)$, category $\mr{Ret}_{\Omega^\bt(M,\g)}$ contains interesting intermediate objects $C^\bt(\Xi,\g)$ --- complexes of $\g$-valued cochains on a triangulation or on a cubical cell decomposition $\Xi$ of the manifold $M$. For each $\Xi$ there is a standard morphism
$$\Omega^\bt(M,\g)\xra{(\iota_\Xi,r_\Xi,K_\Xi)}C^\bt(\Xi,\g)$$
In case when $\Xi$ is a triangulations, it is given by embedding of cochains as Whitney forms, retraction to the cochains by integrals over simplices and the chain homotopy is glued from Dupont's operators for simplices. Respectively, in case when $\Xi$ is a cubical cell complex, the induction data are: embedding of cochains as cubical Whitney forms, retraction to the cochains by integrals over cubical cells, the chain homotopy is glued from symmetrized tensor powers (\ref{cube K}) of the Dupont's operator for interval. The $BF_\infty$ action, corresponding to the $qL_\infty$ structure, transported from de Rham algebra along the standard morphism $(\iota_\Xi,r_\Xi,K_\Xi)$, is called the discrete (simplicial or cell) $BF$ action for $\Xi$. Also, for any triangulation or any cubical cell decomposition $\Xi$ there is a canonical morphism
$$C^\bt(\Xi,\g)\xra{(\iota_{\Xi\ra H^\bt},r_{\Xi\ra H^\bt},K_{\Xi\ra H^\bt})}H^\bt(M,\g)$$
that is constructed analogously to the Hodge morphism  $\Omega^\bt(M,\g)\ra H^\bt(M,\g)$, but does not require any additional structure (while the Hodge morphism depends on Riemannian metric), and uses the fact that $C^\bt(\Xi)$ has a canonical basis, associated to the cells. Namely, instead of $d^*$ we can use the operator $d^T$, whose matrix in the basis of cells is the transposed matrix of the differential. Then we introduce the analog of Hodge Laplacian $d\,d^T+d^T d:C^\bt(\Xi)\ra C^\bt(\Xi)$. Embedding $\iota_{\Xi\ra H^\bt}$ sends cohomology into ``harmonic'' cochains $\mr{Harm}^\bt(\Xi)=\ker (d\,d^T+d^T d)$, retraction $r_{\Xi\ra H^\bt}$ projects to the harmonic cochains in decomposition
$$C^\bt(\Xi)=\mr{Harm}^\bt(\Xi)\oplus C^\bt_{d-ex}(\Xi)\oplus C^\bt_{d^T-ex}(\Xi)$$
and the chain homotopy is
\be K_{\Xi\ra H^\bt}=d^T/(d\,d^T+d^T d)\label{ret 2}\ee
Therefore, for each $\Xi$ there is a canonical two-step induction from $\Omega^\bt(M,\g)$ to $H^\bt(M,\g)$:
\be \Omega^\bt(M,\g)\xra{(\iota_\Xi,r_\Xi,K_\Xi)}C^\bt(\Xi,\g)\xra{(\iota_{\Xi\ra H^\bt},r_{\Xi\ra H^\bt},K_{\Xi\ra H^\bt})}H^\bt(M,\g) \label{ret 1}\ee
The advantage of calculating the induced $qL_\infty$ structure (effective $BF_\infty$ action) on cohomology of the manifold in this way is that the problem of calculating the functional BV integral defining the induction is reduced to the calculation of simplicial (cell) action for $\Xi$. The latter, due to simplicial (cell) locality, reduces to the standard computation for single simplex (cube). And afterwards we only have to evaluate a finite-dimensional BV integral, to pass from cell cochains to cohomology. Thus the advantage of this method over using, for instance, the Hodge morphism for direct computation of the induced $qL_\infty$ structure on cohomology, is that here the hard part of the problem --- computation of the infinite-dimensional part of BV integral --- is standardized and reduced to a series of universal computations, which we already performed partially (in first orders of perturbation theory for general $D$ and exactly for $D=0,1$): Theorems \ref{interval thm},\ref{thm: simplex perturbative result},\ref{thm: cube factorization}.

Thus in the category of retracts $\mr{Ret}_{\Omega^\bt(M,\g)}$ we distinguish the largest object $\Omega^\bt(M,\g)$, the smallest object --- cohomology $H^\bt(M,\g)$ and certain class of intermediate objects $C^\bt(\Xi,\g)$, associated to triangulations and cubical cell decompositions of $M$. We also distinguish Hodge morphisms $\Omega^\bt(M,\g)\ra H^\bt(M,\g)$, depending on the choice of metric on $M$, and canonical morphisms
$\Omega^\bt(M,\g)\ra C^\bt(\Xi,\g) $ and $C^\bt(\Xi,\g)\ra H^\bt(M,\g)$ for each $\Xi$. Other interesting morphisms are the aggregation morphisms $C^\bt(\Xi',\g)\ra C^\bt(\Xi,\g)$,
where $\Xi'$ is a cell subdivision of $\Xi$, but we will not discuss them here.

\subsection{Special properties of effective $BF$ action on cohomology}
\label{section: S on coh properties}
\subsubsection{Cyclic symmetry of Feynman trees for $S_{H^\bt(M,\g)}^0$ for Hodge induction}
Let manifold $M$ be connected, orientable, compact and have no boundary. We denote its dimension $D=\dim M$. Let it also be endowed with Riemannian metric. We are interested in the effective action on cohomology, defined by the Hodge induction data  $(\iota_H,r_H,K_H)$. Let $\{h_i\}$ be a basis in de Rham cohomology $H^\bt(M)$ and $\{h^i\}$ the dual basis in the dual space (homology) $H_\bt(M)$. Super-fields for the space of fields $\FF_{H^\bt(M,\g)}=T^*[-1](H^\bt(M,\g)[1])$ are
$$\omega=\sum_i h_i\omega^i,\quad p=\sum_i p_i h^i$$
where $\omega^i$ and $p_i$ have values in $\g$ and $\g^*$ respectively and have ghost numbers $\gh(\omega^i)=1-|i|,\; \gh(p_i)=|i|-2$ (by $|i|$ we denote the de Rham degree of the cohomology $h_i$). Tree part of the effective action on cohomology is
\begin{multline*}S_{H^\bt(M,\g)}^0(\omega,p)=\sum_{T\in{\bf{T}}_\mr{nonPl}:\,|T|\geq 2}\frac{1}{|\Aut(T)|}\cdot\\
\cdot\sum_{j,i_1,\ldots,i_{|T|}}<p_{j}r_H^*(h^{j}),
\Iter_{T;-K_H[\bt,\bt];[\bt,\bt]}(\iota_H(h_{i_1})\omega^{i_1},\ldots,\iota_H(h_{i_{|T|}})\omega^{i_{|T|}})>\end{multline*}
Next, $\Omega^\bt(M)$ is endowed with Poincar\'{e} pairing
$$(\alpha,\beta)_P=\int_M \alpha\wedge\beta$$
which induces a non-degenerate pairing on $H^\bt(M)$
$$(A,B)_P=(\iota_H(A),\iota_H(B))_P$$
We denote the matrix of Poincar\'{e} pairing on cohomology by
$$P_{ij}=(h_i,h_j)_P$$
Let also  $\g$ be endowed with invariant inner product $\tr_\g T_a T_b$, which we use to identify $\g^*$ with $\g$.
Let us rewrite the tree part of effective action on cohomology using Poincar\'{e} pairing $(\bt,\bt)_P$ instead of canonical pairing $<\bt,\bt>$:
\begin{multline}S_{H^\bt(M,\g)}^0(\omega,p)=\sum_{T\in{\bf{T}}_\mr{nonPl}:\,|T|\geq 2}\frac{1}{|\Aut(T)|}\cdot\\
\cdot \sum_{j,i_0,i_1,\ldots,i_{|T|}}\tr_\g\left(p_{j}P^{j\, i_0}\iota_H(h_{i_0}),
\Iter_{T;-K_H[\bt,\bt];[\bt,\bt]}(\iota_H(h_{i_1})\omega^{i_1},\ldots,\iota_H(h_{i_{|T|}})\omega^{i_{|T|}})\right)_P
\label{cycl eq1}
\end{multline}
where $P^{ij}$ is the inverse matrix for $P_{ij}$ and we used the fact that IR (harmonic) forms are orthogonal to UV ($d$- and $d^*$-exact) forms to rewrite
$(h_{i_0},r_H\Iter(\cdots))_P$ as $(\iota_H(h_{i_0}),\Iter(\cdots))_P$.

Let us introduce the operation of cyclic rotation (by $k$ steps) for planar rooted trees $\mr{Cycl}_k:\bf{T}_\mr{Pl}\ra \bf{T}_\mr{Pl}$ with $0\leq k\leq |T|$. Tree $T'=\mr{Cycl}_k T$ is constructed from $T$ as follows. Recall that leaves of $T$ are enumerated by numbers from $1$ to $|T|$ counter-clockwise starting from the root. Let us call the root the leaf number $0$. Then operation  $\mr{Cycl}_k$ just shifts the numeration of leaves by  $-k\mod (|T|+1)$, without changing  the underlying (non-oriented) graph. Thus the root of $T'$ is $k$-th leaf of $T$, and we re-orient the edges towards the new root. In particular, $\mr{Cycl}_0$ is the identity operation. Operations $\{\mr{Cycl}_k\}_{k=0}^{|T|}$ define an action of cyclic group $\ZZ_{|T|+1}$ on $\bf{T}_\mr{Pl}$.

The skew-symmetry of Poincar\'{e} pairing
$$(\alpha,\beta)_P=(-1)^{|\alpha|\cdot |\beta|}(\beta,\alpha)_P$$
cyclicity of Poincar\'{e} pairing
$$(\alpha,\beta\wedge\gamma)_P=(-1)^{|\alpha|\cdot (|\beta|+|\gamma|)} (\beta,\gamma\wedge\alpha)_P$$
and self-adjointness of Hodge chain homotopy
$$(K_H\alpha,\beta)=(-1)^{|\alpha|}(\alpha,K_H\beta)$$
allow to make local rearrangements for the de Rham part of the contribution of the tree into $S^0_{H^\bt(M,\g)}$
$$C_{T}(i_0,\ldots,i_{|T|})=\left(\iota_H(h_{i_0}),\Iter_{T;-K_H(\bt\wedge\bt);(\bt\wedge\bt)}(\iota_H(h_{i_1}),\ldots,\iota_H(h_{i_{|T|}}))\right)_P$$
leading to cyclic symmetry:
\be C_{T}(i_0,\ldots,i_{|T|})=\e_{T,k}(|i_0|,\ldots,|i_{|T|}|) C_{\mr{Cycl}_k T}(i_{k},\ldots,i_{|T|},i_{0},\ldots,i_{k-1}) \label{cycl eq4}\ee
Here $0\leq k\leq |T|$ and $\e_{T,k}(|i_0|,\ldots,|i_{|T|}|)=\pm1$ is certain sign depending on $T$, $k$ and degrees of cohomologies, with which leaves and the root are decorated.
For example:
\begin{multline*}C_{(**)}(i_0,i_1,i_2)=(\iota_H(h_{i_0}),\iota_H(h_{i_1})\wedge\iota_H(h_{i_2}))_P=(-1)^{|\iota_0|\cdot (|\iota_1|+|\iota_2|)} (\iota_H(h_{i_1}),\iota_H(h_{i_2})\wedge\iota_H(h_{i_0}))_P\\
=(-1)^{|\iota_0|\cdot (|\iota_1|+|\iota_2|)} C_{(**)}(i_1,i_2,i_0)=(-1)^{|\iota_0|\cdot (|\iota_1|+|\iota_2|)} C_{\mr{Cycl}_1(**)}(i_1,i_2,i_0)
\end{multline*}
Another example:
\begin{multline*}
C_{((**)(**))}(i_0,i_1,i_2,i_3,i_4)=\left(\iota_H(h_{i_0}),-K_H(\iota_H(h_{i_1})\wedge \iota_H(h_{i_2}))\wedge -K_H(\iota_H(h_{i_3})\wedge \iota_H(h_{i_4}))\right)_P\\
=(-1)^{|i_0|\cdot (|i_1|+|i_2|+|i_3|+|i_4|)}
\left(-K_H(\iota_H(h_{i_1})\wedge \iota_H(h_{i_2})), -K_H(\iota_H(h_{i_3})\wedge \iota_H(h_{i_4}))\wedge \iota_H(h_{i_0})\right)_P\\
=(-1)^{|i_0|\cdot (|i_1|+|i_2|+|i_3|+|i_4|)+|i_1|+|i_2|}
\left(\iota_H(h_{i_1})\wedge \iota_H(h_{i_2}), -K_H(-K_H(\iota_H(h_{i_3})\wedge \iota_H(h_{i_4}))\wedge \iota_H(h_{i_0}))\right)_P\\
=(-1)^{|i_0|\cdot (|i_1|+|i_2|+|i_3|+|i_4|)+|i_1|+|i_2|}
\left(\iota_H(h_{i_1}), \iota_H(h_{i_2})\wedge -K_H(-K_H(\iota_H(h_{i_3})\wedge \iota_H(h_{i_4}))\wedge \iota_H(h_{i_0}))\right)_P\\
=(-1)^{|i_0|\cdot (|i_1|+|i_2|+|i_3|+|i_4|)+|i_1|+|i_2|}
C_{(*((**)*))}(i_1,i_2,i_3,i_4,i_0)\\
=(-1)^{|i_0|\cdot (|i_1|+|i_2|+|i_3|+|i_4|)+|i_1|+|i_2|}
C_{\mr{Cycl}_1((**)(**))}(i_1,i_2,i_3,i_4,i_0)
\end{multline*}

Next, since the invariant inner product on $\g$ also has the cyclic property
$$\tr_\g T_a [T_b,T_c]=\tr_g T_b [T_c,T_a]$$
cyclic symmetry for the summand of (\ref{cycl eq1}) takes the form
\begin{multline*}
\tr_\g\left(P^{j\; i_0}p_j\iota_H(h_{i_0}),
\Iter_{T;-K_H[\bt,\bt];[\bt,\bt]}(\iota_H(h_{i_1})\omega^{i_1},\ldots,\iota_H(h_{i_{|T|}})\omega^{i_{|T|}})\right)_P\\
=\e_{T,k}(D-2,1,\ldots,1)\cdot
\tr_\g(\iota_H(h_{i_k})\omega^{i_k},
\Iter_{\mr{Cycl}_kT;-K_H[\bt,\bt];[\bt,\bt]}(\iota_H(h_{i_{k+1}})\omega^{i_{k+1}},\ldots,\\
,\iota_H(h_{i_{|T|}})\omega^{i_{|T|}},P^{j\; i_0}p_j\iota_H(h_{i_0}),\iota_H(h_{i_{1}})\omega^{i_{1}},\ldots,\iota_H(h_{i_{k-1}})\omega^{i_{k-1}}))_P
\end{multline*}
for $1\leq k\leq |T|$.
Sign $\e_{T,k}(D-2,1,\ldots,1)$ appears, since now we have to take into account the total degree $\deg+\gh$ when reshuffling objects, and the total degree is 1 for  $\iota_H(h_{i_{j}})\omega^{i_{j}}$ and is $D-2$ for $P^{j\; i_0}p_j\iota_H(h_{i_0})$.
Therefore we can rewrite (\ref{cycl eq1}) as
\begin{multline}
S^0_{H^\bt(M,\g)}(\omega,p)
=\sum_{n=2}^\infty  \sum_{T\in{\bf{T}}_\mr{nonPl}:\,|T|=n}\frac{1}{(n+1)}\;\frac{1}{|\Aut(T)|}\cdot\\
\cdot \sum_{j,i_0,\ldots,i_n}
\left(\tr_\g\left(P^{j\; i_0}p_j\iota_H(h_{i_0}),
\Iter_{T;-K_H[\bt,\bt];[\bt,\bt]}(\iota_H(h_{i_1})\omega^{i_1},\ldots,\iota_H(h_{i_{n}})\omega^{i_{n}})\right)_P+\right. \\
+\sum_{k=1}^n
\e_{T,k}(D-2,1,\ldots,1)\cdot
\tr_\g(\iota_H(h_{i_k})\omega^{i_k},
\Iter_{\mr{Cycl}_kT;-K_H[\bt,\bt];[\bt,\bt]}(\iota_H(h_{i_{k+1}})\omega^{i_{k+1}},\ldots,\\
\left.,\iota_H(h_{i_{n}})\omega^{i_{n}},P^{j\; i_0}p_j\iota_H(h_{i_0}),\iota_H(h_{i_{1}})\omega^{i_{1}},\ldots,\iota_H(h_{i_{k-1}})\omega^{i_{k-1}}))_P\right)
\label{cycl eq2}
\end{multline}
In case the dimension $D$ of $M$ is odd, the cyclic property of Feynman trees may be elegantly formulated as the existence of function  $F(\omega)$, such that $S^0_{H^\bt(M,\g)}(\omega,p)$ can be written as
\be S^0_{H^\bt(M,\g)}(\omega,p)=\left(\tr_g \sum_{i,j}p_i P^{ij}\frac{\dd}{\dd\omega^{j}}\right)F(\omega) \label{cycl eq3}\ee
where
\begin{multline*}F(\omega)=\sum_{n=2}^\infty \sum_{T\in{\bf{T}}_\mr{nonPl}:\,|T|=n}\frac{1}{(n+1)}\;\frac{1}{|\Aut(T)|}\cdot\\
\cdot
\sum_{i_0,\ldots,i_n}\tr_\g \left(\iota_H(h_{i_0})\omega^{i_0}, \Iter_{T;-K_H[\bt,\bt];[\bt,\bt]}(\iota_H(h_{i_1})\omega^{i_1},\ldots,\iota_H(h_{i_{n}})\omega^{i_{n}})\right)_P\\
=\sum_{n=2}^\infty \sum_{T\in{\bf{T}}_\mr{nonPl}/\ZZ_{n+1}:\;|T|=n}\frac{1}{|\Aut_\mr{cycl}(T)|}\cdot\\ \cdot
\sum_{i_0,\ldots,i_n}\tr_\g \left(\iota_H(h_{i_0})\omega^{i_0}, \Iter_{T;-K_H[\bt,\bt];[\bt,\bt]}(\iota_H(h_{i_1})\omega^{i_1},\ldots,\iota_H(h_{i_{n}})\omega^{i_{n}})\right)_P\\
=\frac{1}{6}\tr_\g(\iota_H(h_{i_0})\omega^{i_0},[\iota_H(h_{i_1})\omega^{i_1},\iota_H(h_{i_2})\omega^{i_2}])_P+\\+
\frac{1}{8}\tr_\g(\iota_H(h_{i_0})\omega^{i_0},[\iota_H(h_{i_1})\omega^{i_1},-K_H[\iota_H(h_{i_2})\omega^{i_2},\iota_H(h_{i_3})\omega^{i_3}]])_P+\\
+\frac{1}{8}\tr_\g(\iota_H(h_{i_0})\omega^{i_0},[-K_H[\iota_H(h_{i_1})\omega^{i_1},\iota_H(h_{i_2})\omega^{i_2}], -K_H[\iota_H(h_{i_3})\omega^{i_3}, \iota_H(h_{i_4})\omega^{i_4}]])_P+\cdots
\end{multline*}
Here ${\bf{T}}_\mr{nonPl}/\ZZ_{n+1}$ denotes the set of equivalence classes of planar trees, where two trees are considered equivalent, if they are connected by a combination of an isomorphism of rooted trees and operation $\mr{Cycl}_k$ for some $k$. We call these equivalence classes ``cyclic non-planar trees'' --- root does not play special role here and is regarded as an additional leaf. The group of automorphisms of a cyclic tree $\Aut_\mr{cycl}(T)$ is $\ZZ_{n+1}\ltimes \Aut(T)$. Oddness of $D$ is important first because the signs in (\ref{cycl eq2}) $\e_{T,k}(D-2,1,\ldots,1)=+1$ for odd $D$ and any tree $T$ and any $k$, and second because the differential operator
$\tr_g \sum_{i,j}p_i P^{ij}\frac{\dd}{\dd\omega^{j}}$ is even for $D$ odd.

In case $D=3$ function $F(\omega)\in\Fun(H^*(M,\g)[1])$ can be understood as the tree part of effective action for Chern-Simons theory on $M$, induced on cohomology  $H^\bt(M,\g)$.

In terms of induced $L_\infty$ structure on cohomology, property (\ref{cycl eq3}) means that the structure constants of $L_\infty$ operation $l_{(n)}:\Lambda^n H^\bt(M,\g) \ra H^\bt(M,\g)$ can be written as
$$l_{(n)}(h_{i_1},\ldots,h_{i_n})=\sum_{i,i_0}P^{i\;i_0}h_i l^\mr{cycl}_{(n+1)}(h_{i_0},h_{i_1},\ldots,h_{i_n})$$
or equivalently
$$\left(h_{i_0},l_{(n)}(h_{i_1},\ldots,h_{i_n})\right)_P=l^\mr{cycl}_{(n+1)}(h_{i_0},h_{i_1},\ldots,h_{i_n})$$
where $l^\mr{cycl}_{n+1}:\Lambda^{n+1} H^\bt(M,\g)\ra \RR$ is some $(n+1)$-linear super-antisymmetric map of degree $\deg l_{(n+1)}^\mr{cycl}= 2-D-n$. Thus cyclic symmetry of operations $l_{(n)}$ means that reversing the output by Poincar\'{e} pairing increases the symmetry group of the operation from $S_{n}$ to $S_{n+1}$. Operations $l_{n+1}^\mr{cycl}$ are related to the Taylor expansion for $F(\omega)$:
$$F(\omega)=\sum_{n=2}^\infty \frac{1}{(n+1)!}\;l_{n+1}^\mr{cycl}(\underbrace{\omega,\ldots,\omega}_{n+1})$$
and are calculated as sums over cyclic trees:
\begin{multline*}
l_{n+1}^\mr{cycl}(h_{i_0},\ldots,h_{i_n})=(-1)^{(|i_{n-1}|+1)\cdot|i_n|+(i_{n-2}+1)\cdot (|i_{n-1}|+|i_n|)+\cdots+ (|i_0|+1)\cdot (|i_1|+\cdots+|i_n|)} \frac{\dd}{\dd\omega^{i_n}}\cdots\frac{\dd}{\dd\omega^{i_0}}F(\omega)\\
=\sum_{T\in{\bf{T}}_\mr{nonPl}/\ZZ_{n+1}:\,|T|=n}\frac{1}{|\Aut_\mr{cycl}(T)|}
\sum_{\pi\in S_{n+1}}
\pm \left(\iota(h_{i_{\pi(0)}}),\Iter_{T;-K_H(\bt\wedge\bt);(\bt\wedge\bt)}(\iota(h_{i_{\pi(1)}}),\ldots,\iota(h_{i_{\pi(n)}}))\right)_P
\end{multline*}
Signs here depend on degrees of cohomologies the operation is evaluated on, and also on particular choice of planar rooted representative for a cyclic tree.

\subsubsection{Estimates on degrees of cohomologies in Feynman diagrams for $S_{H^\bt(M,\g)}$}
Here we will consider general (not necessary Hodge) induction $\Omega^\bt(M,\g)\xra{(\iota,r,K)} H^\bt(M,\g)$, which is still assumed to be trivial in $\g$-coefficients. The manifold $M$ is allowed to be non-orientable and/or have boundary (unless otherwise stated). Let us represent the effective action on cohomology as a sum of contributions of Feynman diagrams
$$S_{H^\bt(M,\g)}(\omega,p)=\sum_{T\in{\bf{T}}_\mr{nonPl}:\,|T|\geq 2}S_{H^\bt(M,\g),T}(\omega,p)+\hbar \sum_{L\in{\bf{L}}_\mr{nonPl}}S_{H^\bt(M,\g),L}(\omega)$$
where
\begin{multline}
S_{H^\bt(M,\g),T}(\omega,p)=\frac{1}{|\Aut(T)|}\,<r^*p,\Iter_{T;-K[\bt,\bt];[\bt,\bt]}(\iota(\omega),\ldots,\iota(\omega))>=\\=
\frac{1}{|\Aut(T)|}\sum_{j,i_1,\ldots,i_{|T|}}\e_{T}(|i_1|,\ldots,|i_{|T|}|) C_T(j;i_1,\ldots,i_{|T|})\; <p_j,\Iter_{T;[\bt,\bt];[\bt,\bt]}(\omega^{i_1},\ldots,\omega^{i_{|T|}})>_\g
\label{deg estimates 3}
\end{multline}
\begin{multline*}
S_{H^\bt(M,\g),L}(\omega)=\frac{1}{|\Aut(L)|}\,\Loop_{L;-K[\bt,\bt]}(\iota(\omega),\ldots,\iota(\omega))=\\
=\frac{1}{|\Aut(L)|}\sum_{i_1,\ldots,i_{|L|}} \e_L(|i_1|,\ldots,|i_{|L|}|) C_L(i_1,\ldots,i_{|T|})\; \Loop_{L;[\bt,\bt]}(\omega^{i_1},\ldots,\omega^{i_{|L|}})
\end{multline*}
and de Rham parts of diagrams are
\begin{eqnarray*}
C_T(j;i_1,\ldots,i_{|T|})&=&<r^*h^j,\Iter_{T;-K(\bt\wedge\bt);(\bt\wedge\bt)}(\iota(h_{i_1}),\ldots,
\iota(h_{i_{|T|}}))>\\
C_L(i_1,\ldots,i_{|L|})&=&\Loop_{L;-K(\bt\wedge\bt)}(\iota(h_{i_1}),\ldots,\iota(h_{i_{|T|}}))
\end{eqnarray*}
As before, we denote $\{h_i\}$ the basis in cohomology $H^\bt(M)$ and $\{h^i\}$ the dual basis in homology $H_\bt(M)$. Signs $\e_T, e_L=\pm1$ arise from carrying fields $\omega^{i}$ to the right, and depend on the graph $T$ or $L$ and on degrees of cohomologies.

\begin{statement}
\begin{itemize}
\item For any tree $T$ with $|T|\geq 3$ leaves,
$$C_T(j;i_1,\ldots,i_{|T|})\neq 0$$
implies
\begin{eqnarray}|i_1|\geq 1,\ldots,|i_{|T|}|\geq 1 \label{deg estimates 1}\\
(|i_1|-1)+(|i_2|-1)+\cdots+(|i_{|T|}|-1)=|j|-2 \label{deg estimates 2}
\end{eqnarray}
The latter condition implies in particular the estimate
\be (|i_1|-1)+(|i_2|-1)+\cdots+(|i_{|T|}|-1)\leq D-2 \label{deg estimates 4}\ee
In case of Hodge induction (i.e. for $M$ orientable and without boundary, $(\iota,r,K)=(\iota_H,r_H,K_H)$) the estimate is improved:
\be (|i_1|-1)+(|i_2|-1)+\cdots+(|i_{|T|}|-1)\leq D-3 \label{deg estimates 5}\ee
\item For any one-loop graph $L$,
$$C_L(i_1,\ldots,i_{|L|})\neq 0$$
implies
\be |i_1|=\cdots=|i_{|L|}|=1 \label{deg estimates 6}\ee
\item If $M$ has trivial 1-cohomology $H^1(M)=0$, then the one-loop part of action on cohomology vanishes
\be S^1_{H^\bt(M,\g)}(\omega)=0 \label{deg estimates 7}\ee
and the tree part is a polynomial in fields $\omega^i$ of degree
\be\deg_\omega S^0_{H^\bt(M,\g)} \leq\max(2,D-2)\label{deg estimates 8}\ee
In Hodge case the last estimate is improved:
\be\deg_\omega S^0_{H^\bt(M,\g)} \leq\max(2,D-4)\label{deg estimates 9}\ee
\end{itemize}
\end{statement}

\textbf{Proof.} Condition (\ref{deg estimates 1}) is proved by the following observation: 0-cohomology is necessarily embedded into $\Omega^\bt(M)$ as locally-constant functions, and multiplication by a locally-constant function maps IR forms into IR forms and $K$-exact forms into $K$-exact forms. Therefore any diagram other than $(**)$, with at least one leaf decorated with 0-cohomology, contains either an internal edge where $K$ is applied to an IR form or the expression $K^2$. Hence such diagrams vanish.

Next, (\ref{deg estimates 2}) is a reformulation of the fact that classical operations $l_{(n)}$ have degree $2-n$. Otherwise one can say that (\ref{deg estimates 2}) follows from the fact that expression (\ref{deg estimates 3}) has ghost number 0. Or even more explicitly: multiplication of forms leads to addition of degrees, while application of $K$ decreases the degree by 1. A tree with  $|T|$ leaves contains $|T|-2$ internal edges. Therefore
$$|\Iter_{T;-K(\bt\wedge\bt);(\bt\wedge\bt)}(\iota(h_{i_1}),\ldots,
\iota(h_{i_{|T|}}))|=|i_1|+\cdots+|i_{|T|}|-(|T|-2)=(|i_1|-1)+\cdots+(|i_{|T|}|-1)+2$$
Thus non-vanishing of $C_T(j,i_1,\ldots,i_{|T|})$ implies the condition (\ref{deg estimates 2}).

Estimate (\ref{deg estimates 4}) is a straightforward consequence of (\ref{deg estimates 1},\ref{deg estimates 2}) and the fact that degree of cohomology is bounded by dimension $D$ of the manifold. Improvement (\ref{deg estimates 5}) of this estimate in Hodge case occurs because if $h^j\in H_D(M)$, then $$C_T(j;i_1,\ldots,i_{|T|})=\sum_{i_0}P^{j\;i_0}C_T(i_0,\ldots,i_{|T|})=0$$ due to cyclic symmetry (\ref{cycl eq4}) and because the Poincar\'{e} dual cohomology for $h^j$ belongs to $H^0(M)$, and we return to the argument (\ref{deg estimates 1}).

Condition (\ref{deg estimates 6}) is proved by the following argument. Degree counting for $C_L(i_1,\ldots,i_{|L|})$ implies that this contribution vanishes automatically unless   $$|i_1|+\cdots+|i_{|L|}|=|L|$$
Therefore unless all inbound cohomologies are of degree 1, there is at least one $k$ for which $|i_k|=0$. Then the super-trace $C_L(i_1,\ldots,i_{|L|})$ necessarily contains one of the two structures:
\begin{eqnarray*}
K(\iota(h_{i_k})\wedge \iota(h_{i_j}))=0\\
K(\iota(h_{i_k})\wedge K(\bt))=0
\end{eqnarray*}
This proves (\ref{deg estimates 6}).

Condition (\ref{deg estimates 7}) follows from (\ref{deg estimates 6}) straightforwardly: if there is no 1-cohomology, then all one-loop Feynman diagrams vanish automatically. Estimate (\ref{deg estimates 8}) follows from (\ref{deg estimates 4}) and the fact that $i_k-1\geq 1$ for all $1\leq k\leq |T|$. Its improvement (\ref{deg estimates 9}) follows from (\ref{deg estimates 5}) and the fact that $H_{D-1}(M)=0$ by Poincar\'{e} duality (which implies that equality in (\ref{deg estimates 5}) cannot be achieved).
\\$\Box$

Thus the case $H^1(M)=0$ is the simplest for the problem of calculation of effective $BF$ action on cohomology: all one-loop diagrams vanish and only finite number of tree diagrams contribute. Even greater simplification occurs if in addition to $H^1(M)=0$ manifold  $M$ is formal: there exists an embedding of cohomology $\iota: H^\bt(M)\ra\Omega^\bt(M)$, such that  $\iota(H^\bt(M))$ is a subalgebra in $\Omega^\bt(M)$. Then the only contribution to $S_{H^\bt(M,\g)}$ is due to diagram $(**)$:
\be S_{H^\bt(M,\g)}=S_{H^\bt(M,\g),(**)}=\frac{1}{2}\sum_{i,j,k} (-1)^{(|j|+1)\cdot |k|} m^i_{jk} <p_i,[\omega^j,\omega^k]>_\g \label{formal mfd: action on coh}\ee
where
$$m^i_{jk}=C_{(**)}(i;j,k)=<r^*h^i,\iota(h_j)\wedge\iota(h_k)>$$
are the structure constants of multiplication on cohomology.

In case of general formal manifold $M$ (not supposing $H^1(M)=0$), since $\iota(H^\bt(M))$ is closed under multiplication, the only contributions to $S_{H^\bt(M,\g)}$ are due to diagram $(**)$ and one-loop diagrams of form $(*(*\cdots(*\bt)\cdots))$ (wheels). Therefore the effective action on cohomology is
\be S_{H^\bt(M,\g)}(\omega,p)=\frac{1}{2}<r^*p,[\iota(\omega),\iota(\omega)]>+\hbar\; \Str_{\Omega^\bt(M,\g)}\log (1+K[\iota(\omega^{(1)}),\bt])\label{formal mfd 2}\ee
(this is a special case of formula (\ref{S' via U and I}), moreover for this case the $L_\infty$ morphism $U$ is linear and coincides with embedding $U=\iota$). We denoted $\omega^{(1)}=\sum_{i:\,|i|=1}h_i\omega^i$ the part of super-field $\omega$, corresponding to 1-cohomology. This result is not very explicit indeed: computing the super-trace in (\ref{formal mfd 2}) might still be a hard challenge.

Another obvious corollary of the estimate (\ref{deg estimates 9}) is the absence of Massey operations for simply connected orientable compact manifolds without boundary of dimension $D<7$.

\subsubsection{Effective action on cohomology of the product of manifolds}
Let $M_1$ and $M_2$ be two compact connected manifolds. We are interested in the effective $BF$ action on the cohomology of the product $H^\bt(M_1\times M_2,\g)=\g\otimes H^\bt(M_1)\otimes H^\bt(M_2)$. We assume that two sets of induction data are given:
$\Omega^\bt(M_1)\xra{(\iota_1,r_1,K_1)} H^\bt(M_1)$ and $\Omega^\bt(M_2)\xra{(\iota_2,r_2,K_2)} H^\bt(M_2)$, and we consider the family of induction data for the product:
$$\Omega^\bt(M_1\times M_2)\xra{(\iota_1\otimes\iota_2,r_1\otimes r_2, K^\xi)}H^\bt(M_1\times M_2)$$
where \be K^\xi=K_1 \otimes ((1-\xi)\id+\xi \PP'_2)+ (\xi\id+(1-\xi)\PP'_1)\otimes K_2 \label{coh of product 1}\ee
is parameterized by $\xi\in\RR$.
Let $\{h^1_i\}$ and $\{h^2_i\}$ be bases in $H^\bt(M_1)$ and $H^\bt(M_2)$ respectively. Denote $h^1_{\circ}$ and $h^2_\circ$ the basis elements in $H^0(M_1)$ and $H^0(M_2)$ respectively. We assume that $h^1_\circ, h^2_\circ$ are embedded into $\Omega^\bt(M_1)$, $\Omega^\bt (M_2)$ as unit functions. Super-fields for the space of fields $$\FF_{H^\bt(M_1\times M_2,\g)}=T^*[-1](H^\bt(M_1\times M_2,\g)[1])$$
are
$$\omega=\sum_{i,j}h^1_i\otimes h^2_j\;\omega^{ij},\quad p=\sum_{i,j}p_{ij}\;h^{2j}\otimes\; h^{1i}$$
where $\omega^{ij},p_{ij}$ are in $\g$ and $\g^*$ respectively, and have ghost numbers $\gh(\omega^{ij})=1-|i|-|j|$, $\gh(p_{ij})=|i|+|j|-2$.

\begin{statement}
\label{statement: coh of product}
\begin{itemize}
\item If $H^1(M_2)=0$, then one-loop parts of effective actions on cohomology for $M_1$ and $M_1\times M_2$ are related by
\be S^1_{H^\bt(M_1\times M_2,\g)}=\chi(M_2)\cdot S^1_{H^\bt(M_1,\g)}|_{\omega^i\mapsto \omega^{i\circ}} \label{coh of product 2}\ee
for any value of $\xi$ in (\ref{coh of product 1}). Here $\chi(M_2)$ is the Euler characteristic of the manifold $M_2$.
\item If cohomology of $M_2$ coincides with cohomology of the point: $H^n(M_2)=\RR^{\delta_{n,0}}$, then effective actions on cohomology for $M_1$ and $M_1\times M_2$ coincide up to the substitution $\omega^i\mapsto \omega^{i\circ},\; p_i\mapsto p_{i\circ}$:
    \be S_{H^\bt(M_1\times M_2,\g)}=S_{H^\bt(M_1,\g)}|_{\omega^i\mapsto\omega^{i\circ}, p_i\mapsto p_{i\circ}}\label{coh of product 3}\ee
    for any $\xi$ in (\ref{coh of product 1}).
\item If $M_2=\s^1$ is a circle with basis $h^2_+\in H^0(\s^1),h^2_I\in H^1(\s^1)$ in cohomology (we assume that $\iota_2(h^2_+)=1, \iota_2(h^2_I)=dt$ and use the standard chain homotopy $K_I$ for $\s^1$), then effective actions on cohomology for $M_1$ and $M_1\times \s^1$ are related by
\be S_{H^\bt(M_1\times \s^1,\g)}=\int_0^1\left(S^0_{H^\bt(M_1,\g)}|_{\omega^{i}\mapsto\omega^{i+}+d\nu\, \omega^{iI},\;
p_i\mapsto (-1)^{|i|+1}p_{iI}+d\nu\, p_{i+}}\right)+\hbar\; \chi(M_1)\cdot \tr_g \log\left(\frac{\sinh\frac{\ad_{\omega^{\circ I}}}{2}}{\frac{\ad_{\omega^{\circ I}}}{2}}\right) \label{coh of product 4}\ee
for $\xi=1$ in (\ref{coh of product 1}), i.e. for the asymmetric chain homotopy $K_R=K_1\otimes\PP'_I+\id\otimes K_I$. Here $\chi(M_1)$ is the Euler characteristic of $M_1$ and $\nu\in[0,1]$  is an auxiliary variable.
\end{itemize}
\end{statement}

\textbf{Proof.} Let us prove (\ref{coh of product 2}). Let us use the fact that $S^1_{H^\bt(M_1,\times M_2,\g)}$ depends only on the part of super-fields $\omega$, corresponding to 1-cohomology (\ref{deg estimates 6}). Moreover, we have $H^1(M_1\times M_2)=H^1(M_1)\otimes H^0(M_2)=H^1(M_1)\otimes h^2_\circ$, since $M_2$ has no 1-cohomology. Therefore
\begin{multline*}S^1_{H^\bt(M_1\times M_2,\g)}\\=-\sum_{L\in\bf{L}_\mr{nonPl}}\frac{1}{|\Aut(L)|} \sum_{i_1,\ldots,i_{|L|}}\Loop_{L;-K^\xi[\bt,\bt]; \Omega^\bt(M_1\times M_2,\g)}(\iota_1(h^1_{i_1})\otimes 1\;\omega^{i_1\circ},\ldots,\iota_{1}(h^1_{i_{|L|}})\otimes 1\;\omega^{i_{|L|}\circ})\end{multline*}
The only contribution to each term is due to the decoration of all internal edges by the part $K_1\otimes ((1-\xi)\id+\xi \PP'_2)=K_1\otimes (\PP'_2+(1-\xi)\PP''_2)$ of chain homotopy $K^\xi$ (since the operator under super-trace has to have degree $(0,0)$ in the bigrading on $\Omega^\bt(M_1)\otimes \Omega^\bt(M_2)$, otherwise super-trace vanishes automatically). Therefore
\begin{multline*}
S^1_{H^\bt(M_1\times M_2,\g)}
=-\sum_{L\in\bf{L}_\mr{nonPl}}\frac{1}{|\Aut(L)|}\cdot\\ \cdot \sum_{i_1,\ldots,i_{|L|}} \Loop_{L;-K_1[\bt,\bt];\Omega^\bt(M_1,\g)} (\iota_1(h^1_{i_1})\omega^{i_1\circ},\ldots,\iota_1(h^1_{i_{|L|}})\omega^{i_{|L|}\circ})\cdot \Str_{\Omega^\bt(M_2)} (\PP'_2+(1-\xi)\PP''_2)^{[L]}\\
=S^1_{H^\bt(M_1,\g)}|_{\omega^i\mapsto\omega^{i\circ}}\cdot\Str_{\Omega^\bt(M_2)} (\PP'_2+(1-\xi)^{[L]}\PP''_2)
=\chi(M_2)\cdot S^1_{H^\bt(M_1,\g)}|_{\omega^i\mapsto\omega^{i\circ}}
\end{multline*}
where $[L]$ is the length of cycle in $L$. Thus (\ref{coh of product 2}) is proved.

Now let us turn to (\ref{coh of product 3}). Equality for one-loop parts is implied by (\ref{coh of product 2}), so let us consider the tree part. Since $H^\bt(M_2)=\RR h^2_\circ$ and $H^\bt(M_1\times M_2)=H^\bt(M_1)\otimes h^2_\circ$, the super-fields are
$$\omega=\sum_i h^1_i\otimes h^2_\circ\;\omega^{i\circ},\quad p=\sum_i p_{i\circ} h^{2\circ} \otimes h^{1 i}$$
Therefore
\begin{multline*}S^0_{H^\bt(M_1\times M_2,\g)}
=\sum_{T\in{\bf{T}}_\mr{nonPl}:\;|T|\geq 2}\frac{1}{|\Aut(T)|}\cdot\\ \cdot\sum_{i_0, i_1,\ldots,i_{|T|}} <p_{i_0\circ} h^{2\circ}\otimes h^{1i_0},r_1\otimes r_2\;\Iter_{T;-K^\xi[\bt,\bt];[\bt,\bt]}(\iota_1(h^1_{i_1})\omega^{i_1\circ},\ldots, \iota_1(h^1_{i_{|T|}})\omega^{i_{|T|}\circ})>
\end{multline*}
Here again the only contribution is due to the decoration of all internal edges with the part $K_1\otimes (\PP'_2+(1-\xi)\PP''_2)$ of chain homotopy $K^\xi$ (otherwise the second factor would contain the structure $K_2 1=0$). Therefore
\begin{multline*}
S^0_{H^\bt(M_1\times M_2,\g)}
=\sum_{T\in{\bf{T}}_\mr{nonPl}:\;|T|\geq 2}\frac{1}{|\Aut(T)|}\cdot\\ \cdot \sum_{i_0, i_1,\ldots,i_{|T|}} <p_{i_0\circ} h^{1i_0},r_1 \;\Iter_{T;-K_1[\bt,\bt];[\bt,\bt]}(\iota_1(h^1_{i_1})\omega^{i_1\circ},\ldots, \iota_1(h^1_{i_{|T|}})\omega^{i_{|T|}\circ})>\cdot\\
\cdot <h^{2\circ},r_2\;\Iter_{T;(\PP'_2+(1-\xi)\PP''_2)(\bt\wedge\bt);(\bt\wedge\bt)}(\underbrace{1,\ldots,1}_{|T|})>=
S^0_{H^\bt(M_1,\g)}|_{\omega^i\mapsto\omega^{i\circ},p_i\mapsto p_{i\circ}}\cdot 1
\end{multline*}
Thus (\ref{coh of product 3}) is proved.

Finally, let us prove (\ref{coh of product 4}). First consider the tree part of effective action:
\begin{multline*}
S^0_{H^\bt(M_1\times \s^1,\g)}=\sum_{T\in{\bf{T}}_\mr{nonPl}:\;|T|\geq 2}\frac{1}{|\Aut(T)|}\sum_{i_0,i_1,\ldots,i_{|T|}} \sum_{j_0,j_1,\ldots,j_{|T|}\in\{+,I\}} <p_{i_0 j_0} h^{2 j_0}\otimes h^{1 i_0} ,\\
,r_1\otimes r_2\; \Iter_{T;-K_R[\bt,\bt];[\bt,\bt]}(\iota_1(h^1_{i_1})\otimes \iota_2(h^2_{j_1})\;\omega^{i_1 j_1}, \ldots, \iota_1(h^1_{i_{|T|}})\otimes \iota_2(h^2_{j_{|T|}})\;\omega^{i_{|T|} j_{|T|}})>
\end{multline*}
Since Whitney forms on the circle $\mr{Span}(1,dt)$ are closed w.r.t. wedge product, the only contributions are due to decoration of all internal edges of trees with the part $K_1\otimes\PP'_I$ of chin homotopy $K_R$, and the projector $\PP'_I$ may be substituted with identity operator:
\begin{multline*}
S^0_{H^\bt(M_1\times \s^1,\g)}=\sum_{T\in{\bf{T}}_\mr{nonPl}:\;|T|\geq 2}\frac{1}{|\Aut(T)|}\sum_{i_0,i_1,\ldots,i_{|T|}} \sum_{j_0,j_1,\ldots,j_{|T|}\in\{+,I\}} <p_{i_0 j_0} h^{2 j_0}\otimes h^{1 i_0} ,\\
,r_1\otimes r_2\; \Iter_{T;-K_1\otimes\id [\bt,\bt];[\bt,\bt]}(\iota_1(h^1_{i_1})\otimes \iota_2(h^2_{j_1})\;\omega^{i_1 j_1}, \ldots, \iota_1(h^1_{i_{|T|}})\otimes \iota_2(h^2_{j_{|T|}})\;\omega^{i_{|T|} j_{|T|}})>\\
=\sum_{T\in{\bf{T}}_\mr{nonPl}:\;|T|\geq 2}\frac{1}{|\Aut(T)|}\sum_{i_0,i_1,\ldots,i_{|T|}}\left( <p_{i_0 +} h^{2 +}\otimes h^{1 i_0},\right.\\
,r_1\otimes r_2\; \Iter_{T;-K_1\otimes\id [\bt,\bt];[\bt,\bt]}(\iota_1(h^1_{i_1})\otimes \iota_2(h^2_{+})\;\omega^{i_1 +}, \ldots, \iota_1(h^1_{i_{|T|}})\otimes \iota_2(h^2_{+})\;\omega^{i_{|T|} +})>+\\
+\sum_{k=1}^{|T|}<p_{i_0 I} h^{2 I}\otimes h^{1 i_0},r_1\otimes r_2\;\Iter_{T;-K_1\otimes\id [\bt,\bt];[\bt,\bt]} (\iota_1(h^1_{i_1})\otimes \iota_2(h^2_{+})\;\omega^{i_1 +}, \ldots, \iota_1(h^1_{i_{k-1}})\otimes \iota_2(h^2_{+})\;\omega^{i_{k-1} +},\\
\left. ,\iota_1(h^1_{i_k})\otimes \iota_2(h^2_{I})\;\omega^{i_k I}, \iota_1(h^1_{i_{k+1}})\otimes \iota_2(h^2_{+})\;\omega^{i_{k+1} +},\ldots, \iota_1(h^1_{i_{|T|}})\otimes \iota_2(h^2_{+})\;\omega^{i_{|T|} +})>\right)\\
=\sum_{T\in{\bf{T}}_\mr{nonPl}:\;|T|\geq 2}\frac{1}{|\Aut(T)|}\cdot \\ \cdot\sum_{i_0,i_1,\ldots,i_{|T|}}\left( <p_{i_0 +} h^{1 i_0}
,r_1\; \Iter_{T;-K_1 [\bt,\bt];[\bt,\bt]}(\iota_1(h^1_{i_1})\;\omega^{i_1 +}, \ldots, \iota_1(h^1_{i_{|T|}})\;\omega^{i_{|T|} +})>+\right.\\
\left. +<p_{i_0 I} h^{1 i_0},  (-1)^{i_0}\int_0^1 r_1\;\Iter_{T;-K_1[\bt,\bt];[\bt,\bt]}(\iota_1(h^1_{i_1})\;(\omega^{i_1 +}+d\nu\, \omega^{i_1 I}), \ldots, \iota_1(h^1_{i_{|T|}})\;(\omega^{i_{|T|} +}+d\nu\,\omega^{i_{|T|} I}) )
\right)\\
=\int_0^1\sum_{T\in{\bf{T}}_\mr{nonPl}:\;|T|\geq 2}\frac{1}{|\Aut(T)|}\sum_{i_0,i_1,\ldots,i_{|T|}}
<(d\nu\, p_{i_0+}+(-1)^{|i_0|+1}p_{i_0 I})\;h^{1 i_0},\\
,r_1\;\Iter_{T;-K_1[\bt,\bt];[\bt,\bt]}\left(\iota_1(h^1_{i_1})\;(\omega^{i_1 +}+d\nu\, \omega^{i_1 I}), \ldots, \iota_1(h^1_{i_{|T|}})\;(\omega^{i_{|T|} +}+d\nu\, \omega^{i_{|T|} I}) \right)>\\
=\int_0^1 \left(S^0_{H^\bt(M_1,\g)}|_{\omega^i\mapsto \omega^{i+}+d\nu\,\omega^{iI},\; p_i\mapsto (-1)^{|i|+1} p_{iI}+ d\nu\, p_{i+}}\right)
\end{multline*}
Therefore the tree part of (\ref{coh of product 4}) is proved. Next consider the one-loop part:
\begin{multline*}
S^1_{H^\bt(M_1\times \s^1,\g)}=-\sum_{L\in\bf{L}_\mr{nonPl}}\frac{1}{|\Aut(L)|}\cdot\\
\cdot\sum_{i_1,\ldots,i_{|L|}}\sum_{j_1,\ldots,j_{|L|}\in\{+,I\}}
\Loop_{L;-K_R[\bt,\bt];\Omega^\bt(M_1\times \s^1,\g)}(\iota_1(h^1_{i_1})\otimes \iota_2(h^2_{j_1})\;\omega^{i_1 j_1},\ldots, \iota_1(h^1_{i_{|L|}})\otimes \iota_2(h^2_{j_{|L|}})\;\omega^{i_{|L|} j_{|L|}})
\end{multline*}
Since  $\mr{Span}(1,dt)$ is closed w.r.t. wedge product, the only those decorations of terms can contribute where all internal edges, not belonging to the cycle, are decorated with the part $K_1\otimes \PP'_I$ of chain homotopy $K_R$. On the other hand, due to Lemma \ref{lemma: str on s^1 with P',K}, all edges of the cycle have to be decorated with $\id\otimes K_I$. Next, (\ref{deg estimates 6}) implies that decoration of leaves of $L$ should be such that $|i_k|+|j_k|=1$ for all leaves $k$, i.e. only decorations $i+$ and $\circ I$ are allowed for leaves. The following argument shows that only the wheels $L=(*(*\cdots(*\bt)\cdots))$ contribute: otherwise, let there be some tree plugged into the cycle. Then exactly one leaf of this tree is decorated by $\circ I$ and all others are decorated by $i+$ (as implied by \ref{lemma: str on s^1 with P',K}). Therefore one of the following structures arises in the first factor:
\begin{eqnarray*}
K_1(\iota_1(h^1_i)\wedge \iota_1(h^1_\circ))=0\\
K_1(\iota_1(h^1_\circ)\wedge K_1(\bt))=0
\end{eqnarray*}
Therefore only wheels contribute to the one-loop action:
\begin{multline*}
S^1_{H^\bt(M_1\times \s^1,\g)}=-\sum_{n=2}^\infty (-1)^n \frac{1}{n}\Str_{\Omega^\bt(M_1\times \s^1,\g)}(K_I[dt\;\omega^{\circ I},\bt])^n\\
=-\sum_{n=2}^\infty (-1)^n \frac{1}{n}\Str_{\Omega^\bt(M_1)}1\cdot\Str_{\Omega^\bt(\s^1)}(K_I(dt\wedge\bt))^n\cdot\tr_\g(\ad_{\omega^{\circ I}})^n=
\chi(M_1)\cdot\sum_{n=2}^\infty \frac{B_n}{n\cdot n!}\;\tr_\g(\ad_{\omega^{\circ I}})^n\\
=\chi(M_1)\cdot\tr_g\log\left(\frac{\sinh\frac{\ad_{\omega^{\circ I}}}{2}}{\frac{\ad_{\omega^{\circ I}}}{2}}\right)
\end{multline*}
Thus (\ref{coh of product 4}) is proved.
\\$\Box$

\subsection{Examples}
\label{section: S on coh examples}
\subsubsection{Circle, torus, sphere}
\label{section: S on coh examples: circle, torus, sphere}
The point $pt$ is a trivial example of the induction on cohomology, since $H^\bt(pt,\g)\cong \Omega^\bt(pt,\g)\cong\g$. If we denote the basis in 0-cohomology by $h_\circ\in H^0(pt)$ and assume that it is identified with the unit in $\Omega^\bt(pt)$ (normalization condition), then
$$S_{H^\bt(pt,\g)}=<p_\circ,\frac{1}{2}[\omega^\circ,\omega^\circ]>_\g$$
(cf. the discussion of simplicial $BF$ action for 0-simplex in the beginning of section \ref{section: interval}).

First nontrivial example of induction on cohomology is the circle $\s^1$. We know the result for circle already, since in section \ref{section: gluing, bc examples} we obtained an explicit expression  (\ref{circle}) for the effective action for cell decomposition of the circle
$\Xi=\{[+],[01]\}$ with one 0-cell and one 1-cell. Since the differential on cochains $C^\bt(\Xi,\g)$ of this cell decomposition is identically zero, we can identify $H^\bt(\s^1,\g)\cong C^\bt(\Xi,\g)$, identifying the basis cohomologies $h_+,h_I$ with the basis cochains $e_+,e_{01}$. Therefore the effective action on cohomology of the circle is
\be S_{H^\bt(s^1,\g)}=<p_+,\frac{1}{2}[\omega^+,\omega^+]>_\g+<p_I,[\omega^I,\omega^+]>_\g+\hbar\; \tr_\g\log\left(\frac{\sinh\frac{\ad_{\omega^I}}{2}}{\frac{\ad_{\omega^I}}{2}}\right) \label{circle coh}\ee
The construction of section \ref{section: gluing, bc examples}, which lead us to this result, corresponds to the following choice of induction data
$\Omega^\bt(\s^1,\g)\xra{(\iota,r,K)} H^\bt(\s^1,\g)$ (the data glued from the induction data for the interval, from differential forms to cochains of standard triangulation):
\begin{eqnarray*}\iota:&& \alpha^+ h_+ + \alpha^I h_I\mapsto \alpha^+\cdot 1 + \alpha^I dt\\
r:&& f(t)+g(t)dt\mapsto f(0)+\left(\int_{\s^1}g(\tilde{t})d\tilde{t}\right)dt\\
K=K_I:&& f(t)+g(t)dt\mapsto \int_0^t g(\tilde{t})d\tilde{t}-t \int_{\s^1} g(\tilde{t})d\tilde{t}
\end{eqnarray*}

We can also demonstrate more explicitly, how the construction (\ref{ret 1},\ref{ret 2}) works in the case of the circle $M=\s^1$ and its cell decomposition (triangulation) $\Xi=\{[0]=[2],[1],[01],[12]\}$ with two 0-simplices $[0],[1]$ and two 1-simplices $[01],[12]$ (meaning that $[2]$ is the alternative label of 0-simplex $[0]$). The simplicial action for $\Xi$ is obtained from the known results for reduced actions for 0-simplex and 1-simplex, using  (\ref{simplicial locality decomp}):
\begin{multline}S_\Xi=<p_0,\frac{1}{2}[\omega^0,\omega^0]>_\g+<p_1,\frac{1}{2}[\omega^1,\omega^1]>_\g+\\
+<p_{01},\frac{1}{2}[\omega^{01},\omega^0+\omega^1]+
\left(\frac{\ad_{\omega^{01}}}{2}\coth\frac{\ad_{\omega^{01}}}{2}\right)\circ(\omega^1-\omega^0)>_\g+
\hbar\; \tr_\g\log\left(\frac{\sinh\frac{\ad_{\omega^{01}}}{2}}{\frac{\ad_{\omega^{01}}}{2}}\right)+\\
+<p_{12},\frac{1}{2}[\omega^{12},\omega^1+\omega^0]+
\left(\frac{\ad_{\omega^{12}}}{2}\coth\frac{\ad_{\omega^{12}}}{2}\right)\circ(\omega^0-\omega^1)>_\g
+\hbar\; \tr_\g \log\left(\frac{\sinh\frac{\ad_{\omega^{12}}}{2}}{\frac{\ad_{\omega^{12}}}{2}}\right)
\label{discretized circle action}
\end{multline}
The differential on $C^\bt(\Xi,\g)$ is
$$d:\quad \alpha^0 e_0+\alpha^1 e_1+\alpha^{01} e_{01}+\alpha^{12} e_{12}\mapsto (\alpha^1-\alpha^0) e_{01}+(\alpha^0-\alpha^1) e_{12}$$
The matrix of differential $d$ (in the basis of cells, with ordering $\{e_0,e_1,e_{01},e_{12}\}$), its transpose (analog of $d^*$ for the Hodge case) and the matrix of discrete Laplacian $d\, d^T+d^T d$ are
$$d=\left(\begin{array}{rrrr}0&0&0&0\\0&0&0&0\\-1&1&0&0\\1&-1&0&0\end{array}\right),\quad
d^T=\left(\begin{array}{rrrr}0&0&-1&1\\0&0&1&-1\\0&0&0&0\\0&0&0&0\end{array}\right),\quad
d\,d^T+d^T d=\left(\begin{array}{rrrr}2&-2&0&0\\-2&2&0&0\\0&0&2&-2\\0&0&-2&2\end{array}\right)$$
Therefore the Hodge decomposition for the cochains of triangulation $\Xi$ is
$$C^\bt(\Xi)=\underbrace{\mr{Span}(e_0+e_1,e_{01}+e_{12})}_{\mr{Harm}^\bt(\Xi)}\oplus \underbrace{\mr{Span}(e_{01}-e_{12})}_{C^\bt_{d-ex}(\Xi)} \oplus \underbrace{\mr{Span}(e_1-e_0)}_{C^\bt_{d^T-ex}(\Xi)}$$
Induction data from the cochains of triangulation to cohomology \\$C^\bt_{\Xi,\g}\xra{(\iota_{\Xi\ra H^\bt},r_{\Xi\ra H^\bt},K_{\Xi\ra H^\bt})} H^\bt(\s^1,\g)$ is defined uniquely (up to the normalization of basis $h_+,h_I$ in cohomology) by the Hodge decomposition:
\begin{eqnarray*}
\iota_{\Xi\ra H^\bt}:&& \alpha^+ h_+ + \alpha^I h_I\mapsto \alpha^+ (e_0+e_1)+\alpha^I \frac{1}{2}(e_{01}+e_{12})\\
r_{\Xi\ra H^\bt}:&& \alpha^0 e_0+\alpha^1 e_1+\alpha^{01} e_{01}+\alpha^{12} e_{12}\mapsto \frac{1}{2}(\alpha^0+\alpha^1) h_+ +
(\alpha^{01}+\alpha^{12}) h_I\\
K_{\Xi\ra H^\bt}:&& \alpha^0 e_0+\alpha^1 e_1+\alpha^{01} e_{01}+\alpha^{12} e_{12}\mapsto \frac{1}{4}(\alpha_{01}-\alpha_{12})(e_{1}-e_{0})
\end{eqnarray*}
Effective action on cohomology $H^\bt(\s^1,\g)$ may be computed using (\ref{S' via U and I}). It is easy to see that the $L_\infty$ morhism (\ref{L_infty morph pert series}) from cohomology to the cochains is linear (since harmonic cochains form a subalgebra in the $L_\infty$-algebra of cochains, generated by the tree part of (\ref{discretized circle action})):
\be U(\omega^+ h_+ +\omega^I h_I)=\iota_{\Xi\ra H^\bt}(\omega^+ h_+ +\omega^I h_I)=(e_0+e_1)\omega^+ + \frac{1}{2}(e_{01}+e_{12})\omega^I \label{circle U}\ee
The operator (\ref{I}) (which should be understood as a part of the ``covariant derivative'' operator $d+\mc{I}(\omega_\Xi)$ on cochains in the background of discrete super-connection $\omega_\Xi$) is calculated by differentiation of the tree part of action (\ref{discretized circle action}):
\begin{multline*}
\mathcal{I}(\omega_\Xi): \alpha^0 e_0+\alpha^1 e_1+\alpha^{01} e_{01}+\alpha^{12} e_{12}\mapsto e_0 [\omega^0,\alpha^0]+ e_1[\omega^1,\alpha^1]+ \\ +e_{01}\left(\frac{1}{2}[\omega^{01},\alpha^0+\alpha^1]+\frac{1}{2}[\alpha^{01},\omega^0+\omega^1]+
\left(\frac{\ad_{\omega^{01}}}{2}\coth\frac{\ad_{\omega^{01}}}{2}-1\right)\circ (\alpha^1-\alpha^0)+\right.\\
\left.+\sum_{n=2}^\infty \frac{B_n}{n!}\sum_{k=1}^n (\ad_{\omega^{01}})^{k-1}\ad_{\alpha^{01}}(\ad_{\omega^{01}})^{n-k}\circ (\omega^1-\omega^0)\right)+\\
+e_{12}\left(\frac{1}{2}[\omega^{12},\alpha^1+\alpha^0]+\frac{1}{2}[\alpha^{01},\omega^0+\omega^1]+
\left(\frac{\ad_{\omega^{12}}}{2}\coth\frac{\ad_{\omega^{12}}}{2}-1\right)\circ (\alpha^0-\alpha^1)+\right.\\
\left.+\sum_{n=2}^\infty \frac{B_n}{n!}\sum_{k=1}^n (\ad_{\omega^{12}})^{k-1}\ad_{\alpha^{12}}(\ad_{\omega^{12}})^{n-k}\circ (\omega^0-\omega^1)
\right)
\end{multline*}
In the case when super-connection $\omega_\Xi$ comes from the super-connection on cohomology $\omega_\Xi=U(\omega_{H^\bt})$ (this is the case needed for (\ref{S' via U and I})), the expression for $\mc{I}$ simplifies:
\begin{multline*}
\mc{I}(U(\omega^+ h_+ + \omega^I h_I)):\quad \alpha^0 e_0+\alpha^1 e_1+\alpha^{01} e_{01}+\alpha^{12} e_{12}\mapsto
e_0 [\omega^+,\alpha^0]+ e_1 [\omega^+,\alpha^1]+\\
+ e_{01}\left(\frac{1}{4}[\omega^I,\alpha^0+\alpha^1]-[\omega^+,\alpha^{01}]+
\left(\frac{\ad_{\omega^{I}}}{4}\coth\frac{\ad_{\omega^{I}}}{4}-1\right)\circ (\alpha^1-\alpha^0) \right)+\\
+e_{12}\left(\frac{1}{4}[\omega^I,\alpha^1+\alpha^0]-[\omega^+,\alpha^{12}]+
\left(\frac{\ad_{\omega^{I}}}{4}\coth\frac{\ad_{\omega^{I}}}{4}-1\right)\circ (\alpha^0-\alpha^1) \right)
\end{multline*}
Next,
\begin{multline*}
K_{\Xi\ra H^\bt}\mc{I}(U(\omega^+ h_+ + \omega^I h_I)):\quad \alpha^0 e_0+\alpha^1 e_1+\alpha^{01} e_{01}+\alpha^{12} e_{12}\mapsto
\\ \mapsto (e_1-e_0)\left(-\frac{1}{4}[\omega^+,\alpha^{01}-\alpha^{12}]+
\frac{1}{2}\left(\frac{\ad_{\omega^{I}}}{4}\coth\frac{\ad_{\omega^{I}}}{4}-1\right)\circ (\alpha^1-\alpha^0)\right)
\end{multline*}
Therefore, the logarithm of torsion  (\ref{torsion}) is
\begin{multline}\log\tau(\omega_{H^\bt})=\Str_{C^\bt(\Xi,\g)}\log(1+K_{\Xi\ra H^\bt}\mc{I}(U(\omega_{H^\bt})))\\=
\Str_{C^\bt_{d^T-ex}(\Xi,\g)}\log(1+K_{\Xi\ra H^\bt}\mc{I}(U(\omega_{H^\bt})))
=\tr_\g\log\left(\frac{\ad_{\omega^{I}}}{4}\coth\frac{\ad_{\omega^{I}}}{4}\right) \label{circle tau}
\end{multline}
Finally, substituting (\ref{discretized circle action},\ref{circle U},\ref{circle tau}) into (\ref{S' via U and I}), we obtain
\begin{multline*}
S_{H^\bt(\s^1,\g)}=S_{C^\bt(\Xi,\g)}|_{\omega_\Xi\mapsto U(\omega_{H^\bt}),p_\Xi\mapsto r_{\Xi\ra H^\bt}^*p_{H^\bt}}+\hbar\;\log\tau\\
=<p_+,\frac{1}{2}[\omega^+,\omega^+]>_\g+<p_I,[\omega^I,\omega^+]>_\g+
\hbar\; 2\,\tr_\g\log\left(\frac{\sinh\frac{\ad_{\omega^{I}}}{4}}{\frac{\ad_{\omega^{I}}}{4}}\right)+\hbar\;
\tr_\g\log\left(\frac{\ad_{\omega^{I}}}{4}\coth\frac{\ad_{\omega^{I}}}{4}\right)\\
=<p_+,\frac{1}{2}[\omega^+,\omega^+]>_\g+<p_I,[\omega^I,\omega^+]>_\g+
\hbar\;\tr_\g\log\left(\frac{\sinh\frac{\ad_{\omega^{I}}}{2}}{\frac{\ad_{\omega^{I}}}{2}}\right)
\end{multline*}
And again we come to the result (\ref{circle coh}). Notice that a miracle is that two different inductions give precisely the same results, not just equivalent modulo canonical transformation.

Another series of examples of exactly calculable effective action on cohomology is given by tori $\TT^D$ of dimensions $D\geq 2$ with asymmetric chain homotopy (consecutively contracting circles in $\TT^D$ in arbitrary order). This computation was done in section \ref{section: exact results for cell action}:  we computed the effective action (\ref{torus asym cell result}) for the cell decomposition $\Xi=\{+,I\}^{\times D}$ with zero differential on cochains, and there is an isomorphism $C^\bt(\Xi,\g)\cong H^\bt(\TT^D,\g)$. Therefore we can identify the cell action (\ref{torus asym cell result}) with the action on cohomology and identify the super-fields for cohomology with the super-fields for cell decomposition: $\omega_{H^\bt}=\omega_\Xi,\; p_{H^\bt}=p_\Xi$. Also, the result (\ref{torus sym cell result}) may be interpreted as the action on cohomology of $\TT^2$, obtained using the symmetric chain homotopy.

Next, effective action on cohomology for the sphere $\s^D$ of dimension $D\geq 2$ is calculated trivially, using (\ref{formal mfd: action on coh}): the sphere is a formal manifold with trivial 1-cohomology. Denoting the basis in cohomology of sphere by $h_\circ\in H^0(\s^D),\; h_\Phi\in H^D(\s^D)$, we obtain from (\ref{formal mfd: action on coh}) the result:
$$S_{H^\bt(\s^D,\g)}=<p_\circ,\frac{1}{2}[\omega^\circ,\omega^\circ]>_\g+<p_\Phi,[\omega^\Phi,\omega^\circ]>_\g$$
I.e. for the sphere the action on cohomology generates the usual graded Lie algebra structure on $H^\bt(\s^D,\g)$: all classical and quantum operations vanish, except for the operation $l_{(2)}$ (as for the point and as for the torus).

One can obtain a lot of examples of exactly calculable effective action on cohomology from Statement \ref{statement: coh of product}. For example, for direct product of the circle and the 2-sphere $M=\s^1\times \s^2$ the tree part is calculated via (\ref{formal mfd 2}) and the one loop part --- via (\ref{coh of product 2}). Using the basis $h_+,h_I$ in $H^\bt(\s^1)$ and basis $h_\circ,h_\Phi$ in $H^\bt(\s^2)$, we obtain
\begin{multline*}
S_{H^\bt(\s^1\times\s^2,\g)}=<p_{+\circ},\frac{1}{2}[\omega^{+\circ},\omega^{+\circ}]>_\g+ <p_{I\circ},[\omega^{I\circ},\omega^{+\circ}]>_\g+ <p_{+\Phi},[\omega^{+\Phi},\omega^{+\circ}]>_\g+\\+
<p_{I\Phi},[\omega^{I\Phi},\omega^{+\circ}]+[\omega^{I\circ},\omega^{+\Phi}]>_\g+
\hbar\;2\,\tr_g\log\left(\frac{\sinh\frac{\ad_{\omega^{I\circ}}}{2}}{\frac{\ad_{\omega^{I\circ}}}{2}}\right)
\end{multline*}
The factor 2 in the one-loop part comes from the Euler characteristic of the sphere $\s^2$.

\subsubsection{Klein bottle}
\label{section: S on coh examples: KB}
Effective action on cohomology of the Klein bottle may be calculated using construction (\ref{ret 1},\ref{ret 2}) and using the result (\ref{KB cell result}) for the cell action for Klein bottle with cell decomposition $\Xi=\{++,-I,I+,II\}$ obtained in section \ref{section: exact results for cell action}.

The differential on cell cochains $C^\bt(\Xi,\g)$ is
$$d:\quad \alpha^{++}e_{++}+\alpha^{-I}e_{-I}+\alpha^{I+}e_{I+}+\alpha^{II}e_{II}\mapsto 2\alpha^{-I} e_{II}$$
Matrix of the differential $d$, its transpose $d^T$ and matrix of the cell Laplacian $d\,d^T+d^T d$ written in the basis $\{e_{++},e_{-I},e_{I+},e_{II}\}$ are
$$d=\left(\begin{array}{rrrr}0&0&0&0\\0&0&0&0\\0&0&0&0\\0&2&0&0\end{array}\right),\quad
d^T=\left(\begin{array}{rrrr}0&0&0&0\\0&0&0&2\\0&0&0&0\\0&0&0&0\end{array}\right),\quad
d\,d^T+d^T d=\left(\begin{array}{rrrr}0&0&0&0\\0&4&0&0\\0&0&0&0\\0&0&0&4\end{array}\right)$$
Therefore the Hodge decomposition for the space of cell cochains of $\Xi$ is
$$C^\bt(\Xi)=\underbrace{\mr{Span}(e_{++},e_{I+})}_{\mr{Harm}^\bt(\Xi)}\oplus\underbrace{\mr{Span}(e_{II})}_{C^\bt_{d-ex}(\Xi)} \oplus\underbrace{\mr{Span}(e_{-I})}_{C^\bt_{d^*-ex}(\Xi)}$$
Denote $h_{++},h_{I+}$ the basis 0- and 1-cohomology of the Klein bottle, with the normalization condition that they are identified with $e_{++}$ and $e_{I+}$ by the embedding into cochains. Then the induction data are defined uniquely by the  Hodge decomposition:
\begin{eqnarray*}
\iota_{\Xi\ra H^\bt}:&& \alpha^{++}h_{++}+\alpha^{I+}h_{I+}\mapsto \alpha^{++}e_{++}+\alpha^{I+}e_{I+}\\
r_{\Xi\ra H^\bt}:&& \alpha^{++}e_{++}+\alpha^{-I}e_{-I}+\alpha^{I+}e_{I+}+\alpha^{II}e_{II}\mapsto \alpha^{++}h_{++}+\alpha^{I+}h_{I+}\\
K_{\Xi\ra H^\bt}:&& \alpha^{++}e_{++}+\alpha^{-I}e_{-I}+\alpha^{I+}e_{I+}+\alpha^{II}e_{II}\mapsto \frac{1}{2} \alpha^{II} e_{-I}
\end{eqnarray*}
Next, $\g e_{++}\oplus \g e_{I+}$ is a subalgebra of the $L_\infty$ algebra on cochains of $\Xi$, generated by the tree part of action (\ref{KB cell result}). Therefore the $L_\infty$ morphism from cohomology to cochains of $\Xi$ is linear:
$$U(\omega_{H^\bt})=\iota_{\Xi\ra H^\bt}(\omega_{H^\bt})=e_{++}\omega^{++}+e_{I+}\omega^{I+}$$
Operator (\ref{I}) is calculated by differentiating the tree part of (\ref{KB cell result}):
\begin{multline*}
\mc{I}(\omega_\Xi):\quad \alpha^{++}e_{++}+\alpha^{-I}e_{-I}+\alpha^{I+}e_{I+}+\alpha^{II}e_{II}\mapsto
e_{++}[\omega^{++},\alpha^{++}]+e_{-I}([\omega^{-I},\alpha^{++}]-[\omega^{++},\alpha^{-I}])+\\ +e_{I+}([\omega^{I+},\alpha^{++}]-[\omega^{++},\alpha^{I+}])+ e_{II}([\omega^{II},\alpha^{++}]+[\omega^{++},\alpha^{II}])+\\
+2 e_{II}\left(\left(\frac{\ad_{\omega^{I+}}}{2}\coth \frac{\ad_{\omega^{I+}}}{2}-1\right)\circ \alpha^{-I}
+\sum_{n=2}^\infty\frac{B_n}{n!}\sum_{k=1}^n (\ad_{\omega^{I+}})^{k-1}\ad_{\alpha^{I+}}(\ad_{\omega^{I+}})^{n-k}\circ\omega^{-I} \right)
\end{multline*}
Therefore
\begin{multline*}
K_{\Xi\ra H^\bt}\mc{I}(U(\omega_{H^\bt})):\quad \alpha^{++}e_{++}+\alpha^{-I}e_{-I}+\alpha^{I+}e_{I+}+\alpha^{II}e_{II}\mapsto\\
\mapsto e_{-I}\left(\frac{1}{2}[\omega^{++},\alpha^{II}]+\left(\frac{\ad_{\omega^{I+}}}{2}\coth \frac{\ad_{\omega^{I+}}}{2}-1\right)\circ \alpha^{-I}\right)
\end{multline*}
Hence the logarithm of torsion is
\begin{multline*}
\log\tau(\omega_{H^\bt})=\Str_{C^\bt_{d^T-ex}(\Xi,\g)}\log (1+K_{\Xi\ra H^\bt}\mc{I}(U(\omega_{H^\bt})))=
-\tr_\g\log\left(\frac{\ad_{\omega^{I+}}}{2}\coth \frac{\ad_{\omega^{I+}}}{2}\right)
\end{multline*}
(sign is due to the fact that the operator under super-trace is non-trivial on 1-forms). Thus the effective action on cohomology of Klein bottle is
\begin{multline}
S_{H^\bt(\mr{KB},\g)}=S_{C^\bt(\Xi,\g)}|_{\omega_\Xi\ra U(\omega_{H^\bt}),\,p_\Xi\ra r_{\Xi\ra H^\bt}^* p_{H^\bt}}+\hbar\; \log\tau\\
=<p_{++},\frac{1}{2}[\omega^{++},\omega^{++}]>_\g+<p_{I+},[\omega^{I+},\omega^{++}]>_\g-\hbar\;\tr_\g\log\left(\frac{\ad_{\omega^{I+}}}{2}\coth \frac{\ad_{\omega^{I+}}}{2}\right) \label{KB coh result}
\end{multline}

It is interesting to compare result (\ref{KB coh result}) with the result (\ref{circle coh}): cohomology for circle and Klein bottle are isomorphic $H^\bt(\s^1,\g)\cong H^\bt(\mr{KB},\g)$ and the tree effective actions on cohomology (\ref{circle coh},\ref{KB coh result}) coincide, if we identify $h_+=h_{++},h_I=h_{I+}$ (in other words, the induced $L_\infty$ structures on cohomology, or Massey operations, coincide). On the other hand the one-loop effective actions on cohomology (quantum Massey operations) are different. Thus we have an example of two manifolds with the same $L_\infty$ structure on cohomology, distinguished by quantum operations on cohomology. This means that the homotopy type of de Rham algebra of a manifold as a $qL_\infty$ algebra (in the sense of Definition \ref{def: homotopy of qL_infty algebras}), is a stronger invariant of manifolds than the homotopy type of de Rham algebra as a classical $L_\infty$ algebra.

One could ask, whether actions (\ref{KB coh result}) and (\ref{circle coh}) are equivalent, i.e. maybe there is a canonical transformation from one to another? The answer is negative due to the following argument: consider the densities of measures on respective $qL_\infty$ algebras
\begin{eqnarray*}\rho_{H^\bt(\s^1,\g)}&=&e^{S^1_{H^\bt(\s^1,\g)}}=\mr{det}_\g\left(\frac{\sinh\frac{\ad_{\omega^I}}{2}}{\frac{\ad_{\omega^I}}{2}}\right) , \\
\rho_{H^\bt(\mr{KB},\g)}&=&e^{S^1_{H^\bt(\mr{KB},\g)}}=\mr{det}_\g\left(\frac{\ad_{\omega^{I+}}}{2}\coth \frac{\ad_{\omega^{I+}}}{2}\right)^{-1}\\ &=&\rho_{H^\bt(\s^1,\g)}|_{\omega^I\mapsto\omega^{I+}}\cdot\left(\mr{det}_\g\cosh\frac{\ad_{\omega^{I+}}}{2}\right)^{-1}
\end{eqnarray*}
Notice that the density for circle is a regular expression in $\omega^I$, vanishing if $\ad_{\omega^{I}}$ has (at least one) eigenvalue $2\pi i k$ for some integer $k\neq 0$. On the other hand, the density for Klein bottle is singular if $\ad_{\omega^{I+}}$ has an eigenvalue $\pi i (2k+1)$ for some $k\in\ZZ$. Therefore the actions for circle and Klein bottle cannot be related by a canonical transformation with the generator regular in $\omega$.

The fact that $\rho_{H^\bt(\mr{KB},\g)}$ is singular for some values of $\omega^{I+}$ (far from zero) corresponds to the fact that the moduli space of flat connections on Klein bottle $\Hom(\pi_1(\mr{KB}),G)$ is singular. On the other hand, the moduli space of flat connections on circle $\Hom(\pi_1(\s^1),G)=\Hom(\ZZ,G)\cong G$ is a smooth manifold, which corresponds to the regular behaviour of density $\rho_{H^\bt(\s^1,\g)}$.

\subsubsection{$S_D\ltimes\ZZ_2^D$-bundles over circle with fiber $\TT^D$}
\label{section: M_gamma coh}
Next, we can generalize the result for Klein bottle (\ref{KB coh result}) to manifolds $M_\gamma$, introduced in section \ref{section: M_gamma}, on basis of the result (\ref{M_gamma cell action 1}) for the cell action for $M_\gamma$ with cell decomposition $\Xi=\{I\zeta,\bar{1}\zeta\}_{\zeta\in\{+,I\}^D}$, glued from standard cell decomposition $\{0,1,I\}\times \{+,I\}^D$ for cylinder $I\times \TT^D$.

Let us fix some element $\gamma\in S_D\ltimes\ZZ_2^D$. This element acts on cell cochains $C^\bt(\TT^D)$ of standard cell decomposition of torus in obvious way (by permuting circles and changing orientation of some of them) and generates the splitting:
$$C^\bt(\TT^D)=[C^\bt(\TT^D)]^\gamma\oplus \widehat{C^\bt(\TT^D)}$$
where the first term is the $\gamma$-invariant part, and the second term is the sum of all eigenspaces of $\gamma$-action on $C^\bt(\TT^D)$ corresponding to eigenvalues $\neq 1$. Next, we define a special Hodge decomposition for cell cochains of $M_\gamma$:
\be C^\bt(M_\gamma)=\underbrace{\Span(e_+,e_I)\otimes [C^\bt(\TT^D)]^\gamma}_{\mr{Harm}^\bt(\Xi)} \oplus \underbrace{\{e_I\otimes\tau\,|\,\tau\in \widehat{C^\bt(\TT^D)}\}}_{C^\bt_{d-ex}(\Xi)}\oplus \underbrace{\{e_1\otimes \tau+e_0\otimes \gamma\circ\tau \,|\,\tau\in \widehat{C^\bt(\TT^D)}\}}_{C^\bt_{K-ex}(\Xi)} \label{M_gamma Hodge decomp}\ee
(we use the identification of cell cochains of $M_\gamma$ with their image in cell cochains of the cylinder).

It is obvious from definition of $\mr{Harm}^\bt(\Xi)$ in (\ref{M_gamma Hodge decomp}) that cohomology of $M_\gamma$ is embedded into differential forms of the cylinder $I\times \TT^D$ as constant forms on interval times $\gamma$-invariant constant (Whitney) forms on the torus, which is indeed a subalgebra of $\Omega^\bt(I\times \TT^D)$. This means that manifold $M_\gamma$ is formal and the $L_\infty$ morphism from cohomology of $M_\gamma$ to cell cochains of $M_\gamma$ is linear: $U(\omega_{H^\bt})=\iota_{\Xi\ra H^\bt}(\omega_{H^\bt})$. Thus the action on cohomology has the form
\be S_{H^\bt(M_\gamma)}=S_{H^\bt(M_\gamma),(**)}(\omega_{H^\bt},p_{H^\bt})+\hbar\; \underbrace{\Str_{C^\bt_{K-ex}(\Xi,\g)}\log(1+K_{\Xi\ra H^\bt}\mc{I}(\iota_{\Xi\ra H^\bt}(\omega_{H^\bt})))}_{S^1_{H^\bt(M_\gamma)}} \label{M_gamma coh action}\ee
where the tree part is given by the contribution of tree $(**)$, generating the multiplication on cohomology, while the one-loop part (given by sum of wheels with multivalent cohomology insertions) is expressed as a super-trace.

Let $\{\tau_{(n),k}\}$ denote the eigenvectors of action of $\gamma$ on $C^n(\TT^D)$, with $\eta_{(n),k}\in \mathbb{C}$ the corresponding eigenvalues, so that $\tau_{(n),k}\in C^n(\TT^D,\mathbb{C})$ with $n=0,1,\ldots,D$ and $k=1,2,\ldots,\binom{D}{n}$, and $\gamma\circ \tau_{(n),k}=\eta_{(n),k}\tau_{(n),k}$. Let also denote $\{h_A\}$ the set eigenvectors corresponding to eigenvalue 1:
$\{h_A\}=\{\tau_{(n),k}\,|\, \eta_{(n),k}=1\}$. In other words, $\{h_A\}$ is a basis in $[C^\bt(\TT^D)]^\gamma$. Thus cohomology of $M_\gamma$ (which we identify with harmonic cell cochains of $\Xi$) is
$$H^\bt(M_\gamma)\cong \mr{Harm}^\bt(\Xi)=\Span(\{e_+\otimes h_A, e_I\otimes h_A\})$$
and the super-fields, associated to cohomology are
$$\omega_{H^\bt}=\sum_A e_{+}\otimes h_A \; \omega^{+A} + \sum_I e_{+}\otimes h_A \; \omega^{IA} ,\quad p_{H^\bt}=\sum_A p_{+A}\; h^A\otimes e^+ + \sum_A p_{IA}\; h^A\otimes e^I$$
We will also use notation $h_\circ:=e_{+\cdots +}$ for the basis 0-cochain on torus (which is obviously $\gamma$-invariant).

Let $m_{BC}^A:=<h^A,l_{(2),C^\bt(\TT^D)}(h_B,h_C)>$ be the structure constants of multiplication on $[C^\bt(\TT^D)]^\gamma$ (which is the same as wedge multiplication of $\gamma$-invariant Whitney forms on torus) in terms of basis $\{h_A\}$. Then the first term (tree action) in (\ref{M_gamma coh action}) is
\begin{multline}
S_{H^\bt(M_\gamma)}^0=S_{H^\bt(M_\gamma),(**)}=\\
=\frac{1}{2} \sum_{A,B,C} (-1)^{(|B|+1)\cdot |C|}m^A_{BC}<p_{+A},[\omega^{+B},\omega^{+C}]>_\g+ \sum_{A,B,C} (-1)^{|B|\cdot |C|}m^A_{BC}<p_{IA},[\omega^{IB},\omega^{+C}]>_\g
\label{M_gamma coh tree action}
\end{multline}
The second term (one-loop action) of (\ref{M_gamma coh action}) is straightforwardly computed in basis $\{(e_1+\eta_{(n),k}e_0)\otimes\tau_{(n),k}\}$ in $C^\bt_{K-ex}(\Xi)$:
first we use that one-loop action depends only on 1-cohomology to write
\begin{multline}
\mc{I}(\iota_{\Xi\ra H^\bt}(\omega_{H^1}))\circ\;((e_1+\eta_{(n),k}e_0)\otimes\tau_{(n),k}\cdot \alpha) = \\
=\sum_{m=1}^\infty \frac{1}{m!}l_{(m)}(e_I\otimes h_\circ\cdot \omega^{I\circ}+ \sum_{A_1}e_+\otimes h_{A_1}\cdot \omega^{+A_1},\ldots,e_I\otimes h_\circ\cdot \omega^{I\circ}+ \sum_{A_1}e_+\otimes h_{A_1}\cdot \omega^{+A_1},  (e_1+\eta_{(n),k}e_0)\otimes\tau_{(n),k}\cdot \alpha)= \\
= \frac{1+\eta_{(n),k}}{2}\;e_I\otimes \tau_{(n),k}\cdot [\omega^{I\circ},\alpha] +\sum_{m=2}^\infty \frac{B_m}{m!} (1-\eta_{(n),k})\; e_I\otimes \tau_{(n),k}\cdot (\ad_{\omega^{I\circ}})^m\circ \alpha + e_1\otimes (\cdots) + e_0\otimes (\cdots)
\label{M_gamma 1}
\end{multline}
for $\alpha\in\g$. Here index $A_1$ runs over the basis of $[C^1(\TT^D)]^\gamma$ and operations $l_{(m)}$ are $L_\infty$ operations on cell cochains of cylinder, generated by action (\ref{cylinder cell result}). Last two terms in (\ref{M_gamma 1}) are annihilated by $K_{\Xi\ra H^\bt}$ and therefore can be neglected. Since differential maps $(e_1+\eta_{(n),k}e_0)\otimes\tau_{(n),k}\mapsto (1-\eta_{(n),k})\cdot e_I\otimes \tau_{(n),k}$, we have
$$K_{\Xi\ra H^\bt}(e_I\otimes \tau_{(n),k})=\frac{1}{1-\eta_{(n),k}}\;(e_1+\eta_{(n),k}e_0)\otimes\tau_{(n),k}$$
Therefore
\begin{multline*}
\left(1+K_{\Xi\ra H^\bt}\mc{I}(\iota_{\Xi\ra H^\bt}(\omega_{H^1}))\right)\circ ((e_1+\eta_{(n),k}e_0)\otimes\tau_{(n),k}\cdot \alpha)=\\
=(e_1+\eta_{(n),k}e_0)\otimes \tau_{(n),k}\cdot \left(1+\frac{\eta_{(n),k}+1}{\eta_{(n),k}-1}\;\frac{\ad_{\omega^{I\circ}}}{2}+\left(\frac{\ad_{\omega^{I\circ}}}{2}\coth \frac{\ad_{\omega^{I\circ}}}{2}-1\right)\right)\circ \alpha=\\
=(e_1+\eta_{(n),k}e_0)\otimes \tau_{(n),k}\cdot \frac{\ad_{\omega^{I\circ}}}{2}\cdot\left(\coth\frac{\ad_{\omega^{I\circ}}}{2}+\frac{1+\eta_{(n),k}}{1-\eta_{(n),k}}\right) \circ \alpha
\end{multline*}
Thus we obtain explicit (up to values of $\eta_{(n),k}$, which can be easily worked out for any given $\gamma$) formula for one-loop action on cohomology of $M_\gamma$:
\be S^1_{H^\bt(M_\gamma)}=\sum_{n=1}^D (-1)^n \sum_{1\leq k\leq \binom{D}{n},\, \eta_{(n),k}\neq 1} \tr_\g \log\; \left(\frac{\ad_{\omega^{I\circ}}}{2}\cdot\left(\coth\frac{\ad_{\omega^{I\circ}}}{2}+\frac{1+\eta_{(n),k}}{1-\eta_{(n),k}}\right)\right) \label{M_gamma coh loop action}\ee

Together with (\ref{M_gamma coh tree action}) this gives complete result for effective action on cohomology of $M_\gamma$. The $qL_\infty$ structure on cohomology it generates has trivial classical part, consisting only of the binary product on cohomology (generated by (\ref{M_gamma coh tree action})) and no higher Massey operations, but the quantum part is non-trivial and quantum Massey operations $q_{(2)}, q_{(4)}, q_{(6)},\ldots$ are generated by (\ref{M_gamma coh loop action}). These quantum operations are only non-vanishing on the cohomology $e_I\otimes h_\circ$ (plugged into all inputs).

Before proceeding to examples, we would like to make two more remarks. First, if $\gamma,\gamma'\in S_D\ltimes \TT^D$ are conjugate elements, i.e. $\gamma'=\beta \gamma \beta^{-1}$ for some $\beta \in S_D\ltimes \TT^D$, then manifolds $M_\gamma$ and $M_{\gamma'}$ are diffeomorphic. The diffeomorphism is given by fiberwise action of $\beta$ (regarding these manifolds as bundles over circle).

Second, if $\dim H^1(M_\gamma) > 1$ then one-loop action (\ref{M_gamma coh loop action}) for $M_\gamma$ vanishes. The argument is as follows: $\dim H^1(M_\gamma) > 1$ is equivalent to $\dim [C^1(\TT^D)]^\gamma > 0$, so that one of eigenvalues $\eta_{(1),1},\ldots, \eta_{(1),D}$ is equal to 1; assume that it is $\eta_{(1),D}=1$. Next, the set $\{\eta_{(n),k}\}$ for given $n$ can be described as the set of $n$-fold products of eigenvalues in degree 1:
$$\{\eta_{(n),1},\ldots,\eta_{(n),\binom{D}{n}}\}=\{\eta_{(1),k_1}\cdots \eta_{(1),k_n}\}_{1\leq k_1<\cdots<k_n\leq D}$$
Let us introduce the subset $\{\tilde\eta_{(n),k}\}$ of the set of eigenvalues as given by products, not containing $\eta_{(1),D}$:
$$\{\tilde\eta_{(n),1},\ldots,\tilde\eta_{(n),\binom{D-1}{n}}\}=\{\eta_{(1),k_1}\cdots \eta_{(1),k_n}\}_{1\leq k_1<\cdots<k_n\leq D-1}$$
Then we obviously have
$$\{\eta_{(n),1},\ldots,\eta_{(n),\binom{D}{n}}\}=\{\tilde\eta_{(n),1},\ldots,\tilde\eta_{(n),\binom{D-1}{n}}\}\cup\{\tilde\eta_{(n-1),1},\ldots,\tilde\eta_{(n-1),\binom{D-1}{n-1}}\}$$
where we just split the $n$-fold products of $\eta_{(1),k}$ into those containing the factor $\eta_{(1),D}$ (second term) and products not containing it (first term). Thus we see that the whole set of eigenvalues $\{\eta_{(n),k}\}$ (for all $n$) splits into two subsets, consisting of the same eigenvalues, but in degrees differing by 1. This implies that contributions of these two subsets in (\ref{M_gamma coh loop action}) cancel each other, yielding zero.

\textbf{Examples of $M_\gamma$.} The trivial example is $\gamma=((1,\ldots,D);1,\ldots,1)$ --- the unit element of $S_D\ltimes \ZZ_2^D$. Then $M_\gamma=\TT^{D+1}$ is the torus, $[C^\bt(\TT^D)]^\gamma=C^\bt(\TT^D)$, $\widehat{C^\bt(\TT^D)}=0$ and all eigenvalues $\eta_{(n),k}$ are equal to 1. Thus the one-loop action (\ref{M_gamma coh loop action}) vanishes and the tree part (\ref{M_gamma coh tree action}) gives the result (\ref{torus asym cell result}).

Next example $\gamma=((1);-1)\in S_1\ltimes\ZZ_2$ corresponds to $M_\gamma$ the Klein bottle. Here the eigenvectors of $\gamma$-action on $C^\bt(\s^1)$ are
$$\{\tau_{(n),k}\}=\{\underbrace{e_+}_{n=0}; \underbrace{e_I}_{n=1}\}$$
and the corresponding eigenvalues are
$$\{\eta_{(n),k}\}=\{\underbrace{1}_{n=0};\underbrace{-1}_{n=1}\}$$
and we obtain the result (\ref{KB coh result}).

Next, take $\gamma=((1,2);-1,-1)\in S_2\ltimes \ZZ_2^2$. The corresponding $M_\gamma$ is geometrically a bundle over the circle with fiber $\TT^2$, with transition function given by rotation by angle $\pi$. Eigenvectors of $\gamma$-action on cochains of the fiber torus are
$$ \{\tau_{(n),k}\}= \{\underbrace{e_{++}}_{n=0}; \underbrace{e_{I+}, e_{+I}}_{n=1}; \underbrace{e_{II}}_{n=2}\} $$
with eigenvalues
$$ \{\eta_{(n),k}\}= \{\underbrace{1}_{n=0}; \underbrace{-1, -1}_{n=1}; \underbrace{1}_{n=2}\} $$
which gives
\begin{multline}
S_{H^\bt(M_\gamma)}
=<p_{+++},\frac{1}{2}[\omega^{+++},\omega^{+++}]>_\g+<p_{+II},[\omega^{+++},\omega^{+II}]>_\g+ \\
+<p_{I++},\frac{1}{2}[\omega^{I++},\omega^{+++}]>_\g+<p_{III},[\omega^{I++},\omega^{+II}]+[\omega^{III},\omega^{+++}]>_\g-\\
-\hbar\; 2\cdot\tr_g\log \left(\frac{\ad_{\omega^{I++}}}{2}\cdot\coth\frac{\ad_{\omega^{I++}}}{2}\right)
\label{M_gamma example 1}
\end{multline}

Now consider another 3-dimensional example: $\gamma=((2,1);1,-1)$, which corresponds to $M_\gamma$ the bundle over circle with fiber $\TT^2$ and transition function given by rotation by angle $\pi/2$. Here
$$ \{\tau_{(n),k}\}= \{\underbrace{e_{++}}_{n=0};\, \underbrace{e_{I+}-i e_{+I},\, e_{I+}+ i\e_{+I}}_{n=1};\, \underbrace{e_{II}}_{n=2}\} $$
and
$$ \{\eta_{(n),k}\}= \{\underbrace{1}_{n=0}; \underbrace{i, -i}_{n=1}; \underbrace{1}_{n=2}\} $$
and the action on cohomology is
\begin{multline}
S_{H^\bt(M_\gamma)}
=<p_{+++},\frac{1}{2}[\omega^{+++},\omega^{+++}]>_\g+<p_{+II},[\omega^{+++},\omega^{+II}]>_\g+ \\
+<p_{I++},\frac{1}{2}[\omega^{I++},\omega^{+++}]>_\g+<p_{III},[\omega^{I++},\omega^{+II}]+[\omega^{III},\omega^{+++}]>_\g-\\
-\hbar\; \tr_g\log \left(\left(\frac{\ad_{\omega^{I++}}}{2}\right)^2\cdot\left(\left(\coth\frac{\ad_{\omega^{I++}}}{2}\right)^2+1\right)\right)
\label{M_gamma example 2}
\end{multline}

Comparing (\ref{M_gamma example 1}) with (\ref{M_gamma example 2}) we see that $M_{(1,2);-1,-1}$ and $M_{(2,1);1,-1}$ are another pair of manifolds with coinciding classical $L_\infty$ structure on cohomology, but distinguished by quantum operations. In a sense, this example is better than circle vs. Klein bottle, since here we have two orientable manifolds of same dimension. One can, indeed, find a lot of other pairs like this among manifolds $M_\gamma$.


\end{document}